\documentclass{article}
\usepackage[utf8]{inputenc}

\usepackage[a4paper, total={7in, 9in}]{geometry}
\usepackage{graphicx,color}
\usepackage{amsmath,amssymb}
\usepackage{authblk}
\usepackage{enumitem}   
\usepackage{bm}
\usepackage{tikz}
\usepackage[authoryear,round]{natbib}
\usepackage{hyperref}
\usepackage[labelfont=bf,small]{caption}
\usepackage{subfig}
\usepackage{longtable}
\usepackage{pbox,makecell}

\graphicspath{{./Figures/}}

\makeatletter
\newcommand{\subsubsubsection}{\@startsection{paragraph}{4}{\z@}%
    {-2.5ex\@plus -1ex \@minus -.25ex}%
    {1.25ex \@plus .25ex}%
  {\reset@font\sffamily\normalsize}
}
\makeatother
\usepackage[OT2,T1]{fontenc}
\DeclareSymbolFont{cyrletters}{OT2}{wncyr}{m}{n}
\DeclareMathSymbol{\Sha}{\mathalpha}{cyrletters}{"58}
\DeclareMathSizes{12}{20}{4}{10}

\hypersetup{
    colorlinks = true,
    linkcolor  = blue,
    citecolor  = blue,
    urlcolor   = blue
  }
  
\providecommand{\eprint}[1]{\href{http://arxiv.org/abs/#1}{#1}}
\providecommand{\adsurl}[1]{\href{#1}{ADS}}
\usepackage{ifthen}\def\eprinttmp@#1arXiv:#2 [#3]#4@{\ifthenelse{\equal{#3}{x}}{\href{http://arxiv.org/abs/#1}{#1}}{\href{http://arxiv.org/abs/#2}{arXiv:#2} [#3]}}
\renewcommand{\eprint}[1]{\eprinttmp@#1arXiv: [x]@}

\newcommand{\be}{\begin{equation}}
\newcommand{\ee}{\end{equation}}
\newcommand{\bea}{\begin{eqnarray}}
\newcommand{\eea}{\end{eqnarray}}
\newcommand{\bdm}{\begin{displaymath}}
\newcommand{\edm}{\end{displaymath}}

\newcommand{\ie}{{\em i.e.}~}

\newcommand{\Mpc}{\, h^{-1} \, {\rm Mpc}}

\newcommand\mbar{\ensuremath{\overline{m}}}
\newcommand\mIbar{\ensuremath{\overline{m}_I}}
\newcommand\Mbar{\ensuremath{\overline{M}}}
\newcommand\MIbar{\ensuremath{\overline{M}_I}}
\newcommand\fbar{\ensuremath{\bar{f}}}
\newcommand\Lbar{\ensuremath{\bar{L}}}
\newcommand{\gsim}{\ \raise-2.truept\hbox{\rlap{\hbox{$\sim$}}\raise 5.truept\hbox{$>$}\ }}
\newcommand{\ksm}{km~s$^{-1}$~Mpc$^{-1}$}
\newcommand\arcsec{\mbox{$^{\prime\prime}$}}%
\newcommand\micron{\mbox{$\mu$m}}%

\newcommand\VI{\ensuremath{V{-}I}}

\DeclareRobustCommand{\ion}[2]{%
\relax\ifmmode
\ifx\testbx\f@series
{\mathbf{#1\,\mathsc{#2}}}\else
{\mathrm{#1\,\mathsc{#2}}}\fi
\else\textup{#1\,{\mdseries\textsc{#2}}}%
\fi}

\newcommand{\hi}{H\textsc{i}\ }
\newcommand{\hinospace}{\textrm{H\textsc{i}}}

\newcommand{\wzero}{$w_0$}
\newcommand{\wa}{$w_a$}
\newcommand{\omegam}{$\Omega_{\rm m}$}
\newcommand{\omegade}{$\Omega_{\rm de}$}
\newcommand{\omegal}{$\Omega_{\Lambda}$}
\newcommand{\epi}{$E_{\rm p,i}$}
\newcommand{\eiso}{$E_{\rm iso}$}
\newcommand{\epeiso}{$E_{\rm p,i}$--$E_{\rm iso}$}

\newcommand{\Hunit}{km s$^{-1} {\rm Mpc}^{-1}$}
\newcommand{\Ho}{$H_0$}

\usepackage{soul}

\begin{document}
\title{Unveiling the Universe with Emerging Cosmological Probes}

\author{Michele Moresco$^{1,2}$,
Lorenzo Amati$^{2}$,
Luca Amendola$^{3}$,
Simon Birrer$^{4,5}$,
John P.~Blakeslee$^{6}$,
Michele Cantiello$^{7}$,
Andrea Cimatti$^{1,8}$, 
Jeremy Darling$^{9}$,
Massimo Della Valle$^{10}$,
Maya Fishbach$^{11}$,
Claudio Grillo$^{12,13}$,
Nico Hamaus$^{14}$,
Daniel Holz$^{15,16}$,
Luca Izzo$^{17}$,
Raul Jimenez$^{18,19}$,
Elisabeta Lusso$^{20,8}$,
Massimo Meneghetti$^{2,21}$,
Ester Piedipalumbo$^{22,23}$,
Alice Pisani$^{24,25,26}$,
Alkistis Pourtsidou$^{27,28,29}$,
Lucia Pozzetti$^{2}$,
Miguel Quartin$^{30,31,3}$,
Guido Risaliti$^{20,8}$,
Piero Rosati$^{32,2}$,
Licia Verde$^{18,19}$}

\affil[]{\footnotesize michele.moresco@unibo.it\\ (Extended author information available at the end of the article)}

\date{}

\maketitle

\begin{abstract}
The detection of the accelerated expansion of the Universe has been one of the major breakthroughs in modern cosmology. Several cosmological probes (Cosmic Microwave Background, Supernovae Type Ia, Baryon Acoustic Oscillations) have been studied in depth to better understand the nature of the mechanism driving this acceleration, and they are being currently pushed to their limits, obtaining remarkable constraints that allowed us to shape the standard cosmological model. In parallel to that, however, the percent precision achieved has recently revealed apparent tensions between measurements obtained from different methods. These are either indicating some unaccounted systematic effects, or are pointing toward new physics.  Following the development of CMB, SNe, and BAO cosmology, it is critical to extend our selection of cosmological probes. Novel probes can be exploited to validate results, control or mitigate systematic effects, and, most importantly, to increase the accuracy and robustness of our results. 

This review is meant to provide a state-of-art benchmark of the latest advances in emerging ``beyond-standard'' cosmological probes. We present how several different methods can become a key resource for observational cosmology. In particular, we review cosmic chronometers, quasars, gamma-ray bursts, standard sirens, lensing time-delay with galaxies and clusters, cosmic voids, neutral hydrogen intensity mapping, surface brightness fluctuations, stellar ages of the oldest objects, secular redshift drift, and clustering of standard candles.
The review describes the method, systematics, and results of each probe in a homogeneous way, giving the reader a clear picture of the available innovative methods that have been introduced in recent years and how to apply them. The review also discusses the potential synergies and complementarities between the various probes, exploring how they will contribute to the future of modern cosmology.
\end{abstract}

\tableofcontents

\newpage
{\Large Credits}
\vspace{1cm}

{\bf Cosmic Chronometers (CC)}: Michele Moresco, Andrea Cimatti, Raul Jimenez, Licia Verde, Lucia Pozzetti

{\bf Quasars (QSO)}: Elisabeta Lusso, Guido Risaliti

{\bf Gamma-Ray Bursts (GRB)}: Lorenzo Amati, Massimo Della Valle, Ester Piedipalumbo, Luca Izzo

{\bf Standard Sirens (SS)}: Maya Fishbach, Daniel Holz

{\bf Time delay Cosmography (TDC)}: Simon Birrer

{\bf Cosmography with Cluster Strong Lensing (CCSL)}: Claudio Grillo, Piero Rosati, Massimo Meneghetti

{\bf Cosmic Voids (CV)}: Nico Hamaus, Alice Pisani

{\bf Neutral Hydrogen Intensity Mapping (NHIM)}: Alkistis Pourtsidou

{\bf Surface Brightness Fluctuations (SBF)}: Michele Cantiello, John Blakeslee

{\bf Stellar Ages (SA)}: Raul Jimenez

{\bf  Redshift Drift (RD)}: Jeremy Darling

{\bf  Clustering of Standard Candles (CSC)}: Luca Amendola, Miguel Quartin
    
\newpage


\section{Introduction}
\label{sec:intro}

The discovery of the accelerated expansion of the Universe \citep{perlmutter1998,perlmutter1999,riess1998} has been one of the major breakthrough in modern cosmology, and also in physics in general. The general framework established in the previous century, where the entire evolution of the Universe was thought to be dominated by matter and radiation, needed to readjust to make space for a new form of energy with negative pressure that can be responsible for this acceleration (that was named {\it dark energy}), or, alternatively, to account for some breaking of the well-known general relativity at very large scales. 
Driven by these pioneering results, in the subsequent decades the scientific and technical efforts of the scientific community were dedicated to the study of methods to measure and characterize this accelerated expansion, and to the development of large facilities providing massive datasets to be analyzed. In this process, a few of these methods, also referred to as cosmological probes, have become standard approaches in the cosmological analysis given the large efforts spent in measurements, theoretical analyses, systematics characterization, and also investments.

A comprehensive review on these methods is provided in \cite{huterer2018}. Here we just recall that most of these approaches are based on the determination of some standard properties of astrophysical objects that can be used to calibrate observations and measure the expansion history of the Universe. In particular, it was discovered that the peculiar physical characteristics of some objects allow us to infer a-priori their absolute luminosity, making them {\it standard candles} (or {\it standardizable candles}) with which it became possible to measure their luminosity distance. Locally, it was found that some stars have  a variable luminosity (Cepheids, RR-Lyrae) whose period of variability can be used to determine precisely their absolute luminosity; detached eclipsing binaries have been also used as local distance indicators to determine the distance to the Large Magellanic Cloud (LMC) to $<$1\% \citep{pietrzynski19}. At larger distances, it was discovered that also the stars at the Tip of the Red Giant Branch (TRGB), easily identifiable in the upper part of the the Hertzsprung–Russell diagram, can be used as standard candles, having an almost constant I-band magnitude \citep{lee1993}.
Finally, at cosmological distances, Type Ia Supernovae (SNe) have been found to be ideal standardizable candles, since their peak luminosity is found to strictly correlate with their absolute luminosity after a proper calibration \citep{phillips1993}, allowing us to probe the Universe with precise distance indicators up to $z\sim1.5$. Similarly, the analysis of large-scale structures in the Universe highlighted, among other features, the presence of correlated over-densities in the matter distribution at a specific separation of $r\sim100$ Mpc/h. This effect, known as Baryon Acoustic Oscillations (BAO), was clearly seen both as wiggles in the power spectrum of galaxies and as a peak in the two-point correlation function \citep{percival2001,cole2005,eisenstein2005}, and can be interpreted as the imprint of the sound horizon in the  original fluctuations in the photo-baryonic fluid present in the very early Universe. These oscillations have been in particular used as a {\it standard ruler} to study the expansion history of the Universe. 
While the BAO is the most direct probe of the expansion history from  large scale-stucture, the massive galaxies and quasars surveys that have enabled the BAO success,  enclose also  additional signals of great cosmological interest. These analyses are also well established, and, while they might not have reached their full potential yet, they do not qualify as ``emerging''.

As a parallel effort, the observation and study of the first light emitted in the Universe, the Cosmic Microwave Background (CMB) radiation, done with several ground- and space-based missions \citep{COBE,WMAP,Planck14,ACT,SPT} gave us a privileged view on the early Universe, providing fundamental insights on the process of formation and on the main components in that early times. In addition to those, other cosmological probes have been widely used in the past decades to constrain the expansion of the Universe and the evolution of the matter within it. Amongst the most important ones, here we just mention the weak gravitational lensing \citep[see, e.g.,][]{bartelmann2001}
and the properties of massive clusters of galaxies, in particular the cluster counts \citep[see, e.g.,][]{allen2011}.
Weak gravitational lensing, while being a younger field than CMB or galaxy surveys, has matured tremendously in the past two decades;  efforts in this direction have culminated recently with the DES analysis \citep{Abbott:2016ktf, DES2021} and weak lensing  is one of the science driver of future surveys such as the Legacy Survey of Space and Time (LSST) on the Vera Rubin Observatory\footnote{\url{https://www.lsstcorporation.org}}.

While CMB, BAO, SNe, and the other previously quoted probes have increasingly gained interest in these years in the cosmological community, it soon became clear also that a single probe is not sufficient to constrain accurately and precisely the properties of the components of the Universe. Ultimately, each probe has its own strengths and weaknesses, being sensitive to specific combinations of cosmological parameters, to specific physical processes, specific range of cosmic time, and affected by specific set of systematics. In the end, the only road to move forward in our knowledge of the Universe is found to reside in the combination between complementary cosmological probes, allowing us to break degeneracies between the estimate of parameters, and also to keep under control systematic effects \citep[see, e.g.,][]{scolnic2018}. This point was clearly first highlighted in the Dark Energy Task Force report \citep{albrecht2006}, and since then the effort of the scientific community proceeded towards that direction, also with space missions specifically designed to take advantage of the synergy between different probes\footnote{As an example, the ESA space mission Euclid \citep{Laureijs:2011gra} will study the expansion history of the Universe and the growth of the structures within, taking advantage from the combination of two cosmological probes, galaxy clustering and weak lensing.}. 

With the development of these cosmological probes, it soon begun the era of precision cosmology, where the advances in the instrumental technology,  supported by a more mature assessment and reduction of systematic uncertainties and by an increasing volume of data, led to percent and sub-percent measurements of cosmological parameters. However, instead of eventually closing all the questions related to the nature of the accelerated expansion of our Universe and of its constituents, this newly achieved accuracy actually opened even more the Pandora's box. One of the most pressing issues is that the Hubble constant \Ho~as determined from early-Universe probes (CMB) appears to be in significant disagreement with respect to the estimates provided by late Universe (Cepheids, TRGB, masers, ...). Many analyses addressed whether this might be due to some systematics hidden in either measurement, but, as of the current status, this seems disfavored \citep{Riess2011,Riess2016,Bernal2016,DiValentino2016,Efstathiou2020,Riess2020,DiValentino2021,Efstathiou2021,Riess:2021,Dainotti2021, Riess2021b}. At the same time, smaller  and less statistically significant differences started arising also in other cosmological parameters as estimated from early- and late-Universe probes, such as the tension in the estimate of dark matter energy density \omegam~and of $\sigma_8$, the matter power spectrum normalization at 8$\Mpc$, often summarized in the quantity $S_8\equiv\sigma_8\sqrt{\Omega_{\rm m}/0.3}$ \citep{Heymans2013,MacCrann2015,Joudaki2017,Hildebrandt2017,Asgari2020,Park2020,Joudaki2020,Troster2021,Asgari2021,Heymans2021,Amon2021,Secco2021,DES2021}. All these constraints are pointing toward significant differences of the order of 4-5 $\sigma$, and in the case (if confirmed) this is not attributable to some problems with the data, this may open the road to new physics with which to explain such discrepancies in the measurement of the same quantity probing different cosmic times.

Now that the precision in many standard cosmological probes is close to reaching its maximum, given the current analyses or the ones planned in the near future, a way to take a step forward in our understanding of the Universe is to look for new independent cosmological probes \citep[as also highlighted by][]{verde2019,DiValentino2021}, that could either confirm the discrepancies found, pointing us toward the need of new models, or deny those, helping us to understand better possible systematics, or unknown unknowns. Moreover, the synergy and complementarity between different probes can also help to reduce, when different probes are combined, the uncertainty on cosmological parameters. In general, the diversity between different methods will not only enrich the panorama of ways to look at and study our Universe, but also possibly open new observational and theoretical windows, as happened in the past with the study of CMB, SNe, and BAO.

This is an exciting time for cosmology, and in this review we aim to provide a state-of-art review of the new emerging cosmological probes, discussing how to apply them, the systematics involved, the measurements obtained, and the forecasts of how they could contribute to understand the evolution of the Universe. In particular, we will review cosmic chronometers, quasars, gamma-ray bursts, gravitational waves as standard sirens, time-delay cosmography, cluster strong lensing, cosmic voids, neutral hydrogen intensity mapping, surface brightness fluctuations, stellar ages, secular redshift drift, and clustering of standard candles. In Sect.~\ref{sec:introcosmo} we will provide a general overview of the basic notation and fundamental equations assumed in the review, in Sect.~\ref{sec:probes} we will discuss separately each emerging cosmological probe, in Sect.~\ref{sec:synergy} we will discuss the synergy and complementarity between the various described cosmological probes, and in Sect.~\ref{sec:conclusions} we will draw our conclusions. 


\section{Notations and fundamental equations}
\label{sec:introcosmo}

One of the main assumptions in modern cosmology is the {\it cosmological principle}, which describes our Universe at very large scales based on two main premises: the homogeneity (the Universe is the same in every positions) and isotropy (there is no preferential spatial direction). Under this principle, the space-time metric can be described by the 
Friedmann-Lema\^{i}tre-Robertson-Walker (FLRW) metric:
\begin{equation}
ds^2=-c^2 dt^2+a(t)^2\left(\frac{dr^2}{1-kr^2}+r^2d\theta^2+r^2\sin^2\theta d\phi^2\right) \;\; ,
\label{eq:flrw}
\end{equation}
where $a(t)$ is the scale factor, that describes how the universe is expanding relating physical and comoving distances as $R(t)=a(t) r$, $c$ is the speed of light, $\theta$ and $\phi$ are the angles describing the spherical coordinates, and $k$ is the parameter describing the curvature of space; in particular, a $k=0$ corresponds to a flat universe described by an Euclidean geometry, a positive $k>0$ to a closed universe with a spherical geometry, and a negative $k<0$ to an open universe with a hyperbolic geometry. Within a FRLW metric, it is also possible to relate the scale factor with the redshift $z$, having:
\begin{equation}
a(t)=\frac{1}{1+z} \;\; .
\label{eq:scalez}
\end{equation}

If we define the expansion rate of the universe $H(t)$ as the rate with which the scale factor evolves with time, $H(t)\equiv\left(\frac{\dot{a}}{a}\right)$, we can describe how it evolves with cosmic time $t$ through the Friedmann equations:
\begin{eqnarray}
\left(\frac{\dot{a}}{a}\right)^2&=&\frac{8\pi G\rho}{3}-\frac{k}{a^2}+\frac{\Lambda}{3} \label{eq:Fried1} \;\; ,\\
\frac{\ddot{a}}{a}&=&-\frac{4\pi G}{3}(\rho+3p)+\frac{\Lambda}{3} \;\; , \label{eq:Fried2}
\end{eqnarray}
where $G$ is the gravitational constant, $\rho$ and $p$ are the total energy density and pressure, $\Lambda$ is the cosmological constant, and the dot indicates a derivative with respect to time. Historically, a critical value of density producing a flat universe has been defined by equating, in the absence of a $\Lambda$ term, Eq.~\ref{eq:Fried1} to zero, obtaining $\rho_{\rm crit}=\frac{3H^2}{8\pi G}$. This quantity has proven to be extremely useful to define adimensional density parameters for the various constituents of the universe as $\Omega_i=\frac{\rho_i}{\rho_{\rm crit}}$.
This allows us to write the total energy density of the universe as the sum of the contribution of various components, namely matter and radiation; analogously, considering the terms on the right-hand side of Eq.~\ref{eq:Fried1}, we can define an energy density for the curvature $\Omega_k\equiv\frac{k}{H^2}$ and for dark energy (in the case of a Cosmological Constant) $\Omega_{\rm \Lambda}\equiv\frac{\Lambda}{3H^2}$. In this way, we have:
\begin{equation}
1=\sum_i\Omega_i(z)=\Omega_{\rm m}(z)+\Omega_{\rm r}(z)+\Omega_{k}(z)+\Omega_{\Lambda}(z) \;\; ,
\label{eq:om_dens}
\end{equation}
where the density parameters are here defined at any given time, so as a function of redshift $z$.
In this context, it is also useful to define the equation of state (EoS) parameter of a generic component as the value $w$ relating its pressure and density, $w=p/\rho$. In general, we can express the evolution of the energy density as:
\begin{equation}
\rho_{\rm i}(z)=\rho_{\rm i,0}\exp\left\{\int_0^z \frac{3[1+w_i(z')]}{1+z'}dz'\right\} \;\; .
\label{eq:rhoz}
\end{equation}
While the EoS could depend on time, we recall here that the different components have different EoS parameters, namely $w=1/3$ for radiation, $w=0$ for matter, and $w=-1$ for the term we referred as to dark energy (in the case it is a Cosmological Constant). If we consider Eq.~\ref{eq:rhoz} in the case of a constant $w_i$, it simplifies to:
\begin{equation}
\rho_{\rm i}(z)=\rho_{\rm i,0}(1+z)^{3(1+w_i)}\;\; .
\label{eq:rhoz_simpl}
\end{equation}
 
Combining Friedmann equations \ref{eq:Fried1} and \ref{eq:Fried2} with Eqs.~\ref{eq:om_dens} and \ref{eq:rhoz_simpl}, it is possible to express the expansion rate of the universe as a function of the evolution with redshift of its main components:
\begin{equation}
H(z) = H_0 \left[\Omega_{\rm r}(1+z)^4+\Omega_{\rm m}(1+z)^3 + \Omega_k(1+z)^2 + \Omega_{\rm de}(1+z)^{3(1+w)}\right]^{1/2} \;\; ,
\label{eq:Hz1}
\end{equation}
where each component evolves with a different power of $(1+z)$ due to the different EoS parameter of each term; here, we implicitly assumed the density parameters defined as constant, referred to as today's values $\Omega_{i,0}$. We will assume this convention throughout the review, unless otherwise specified. In Eq.~\ref{eq:Hz1} we also introduced the dark energy density as $\Omega_{\rm de}$, since in this case its EoS parameter is considered having a generic value $w$. 
While, in principle, one could take into account also the contribution of radiation $\Omega_{\rm r}$ that scales as $(1+z)^4$, typically this is not considered given the current constraint $\Omega_{\rm r}\sim2.47\;10^{-5}h^{-2}$ \citep{fixsen2009}, and in the following we will neglect its contribution.

So far, we have considered the dark energy as having a constant EoS parameter $w=-1$; however, to be more generic, we can allow it to vary with cosmic time, as different cosmological model would actually suggest. The most widely used way to parameterize this evolution is the Chevallier, Polanski, Linder (CPL) parameterization \citep{chevallier2001,linder2003}, where:
\begin{equation}
w(z)=w_0+w_a\left(\frac{z}{1+z}\right) \;\; .
\label{eq:de_CPL}
\end{equation}
Considering Eqs.~\ref{eq:rhoz} and \ref{eq:de_CPL}, we can therefore generalize Eq.~\ref{eq:Hz1} as follows:
\begin{equation}
H(z) = H_0 \left[\Omega_{\rm m}(1+z)^3 + \Omega_k(1+z)^2 + \Omega_{\rm de}(1+z)^{3(1+w_0+w_a)}e^{(-3w_a(z/(1+z))}\right]^{1/2} \;\; .
\end{equation}
From this more general formulation where most of the cosmological parameters are let free to vary (which we will refer as to open $w_{0}w_{a}$CDM model, o$w_{0}w_{a}$CDM), it is possible to derive more specific cases. In the case we fix the curvature of the universe to be flat ($\Omega_k=0$), we will have a flat $w_{0}w_{a}$CDM model (f$w_{0}w_{a}$CDM); in case we also fix the time evolution of the dark energy EoS to be null ($w_{a}=0$), we will have a flat $w$CDM model (f$w$CDM); finally, if we assume the dark energy EoS to be constant and equal to $w=-1$, we will obtain the standard $\Lambda$CDM model.
In this context, it is also useful to define the normalized Hubble parameter as:
\begin{equation}
E(z) = H(z)/H_0 \;\; .
\label{eq:Ez}
\end{equation}

The previously discussed equations describe how the cosmological background evolves. From these, we can introduce several additional quantities that will be extremely relevant in describing astrophysical phenomena, namely distances and times. Following \cite{huterer2018}, the comoving distance can be defined as:
\begin{equation}
D(z)=\frac{c}{H_{0}\sqrt{|\Omega_k|}}S{\left[\sqrt{|\Omega_k|}\int_0^z\frac{H_{0}dz'}{H(z')}\right]}\;\;\;\;\;\;\;\;{\rm where}\;\;\;S(x)= \begin{cases}
\sinh(x),\;\; & \Omega_k>0\\
x,\;\; & \Omega_k=0\\
\sin(x),\;\; & \Omega_k<0\\
\end{cases} \;\; .
\label{eq:dist}
\end{equation}
It is interesting to notice that in the case of a standard flat $\Lambda$CDM cosmology, this equation can be significantly simplified to:
\begin{equation}
D(z)=c\int_0^z\frac{dz'}{H(z')} \;\; .
\label{eq:dist_lcdm}
\end{equation}
From this equation, we can define two fundamental quantities in astrophysics, namely the luminosity distance $D_{\rm L}(z)$ and the angular diameter distance $D_{\rm A}(z)$ as:
\begin{equation}
D_{\rm L}(z)=(1+z)D(z)\;\;\;\;\;\;\;\; ; \;\;\;\;\;\;\;\;D_{\rm A}(z)=\frac{1}{(1+z)}D(z) \;\; ,
\label{eq:lumdist}
\end{equation}
where we have assumed that the Etherington relation holds, and therefore we rely on assumptions such as a metric theory and photon number conservation.  
Similarly, considering the previous definition of $H(t)$ and considering Eq.~\ref{eq:scalez}, we can write:
\begin{equation}
H(z)=\frac{\dot{a}}{a}=-\frac{1}{(1+z)}\frac{dz}{dt} \;\; ,
\label{eq:CC1}
\end{equation}
and by integrating it we obtain the expression of the age of the universe as a function of redshift:
\begin{equation}
t(z)=\int_0^z \frac{dz'}{H(z')(1+z)} \;\; .
\end{equation}

\clearpage

\section{Cosmology with emerging cosmological probes}
\label{sec:probes}

All the new emerging cosmological probes are presented following a common scheme, introducing at the beginning of each section the basic idea of the method and its main equations, describing how to optimally select each probe, discussing how it can be (and has been) applied, reviewing the current status of the art of the measurements, and providing forecasts on how the method is expected to improve its performance in the near future. A fundamental part is dedicated, in particular, to the presentation of the systematics involved in each probe, discussing how they impact the measurements and possible strategies to handle and mitigate them.

\subsection{Cosmic Chronometers}
\label{sec:CC}

The age of the Universe has been an important (derived) cosmological parameter, being closely related to  the Hubble constant and the background parameters governing Universe's expansion history.  Determinations of the age of the Universe today from the age of old cosmological objects at $z\sim0$ (see e.g. the reviews by \citealp{Catelan,Soderblom,Bolte+} and recent determinations by \citealp{OMalley,Valcin1,Valcin2}) and of the look-back time at higher redshifts \citep{Dunlop,Spinrad} have been very  influential in the establishment of the (now) standard cosmological model. 

The age of the Universe or the look-back time, being an integrated quantity of $H(z)$, has some limitations (both in terms of statistical power and in terms of susceptibility to systematics) that the {\it cosmic chronometers} approach attempts to overcome.

\subsubsection{Basic idea and equations}
\label{sec:CCidea}

The accurate determination of the expansion rate of the Universe, or Hubble parameter $H(z)$ has become in recent years one of the main drivers of modern cosmology, since it can provide fundamental information about the energy content and on the main physical mechanisms driving its current acceleration. Its measurement is, however, very challenging, and while many works have focused on the estimate of its local value at $z=0$ (the Hubble constant \Ho, see Sect.~\ref{sec:introcosmo}), we have nowadays few determinations of $H(z)$, and mainly based on few methods \citep[e.g., on the detection of the BAO signal in the clustering of galaxies and quasars, or on the analysis of SN data, see][]{font-ribera2014,delubac2015,Alam2017, riess2018b, scolnic2018,bautista2021,hou2021,raichoor2021,Riess:2021}. These measurements, while having their own strengths,  
rely on  the adoption  of a cosmological scenario such as assumption of flatness, on early physics assumptions (in the case of BAO) and on calibration of the cosmic distance ladder (in the case of SNe); without these assumptions, these probes yield the determination of the normalized expansion $E(z)$ instead of $H(z)$.

In this context, it is very important to explore alternative ways to determine the Hubble parameter, that can be compared, and eventually combined, with other determinations. The \textit{cosmic chronometers} method is a novel cosmological probe able to provide a direct and cosmology-independent estimate of the expansion rate of the Universe. The main idea, introduced by \citet{Jimenez_2002}, is based on the fact that in a universe described by a FLRW metric the scale factor $a(t)$ can be directly related with the redshift $z$ as in Eq.~\ref{eq:scalez}. With this minimal assumption, it is therefore possible to directly express the Hubble parameter as a function of the differential time evolution of the universe $dt$ in a given redshift interval $dz$, as provided by Eq.~\ref{eq:CC1}:
\begin{equation}
H(z)=-\frac{1}{(1+z)}\frac{dz}{dt}\,.\nonumber
\end{equation}
Here $dt/dz$ can be taken to be the look-back time differential change with redshift. Since redshift is a direct observable, the challenge is to find a reliable estimator for look-back time, or age, over a range of redshifts, i.e. to find cosmic chronometers (CC). 

The novelty and added value of this method with respect to other cosmological probes is that it can provide a direct estimate of the Hubble parameter without any cosmological assumption \citep[beyond that of an FLRW metric, see also][]{koksbang2021}. From this point of view, the strength  of this method is its (cosmological) model independence: no assumption is made about the functional form of the expansion history or about spatial geometry; it only assumes homogeneity and isotropy, and a metric theory of gravity. Constrains obtained with this method, therefore, can be used under extremely varied cosmological models.

There are three main ingredients at the basis of the CC method:
\begin{enumerate}
\item {\bf the definition of a sample of optimal CC tracers.} As highlighted in Eq.~\ref{eq:CC1},  a sample of objects able to trace, at each redshift, the differential age evolution of the Universe is needed. It is fundamental that this sample of {\it cosmic chronometers} is homogeneous as a function of cosmic time (i.e., the chronometers started ticking in a synchronized way independently of the redshift they are observed at), and optimized in order to minimize the contamination due to outliers. The optimal selection process will be described in detail in Sect.~\ref{sec:CCselection}.
\item {\bf the determination of the differential age $dt$.} The CC method is typically applied on tracers identified through spectroscopic analysis, where the redshift determination is extremely accurate \citep[$\delta z/(1+z)\lesssim0.001$, see e.g.][]{moresco2012}. As a consequence, as can be seen from Eq.~\ref{eq:CC1}, the only remaining unknown is the differential age  $dt$. Different techniques have been explored to obtain robust and reliable differential age estimates for CC, to estimate statistical and systematic uncertainties, and they will be presented in Sect.~\ref{sec:CCage}.
\item {\bf the assessment of the systematic effects.} As any other cosmological probe, one of the fundamental issues to be assessed is the sensitivity  of the method to effects that can systematically bias the measurement. All the various systematic effects will be examined in Sect.~\ref{sec:CCsys}.
\end{enumerate}

\subsubsection{Sample selection}
\label{sec:CCselection}

Cosmic chronometers are objects that should allow us to trace robustly and precisely the differential age evolution of the Universe across a wide range of cosmic times. For this reason,  the most useful astrophysical objects are galaxies: with current ground and  space-based facilities, these objects can be observed with reasonably high signal-to-noise over a wide area and range of redshifts.
Two different approaches have been explored.

Imagine to select, in a given redshift range, a complete sample of galaxies, independently of their properties, and estimate their age as to  homogeneously populate the $age(z)$ plane. With enough statistics, it becomes possible to estimate the {\it upper envelope} (also called red envelope) of the $age(z)$ distribution. Under the assumption that all galaxies formed at the same time independently of the observed  redshift (which relies on the Copernican principle)  and that the sample is complete, the envelope can be used  to measure the differential age of the Universe. The advantage of this kind of approach is that the selection of the sample is very straightforward, at the cost of being significantly demanding, since to determine robustly the ``edge'' of the distribution and its associated error, very high statistics are needed in order not to be biased by random fluctuations in the determination of the ages of the population \citep[e.g., see][where over 11000 massive and passive galaxies have been selected to apply this method]{jimenez2002,jimenez2003,simon2005,moresco2012}.

A more practical solution, therefore, is to (pre-)select an homogeneous population representing at each redshift the oldest objects in the Universe. The best cosmic chronometers that have been  identified are extremely massive ($\log(M/M_{\odot})>$10.5--11) and passively evolving galaxies (sometime also inappropriately referred as early-type galaxies). These objects represent the most extreme tails in the mass function (MF) and luminosity function (LF), from the local Universe \citep{baldry2004,baldry2006,baldry2008,peng10} up to high redshift \citep{pozzetti2010,ilbert2013,zucca2009,davidzon2017}. Many recent studies \citep[e.g.,][]{daddi2004,fontana2006,ilbert2006,wiklind2008,caputi2012,castro2012,muzzin2013,stefanon2013,nayyeri2014,straatman2014,wang2016,mawatari2016,deshmukh2018,merlin2018,merlin2019,girelli19} have identified a population of massive quiescent galaxies at high redshift ($z \gtrsim 2.5$). There is a large literature supporting the scenario in which these systems have built up their mass very rapidly \citep[$\Delta t<0.3$ Gyr,][]{thomas2010,mcdermid2015,citro2017,carnall2018} and at high redshifts \citep[$z>2-3$,][]{daddi2005,choi2014,mcdermid2015,pacifici2016,carnall2018,estrada2019,carnall2019}, having quickly exhausted their gas reservoir and being then evolving passively. For this reason, such objects constitute a very homogeneous population also in terms of metal content, having  been found to have a solar to slightly oversolar metallicity from  $z\sim0$ up to $z\sim2$ \citep{gallazzi2005,onodera2012,gallazzi2014, conroy2014,onodera2015,mcdermid2015,citro2016,comparat2017,saracco2019,morishita2019,estrada2019,kriek2019}. The mere existence of a population of passive and massive galaxies already at $z\sim2$ further supports  this scenario \citep{franx2003,cimatti2004,onodera2015,kriek2019,belli2019}. A clear pattern has also been found strictly connecting the mass, the star formation history (SFH), and the redshift of formation of these galaxies; within this scenario, referred to  to as {\it mass downsizing}, more massive galaxies are found to have been formed earlier, to have experienced a more intense, even if short, episode of star formation, and to have a very homogeneous SFH \citep{Heavens2004,cimatti2004,thomas2010}.
To summarize, these galaxies represent a population where the age difference $dt$ between two suitable separated (and suitably narrow) redshift bins is significantly larger than their internal time-scale evolution, making them optimal chronometers.
For a more detailed review on massive and passive galaxies, we refer to \cite{renzini2006}.

\begin{figure}[t!]
\centering
\mbox{\includegraphics[width=0.53\textwidth]{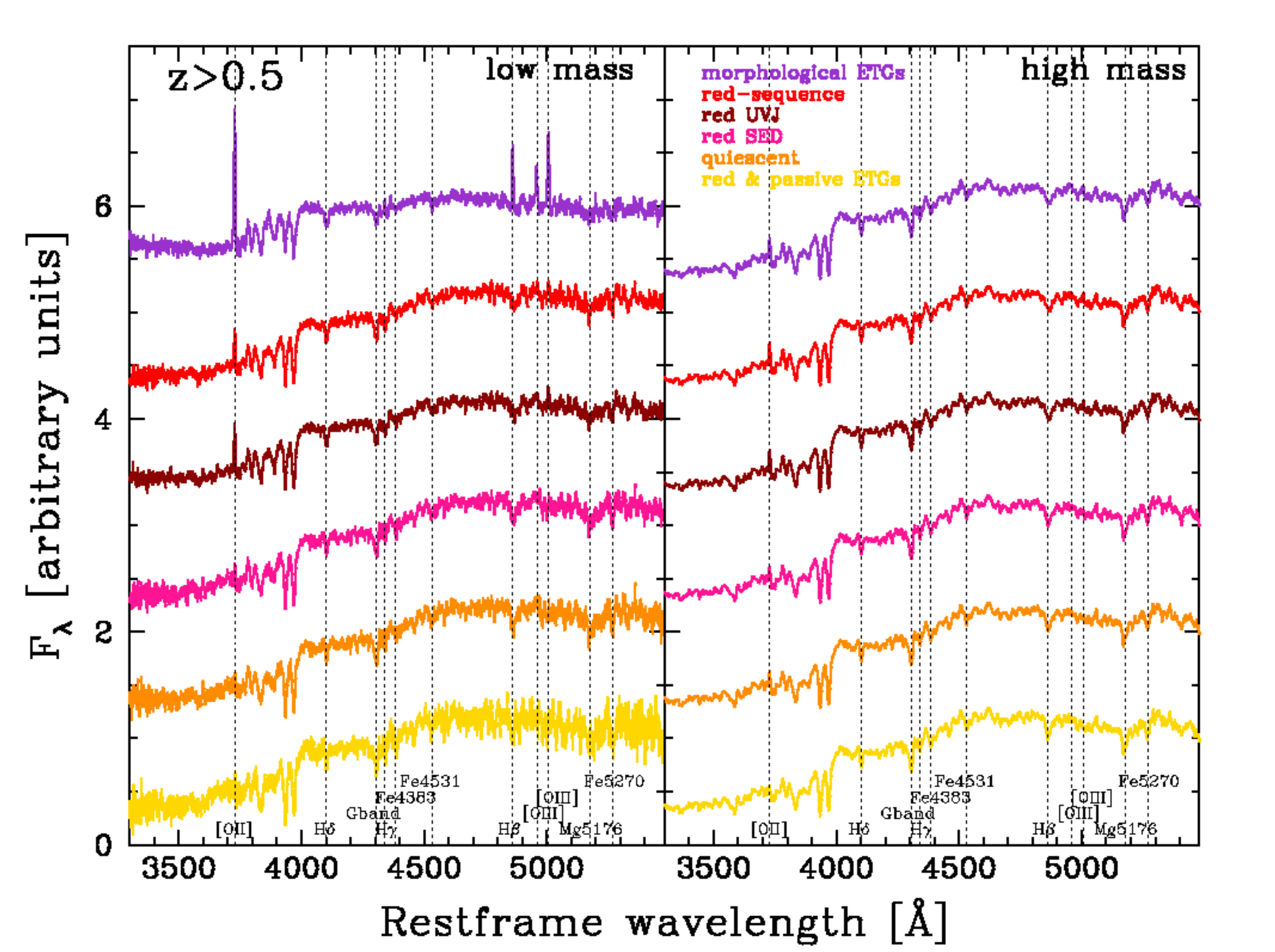}
\includegraphics[width=0.45\textwidth]{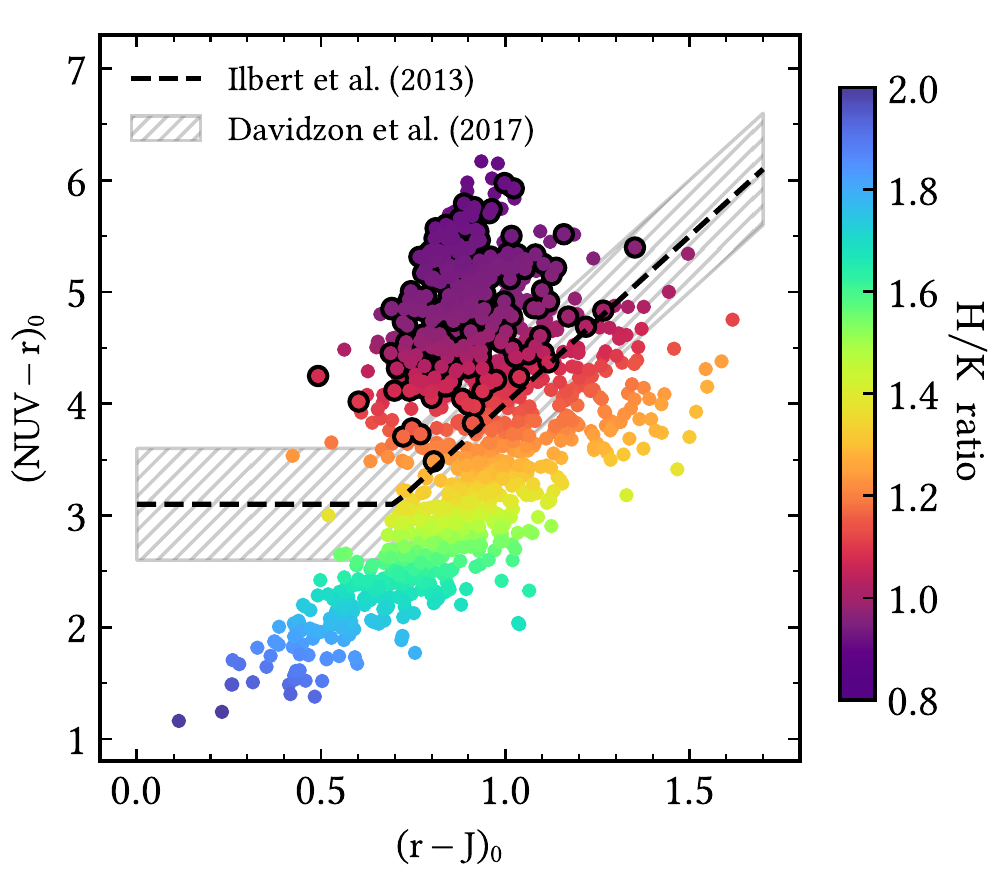}}
\caption{Impact of selection criteria on the purity of CC samples. Left panel: stacked spectra of differently selected samples of passive galaxies from the zCOSMOS survey in two different mass bins ($\log(M/M_{\odot})<10.25$ and  $\log(M/M_{\odot})>10.75$), showing how, in many selection criteria, the contamination by significant emission lines is still clearly evident, especially in the low mass bin. Note that in the high mass bin emission lines are not visible indicating much reduced contamination of the sample. Right panel: NUVrJ diagram for galaxies from the LEGA-C survey. The points have been colored by their H/K ratio, where the dashed line shows the division between passive and star-forming objects \citep{ilbert2013} \citep[the shaded region identifies the green valley][]{davidzon2017}, and the points highlighted in black the selected CC. Images reproduced with permission from \cite{Moresco2013} and \cite{borghi2021}, copyright by Astronomy \& Astrophysics and Astrophysical Journal.}
\label{fig:CC_sel}
\end{figure}

Many different  prescriptions  have been suggested in the literature to select passive galaxies, based on rest-frame colors \citep{williams2009,ilbert2010,ilbert2013,arnouts2013}, the shape of the spectral energy distribution (SED) \citep{zucca2009,ilbert2010}, star formation rate (SFR) or specific SFR (sSFR) \citep[see, e.g.,][]{ilbert2010,ilbert2013,pozzetti2010}, presence or absence of emission lines \citep[see, e.g.,][]{mignoli2009,wang2018}, and even morphology. The important question in this context is whether these different selection criteria are all equivalent to select CC. The short answer is no. In several papers \citep{franzetti2007,Moresco2013,belli2017,schreiber2018,fang2018,merlin2018,leja2019,diaz2019} it has been found that a simple criterion is not able per-se to select a pure sample of passively evolving galaxies, and that, depending on the criterion, a conspicuous number of contaminants might remain. This is clearly shown in the left panel of Fig.~\ref{fig:CC_sel}, reproduced from  \cite{Moresco2013}. The reference and the figure highlight how passive galaxies selected with  several different criteria still shows evidence of emission lines, with a residual contamination by blue/star-forming objects that, depending on the criterion, can be as high as 30-50\%. In the same work, as also reported by the figure, it was also shown that a cut in stellar mass is helpful to increase the purity of the sample, and that, at fixed criterion, the contamination is significantly smaller at high masses (decreasing by a factor 2-3 from $\log(M/M_{\odot})<10.25$ to $\log(M/M_{\odot})>10.75$).

Both in \cite{Moresco2013} and in \cite{borghi2021} it has been demonstrated that, in order to maximize the purity of the sample and to select the best possible sample of CC, different criteria should be combined (photometric, spectroscopic, stellar mass/velocity dispersion cut, potentially morphological). In \cite{Moresco2018}, a detailed selection workflow has been proposed, which can be summarized in the following three criteria: 
\begin{enumerate}[label=\roman*)]
    \item a {\it photometric criterion} to select the reddest objects, based on the available photometric data. Among the best ones there is the one based on the NUVrJ diagram \citep{ilbert2013}, but other alternatives are the UVJ diagram \citep{williams2009}, or the NUVrK \citep{arnouts2013}, or also selections based on full SED modeling \citep[e.g., see][]{ilbert2009,zucca2009}. It is important to underline, however, that having information about the UV flux is proven to be very important to discard the contamination by a young (0.1-1 Gyr) population, and that the NUVrJ diagram has been demonstrated to be the most robust one to distinguish star-forming and passive populations.
    \item a {\it spectroscopic criterion}, in order to check that no residual emission lines, that might trace the presence of on-going star formation, are present in the spectrum. Depending on the redshift and on the wavelength coverage of the data, the most important emission lines to be checked are [OII]$\lambda$3727, H$\beta$ ($\lambda=4861$\AA), [OIII]$\lambda$5007, and H$\alpha$ ($\lambda=6563$\AA), and different kind of cuts can be adopted, based on the equivalent width (EW) of the line \citep[e.g., EW$<$5\AA][]{mignoli2009, moresco2012, borghi2021}, on its signal-to-noise ratio \citep[S/N, e.g.,][]{moresco2016,wang2018}, or a combination of these. In general, it  is important  that the selected spectra do not show any sign of emission lines (as an example, see Fig.~\ref{fig:CC_stack}).
    \item a {\it cut in stellar mass}, or, equivalently, in stellar velocity dispersion $\sigma_{\star}$. As discussed above, the more massive a system is, the oldest, more coeval, and less contaminated it is. Therefore, typically a cut around $\log(M/M_{\odot})>$10.6--11 is adopted.
\end{enumerate}

\begin{figure}[t!]
\centering
\includegraphics[width=0.95\textwidth]{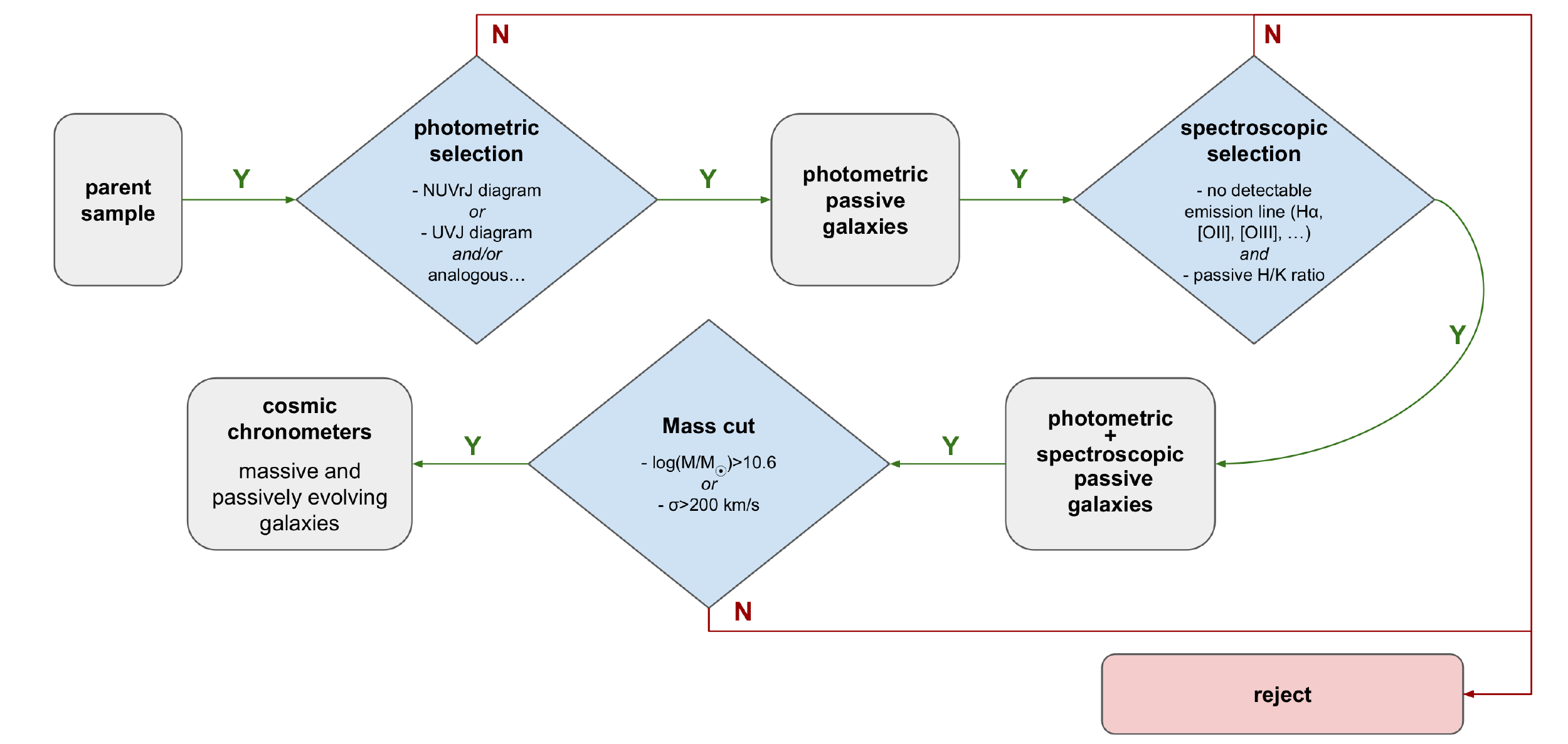}
\caption{Selection workflow for CC \citep[adapted from][]{Moresco2018}. The rounded boxes show the selected samples at the different steps, from the parent sample to the final CC sample, while the blue diamond boxes represent the incremental selection criteria adopted, with a green (red) arrow indicating when the criterion is met (or not) and the galaxy included (or excluded) from the sample. Each criterion is fundamental to maximize the purity of the sample, removing star-forming contaminants at different typical young ages \citep[see][]{Moresco2018}.}
\label{fig:CC_wf}
\end{figure}

Any other less stringent selection criteria will yield a sample with a residual degree of contamination by star-forming objects, which we will address in Sect.~\ref{sec:CCsys}.
It is interesting to notice that recently some alternative estimator has been suggested that can help to track the purity of the sample. In \cite{Moresco2018}, the ratio between the CaII H ($\lambda=3969$\AA) \& K ($\lambda3934$\AA) lines has been introduced as a novel way to trace the degree of contamination by a star-forming component. The reason is that, while for a passive population typically the ratio $H/K$ is larger than one (being the K line deeper than the H line), the presence of a young component affects this quantity, being characterized by non-negligible Balmer line absorptions, and in particular by the presence of H$\epsilon$ ($\lambda=3970$\AA) that get summed with the CaII H line, inverting the ratio. This new diagnostic has been demonstrated to be extremely powerful, since it correlates extremely well with almost all other indicators of ongoing star formation, as shown in the right panel of Fig.~\ref{fig:CC_sel} \citep[NUV and optical colors, SFR, emission lines, see][]{borghi2021}, and can be an useful independent indicator of the presence of a residual ongoing star formation.
The workflow for the selection criteria is summarized in Fig.~\ref{fig:CC_wf}.

\subsubsection{Measurements}
\label{sec:CCage}

Measuring the age of a stellar population presents several challenges.
One of the main issues is the existence of degeneracies between the physical parameters, so that the spectral energy distribution (SED) of a galaxy can be  approximately reproduced with quite different combinations of age and other parameters. The most well-known one is the {\it age-metallicity degeneracy} \citep{worthey1994,ferreras1999}, and it is connected to the fact that both an older age and an higher metallicity produce a reddening of galaxies spectra; in particular, it has been found from synthetic stellar population models that the optical colors of early-type galaxies obtained by changing their ages and metallicities while keeping the ratio $\Delta{\rm age}/\Delta [Z/H]\sim3/2$ are almost the same. The degeneracy between the age of a galaxy and its star formation history (SFH) \citep{gavazzi2002} or the dust content should also be mentioned \citep[even though we note that the second one is typically negligible at most for accurately selected passive galaxies, due to their low contamination by dust, see][]{pozzetti2000}. Therefore, while age estimates for galaxies obtained from multi-band SED-fitting are quite common in the literature,  they are not suitable for this purpose.

With the advent of high-resolution spectroscopy over a wide wavelength range and for large galaxy samples, and more accurate stellar model and fitting methods, it has become possible to lift these degeneracies and estimate the ages of stellar population of galaxies much more accurately and precisely.
Moreover the main strength of the CC method is that it is a differential approach, where the quantity to be measured is the differential age $dt$, and not the absolute age $t$. The advantage is that any systematic effect that might be introduced by any method in the estimate of $t$ is significantly minimized in the measurement of $dt$;  any systematic offset in the absolute age estimation will not impact the determination of $dt$. This is confirmed also by independent analysis \citep[e.g., see][]{marinfranch2009}, demonstrating  that the accuracy  reached in the determination of relative ages is much higher than the one on absolute ages.

Different methods have been proposed in the literature to obtain a robust estimate of $dt$ from galaxy spectra. These can be roughly  classified in two ``philosophies'': using the full spectral information versus selecting only specific features sensitive to the age and well localized in  wavelength. Using the full spectral information extracts the maximal amount of information possible (minimizes statistical errors) but is more sensitive to systematics, i.e., other physical process than age that leave their imprint on the spectrum, and exhibit some dependence of the age estimate on evolutionary stellar population synthesis models. Using localized features attempts to mitigate that, at the expense of possibly larger statistical errors. To keep systematic errors well below the statistical ones, the preferred methodology might change depending on the statistical power of the datasets available. With very large, high statistics datasets becoming available, the focus has shifted from full spectral fitting to using only specific features.\\
The main methods to measure $dt$ from galaxy spectra can be summarized as follows.

\subsubsubsection{Full-spectrum fitting} 

\begin{figure}[t!]
\centering
\includegraphics[width=0.95\textwidth]{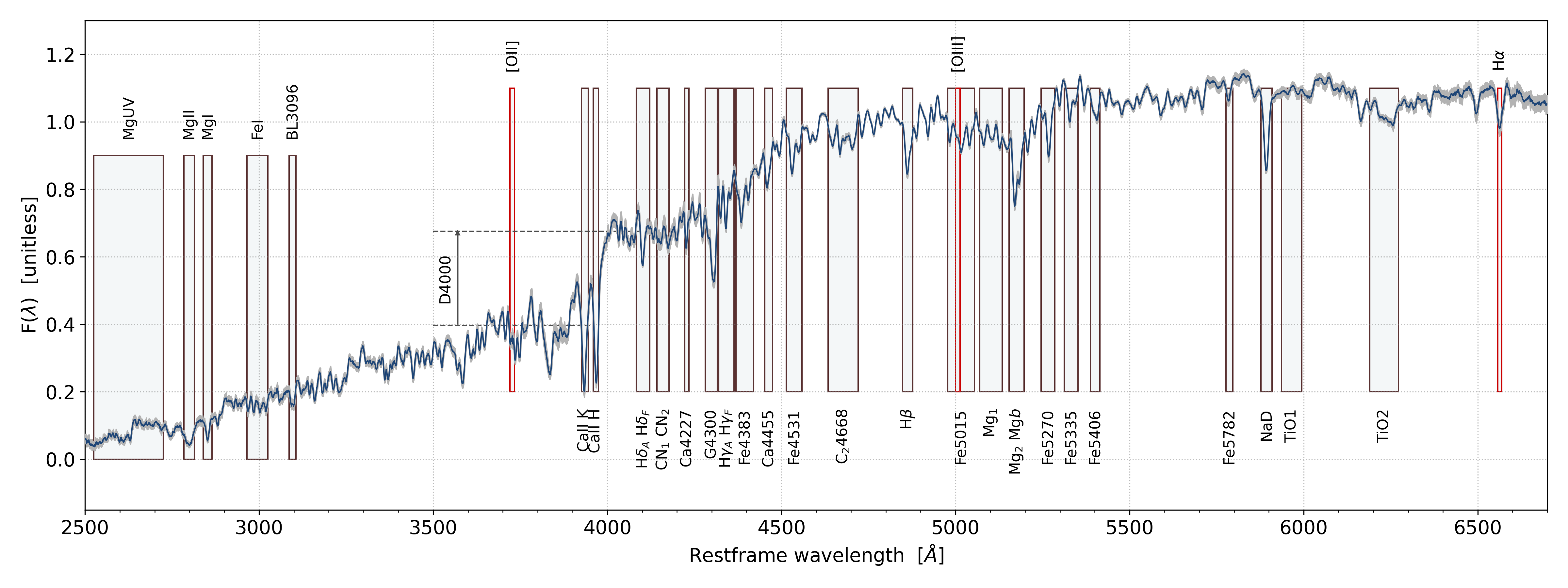}
\caption{Stacked spectrum of $\sim$100,000 massive and passive CC selected from SDSS DR12. It is possible to see clearly how it is characterized by a red continuum, several absorption lines (identified by the black boxes), and by the absence of significant emission lines (whose position is highlighted by the red boxes).}
\label{fig:CC_stack}
\end{figure}

The most straightforward approach is to take advantage of the full spectroscopic information available by fitting the entire spectrum with theoretical models. Different components, obtained from stellar population synthesis models, are typically combined with a mixture of different physical properties (age, metal content, mass), and properly weighted to reproduce the observed spectrum in a given wavelength window (usually within the optical range). The strength of this approach is therefore to be able to reconstruct, together with the age and metallicity of the population, also its star formation history, either in a parametric or non-parametric way.
Currently, several codes have been developed and are publicly available to perform a full spectrum fitting, differing slightly for the model implemented, how the SFH is reconstructed, and on the statistical methods. The first such method that started the field is the \texttt{MOPED} algorithm \citep{MOPEDI,MOPEDII}; after that, amongst the most used we can find \texttt{STARLIGHT} \citep{cidfernandes2005}, \texttt{VESPA} \citep{tojeiro2007}, \texttt{ULySS} \citep{koleva2009}, \texttt{BEAGLE} \citep{chevallard2016}, \texttt{FIREFLY} \citep{wilkinson2017}, \texttt{pPXF} \citep{cappellari2017}, and \texttt{BAGPIPES} \citep{carnall2018}. In Fig.~\ref{fig:CC_stack} we show as an example the typical spectrum of a passively evolving population obtained by stacking roughly 100000 spectra extracted from the Sloan Digital Sky Survey Data Relase 12 (SDSS-DR12). The figure also  highlights the locations of relevant spectral features. 

\subsubsubsection{Absorption features (Lick indices) analysis} 

Another approach is to analyze, instead of the full spectrum, only some specific regions characterized by well understood absorption features, also known as Lick indices. These indices, originally introduced by \cite{worthey1994,worthey1997} are characterized by a strength that can be directly linked to a variation of the property of the stellar population; some indices are more useful to trace to the age of the population (typically the Balmer lines), others the stellar metallicity (typically Fe lines), and others the alpha-enhancement (e.g. Mg lines). Also in this case, public codes exist to measure Lick indices (see, e.g., \texttt{indexf} \citealp{cardiel2010} and \texttt{pyLick} \citealp{borghi2021}). The specific dependence of each index (shown in Fig.~\ref{fig:CC_stack}) on physical properties has been at first assessed in \cite{worthey1994}. A significant step forward in their use to quantitatively determine the age of a stellar population has been done by \cite{thomas2011}; this consists in  constructing stellar population models specifically suited for modeling Lick indices, including a variable element abundance ratio, that can be compared with the data (e.g., with a Bayesian approach). This step is fundamental since it overcomes the limitation of the full-spectrum fitting, allowing also the possibility to  determine, together with the age and metallicity, also the alpha-enhancement of a stellar population.
It is worth noting that more recently other models with variable element ratios that could be used for this purpose have also been proposed by \cite{conroy2012} and \cite{vazdekis2015}.

\subsubsubsection{Calibration of specific spectroscopic features} 

Finally, one of the more commonly adopted approach in the CC works, is to focus on a single spectroscopic feature found to have a tight correlation with the age of the population. This approach was introduced by \cite{moresco2012}, who proposed to use the break in the spectrum at 4000~\AA~rest-frame ($D4000$, one of the main characteristic of the spectrum of a passive galaxy, as also shown in Fig.~\ref{fig:CC_stack}). The $D4000$ has been demonstrated to correlate extremely well with the stellar age (at fixed metallicity). Moreover it has been shown that the dependence of $D4000$ on the two quantities (age and metallicity $Z$) can be described by a simple (piece-wise) linear relation in the range of interest for the analysis:
\begin{equation}
D4000=A(Z, SFH)\times{\rm age}+B\; ,     
\label{eq:D4000age}
\end{equation}
where $B$ is a constant and  $A(Z, SFH)$ is a parameter, which for a broad age range depends only on the metallicity $Z$ and on the SFH, and can be calibrated on stellar population synthesis (SPS) models. By differentiating Eq.~\ref{eq:D4000age}, it is possible to derive the relation between the differential age evolution of the population $dt$ and the differential evolution of the feature, $dD4000$, in the form $dD4000=A(Z, SFH)\times{\rm dt}$. This allows us to rewrite Eq.~\ref{eq:CC1} as:
\begin{equation}
H(z)=-\frac{A(Z, SFH)}{1+z}\frac{dz}{dD4000}
\label{eq:CC2}
\end{equation}
with the advantage of having decoupled statistical (all included in the observationally measurable term $dz/dD4000$) from systematic effects (captured by the coefficient $A(Z, SFH)$). We note here that different definitions have been proposed in the literature to measure the $D4000$, which is the ratio between the average flux $F(\nu)$ in two windows adjacent to 4000~\AA \,rest-frame, one assuming wider bands ($D4000_w$, [3750-3950]~\AA~ and [4050-4250]~\AA, \citealt{bruzual1983}) and one with narrower ones ($D4000_n$, [3850-3950]~\AA~ and [4000-4100]~\AA, \citealt{balogh1999}); in the following, we will consider $D4000_n$, since it has been shown that it has been demonstrated to have a significant smaller dependence on potential reddening effects \citep{balogh1999}.

\begin{figure}[t!]
\centering
\includegraphics[width=0.95\textwidth]{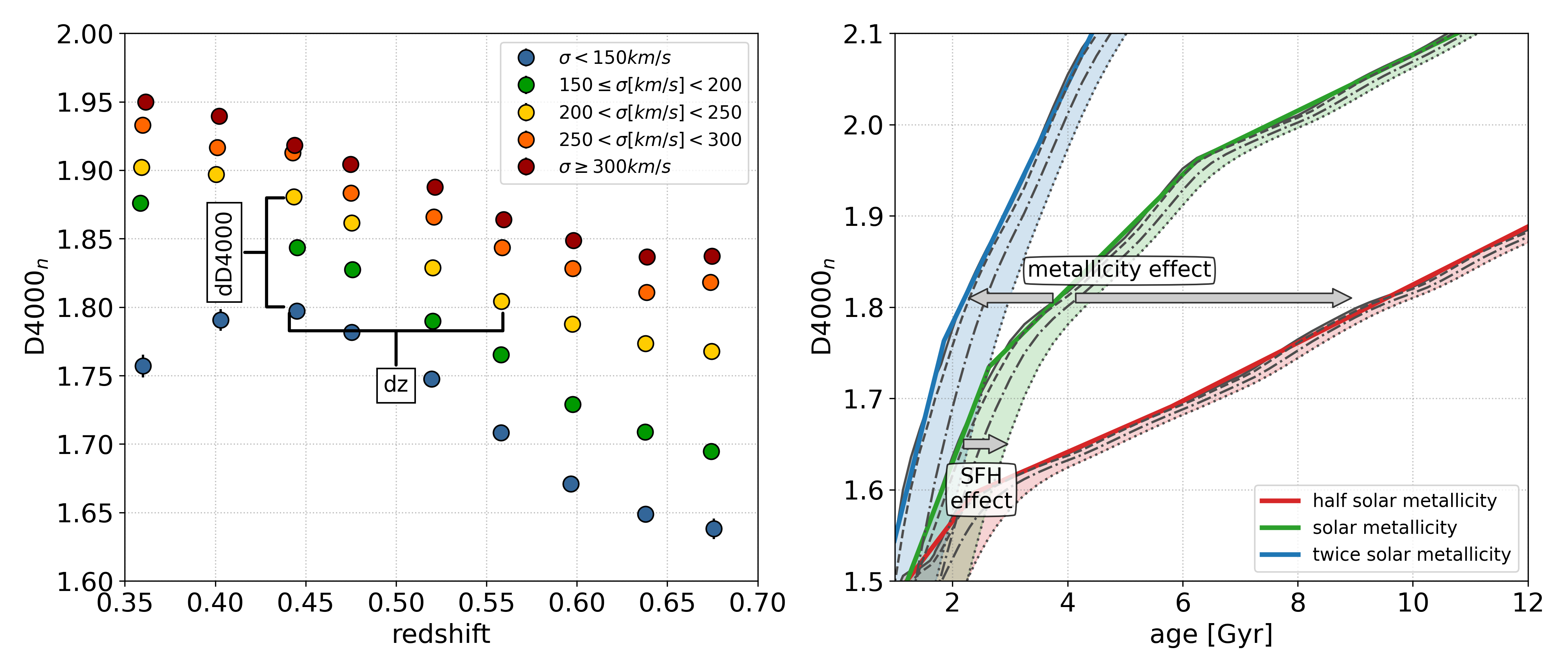}
\caption{Application of the CC method. In the left panel is shown an example of averaged $D4000-z$ (with uncertainties smaller than the  symbol size, thus differences can be robustly computed) relations for a CC sample extracted from SDSS-DR12, in different velocity dispersion bins as shown in the label. Each point has been estimated from a stacked spectrum of $\sim1000$ objects, and its uncertainty is the error on $D4000$ measured from the stacked spectrum.
It is clearly evident a downsizing pattern, for which more massive (with higher $\sigma$) galaxies have also larger $D4000$ values, corresponding to higher ages; it also shows the expected decrease of $D4000$ with redshift. The brackets show for an illustrative couple of points the calculation of $dD4000$ and $dz$. The right panel shows theoretical $D4000$-age relations obtained with SPS models by \cite{maraston2011} used to calibrate Eq.~\ref{eq:CC2}. Lines from the upper to the lower ones show different stellar metallicities, from twice as solar, to solar and half solar. Different lines present, at fixed metallicities, different SFH, namely with $\tau=[0.05,0.1,0.2,0.3]$ Gyr (from left to right). The colored lines show, for one SFH for each metallicity, the best fit obtained with a piece-wise linear relation.
The arrows indicate how different parameters affect the $D4000$-age relations, being important to keep in mind that the calibration parameter $A(Z,SFH)$ is the slope of the relation.}
\label{fig:CC_D4000age}
\end{figure}

\noindent
To apply the improved CC method as described by Eq.~\ref{eq:CC2}, it is therefore necessary to measure the following quantities:
\begin{enumerate}
    \item {\it the differential $\Delta D4000$ of a sample of CC over a redshift interval $\Delta z$}. Since this process involves the estimate of a derivative, to increase its accuracy and minimize the noise due to statistical fluctuation of the signal, it should be done both averaging the $D4000$ of galaxies in redshift slices and then estimating $dD4000$, or stacking multiple spectra of CC to increase the spectral S/N, and measuring the $D4000$ on the stacked spectra, as shown in Fig.~\ref{fig:CC_D4000age}. Equation~\ref{eq:CC2}  disentangles  observational errors from the systematic errors associated with the interpretation (such as dependence on the SSP model, degeneracies with metallicity, etc.). The $D4000$ is a purely observational quantity, and thus, barring observational systematics such as wavelength calibration or instrument response, its measurement is affected only by statistical uncertainty which can be reduced by increasing  the number of objects with spectra and/or increasing the $S/N$ per spectrum.
    \item {\it the metallicity $Z$ and SFH of the selected sample}. As a result of the strict selection criteria (see Sect.~\ref{sec:CCselection}), selected galaxies are characterized by a SFH with a very small duration:  $\tau<0.5$ Gyr (in many cases $<0.2$ Gyr) when parameterized with an exponentially declining SFH with $\tau$ the formation time scale (in Gyr). Nevertheless, SFH should be taken into account and correctly propagated in the measurement, since despite the selection, describing those system as single stellar population (SSP) would be over-simplistic. The method to estimate the SFH are mostly based on SED-fitting or on full-spectrum fitting, or on a combination of those \citep[see, e.g,][]{tojeiro2007,chevallard2016,citro2016,carnall2018,carnall2019}. Despite the fact that by construction the CC population is very homogeneous  also in metal content, and it has been observed to have a solar to slightly over-solar metallicity over a very wide range of cosmic times (see Sect.~\ref{sec:CCselection}), the stellar metallicity $Z$ need to be determined too. Also in this case, different approaches are viable, from considering a data-driven prior on it \citep{moresco2012}, estimating it with full-spectrum fitting considering different codes and models \citep{moresco2016}, or measuring it from Lick index analysis \citep{gallazzi2005,borghi2021}. 
    \item {\it the calibration parameter A(Z/SFH)} to connect variations in $D4000$ to variations in the age of the stellar population, assuming different SPS models. This involves generating several $D4000$-age relations exploring different metallicities and SFH, and adopting several different SPS models.

    As already discussed, these relation can be well approximated to be  linear (or, better,  piece-wise linear, as shown in Fig.~\ref{fig:CC_D4000age}), whose slopes are the parameter $A(Z,SFH)$ in the regime of interest. 
    At fixed metallicity and in a given $D4000$ regime, it is then possible to estimate the spread in the slopes obtained by varying the SFH within the observed ranges, and use this as associated uncertainty to the calibration parameter, i.e. $A(Z,SFH)=A(Z)\pm\sigma_A(SFH)$. These measurements, available from models at given metallicities (e.g. $Z/Z_{\odot}=0.5,1,2$ for the example in Fig.~\ref{fig:CC_D4000age}), can be afterwards interpolated, to obtain a value with its error for any given metallicity. The correct calibration parameter for each point will be therefore estimated from the measured (or assumed) metallicity, together with its error, for a global $A\pm\sigma_A$ that takes into account both the uncertainty on SFH and on metallicity. We will explore the impact of the SPS model choice on the systematic error budget in Sect.~\ref{sec:CCsys}.
\end{enumerate}
All these quantities will be combined in Eq.~\ref{eq:CC2} to obtain an estimate of the Hubble parameter $H(z)$ and of its uncertainty.

A final, yet important point to keep in mind is that, in order to be  cosmology-independent, the CC approach must rely on age estimates that do not assume any cosmological prior. This is a very important point, since in many (if not in most) analyses, a cosmologically-motivated upper prior on age is adopted in order to break or minimize the previously discussed degeneracies.
Of course, for the CC method  to be used as a test for cosmology, it is of paramount importance to obtain a robust age estimate without introducing any (prior) dependence on a cosmological model, in order to avoid circularity and, basically, retrieve the cosmological model used as a prior.

\subsubsection{Systematic effects}
\label{sec:CCsys}

In this section, we give an overview of the possible systematic effects that can affect the CC method, discussing approaches  to minimize them  and propagate them  to the total covariance matrix. We begin by discussing effects and assumptions that have a direct impact on the uncertainty on $H(z)$, and conclude presenting additional possible issues that might impact on the measurement, but that turn out to be negligible.

The main systematic effects can be divided into four components, and are summarized below. Each one of those will provide a contribution in the total systematic covariance matrix ${\rm Cov}_{ij}^{\rm sys}$.

\noindent
{\bf Error in the CC metallicity estimate ${\rm Cov}_{ij}^{\rm met}$.} The metallicity estimate enters in Eq.~\ref{eq:CC2} by changing the calibration parameter $A(Z,SFH)$. An error in its value, therefore, directly affects the $H(z)$ measurement and its associated error budget. In \cite{Moresco2020}, this issue has been addresses by performing a Monte Carlo simulation of SSP-generated galaxy spectra considering a variety of SPS models, with metallicities spanning different ranges ($\pm$10\%,5\%,1\%) and estimating the Hubble parameter.
In this way, it was estimated that the error induced on $H(z)$ scales almost linearly with the uncertainty on the stellar metallicity, which is corroborated observationally by the analysis in \cite{moresco2016}, where a 10\% error on the metallicity was found to correspond  to a 10\% error on the Hubble parameter. 
Hence, the uncertainty on  stellar metallicity (if known and quantified correctly) can be quantitatively propagated to an error on $H(z)$ following the procedure highlighted in Sect.~\ref{sec:CCage}. This contribution does not introduce off-diagonal terms in the covariance matrix because it depends on the stellar metallicity of each spectra (be of an individual object or a co-add) and  does not correlate different spectra.

\noindent 
{\bf Error in the CC SFH ${\rm Cov}_{ij}^{\rm SFH}$.} 
Even if CC have SFH characterized by very short timescales, assuming that the entire SFH is concentrated in a single burst (SSP) introduce a systematic error which must be accounted for as described in Sect.~\ref{sec:CCage}. This is typically a systematic contribution of the order of 2-3\%; as an example, in \cite{moresco2012}, where the estimated uncertainty on the SFH timescale was $0<\tau<0.3$ Gyr, the contribution to the final error on $H(z)$ was of $\sim$2.5\%. Also this contribution to the covariance matrix is taken to be purely diagonal.

\noindent 
{\bf Assumption of SPS model ${\rm Cov}_{ij}^{\rm model}$.} 
The major source of systematic uncertainty in the CC method, independently of the process adopted to estimate $dt$, is the assumption of the SPS model. This is also by definition a term that  introduces non-diagonal elements in the total covariance matrix, as the errors are highly correlated across different spectra. The estimation of its impact on the $H(z)$ error was assess in \cite{Moresco2020}. In this work, a wide combination of models was studied, including a variety of SPS models (BC03 and BC16 \citealp{bruzual2003}, M11 \citealp{maraston2011}, FSPS \citealp{conroy2009,conroy2010}, and E-MILES \citealp{vazdekis2016}), initial mass functions (IMF, including Salpeter \citealp{salpeter1955}, Kroupa \citealp{kroupa2001} and Chabrier \citealp{chabrier2003}), and stellar libraries (STELIB \citealp{leborgne2003} and MILES \citealp{sanchezblazquez2006}). These models have then been used with a MC approach by simulating a measurement assuming a model and measuring the Hubble parameter with all the other ones, estimating in this way the contribution to the total covariance matrix due to the assumption of a specific SPS model, IMF and stellar library. It was demonstrated that the error introduced on $H(z)$ is, on average, smaller than 0.4\% for the IMF contribution, and of the order of 4.5\% for the SPS model contribution.  

The component due to stellar library is slightly higher, however this estimate is overly-conservative as the effect is driven by the inclusion of a stellar library model that has now been superseded.
More importantly, it has been found that this uncertainty is also redshift dependent, and an explicit estimate for each component is provided as a function of $z$.

\noindent 
{\bf Rejuvenation effect ${\rm Cov}_{ij}^{\rm young}$.} 
Another possible bias to take into account is if the CCs selected present a residual contamination by a young component. We can divide this systematic effect into two cases. On the one side, we can have a part of the selected CC population composed by star-forming or intermediate systems; this event should be avoided, or maximally mitigated, by the accurate and combined selection process described in Sect.~\ref{sec:CCselection}. On the other side, despite the accurate selection we could have that the population of a single CC, even if dominated by an old component, still have a minor contribution by a young underlying component of stars. This effect can bias the $H(z)$ determination because it influences the overall shape of the spectrum due to the bluer color of younger stars, causing the measurement of younger ages and hence a biased $dt$. This issue has been studied in detail in \cite{Moresco2018}, where several indicators have been explored and proposed to trace the eventual presence of o residual young sub-population, from the UV flux \citep{kennicutt1998} to the presence of emission lines \citep[see, e.g.,][]{magris2003} or of strong absorption higher-order Balmer lines \citep[like H$\delta$][]{leborgne2006}. In particular, by studying theoretical SPS models, the previously discussed CaII H/K indicator was proposed to quantitatively trace the percentage level of contamination, taking advantage of the fact that the H$\epsilon$ line, characteristic of a young stellar component, directly affects the CaII H line, and therefore the ratio. It was then assessed, given a certain degree of contamination, how much the $D4000$ would be decreased, and, therefore, how much the estimate of $H(z)$ is impacted, giving in this was a direct recipe between the measured CaII H/K (or upper limit due to non-detection) and an additional error on the Hubble parameter. A contamination by a star-forming young component of 10\% (1\%) of the total light was found to propagate to an $H(z)$ error of 5\% (0.5\%); in particular, for the CC samples analyzed so far \citep{moresco2012,moresco2015,moresco2016,borghi2021}, it has been found this contamination to be below the detectable threshold, with an eventual additional error on $H(z)$<0.5\%. In case of a lack of detection and given the stringent upper limit on a possible residual contamination this contribution to the covariance is also taken to be diagonal.

Following \citet{Moresco2020}, the total covariance matrix for CC is defined as the combination of the statistical and systematic part as:
\begin{equation}
{\rm Cov}_{ij}= {\rm Cov}_{ij}^{\rm stat}+ {\rm Cov}_{ij}^{\rm syst} \;\; ,
\end{equation}
where ${\rm Cov}_{ij}^{\rm syst}$, for simplicity and transparency, is decomposed the several contributions discussed above:
\begin{equation}
{\rm Cov}_{ij}^{\rm syst}= {\rm Cov}_{ij}^{\rm met}+ {\rm Cov}_{ij}^{\rm young}+ {\rm Cov}_{ij}^{\rm model} \;\; ,
\end{equation}
where the latest component can be further decomposed in:
\begin{equation}
{\rm Cov}_{ij}^{\rm model}={\rm Cov}_{ij}^{\rm SFH}+{\rm Cov}_{ij}^{\rm IMF}+{\rm Cov}_{ij}^{\rm st. lib.}+{\rm Cov}_{ij}^{\rm SPS} \;\; .
\end{equation}

As discussed above, ${\rm Cov}_{ij}^{\rm met}$, ${\rm Cov}_{ij}^{\rm SFH}$ and ${\rm Cov}_{ij}^{\rm young}$ are purely diagonal terms, since they are related to the estimate of physical property of a galaxy (the stellar metallicity, and the eventual contamination by a younger subdominant population) uncorrelated for objects at different redshifts. ${\rm Cov}_{ij}^{\rm model}$, instead, has been conservatively estimated as the contribution from different redshifts are fully correlated.
In the published analyses  of currently available datasets, the contributions ${\rm Cov}_{ij}^{\rm met}$, ${\rm Cov}_{ij}^{\rm SFH}$ and ${\rm Cov}_{ij}^{\rm young}$ are already included in the errors provided  (and discussed later in Sect.~\ref{sec:CCres} and Tab.~\ref{tab:CC1}); the other terms have instead to be included following these recipes\footnote{To expand the analysis taking into account also the other systematic effects, a tutorial with dedicated jupyter notebooks is provided at \url{https://gitlab.com/mmoresco/CC_covariance}.}. 

Other effects, which have been demonstrated to o have a negligible impact on the measurement, but which should be mentioned are the following:
\begin{itemize}
\item {\bf progenitor bias.} 
A common observational effect that can introduce biases in the analysis of early-type galaxies is the  so called progenitor bias \citep{franx1996,vandokkum2000}: a given selection criterion might be effectively more stringent when applied at high redshift than at low redshift. In particular, high redshift objects that pass the sample selection might be older and more massive than those selected at low redshift, effectively representing the progenitor population of the low redshift sample. 
This bias becomes increasingly relevant when comparing objects spanning a wide range of redshifts, and, if not properly taken into account, could significantly affect the CC approach, since by definition it flattens the $age-z$ relations, changing its slope and hence producing a biased $H(z)$.  The differential approach at the basis of CC by definition acts to minimize this effect, since in all cases galaxies being compared span a very small range of redshifts. A quantitative estimate of its impact on the CC approach has been done in \cite{moresco2012} with two different methods. On the one side, the analysis has been repeated considering only the upper envelope of the $age-z$ distribution, that, by definition, could not be biased by the progenitor bias effect. The resulting $H(z)$ obtained is in perfect agreement with the baseline analysis, even if with larger error-bars due to the lower statistics on which the upper envelope approach is based on (see Sect.~\ref{sec:CCselection}). On the other side, the expected change in slope of the $age-z$ relation, assuming a very conservative change in formation times for the CCs considered, has also been estimated. In this conservative estimate, it was found that the error induced on the estimated $H(z)$ is $\sim$1\% on average, which is negligible considering the rest of the error-budget.
\item {\bf mass-dependence.} A final effect to be further explored is if the results have some mass-dependent bias. This effect has been explored thoroughly in many analyses \citep{moresco2012,moresco2016,borghi2021b}, and in all cases the $H(z)$ measured in different mass (or velocity dispersion) bins have been found to be mutually consistent, and with no systematic trends. This is in agreement with the expectation  since CC are selected to be already very massive galaxies ($\log(M/M_{\odot})\gtrsim11$), comprising very homogeneous systems, as discussed in Sect.~\ref{sec:CCselection}.
\end{itemize}

\subsubsection{Main results}
\label{sec:CCres}

The first  measurement with the CC method dates back to \cite{simon2005}, where they analyzed a sample of passively evolving galaxies from the luminous red galaxy (LRG) sample from SDSS early data release combined with higher redshift data from GDDS survey and archival data. The ages of these objects have been estimated with a full-spectrum fitting using SPEED models \citep{jimenez2004} estimating the age of the oldest components marginalizing over metallicity and SFH. Applying then the CC approach, 8 $H(z)$ measurements were obtained in the range $0<z<1.75$\footnote{Because of the high-quality of the spectra analysed in \citet{jimenez2004}, it was possible to compute {\em relative} ages with few percent accuracy in between the different redshift bins. Here the clue is once again {\em relative} ages and not the absolute ones plotted in Fig.~1 of \citep{jimenez2004}.}. 
\begin{figure}[t!]
\includegraphics[width=0.95\textwidth]{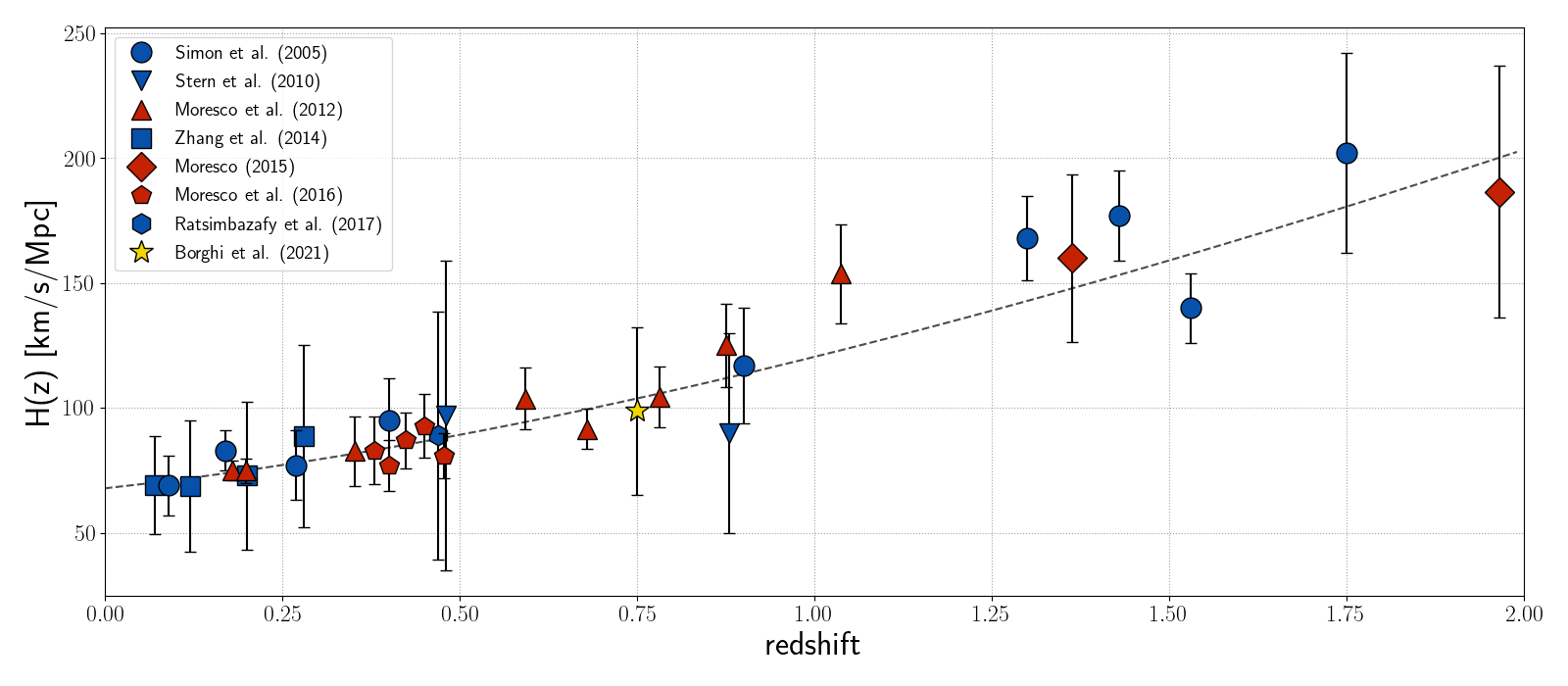}
\caption{Hubble parameter measurements obtained with the CC method. Different colors refer to different methods adopted to estimate $dt$, as presented in Tab.~\ref{tab:CC1}. The dashed line shows the flat $\Lambda$CDM cosmological model from \cite{planck2018} as a pure illustrative reference.}
\label{fig:HzCC}
\end{figure}
Similarly, also \cite{zhang2014} and \cite{ratsimbazafy2017} determined new values of the Hubble parameter measuring $dt$ with a full-spectrum fitting technique. They studied a sample of $\sim$17,000 LRGs from SDSS Data Release Seven (DR7) and of $\sim$13,000 LRGs from 2dF–SDSS LRG and QSO catalog, respectively, both extracting differential age information for their sample using the \texttt{UlySS} code and BC03 models, obtaining four  additional estimates of $H(z)$ at $z<0.3$  and  one at $z\sim0.47$, respectively.\\
The results by \cite{moresco2012}, \cite{moresco2015}, and \cite{moresco2016} are instead based on the analysis of the $D4000$ feature described in Sect.~\ref{sec:CCage}. The first paper examined a compilation of very massive and passively evolving galaxies extracted from SDSS Data Release 6 Main Galaxy Sample and Data Release 7 LRG sample and from a combination of spectroscopic surveys at higher redshifts (zCOSMOS, K20, UDS), comprising in total $\sim$11,000 galaxies in the range $0.15<z<1.3$. The second paper analyzed a significantly smaller sample (29 objects) of massive and passive galaxies available in the literature at very high redshifts $z>1.4$. Finally, in the last paper considers the SDSS BOSS Data Release 9, selecting a sample of more than 130000 CC in the range $0.3<z<0.55$. In total, 15 additional $H(z)$ estimates are presented in the range $0.18<z<2$.\\
Most recently, in \cite{borghi2021} a new approach was explored, using a Lick-indices-based analysis applied on CC extracted from the LEGA-C survey to derive information of the physical properties (age, metallicity and $\alpha$-enhancement) of the population, and in \cite{borghi2021b} the resulting $dt$ measurements were used to obtain a new estimate of the Hubble parameter.

The current, most updated compilation of $H(z)$ measurements obtained with CC is shown in Fig.~\ref{fig:HzCC}, and provided in Tab.~\ref{tab:CC1}. All these measurements have been obtained assuming a SPS model \citep[BC03,][]{bruzual2003}, except from the measurements from \cite{moresco2012}, \cite{moresco2015}, and \cite{moresco2016}, that are available also with a different set of SPS models \citep[M11,][]{maraston2011}.
Since, as discussed above, one of the main source of systematic uncertainties is the SPS model assumed, for a coherent analysis the systematic off-diagonal component to the covariance has to be added following the recommendations of Sect.~\ref{sec:CCsys}, and with the recipes presented in \cite{Moresco2020}.

These data have been widely used in the literature in a variety of applications, which we proceed to present below. 
\begin{table}[t!]
\begin{center}
\begin{tabular}{|llllr|}
\multicolumn{5}{c}{{}}\\
\hline \hline
$z$ & $H(z)$ & $\sigma_{H(z)}$ & M & reference\\
\hline
0.07 & 69.0 & 19.6 & F & \cite{zhang2014}\\
0.09 & 69 & 12 & F & \cite{simon2005}\\
0.12 & 68.6 & 26.2 & F & \cite{zhang2014}\\
0.17 & 83 & 8 & F & \cite{simon2005}\\
0.179 & 75 & 4 & D & \cite{moresco2012}\\
0.199 & 75 & 5 & D & \cite{moresco2012}\\
0.20 & 72.9 & 29.6 & F & \cite{zhang2014}\\
0.27 & 77 & 14 & F & \cite{simon2005}\\
0.28 & 88.8 & 36.6 & F & \cite{zhang2014}\\
0.352 & 83 & 14 & D & \cite{moresco2012}\\
0.38 & 83 & 13.5 & D & \cite{moresco2016}\\
0.4 & 95 & 17 & F & \cite{simon2005}\\
0.4004 & 77 & 10.2 & D & \cite{moresco2016}\\
0.425 & 87.1 & 11.2 & D & \cite{moresco2016}\\
0.445 & 92.8 & 12.9 & D & \cite{moresco2016}\\
0.47 & 89.0 & 49.6 & F & \cite{ratsimbazafy2017}\\
\hline \hline
\end{tabular}
\begin{tabular}{|llllr|}
\multicolumn{5}{c}{{\it continues}}\\
\hline \hline
$z$ & $H(z)$ & $\sigma_{H(z)}$ & M & reference\\
\hline
0.4783 & 80.9 & 9 & D & \cite{moresco2016}\\
0.48 & 97 & 62 & F & \cite{stern2010}\\
0.593 & 104 & 13 & D & \cite{moresco2012}\\
0.68 & 92 & 8 & D & \cite{moresco2012}\\
0.75 & 98.8 & 33.6 & L & \cite{borghi2021b}\\
0.781 & 105 & 12 & D & \cite{moresco2012}\\
0.875 & 125 & 17 & D & \cite{moresco2012}\\
0.88 & 90 & 40 & F & \cite{stern2010}\\
0.9 &  117 &  23 & F & \cite{simon2005}\\
1.037 & 154 & 20 & D & \cite{moresco2012}\\
1.3 & 168 & 17 & F & \cite{simon2005}\\
1.363 & 160 & 33.6 & D & \cite{moresco2015}\\
1.43 & 177 & 18 & F & \cite{simon2005}\\
1.53 & 140 & 14 & F & \cite{simon2005}\\
1.75 & 202 & 40 & F & \cite{simon2005}\\
1.965 & 186.5 & 50.4 & D & \cite{moresco2015}\\
\hline \hline
\end{tabular}
\caption{$H(z)$ measurements (in units of [\Hunit]) obtained with the CC method and their associated errors. The error reported in the table represent only the diagonal part of the covariance matrix; in order to appropriately use these data, the full covariance has to be taken into account, as discussed in Sect.~\ref{sec:CCsys}. The last two columns report the method (M) used to derive the differential age $dt$ (full-spectrum fitting F, Lick indices L, calibrated $D4000$ D) and the corresponding reference. We note that all these measurements are independent, since they consider different datasets.}
\label{tab:CC1}
\end{center}
\end{table}

\subsubsubsection{Independent estimates of the Hubble constant \Ho.} 

In the framework of the well-established tension between early- and late-Universe-based determinations of  the Hubble constant \citep{verde2019,DiValentino2021}, obtaining independent estimates of \Ho~is  of great importance as it can provide additional information to test or constrain the underlying cosmological models. By providing cosmology-independent estimates of $H(z)$, whose calibration does not depend on early-time physics or on the traditional cosmic distance ladder, CCs are of value and, by extrapolating $H(z)$ to $z=0$, could inform the current debate over the Hubble tension. 

This analysis can be done either by directly fitting CC data with a cosmological model \citep{moresco2011,moresco2012b,moresco2016b}, or to take full advantage of the cosmology-independent approach, employing extrapolation techniques that do not rely on cosmological models, such as Gaussian Processes or Pade' approximation \citep{protopapas2014,montiel2014,haridasu2018,gomezvalent2018,capozziello2019,sun2021,bonilla2021,colgain2021b}, or also based on alternative diagnostics \citep[e.g., see][]{sapone2014,krishnan2021}. 
For currently published analyses using CC alone, the size of the error-bars on \Ho~including systematic uncertainties is still too large to weigh in on the tension.

\subsubsubsection{Comparison with independent probes.} 

With respect to other probes, one of the strengths of CC method is that it is a direct probe of the Hubble parameter $H(z)$, instead of one of its integrals (see, e.g., Eqs.~\ref{eq:lumdist}). As a consequence, as highlighted in \cite{jimenez2002}, it is more sensitive to cosmological parameters which affect the evolution of the expansion history, where a difference in luminosity distance of 5\% correspond to a difference in $H(z)$ of 10\%.  In several works the performance of CC in constraining cosmological parameters  has been compared with that other probes. In \cite{moresco2016} constraints from  CC have been compared with the ones from SNe Ia and BAO considering different cosmological models, finding that for a flat $w_{0}w_{a}$CDM model, the accuracy on cosmological parameters that can be obtained from CC and BAO are comparable, and that in comparison with other probes CC are in particular useful to measure \Ho~and \omegam. Similar conclusions are also found by \cite{vagnozzi2021} and \cite{gonzalez2021}, where the results from CC are found in good agreement with the ones of BAO and SNe over a wide range of cosmological models. \cite{lin2020} focused the comparison in particular on \Ho~and \omegam, confirming a good consistency between CC and an even broader collection of cosmological probes, and also highlighting the crucial synergy between the various probes. 

\subsubsubsection{Constraints on cosmological parameters using CC alone and in combination with independent probes.} 

CC are a very attractive probe to study non-standard cosmological models, since no cosmological assumption is made in the derivation of $H(z)$. For this reason, several works have explored how they can be used to put constraints and provide evidences in favor or against various cosmological models, from reconstructing the expansion history of the Universe with a cosmographic approach \citep{capozziello2018, capozziello2019b}, to testing the consistency with concordance models \citep{seikel2012} or the spatial curvature of the Universe \citep{vagnozzi2021,arjona2021}, to exploring more exotic cosmological models \citep[such as interacting dark energy models, but not only, see e.g.][]{bilicki2012,nunes2016,colgain2019,vonmarttens2019,yang2019,benetti2019,aljaf2021,ayuso2021,reyes2021,benetti2021}, to directly measuring cosmological parameters (see, e.g., Sect.~\ref{sec:CCforecast}). In particular, it has been found that CC are extremely useful in combination with other cosmological probes (SNe, BAO, CMB) to increase the accuracy on cosmological parameters \citep[such as $\Omega_k$, \omegam~and \Ho, see, e.g.,][]{haridasu2018,gomezvalent2018,lin2021}, to determine the time evolution of the dark energy EoS \citep{moresco2016b,zhao2017,divalentino2020,colgain2021a}, and also to provide tighter constraints on the number of existing relativistic species and on the sum of neutrino masses by breaking the existing degeneracies between parameters \citep{moresco2012b,moresco2016b}. As suggested by \cite{linder2017}, the measured $H(z)$ data have also been used in combination with the growth rate of cosmic structures to construct a new diagram to disentangle cosmological models \citep{moresco2017,basilakos2017,bessa2021}. Finally, the CC data, in combination with BAO and SNe, have proven to be extremely useful also to test the distance-duality relation and measure the transparency (or equivalently, the opacity) of the Universe \citep{holanda2013,santos2015,chen2016,vavrycuk2020,bora2021,mukherjee2021,renzi2021}.

\subsubsection{Forecasting the future impact of cosmic chronometers}
\label{sec:CCforecast}

Currently, there are two main limitations in the CC method: {\it i)} the error-bars are dominated by the uncertainty due to metallicity and SPS model, and {\it ii)} the absence of a dedicated survey (such as for SNe or BAO) to obtain a statistically significant sample of CC with high spectral S/N and resolution.
For the first one, as highlighted in \cite{Moresco2020}, there is a clear path to make progress, which involves a meticulous and detailed analysis and comparison of the  various models with high-resolution and high S/N observations of CC spectra and SEDs. This program appears to be feasible, enabled by current or forthcoming observational instruments and facilities (e.g., X-Shooter, MOONS) possibly combined with some dedicated observations.

On the other hand, large campaigns to detect massive and passive galaxies with spectra at high S/N and resolutions are not directly foreseen at the moment, and for this science case one should rely on legacy data coming from other planned surveys. Nevertheless, future missions, either already planned \citep[like Euclid,][]{Laureijs:2011gra}, under study \citep[ATLAS probe,][]{Atlasmission}, or large data sets yet not fully exploited \citep[SDSS BOSS Data Release 16,][]{boss16}, could provide  significant large statistics of massive and passive galaxies either in redshift ranges previously poorly mapped ($1.5<z<2$) or previously exploited with  significantly lower statistics ($0.2<z<0.8$).

In the following, we therefore explore two different scenarios, constructing their corresponding simulations and extracting forecasts on the expected performance of CC with future data.
In the first scenario, we will assume to be able to exploit the available spectroscopic surveys at redshifts $0.2<z<0.8$ ({\it low-$z$}, e.g. BOSS DR16), and to be able to obtain a sample large enough to measure 10 $H(z)$ points with a statistical error of 1\%, and including in the systematic error budget both the contribution of IMF and SPS models \citep[as suggested by][]{Moresco2020}; note that already in the analysis by \cite{moresco2016} the statistical error was of the order of 2-3\%.
In the second scenario, we perform a simulation of CC measurements as they will be enabled by future spectroscopic surveys at higher redshifts ({\it high-$z$}), producing 5 $H(z)$ points with a statistical error of 5\% at $1.5<z<2.1$; as an example, Euclid is expected to provide, especially with its Deep Fields, up to a few thousands very massive and passive galaxies in this redshift range, increasing by 2 orders of magnitude the currently available statistics \citep{Laureijs:2011gra,Atlasmission} 
As a final step, we will analyze the {\it combined} measurements, and also a more {\it optimistic} scenario where the systematic error component is assumed to be minimized following the recipes described in Sect.~\ref{sec:CCsys} (and in particular considering the uncertainty due to SPS models resolved, remaining just with the covariance due to the IMF contribution). 

\begin{figure}[t!]
\centering
\includegraphics[width=0.95\textwidth]{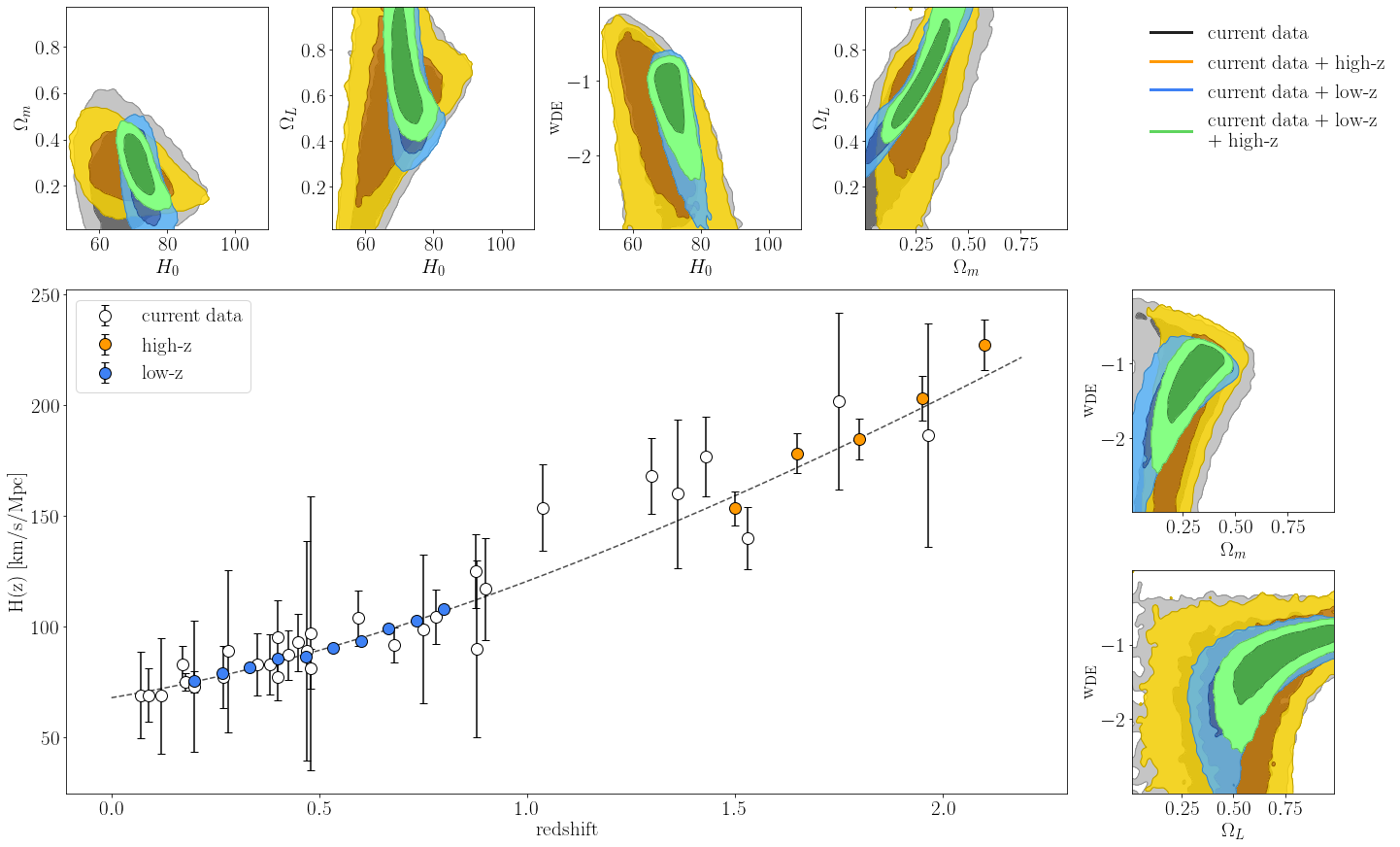}
\caption{Forecast of CC measurements with future surveys. In the bottom left panel, current CC data are shown with white points, while blue and yellow points present forecasts on the expected accuracy with the CC approach respectively at low redshift (with an accurate re-analysis of current surveys, e.g. SDSS) and from future surveys, like the ESA Euclid mission \citep{Laureijs:2011gra} or the ATLAS probe mission \citep{Atlasmission}. For the blue points, the error-bars are smaller than the points. The outer plots shows the constraints for an open $w$CDM cosmology that can be obtained with current data (gray contours), and with different combinations of the simulated datasets.}
\label{fig:HzCC_forecasts}
\end{figure}

The $H(z)$ simulated data are generated with a given error (uncorrelated across data points) assuming cosmological parameters for the $\Lambda$CDM model from \cite{planck2018}, and are shown, together with the current CC measurements, in the larger panel of Fig.~\ref{fig:HzCC_forecasts}. The associated covariance matrix is, then, calculated as presented in Sect.~\ref{sec:CCsys}, considering the contributions previously discussed. To assess the capability of the CC method to constrain cosmological parameters, we explore the constraints current and future data can provide on an open $w$CDM cosmology, where both the spatial curvature density $\Omega_k$ and the dark energy EoS are let free to vary. We considered flat priors on [\Ho, \omegam, \omegade, $w_0$] (the free parameters in our fit), and analyzed the data in a Bayesian framework with a MonteCarlo Markov Chain (MCMC) approach using the public \texttt{emcee} \citep{foreman2013} python code. The results are shown in Fig.~\ref{fig:HzCC_forecasts} and in Tab.~\ref{tab:CC2}.

\begin{table}[t!]
\centering
\begin{tabular}{|c|cr|cr|cr|cr|}
\hline \hline
 & \Ho & \% acc & \omegam & \% acc & \omegade & \% acc & $w$ & \% acc \\ 
 & [\Hunit] & & & & & & &\\
\cline{2-9}
 & \multicolumn{8}{c|}{open $w$CDM} \\
\cline{2-9}
current dataset & $67.8^{+8.7}_{-7.2}$ & 11.7\% & $0.22^{+0.15}_{-0.12}$ & 61\% & $0.51^{+0.24}_{-0.26}$ & 49\% & $-1.6^{+0.8}_{-0.9}$ & 52\%\\
low$-z$ & $72.5^{+3.5}_{-2.9}$ & 4.4\% & 
$0.23^{+0.14}_{-0.13}$ & 59\% &
$0.59^{+0.24}_{-0.16}$ & 34\% & $-1.3^{+0.3}_{-0.6}$ & 38\%\\
high$-z$ & $67.8^{+8.5}_{-7.4}$ & 11.7\% & $0.26^{+0.11}_{-0.09}$ & 40\% & $0.57^{+0.21}_{-0.24}$ & 40\% & $-1.5^{+0.6}_{-0.9}$ & 50\%\\
combined & $71.6^{+3.1}_{-2.7}$ & 4\% & $0.29^{+0.09}_{-0.09}$ & 31\% & $0.67^{+0.19}_{-0.13}$ & 24\% & $-1.w^{+0.3}_{ -0.4}$ & 29\%\\
optimistic & $70.7^{+2.7}_{-1.9}$ & 3.3\% & $0.3^{+0.08}_{-0.08}$ & 27\% & $0.68^{+0.17}_{-0.12}$ & 21\% & $-1.2^{+0.3}_{-0.4}$ & 28\%\\
\hline
 & \multicolumn{8}{c|}{flat $\Lambda$CDM} \\
\cline{2-9}
current dataset & $66.5\pm5.4$ & 8.1\% & $0.34^{+0.08}_{-0.06}$ & 20.6\% & -- & -- & -- & --\\
optimistic & $69.0\pm2.1$ & 3\% & $0.3\pm0.01$ & 3.3\% & -- & -- & -- & --\\
\hline \hline
\end{tabular}
\caption{Constraints with current and future CC measurements in an open $w$CDM (upper rows) and in a flat $\Lambda$CDM cosmology (lower rows).}
\label{tab:CC2}
\end{table}

As a first comment, we note that the reduced $\chi^2$ ($\chi^2_{red}$) of the analysis of the current dataset in the flat $\Lambda$CDM model is smaller than expected, with a value $\sim$ 0.5. This effect is driven by the fact that in some of the CC analyses (and hence for some of the $H(z)$ points), some sources of error have been estimated a bit too conservatively. This is true in particular for the diagonal part of the covariance matrix, where the error on the metallicity (which is driving in most cases the total error) has been overestimated in some works, either for a large prior assumed due to the fact that a accurate measurement was not feasible \citep{moresco2015}, or due to a propagation in that error of additional error contributions \citep[e.g., the one due to SPS modeling or SFH estimate,][]{moresco2016,borghi2021b} that in this way would have been counted twice, after the full covariance matrix has been considered. This is particularly evident since this effect disappears for the analyses where a full metallicity estimate was available and in which the error has not been overestimated, providing in those cases reasonable values of $\chi^2_{red}$, like, e.g., in \cite{simon2005} dataset where $\chi^2_{red}\sim1.1$, or in \cite{moresco2012} where $\chi^2_{red}\sim0.75$. This effect, however, does not have a significant impact on the results, since the points with larger errors (driving $chi^2_{red}$ to smaller values) are also the less relevant for the cosmological constraints, and the cosmological analyses of different CC subsamples provide compatible results.

As discussed in \cite{moresco2016b}, $H(z)$ measurements at low redshift are crucial to better constrain the intercept the Hubble parameter at $z\sim0$, while measurements at higher redshift become more and more important to determine the shape of the $H(z)$ evolution, critically dependent on dark energy and dark matter parameters. As expected, the simulated CC data at low$-z$ significantly improve the current accuracy on the estimated Hubble constant by a factor of $\gtrsim2$ by increasing the precision on the extrapolation of  $H(z)$ to $z\sim0$. On the other hand, the high$-z$ simulated data become fundamental to determine the 
dark energy EoS especially when combined with lower redshift data, improving the accuracy on $w$ from 38\% to 29\% and on \omegam~from 59\% to 31\%. When considering the {\it optimistic} scenario, CC data will enable an accuracy on \Ho~to the 3\% level, and on \omegam~and $w$ to $\sim$30\%.

Clearly, as the dimensionality of the problem decreases, the accuracy on the derived parameter increases. As a comparison, in Tab.~\ref{tab:CC2} we show also, for the {\it current dataset} and the {\it optimistic} scenarios, the constraints on \Ho~and \omegam~achievable in a flat $\Lambda$CDM model. In this regime, we observe a particular improvement in the accuracy on \omegam~up to the 3\% level.

\clearpage

\subsection{Quasars}
\label{sec:QSO}
There have been numerous proposals in the literature for standardising the emission of quasars \citep[e.g.][]{watson2011,lafranca2014,ss2022}; in the following we will focus on the one presented by \citet{rl15}.
Quasars are the most luminous persistent objects in the Universe, with integrated luminosities of $10^{44-48}$ erg s$^{-1}$ over the ultra-violet (UV) to the X-ray energy range. The UV emission is interpreted as the radiation produced by the material flowing towards the supermassive black hole, located in the center of a galaxy, in the form of an accretion disc, and it makes up to roughly 90\% of the quasar bolometric budget \citep{ss1973}. The rest is released as X-rays, which are thought to originate in a hot plasma of relativistic electrons \citep{sz1994}, called corona for analogy with the Sun, that Compton up-scatter photons coming from the disk. The UV and X-ray fluxes have long been known to obey a non-linear relation between their UV (at the rest frame 2500 \AA, $L_{\rm UV}$) and X-ray (at the rest frame 2 keV, $L_{\rm X}$) emission (e.g., \citealt{avnitananbaum79,zamorani81,avnitananbaum82}, parameterized as $L_{\rm X}\propto L_{\rm UV}^\gamma$, with $\gamma\simeq0.6$), yet how the gravitational energy is partly transferred from the disc to the corona, preventing its fast cooling via the production of X-ray photons through the inverse Compton process, is unknown.

\subsubsection{Basic idea and equations}
\label{sec:QSObasics}

The technique that makes use of quasars as cosmological probes hinges on the non-linear relation mentioned above to provide an independent measurement of their distances, thus turning quasars into standardizable candles and extending the distance modulus-redshift relation (or the so-called Hubble-Lema\^itre diagram) of supernovae Ia to a redshift range that is still poorly explored \cite[$z>2$;][]{rl15}. 
The applicability of this methodology is based on two key points. Firstly, the understanding that most of the observed dispersion in the $L_{\rm X}-L_{\rm UV}$ relation is not intrinsic to the relation itself but due to observational issues, such as gas absorption in the X-rays, dust extinction in the UV, calibration uncertainties in the X-rays \cite[e.g.][]{lusso2019a}, variability, and selection biases associated with the flux limits of the different samples. In fact, with an optimal selection of clean sources (i.e. where the intrinsic UV and X-ray quasar emission can be measured), the observed dispersion drops from 0.4 dex to $\simeq$0.2 dex \citep{lr16,lr17}. The interested reader should refer to \citet{lr16} and \citet{lusso2020} for further details on the sample selection. Specifically, \citet{lr16} determined how both slope and dispersion vary depending upon a given selection criterion by also including censored data at X-ray energies (see their Table 3). They also discussed the additional effect of X-ray variability and measurement uncertainties on the determination of the slope and the dispersion.
Secondly, the slope of the $L_{\rm X}-L_{\rm UV}$ relation does not evolve with redshift up to $z\simeq4$ (i.e. the highest redshift where the source statistics is currently sufficient to verify any possible dependence of the slope with distance).
This point has been recently  discussed also by \citet{sacchi2022arXiv}, who  demonstrated that a one-by-one spectral analysis of a sample of quasar at redshift higher than 2.5, with high-quality X-ray and UV observations, further reduces the dispersion from 0.2 dex (by employing photometric data only) to 0.12 dex, whilst the observed slope of the relation is still around 0.6. \cite{sacchi2022arXiv} also showed that the composite X-ray and UV spectra of these high-redshift quasars do not show any peculiar spectral feature or systematic difference with respect to the average spectra of quasars at lower redshifts. The absence of any spectral variance between high- and low-redshift quasars, combined with the tightness of the X-ray to UV relation, suggests that no evolutionary effects are present in the relation itself.
A key consequence is that the $L_{\rm X}-L_{\rm UV}$ relation must be the manifestation of a universal mechanism at work in the quasar engines. 

To fit the Hubble diagram, the distance modulus for each object should be computed first. The method is based on the non-linear relation between $L_{\rm X}$ and $L_{\rm UV}$
\begin{equation}
\label{eq:luvlx}
\log L_{\rm X}=\beta+\gamma\log L_{\rm UV},
\end{equation}
from which the luminosity distance \cite[e.g., see][]{rl15,rl19} can be derived as:
\begin{equation}
\label{dl}
\log D_{\rm L} = \frac{\left[\log F_{\rm X} -\beta -\gamma(\log F_{\rm UV}+27.5) \right]}{2(\gamma-1)}-\frac{1}{2}\log(4\pi) + 28.5,
\end{equation}
assuming that $F=L/4\pi D_{\rm L}^2$, where $F_{\rm X}$ and $F_{\rm UV}$ represent the flux densities (in erg s$^{-1}$ cm$^{-2}$ Hz$^{-1}$) at X-ray and UV energies, respectively. 
$F_{\rm UV}$ is normalized to the (logarithmic) value of 27.5 in the equation above, whilst $D_{\rm L}$ is in units of cm and is normalized to 28.5 (in logarithm)\footnote{The values of these normalisations are representative of the average luminosity and distance probed by the Lusso et al.~quasar sample and should be tailored to other data sets accordingly.}. 
The slope of the $F_{\rm X}-F_{\rm UV}$ relation, $\gamma$, is a free parameter, and so is the intercept $\beta$. The intercept $\beta$ of the $L_{\rm X}-L_{\rm UV}$ relation is related to the one of the $F_{\rm X}-F_{\rm UV}$ relation, $\hat\beta$, as $\hat\beta(z)=2(\gamma-1)\log D_{\rm L}(z) + (\gamma-1)\log 4\pi + \beta$. The distance modulus, $DM$, is thus:
\begin{equation}
DM= 5 \log D_{\rm L} - 5 \log (10\,{\rm pc}) \;\; ,
\end{equation}
and the uncertainty on $DM$, $d DM$, is:
\begin{equation}
\label{dDM}
d DM = \frac{5}{2(\gamma-1)} \left[ \left(d\log F_{\rm X}\right)^2 + \left(\gamma d\log F_{\rm UV}\right)^2 + \left(d\beta\right)^2 + \left( \frac{d\gamma \left[\beta+\log F_{\rm UV}+27.5-\log F_{\rm X}\right]}{\gamma-1} \right)^2\right]^{1/2} \;\; ,
\end{equation}
where $d\log F_{\rm X}$ and $d\log F_{\rm UV}$ are the logarithmic uncertainties on $F_{\rm X}$ and $F_{\rm UV}$, respectively. Equation~\ref{dDM} assumes that all the parameters are independent, and takes into account also the uncertainties on $\beta$ and $\gamma$.
The fitted likelihood function, $\mathcal{L}$, is then defined as:
\begin{equation}
\label{lf}
\ln \mathcal{L} = - \frac{1}{2} \sum_i^N\left(\frac{(y_i-\psi_i)^2}{s_i^2} - \ln s^2_i\right) \;\; ,
\end{equation}
where $N$ is the number of sources, $s_i^2 = d y_i^2 +\gamma^2 d x_i^2 + \exp(2\ln\delta)$ takes into account the uncertainties on both the $x_i$ ($\log F_{\rm UV}$) and $y_i$ ($\log F_{\rm X}$) parameters of the fitted relation. The parameter $\delta$ represents what is left in the scatter of the relation once it is marginalized over all the parameters and thus it can be considered a proxy of the \textit{intrinsic} dispersion under the assumption that all the systematics have been taken into account\footnote{$\delta=0$ means that all the observed dispersion is intrinsic.}. The variable $\psi$ is the modeled X-ray monochromatic flux ($F_{\rm X,\,mod}$), defined as:
\begin{equation}
\label{model}
\psi = \log F_{\rm X,\,mod} = \beta + \gamma(\log F_{\rm UV}+27.5) +2(\gamma-1)(\log D_{\rm L,\,mod} -28.5) \;\; ,
\end{equation}
and it is dependent upon the data, the redshift and the model (cosmological or parametric) assumed for the distances (e.g., $\Lambda$CDM, $w$CDM or a polynomial function). 

In the case of a parametric (cosmology independent) approach, the data are fitted with a luminosity distance described by a fifth-grade polynomial of $\log(1+z)$, where the cosmographic function is:
\begin{equation}
  \label{dl_mod}
D_{\rm L,\,mod}(z)=k \ln(10)\frac{c}{H_0}\sum^5_{i=1} a_i\log^i(1+z) \;\; ,
\end{equation}
where $k$ and $a_i$ ($a_1$ is fixed to 1 to reproduce the local Hubble law) are free parameters. The polynomial order is chosen depending upon the range of redshift spanned by the quasars to ensure convergence \cite[see][]{bargiacchi2021}.

For any analysis that involves a detailed test of cosmological models, the quasar distances should be cross-calibrated by making use of the distance ladder through supernovae Ia. In fact, the $DM$ values of quasars are not absolute, thus a cross-calibration parameter ($k$) is needed. The parameter $k$ should be fitted simultaneously for supernovae Ia and quasars (i.e. $k$ is a rigid shift of the quasar Hubble diagram to match the one of supernovae). 

The slope of the $L_{\rm X}-L_{\rm UV}$ relation can be kept fixed in the procedure above. Yet, it is better to marginalize over $\gamma$ to check whether any degeneracy of the slope with the other parameters is present, and whether the statistical significance of any deviation from a cosmological model can be affected by the assumption of a $\gamma$ value that slightly deviates from the true one. The marginalization on $\gamma$ is a more conservative procedure, as it reduces the significance of any observed deviation with respect to the same MCMC analysis with $\gamma$ fixed. Therefore, if a statistical deviation persists with respect to a cosmological model even allowing for a variable $\gamma$, its significance should be considered as an indicative lower limit with respect to the case where $\gamma$ is fixed. 
Finally it should be noted that the Hubble constant \Ho~in Eq.~\ref{dl_mod} is degenerate with the $k$ parameter, so it can assume any arbitrary value. In the following, the Hubble constant is assumed to be fixed to \Ho=70 \Hunit \citep[see also][]{lusso19b,lusso2020,bargiacchi2021}.

\subsubsection{Sample selection}
\label{sec:qsosampleselection}
To build a quasar sample that can be utilized for cosmological purposes, both X-ray and UV data are required to cover the rest-frame 2 keV and 2500 \AA. The most up-to-date broad-line quasar sample considered for cosmological purposes has been assembled by combining seven different samples from both the literature and the public archives \cite[][]{lusso2020}. The former group includes the samples at $z\simeq3.0-3.3$ by \citet{nardini2019}, $4<z<7$ by \citet{salvestrini2019}, $z>6$ by \citet{vito2019}, the XMM-XXL North quasar sample published by \citet{menzel2016}, and one new optically-selected SDSS quasar at $z=4.109$, J074711.14$+$273903.3, whose X-ray observation was obtained as part of a proposed large programme with XMM-\textit{Newton} (cycle 18, proposal ID: 084497, PI: Lusso). 
This collection is complemented by including quasars from a cross-match of optical (i.e. the Sloan Digital Sky Survey) and X-ray public catalogs (i.e. XMM-\textit{Newton} and \textit{Chandra}), which will be labeled as SDSS-4XMM and SDSS-\textit{Chandra} samples hereafter \citep{bisogni2021}. A local subset of active galactic nuclei (AGN) with UV (i.e. {\it International Ultraviolet Explorer}) data and X-ray archival information was also added to improve the sampling at very low redshifts. The reader interested on the description of the different subsets should refer to \cite{lusso2020}. The main parent sample is composed by $\sim$19,000 objects, from local up to $z=7.52$, where quasars with bright radio jets and broad absorption lines (BALs) have been removed. 
In fact, an excess of X-rays due to synchrotron emission is observed in bright radio quasars due to the presence of the jet, whilst the strong absorption features observed in BALs, and usually attributed to winds/outflows, hamper a robust measurement of the quasar continuum in the UV.

\begin{figure}[t!]
   \centering
   \includegraphics[width=0.6\textwidth]{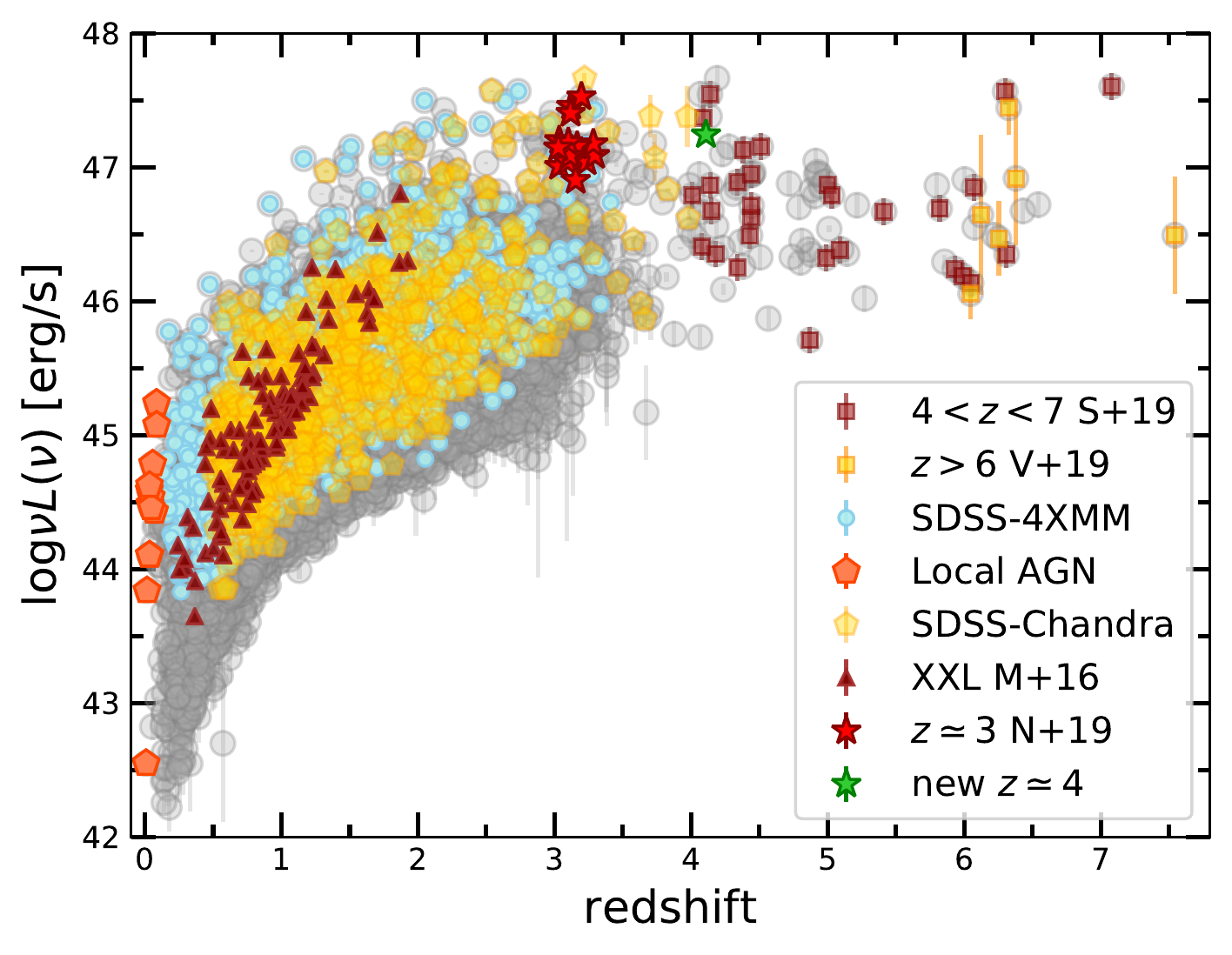}
   \caption{Distribution of luminosities at rest-frame 2500 \AA\ as a function of redshift for the main (grey points, $\simeq19,000$ objects) and the selected (cleaned) samples \citep{lusso2020}. Brown and yellow squares show the high-$z$ sample \citep{salvestrini2019,vito2019}, cyan points the SDSS-4XMM one, brown triangles the XMM-XXL one \citep{menzel2016}, orange pentagons the local AGN sample, red stars the $z\simeq3$ quasar sample \citep{nardini2019}, the green star represents the new $z\simeq4$ quasar from a dedicated XMM programme (see text for details), and gold pentagons the SDSS-Chandra one \citep{bisogni2021}. Image reproduced with permission from \cite{lusso2020}, copyright by Astronomy \& Astrophysics.}
   \label{loz}
\end{figure}

To select a sub-sample with accurate estimates of $F_{\rm X}$ and $F_{\rm UV}$, systematic effects should be taken into account and low-quality measurements should be neglected. A minimum signal-to-noise (S/N) of 1 on the soft and hard X-ray band fluxes should be considered, whilst no such a filter is required in the UV since the S/N at these wavelengths is typically significantly higher than 1. 
The main possible sources of contamination or systematic error that may affect the flux measurements are: dust reddening and host-galaxy contamination in the optical/UV, gas absorption in the X-rays, and the Eddington bias associated with the flux limit of the X-ray observations. 

Regarding the latter, any flux limited sample is biased towards brighter sources at high redshifts and this should be more relevant to the X-rays, since the relative observed flux range is narrower than in the UV.
Specifically, AGN with an average X-ray intensity close to the flux limit of the observation will be observed only in case of a positive fluctuation. This introduces a systematic, redshift-dependent bias towards high fluxes, known as {\it Eddington bias}, which has the effect to flatten the $F_{\rm X}-F_{\rm UV}$ relation. 
Samples with datasets of only detections might thus be affected by such a bias. One possibility is to include censored data in the analysis. Yet, the investigation of both the $F_{\rm X}-F_{\rm UV}$ and the distance modulus-redshift relations is far from trivial, since it strongly depends on the weights assumed in the fitting algorithm. Therefore, one needs to find an alternative method to obtain an (almost) unbiased sample. 

To minimize this bias, one possible approach is to neglect all X-ray detections below a threshold defined as $\kappa$ times the intrinsic dispersion of the $F_{\rm X}-F_{\rm UV}$ relation ($\delta$) computed in narrow redshift intervals (\citealt{lr16,rl19}), specifically:
\begin{equation}
\label{fthr}
\log F_{2\,\rm keV,\,exp} - \log F_{\rm min} < \kappa \delta \;\; ,
\end{equation}
where $F_{2\,\rm keV,\,exp}$ is the monochromatic flux at 2 keV expected from the observed rest-frame quasar flux at 2500 \AA\ with the assumption of a {\it true} $\gamma$ of 0.6; it is calculated as follows:
\begin{equation}
\label{fexp}
\log F_{2\,\rm keV,\,exp} =(\gamma-1)\log(4\pi) + (2\gamma-2)\log D_{\rm L} + \gamma\log F_{\rm UV} + \beta \;\; ,
\end{equation}
where $D_{\rm L}$ is the luminosity distance calculated for each redshift with a fixed cosmology, and the parameter $\beta$ represents the pivot point of the non-linear relation in luminosities, $\beta=26.5-30.5\gamma\simeq8.2$\footnote{We note again that the value of the luminosity normalizations should be chosen based on the average values for the entire sample.}.
\begin{figure}[t!]
   \centering
   \includegraphics[width=0.6\linewidth]{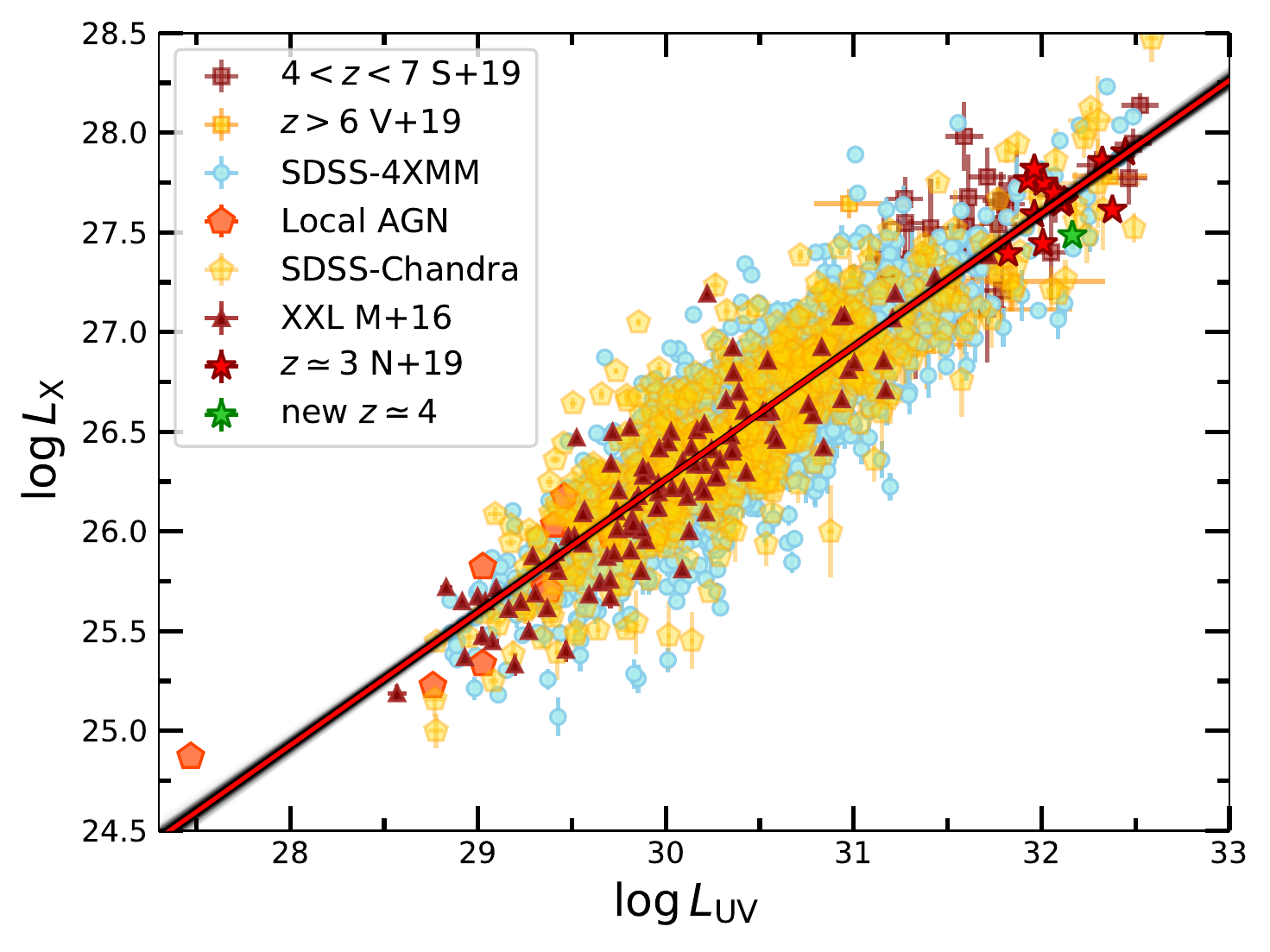}   
   \caption{The $L_{\rm X}-L_{\rm UV}$ relation for the $\simeq2400$ quasars published by \citet{lusso2020}. Symbol keys are the same as in Figure~\ref{loz}. The red line represents the linear regression fit of the data obtained through the hierarchical Bayesian model \texttt{linmix} \citep{kelly2007}. The light black lines represent some random realisations of the $L_{\rm X}-L_{\rm UV}$ relation. The resulting slope and intercept of the best-fit regression line are $\gamma=0.667\pm0.007$ and $\beta=6.25\pm0.23$. The observed dispersion along the $L_{\rm X}-L_{\rm UV}$ relation is 0.24 dex. Luminosity values are computed by assuming a flat $\Lambda$CDM model with $\Omega_{\rm M}=0.3$.}
  \label{lolx}
\end{figure}
The parameter $F_{\rm min}$ in the Eq.~\ref{fthr} represents the flux limit of a given observation or survey, whilst the product $\kappa \delta$ is a value that should be estimated for all the sub-samples constructed from archives (e.g., SDSS-4XMM, SDSS-\textit{Chandra}) or surveys (XXL).
The Eddington bias is then reduced by including only X-ray detections for which the minimum detectable flux $F_{\rm min}$ in that given observation is lower than the expected X-ray flux $F_{2\,\rm keV,\,exp}$ by a factor that is proportional to the dispersion in the $F_{\rm X}-F_{\rm UV}$ relation in narrow redshift bins (see Appendix~A in \citealt{lr16} and \citealt{rl19}).

A complete description and implementation of these filters to obtain the final {\it best} sample for a cosmological analysis is presented in \citet[][see their Section~5]{lusso2020}. 
The most up-to-date quasar sample is composed by 2,421 quasars spanning a redshift interval $0.009\leq z\leq7.52$, with a mean (median) redshift of 1.442 (1.295)  and it is shown in Fig.~\ref{loz}. 
Figure~\ref{lolx} presents the $L_{\rm X}-L_{\rm UV}$ relation for this sample, where the best fit regression line is obtained through the hierarchical Bayesian model \texttt{linmix} \citep{kelly2007}.

\subsubsection{Measurements}

Ideally, spectroscopy can deliver cleaner measurements of the relevant parameters (i.e. the X-ray and UV rest frame fluxes), but since a detailed spectroscopic UV and X-ray analysis can be carried out only for a relatively small number of sources, the currently published quasar sample also still heavily relies on broadband photometry in both UV and X-rays to compute the monochromatic UV and X-ray fluxes, as well as the UV colors and X-ray slopes. These parameters are thus derived from the photometric AGN spectral energy distribution (SED).  

To compile the quasar SEDs, multi-wavelength data from radio to UV should be considered, such as the FIRST survey in the radio \cite{becker1995}, the Wide-Field Infrared Survey \cite[WISE][]{write2010} in the mid-infrared, the Two Micron All Sky Survey \cite[2MASS][]{cutri2003,2006AJ....131.1163S} and the UKIRT Infrared Deep Sky Survey \cite[UKIDSS][]{lawrence2007} in the near-infrared, SDSS in the optical and the Galaxy Evolution Explorer \cite[GALEX][]{martin2005} survey in the UV. Most of the relevant broadband information, as well as the spectroscopic redshifts, are compiled in the SDSS quasar catalogs.
Galactic reddening must be taken into account by utilizing the selective attenuation of the stellar continuum $k(\lambda)$ \cite[e.g.][]{F99}, along with the relative Galactic extinction  \cite[e.g.][]{schlegel98} for each object.
For each source, the observed flux and the corresponding frequency in all the available bands should be computed. The data used in the SED computation are then blue-shifted to the rest-frame (with no K-correction). All the rest-frame luminosities are then determined from a first-order polynomial between two adjacent points. 
At wavelengths bluer than about 1400 \AA,  significant absorption by the intergalactic medium (IGM) is expected in the continuum \cite[$\sim$10\% between the \ion{Ly}{$\alpha$} and \ion{C}{iv} emission lines, see ][for details]{lusso2015}. 
Hence, when computing the relevant parameters, all the rest-frame data at $\lambda<1500$\AA\ should be excluded from the SED (or corrected for such an absorption if possible).

By compiling a broad photometric coverage, the rest-frame luminosity at 2500 \AA\ can be computed via interpolation for the majority of the quasars whenever the reference frequency is covered by the photometric SED. Otherwise, the value can be extrapolated by considering the slope between the luminosity values at the closest frequencies. Uncertainties on monochromatic luminosities ($L_\nu\propto \nu^{-\gamma}$) from the interpolation (extrapolation) between two values $L_1$ and $L_2$ are computed as:
\begin{equation}
\label{uncertainties}
\delta L = \sqrt{\left( \frac{\partial L}{\partial L_1}\right)^2 (\delta L_1)^2 + \left( \frac{\partial L}{\partial L_2}\right)^2 (\delta L_2)^2} \;\; .
\end{equation}  

To obtain the rest-frame luminosities at 2 keV, a detailed X-ray spectral analysis of all the quasars is impractical, given the overall large number of sources, while a photometric approach is a viable solution \citep{rl19,lusso2020}.
Briefly, for sources having an entry in the 4XMM-DR9 serendipitous source catalog\footnote{http://xmm-catalog.irap.omp.eu/}, the rest-frame 2 keV fluxes and the relative (photometric) photon indices, $\Gamma_X$ (along with their 1$\sigma$ uncertainties), can be derived from the tabulated 0.5-2 keV (soft, $F_{\rm S}$) and 2-12 keV (hard, $F_{\rm H}$) fluxes. These band-integrated fluxes are blue-shifted to the rest-frame by considering a pivot energy value of 1 keV ($E_{\rm S}$) and 3.45 keV ($E_{\rm H}$), respectively, and by assuming the same photon index used to derive the fluxes in the 4XMM catalog (i.e. $\Gamma_X=1.42$, \citealt{nb2020}). For the soft band, the monochromatic flux at $E_{\rm S}$ is then:
\begin{equation}
\label{soft}
F_E(E_S)=F_{\rm S}\frac{(2-\Gamma_X) E_{\rm S}^{1-\Gamma_X}}{(2\,\rm keV)^{2-\Gamma_X}-(0.5\,\rm keV)^{2-\Gamma_X}} \;\; ,
\end{equation}
in units of erg s$^{-1}$ cm$^{-2}$ keV$^{-1}$. An equivalent expression holds for the hard band, with the obvious modifications. Flux values must be corrected for Galactic absorption. The photometric photon index is then estimated from the slope of the power-law connecting the two soft and hard monochromatic fluxes at the rest-frame energies corresponding to the observed pivot points. 
The rest-frame photometric 2 keV flux (and its uncertainty) is interpolated (or extrapolated) based on such a power-law. 
A similar approach can be adopted for any X-ray catalog (e.g., the \textit{Chandra} source catalog\footnote{https://cxc.cfa.harvard.edu/csc/}, see \citealt{bisogni2021}). 

\subsubsection{Systematic effects}

This method may still have several shortcomings, thus it is mandatory to demonstrate that the observed deviation from $\Lambda$CDM at a redshift $>2$ is neither driven by systematics in the quasar sample selection nor by the procedure adopted to fit the quasar Hubble-Lema\^itre diagram. Potential convergence issues may arise from the use of the polynomial expansion (Eq.~\ref{dl_mod}) to fit the Hubble diagram when observational data go beyond $z \simeq 1$ (see \citealt{bargiacchi2021} for an in-depth discussion). Moreover, the choice of these monochromatic luminosities is rather arbitrary, and mostly based on historical reasons. It is possible that the $L_{\rm X}-L_{\rm UV}$ relation is tighter with a different choice of the indicators of UV and X-ray emission \cite[see e.g.][]{young2010}. A careful analysis of this issue may also provide new insights as to the physical process responsible for this relation. A small fraction of moderate/bright radio sources may still be present in the sample. Deep all-sky radio surveys and multi-wavelength approaches \citep{mingo2016} are necessary to better remove these sources from the clean samples.

One serious issue that could affect the precision of the flux estimates at X-rays is gas absorption. Previous studies based on large AGN surveys show that about 25\% of optically selected un-obscured AGN display some levels of X-ray absorption \citep{merloni2014} in excess of the Galactic value. 
If not corrected for, this absorption leads to an underestimate of the X-ray flux, and an overestimate of the distance. As absorption mostly affects the low energy part of the X-ray spectrum, this bias is expected to be more relevant at low redshift ($z<1$). Nonetheless, the global effect on the Hubble diagram will be a decrease of the ratio between high-redshift and low-redshift distances, i.e., qualitatively, this effect may lead a discrepancy with the concordance model. In fact, including AGN with $\Gamma_X<1.5$ produces a flattening of the Hubble diagram, as expected if absorbed sources start to contaminate the sample. A conservative threshold should thus be $\Gamma_X>1.7$. Therefore, sources with an X-ray photon index below that value are removed from the sample.

Work still needs to be done regarding the effect of the X-ray and UV variability on the relation \citep{lr16}. Variations in the UV brightness are on the order of about 10\% (i.e. 0.04 dex in logarithmic units) on time scales of months to years \citep[e.g.][]{vandenberk2001}. The X-ray variability is on the order of 5\% on long time scales at high luminosity and somewhat larger at lower luminosity \citep[e.g.][]{zheng2017} and it represents about 30\% on the dispersion of the X-ray/UV relation overall \citep[about 0.12 dex compared to the observed 0.24 dex, see][for details]{lr16}. Moreover, it is well known that the UV and X-ray variability are not correlated on short timescales (e.g., NGC5548 \citealt{edelson2015}), so the intrinsic variance on the relation could be even lower than 0.1 dex. 
Yet, regarding the X-ray variability, the increase of dispersion due to variability does not modify the slope of the relation \citep{lr16}, even when using simultaneous datasets \citep{grupe2010,wu2012,lr16}. Although in the case of low fluxes, X-ray and UV variability may bias our data towards brighter states, both X-ray and UV variability have the only effect of producing higher uncertainties on the final computation of the parameters, without introducing any major systematic. 

Another key issue that could affect the analysis of the distance modulus-redshift relation is the correction for the Eddington bias, which flattens the $F_{\rm X}-F_{\rm UV}$ relation and thus the Hubble diagram, especially at high redshifts. At present, such a correction is at the expenses of the sample statistics. Depending on the flux limit of the given observation/survey, the sample statistics of the parent sample may drop by more than 50\%. Additionally, the assumption that the {\it true} slope of the $F_{\rm X}-F_{\rm UV}$ is $\gamma=0.6$ may leave some hidden trends in the residuals of the Hubble diagram as a function of redshift. Nonetheless, the analysis of the residuals of the Hubble diagram as a function of redshift and $\gamma$ for different values of the threshold $\kappa\delta$ does not show any obvious trend (see Sect.~9.1 by \citealt{lusso2020} and appendix~A in \citealt{lr16}).

The presence of an additional contribution of dust reddening in the UV band should be considered amongst the possible residual (and redshift-dependent) observational systematics in the Hubble diagram. Going to higher redshifts, the rest-frame optical/UV spectra shift to higher (shorter) frequencies (wavelengths), where the dust absorption cross-section is higher. This might underestimate $F_{\rm UV}$ measurements, which would imply an intrinsically larger value of the luminosity distance (and thus the distance modulus) than the measured one (see Section~9.4 by \citealt{lusso2020} for details). 

The results on the quasar Hubble diagram crucially depend on the assumption of the non-evolution of the relation, which has been verified by the constancy of the slope across a wide redshift interval (see e.g. Figure~8 in \citealt{lusso2020}) and by the agreement with supernovae (and, hence, with the flat $\Lambda$CDM model) up to $z$\,$\sim$\,1.5. Both findings suggest a non-evolution as also implied by the analysis of \citet[][and references therein]{dainotti2022}. The standardization of quasars through the $L_{\rm X}-L_{\rm UV}$ relation is a rather young technique and it is still subject to extensive tests by several independent groups ( e.g. \citealt{dainotti2022,colgain2022a,colgain2022b}, see also \citealt{kr2020a,kr2020b,kr2021}).

Finally, we note that, in a Bayesian framework, handling data with uncertainties on both the $x$ and $y$ parameters is quite subtle, and can lead to biases unless a hierarchical model is adopted where the \textit{true} value of a certain parameter is considered in the fitting procedure and then marginalized over. As uncertainties on the X-ray fluxes are on average higher than the ones on the UV ones, one can fit the data by including uncertainties on the X-ray emission only. To alleviate possible issues, the priors should incorporate as much information as possible in the Bayesian formulation in order to constrain the set of plausible solutions. Something along this line was performed by \citet{lr16}, who considered the hierarchical Bayesian model for linear regression by \citet[][see also Figure~\ref{lolx}]{kelly2007}. They find statistically consistent results with the Bayesian analysis performed with \texttt{emcee} in the case of the fit of the $L_{\rm X}-L_{\rm UV}$ relation, so the implementation of the latter algorithm does not seem to show any clear bias. Nonetheless, the use of hierarchical models for the analysis of the quasar Hubble diagram is surely a point that should be investigated further.

\subsubsection{Main results and forecasts}

\begin{figure}[t!]
   \centering
   \includegraphics[width=0.6\linewidth]{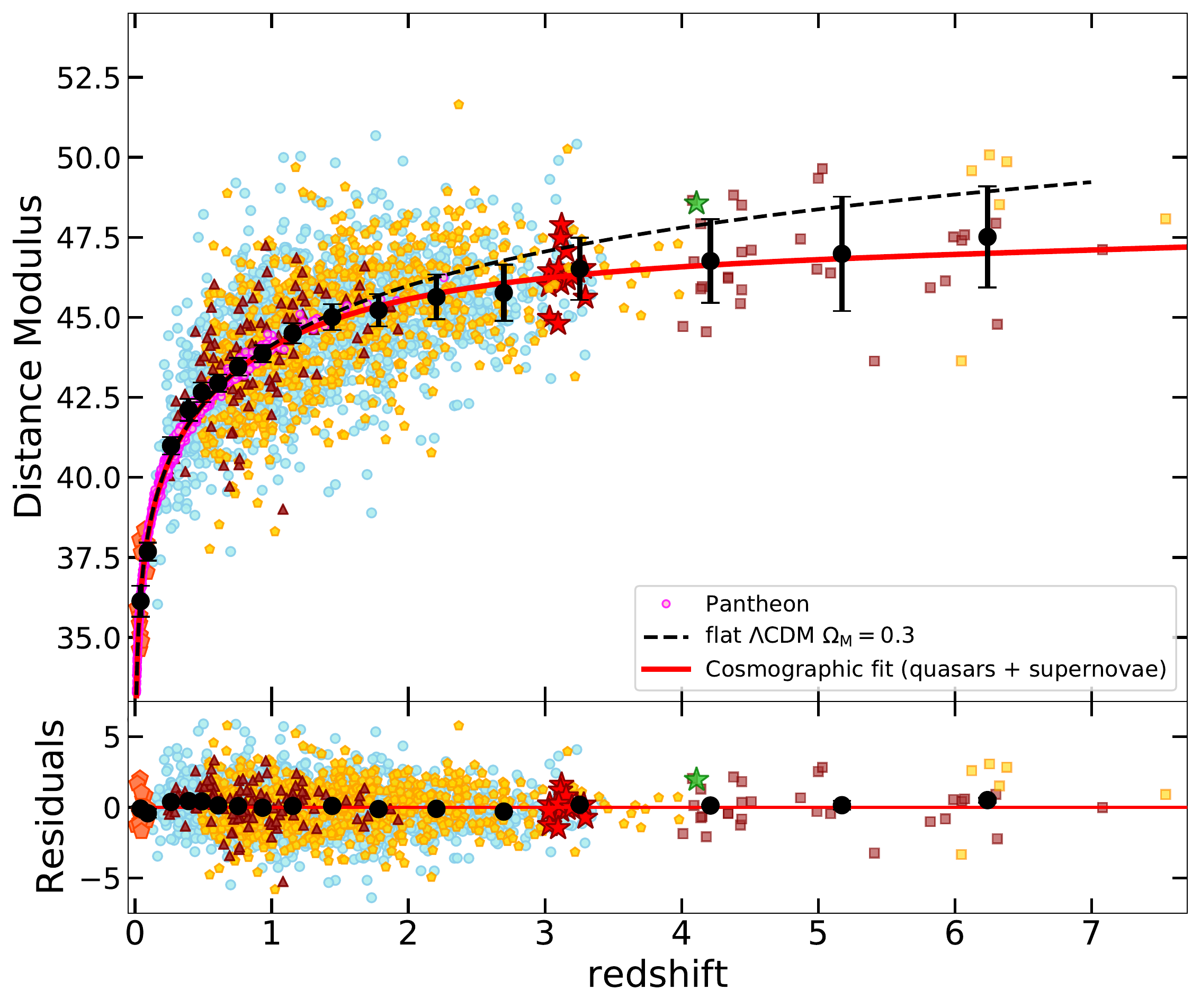}   
   \caption{Distance modulus-redshift relation (Hubble diagram) for the clean quasar sample and Type Ia supernovae ({\it Pantheon}, magenta points). Symbol keys are the same as in Figure~\ref{loz}. The red line represents a fifth order cosmographic fit of the data, whilst the black points are averages (along with their uncertainties) of the distance moduli in narrow (logarithmic) redshift intervals. The dashed black line shows a flat $\Lambda$CDM model fit with \omegam$=0.3$. The bottom panel shows the residuals with respect to the cosmographic fit and the black points are the averages of the residuals over the same redshift intervals. Image reproduced with permission from \cite{lusso2020}, copyright by Astronomy \& Astrophysics.}
  \label{hubbleclean}
\end{figure}

Quasars have been now extensively used to determine cosmological constraints by fitting their Hubble diagram in combination with the one of supernovae Ia, as discussed in Sect.~\ref{sec:QSObasics} \citep[e.g.][]{corr2016,brl2017,lusso2019b,melia2019,wm2020,demianski2020,2021EPJC...81..694Z,2021MNRAS.507..919L,bargiacchi2021arXiv,2021arXiv211201492L}. Amongst the main results, it has been found that the expansion rate of the Universe based on the combined quasar and supernovae Ia Hubble diagram shows a deviation from the concordance model at high redshifts ($z>1.4$), with a statistical significance of $\sim3-4\sigma$. Figure~\ref{hubbleclean} presents the Hubble diagram for the most up-to-date samples of quasars \citep{lusso2020} and Type Ia supernovae from the {\it Pantheon} survey \citep{scolnic2018}. 
The best MCMC cosmographic fit (see Eq.~\ref{dl_mod}) is shown with a red line, whilst black points are the means (along with the uncertainty on the mean) of the distance modulus in narrow (logarithmic) redshift intervals, plotted for visualization purposes only. The residuals are displayed in the bottom panel with the same symbols, and do not reveal any apparent trend with redshift. 
The MCMC fit assumes uniform priors on the parameters \citep[see][for more details on the cosmographic technique employed]{bargiacchi2021}.

\begin{figure}[t!]
   \centering
   \includegraphics[width=0.44\linewidth]{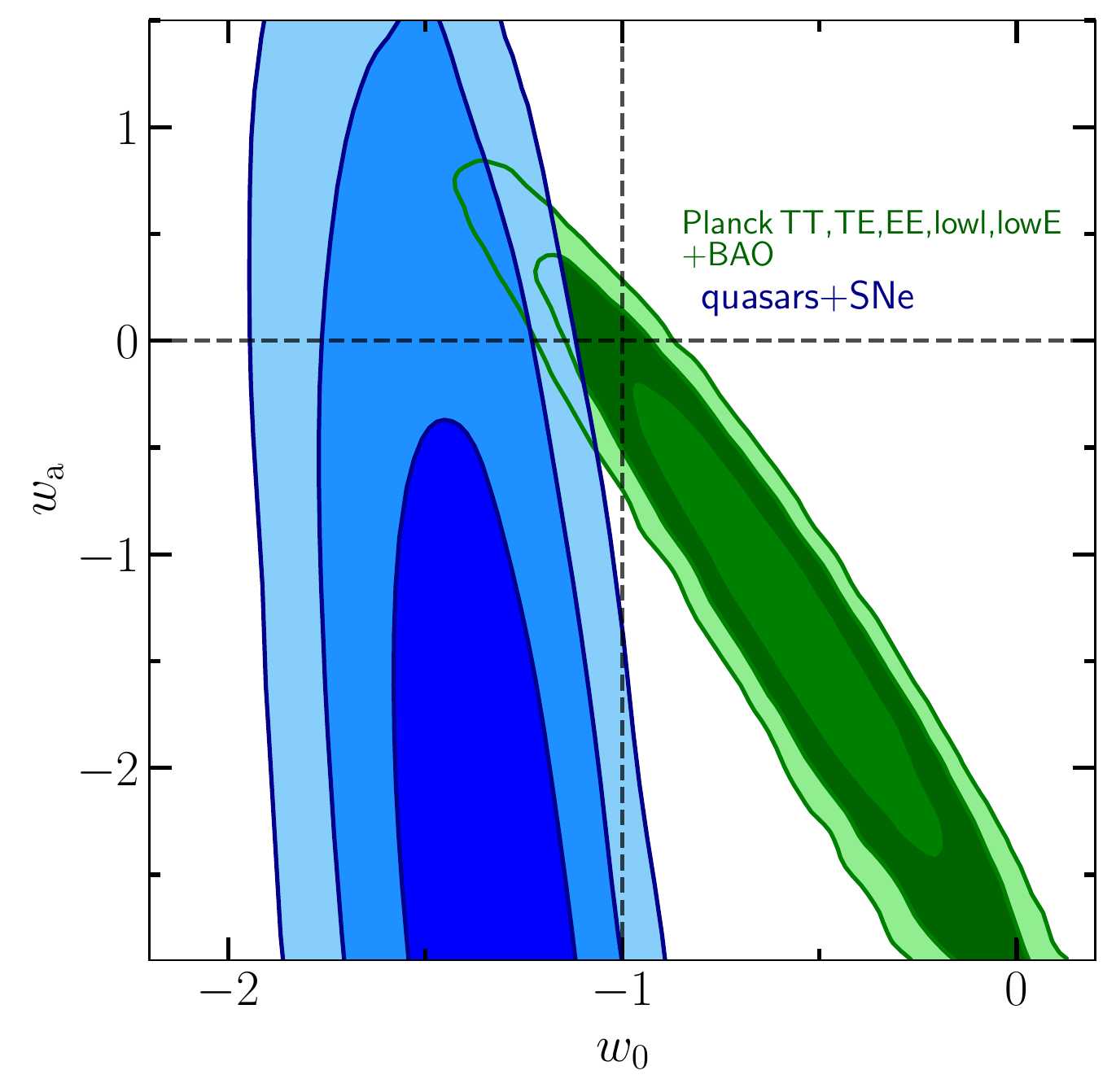}   
   \includegraphics[width=0.44\linewidth]{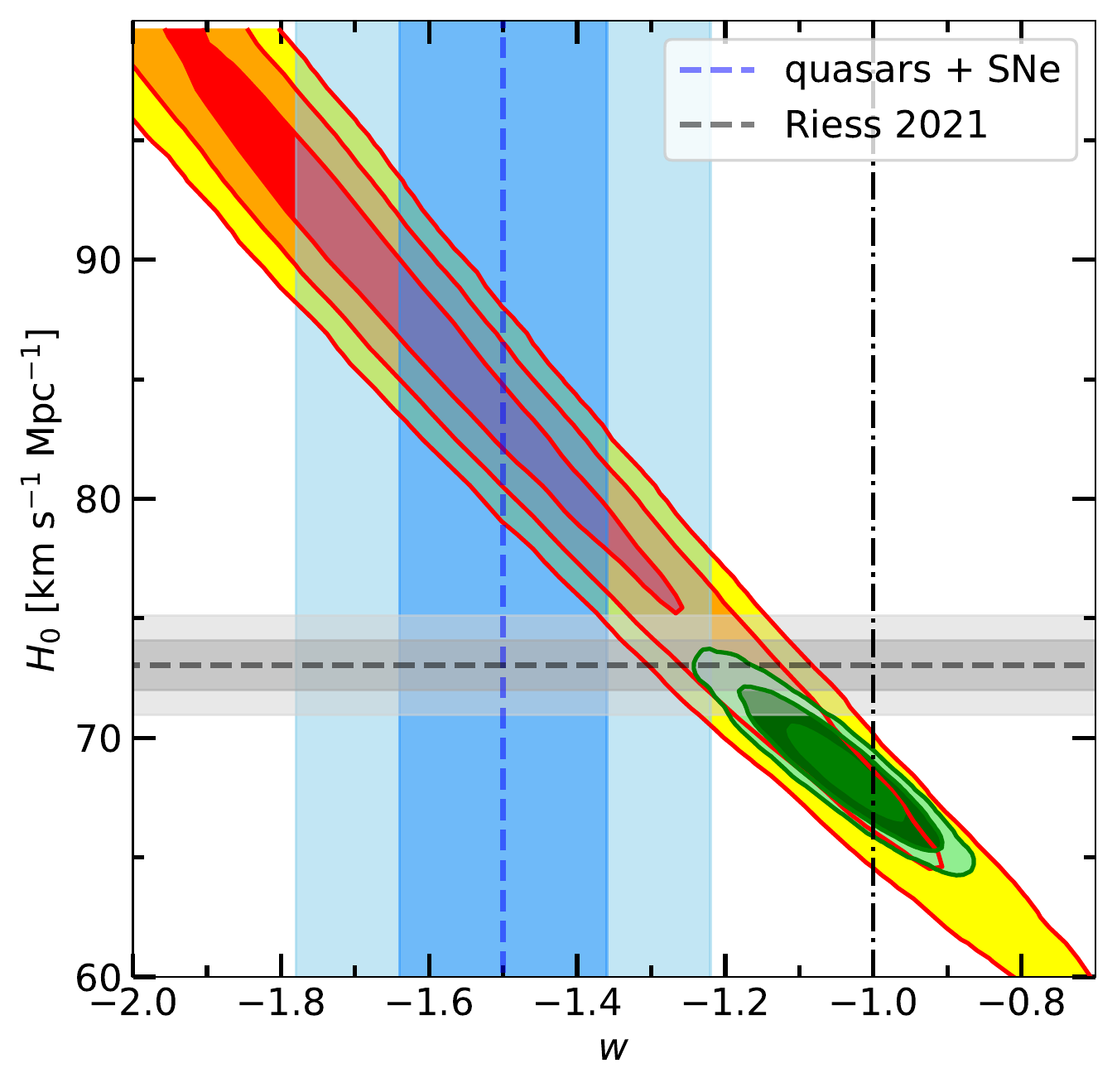}   
   \caption{\textit{Left panel}. Marginalized posterior distributions (1, 2 and 3$\sigma$) of the ($w_0$,$w_{\rm a}$) parameters for the combined quasars \citep{lusso2020} and supernovae Ia \citep{scolnic2018} samples (blue contours). The constraints from the combination of \textit{Planck} TT,TE,EE+lowE+lowl + BAO are also shown (green contours, \citealt{planck2018}). The dashed lines mark the point corresponding values to the $\Lambda$CDM model. The resulting ($w_0$,$w_{\rm a}$) for the combined quasars + SNe are statistically consistent with the phantom regime ($w<-1$) and at variance with the $\Lambda$CDM model at more than the $3\sigma$ statistical level. \textit{Right panel.} Marginalized posterior distributions (1, 2 and 3$\sigma$) of the (\Ho,$w$) parameters. The green contours are the same as the left panel, whilst the red, orange and yellow contours represents the constraints from \textit{Planck} TT,TE,EE+lowE+lowl only (i.e. excluding BAO). The dashed grey line marks the \Ho\ value resulting from the baseline model for the Cepheid–supernovae Ia sample along with the 1$\sigma$ and 2$\sigma$ uncertainty (i.e. \Ho$=73.04\pm1.04$ km s$^{-1}$ Mpc$^{-1}$;  \citealt{Riess2021b}). The dashed blue line marks the best fit $w$ value in a flat $w$CDM model for the combined quasars and supernovae Ia \citep[i.e. $w=-1.49\pm0.14$;][]{bargiacchi2021arXiv}.}
   \label{w0wa}
\end{figure}
The constraints on $w_0$ and $w_a$ in a $w_0w_a$CDM cosmological model combining the latest quasar and supernovae samples are shown in Fig.~\ref{w0wa}. The constraints from the combination of Planck18 \citep{planck2018} TT,TE,EE+lowE+lowl + BAO are also shown for reference. The dashed lines mark the point corresponding to the flat $\Lambda$CDM model for $w_0=-1$ and $w_a=0$. The resulting ($w_0$,$w_a$) for the combined quasars+SNe are statistically consistent with the phantom regime ($w<-1$) and at variance with the $\Lambda$CDM model at more than the $3\sigma$ statistical level. A summary of the cosmological fits to the combined quasar and supernovae samples are presented in Tab.~\ref{tab:qsosn}.
The detailed discussion of the cosmological implications of this deviation and its statistical significance is discussed at length by \citet{rl19,lusso19b}.
Figure~\ref{w0wa} also presents the marginalized posterior distributions for the \Ho\ and $w$ parameters. The red, orange and yellow contours represents the 1, 2 and 3$\sigma$ constraints from \textit{Planck} TT,TE,EE+lowE+lowl only (i.e. base $w$ model excluding BAO). The dashed grey line marks the \Ho\ value resulting from the baseline model for the Cepheid–supernovae Ia sample along with the 1$\sigma$ and 2$\sigma$ uncertainty (i.e. \Ho$=73.04\pm1.04$ km s$^{-1}$ Mpc$^{-1}$;  \citealt{Riess2021b}), whilst the dashed blue line marks the best fit $w$ value in a flat $w$CDM model for the combined quasars and supernovae Ia \citep[i.e. $w=-1.49\pm0.14$; see Table 2 by][]{bargiacchi2021arXiv}. As discussed in Section~\ref{sec:QSObasics}, the technique presented here is degenerate on \Ho\ but it provides constraints on the $w$ parameter.
Notably, CMB alone predicts high values on \Ho\ (i.e. \Ho$=87^{+8}_{-11}$ km s$^{-1}$ Mpc$^{-1}$) and $w$ constraints in the \textit{phantom} regime (i.e. $w=-1.6^{+0.3}_{-0.2}$), and it is only when BAO are included that \textit{Planck} becomes consistent with the concordance model (we refer the interested reader to the Section 7.4.1 by \citealt{planck2018}).

Concerning the deviation from the concordance model, \citet{bargiacchi2021arXiv} also presented a detailed analysis of BAO, SNe, and quasar data to understand their compatibility as well as their implications for extensions of the standard cosmological model. Specifically, they considered a flat and non-flat $\Lambda$CDM  cosmology, a flat and non-flat dark energy model with a constant dark energy equation of state parameter, and four flat dark models with variable $w$. They find that a joint analysis of quasars and SNe with BAO is only possible in the context of a flat Universe. BAO confirm the flatness condition assuming a curved geometry, whilst SNe+QSO show evidence of a closed space. They also find $\Omega_{\rm M}=0.3$ in all data sets assuming a flat $\Lambda$CDM model. Yet, all the other models show a statistically significant deviation at $2-3 \sigma$ with the combined SNe+quasars+BAO data set. In the models where the dark energy density evolves with time, SNe+QSO+BAO data always prefer $\Omega_{\rm M}> 0.3$, $w_0<-1$ and $w_{\rm a}>0$. They finally argue that this phantom behaviour is mainly driven by SNe+QSO, while BAO are closer to the flat $\Lambda$CDM model.
Recently, \citet{ss2022} have also presented a combined Type Ia SNe and quasar Hubble diagram in the redshift interval $z\simeq0.5-3.5$, by making use of a variability-absolute magnitude relation in quasar light curves. Their analysis seems to show a similar discrepancy with the $\Lambda$CDM at redshift higher than 2. Type Ia SNe at redshift $z=1-2$ also appear to show a similar trend (see Figure~6 in \citealt{bargiacchi2021}), although only $\sim$23 Type Ia SNe are currently observed in that redshift range. Future surveys that will target a higher number of Type Ia SNe at high redshift could provide compelling evidences that the discrepancy may indeed be confirmed by a completely independent method.

\renewcommand{\arraystretch}{1.2}
\begin{table}[t!]
\begin{center}
\begin{tabular}{|l|cccc|}
\multicolumn{5}{c}{{}}\\
\hline \hline
Model & \omegam & \omegade & $w_0$ & $w_{\rm a}$ \\
\hline
flat $\Lambda$CDM  & $0.295^{+0.013}_{-0.012}$ & -- & -- & -- \\
o$\Lambda$CDM  & $0.51^{+0.03}_{-0.03}$ & $1.10^{+0.05}_{-0.05}$ & -- & -- \\
flat $w_0-w_a$CDM & $0.45^{+0.02}_{-0.03}$ & -- & $-1.3^{+0.2}_{-0.2}$ & $-4.0^{+2.3}_{-2.7}$
\\
\hline \hline
\end{tabular}
\caption{Summary of the cosmological constraints for the combined quasars \citep{lusso2020} and supernovae Ia \citep{scolnic2018} sample for three different cosmological models: flat $\Lambda$CDM, open $\Lambda$CDM (o$\Lambda$CDM) and flat $w_0-w_a$CDM \citep[see][for more details]{bargiacchi2021arXiv}}
\label{tab:qsosn}
\end{center}
\end{table}
\renewcommand{\arraystretch}{1.}

With currently operating facilities, dedicated observations of well-selected high-$z$ quasars will greatly improve the test of the cosmological model and the study of the dispersion of the $L_{\rm X}-L_{\rm UV}$ relation, especially at $z\simeq4$ and beyond. 
The {\it extended Roentgen Survey with an Imaging Telescope Array} (eROSITA, \citealt{2012SPIE.8443E..1RP, 2012arXiv1209.3114M}), flagship instrument of the ongoing Russian {\it Spektrum-Roentgen-Gamma} (SRG) mission, will represent a powerful and versatile X-ray observatory in the next decade. The eROSITA sky will be dominated by the AGN population, with $\sim$3 million AGN with a median redshift of $z\sim1$ expected by the end of the nominal 4-year all-sky survey at the sensitivity of $F_{0.5-2\,\rm keV} \simeq 10^{-14}$ erg s$^{-1}$ cm$^{-2}$ and for which extensive multi-wavelength follow-ups are already planned. 
Concerning the constraints on the cosmological parameters (such as \omegam, \omegade, and $w$) through the Hubble diagram of quasars, the 4-year eROSITA (launched from Baikonur on July 13, 2019) all-sky survey alone, complemented by redshift and broadband photometric information, will supply the largest quasar sample at $z<2$ (average redshift $z\simeq1$). Nonetheless, a relatively small population should survive the Eddington bias cut at higher redshifts (see, e.g., \citealt{medvedev2020} for the highest redshift radio bright quasar), thus being available for cosmology as eROSITA samples the brighter end of the X-ray luminosity function \citep[][but see also section 6.2 in \citealt{comparat2020}]{lusso2020f}.
The large number of eROSITA quasars at $z\simeq1$ will be essential for both a better cross-calibration of the quasar Hubble diagram with supernovae and a more robust determination of \omegade, which is sensitive to the shape of the low redshift part of the distance modulus-redshift relation \citep[see Figure~2 in][]{lusso2020f}. In the mid and long term, surveys from {\it Euclid} (planned launch in 2023) and LSST (first light in July 2023, with the start of operations beginning of 2024) in the optical and UV, and {\it Athena} (currently in phase B1 study) in the X-rays, will also provide statistical samples of millions of quasars. 
With these datasets, it will be possible to obtain constraints on the observed deviations from the standard cosmological model, which will rival and complement those available from the other cosmological probes.

\clearpage

\subsection{Gamma-Ray Bursts}
\label{sec:GRB}

\begin{figure}[b!]
\centerline {\includegraphics[width=12cm]{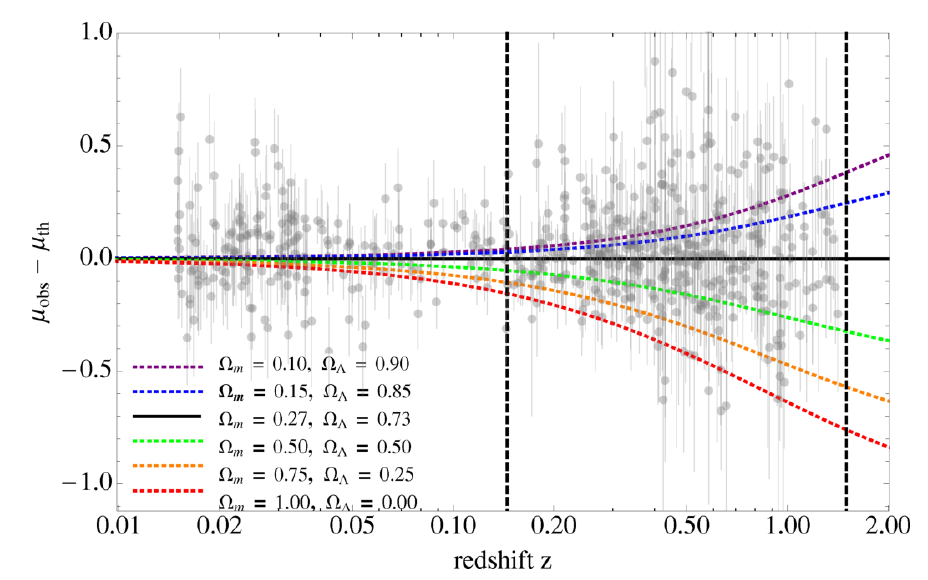}} 
\caption{Residual distance modulus for different values of the density cosmological parameters up to z = 2.0. We consider the best fit to be the standard $\Lambda$CDM model, where \omegam=0.27, \omegal=0.73, and \H0=71 \Hunit (black line). Union2 SNe Ia data residuals are shown in grey. The large spread (more than 1 mag) shown by $\mu$ at z = 1.5 and at z = 0.145 (the two vertical dashed lines) where the scatter is almost 0.2 mag is clearly evident. Image reproduced with permission from \cite{Izzo15}, copyright by ESO.} 
\label{izzo15}
\end{figure}

Observations of SNe Ia obtained at the end of 90’s by two different teams \citep{perlmutter1998,perlmutter1999,riess1998,schmidt1998} found that starting from $z\sim0.5$ SNe Ia appeared dimmer by $\sim$0.25 mag. Given the nature of {\it standard candles} of SNe Ia \citep{phillips1993} this result suggested that we are living in an Universe characterized by an accelerated expansion. In the following decades, other cosmological probes (e.g. CMB and BAO) provided further support to the existence of an unknown form of ``dark energy'' propelling the acceleration. 
By combining SNe data with the constraints from CMB measurements, several groups \citep[e.g.,][]{Riess:2004nr} found $w_0 \sim-1$ and $w_a \sim0$. This result might identify the dark energy as originated from a genuine cosmological constant. In subsequent years, new SN surveys have shown that the Hubble diagram does not exploit the growing number of SN discoveries (Fig.~\ref{izzo15}) in terms of the accuracy of cosmological parameter measurements. This is likely due to the fact that SN observations are affected by numerous sources of systematic effects, such as different classes of progenitor systems and different explosion mechanisms, anomalous reddening law, contamination of the Hubble diagram by non standard SNe Ia and/or bright SNe Ibc. Taking advantage from the existence of this ``systematic wall'' some authors \citep[e.g.,][]{nielsen2016} have questioned, on statistical basis, the evidence for cosmic acceleration from SNe Ia. In fact, SNe Ia detected in the Supernova Legacy Survey \citep[e.g.,][]{astier2006,guy2010} confirm the acceleration, although their measurements suggest different values for the cosmological parameters. The cosmological interpretation of SN Ia peaks decreased by 0.25 mag is based on the lack of evolutionary effects of their progenitors. 

Gamma-ray bursts (GRBs) are the brightest cosmological sources in the Universe, and detectable up to the first hundred millions years after the Big-Bang thanks to the enormous energy that they release in the X/gamma-rays (the isotropic radiated energy, $E_{\rm iso}$, can reach $ \simeq 10^{54}$ erg released typically in a few tens or hundreds of seconds). Their redshift distribution extends from 0.0085 (GRB 980425) up to $\sim9.4$  (GRB 090429B). In addition, they emit most of their radiation in the hard X-rays, so that they do not suffer for dust absorption. These phenomena are not standard candles, given that their total radiated energies or peak luminosities span several orders of magnitude, but the discovery and intensive study of empirical correlations between distance-dependent quantities and rest-frame observables  has opened to us the possibility of standardizing these sources as cosmological probes, and extend the Hubble diagram in a previously unexplored range of redshift. The use of GRBs for cosmology through the \epeiso~(``Amati'') relation and other correlations involving radiated energy or luminosity of the prompt and/or afterglow emission has been subject of tough investigation by many research groups worldwide since more than one decade ago. The main power of GRBs as cosmological probes lies in their huge brightness in the X- and soft gamma-rays domain, which makes them detectable up to redshift 10 or more, combined to the huge follow-up efforts in the optical/NIR follow-up, allowing the redshift measurement. This allows to extend the Hubble diagram substantially beyond the redshift range of Type-Ia SNe and even BAO, in a regime where only high-z AGNs can partly compete. As demonstrated by analysis and simulations, this redshift extension is fundamental for testing DE models and, more in general, cosmological scenarios alternative to the standard $\Lambda$CDM.
In the following, we describe three methods, based on gamma-ray bursts, to measure \omegam independently of SNe Ia, and to constrain the dark energy EoS aimed at describing the expansion history of the Universe.

\subsubsection{Basic idea and equations}

GRBs are very promising probes for investigating the history and evolution of the Universe, understanding the nature and evolution of dark energy, and testing alternative cosmological models. For recent general reviews on the GRB phenomenon, we refer to \citet{Meszaros02,Zhang14,Kumar15,Peer15}.
Although GRBs are not standard candles, as their peak luminosity and radiated energy span several orders of magnitude, some empirical correlations between distance-dependent quantities and rest-frame observables have opened up the possibility of using GRBs as distance indicators \citep[see, for instance,][]{Amati08,Amati13b,Lin15, Lin16a, Lin16b,Wei17,Si18,Fana19,Khadka20,Zhao20,Cao21,Khadka21,Cao22}. 
Actually, from a phenomenological point of view GRBs show a prompt emission, consisting of $\gamma$-rays and hard X-rays high-energy photons, and an afterglow emission, which is a long-lasting multi-wavelength emission from X-ray, to infrared and sometimes also radio, which follows the prompt emission and shows a typical power-law decay \citep[e.g.,][]{Gehrels09}. In addition, GRBs can be generally classified into short (with duration $T_{90}<2\,{\rm s}$, SGRBs) and long (with $T_{90}>2\,{\rm s}$, LGRBs \citealt{kouveliotou93}), where $T_{90}$ is the time interval in which $ 90\%$ of the GRB burst fluence is accumulated, starting from the time at which $5\%$ of the total fluence was detected. The classification is very important for standardizing GRBs since most of these correlations hold for long GRBs only. In Tab.~\ref{tab1} we list some of the correlations widely investigated in the literature, based on both prompt and afterglow emission properties (see references above for the definitions of the parameters mentioned in the Table).
 
Throughout this Section, we mostly focus on the $E_{\rm p ,i}$--$E_{\rm iso} $ correlation for measuring cosmological parameters and investigating dark energy properties and evolution.
In addition, as an example of the potentiality of combining prompt and afterglow emission properties, we will also discuss the perspectives for cosmology of the so called Combo-relation \citep{Izzo15}, obtained by combining  the $E_{\gamma,iso}$--$E_{X,iso}$--$E_{p,i}$, the $E_{\rm p ,i}$--$E_\gamma$ correlations, and the analytical formulation of the X-ray afterglow component given in \citep{Ruffini2014}.   

\begin{table*}
\begin{center}
\begin{tabular}{|c|c|}
\hline
{\bf Correlation} & {\bf Reference}\\
\hline
 $E_{\rm p ,i}$--$E_{\rm iso} $ & \cite{Amati02} \\
 $E_{\rm p ,i}$-- $E_\gamma$  &\cite{Ghirlanda04} \\
 $E_{\rm p ,i}$--$L_{\rm iso} $ &\cite{Yonetoku04} \\
 $L_{\rm peak}-\tau_{\rm lag}$& \cite{Azzam2012} \\
$L_{\rm iso}-V$ & \cite{Fenimore2000}\\
$L_{\rm iso}-E_{\rm p,i}-T_{0.45}$ & \cite{Firmani06}\\
 $L_{\rm iso}-E_{\rm p,i}-t_{\rm break}$&  \cite{Liang05}\\
$L_X-T_a$ & \cite{Dainotti08}\\
 $E_{X,iso}-E_{\gamma,iso}-E_{pk}$& \cite{Bernardini12}\\
 $E_{\gamma,iso}-E_{X,iso}-E_{pk}$& \cite{Izzo15}\\
\hline
\end{tabular}
\end{center}
\caption{List of the most investigated GRB correlations. } \label{tab1}
\label{tab-grb-correlations}
\end{table*}

\subsubsubsection{The $E_{\rm p ,i}$--$E_{\rm iso} $ (``Amati'') correlation}

GRBs show non thermal spectra which can be empirically modeled with the  Band function \citep{Band93}, which is a smoothly broken power law with parameters  $\alpha$, the low-energy spectral index, $\beta$, the high energy spectral index and the {\it roll-over} energy $E_0$:

\[N(E)=\left\{
\begin{array}{ll}
 A \left(\frac{E}{100keV}\right)^{\alpha} \exp{\left(-{\frac{E}{E_0}}\right)} & \left(\alpha-\beta\right)E_0\geq E \;\; ,\\
 A \left(\frac{\left(\alpha-\beta\right)E}{100keV}\right)^{\alpha-\beta} \exp{\left(\alpha-\beta\right)\left(\frac{E}{100keV}\right)^{\beta}} & \left(\alpha-\beta\right)E_0\leq E \;\; .\\
\end{array}
\right. \]

\begin{figure}
\centerline {\includegraphics[width=12cm]{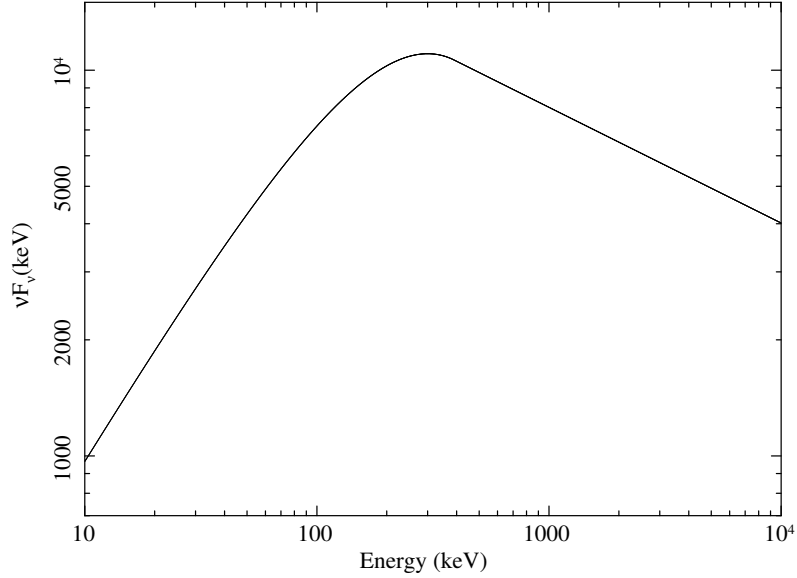}} 
\caption{A typical $\nu$F$_\nu$ spectrum of a GRB.} 
\label{grbspec}
\end{figure}

Given that $\beta$ is almost always found to be $<$-2, GRB $\nu$F$\nu$ spectra show a peak corresponding to a value of the photon energy $E_{\rm p} = E_0 (2 + \alpha)$ (Fig.~\ref{grbspec}), ranging typically from $\sim$5-10 up to 1000-5000 keV \citep[see, e.g.,][]{Zhang14}. For those GRBs with well measured prompt emission spectrum and redshift, it is possible to evaluate the ``intrinsic'' (i.e., in the cosmological rest-frame) peak energy, $E_{\rm p,i} = E_{\rm p} (1 + z)$ and the isotropic-equivalent radiated energy, defined as:
\begin{equation}
E_{\rm iso}= 4 \pi D_{\rm L}^2(z,{\mathrm \theta}) \left(1+z\right)^{-1}\int^{10^4/(1+z)}_{1/(1+z)} E N(E)
dE \;\;,
\label{eqEiso}
\end{equation}
or equivalently
\begin{equation}\label{eiso-sbolo}
E_{\rm iso}= 4 \pi D_{\rm L}^2(z,{\mathrm \theta}) \left(1+z\right)^{-1}S_{bolo}\,,
\end{equation}
being $S_{bolo}$ the bolometric fluence\,.

The quantity $E_{\rm iso} $ spans several orders of magnitude, typically ranging from 10$^{50}$ to 10$^{54}$ erg. It is important to note that, while there are observational and theoretical evidences suggesting that the GRB emission is collimated within a few tens of degrees or less, we are still lacking a firm and reliable method for estimating the jet opening angle of single GRBs. This is why, conservatively, $E_{\rm iso}$, or the isotropic-equivalent peak luminosity, $L_{\rm iso}$, are still used as indicators of the GRB ``brightness''.

The existence of a strong correlation between $E_{\rm p,i} $ and $E_{\rm iso}$ of long GRBs was inferred more than 20 years ago based on the systematic analysis of GRB spectra and fluences \citep{Lloyd00}, and was actually discovered in 2002 \citep{Amati02} based on the first sample of BepppoSAX GRBs with measured redshift. The \epeiso~(``Amati'') correlation was then confirmed by later measurements by several different GRB detectors and can be modeled as a linear relation between the logarithms of the two quantities:

\begin{equation}
\log \left[\frac{E_{\mathrm{p, i}}}{\mathrm{keV}}\right]=b+a \log \left[\frac{E_{\rm iso}}{10^{52}\;\mathrm{erg}}\right] \;\; ,
 \label{correlation_amati}
\end{equation}

\begin{figure}
\centerline{\includegraphics[width=12cm]{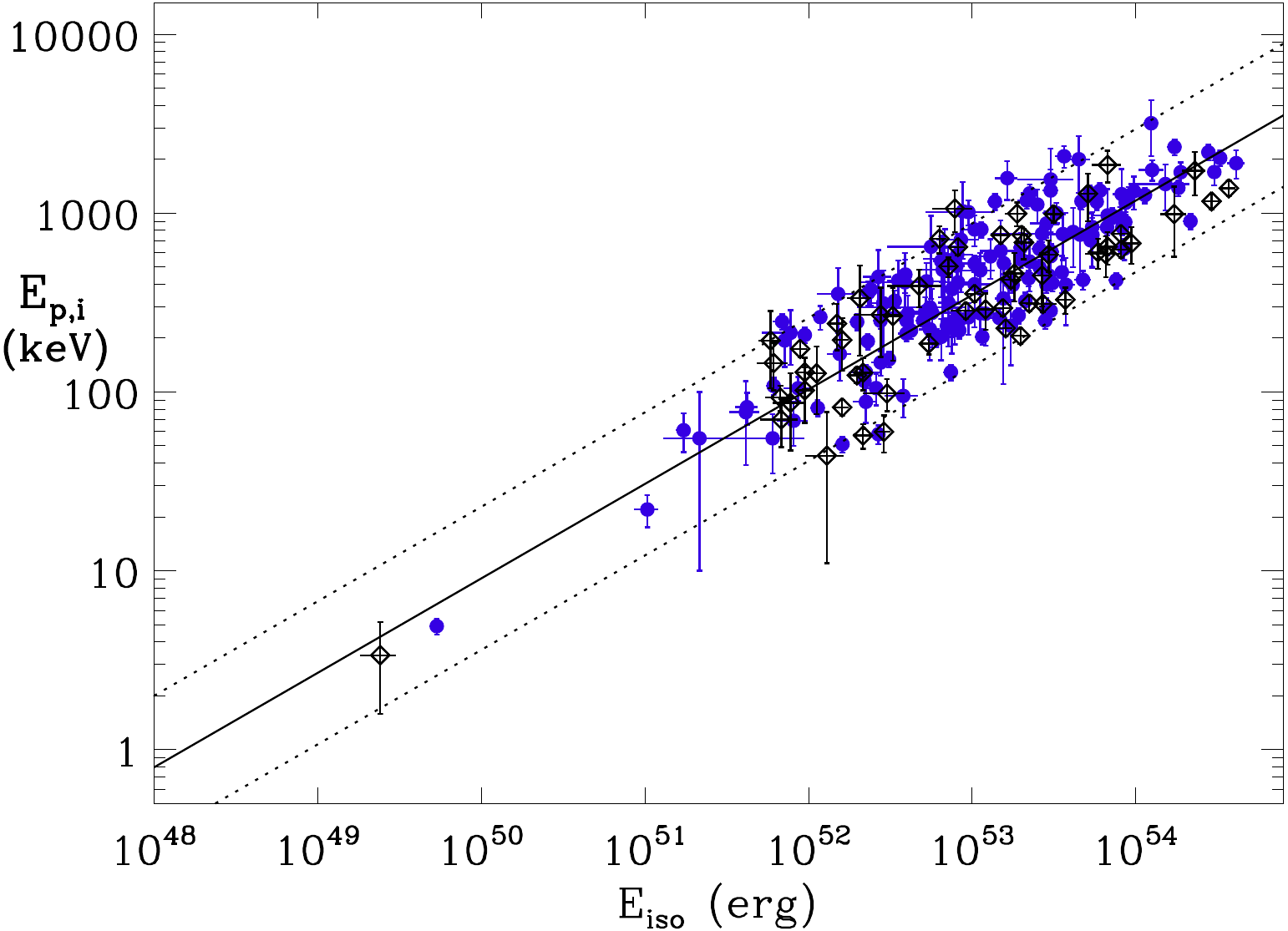}} 
\caption{The \epeiso~correlation for long GRBs based on the updates sample of 208 events used for this review. Blue points indicate GRBs detected and localized by the Swift satellite.}
\label{grbepeiso}
\end{figure}

The \epeiso~correlation (see Fig.~\ref{grbepeiso}) is characterized by an intrinsic additional extra-Poissonian scatter, $\sigma_{int}$, around the best-fit line that has to be taken into account and determined together with $(a, b)$ by the fitting procedure. A commonly used method is the maximization of the likelihood implemented by \citet{Reichart01}. According to this method the  data $\left(x_i , y_i \right)$ are correlated by a linear function $y = a x + b $ with the addition of an extrinsic scatter $\sigma_{int}$, and the best fit value of the parameters $\left(a, b and \sigma_{int} \right)$ are obtained by minimizing the -log(likelihood) function, in which the uncertainties,$ \sigma_{x,i}$ and $\sigma_{y,i}$ on both $\left(x_i , y_i \right)$  are taken into account. The general log(likelihood) is:
\begin{eqnarray}
\displaystyle \log\; \mathcal{L}_{\rm Reichart}(a, b, \sigma_{int})=-\frac{1}{2}\,\sum_{i=1}^N \Big[\log{\Big(\frac{1+a^2}{2 \pi (\sigma_y^2 + a^2\,\sigma_{x}^2 + \sigma_{y,i}^2 + a^2\,\sigma_{x,i}^2)}\Big)}-\frac{(y_i - a\,x_i - b)^2}{\sigma_y^2 + a^2\,\sigma_{x}^2 + \sigma_{y,i}^2 + a^2\,\sigma_{x,i}^2}\Big]\,,
\label{eqreich}
\end{eqnarray}
where $x={\rm log}(E_{\rm p,i})$ or $x={\rm log}(E_{\rm iso})$ (depending on whether one wants to investigate the correlation in the form  $E_{\rm p ,i}$--$E_{\rm iso}$ or $E_{\rm iso}$--$E_{\rm p ,i}$),   $\sigma_x$ = 0 and $\sigma_y =\sigma_{int}$. 
Here the sum is over the $N$ objects in the sample. We note that this maximization can actually be performed in the
two-parameter space $(a, \sigma_{int})$ only, since $b$ may be calculated analytically by solving the equation  $\displaystyle
{\frac{\partial }{\partial b}L(a, b, \sigma_{int})=0}$: 
\begin{equation}
b = \left[ \sum{\frac{y_i - a x_i}{\sigma_{int}^2 + \sigma_{y_i}^2
+ a^2 \sigma_{x_i}^2}} \right ] \left [\sum{\frac{1}{\sigma_{int}^2 + \sigma_{{y_i}}^2 + a^2 \sigma_{x_i}^2}} \right ]^{-1} \;\; . \label{eq:calca}
\end{equation}.

The values of the normalization, slope and intrinsic dispersion of the \epeiso~correlation in the logarithmic form expressed above are found to be $\sim$2, $\sim$0.5 and $\sim$0.2 dex, respectively, with slight variations depending on the sub-sample considered \citep[e.g.,][]{Amati02,Ghirlanda04,Amati06,Amati08,Amati13b,demianski17}. 

The existence and properties of this correlation have been widely investigated by many research groups in the last twenty years, because of its key role for the understanding of the GRB prompt emission physics, jet structure and geometry and viewing angle effects, as well as for the identification and nature of different sub-classes of these events, such as: short vs. long, X-Ray Flashes and under-luminous GRBs, ultra-long GRBs, etc. \citep[see, e.g., ][]{Zhang02, Amati06, Zhang14, Kumar15, Peer15}. 

\subsubsubsection{Independent measurements of cosmological parameters through the \epeiso~correlation of GRBs}\label{independent}

The ``Amati'' relation becomes a distance indicator through the measurement of $E_{\rm iso}$ that is derived from the observed fluence, which in turns depends on the geometry and expansion rate of our Universe through the so-called luminosity distance. 
Unlike historical ``standardized'' candles as SNe Ia that can be calibrated via Cepheids \citep[e.g.,][]{Riess:2021}, we don't have a statistically significant sample of GRBs at low redshift allowing us to determine the parameters of the correlation in a cosmology-independent way. Which means that the existence and properties of the correlation were found by assuming a fiducial cosmological model. Thus, if we wish to use it for measuring cosmological parameters we are obviously affected by a circularity problem. The most straight way to get rid of it is to simultaneously constrain the calibration parameters $(a, b, \sigma_{int})$ and the set of cosmological parameters by considering a chosen likelihood function. In practice, this task consists in determining the multi-dimensional probability distribution function (PDF) of the parameters $\{ a,b,\sigma_{\rm int} , {\mathbf p} \} $, where ${\mathbf p}$ is the $N$-dimensional vector of the cosmological parameters.

\begin{figure}
\centerline{\includegraphics[width=12cm]{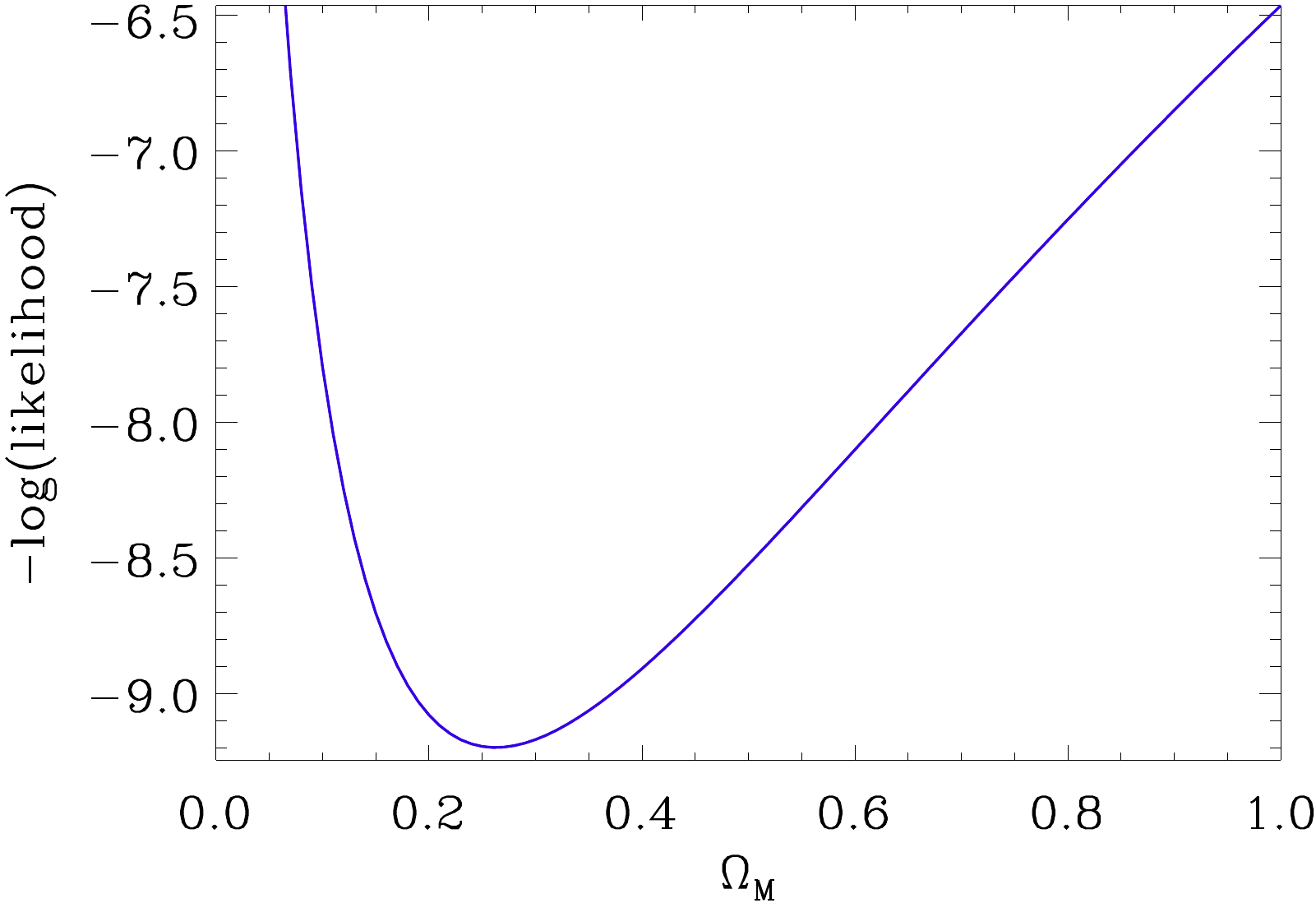}}
\caption{Goodness of fit (in terms of $-\log$(likelihood) of the \epeiso~correlation of long GRBs (based on the updates sample of 208 events used for this review) as a function 
of the value of \omegam~assumed in the computation of the $E_{\rm iso}$ values by assuming a flat $\Lambda$CDM cosmology.} 
\label{grb-omegam}
\end{figure}

This is the method adopted by \citet{Amati08} in the first work aimed at verifying if the \epeiso~correlation could actually be used for cosmology. By assuming a flat $\Lambda$CDM cosmology it was found that, actually, the goodness of fit of the correlation varied as a function of \omegam~following a nice parabolic shape with a minimum at about 0.2-0.3, as shown in Fig.~\ref{grb-omegam}. The analysis performed on larger samples in the following years made this result more and more reliable and accurate \citep[see, e.g., ][]{Amati13b}, showing that GRBs provide - in the framework of $\Lambda$CDM cosmology -  a firm and independent evidence for the case of an accelerating Universe with \omegam$\sim$0.3. This result is further confirmed by releasing the flat universe assumption, i.e., by letting both \omegam~and \omegal~free to vary (see next sections).

\subsubsubsection{Calibrating the \epeiso~correlation with SNe Ia and other probes}

In addition to the clean and independent approach described above, different and alternative techniques for getting rid of the circularity issue when using the \epeiso~($L_{\rm iso}$) correlation for cosmology have been developed and presented in literature \citep[see for instance][]{montiel21,Amati19,muccino21,Izzo15,Wang15, Liang08,kodama08,Wei10,Lin15}.

As anticipated, most of these methods use SNe Ia for calibrating the correlation for those GRBs at redshift lower than about 1.5 using the luminosity distances derived from SNe Ia \citep[see][]{kodama08, Liang08,demianski17}.
It is worth pointing out that the use of GRBs as distance indicators has an advantage over SNe Ia: they can explore a broader range of redshifts, extending to $z \sim 10$ instead of $z \sim 2$.
However, if we calibrate the GRBs with SNe Ia, the GRBs are no longer independent distance indicators. In the ``distance scale'' jargon, the GRBs have become ``tertiary'' indicators because in turn the SNe Ia are calibrated with the Cepheids. This is a different approach from that described in Sect.~\ref{independent}, where GRBs remain independent cosmological probes.

The typical regression procedure adopted in these approach can be schematically sketched as follows:
\begin{enumerate}

\item[1.]{set the redshift range where the modulus of distance, $\mu(z)$, has to be reconstructed; }

\item[2.]{sort the SNe Ia sample by increasing value of $|z - z_i|$ and
select the first $n = \alpha N_{\rm SNe Ia}$, where $\alpha$ is a user-selected value and $N_{\rm SNe Ia}$ the total number of SNe Ia;  }

\item[3.]{apply the weight function
\begin{equation}
W(u) = \left \{
\begin{array}{ll}
(1 - |u|^2)^2 & |u| \le 1 \\ ~ & ~ \\ 0 & |u| \ge 1
\end{array} \;\; ,
\right .
\label{eq: wdef}
\end{equation}
where $u = |z - z_i|/\Delta$ and $\Delta$ is the highest value of the $|z -z_i|$ over the previously selected subset;}

\item[4.]{fit a first-order polynomial to the data previously selected and weighted, and use the zeroth-order term as the best-fit value of the modulus of distance $\mu(z)$;}

\item[5.]{evaluate the error $\sigma_{\mu}$ as the root mean square of the weighted residuals with respect to the best-fit value. }

\end{enumerate}
Therefore, we  use the reconstructed $\mu(z)$ to obtain the luminosity distance, and fit the \epeiso~correlation relation (i.e. determine the parameters $(a, b, \sigma_{int})$) as expressed by Eq.~\ref{correlation_amati}, without assuming any particular cosmological model. We actually considered only GRBs with $z \le 1.414$ to cover the same redshift range spanned by the SNe Ia data.

\begin{figure}
\centerline{\includegraphics[width=12 cm]{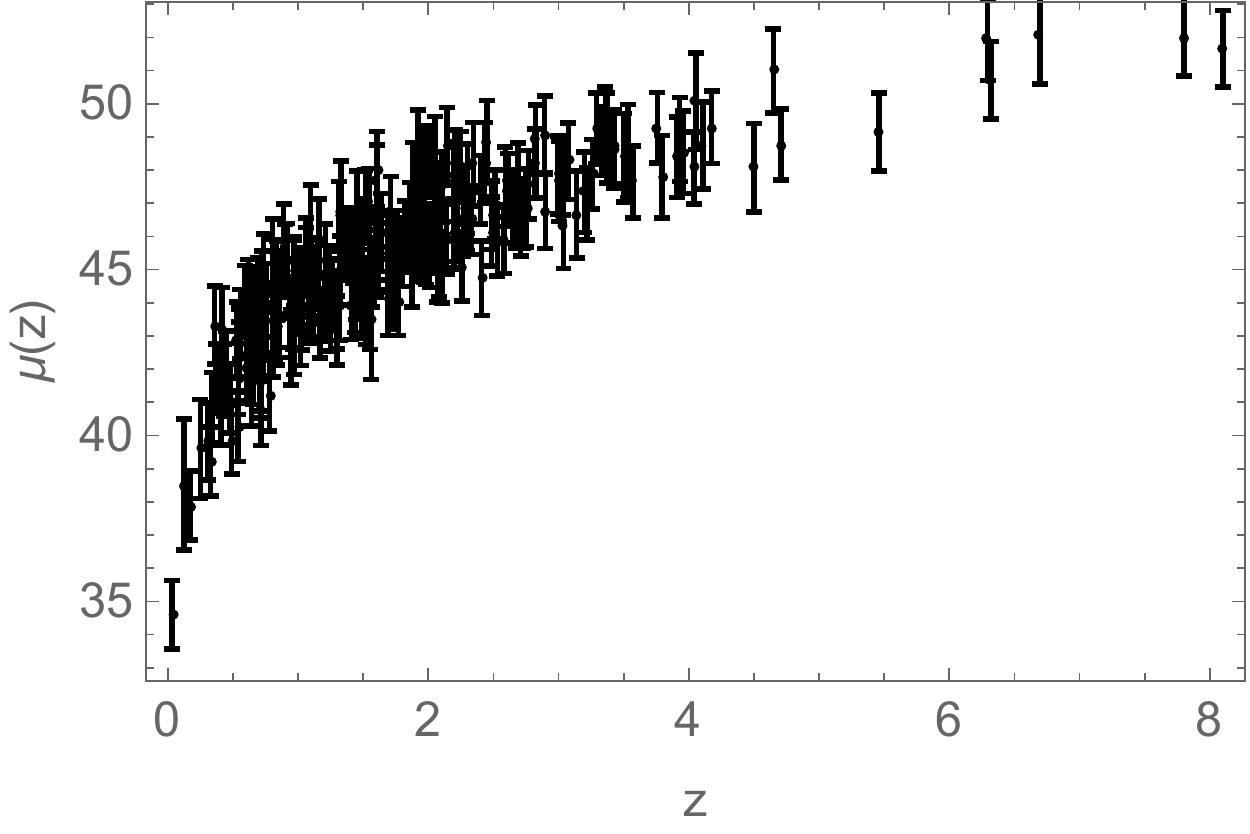}} 
\caption{GRB Hubble diagram build up by calibrating the E$ _{p,i}$--E$_{iso}$ correlation for the updated sample of 208 GRBs used for this review.} 
\label{GRBHD}
\end{figure}

After that, the values $(a, b)$ have been estimated through the calibration, and if we further assume that the \epeiso~correlation do not evolve with redshift, we obtain the energy $E_{\rm iso}$ of each burst at high redshift  through Eq.~\ref{correlation_amati}. We finally  obtain the luminosity distance, $D_{\rm L}(z)$, and construct the GRB Hubble diagram:
\begin{equation}\label{lumdist}
D_{\rm L}(z) = \left( \frac{E_{\rm iso}(1 + z)}{4 \pi  S_{bolo}}\right)^{1/2} \;\; .
\end{equation}
The uncertainty of $D_L(z)$ was estimated through the propagation of the measurement errors of the pertinent quantities. It turns out that
\begin{equation}
5 \log{D_L(z)} = \left(\frac{5}{2}\right)\left\{b+a\log
\left[\frac{E_{\mathrm{p,i}} }{300\;\mathrm{keV}}\right]-\log\left(4 \pi S_{bolo}\right)+\mu_0 \right\} \;\; ,
\end{equation}
where $\mu_0$ is a normalization parameter, due to the fact that the distance moduli of GRBs are not absolute; thus, this cross-calibration parameter is needed to match the GRB Hubble diagram and the one of SNe Ia \citep[see for instance][]{demianski21}. In Fig.~\ref{GRBHD} we plot the GRB Hubble diagram obtained for a new sample of 212 objects. 
It is worth noting that the calibration technique of the \epeiso~correlation, and its impact on reliability of the GRBs as distance indicators, deserves any possible attention, and for this reason in our analysis we tested the results also with different calibration techniques based on an approximated luminosity distance able to reproduce the exact function in different models \citep{demianski21}. Moreover, other approaches have been presented in literature, which exploit different interpolation of the luminosity distance and employ different data samples at intermediate redshifts \citep[including BAO datasets, see for instance][]{Amati19,muccino21}.
As anticipated, it turns out that the price of applying all these calibration techniques is that GRBs are a cosmological probe not fully independent. 
On the other hand, if we simultaneously constrain the calibration parameters $(a, b, \sigma_{int})$ and the set of cosmological parameters, it turns out that the parameters of the correlations depend on the cosmological model and are coupled to the cosmological parameters. 
When future GRB missions will substantially increase the number of GRBs available to construct the \epeiso~correlation, they may shed new light on the properties of this important correlation.

\subsubsection{Measurements and sample selection}

The use of GRBs for measuring cosmological parameters through the \epeiso~correlation, or other correlations involving the spectral peak energy  $E_{\rm p,i}$ (see, e.g., Tab.~\ref{tab-grb-correlations}), requires {\it i)} the measurement of the redshift, through either absorption spectroscopy of the optical/NIR afterglow spectrum or emission line spectroscopy of the host galaxy, and {\it ii)} the measurement of the prompt emission spectrum over a broad energy band and for most of the duration of the event, to allow an accurate characterization of the spectral continuum curvature. These combined requirements reduce the size of the sample from the several thousands of GRBs detected since the '70s to less than three hundreds nowadays. For instance, while GRB broad band spectroscopy from 10-20 keV up to a few MeVs was already available in the '80s and '90s, thanks, e.g., to CGRO/BATSE and Konus-WIND GRB detectors, it was possible to discover GRB afterglow emission and hence get the first redshift measurements only in the late '90s. On the other hand, the Swift mission, while providing great and fast localization of GRB prompt and afterglow emissions, thus substantially improving the efficiency in the follow-up process leading to redshift determination, is limited by the narrow energy band (15-150 keV) of its GRB detector.  

The samples used up to now for this line of investigation \citep[e.g.,]{Amati08,Amati13b,demianski17, Amati19, demianski21} include GRBs with measured redshift for which detection, localization, and spectral measurements come from the following main GRB missions: BATSE, BeppoSAX, HETE--2, Konus--WIND, Fermi/GBM, Swift/BAT. For this work, we consider a slightly updated sample wit respect to that used by \citet{Amati19, demianski21}, comprising a total number of 208 GRBs. This update is based on events for which redshift and spectral measurements became available in 2017 and 2018. A substantially updated sample including data form very recent Konus-WIND, Konus-WIND + Swift/BAT and Fermi/GBM spectral catalogs will be presented and analyzed in Amati et al. (in prep.), as well as in the next version of this review.

As discussed, e.g., in \citet{demianski17} and \citet{demianski21}, the criteria behind selecting the measurements from a particular mission are based on objective conditions aimed at minimizing selection and systematic effects (see also Sect.~\ref{sec:GRB_sys}):
\begin{itemize}
    \item given the broad energy band and good calibration, spectral measurements by Konus-WIND and Fermi/GBM are preferably chosen whenever available. The SWIFT BAT observations were chosen when no other preferred mission (Konus-WIND, Fermi/GBM) was able to provide information. They were considered only for GRBs with the  observed value of $E_{\rm p,i}$ within the energy band of the instrument. 
    \item in order to minimize biases due to event spectral evolution,and hence possible systematics on $E_{\rm p,i}$,
    only GRBs for which the exposure time was at least 2/3 of the whole event duration are selected (this condition is satisfied by about 80\% of the publicly available spectral catalogs); 
    \item those GRBs usually classified as ``under-luminous events'', for which there is significant possibility that their radiated energy, luminosity and spectral parameters are strongly biased by off-axis viewing effects or very long-to-soft spectral evolution \citep[see, e.g.,][]{Amati06,Martone17}, as well as being a different class of events with respect to classical cosmological long GRBs, are not included in the sample.
\end{itemize}

In the estimates of $E_{\rm p,i}$ and $E_{\rm iso}$, the values and uncertainties of all the observations are taken into account. When the observations were to be included in the data sample, it has been checked that the uncertainty on any value is not below 10 per cent in order to account for the instrumental capabilities. When the error was lower, it has been assumed to be 10\%, which is a reliable level of accuracy in the calibration of these kind of detectors. When available, the Band model \citep{Band93} was considered since the cut-off power-law tends overestimate the value of $E_{\rm p,i}$. 

\subsubsection{Systematic effects}
\label{sec:GRB_sys}

Given their relevance for shedding light on the emission processes, on the jet properties (e.g., structure, degree of magnetization), on the identification and understanding of different sub-classes of GRBs (long, short, under-luminous, ultra-long, GRB-SN connection), and as well for their great potential for GRB cosmology, the \epeiso~and other main correlations involving prompt and afterglow emission properties have been subject of many tough investigations aimed at identifying, understanding, and overcoming, possible selection effects and systematic (see, e.g., \citealt{Dainotti18} for an exhaustive review). 

\noindent
{\bf Reliability of the \epeiso~correlation.}
Different GRB detectors are characterized by different thresholds and spectroscopic sensitivity, therefore they can spread relevant selection effects and biases in the observed \epeiso~correlation. In the past, there were claims that a high fraction (70-90\%) of BATSE GRBs without redshift would be inconsistent with the correlation for any redshift \citep{Band05,Nakar05}.
However, this ``peculiar'' conclusion was refuted by other authors \citep{Ghirlanda05,Bosnjak08,Ghirlanda08,Nava11} who show that, in fact,  most BATSE GRBs with unknown redshift were well consistent with the \epeiso~correlation. We also note that the inconsistency of such a high percentage of GRBs of unknown redshift would have implied that most GRBs with known redshift should also be inconsistent with the \epeiso~relation, and this fact was never observed. Moreover, \cite{Amati09} showed that the normalization of the correlation varies only marginally using GRBs measured by individual instruments with different sensitivities and energy bands, while \cite{Ghirlanda10} show that the parameters of the correlations ($m$ and $q$) are independent of redshift.

Furthermore, the Swift satellite, thanks to its capability of providing quick and accurate localization of GRBs, thus reducing the selection effects in the observational chain leading to the estimate of GRB redshift, has further confirmed the reliability of the \epeiso~correlation \citep{Amati09,Ghirlanda10,Sakamoto11}.

Finally, based on time-resolved analysis of BATSE, BeppoSAX, and Fermi GRBs, it was found that the \epeiso~correlation also holds within each single GRB with normalization and slope consistent with those obtained with time-averaged spectra and energetic/luminosity \citep{Ghirlanda10,Lu12,Frontera12,Basak13}. This ultimate test confirms the physical origin of the correlation, also providing clues to its explanation.

\noindent
{\bf Possible evolutionary effects of the \epeiso~correlation.}
Possible evolutionary effects that may affect the correlation and have been investigated by several authors. By dividing the GRB sample into subsets with different redshift ranges (e.g., 0.1 $<$ z $<$ 1, 1 $<$ z $<$ 2, etc.), it is found that slope, normalization, and dispersion of the correlation do not change significantly. This result also implies that Malmquist–like selection effects are negligible.

In any case, to take into account possible evolutionary effects due, for instance, to the effects of local inhomogeneities distribution along the GRB line of sight \citep[see, for instance,][]{Shirokov20,demianski21}, it is also possible to consider a sort of {\it extended} \epeiso~correlation, introducing terms representing the redshift evolution, in the form of power-law functions: $g_{iso}(z)=\left(1+z\right)^{k_{iso}}$ and
$g_{p}(z)=\left(1+z\right)^{k_{p}}$, so that $E_{\rm iso}^{'}
=\displaystyle\frac{E_{\rm iso}}{g_{iso}(z)}$ and $E_{\rm p,i}^{'} =\displaystyle\frac{E_{\rm p,i}}{g_{p}(z)}$ are the
new fitting quantities \citep[see also][]{Shirokov20,demianski21}. In this approach, we consider a correlation with three parameters $a$, $b$, and $k_{iso} - ak_{p}$:
\begin{equation}
\log \left[\frac{E_{\rm iso}}{1\;\mathrm{erg}}\right] = b+a \log  \left[
    \frac{E_{\mathrm{p,i}} }{300\;\mathrm{keV}} \right]+\left(k_{iso} - a k_{p}\right)\log\left(1+z\right) \;\; .
\label{eqamatievol}
\end{equation}
The redshift dependence term in Eq.~\ref{eqamatievol} can be expressed by a single average coefficient $\gamma$:
\begin{equation}
\log \left[\frac{E_{\rm iso}}{1\;\mathrm{erg}}\right] = b+a \log  \left[
    \frac{E_{\mathrm{p,i}} }{300\;\mathrm{keV}} \right]+ \gamma\log\left(1+z\right) \;\; .
\label{eqamatievol2}
\end{equation}
To calibrate this 3D relation we have to fit the coefficients $a$, $ b$, $\gamma$, and the intrinsic scatter $\sigma_{int}$. It turns out that low values of $\gamma$ would indicate negligible evolutionary effects. Therefore it is possible to consider a 3D Reichart  general log(likelihood), which is:
\begin{equation}
\log\; \mathcal{L}^{3D}_{\rm Reichart}(a,  \gamma, b,  \sigma_{int}) = - \frac{1}{2} \sum{\log{\Big(\frac{(1+a^2)}{2\pi(\sigma_{int}^2 + \sigma_{y_i}^2 + a^2
\sigma_{x_i}^2)}\Big)}}\,-\frac{1}{2} \sum{\frac{(y_i - a x_i -\gamma z_i-b)^2}{\sigma_{int}^2 + \sigma_{x_i}^2 + a^2
\sigma_{x_i}^2}} \;\; .
\label{reich3dl}
\end{equation}
This likelihood can be maximized with respect to $a$ and $\gamma$, since $b$ can be evaluated analytically by solving the equation:
\begin{equation}
{\frac{\partial }{\partial b}L^{3D}_{\rm Reichart}(a, k_{iso},
\alpha, b, \sigma_{int})=0 \;\; .}
\end{equation}
Actually, it turns out that:
\begin{equation}
b = \left [ \sum{\frac{y_i - a x_i-\gamma z_i}{\sigma_{int}^2 + \sigma_{y_i}^2
+ a^2 \sigma_{x_i}^2}} \right ] \left [\sum{\frac{1}{\sigma_{int}^2 + \sigma_{{y_i}}^2 + a^2 \sigma_{x_i}^2}} \right ]^{-1} \;\;. 
\label{eq:calca1}
\end{equation}

\subsubsection{Main results and forecasts}

In this section we show and discuss the current and perspective potentiality of the three methods described above for using GRBs as probes of the expansion rate and geometry of the Universe. The main results and forecasts reported are based on the partially updated sample of 208 GRBs described above and a sample of 208 real + 292 simulated GRBs which may be expected from future dedicated space missions, as described below (giving a sample of 500 GRBs in total), respectively. The latter sample was produced following the procedure and assumptions detailed in \citet{Amati13b}.

\subsubsubsection{GRBs as independent probes}

\begin{figure}
\centerline{\includegraphics[width=8cm]{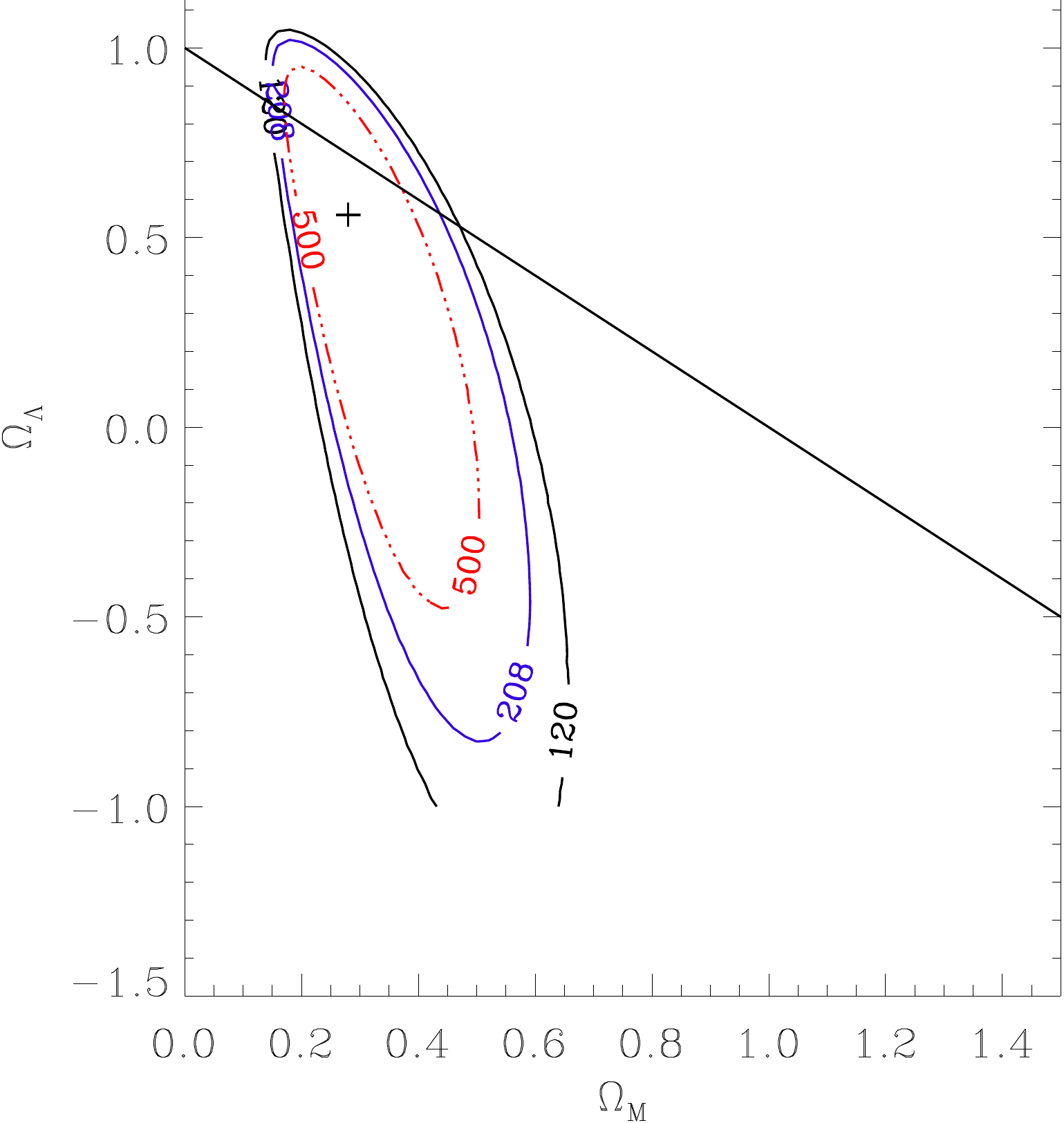}\hspace{1cm}\includegraphics[width=8cm]{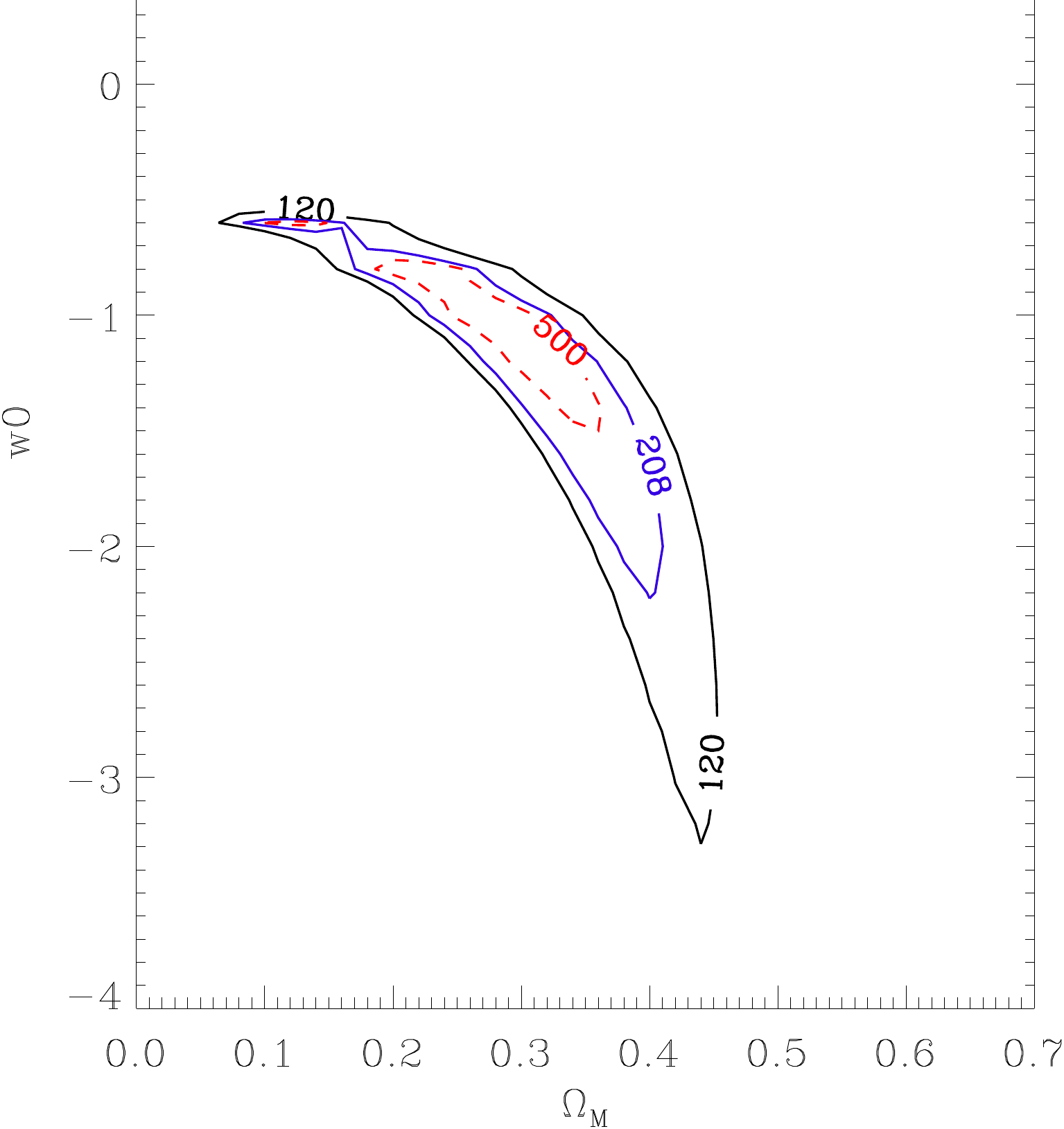}}
\caption{Left: 68\% confidence level contour in the \omegam-\omegade~plane obtained by releasing the flat universe assumption with the sample of 208 GRBs considered in this work (red contour) compared to those obtained with a sub-sample of 120 GRBs and what expected in the next years with the increasing of GRBs in the sample (500 GRBs, blue). Right: 68\% confidence level contour in the \wzero-\omegam~plane for a flat FLRW universe with \omegam=0.3 obtained for the same samples as for the left panel. As for the results and simulations reported in Tab.~\ref{ta1}, for the dark energy equation of state $w_a=0.5$ was assumed.}
\label{grb-de}
\end{figure}

In Tab.~\ref{ta1}, we show the 68\% confidence level intervals for \omegam~and \wzero~in a flat FLRW universe derived with the 70 GRBs of \cite{Amati08}, the partially updated sample of 208 GRBs and the partially simulated sample of 500 GRBs  
These values were obtained with the same approach as \cite{Amati08}, but using the likelihood function proposed by \citep{Reichart01}, which has the advantage of not requiring the arbitrary choice of an independent variable among \epi~and \eiso. Interesting enough, we note that, after increasing the number of GRBs from 70 to 156, the accuracy of the estimate of \omegam~improves by a factor of $\sim \sqrt{N_2/N_1}$. The accuracy of these measurements is still lower than that obtained with supernova data, but promising in view of the increasing number of GRBs with measured redshift and spectra (see also Fig.~\ref{grb-omegam}, Fig.~\ref{grb-de}, and Sect.~\ref{sec:GRB_SNe}).

\begin{table}[b!]
{\centerline{
\begin{tabular}{|lcc|}
\hline
 Number of GRBs & \omegam & \wzero \\
 &  (flat)  & (flat, \omegam=0.3,\wa=0.5)  \\
\hline
 70 (real) GRBs \citep{Amati08} & 0.27$_{-0.18}^{+0.38}$  & $<$$-$0.3 (90\%)   \\
 208 (real) GRBs (this work)  & 0.26$_{-0.12}^{+0.23}$  & $-$1.2$_{-1.1}^{+0.4}$  \\
 500 (208 real + 292 simulated) GRBs &  0.29$_{-0.09}^{+0.10}$  & $-$0.9$_{-0.8}^{+0.2}$ \\
\hline
 208 (real) GRBs, calibration &  0.30$_{-0.06}^{+0.06}$ &  $-$1.1$_{-0.30}^{+0.25}$ \\
 500 (208 real + 292 simulated) GRBs, calibration &  0.30$_{-0.03}^{+0.03}$ & $-$1.1$_{-0.15}^{+0.12}$ \\
\hline
\end{tabular}
}}
\caption{Comparison of the 68\% confidence intervals on \omegam{} and \wzero{}
(\omegam=0.3, \wa=0.5) for a flat FLRW
universe obtained with the sample of 70 GRBs \citet{Amati08}, the
updated sample of 208 GRBs considered in this work and
simulated sample of 500 GRBs (see text).  In the last two lines we also show the results obtained
for the same samples by assuming that the slope and normalization of the \epeiso~correlation
are known with a 10\% accuracy based, e.g., on calibration against SNe Ia or self-calibration
with a large enough number of GRBs at similar redshift.}
\label{ta1}
\end{table}

In the last 3 lines of Tab.~\ref{ta1}, we report the estimates of \omegam~and \wzero~derived from the present and expected future samples by assuming that the \epeiso~correlation is calibrated with a 10\% accuracy by using, e.g., the luminosity distances provided by SNe Ia, GRBs self-calibration, or the other methods shortly described below.
The perspectives of this method for improving estimates of \omegam and the investigation of the properties of dark energy, combined with the expected increase of the number of GRBs in the sample, are shown in Fig.~\ref{grb-de}. In particular, as an example, we are showing the current and expected accuracy on \wzero~in case of an evolving dark energy with \wa $\sim$0.5.

It is important to note that, as the number of GRBs in each $z$-bin increases, also the feasibility and accuracy of the self-calibration of the \epeiso~correlation will improve. Thus, the expected results shown in the last part of Tab.~\ref{ta1} and in Fig.~\ref{grb-de} may be obtained even without the need of calibrating GRBs against other cosmological probes.

The results presented in Tab.~\ref{ta1} show a sharp increase of the accuracy of \omegam~as a consequence of the increasing number of GRBs in the \epeiso~plane. Currently, the main contribution to enlarge the GRB sample comes from joint detections by Swift, Fermi/GBM or Konus-WIND. Hopefully, these missions will continue to operate in the next years, then providing us with an ``actual'' rate of $\sim$15-20 GRB/year. However, a real breakthrough in this field should come from next generation missions capable of promptly pinpointing the GRB localization and of carrying out broad-band spectroscopy. We build our hopes on the Chinese-French mission SVOM \citep{Cordier19}, for the very near future, and on mission concepts like THESEUS \citep{Amati18} for the next decade.

In Fig.~\ref{grb-de} we show the confidence level contours in the \omegam-\omegade~and \omegam-\wzero~planes  by using the real data, and by adding to them the 292 simulated GRBs (resulting a sample of 500 GRBs in total, respectively). The simulated dataset was obtained via Monte Carlo techniques by taking into account the slope, normalization and dispersion of the observed \epeiso~correlation, the observed redshift distribution of GRBs and the distribution of the uncertainties in the measured values of \epi~and \eiso.
These simulations indicate that with a sample of 500 GRBs (achievable within a few years from now) the accuracy in measuring \omegam~will be comparable to that currently provided by SNe data.

\subsubsubsection{Use of GRBs calibrated against SNe Ia}
\label{sec:GRB_SNe}

\begin{figure}
\centerline 
\mbox{\includegraphics[width=0.485\linewidth]{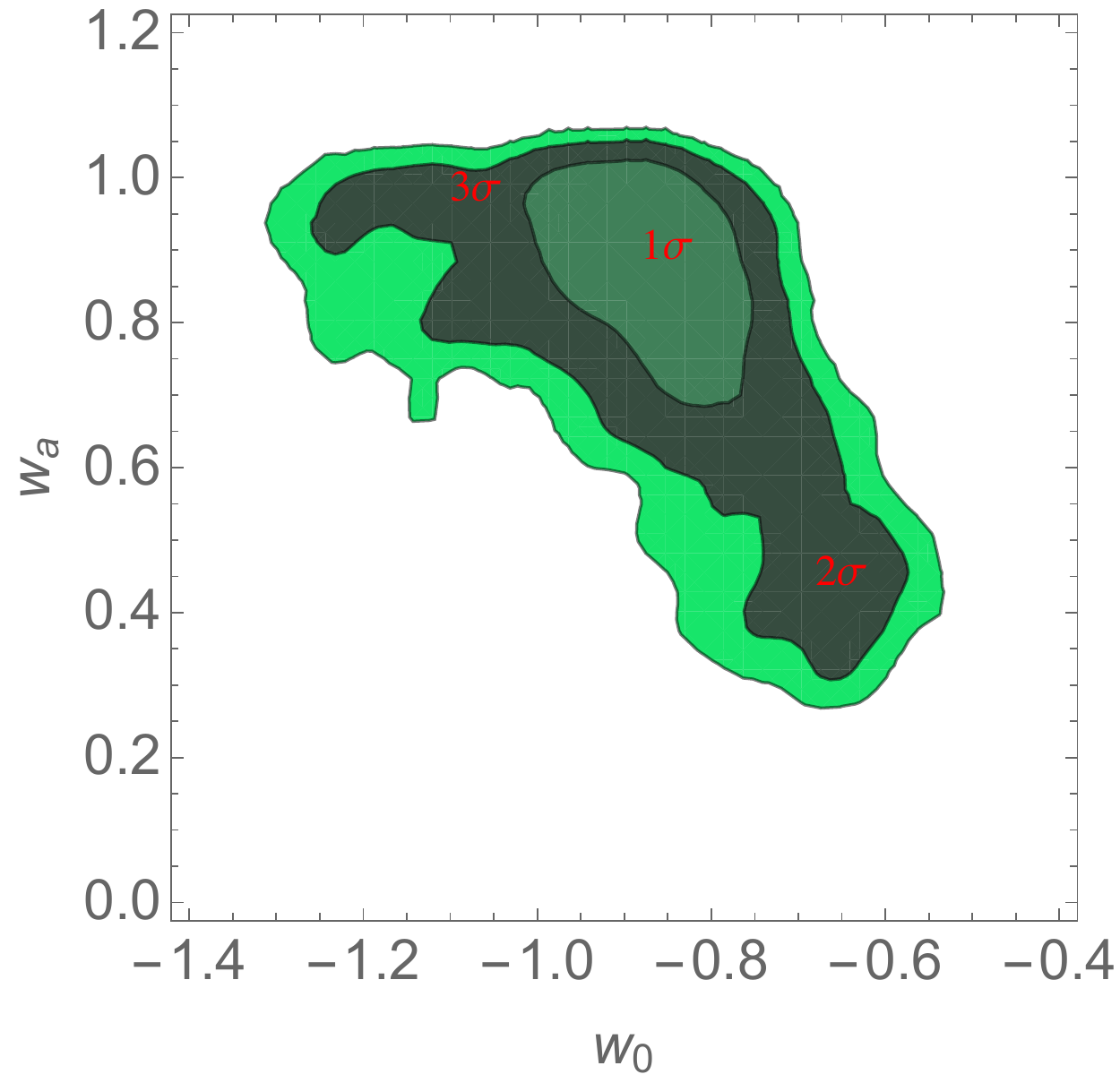}
\includegraphics[width=0.47\linewidth]{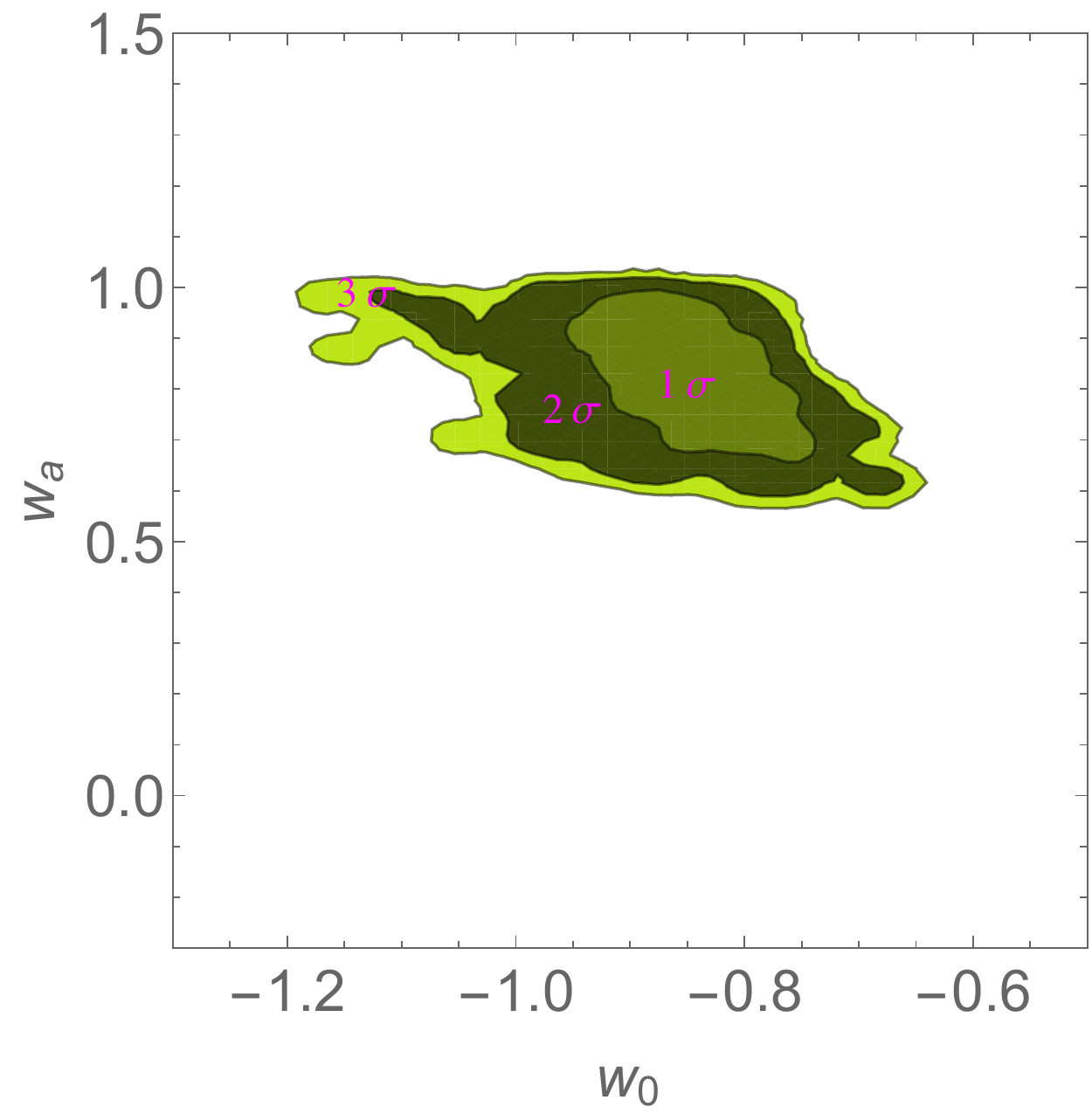}} 
\caption{2D confidence regions in the $w_0-w_a$ plane for the CPL model, obtained from a simulated (right panel) and real (left panel) GRBs Hubble diagram.} 
\label{plotcpl}
\end{figure}

To test different cosmological models, we use a Bayesian approach based on the MCMC method. In order to set the starting points for our chains, we first performed a preliminary and standard fitting procedure to maximize the likelihood function ${\mathcal{L}}({\bf p})$. We sample the space of parameters by running five parallel chains and use the Gelman-Rubin diagnostic approach to test the convergence. As a test probe, it uses the reduction factor $R$, which is the square root of the ratio of the variance between-chains and the variance within-chain. A large $R$ indicates that the between-chains variance is substantially greater than the within-chain variance, so that a longer simulation is needed. We require that $R$  converges to 1 for each parameter. We set $R - 1$ of order $0.05$, which is more restrictive than the often used and recommended value $R - 1 < 0.1$ for standard cosmological investigations. After that, we ran multiple chains in parallel, we discarded the first $30\%$ of the point iterations, and finally extracted the constrains on cosmological parameters by co-adding the thinned chains. The histograms of the parameters from the merged chains were then used to infer median values and confidence ranges. 
As a simple example, let us consider the CPL parameterization of the dark energy EoS described in Eq.~\ref{eq:de_CPL}. In Fig.~\ref{plotcpl} we plot the 2D confidence regions in the $w_0-w_a$ plane for the CPL model, obtained from real (upper panel) and a simulated (bottom panel) GRBs Hubble diagram.
 
We join our sample of 208 GRBs to a simulated sample of 792 objects. These simulated data have been obtained by implementing a  Monte Carlo approach and taking into account the slope, normalization, dispersion of the observed \epeiso~correlation. It is worth noting that the $\Lambda$CDM model, which in the CPL parameterization corresponds to $w_0=-1$\, and $w_a=0$, is disfavoured with respect to a dynamical model of dark energy.

\subsubsubsection{The ``Combo'' relation: shedding light on the evolution of dark energy}
\label{sec:GRB_combo}

As discussed by \citet{Izzo15} and \citet{muccino21}, an important step forward in this line of investigation may be provided by the use of the ``Combo'' relation, which extends the ``Amati'' relation through the inclusion of X-ray afterglow observables like the initial luminosity, the rest-frame duration of the shallow phase, and the index of the late power-law decay, combined with an innovative calibration method minimizing the dependence on the systematics possibly affecting SNe Ia. The main novelty provided with the Combo relation consists in the afterglow X-ray light-curve fitting procedure through a piece-wise function, first introduced by \citet{Willingale2007}, that is capable to model the very early power-law decay and the following ``plateau'' emission \citep{Izzo15}, getting rid of X-ray flaring emission over-imposed to the underlying afterglow behavior \citep{Zaninoni2014}. This procedure, similar to the analysis currently developed for SNe Ia, allows to measure with great accuracy the main observables of the Combo relation: indeed, among the entire sample of Swift long GRBs showing a complete light curve in X-ray, and characterized by a known peak energy of the corresponding prompt emission, no outliers have been found so far \citep{muccino21,Xu2021,Wang2021}.

\begin{figure}
\centerline{\includegraphics[width=1.0\linewidth]{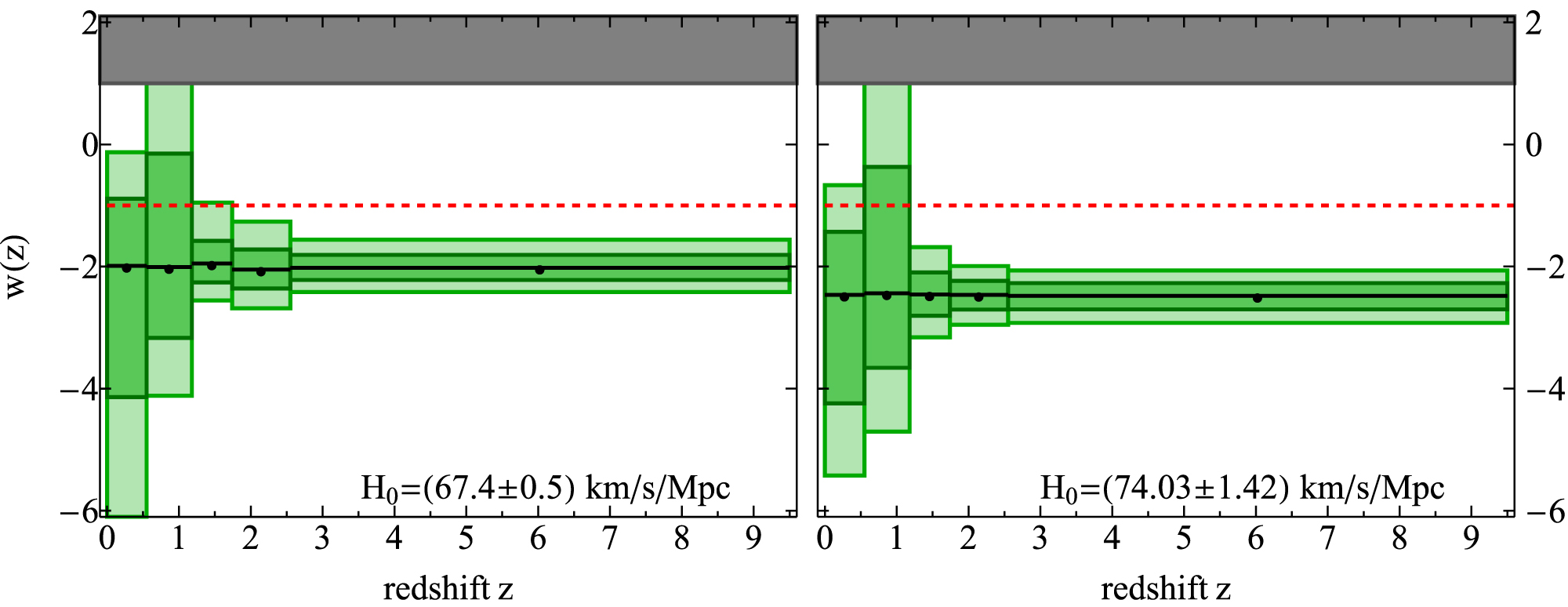}}
\caption{The DE EoS reconstructed evolution through the redshift-binned parameterization of $w(z)$ (1 and 2$\sigma$ from the inner/darker to the outer/lighter) for the selected \Ho. The dashed red lines mark the value $w=-1$ in the flat $\Lambda$CDM model. The darker region shows un-physical EoS, i.e., exceeding the stiff matter regime. Image reproduced with permission from \cite{muccino21}, copyright by the American Astronomical Society.}
\label{fig-muccino20-de}
\end{figure}

In a preliminary analysis on a sample of 60 GRBs with well measured parameters of both prompt and early X-ray afterglow emission, \citet{Izzo15} showed that actually the Combo relation could provide a value of \omegam= 0.29$_{-0.15}^{+0.23}$. By applying the Combo relation to an updated sample of 174 gamma-ray bursts, \citet{muccino21} could obtain tighter bounds on \omegam, and investigate the possible evidence of evolving dark energy parameter $w(z)$. As shown in Fig.~\ref{fig-muccino20-de}, the $w(z)$ evolution was studied by binning the GRB Hubble diagram in seven redshift intervals and assuming two priors over the Hubble constant in tension at 4.4$\sigma$, i.e., H0 = (67.4$ \pm$ 0.5) \Hunit and H0 = (74.03 $\pm$ 1.42) \Hunit. It was found that at $z\leq1.2$ $w(z)$ agrees within 1$\sigma$ with the standard value $w=-1$, whereas at larger z the $w(z)$ estimated from GRBs seem to deviate from $w=-1$ at 2$\sigma$ and 4$\sigma$ level, depending on the redshift bins (Fig.~\ref{fig-muccino20-de}). These results indicate that dark energy equation of state parameter can be different from the $\Lambda$CDM value $w = -1$ at larger $z$, although its contribution to the energy budget of the Universe is still negligible\footnote{Indeed, it is worth noting that the dependence of the luminosity distance -- that is the cosmological observable involved in the GRB datasets -- on the dark energy equation of state parameters is mediated by a redshift integral, which introduces a kind of cumulative effect, and partially compensates for the decreasing contribute of the dark energy to the total energy budget at high redshifts.}, and also confirm the Combo relation as a powerful tool to investigate cosmological evolution of dark energy.

In view of the increasing size of the GRB database, thanks to incoming future missions, the Combo-relation is a promising tool for measuring \omegam~with an accuracy comparable to that exhibited by SNe Ia, and to investigate a possible evolution of the dark energy up to $z\sim10$.

\subsubsubsection{The promises of correlations involving $L_X$ and $T_a$}

As discussed at the beginning of this section, and shown in Tab.~\ref{tab-grb-correlations}, the quest for correlations between GRB properties, aimed at shedding light on the emission processes and at enabling the use of these phenomena for measuring cosmological parameters, involved not only the X/Gamma-ray prompt phase but also the early X-ray afterglow emission. Among these, the most investigated are those involving the duration of the ``plateau'' phase, $T_a$ and the luminosity at the end of this phase, usually referred to as $L_X$. Indeed, as shown and discussed by several authors (e.g. \citealt{Cardone09, Dainotti20} and references therein, \citealt{Hu21}), there exists a significant correlation between these two quantities, as well as a 3D correlation obtained by including the peak luminosity of the prompt emission, $L_p$. In particular, it has been found that these correlations become tight for sub-samples selected based on other characteristics, including the nature of the progenitor and multi-wavelength properties. This method, while still affected by the relatively low number of events that can be used for each sub-sample and sample selection effects, seems promising for the purpose of GRB cosmology, especially in view of the wealth of new data on GRB prompt and afterglow emission expected in the near future thanks to the continuing operation of Swift, Fermi, Konus-WIND and other GRB experiments, as well as increased efficiency of follow-up with ground facilities.

\clearpage

\subsection{Standard Sirens}
\label{sec:ss}

As first pointed out by~\citet{1986Natur.323..310S}, merging black holes and neutron stars, when observed in gravitational waves (GWs), can serve as powerful cosmological probes~\citep{2005ApJ...629...15H,2006PhRvD..74f3006D}. These merging binaries emit GW signals that directly encode the luminosity distance to the binary $D_{\rm L}$, calibrated by the theory of general relativity. There are three primary approaches to standard siren cosmology: ``bright'', ``dark'', and ``spectral'' sirens, as detailed below. LIGO \citep{2015CQGra..32g4001L}, Virgo \citep{2015CQGra..32b4001A}, and KAGRA \citep{2020arXiv200505574A} are observing a growing catalog of gravitational-wave events, with hundreds to thousands more detections expected in the coming years. Most standard siren measurements to date have relied on the closest standard sirens, with luminosity distances $D_L \lesssim 400\ \mathrm{Mpc}$, and thus probe the local distance-redshift relation through the Hubble constant \Ho. However, these analyses are starting to take advantage of the full gravitational-wave catalog, which extends to $D_L \gtrsim 5\ \mathrm{Gpc}$ with the current LIGO and Virgo detections, and will extend past 10 Gpc with upgrades to the gravitational-wave detector network over the next few years. Standard sirens are therefore starting to provide measurements of the expansion history out to $z > 1$ in addition to measuring the Hubble constant. Furthermore, standard sirens are unique probes of modified gravitational wave propagation, a prediction of many cosmological modified gravity and dark energy theories. 

\subsubsection{Basic idea and equations}

When two compact objects, such as black holes and/or neutron stars, orbit each other, the time-varying mass quadrupole sources space-time perturbations, or GWs. At sufficiently tight orbital separations, the energy and angular momentum radiated by GWs shrinks the orbit until the two objects merge, forming a bigger black hole or neutron star. Such sources of GWs are known as ``compact binary coalescences''. A passing GW signal stretches and squeezes space-time, creating a relative change in length $\Delta L / L$, known as the \emph{strain} $h$, or GW amplitude. The typical strain for a GW signal sourced by a compact binary coalescence is $10^{-21}$. This stretching and squeezing of space-time happens at a certain frequency. The frequency of a GW from a compact binary coalescence is twice the orbital frequency, and it therefore evolves with time as the orbit shrinks. The frequency evolution is driven by a combination of the masses of the two compact objects known as the \emph{chirp mass}.

For compact binary coalescences, the GW strain as a function of time $h(t)$ scales inversely with the luminosity distance $D_L$. To first order:
\begin{equation}
\label{eq:hoft}
h(t) = \frac{\mathcal{M}_z^{5/3}f(t)^{2/3}}{D_L}F(\mathrm{angles})\cos(\Phi(t)) \;\; ,
\end{equation}
where $f(t)$ is the GW frequency, $F(\mathrm{angles})$ is a function of the source's position on the sky, inclination and polarization, and $\Phi(t)$ is the orbital phase.
The ``intrinsic loudness'' of the GW depends on the redshifted chirp mass $\mathcal{M}_z$:
\begin{equation}
    \mathcal{M}_z = (1 + z) \frac{(m_1 m_2)^{3/5}}{(m_1 + m_2)^{1/5}} \;\; ,
\end{equation}
for binary component masses $m_1$ and $m_2$, measured in the source-frame; the factor of $(1+z)$ converts between source-frame and detector-frame quantities. 
Interestingly, this same combination of masses governs the GW frequency evolution, $f(t)$ and its derivative $\dot{f}(t)$:
\begin{equation}
    \mathcal{M}_z = \left( \frac{5}{96}\pi^{-8/3}\left( f(t) \right)^{-11/3}\dot{f}(t) \right)^{3/5} \;\; ,
\end{equation}
so that by measuring both the amplitude and frequency evolution of the GW signal, the luminosity distance can be derived. Note that the amplitude also depends on source geometry encoded in $F(\mathrm{angles})$. For example, a face-on binary will emit a louder GW signal than an edge-on binary. We also see that while the cosmological redshift $z$ affects the measured GW frequency, this effect is degenerate with the binary's mass; only redshifted masses appear in the equations describing the amplitude and frequency of the GW signal. In order to do cosmology with GW sources, we must identify external sources of redshift information. Matching GW source distances with their redshifts allows us to probe the cosmological parameters with the usual distance--redshift relation: 
\begin{equation}
D_{\rm L}= c (1+z) \int_0^z\frac{dz'}{H_0 E(z')}
\end{equation}
We discuss methods for measuring the redshift of GW sources in the following Sect.~\ref{sec:ss-samplesel}.

\subsubsection{Sample selection}
\label{sec:ss-samplesel}

In order to use GW sources as cosmological indicators and standard sirens, the required ingredients are {\it i)} estimating the GW distances, and {\it ii)} assigning redshifts to the GW sources.

{\bf Gravitational-wave distances.} Every GW detection of a compact binary coalescence provides a measurement of the source's luminosity distance. For a given source, the accuracy of the GW luminosity distance measurement is typically $\mathcal{O}(10\%)$, depending on the parameters of the source and its signal-to-noise ratio. For some systems, the distance constraints are much tighter because the distance-inclination degeneracy, which stems from the $F(\mathrm{angles})/D_L$ factor in Eq.~\ref{eq:hoft}, can be broken. This occurs for binaries with misaligned spins leading to measurable orbital precession and binaries with asymmetric mass ratios that emit measurable higher-order GW harmonics \citep{2018PhRvL.121b1303V,2020ApJ...896L..44A,2020arXiv200702883B,2021ApJ...912L..10C}. Occasionally, electromagnetic observations of the same source (for example, observations of beamed emission from binary neutron star mergers) can be used to independently measure the source inclination, resulting in a tighter GW distance measurement \citep{2018Natur.561..355M,2020MNRAS.494.2449D}. However, this introduces layers of astrophysical modeling, and in this case the standard siren is not calibrated by general relativity alone.

{\bf Assigning redshifts to gravitational-wave sources.} The challenge for standard siren cosmology is to identify the redshifts of GW sources. 
Multi-messenger observations, such as neutron star mergers with electromagnetic counterparts like short gamma-ray bursts or kilonovae, provide the most straightforward measurement \citep{2005ApJ...629...15H,2006PhRvD..74f3006D}. An electromagnetic counterpart like a kilonova can typically be pinpointed to a specific galaxy, thereby identifying the host galaxy of the GW merger. The GW signal provides the distance to the host galaxy, while its electromagnetic spectrum provides the redshift. {These sources are typically referred to as {\it bright sirens}.}

Without an electromagnetic counterpart, the GW event is usually too poorly localized on the sky to allow for a unique host galaxy identification \citep{2018LRR....21....3A}. Only the loudest, best-localized GW events (1 per several hundred events) are expected to have only a single galaxy in their localization volumes~\citep{2016arXiv161201471C}.
Nevertheless, if a sufficiently complete galaxy catalog is available, one can consider all of the galaxies within the GW localization volume as potential host galaxies, and statistically marginalize over them. This was the original proposal by \citet{1986Natur.323..310S}, and the method was further developed in a Bayesian context by \citet{2012PhRvD..86d3011D,2018Natur.562..545C}. {These sources are often called {\it dark sirens}.}
At the typical distances of GW events (greater than several hundred Mpc), spectroscopic galaxy catalogs are rare, although photometric galaxy catalogs (with redshifts inferred by photometry rather than spectra) can be useful when they overlap with the GW skymap \citep{2019ApJ...876L...7S,2020ApJ...900L..33P}. New and upcoming large-scale spectroscopic galaxy surveys like DESI, Taipain, SDSS-V, and 4MOST may provide useful galaxy catalogs for statistical GW standard siren analyses, either by cataloging a large fraction of the sky or through targeted follow-up of GW event localizations.

In the absence of counterparts or galaxy catalogs, alternative sources of redshift information have been proposed. If galaxy catalogs are incomplete but GW events are well-localized, matching the spatial clustering of GW sources as a function of distance to the clustering of galaxies as a function of redshift can constrain cosmological parameters~\citep{2008PhRvD..77d3512M,Oguri:2016dgk,Mukherjee:2018ebj,2020arXiv200501111V,2020ApJ...902...79B,2021PhRvD.103d3520M}.  

Another extension of the statistical dark standard siren method is to use prior knowledge of the merger redshift distribution, derived from external measurements of the star formation rate and time delay distribution of binary mergers, to compare against the observed gravitational-wave distance distribution~\citep{2019JCAP...04..033D,2021arXiv210314038Y,2021arXiv210907537L}.
Finally, a particularly promising avenue for gravitational-wave only standard siren analyses is to use known features in the source population to directly extract the redshift and distance from the gravitational-wave signal alone. These sources have been dubbed ``spectral sirens''~\citep{2022arXiv220208240M}. If information about the source-frame frequency is available, the redshift can be derived from the observed GW frequency. This source-frame GW frequency information can come from features in the source-frame mass distribution \citep{1993ApJ...411L...5C,2012PhRvD..85b3535T,2012PhRvD..86b3502T,2019ApJ...883L..42F,2021ApJ...908..215Y,2021ApJ...909L..23E,2022arXiv220208240M} as well as tidal effects in neutron star mergers \citep{2012PhRvL.108i1101M,2017PhRvD..95d3502D,2021PhRvD.104h3528C}. 

As \citet{2019ApJ...883L..42F} showed, an especially promising feature in the black hole mass distribution is the lower edge of the pair-instability mass gap: a steep drop-off in the black hole mass distribution at $\sim 40$--$65\,M_\odot$, which may be accompanied by a pile-up of black holes immediately below the gap at $\gtrsim 35\,M_\odot$. Stellar models \citep{1964ApJS....9..201F,1967ApJ...150..131R,Fryer:2000my,Heger:2001cd} show that when the black hole progenitor Helium star is in the mass range $\sim40$--$120\,M_\odot$, after the helium burning stage, unstable electron-positron pair production occurs in the carbon-oxygen core. This pair production reduces the photon pressure in the stellar core, and causes oxygen to explosively ignite. This explosive oxygen burning generates an energetic outwards pulse, which can disrupt the star entirely, leaving behind no stellar remnant, or shed off enough mass so that when the star collapses to a black hole, its mass is below the mass gap. Because the physics of pair instability depends primarily on the mass of the carbon-oxygen core, the location of the lower and upper edge of the gap are expected to be independent of redshift \citep{2019ApJ...887...53F}. By observing the {\it redshifted} mass distribution as a function of luminosity distance in gravitational waves, the location of the pair-instability feature(s) can be jointly inferred together with the redshift-distance relation \citep{2019ApJ...883L..42F,2021PhRvD.104f2009M}. Gravitational-wave observations of binary black holes support the existence of bump, followed by a steepening of the black hole mass distribution at $\sim 40\,M_\odot$~\citep{2017ApJ...851L..25F,2021ApJ...913L...7A,2021arXiv211103634T}. The interpretation of this feature as the imprint of pair-instability supernovae is still uncertain; however, as black hole population models improve, such features in the black hole mass distribution can be theoretically calibrated and reach their potential as robust cosmological probes. 
It is to be emphasized that all features in the mass distribution, including properties around the putative NS-BH lower mass gap, can be used as spectral sirens; the combination of these many features can be used to self-calibrate and control potential bias from systematic errors~\citep{2022arXiv220208240M}.

\subsubsection{Measurements}
\label{sec:ssmeasurements}

While GWs directly provide the luminosity distance to the source, there are multiple ways to estimate its redshift. As discussed in the previous section, standard siren redshift measurements fall under three main categories: electromagnetic counterparts, galaxy catalogs, and features in the GW source population.  

\subsubsubsection{Electromagnetic counterparts} 

The multi-messenger binary neutron star detection, GW170817, provided the first standard siren measurement of the Hubble constant \citep{2017PhRvL.119p1101A,2017Natur.551...85A}. Gravitational-wave parameter estimation provided a luminosity distance of $43.8^{+2.9}_{-6.9}$ Mpc. The kilonova optical counterpart allowed for the identification of a unique host galaxy NGC4993. Because this event was relatively nearby, the measured redshift of NGC4993 is significantly affected by its peculiar (non-Hubble flow) velocity. In this case, the peculiar velocity is large ($\sim 300$ km/s) because NGC4993 is near to the Great Attractor. Correcting for inter-group and bulk flow velocities, the Hubble flow velocity is $3017\pm 166$ km/s. At $z \sim 0.01$, this event is only sensitive to the first-order linear redshift-distance relation, and the resulting Hubble constant measurement is \Ho$=70^{+12}_{-8}$ \Hunit~(maximum a-posteriori value and 68.3\% highest density credible interval, taking a flat-in-log prior on \Ho). With improved analysis of the gravitational-wave signal and slightly updated distance measurement, the Hubble constant measurement was updated to \Ho$=70^{+13}_{-7}$ \Hunit~\citep{2019PhRvX...9a1001A}. In addition to measuring the Hubble constant, GW170817 and its electromagnetic counterpart enabled impressively tight constraints on cosmological modified gravity theories, including the speed of gravity and gravitational-wave friction \citep{Abbott2017gamma,Amendola:2017orw,Ezquiaga2017,Sakstein2017,Creminelli2017,Baker2017,Crisostomi:2017pjs,Boran:2017rdn,2019PhRvD..99h3504L,2018JCAP...07..048P,2019PhRvL.123a1102A}. 

\subsubsubsection{Galaxy catalogs} 

To date, the only gravitational-wave event with a confident electromagnetic counterpart is GW170817. A possible AGN flare association was identified for the binary black hole event GW190521 \citep{2020PhRvL.124y1102G}, but the association is debatable \citep{Ashton:2020kyr,DePaolis:2020onl,Palmese:2021wcv}. However, the statistical galaxy catalog method has been applied to several gravitational-wave events. As a proof of concept, \citet{2019ApJ...871L..13F} demonstrated the statistical method with GW170817, marginalizing over galaxies in the GLADE catalog \citep{2018MNRAS.479.2374D}, rather than using the uniquely identified host galaxy NGC4993. Because GW170817 was exceptionally loud and close-by, and all three detectors of the LIGO-Virgo network were operational, it was localized to only 16 deg$^2$ with 90\% credibility \citep[215 Mpc$^3$ assuming standard cosmological parameters from][]{PlanckCollaboration2015}. This small localization volume contains only one large group of galaxies (the group containing NGC4993) at $z \sim 0.01$, and so the statistical standard siren measurement of \Ho~from GW170817 is almost as informative as the counterpart measurement. In most cases, the gravitational-wave localization volume contains $\mathcal{O}(10^4$--$10^5)$ potential host galaxies, and so the statistical standard siren method would be substantially less informative even if we had complete galaxy catalogs with well-measured redshifts. 
The two best statistical standard sirens, excluding GW170817, are the binary black hole event GW170814 \citep{2017PhRvL.119n1101A} and the (probable) binary black hole event GW190814 \citep{2020ApJ...896L..44A}. (The secondary mass of GW190814 is ambiguous, and GW190814 may be a neutron star--black hole system.) Both of these events lack electromagnetic counterparts, but their sky position and gravitational-wave location are ideal for the statistical galaxy catalog method. Not only are they the best-localized events from the first three observing runs (other than the binary neutron star event GW170817), but they also both fall within the footprint of the Dark Energy Survey \citep[DES,][]{Abbott:2016ktf}. 

GW170814 was the first three-detector gravitational-wave event, observed by Virgo in addition to the two LIGO observatories in their second observing run. Using data from all three detectors enabled a 90\% sky localization of only 60 deg$^2$ (compared to 1160 deg$^2$ using only data from the two LIGO detectors). Correlating the gravitational-wave sky map and distance measurement of $540^{+130}_{-210}$ Mpc with the photometric galaxy catalog from DES,
\citet{2019ApJ...876L...7S} performed the first standard siren measurement of the Hubble constant using a binary black hole. With only a single event, the measurement was relatively broad, with the $68\%$ posterior credible interval encompassing $\sim60\%$ of the prior, but nevertheless there was a clear peak at \Ho$\sim75$ \Hunit associated with an over-density of galaxies at $z \sim 0.12$.  

GW190814, detected by LIGO and Virgo in their third observing run, is the best-localized dark standard siren observed to date. It was localized to 18 deg$^2$ (90\% credibility) on the sky. At $241^{+41}_{-45}$ Mpc, it is nearby and has an impressive signal-to-noise ratio of 25. Furthermore, because of its asymmetric masses (mass ratio of approximately 1:10), the gravitational-wave signal contains detectable higher harmonics, which reduce the distance-inclination degeneracy and yields a tighter distance measurement. Combining the gravitational-wave localization with the GLADE galaxy catalog, \citet{2020ApJ...896L..44A} performed a statistical standard siren measurement of the Hubble constant, finding a broad peak at \Ho$=75^{+59}_{-13}$ \Hunit~(with the 68\% highest posterior density interval comprising 60\% of the prior range). Although GW190814 is very nearby for a gravitational-wave event, it is at the limit of where currently-available spectroscopic galaxy catalogs are useful. At these distances, the GLADE catalog is 40\% complete. Meanwhile, like GW170814, GW190814 lies within the DES footprint. Although the DES catalog contains photometric, rather than spectroscopic redshifts, which means larger errors on each galaxy's redshift, it does not suffer from incompleteness. \citet{2020ApJ...900L..33P} used the DES galaxies within the GW190814 sky map to measure the Hubble constant to \Ho$=78^{+57}_{-13}$ \Hunit, consistent with the result of \citet{2020ApJ...896L..44A}. 

\subsubsubsection{Standard siren population} 

In order to achieve competitive cosmological constraints, information must be combined across multiple standard sirens. Analyzing a population of standard sirens requires a careful treatment of measurement uncertainties and selection effects \citep{2018Natur.562..545C,2019MNRAS.486.1086M,2019PhRvD.100j3523M}. The importance of incorporating selection effects can be understood by considering that gravitational-wave detectors are significantly more likely to observe sources at smaller distances, but there are more potential host galaxies at higher redshifts. If the analysis did not account for selection effects, it would tend to overestimate the redshifts of gravitational-wave events and therefore overestimate the Hubble constant. Meanwhile, because the probability of detecting a gravitational-wave source is a strong function of its mass and distance (and, to a lesser degree, the component spins), we must simultaneously fit the gravitational-wave source distribution, particularly the astrophysical mass distribution and distance/redshift distribution, with the cosmological parameters. For example, if the wrong binary black hole mass distribution is assumed in the statistical galaxy catalog method, the recovered cosmological parameters will be biased \citep{2021ApJ...909..218A,2021PhRvD.104f2009M,LIGO2021}. The assumed black hole and neutron star spin distribution can also affect the cosmological inference, both because the binary spin impacts the gravitational-wave detection probability and because of mild degeneracies between the measured binary spin, inclination and luminosity distance. The latter effect was already noted for the GW170817 standard siren measurement; assuming different priors on the neutron star spin magnitudes yielded slightly different posteriors on the Hubble constant \citep{2019PhRvX...9a1001A}.
In addition to the gravitational-wave data, care must be taken in the statistical treatment of the redshift information. If redshifts are supplied from a galaxy catalog, particular attention is required in treating galaxy catalog incompleteness \citep{2019ApJ...871L..13F,2020PhRvD.101l2001G,2021arXiv210112660F,2021arXiv211104629G}. 

\begin{figure}[t!]
\centering
\mbox{\includegraphics[width=0.56\textwidth]{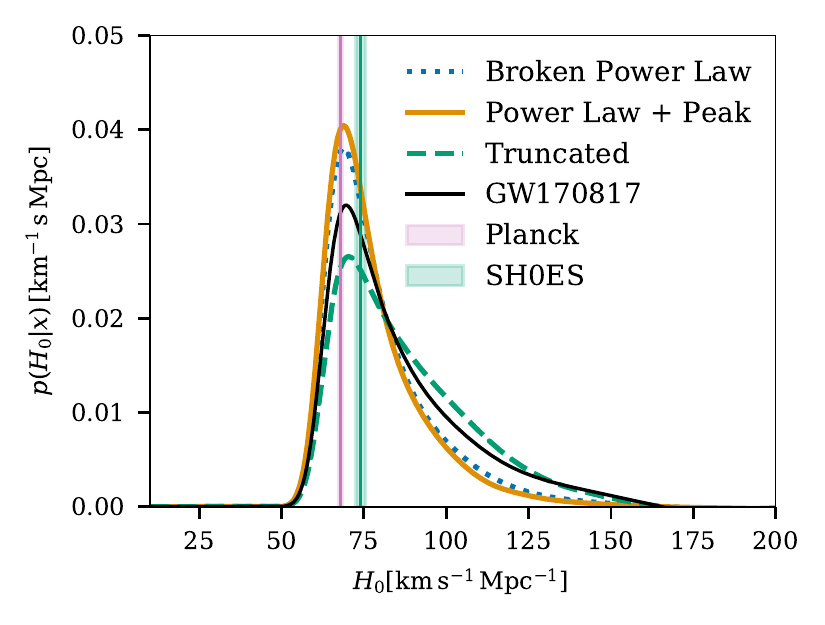}
\includegraphics[width=0.42\textwidth]{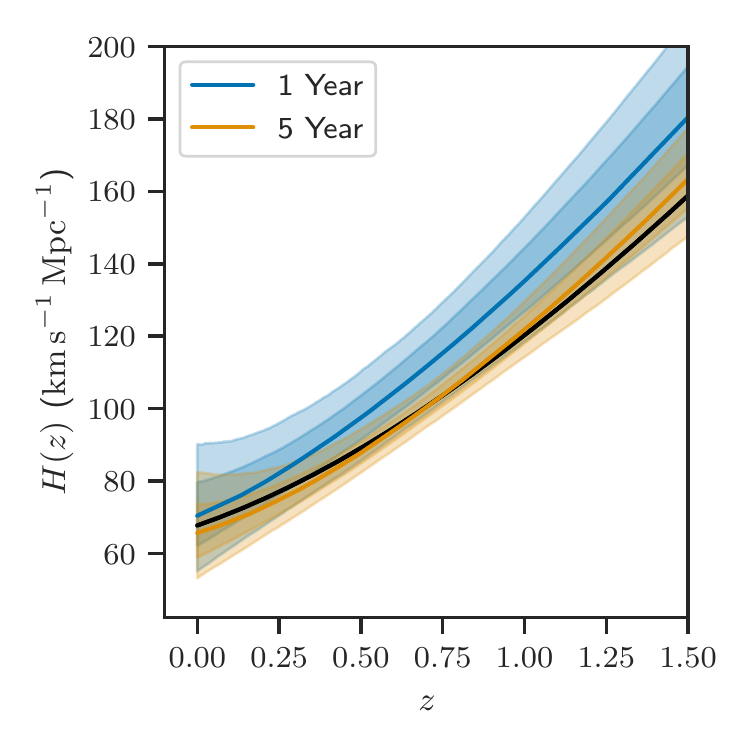}}
\caption{Constraints on $H_0$ and $H(z)$ from GWs used as standard sirens. Left panel: $H_0$ posterior obtained from the combination of the signal of 42 black hole-black hole mergers from GWTC-3 with the detection of GW170817 \cite{LIGO2021}. Right panel: Forecasts on $H(z)$ measurements obtained from the simulation of five years (orange line, one year with the blue line) of detection from the Advanced LIGO and Virgo detectors \citep{2019ApJ...883L..42F}. Images reproduced with permission from \cite{LIGO2021} and \cite{2019ApJ...883L..42F}, copyright by Astrophysical Journal.}
\label{fig:H0Hz_GW}
\end{figure}

The latest gravitational-wave catalog consists of $\sim$90 events from three observing runs of LIGO and Virgo~\citep{2021arXiv211103606T}. Using redshift information either from galaxy catalogs or from the redshifted binary black hole mass spectrum, these events have been used in combination with the counterpart standard siren measurement of GW170817 to constrain the expansion history $H(z)$ and several cosmological modified gravity theories~\citep{2021arXiv210112660F,LIGO2021,2021arXiv211106445P,2021arXiv211205728M}. With the relatively low-redshift sample, the best measured cosmological parameter remains the Hubble constant, and the constraints using all events represent a {$\sim20\%$} improvement over the measurement from GW170817 and its counterpart (see Fig.~\ref{fig:H0Hz_GW}, left panel). 

\subsubsection{Systematic effects}

The limiting systematic uncertainty for standard siren measurements is the detector calibration, specifically the amplitude uncertainty. Each detector's amplitude response uncertainty translates to a systematic distance uncertainty for the GW source, contributing at the few-percent level. For individual events, the statistical distance uncertainty of $\mathcal{O}(10\%)$ dominates calibration uncertainty. But when stacking events to infer cosmological parameters, unlike the statistical distance uncertainty, the calibration uncertainty may not average out. An important prerequisite for reaching a percent-level \Ho~measurement with standard sirens is to reduce the amplitude calibration uncertainty below 1\%~\citep{2021arXiv210700129S}.

As the standard siren catalog continues to grow, other uncertainties in the gravitational-wave distance measurements will become important. One of these uncertainties is the gravitational waveform model. Extracting the distance of the source from the gravitational-wave measurement requires a gravitational waveform model that is not perfectly known, especially for systems with strong matter effects, extreme spins or mass ratios \citep{2021PhRvD.103h3001H}. For standard sirens at larger distances, the gravitational-wave signal may be (de)magnified due to weak gravitational lensing by matter along the line of sight. Most of these uncertainties may be incorporated into the statistical framework and contribute to a statistical rather than systematic uncertainty. For example, if the distribution of lensing magnifications is known, this contribution can be marginalized over in the GW distance likelihood \citep{2005ApJ...629...15H,2010PhRvD..81l4046H,2010CQGra..27u5006S}. As discussed in Sect.~\ref{sec:ssmeasurements}, the astrophysical distributions of the masses, spins and distances of black hole and neutron star mergers must be simultaneously inferred with the cosmological parameters, especially when analyzing a population of standard sirens at cosmological distances. Even compared to the current large statistical uncertainties, fixing the binary black hole mass distribution in the galaxy catalog standard siren analysis results in a significant systematic uncertainty, whereas the joint inference transfers the systematic uncertainty to a statistical uncertainty that converges with many events~\citep{LIGO2021}. 

There are also uncertainties in the redshift measurements that, if not properly understood, can contribute to a systematic uncertainty. The counterpart standard siren method, where the redshift information comes directly from a unique host galaxy identification, is the least susceptible to systematic effects. 
A possible systematic uncertainty in the redshift measurement can come from errors in the peculiar velocity correction, but the statistics of peculiar velocities are well-understood and, especially at typical standard siren distances, contribute a negligible fraction of the uncertainty budget. On the other hand, when galaxy catalogs are used for the redshift information, they introduce more potential sources of systematic uncertainty. Factors such as catalog incompleteness, photometric redshift uncertainties, and the galaxies' probabilities of hosting gravitational-wave sources must be understood. If the redshift information is supplied by features in the source distribution, it is important to check that the population model is not mis-specified. For example, fitting the binary black hole mass distribution to a power law, where the true distribution more closely resembles a mixture model between a power law and a Gaussian, would lead to biased recovery of the mass distribution and the cosmological parameters. In general, the source distribution also needs to be calibrated against theoretical models. If the source mass distribution evolves with redshift~\citep{2021ApJ...912...98F}, theoretical guidance may help disentangle the source mass evolution with cosmological redshift, although analysis of the full distribution may help self-calibrate the sample~\citep{2022arXiv220208240M}. 

\subsubsection{Main results and forecasts}

The current best standard siren constraints are dominated by the Hubble constant measurement from GW170817 and its electromagnetic counterpart, which yielded \Ho$=70^{+13}_{-7}$ \Hunit. However, with $\sim90$ gravitational-wave events detected to date, the population of standard sirens without counterparts is beginning to contribute. Out of the gravitational-wave events without counterparts, a couple of events, namely GW170814 and GW190814, have been particularly well-localized so that comparing their localization posteriors to a galaxy catalog yields only $\sim1$ probable galaxy structure that contains the host galaxy, resulting in a uni-modal, fairly informative Hubble constant measurement. The remaining dozens of events have also been used for standard siren analyses in conjunction with galaxy catalogs, but care is required in the interpretation of these results. Unless the source population, particularly the binary black hole mass distribution, is simultaneously inferred with the cosmological parameters, hidden assumptions about the source population can impact the cosmological inference and result in overly optimistic constraints. So far, the only analyses that simultaneously fit the source population and cosmological parameters do so without incorporating galaxy catalog information~\citep{LIGO2021}. With the latest gravitational-wave catalog GWTC-3, these methods yield a 17\% improvement in the Hubble constant measurement over the measurement from GW170817 and its counterpart \citep[see the left panel of Fig.~\ref{fig:H0Hz_GW};][]{LIGO2021}. 

The most robust standard sirens are gravitational-wave sources with electromagnetic counterparts, typically binary neutron stars, although some neutron star-black hole mergers may also produce electromagnetic emission. With the current ground-based gravitational-wave detectors, these sources will predominantly be sensitive to the Hubble constant, and with $N$ sources with counterparts, we expect the Hubble constant measurement to converge as $15\%/\sqrt{N}$~\citep{2018Natur.562..545C}.

For the majority of gravitational-wave events that lack counterparts, galaxy catalogs can be used for the redshift information. Further work needs to be done to develop galaxy catalogs specifically for the standard siren application, manage catalog incompleteness, and jointly fit the source population together with the cosmological parameters to avoid systematic bias. Another promising method is to use features in the source mass distribution to fit cosmological parameters together with the source population in a gravitational-wave only analysis. \citet{2019ApJ...883L..42F} showed that leveraging the pair-instability feature in the black hole mass distribution can provide percent-level constraints on $H(z)$ at $z = 0.8$ within 5 years of Advanced LIGO observations (see the right panel of Fig.~\ref{fig:H0Hz_GW}). By combining binary black holes, which can be observed at higher redshifts, with nearby binary neutron stars with counterparts, the expansion history can therefore be measured out to $z \sim 1.5$. For this method to provide robust cosmological constraints, further progress is required in theoretical models of the black hole mass distribution. In particular, the redshift evolution of the source mass distribution must be theoretically understood, or controlled through self-calibration~\citep{2022arXiv220208240M}.

Standard sirens are unique cosmological probes in that they simultaneously probe the background cosmology and gravitational perturbations, namely the propagation of gravitational waves. Beyond constraining the Hubble constant, standard sirens are therefore especially promising for constraining dark energy theories both through their effects on the background cosmology and their effects on gravitational-wave propagation.

The era of gravitational-wave cosmology has just begun. The gravitational-wave catalog is growing at an incredible rate, and by the late 2020s, the gravitational-wave detector network of LIGO, Virgo and KAGRA are expected to detect hundreds to thousands of events annually. In the coming decades, the space-based gravitational-wave detector LISA is expected to launch, and the next generation of ground-based gravitational-wave detectors may become a reality, including Cosmic Explorer \citep{Reitze:2019}, Einstein Telescope \citep{Punturo:2010}, Taiji \citep{2020SciBu..65.1340Z}, and  TianQin \citep{2020JCAP...11..012W}. 
The growth of the gravitational-wave dataset is accompanied by new electromagnetic telescopes to hunt counterparts, galaxy surveys to expand redshift catalogs, theoretical developments to model the gravitational-wave source population, and computational techniques to carry out the standard siren inference. Standard siren cosmology is a rapidly growing field with a promising future.

\clearpage

\subsection{Time Delay Cosmography}
\label{sect:tdc}

\begin{figure*}[h!]
\centering
    \includegraphics[width=0.99\textwidth]{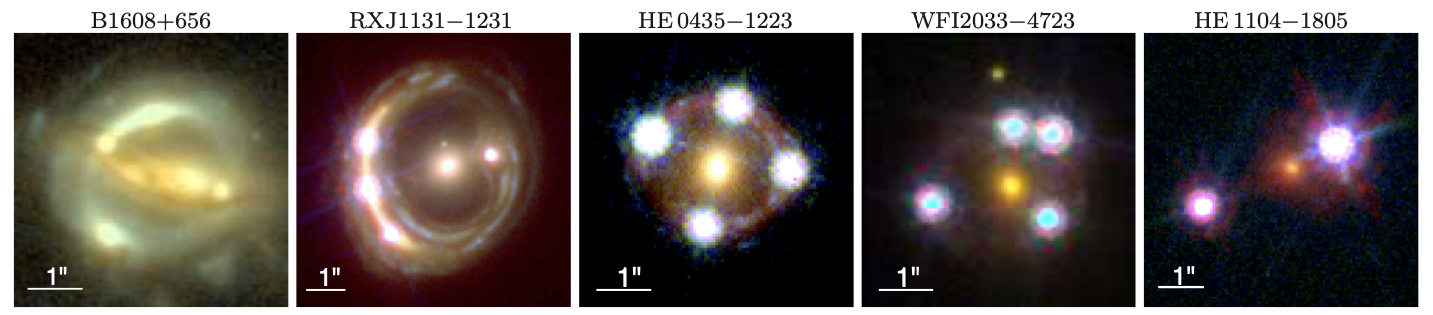}
    \caption{
Four quadruply lensed quasar systems and one doubly lensed quasar system from the H0LiCOW sample. The lens name is indicated above each panel. The color images are composed using 2 (for B1608+656) or 3 (for other lenses) HST imaging bands in the optical and near-infrared. North is up and east is left. Image reproduced with permission from \cite{Suyu:2017}, copyright by Monthly Notices of the Royal Astronomical Society.}
    \label{tdc:fig_lenses}
\end{figure*}

Time-delay cosmography uses measurements of relative arrival times of multiply gravitationally lensed sources to measure an absolute scale of the Universe. The method was originally proposed by \cite{Refsdal:1964} over half a century ago, prior to the discovery of the first extra-galactic gravitational lens. The methodology provides a one-step measurement of the Hubble constant, completely independent of the local distance ladder or probes anchored with sound horizon physics, such as the cosmic microwave background (CMB). 
Figure~\ref{tdc:fig_lenses} illustrates different galaxy-scale gravitational lenses with a multiply imaged quasar in different configurations \citep{Suyu:2017}.

\subsubsection{Basic idea and equations}
\label{tdc_basics}

The phenomena of gravitational lensing can be described by the lens equation, which maps the source plane coordinate $\boldsymbol{\beta}$ to the image plane $\boldsymbol{\theta}$:
\begin{equation} \label{eqn:lens_equation}
  \boldsymbol{\beta} = \boldsymbol{\theta} - \boldsymbol{\alpha}(\boldsymbol{\theta}) \;\; ,
\end{equation}
where $\boldsymbol{\alpha}$ is the angular shift on the sky between the original un-lensed and the lensed observed position of an object.

For a single deflector plane, the lens equation can be expressed in terms of the physical deflection angle $\hat{\boldsymbol{\alpha}}$ as:
\begin{equation} \label{eqn:lens_equation_single_plane}
  \boldsymbol{\beta} = \boldsymbol{\theta} - \frac{D_{\rm ds}}{D_{\rm s}}\hat{\boldsymbol{\alpha}}(\boldsymbol{\theta}) \;\; ,
\end{equation}
where $D_{\rm s}$ and $D_{\rm ds}$ are the angular diameter distance from the observer to the source and from the deflector to the source, respectively.
In the single lens plane regime, we can introduce the lensing potential $\psi$ such that the reduced deflection angle is the gradient of the potential:
\begin{equation}
    \boldsymbol{\alpha}(\boldsymbol{\theta}) = \nabla \psi(\boldsymbol{\theta}) \;\; ,
    \label{eq:anglepot}
\end{equation}
and the lensing convergence as:
\begin{equation}
    \kappa(\boldsymbol{\theta}) =  \frac{1}{2}\nabla^2 \psi(\boldsymbol{\theta}) \;\; .
\end{equation}
Physically, the lensing convergence in this regime corresponds to the projected surface mass density $\Sigma$ normalized to the critical lensing surface density $\Sigma_{\rm crit}$:
\begin{equation}
    \kappa(\boldsymbol{\theta}) = \frac{\Sigma(\boldsymbol{\theta})}{\Sigma_{\rm crit}} \;\; ,
\end{equation}
with the critical lensing surface density:
\begin{equation}
    \Sigma_{\rm crit} = \frac{c^2 D_{\rm s}}{4\pi G D_{\rm d}D_{\rm ds}} \;\; ,
\end{equation}
where $D_{\rm d}$ is the angular diameter distance to the deflector, $c$ is the speed of light and $G$ is the gravitational constant\footnote{The critical lensing surface density is only considering mass relative to the mean background cosmological density.}.

The relative arrival time between two images $\boldsymbol{\theta}_{\rm A}$ and $\boldsymbol{\theta}_{\rm B}$, $\Delta t_{\rm AB}$, originated from the same source is given by:
\begin{equation}\label{eqn:time_delay}
    \Delta t_{\rm AB} = \frac{D_{\Delta t}}{c} \left[\tau(\boldsymbol{\theta}_{\rm A}, \boldsymbol{\beta}) - \tau(\boldsymbol{\theta}_{\rm B}, \boldsymbol{\beta}) \right] \;\; ,
\end{equation}
where:
\begin{equation}\label{eqn:fermat_potential}
    \tau(\boldsymbol{\theta}, \boldsymbol{\beta}) = \left[ \frac{\left(\boldsymbol{\theta} - \boldsymbol{\beta} \right)^2}{2} - \psi(\boldsymbol{\theta})\right]
\end{equation}
is the Fermat potential \citep{Schneider:1985, Blandford:1986}, and:
\begin{equation} \label{eqn:ddt_definition}
    D_{\Delta t} \equiv \left(1 + z_{\rm d}\right) \frac{D_{\rm d}D_{\rm s}}{D_{\rm ds}}
\end{equation}
is the time-delay distance \citep{Refsdal:1964, Schneider:1992, Suyu:2010}.

Constraints on the Fermat potential difference $\Delta \tau_{\rm AB}$ and a measured time delay $\Delta t_{\rm AB}$ allows one to constrain the time-delay distance $D_{\Delta t}$. This absolute physical distance anchors the scale in the Universe within the redshifts involved in the lensing configuration.
The Hubble constant is inversely proportional to the absolute scales of the Universe, and thus scales with $D_{\Delta t}$ as:
\begin{equation} \label{eqn:H0_ddt}
	H_0 \propto D_{\Delta t}^{-1} \;\; ,
\end{equation}
mildly dependent on the relative expansion history from current time ($z=0$) to the redshift of the deflector and source. 

While the time delay $\Delta t_{\rm AB}$ can be directly measured (see Sect.~\ref{tdc:measurement}), the relative Fermat potential $\Delta \tau_{\rm AB}$ is not a direct observable. The primary information to infer $\Delta \tau_{\rm AB}$ are positional constraints and extended distortions from the lensing effect. However, there are degeneracies inherent in gravitational lensing that limit the amount of information accessible by lensing distortions \citep[e.g.,][]{Falco:1985, Gorenstein:1988, Kochanek:2002, Saha:2006, Schneider:2013, Schneider:2014, Birrer:2016, Unruh:2017, Birrer:2021curvedarcs}.

The most prominent lensing degeneracy impacting the time-delay prediction is the mass-sheet degeneracy \citep[MSD,][]{Falco:1985}.
The MSD is a multiplicative transform of the lens equation (Eq.~\ref{eqn:lens_equation}) which preserves image positions (and any higher order relative differentials of the lens equation) under a linear source displacement $\boldsymbol{\beta} \rightarrow \lambda\boldsymbol{\beta}$ 
combined with a transformation of the convergence field:
\begin{equation}\label{eqn:mst}
    \kappa_{\lambda}(\boldsymbol{\theta}) = \lambda \kappa(\boldsymbol{\theta}) + \left( 1 - \lambda\right) \;\; .
\end{equation}
The term $(1 - \lambda)$ in Eq.~\ref{eqn:mst} above describes an infinite sheet of convergence (or mass), and hence the name mass-sheet transform (MST).
Only observables related to the unlensed apparent source size, to the unlensed apparent brightness, or to the lensing potential are able to break this degeneracy.
Thus, the same relative lensing observables can result if the mass profile is scaled by the factor $\lambda$ with the addition of a sheet of convergence (or mass) of $\kappa(\boldsymbol{\theta}) = (1-\lambda)$.

The Fermat potential (Eq.~\ref{eqn:fermat_potential}) scales with $\lambda$ as:
\begin{equation}\label{eqn:fermat_mst}
    \Delta \tau_{\rm AB , \lambda} =  \lambda \Delta \tau_{\rm AB} \;\; ,
\end{equation}
and so does the time delay as:
\begin{equation}\label{eqn:time_delay_mst}
    \Delta t_{\rm AB , \lambda} =  \lambda \Delta t_{\rm AB} \;\; .
\end{equation}
When transforming a lens model with a mass-sheet transformation, the inference of the time-delay distance (Eq.~\ref{eqn:ddt_definition}) from a measured time delay and inferred Fermat potential transforms as:
\begin{equation} \label{eqn:ddt_mst}
    D_{\Delta t , \lambda} = \lambda^{-1}D_{\Delta t} \;\; .
\end{equation}
Thus, the Hubble constant, when inferred from the time-delay distance $D_{\Delta t}$, transforms from Eqn.~\ref{eqn:H0_ddt} as:
\begin{equation} \label{eqn:h0_mst}
H_{0 , \lambda} =  \lambda H_0 \;\; .
\end{equation}

An MSD effect relative to a proposed deflector model might occur either within the mass distribution of the main deflector, referred as internal MSD with $\lambda_{\rm int}$, or being caused due to homogeneities along the line-of-sight (LOS) of the strong lens system.

Mass over- or under-densities along the LOS of the strong lensing system cause, to first order, shear and convergence perturbations. Reduced shear distortions have a measurable imprint on the azimuthal structure of the strong lensing system \citep[see e.g.,][]{Birrer:2021curvedarcs} while the convergence component of the LOS, denoted as $\kappa_{\rm ext}$, is equivalent to an MST, and thus not directly measurable from imaging data. 
The lensing kernel impacting the linear distortions, both shear and $\kappa_{\rm ext}$, is different from the standard weak lensing kernel \citep{McCully:2014, McCully:2017, Birrer:2017los, Birrer:tdcosmoiv, Fleury:2021los}. 

We define $D^{\rm lens}$ as the specific angular diameter distance along the line-of-sight of the lens being corrected by LOS structure and $D^{\rm bkg}$ as the angular diameter distance from the homogeneous background metric without any perturbative contributions. $D^{\rm lens}$ and $D^{\rm bkg}$ are related through the convergence terms as \citep{Birrer:tdcosmoiv}:
\begin{eqnarray}
D^{\rm lens}_{\rm d} &=& (1 - \kappa_{\rm d})D_{\rm d}^{\rm bkg}\\
D^{\rm lens}_{\rm s} &=& (1 - \kappa_{\rm s})D_{\rm s}^{\rm bkg}\\
D^{\rm lens}_{\rm ds} &=& (1 - \kappa_{\rm ds})D_{\rm ds}^{\rm bkg} \;\; ,
\label{eqn:ang_distance_kappa}
\end{eqnarray}
where $\kappa_{\rm d}$ is the weak lensing effect from the observer to the deflector, $\kappa_{\rm s}$ from the observer to the source, and $\kappa_{\rm ds}$ from the deflector to the source, respectively \citep{Birrer:tdcosmoiv}.
The lensing kernel impacting the time delay can be described as the product of three different angular diameter distances entering $D_{\Delta t}$ in Equation~\ref{eqn:ddt_definition} \citep{Birrer:tdcosmoiv, Fleury:2021metric}, 
\begin{equation}\label{eqn:mst_combined}
    1 - \kappa_{\rm ext} = \frac{(1 - \kappa_{\rm d})(1 - \kappa_{\rm s})}{1 - \kappa_{\rm ds}} \;\; .
\end{equation}

MSD uncertainties or biases may also arise relative to assumptions made in the radial density profile of the main deflector galaxy \citep[see, e.g.,][]{Kochanek:2002, Read:2007, Schneider:2013, Coles:2014, Xu:2016, Birrer:2016, Unruh:2017, Sonnenfeld:2018, Kochanek:2020, Blum:2020, Birrer:tdcosmoiv, Kochanek:2021}. Any lensing-only constraints on the radial density profile is over-constrained, and constraints rely on the functional form imposed.

The total MST, i.e. the relevant transform to constrain for an accurate cosmography and \Ho~measurement, is the product of the internal and external MST \citep[e.g.,][]{Schneider:2013, Birrer:2016, Birrer:tdcosmoiv}:
\begin{equation}\label{eqn:lambda_combined}
    \lambda = (1-\kappa_{\rm ext}) \times \lambda_{\rm int} \;\; .
\end{equation}

The external line-of-sight lensing contribution can be estimated by tracers of the large-scale structure, either using galaxy number counts \citep[e.g.,][]{Greene:2013, Rusu:2017}, or weak-lensing measurements \citep{Tihhonova:2018}. These measurements, paired with a cosmological model including a galaxy-halo connection are able to constrain the probability distribution of $\kappa_{\rm ext}$ to few per cent per sight line.

Among those observations that are sensitive to the total MST $\lambda$, stellar kinematics is the most prominent and commonly used one. The dynamics of stars is a direct tracer of the three-dimensional gravitational potential and provides an independent mass estimate.
Joint lensing and dynamics constraints have been used to provide measurements of galaxy mass profiles \citep[e.g.,][]{Grogin:1996, Romanowsky:1999, Treu:2002}.
The modeling of the kinematic observables in lensing galaxies range in complexity from spherical Jeans modeling \citep{BinneyTremaine:2008} to Schwarzschild \citep{Schwarzschild:1979} methods.

Regardless of the approach, the prediction of any $\sigma_{\rm v}$ from any model can be decomposed into a cosmological-dependent and cosmology-independent part as \citep[see e.g.,][]{Birrer:2016, Birrer:2019}:
\begin{equation}\label{eqn:los_sigma_v}
    \sigma_{\rm v}^2 = \lambda \frac{D_{\rm s}}{D_{\rm ds}}c^2 J(\boldsymbol{\xi}_{\rm lens}, \boldsymbol{\beta}_{\rm ani}) \;\; ,
\end{equation}
where $c$ is the speed of light, $J$ is a dimensionless quantity dependent on the deflector model ($\boldsymbol{\xi}_{\rm lens}$), the stellar anisotropy distribution ($\boldsymbol{\beta}_{\rm ani}$) and the observational conditions and luminosity-weighting within the aperture \citep[e.g.,][]{Binney:1982, Treu:2004, Suyu:2010}.

The constraints obtained from joint lensing and dynamics are either able to determine the MST component of the deflector model, or provide additional cosmographic constraints on the relative expansion history through the involved angular diameter distance ratio ($D_{\rm s}/D_{\rm ds}$, Eq.~\ref{eqn:los_sigma_v}). 
When adding a time delay, the joint cosmographic constraints from a combined analysis of time-delay, lensing, and dynamics can be translated into a two-dimensional angular diameter distance plane \citep{Birrer:2016, Birrer:2019}.
When mapped into the $D_{\Delta t}$-$D_{\rm d}$-plane, the projection in $D_{\rm d}$ is invariant under any pure MSD parameter $\lambda$ \citep{Paraficz:2009, Jee:2015, Birrer:2019}\footnote{$D_{\rm d}$ is still dependent on the LOS between observer and lens, $\kappa_{\rm d}$ (Eqn. \ref{eqn:ang_distance_kappa}).}.

An alternative approach to constrain the MSD is with absolute lensing magnifications. The MSD transforms the lensing magnification $\mu$ by:
\begin{equation}
    \mu_{\lambda} = \lambda^{-2}\mu \;\; .
\end{equation}
Thus, a known apparent unlensed brightness of an object $F_{\rm unl}$ with a measured flux $F_{\rm obs}$ can directly measure the target magnification:
\begin{equation}
    \mu_{\lambda} = \frac{F_{\rm obs}}{F_{\rm unl}} \;\; .
\end{equation}
Gravitationally lensed supernovae (glSNe) can provide, in addition to measurable time delays, lensing magnification constraints when knowledge about the unlensed apparent brightness of the explosion is imposed. This measurement does not require an absolute bolometric calibration of the exploding transient, but only relative to an unlensed field \citep[e.g.,][]{Kolatt:1998, Oguri:2003, Foxley-Marrable:2018, Birrer:2021glSNe}.

\subsubsection{Sample selection}
\label{tdc:sample_selection}

The primary requirement to provide an absolute distance measurement is a measured relative time delay between a multiply imaged source. A time delay can only be measured if the source is bright and time-variable, or a transient.
The original proposed source by \cite{Refsdal:1964} were lensed supernovae before the discovery of the strong-lensing phenomena on cosmological scales.
The first extra-galactic lens discovered was a doubly lensed quasar \citep{Walsh:1979}.
Lensed quasars were quickly identified as excellent sources for time-delay cosmography as they are variable on short time scale, making the time-delay measurements possible, and they are sufficiently bright to be observed at cosmological distances. Lensed quasars are typically found at redshift $z_{\rm s} \sim$1-3, lensed by massive early-type galaxies located around redshift $z_{\rm d} \sim$0.2-0.8. This configuration typically produces multiple images separated by 1-3\arcsec.

Strongly lensed quasars are rare objects on the sky. The discovery of currently known lensed quasars followed different paths. Some lenses were serendipitously discovered by visual inspection of astronomical images, in particular in the early days \citep[e.g.,][]{Sluse:2003}.
More recently, with the advent of large ground and space-based imaging surveys, more systematic searches could be conducted, involving astrometric and color selections on post-processed catalogs \citep[][]{Krone-Martins:2018, Agnello:2018, Lemon:2019}, and more recently directly employed machine learning techniques on both catalogs and images. The discovery process is made in phases of certainty of the lensing nature with increased follow-up efforts. The first step with wide-field surveys often results in hundreds of candidates, of which a subset of the highest ranked candidates is followed-up with spectroscopic observations to confirm the identical redshift of the pair or quartet of quasar images, and with deep high-resolution imaging to detect the deflector galaxy and extended lensed features from the quasar host galaxy.

The most prominent lensing system being utilized are galaxy-scale lenses with quadruply imaged quasars. These systems can offer several relative time delays, additional constraints on the lens model from both positional constraints of the quasars and the often Einstein-ring-like lensed structure of the quasar host galaxy. Thus, a significant effort in the search and follow-up work has been spent to find quadruply lensed quasars.
Quadruply lensed quasars are less frequent than doubly lensed quasars by a factor of about $\sim$5 \citep{Oguri:2010}.
The more abundant population of doubly lensed quasars provide less constraints per individual lens, but come with a potential in a population-level analysis.

More recently, the first multiply imaged supernovae were discovered in a galaxy cluster environment \citep{Kelly:2015} and on a galaxy-scale lens \citep{Goobar:2017}. This opens the path, as envisioned by \cite{Refsdal:1964}, to use lensed supernovae as the time-variable source to measure \Ho~and with it the opportunity to utilize an entirely new source population.

\subsubsection{Measurements}
\label{tdc:measurement}

In order to measure the distances $D_{\Delta t}$, or more generally the $D_{\Delta t}$-$D_{\rm d}$ combination, from a time-delay lens system for cosmography, we need the following data products:
\begin{enumerate}
\item discovery of a lens with a time-variable source;
\item spectroscopic redshifts of the lens $z_{\rm d}$ and source $z_{\rm s}$;
\item time delays between the multiple images;
\item lens mass model to determine the Fermat potential; 
\item lens environment studies to constrain external lensing effects related to the mass-sheet degeneracy.
\end{enumerate}

The dataset required for each step are observationally cheap in comparison to other cosmological probes. However, the combined analysis, even of a single lens, requires the coordination of multiple independent observations. The analysis can be impossible or severely limited in its precision and reliability by a single missing ingredient.
For the discovery datasets, we refer to Sect.~\ref{tdc:sample_selection} and references therein.

\noindent
{\bf Spectroscopic redshifts.} The spectroscopic redshifts of the quasar sources $z_{\rm s}$ are often easy to obtain given the frequent emission lines in quasars. The redshift of the lens $z_{\rm d}$ can be challenging since the bright quasar images can outshine the lens galaxy. Getting $z_{\rm d}$ of lensed quasar systems often require spectra taken under good seeing condition, to deblend the lensing galaxy from the quasar. 

\noindent
{\bf Time delays.} Without measurements of a time delay, no constraints on absolute distances involved can be inferred, and thus, regardless of the approach chosen, no direct constraints on the Hubble constant can be achieved.
Relative time delays are measured with monitoring campaigns to extract light curves from individual images.
Lensed quasars with images separated by 1-3\arcsec are sufficient to be resolved with small ground-based telescope. The monitoring of lensed quasars is thus challenging but possible with 1-m or 2-m class telescope.
To perform the measurement, several conditions need to be met: {\it i)} photometric accuracy with few milli-magnitudes are required to catch the low-amplitude variability signal, {\it ii)} a good sampling of the light curves is necessary if one targets the fast variations of small amplitude, and {\it iii)} the duration of the monitoring campaign also need to be sufficient to cover the duration of time delays and to ensure that enough variations of the quasar are recorded. 
Furthermore, seasonal gaps are unavoidable in optical light curves since most lensed quasars are not visible all year long. In addition, extrinsic variations caused mainly by the micro-lensing of the quasar images, but also a variety of other astrophysical effects, are often observed in the light curves. These extrinsic variations and gaps can severely bias time-delay measurements if not appropriately modeled for.
Once well-sampled light-curves have been acquired, the next step consists in identifying features that can be matched in all light curves, and measure the time delays. We refer to \citep{Vuissoz2007, Vuissoz2008, Courbin2011, Tewes2013b, Eulaers2013, Rathna2013, Courbin2018, millon2020c} for recent measurements and methodology taking into account various aspects of model and data uncertainties.

\noindent
{\bf Lens mass model.}
The Fermat Potential (Eq.~\ref{eqn:fermat_potential}) is a crucial component we need to know precisely to be able to use time-delay measurements to probe cosmic distances (Eq.~\ref{eqn:ddt_definition}).
High-resolution imaging of gravitational lenses is a crucial observation to achieve a precise determination of the relative Fermat potential between multiple images of a time-variable source.
Imaging modeling is primarily performed on high-resolution space-based \textit{Hubble Space Telescope} \citep[HST;][]{Suyu:2010, Birrer:2016, Wong:2017, Rusu:2020}, or ground-based adaptive-optics \citep[AO;][]{Chen:2016, Chen:2019, Chen:2021} imaging.
To derive constraints on the lensing deflector from imaging data, all components that affect the imaging data need to be modeled and accounted for simultaneously with the lens model. This includes, but is not limited to, the extended source component of the AGN or transient host that is lensed, the image positions of the time variable source and its resulting point-like flux emission, the surface brightness of the deflector galaxy, differential dust extinction, and any other sources of surface brightness. In addition, instrument effects, such as the point spread function (PSF), noise (both shot-noise and instrumental noise), pixelization, and potential data reduction artifacts need to be accurately taken into account.
Different techniques have been developed to jointly marginalize over a complex and unknown source morphology. These consist of regularized pixelated source reconstruction\citep[e.g.,][]{Suyu:2006, Suyu:2009}, a set of basis functions such as shapelets \citep[e.g.,][]{Birrer:2015, Birrer:lenstronomy}, or parameterized surface brightness profiles, such as Sersic profiles.
The surface brightness amplitude components of all these methods have in common that they create a linear response on the pixels. The maximum likelihood of the data given a proposed model for the amplitude components is thus a linear problem, and the Gaussian covariance matrix of the linear coefficients can be used to analytically marginalize over the prior \citep[e.g.,][]{Suyu:2006, Birrer:2015}.

In the absence of knowledge of an absolute source size or brightness, imaging data constraints can not break the MST (as discussed in Sect.~\ref{tdc_basics}) and its generalization, the Source-Position-Transform \citep[SPT;][]{Schneider:2014}.
The quantity that is constrained by imaging data along the radial direction is \citep{Kochanek:2002, Sonnenfeld:2018, Kochanek:2020, Birrer:2021curvedarcs}:
\begin{equation}\label{eqn:rad_constraint}
	\xi_{\rm rad} \equiv  \frac{\theta_{\rm E} \alpha_{\rm E}^{\prime \prime}}{1-\alpha_{\rm E}^{\prime}} \propto \frac{\theta_{\rm E} \alpha^{\prime\prime}_{\rm E} }{1 - \kappa_{\rm E}} \;\; ,
\end{equation}
where $\alpha^{\prime}_{\rm E}$ is the derivative and $\alpha^{\prime\prime}_{\rm E}$ is the double derivative of the deflection angle at the Einstein radius $\theta_{\rm E}$, respectively, and $\kappa_{\rm E}$ is the convergence at $\theta_E$.
We refer to \cite{Birrer:2021curvedarcs} for a discussion on azimuthal constraints.

The currently used data to break the MST is a measurement of the lens velocity dispersion (see Eq. \ref{eqn:los_sigma_v}). The measurement is performed with high-spectral resolution spectrographs on large ground-based adaptive-optics supported instruments targeting stellar absorption lines in the rest-frame of the lensing galaxy, such as Keck-DEIMOS, Keck-KCWI, or VLT-MUSE.
The velocity dispersion measurement is then a joint fit of the spectra taking into account the observation conditions, including the atmospheric absorption, the stellar templates matching the lensing galaxy type in age distribution and metallicity, and the dispersion width in the stellar distribution on top of the line-spread function.
For measurements of velocity dispersion used in current time-delay cosmography studies we refer to \cite{Koopmans:2003, Suyu:2010, Suyu:2013, Courbin2011, Wong:2017, Agnello:2016, Sluse:2019, Buckley-Geer:2020}.

\noindent
{\bf Line-of-sight and lens environment.}
The contribution of large-scale density perturbations and individual massive objects along the line-of-sight alter the lensing deflections. To first order, these effects can be captured as cosmic shear and convergence. The reduced cosmic shear term is a commonly used model component. The convergence component, however, is equivalent to an external mass-sheet $\kappa_{\rm ext}$ (Eq.~\ref{eqn:mst_combined}), and can not be measured from imaging data.
Higher-order effects from nearby groups or individual groups need to be explicitly modeled. Explicit modeling of individual groups has been done by, e.g., \cite{Fassnacht2002, Momcheva2006, Wilson2016, Sluse2017}. For theoretical aspects of the approximation made and in which regime they hold we refer to \cite{McCully:2014, McCully:2017, Birrer:2017los, Fleury:2021los}.

Typically, methods taking advantage of the knowledge of the galaxy-halo connection are employed, and using luminous tracers of the underlying dark matter distribution. The most commonly used approach adopts galaxy number counts in different weighting schemes \citep[e.g.,][]{Suyu:2010, Greene:2013, Rusu:2017}. The comparison of these weights (summary statistics) with control fields and numerical simulations with an imposed galaxy-halo connection allows the computation of the posterior density in $\kappa_{\rm ext}$.
Weak lensing mass mapping is an alternative and complementary approach \citep{Tihhonova:2018, Tihhonova:2020}.
The required data for galaxy number counts are deep multi-band photometry within several square arc minutes of the deflector, and spectroscopy of the nearby galaxies and group identification \citep[e.g.,][]{Rusu:2017, Buckley-Geer:2020}. For weak lensing, preferentially deep space-based images are used to reduce the shape noise and enhance the signal.

\subsubsection{Systematic effects}
\label{tdc:systematics}

The currently two main uncertainties that, if not properly taken into account, can lead to systematic uncertainties are the mass profile assumptions of the main deflectors and the selection effects of the lens sample used for the analysis.

\noindent
{\bf Mass profile assumptions.}
The dominant uncertainty in the current measurement of the Hubble constant with strong gravitational lensing time delays is attributed to uncertainties in the mass profiles of the main deflector galaxies. 
The currently employed models mitigating the MST effect is parameterized with a pure MST parameter $\lambda$ \citep{Birrer:tdcosmoiv}. This parameterization is purely of mathematical nature, and leaves the physical interpretation \citep[e.g.,][]{Blum:2020} ambiguous, or, in certain regimes even un-physical, with e.g. mass profiles with negative density in the outskirts.
Such a one-parameter extension to the previously considered more simple and rigid mass profiles may also not encompass the necessary flexibility beyond the pure MST that can affect kinematics observations \citep[e.g.,][]{Birrer:tdcosmoiv, Yildirim:2021}.
To make progress, the full degeneracy of the MST needs to be folded into flexible, but physically motivated, mass profile parameters, an approach explored by \citep{Shajib:2021slacs}, but not yet employed for time-delay cosmography.
The kinematics observations add additional potential systematics in the inference of $H_0$ when employed to break the MST. The primary limitation of the kinematics is the mass-anisotropy degeneracy \citep{Binney:1982}, as well as projection effects in the light and mass profile and de-projection assumptions employed, and rotation and ellipticity moments in the data. These assumptions have to be validated sufficiently to quarantee an unbiased interpretation of the mass density profiles and hence $H_0$ form time-delay cosmography.

\noindent
{\bf Selection effects.}
Strong lenses are inherently tracing a narrow and rare distribution of matter in the Universe. Quantifying the selection effects, including the differential selection effects among different samples of lenses, is going to be crucial to maintain accuracy in the years to come.
Selection effects can impact the line-of-sight distribution, the main deflector mass density and ellipticity, the galaxy properties of the deflector as well as of the source, and projection effects. Many of these effects can not precisely quantified on a lens-by-lens basis.\\
There are two approaches to mitigate selection effects. First, one can try to understanding selection from first principles, and explicitly account for the theoretical selection function in the analysis procedure. This approach requires extensive simulations and a reproducible selection function, including the discovery channel and follow-up decision.
Second, one can empirically measure selection functions from a set of observables at hand with assumptions of self-similarity among galaxies and line-of-sights with identical properties, such as stellar mass, morphology, redshift and environment, and explore empirical scaling relation among them.
With the anticipated large number of lenses in the near future, and the more uniform dataset of large and deep surveys, both approach will become feasible and we advocate analyses that take into account the specific discovery channel in the analysis.

The two limiting systematics, the mass profile assumptions and selection effects, result to uncertainties on the combined $H_0$ measurement of few per cent. Pinning down these systematics to sub-percent levels with new observations and methodology is a major current undertaking of the field.

\subsubsection{Main results}
\label{tdc:results}

The H0LiCOW collaboration \citep{Suyu:2017} inferred from the independent analysis of six lensed quasar systems \citep{Suyu:2010, Suyu:2013, Wong:2017, Bonvin:2017, Birrer:2019, Chen:2019, Rusu:2020} a Hubble constant value of \Ho$=73.3^{+1.7}_{-1.8}$ \Hunit, describing deflector mass density profiles by either a power law or stars (constant mass-to-light ratio) plus standard dark matter halos \citep{Wong:2020}.
This is a 2\% precision on H$_0$, in excellent agreement with the local distance ladder measurement by the SH0ES team \citep{Riess:2019, Riess:2021} and more than 3$\sigma$ statistical tension with early-Universe probes \citep[e.g.,][]{Planck2020, Aiola:2020}. 
The STRIDES collaboration presented an additional lens with the most precise single-lens measurement of \Ho$=74.2^{+2.7}_{-3.0}$ \Hunit with the same mass profile assumptions as the H0LiCOW collaboration \citep{Shajib:2020}.
\cite{Millon2020a} found, combining six lenses from H0LiCOW, SHARP and STRIDES, that the previous result are valid when assuming that all lenses are either one or the other of the two previously assumed forms of the mass density profile.
In sum, if the mass density profiles are well described by a power-law or a constant mass-to-light ratio plus a Navarro-Frank-White \citep[NFW,][]{NFW} dark matter halo\footnote{Imposing standard priors on the mass and concentration of the halo.}, and covariant assumptions and priors are negligible, the tension from the strong lensing measurements alone with early-Universe results is significant, corroborating other measurements, and new physics may be required.

The attention thus turned to relaxing the radial profile assumption (see Sect.~\ref{tdc:systematics}) and the covariant treatment of population priors that can not be constrained on a lens-by-lens basis. \cite{Birrer:tdcosmoiv} addressed the issue in the most direct way, by choosing a parameterization of the radial mass density profile that is maximally degenerate with H$_0$, via the MST. With this more flexible parameterization, H$_0$ is only constrained by the measured time delays and stellar kinematics, increasing the uncertainty on H$_0$ from 2\% to 8\% for the TDCOSMO sample of 7 lenses resulting in \Ho$=74.5^{+5.6}_{-6.1}$ \Hunit, without changing the mean inferred value significantly.

\begin{figure}[t!]
\centering
    \includegraphics[width=0.7\textwidth]{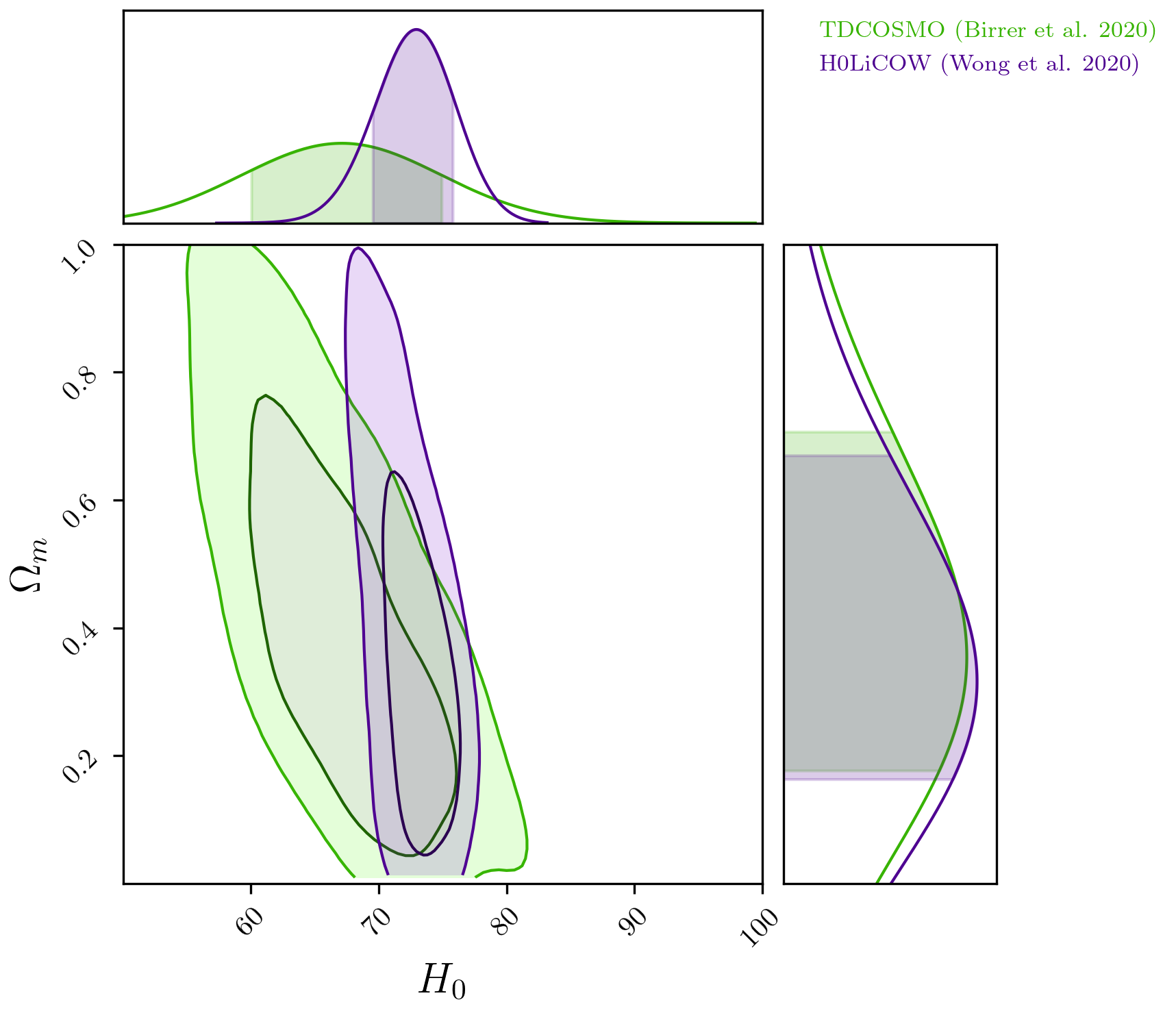}
    \caption{Comparison between current cosmological constraints with strongly lensed quasars on $H_0$-$\Omega_{\rm m}$ from H0LiCOW/TDCOSMO. Purple contours: Results from H0LiCOW by \cite{Wong:2020} based on six lensed quasars with assertive mass profile assumptions
    when averaging power-law and composite NFW plus stars (with constant mass-to-light ratio).
    Green contours: Results from TDCOSMO by \cite{Birrer:tdcosmoiv} with a maximally conservative assumption on the mass density profile constraining the MST solely by kinematics data. The constraints are based on seven lensed quasars (six in common with \cite{Wong:2020}) and one added from STRIDES by \cite{Shajib:2020}, as well as 33 SLACS lenses with imaging and kinematics data \citep{Bolton:2008, Shajib:2021slacs}.}
    \label{fig:TDC_contours}
\end{figure}

\cite{Birrer:tdcosmoiv} introduce a hierarchical framework in which external datasets can be combined with the time-delay lenses to improve the precision. They achieved a 5\% precision measurement on H$_0$ by combining the TDCOSMO lenses with stellar kinematic measurements of a sample of lenses from the Sloan Lens ACS (SLACS) survey with no time-delay information \citep{Bolton:2008, Auger:2009, Shajib:2021slacs}, and measure \Ho$=67.4^{+4.1}_{-3.2}$ \Hunit. The mean of the TDCOSMO+SLACS measurement is offset with respect to the TDCOSMO-only value, in the direction of the CMB value, although still statistically consistent given the uncertainties. The \cite{Birrer:tdcosmoiv} measurements are in statistical agreement with each other. The analysis by \cite{Birrer:tdcosmoiv} can not rule out the mass profile assumptions by earlier H0LiCOW/SHARP/STRIDES measurements with statistical significance. \cite{Birrer:tdcosmoiv} is also consistent, by construction, with the study by \citet{Shajib:2021slacs}, since they share the same measurements for SLACS. \citet{Shajib:2021slacs} concluded that using a mass profile combining an NFW profile for the dark matter component and stars \footnote{using wider priors on mass and concentration than earlier H0LiCOW/SHARP/STRIDES measurements} is a sufficiently accurate description of the mass density profile of the SLACS lenses. However, small departures from those forms are allowed by the data, resulting in the uncertainties quoted by \cite{Birrer:tdcosmoiv}. The shift in the mean could be real or it could be due to an intrinsic difference between the deflectors in the TDCOSMO and SLACS samples, arising from selection effects. For example, the two samples could be well matched in stellar velocity dispersion, but they differ in redshift, or the TDCOSMO sample could be source selected and composed mostly of quadruply imaged quasars, while SLACS is deflector selected and dominated by doubly imaged galaxies. In Fig.~\ref{fig:TDC_contours} is shown the comparison between the constraints on \Ho-\omegam~obtained from the H0LiCOW and TDCOSMO analyses.

\subsubsection{Outlook in the near future}
On the full sky, we expect to exist several 10,000 galaxy-galaxy lenses and several hundred quadruply lensed quasars \citep[e.g.,][]{Oguri:2010, Collett:2015}. With the upcoming wide and deep ground- and space-based surveys, we expect many of those to be discovered within a decade by the Vera Rubin Observatory \citep{LSSTScience:2009jmu}, Nancy Grace Roman Space Telescope \citep{Spergel:2015}, and Euclid \citep{Laureijs:2011gra}. This is an e-folding of the number of lenses possibly suitable for time-delay analyses compared to the current analyses conducted on few lenses (e.g., 7 lenses in case of current TDCOSMO results) and will transform the measurements and approaches in the domain of time-delay cosmography.
The first step in utilizing these lenses is to discover them in large datasets. 
The next step is to acquire all the necessary follow-up information, from monitoring data for a time-delay measurement, high-resolution imaging, to spectroscopic information about the source and lens redshift as well as velocity dispersion of the deflector. This step is going to be challenging with limited resources and there needs to be made decisions which lenses being excessively followed-up and which ones left aside.
Some lenses might require less substantial follow-up in case where Rubin light curves are good enough for a time-delay measurement, and or where high-resolution and sufficiently high signal-to-noise ratio data exists from wide field space surveys, such as Euclid or Roman.

The key to assess the need for follow-up and on which lenses to spend it, is to what extend these datasets impact the precision on \Ho.
Follow-up decisions, besides the limited resources, are currently also impacted by the accessible to adaptive optics (AO) coverage. With next-generation AO instrumentation on both hemispheres, we expect a full sky coverage of instrumentation that allows the community, at least from a technical view point, to target every single gravitational lens on the sky.

\begin{figure}[t!]
\centering
    \includegraphics[width=0.95\textwidth]{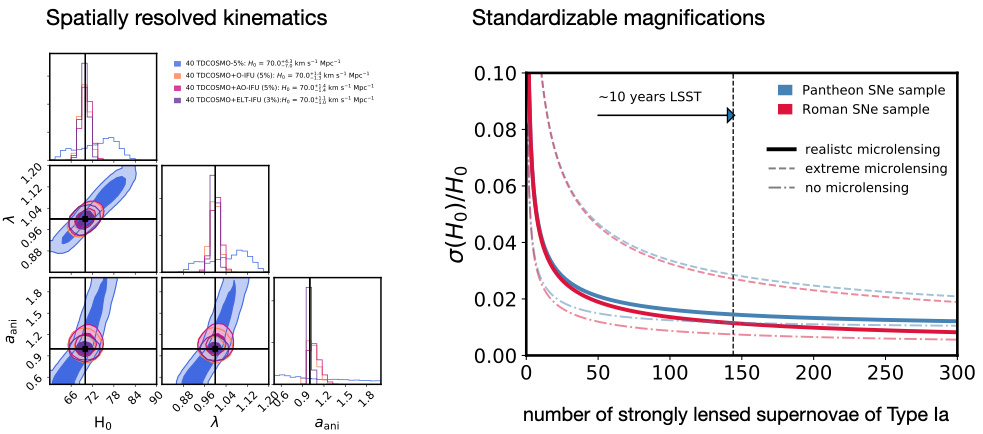}
    \caption{Forecast for \Ho~measurements in the near future with the upcoming ground- and space-based facilities. Left: Spatially resolved kinematics measurements of a sample of 40 time-delay lenses enable a precision on \Ho~of 1.5\% (Figure adopted from \cite{BirrerTreu:2021}). Right: Standardizable magnification measurements of $\sim$144 gravitationally lensed supernovae enable a precision on \Ho~of 1.5\% (Figure adopted from \cite{Birrer:2021glSNe}). Both approaches do constrain the MST with independent observations.}
    \label{fig:forecast}
\end{figure}

The dominant uncertainty in the current measurement of the Hubble constant with strong gravitational lensing time delays is attributed to uncertainties in the mass profiles of the main deflector galaxies.
There are several independent avenues of data available in the near future to approach a 1\% measurement of \Ho~that we focus in this section.

Spatially resolved kinematics of the deflector galaxy  with the next generation space \citep[JWST,][]{Gardner:2006} and ground-based (ELT's) instruments provides precise measurement of the kinematics and have the ability to break the mass-anisotropy degeneracy, a currently limiting systematic when using integrated kinematic measurements. \cite{BirrerTreu:2021} forecasts that with 40 time-delay lenses with exquisite spatially resolved kinematics, a 1.5\% precision on \Ho~can be achieved without relying on mass-density profile assumptions to break the MST, as shown in the left panel of Fig.~\ref{fig:forecast} \citep[see also, e.g.,][]{Yildirim:2020, Yildirim:2021}.
Resolved spectroscopy can also be employed on non time-delay lenses without bright and contaminating quasar images, which can further improve the kinematic measurement precision and enlarge the dataset \citep{Birrer:tdcosmoiv,  BirrerTreu:2021}.

Standardizable magnifications with gravitationally lensed supernovae (glSNe) provide another promising avenue to constrain the MST in the near future with the onset of Rubin.
As reported in the left plot of Fig.~\ref{fig:forecast}, \cite{Birrer:2021glSNe} provides a forecast with glSNe in constraining \Ho~independently of stellar kinematics. They conclude that the standardizable nature enables a 1.5\% \Ho~measurement with a 10 years Rubin survey. On the discovery, expected number of glSNe, the challenges of following them up, and the caveats of micro-lensing, we refer to \cite{Goldstein:2018, Foxley-Marrable:2018, Wojtak:2019, Goldstein:2019, Huber:2021, Birrer:2021glSNe}.

In summary, in the next decade with an increasing of the number of lenses and the improved data quality, a $\sim$1\% measurement of the Hubble constant becomes feasible, when also major efforts in the validation and possible covariant systematics are being invested in.

\clearpage

\subsection{Cosmography with Cluster Strong Lensing}
\label{sec:CCSL}

While Sect.~\ref{sect:tdc} considers strong lensing effects produced by galaxy-scale lenses on intrinsically variable sources, this section focuses on much more massive structures in the Universe, galaxy clusters. In particular, we illustrate the principles of cluster strong lensing cosmography. 

\subsubsection{Basic idea and equations}
\label{CCSL:Biae}

For simplicity, we use the thin-screen approximation, i.e. we assume that the lens total mass distribution is confined on a plane, called the lens plane. In addition, we assume a single lens plane. The equations described in Sect.~\ref{tdc_basics} remain valid in this context. The measurement of relative time-delays between the multiple images of intrinsically variable sources lensed by galaxy clusters can be used to constrain cosmological parameters such as \Ho.

Due to their large mass, galaxy clusters can have large cross sections for strong lensing. The size of these cross sections depends on several properties of the lenses, including their total mass, dynamical state, ellipticity and asymmetry \citep{Torri:2004,Hennawi:2007,Meneghetti:2007,Meneghetti:2010a}. It is not uncommon that massive clusters strongly lens several tens of background sources simultaneously \citep[e.g.,][]{Postman:2012,Lotz:2014,Coe:2019,Steinhardt:2020,Caminha:2017b,Caminha:2019, Lagattuta:2019, bergamini:2021}. In this case, additional constraints on the cosmological parameters can be set, even with sources that are not intrinsically variable and for which relative time-delays cannot be measured. 

Equations~\ref{eqn:lens_equation}, \ref{eqn:lens_equation_single_plane}, and \ref{eq:anglepot} show that the difference between the observed and intrinsic positions of a source whose light is deflected by a gravitational lens is the product of two factors. The first factor is the deflection angle $\hat{\boldsymbol\alpha}(\boldsymbol\theta)$, which is proportional to the two-dimensional gradient of the integral of the lens Newtonian gravitational potential along the line-of-sight:
\begin{equation}
\hat{\boldsymbol{\alpha}}(\boldsymbol{\theta}) = \frac{2}{c^2}\boldsymbol{\nabla}\int \Phi(\boldsymbol\theta,l) {\rm d}l \;\; .
\end{equation} 
Thus, the deflection angle depends on the lens total mass distribution.  

The second factor is the ratio between the angular diameter distances $D_{\rm ds}$ and $D_{\rm s}$. In a flat cosmological model, the angular diameter distance to redshift $z$ is given by:
\begin{equation}
D_{\rm A}(z) = \frac{c}{H_0}\frac{1}{1+z}\int_0^z\frac{{\rm d}z}{[\Omega_{\rm m}(1+z)^3+(1-\Omega_{\rm m})(1+z)^{3(1+w)}]^{1/2}} \;\; ,
\end{equation}
where $w$ is the EoS parameter for the dark energy. Thus, the angular diameter distance depends on the values of cosmological parameters, such as \Ho, \omegam, and $w$. The ratio of two angular diameter distances does not depend on \Ho.

For simplicity, we consider a circular symmetric lens and choose to measure the angular positions $\boldsymbol\theta$ and $\boldsymbol\beta$ with respect to the lens center. The deflection angle for any circular-symmetric mass distribution is:
\begin{equation}
\hat{\boldsymbol{\alpha}}(\boldsymbol\theta) = \frac{4GM(|\boldsymbol\theta|)}{c^2D_{\rm d}|\boldsymbol\theta|^2}\boldsymbol\theta \;\; .
\label{eq:circ_angle}
\end{equation}

\begin{figure}[t!]
    \centering
    \includegraphics[width=.8
    \columnwidth]{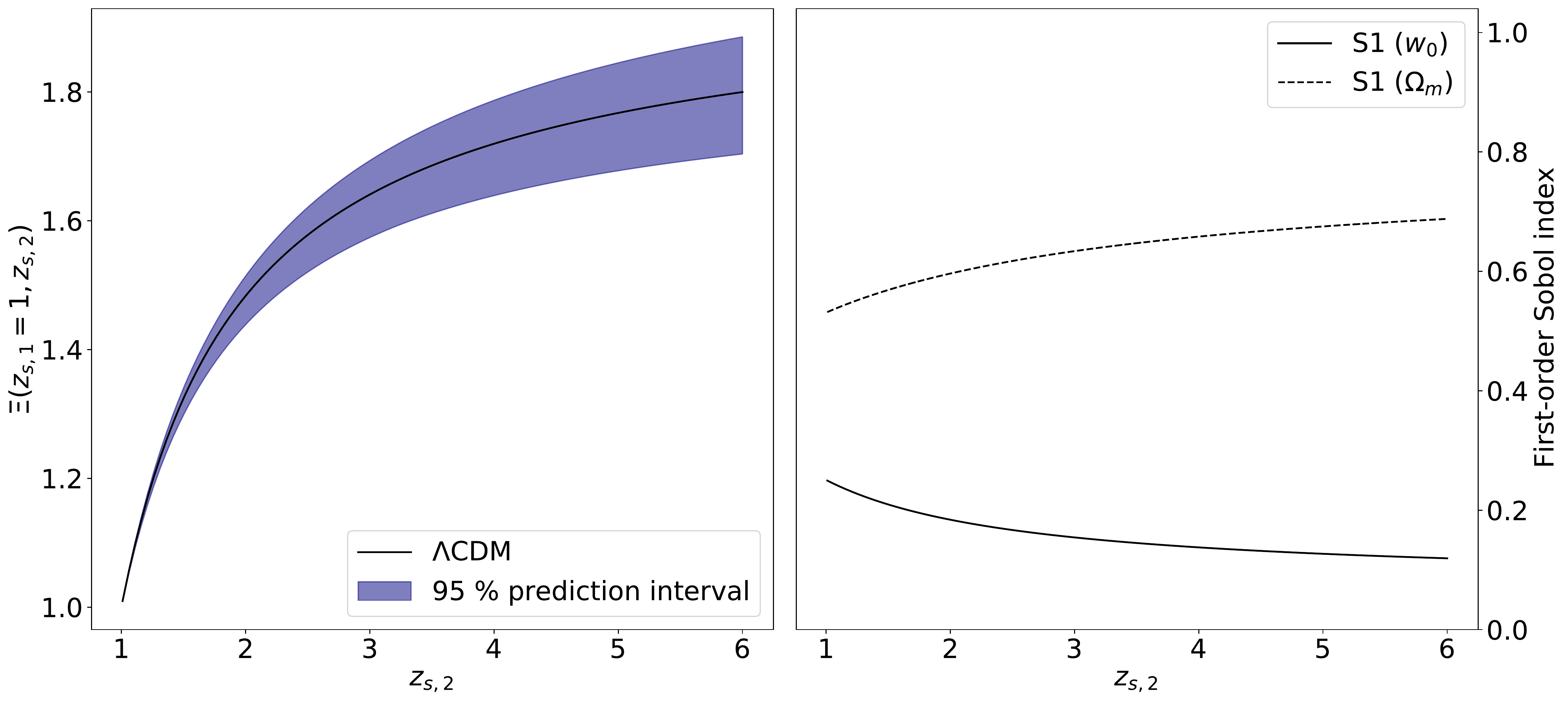}
    \caption{\label{fig:sensitivity_famr} Sensitivity of the family ratio to the values of the cosmological parameters \omegam~and $w_0$. The left panel shows the family ratio for a lens at redshift $z_{\rm d}=0.5$ in a flat $\Lambda$CDM cosmological model with \omegam=0.3, $w(t)=w_0=-1$, and \Ho=70 \Hunit. We assume $z_{\rm s,1}=1$. The solid black curve describes how the family ratio varies as a function of the second source redshift, $z_{\rm s,2}$. The shaded blue region indicates the 95\% prediction interval estimated by sampling the parameter plane \omegam-$w_0$ assuming uniform priors (\omegam$\in [0.1,1]$ and $w_0\in [-2,-0.5]$). The right panel shows the results of the Sobol's sensitivity analysis. We show the first-order Sobol index for both parameters as a function of $z_{\rm s,2}$.
    }
\end{figure}

Inserting Eq.~\ref{eq:circ_angle} into Eq.~\ref{eqn:lens_equation_single_plane}, we obtain that the image of a source perfectly aligned with the lens and the observer ($\boldsymbol\beta = 0$) is a ring, whose angular size is:
\begin{equation}
\theta_{\rm E}(z_{\rm d},z_{\rm s}) =  \sqrt{\frac{4GM[\theta_{\rm E}(z_{\rm d},z_{\rm s})]}{c^2}\frac{D_{\rm ds}}{D_{\rm d}D_{\rm s}}} \;\; .
\end{equation}
This radius is called {\it Einstein radius}. The mass $M(\theta_{\rm E})$ is the projected mass enclosed by the ring.

In the case of two sources at redshifts $z_{\rm s,1}$ and $z_{\rm s,2}$ aligned with the lens and the observer, the ratio of the corresponding Einstein radii is given by:
\begin{equation}
\frac{\theta_{\rm E}(z_{\rm d},z_{\rm s,1})}{\theta_{\rm E}(z_{\rm d},z_{\rm s,2})} = \sqrt{\frac{M[\theta_{\rm E}(z_{\rm d},z_{\rm s,1})]}{M[\theta_{\rm E}(z_{\rm d},z_{\rm s,2})]} \frac{D_{\rm ds}(z_{\rm d},z_{\rm s,1})}{D_{\rm s}(z_{\rm s,1})}\frac{D_{\rm s}(z_{\rm s,2})}{D_{\rm ds}(z_{\rm d},z_{\rm s,2})}} \;\; .
\end{equation} 
The function:
\begin{equation}
\label{family}
    \Xi(z_{\rm d},z_{\rm s,1},z_{\rm s,2}) = \frac{D_{\rm ds}(z_{\rm d},z_{\rm s,1})}{D_{\rm s}(z_{\rm s,1})}\frac{D_{\rm s}(z_{\rm s,2})}{D_{\rm ds}(z_{\rm d},z_{\rm s,2})}
\end{equation}
is called the {\it family ratio}, and depends on the values of cosmological parameters, such as \omegam~and $w$. This result holds also in the case of sources not perfectly aligned with the lens and the observer, or for lenses whose total mass distribution is not circular. The general principle is that the relative positions of multiple image families depend both on the lens mass distribution and the family ratios. 

\begin{figure}[t!]
    \centering
    \includegraphics[width=.8\columnwidth]{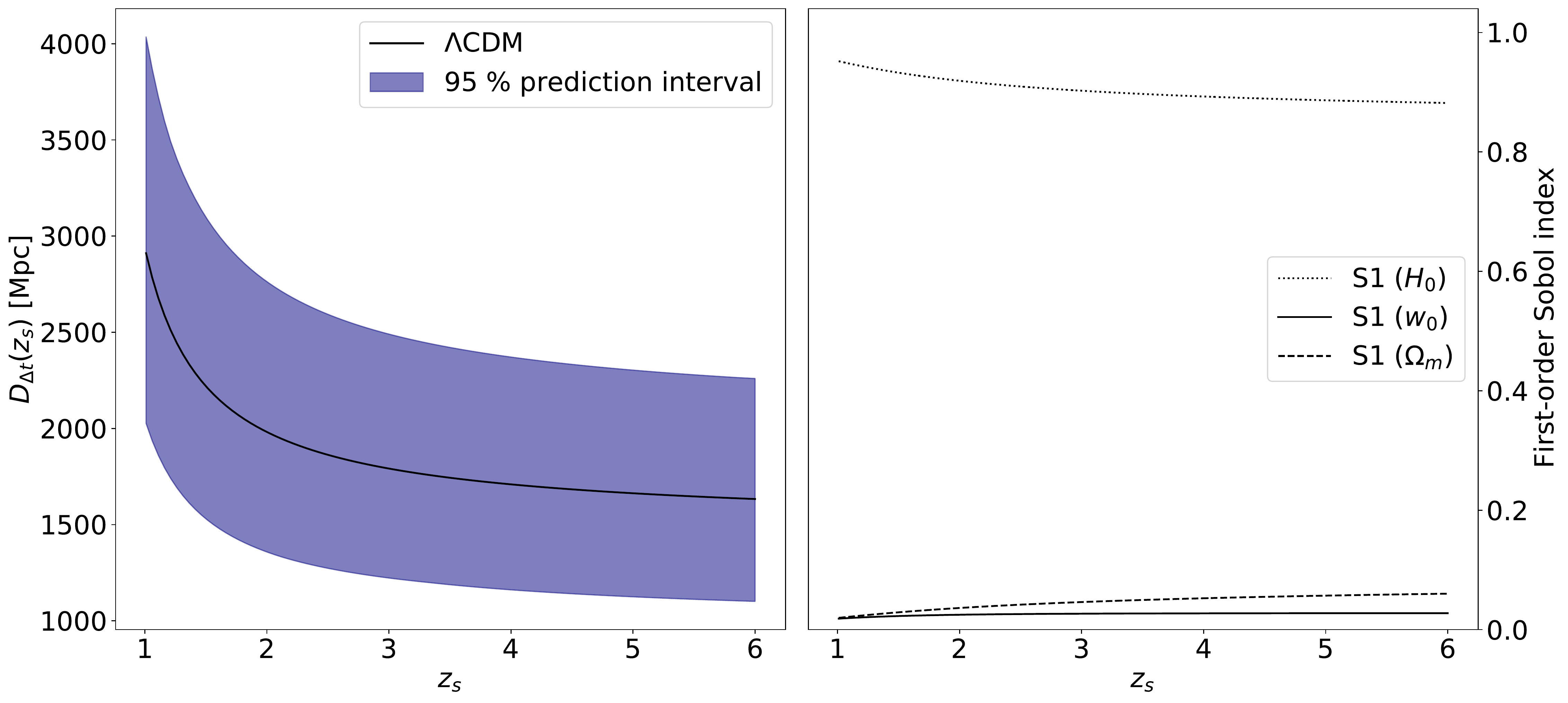}
    \caption{\label{fig:sensitivity_td} Similar to Fig.~\ref{fig:sensitivity_famr}, but showing the sensitivity of the time-delay distance, $D_{\Delta t}(z_{\rm s})$, to the values of the cosmological parameters \Ho, \omegam, and $w_0$.
    }
\end{figure}

In the left panel of Fig.~\ref{fig:sensitivity_famr}, we show the family ratio for a lens at redshift $z_{\rm d}=0.5$ in a flat $\Lambda$CDM cosmological model with \omegam = 0.3, $w(z)=w_0=-1$, and \Ho = 70 \Hunit. We assume $z_{\rm s,1}=1$. The solid black curve describes how the family ratio varies as a function of the second source redshift $z_{\rm s,2}$. The shaded blue region indicates the 95\% prediction interval estimated by sampling the parameter plane \omegam-$w_0$ using the Saltelli's scheme \citep{saltelli:2002}. We assume uniform priors on the cosmological parameters, with \omegam$\in [0.1,1]$ and $w_0\in [-2,-0.5]$. Performing a Sobol's sensitivity analysis \citep{sobol:2001,saltelli:2010}, we find that $\sim 60-70\%$ of the variance of the family ratio is due to the variance of \omegam, as indicated by the first-order Sobol index $S1$ plotted in the right panel. The contribution of the $w_0$ variance amounts to $\sim 10-25\%$, while the remainder of the variance is due to second-order interactions between \omegam~and $w_0$. Thus, the family ratio is primarily sensitive to \omegam, but it is also sensitive to the dark energy equation of state. 

As shown in Fig.~\ref{fig:sensitivity_td}, performing a similar analysis for the time-delay distance (assuming again flat priors on the cosmological parameters, with \Ho $\in [50,100]$ \Hunit, \omegam $\in [0.1,1]$ and $w_0\in [-2,-0.5]$), we find that the variance of this quantity is mostly contributed by the variance of \Ho~($\sim 90-95\%$), while the sensitivity to other cosmological parameters is much weaker. Thus, these results suggest that the family ratio and the time-delay distance are highly complementary cosmological probes.

\begin{figure}[t!]
    \centering
    \includegraphics[width=0.5\columnwidth]{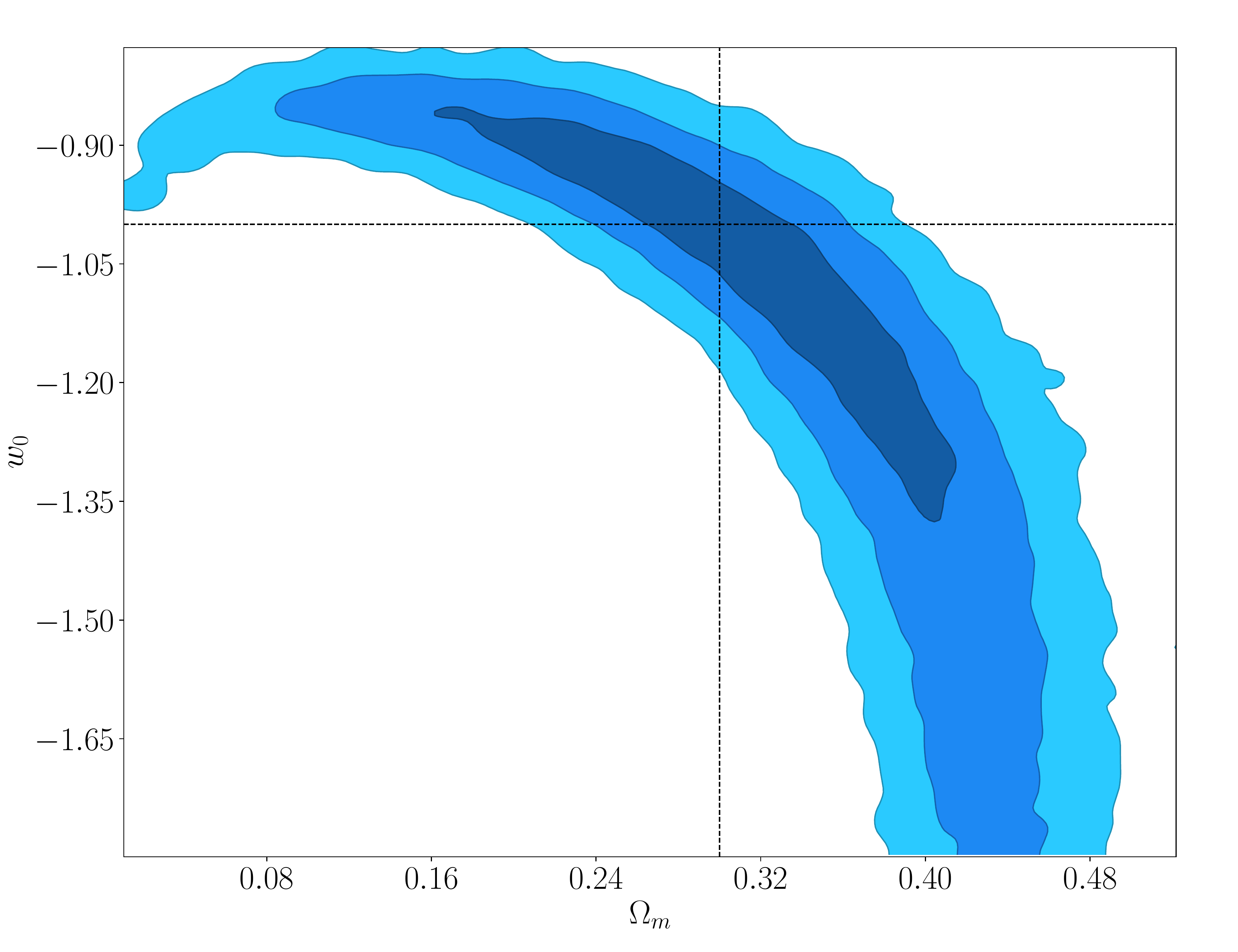}
    \caption{\label{fig:degeneracy} Degeneracy between the parameters $w_0$, and \omegam, derived by fitting 45 family ratios (obtained by combining 10 multiply imaged sources uniformly distributed between $z=1$ and $z=6$). The dashed lines indicate the true values of the cosmological parameters.
    }
\end{figure}

The existing degeneracy between the parameters $w_0$ and \omegam~estimated from the family ratios of several multiply imaged sources is illustrated in Fig.~\ref{fig:degeneracy}. We assume to fit 45 family ratios obtained by combining 10 multiply imaged sources uniformly distributed in redshift between $z=1$ and $z=6$. The confidence contours (at 1, 2, and 3$\sigma$) do not account for the uncertainties related to lens modeling (see the discussion in Sect.~\ref{CCSL:results}). As we see, the degeneracy is strong: we obtain similar family ratios in cosmologies with high value of $w_0$ and low value of \omegam~and vice-versa. Breaking the degeneracy requires increasing the number of constraints by either accumulating a larger number of multiple image families by means of deeper observations of single clusters or stacking multiple lenses \citep{2009MNRAS.396..354G}.

\subsubsection{Sample selection}

Currently, only five lens galaxy clusters with multiple images of time-varying sources (3 QSOs and 2 SNe) and measured time-delays are known. Systematic searches for gravitationally lensed quasars over a range of angular separations have ramped up in the last 20 years with the availability of the Sloan survey. The SDSS Quasar Lens Search \citep[SQLS,][]{SQLS} used a combination of morphological and color selection criteria applied to a SDSS sample of spectroscopically confirmed QSOs to find over 200 candidate strongly lensed QSOs. A few of these were found with angular separations exceeding 10\arcsec, characteristic of group-cluster scale lenses;  remarkable examples are SDSS~J1004$+$4112 \citep{sdss1004}, a five-image lensed QSO with a separation of $\sim\! 15\arcsec$, and SDSS J1029$+$2623 \citep{sdss1029}, a three-image system with the largest known separation to date ($\sim\! 23\arcsec$). The SLOAN Giant Arc Survey \citep{Hennawi08}, a study largely based on visual inspection of strong lensing features around massive clusters, has discovered SDSS J2222$+$2745 \citep{sdss2222}, with six detected images of a QSO at $z=2.2$ with a separation of $\sim\! 15\arcsec$. More advanced methods to search for multiply lensed quasars based on machine-learning techniques, which work directly on image cutouts using neural network pattern recognition methods, have been developed over the last years \citep[e.g.][]{Agnello15,Petrillo2017,Metcalf2019,Canameras2020}, and are being applied to new wide-area optical surveys such as the Dark Energy Survey \citep{Huang2020}. Machine-learning methods have also recently been applied to search for  strongly lensed quasars selected from Gaia catalogs, in combination with near-IR surveys to identify likely lenses \citep[e.g.][]{Stern21}. These techniques have initially been developed to discover lensed QSO systems with small separations (a few arcsec), however they can be easily extended to cluster-scale lenses.   

The Vera C. Rubin Observatory project \citep{LSSTScience:2009jmu} will discover hundreds of new multiply imaged QSOs and SNe (a few tens of which will be lensed by galaxy clusters) and will measure their time-delays \citep{Oguri:2010}. The latter can require time consuming monitor campaigns, particularly in case of cluster-scale lenses. 

By assuming a singular isothermal sphere (SIS) profile for the total mass distribution of the lens (cluster), the time-delay between the two multiple images of the same background source can vary between 0 and a maximum value given by:
\begin{equation}
    \Delta t_{\rm SIS,max} = \frac{1+z_{\rm d}}{c^5}\, \frac{D_{\rm d}\,D_{\rm ds}}{D_{\rm s}}\, 32\pi^2\sigma_{\rm SIS}^4=
    127.5\,(1+z_{\rm d})\left( \frac{D_{\rm d}\,D_{\rm ds}}{D_{\rm s}\,{\rm 1\,Gpc}} \right)\bigg( \frac{\sigma_{\rm SIS}}{\rm 1000\,km\,s^{-1}}\bigg)^4\,{\rm yr} \;\; ,
\end{equation}
where $\sigma_{\rm SIS}$ is the value of the effective velocity dispersion associated to the isothermal total mass profile. 
Alternatively, one can write: 
\begin{equation}
    \Delta t_{\rm SIS,max} = \frac{D_{\Delta t}}{c}\, 2\theta_{\rm E}^2=
    \frac{D_{\Delta t}}{L_H}\, \frac{2}{H_0}\theta_{\rm E}^2 = 
    \left(\frac{D_{\Delta t}}{L_H}\right)\, 0.66\, \theta_{\rm E}^2({\rm arcsec})\; {\rm yr}\;\; , 
\end{equation}
where $D_{\Delta t}$ is the time delay distance (Eq.~\ref{eqn:ddt_definition}), $L_H=c H_0^{-1}$ is the Hubble length, so that $D_{\Delta t}/L_H\lesssim 1$, and $\theta_{\rm E}$ the Einstein radius, ranging from a few arcsec to $\sim\! 15\arcsec$ (\Ho$=70$ \Hunit is adopted).  

Cluster-scale multiply lensed quasars \citep[e.g.][]{Dahle2015,Fohlmeister2013}, as well as the multiple images of SN Refsdal \citep[the first multiply-imaged and spatially resolved supernova,][]{Kelly:2015} and SN Requiem \citep{Rodney:2021} do indeed show model-predicted and measured time-delays spanning from a few days to years and tens of years.

\subsubsection{Measurements}

To constrain cosmology using the strong lensing cosmography approach, one has to simultaneously fit as many strong lensing constraints as possible, using a model that incorporates the cluster mass distribution, and the cosmological parameters. This process is called {\it lens inversion}. There are two general classes of inversion algorithms. A first approach is called free-form, wherein the cluster is subdivided into a mesh on to which the lensing observables are mapped, and which is then transformed into a pixelized mass distribution. Other methods comprise parametric models \citep[e.g.][]{kneib:1996,jullo:2007,jullokneib:2009}, wherein the mass distribution is reconstructed by combining clumps of matter on different scales. One or more large-scale mass components are used to describe the diffuse cluster dark matter halo. They are often positioned where the brightest cluster galaxies are located. The other cluster galaxies are used to trace the cluster substructure. Both the large and small scale mass components have density profiles given by analytic functions. 

Using either of these approaches, the cluster mass distribution is described by a set of parameters (which can be a set of pixel values or parameters describing the shape and density profiles of each mass clump). Let $\mathbf{p}$ be the totality of the parameters used to model the cluster mass distribution, and $\mathbf{p_{\rm cosmo}}$ the cosmological parameters we want to estimate.

The strong lensing constraints are generally in the form of positions of multiple images of several sources. These images are identified based on the morphology and color similarities of the lensed features. Gravitational lensing conserves the source surface brightness, implying that several source properties (e.g., star forming regions, spiral arms, bulges, etc.) can be recognized in all their multiple images. The geometry of the lens, inferred from the spatial distribution of the cluster galaxies, is useful to find counter-images of a given source. Typically, the cluster galaxies in the central regions of galaxy clusters are early-type galaxies, most of which can be recognized because they populate a red sequence in the color-magnitude diagram. More sophisticated methods to identify these galaxies and separate them from foreground and background sources also include deep-learning models trained using multi-band images \citep{angora:2020}. Thus, finding multiple images and cluster members requires high resolution multi-band imaging observations that only the Hubble Space Telescope (HST) can currently deliver. 

Candidate multiple images can be confirmed by verifying that they have similar spectra. Spectroscopy is also crucial to measure the redshifts of lenses and sources. Without redshifts it is impossible to convert angular scales into physically meaningful units. 

The multiple images of the same source form a {\it family}. Each family provides some constraints on the lens deflection field. Indeed, given a source at the intrinsic angular position $\boldsymbol{\beta}$, the positions of its images, $\boldsymbol{\theta}_i$, satisfy the lens equation (Eq.~\ref{eqn:lens_equation_single_plane}). In the case of intrinsically variable sources, such as Supernovae or QSOs, we can derive additional constraints by measuring the relative time-delays between the multiple images. In the following equations, we assume that both positional and time-delay measurements are available for the lensing analysis. Nevertheless, we remark once more that the strong lensing cosmography approach can be used to estimate the values of cosmological parameters, such as those of \omegam, \omegade, and $w$, even without measuring time-delays.

The cluster potential and the cosmological model are constrained simultaneously by maximizing the posterior probability distribution:
\begin{equation}
    P(\boldsymbol{\theta^{\rm obs}}\frown\mathbf{\Delta t^{obs}}|\mathbf{p}\frown \mathbf{p}_{\rm cosmo}) \propto  P(\mathbf{p}\frown \mathbf{p}_{\rm cosmo}|\boldsymbol{\theta^{\rm obs}}\frown\mathbf{\Delta t^{obs}})\cdot P(\mathbf{p} \frown \mathbf{p}_{\rm cosmo}) \;\; ,
\end{equation}
where $\boldsymbol{\theta^{\rm obs}}$ and $\mathbf{\Delta t^{obs}}$ are the observed positions and relative time-delays of the multiple images, respectively. The symbol $\frown$ denotes the concatenation operator. The model likelihood is given by:
\begin{equation}
    \mathcal{L}(\mathbf{p}\frown\mathbf{p}_{\rm cosmo}|\boldsymbol{\theta^{\rm obs}}\frown\mathbf{\Delta t^{obs}}) \propto \exp{\left[-\frac{1}{2}\chi^2(\mathbf{p}\frown\mathbf{p}_{\rm cosmo})\right]} \;\; .
\end{equation}
Since the datasets (positions and time-delays) are independent, the likelihood is separable. Thus the $\chi^2(\mathbf{p}\frown\mathbf{p}_{\rm cosmo})$ function is the sum of two terms. The first quantifies the separation between the observed and the model-predicted multiple image positions:
\begin{equation}
    \label{eq.: chi_lt}
    \chi^{2}_{\rm pos}(\mathbf{p}\frown\mathbf{p}_{\rm cosmo}) = \sum_{i=1}^{N_{fam}} \sum_{j=1}^{n_{i}}\left(\frac{\left\|\boldsymbol{\theta^{\rm obs}}_{i, j}-\boldsymbol{\theta^{\rm pred}}_{i, j}(\mathbf{p}\frown\mathbf{p}_{\rm cosmo})\right\|}{\sigma_{\boldsymbol{\theta}_{i, j}}}\right)^{2} \;\; ,
\end{equation}
where $\boldsymbol{\theta^{\rm obs}}_{i, j}$ and $\boldsymbol{\theta^{\rm pred}}_{i, j}$ are the observed and model-predicted positions of the $j$-th multiple image belonging to the $i$-th family, $N_{fam}$ is the total number of multiple image families, and $n_{i}$ is the number of multiple images belonging to the $i$-th family. The uncertainty on the image positions, $\sigma_{\boldsymbol{\theta}_{i, j}}$, is generally unknown. It depends not only on the effective resolution of the observations (i.e. the pixel scale and the size of the Point-Spread-Function), but also on several properties of the lens not directly accounted for in the lens model (such as unseen substructures in the cluster or along the line-of-sight or asymmetries of the dark matter distribution). Generally, this uncertainty is scaled to obtain a value of reduced $\chi^2$ of $\sim 1$ \citep{bergamini:2021}.

The second term quantifies the difference between the observed and model-predicted relative time-delays:
\begin{equation}
    \chi^2_{\rm td}(\mathbf{p}\frown\mathbf{p}_{\rm cosmo}) = \sum_i^{N_{fam,{\rm td}}}\sum_j^{n_{i,{\rm td}}-1} \left(\frac{\left\| \Delta t_{i,j}^{\rm obs} - \Delta t_{i,j}^{\rm pred}(\mathbf{p}\frown\mathbf{p}_{\rm cosmo}) \right\|}{\sigma_{\Delta t_{i,j}}}\right)^2 \;\; ,
\end{equation}
where $N_{fam,{\rm td}}$ is the number of families of multiple images for measured time-delays, $n_{i,{\rm td}}$ is the number of multiple images of the $i$-th family (note that this implies $n_{i,{\rm td}}-1$ relative time-delays measurements after choosing the $n_{i,{\rm td}}$-th image as reference), $\Delta t_{i,j}^{\rm obs}$ and $\Delta t_{i,j}^{\rm pred}$ are the observed and model-predicted relative time-delays of the $j$-th multiple image belonging to the $i$-th family, and $\sigma_{\Delta t_{i,j}}$ is the error of the time-delay $\Delta t_{i,j}^{\rm obs}$.

If the lens model and the cosmology are constrained by the positions of $N_{im}^{tot}=\sum_{i=1}^{N_{fam}} n_{i}$ observed multiple images and $N_{\rm td}^{tot}=\sum_{i=1}^{N_{fam,td}} (n_{i,{\rm td}}-1)$ relative time-delay measurements, by defining $N_{par}$ as the total number of model free parameters, we can write the number of degrees-of-freedom (DoF) of the lens model as:
\begin{equation}
    \label{eq.: DoF}
    \mathrm{DoF} = 2 \times N_{im}^{tot} + N_{\rm td}^{tot} - 2 \times N_{fam} - N_{par} = N_{con}-N_{par} \;\;.
\end{equation}
The term $2 \times N_{fam}$ stems from the fact that the unknown positions of the $N_{fam}$ background sources (2 coordinates for each of them) are additional free parameters of the model. Thus, $N_{con}$ is the effective number of available constraints.

\subsubsection{Systematic effects}

As described in Sect.~\ref{sect:tdc}, the strong lens time-delay method has been successfully utilized with quasars lensed by galaxies. Several studies \citep[e.g.,][]{Birrer:2016, Treu:2016, Suyu:2017} have recognized that, in addition to the spectroscopic redshifts of the lens and the source, the most important steps toward accurate and precise cosmological measurements are: {\it i)} precise time-delays, {\it ii)} high-resolution images of the lensed sources, {\it iii)} precise stellar kinematics of the lens galaxy, and {\it iv)} detailed information about the lens environment. Long-term monitoring campaigns of lensed quasars with optical, notably by the COSMOGRAIL collaboration \citep[e.g.,][]{Tewes:2013a, Courbin:2018}, or radio \citep[e.g.,][]{Fassnacht:2002} telescopes, together with advances in light-curve analyses \citep[e.g.,][]{Tewes:2013b, Hojjati:2013}, have provided precise time-delays. To convert these delays to cosmologically relevant quantities, an accurate lens mass model is needed, particularly concerning its radial total mass density profile. Steeper profiles yield larger Fermat potential differences between two images, resulting in larger inferred values of \Ho~\citep{Wucknitz:2002, Kochanek:2002}. In addition to the main lens, there could be other mass contributions, associated to galaxies belonging to the same group/cluster of the main lens or to line-of-sight structures. If not properly accounted for, this term represents an important source of systematic error, the so-called {\it mass-sheet degeneracy} \citep{Falco:1985, Schneider:2013}, in the model prediction of the time-delays. This clarifies why the extended reconstruction of multiple images, the use of independent mass diagnostics \citep[e.g., stellar dynamics; see][]{Treu:2002} for the main lens, and a detailed characterization of its environment (i.e., points {\it ii)}, {\it iii)}, and {\it iv)} listed above) are so relevant to a very accurate total mass model of the lens, thus to the success of this cosmological probe \citep[e.g.,][]{Suyu:2014, Birrer:2016, McCully:2017, Rusu:2017, Sluse:2017, Shajib:2018, Tihhonova:2018}.

Despite being more complex than that of an isolated galaxy, the strong lensing modeling of a galaxy cluster presents some advantages. First, the identification of several multiple images, some of which might be very close to the cluster center and radially elongated, provides important information about the slope of the cluster total mass density profile \citep[see, e.g.,][]{Caminha:2017b}. Second, the frequent observations of pairs of angularly close multiple images from sources at different redshifts \citep[see, e.g.,][]{Grillo:2016} locate very precisely the positions of the lens tangential critical curves, thus resulting in precise calibrations of the projected total mass of the cluster within different apertures. These facts reduce the need to rely on different total mass diagnostic, such as stellar dynamics in lens galaxies. Moreover, the large number of secure and spectroscopically confirmed multiple images observed in galaxy clusters allows one to choose the best mass model among the different tested ones \citep[i.e., the best reconstruction of the cluster mass components; see][]{Grillo:2015, Grillo:2016}, according to the value of the minimum $\chi^2$. As shown in \cite{Grillo:2015, Grillo:2016}, it is remarkable that all considered mass models lead to statistical and systematic relative errors of only a few percent for the cluster total mass. Very good agreement has also been found with the measurements from independent total mass diagnostics, e.g. those from weak lensing, dynamical and and X-ray observations \citep[see, e.g.,][]{Grillo:2015, Balestra:2016, Caminha:2017b}. In addition, in a galaxy cluster, the modeling of its different mass components \citep[i.e., extended dark-matter haloes, cluster members, and possibly hot gas; see, e.g.,][]{Bonamigo:2017, Bonamigo:2018, Annunziatella:2017} provides a good first-order approximation of possible additional lensing effects (i.e., of the environment) in the regions adjacent to where the time-delays can be measured. Some recent studies have exploited kinematic data for the cluster members to model more realistically their total mass contribution through scaling relations with non-zero scatter or information from the Fundamental Plane relation \citep[e.g.,][]{Bergamini:2021b, Granata:2021}. In summary, if extensive multi-color and spectroscopic information is available in lens galaxy clusters, robust mass maps can be constructed \citep[see][]{Grillo:2015, Caminha:2017a, Lagattuta:2017}. The feasibility of using the measured time-delays of the first multiply-imaged and spatially-resolved supernova (SN ``Refsdal'') for measuring \Ho~with high statistical precision has been demonstrated \citep{Grillo:2018}, and a full systematic analysis has been performed \citep{Grillo:2020}. Adding to the model a uniform sheet of mass at the cluster redshift or a cluster main mass density profile with a variable slope (optimized together with all the other model parameters), result in \Ho~probability distribution functions that are just slightly broader than those without these extra model parameters. Based on our previous studies \citep[see, e.g.,][on the influence of mass structures along the line of sight on lensing modeling]{Chirivi:2018}, systematic effects in lens galaxy clusters seem to be controlled to a level similar to or even lower than the statistical uncertainties, given the exquisite datasets in hand and soon becoming available, making time-delay cluster cosmography a potentially very competitive method.

Finally, we remark that in any cluster strong lensing model the values of the cosmological parameters and those defining the mass distribution of the lens are not independent, and they cannot be considered separately in obtaining model-predicted quantities (e.g., the time-delays, positions, and flux ratios of the multiple images). Therefore, the results obtained by simplistically keeping the cluster mass distribution fixed are likely to underestimate the uncertainty on the values of the cosmological parameters, and possibly introduce biases, since they neglect the covariance between the cosmological and cluster mass model parameters \citep[see, e.g.,][]{Acebron:2017}. \cite{Zitrin:2014} confirm that the values of the cosmological parameters are biased when they are estimated by applying a fixed cluster mass distribution for correcting the luminosity distances of lensed SNe Ia.

\subsubsection{Main results and forecasts}
\label{CCSL:results}

As detailed in Sect.~\ref{CCSL:Biae}, time-delay distances are primarily sensitive to the value of \Ho, and more mildly to those of other cosmological parameters. In galaxy clusters, usually showing several multiple images, different values of the family ratio (see Eq.~\ref{family}) can be used at the same time to add constraints on the values of the cosmological matter (\omegam) and dark-energy (\omegade) density parameters, defining the global geometry of the Universe. In general, the cosmological contribution is difficult to disentangle from that associated to the total mass of a lens, because of a strong degeneracy between the two. However, when a significant number of multiply lensed sources (with spectroscopic redshifts spanning a wide range) is present, valuable information about the cosmological parameters can be inferred. This technique has been applied without time-delay measurements in the galaxy clusters Abell 2218 \citep{Soucail:2004}, Abell 1689 \citep{Jullo:2010} and, more recently, RXC J2248.7$-$4431 \citep{Caminha:2016}, and in combination with time-delay measurements in MACS J1149.5+2223 for the first time \citep[see Fig.~3 of][]{Grillo:2018}.

In \cite{Caminha:2016}, by exploiting the observed positions of 47 multiple images, 24 of which spectroscopically confirmed, from a total of 16 background sources over the redshift range 1.0-6.1, a comprehensive study of the total mass distribution of the galaxy cluster RXC J2248.7$-$4431 with a set of high precision strong lensing models has resulted into measurements (from lensing only) of the values of \omegam~and $w_0$ with, respectively, between $\sim\,$40\% and $\sim\,$60\%, depending on the adopted cosmological model, and $\sim\,$30\% (1$\sigma$) statistical uncertainties. In \cite{Caminha:2021}, thanks to a sample of five detailed cluster total mass models, it has been demonstrated that strong lensing measurements of the values of the cosmological parameters are complementary and in good agreement with the estimates from the CMB, BAO, and SNe Ia. In particular, the strong lensing cosmographic analysis has allowed to improve the constraints from the CMB on the values of \omegam~and $w_0$ (in a flat $w$CDM model) by factors of 2.5 and 4.0, respectively.

\begin{figure}[b!]
    \centering
    \includegraphics[width=.8\columnwidth]{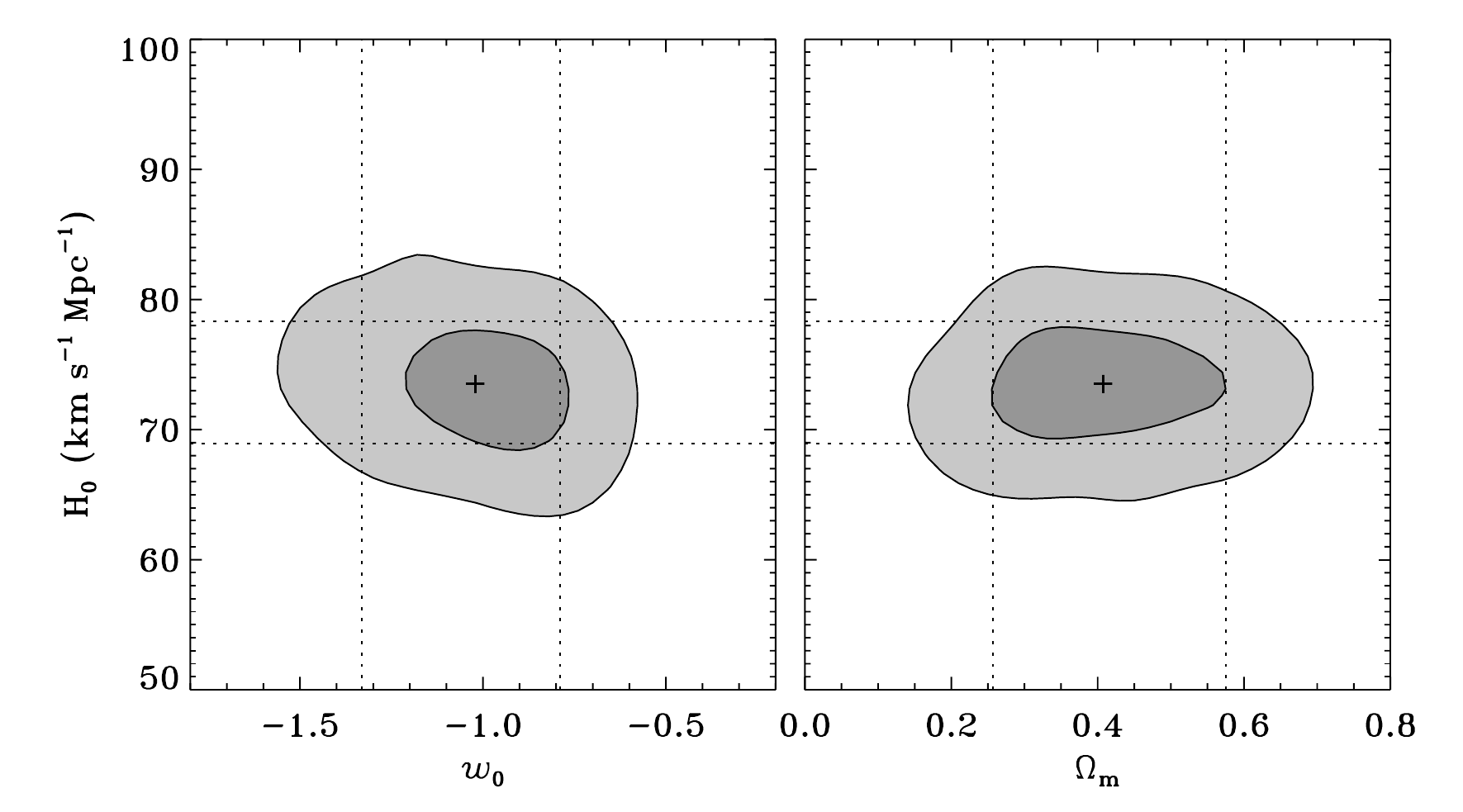}
    \caption{\label{fig:H0_w_Om} Confidence regions (at 1 and 2$\sigma$ levels) and median values (crosses) of $H_{0}$, $w$  and \omegam~obtained from the lensing models of SN Refsdal \citep[adapted from][]{Grillo:2020}. Dotted lines corresponding to the 16th and 84th percentiles for each parameter. A time-delay between SX and S1 of 345 days with a 2\% relative error is adopted. Flat $w$CDM models (\omegam+\omegade=1) with uniform priors on the values of the cosmological parameters (\Ho$\in [20,120]$ \Hunit, \omegam$\in [0,1]$ and $w_0 \in [-2,0]$) are considered. Constraints on the matter density and dark energy EoS parameters are mostly due to the angular diameter distance ratios (Eq.~\ref{family}), whereas those on the Hubble constant are mainly driven by optimizing the measured time-delay of SN Refsdal with the blind mass model by \cite{Grillo:2016}.
    }
\end{figure}

\begin{figure}[t!]
    \centering
    \includegraphics[width=.5\columnwidth]{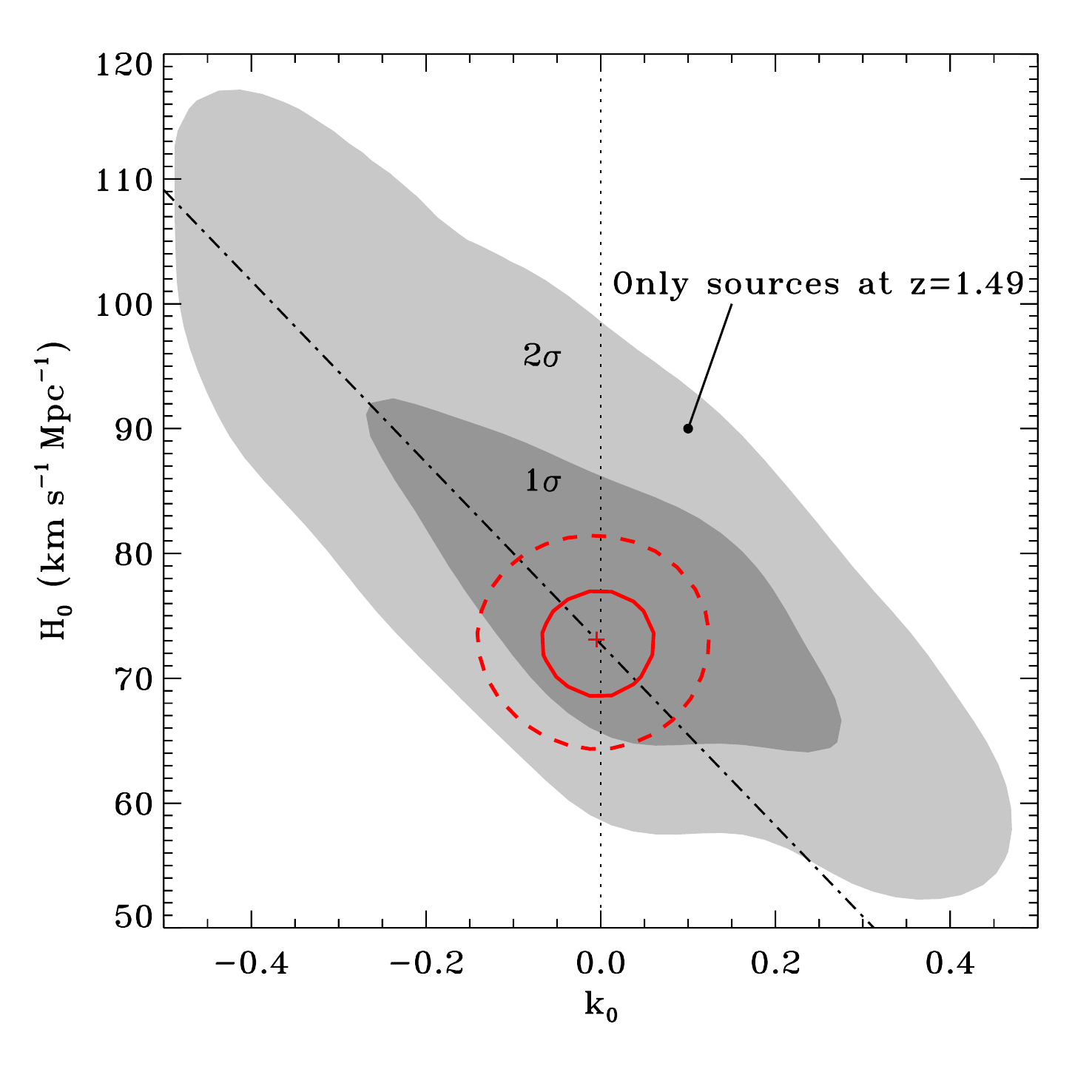}
    \caption{\label{fig:H0_k_MSD} The impact of the mass sheet degeneracy (MSD) on the lensing model of SN Refsdal, where $k_0$ is the value of the convergence of a constant sheet of mass at the cluster redshift \citep{Grillo:2020}. In red: confidence contour levels at 1 and 2$\sigma$ for \Ho~and $k_0$ obtained using all (89) multiple images at different redshifts. 
    A time-delay between SX and S1 of $345\pm 10$ days is adopted.
    In this case the best fit model yields a vanishing mass-sheet ($k_0=0.00_{-0.08}^{+0.06}$, see vertical dotted line). In gray: confidence regions obtained from a model using only those images (63) belonging to SN Refsdal and its host, all at $z=1.49$. 
    The dashed-dotted line illustrates the theoretical effect of the MSD \citep{Schneider:2013}. Flat $\Lambda$CDM models (\omegam+\omegal=1) with uniform priors on the values of the cosmological parameters (\Ho$\in [20,120]$ \Hunit and \omegam$\in [0,1]$) and on the value of k$_{0}$ ($\in [-0.2,0.2]$ or $[-0.5,0.5]$) are considered.
    }
\end{figure}

By using the observed positions of 89 multiple images, with extensive spectroscopic information, from 28 background sources and the measured time-delays between the images S1–S4 and SX of SN Refsdal, \cite{Grillo:2018} have inferred blindly the values of \Ho~and \omegam~with relative (1$\sigma$) statistical errors of, respectively, 6\% (7\%) and 31\% (26\%) in flat (general) cosmological models, assuming a conservative 3\% uncertainty on the final time-delay of image SX and, remarkably,
no priors from other cosmological experiments. Moreover, by investigating separately the impact of a constant sheet of mass at the cluster redshift (see Fig.~\ref{fig:H0_k_MSD}), of a power-law profile for the mass density of the cluster main halo and of some scatter in the cluster member scaling relations, \cite{Grillo:2020} have found that, in a flat $\Lambda$CDM cosmology, these systematic effects do not introduce a significant bias on the inferred values of \Ho~and \omegam, and that the statistical uncertainties dominate the total error budget: a 3\% uncertainty on the time-delay of image SX translates into approximately 6\% and 40\% (including both statistical and systematic 1$\sigma$) uncertainties for \Ho~and \omegam, respectively. They have also presented the interesting possibility of measuring the value of the EoS parameter $w$ of the dark energy density with a 30\% uncertainty (see Fig.~\ref{fig:H0_w_Om}).

By comparing different results of the strong lens time-delay method, with SN Refsdal in MACS J1149.5+2223 \citep{Grillo:2020} and with lensed quasars in the galaxy-scale systems of the H0LiCOW program \citep{Suyu:2017}, we can conclude that {\it i)} the relative error on the inferred value of \Ho~from a single (galaxy or cluster) strong lensing system is similar \citep[mean value of 6.4\% in Fig.~2 of][]{Wong:2020}, {\it ii)} in a single lens cluster, there is the additional possibility of estimating the value of \omegam~(and $w$), thanks to the observations of different multiple-image families with spectroscopically confirmed redshifts and to the measurements of the time-delay value between the multiple images of intrinsically variable sources, and {\it iii)} the observed positions of many spectroscopic multiple images (some of which are key to locating the lens tangential and radial critical curves)
provide precise calibrations of the different mass components (i.e., extended dark-matter halos, cluster members, and hot gas) considered in the model of a galaxy cluster and, thus, also a good approximation of the effect of the ``environment'' where the time-delays are measured.

In particular, it has been tested on models with either the entire sample of 89 multiple images from 28 sources at different redshifts, or only the 63 multiple images from SN Refsdal and its host, all at the same redshift, that in the former case the effect of the so-called ``mass-sheet degeneracy'' is significantly reduced \citep{Grillo:2020}. More quantitatively, this has produced an approximately 9\% difference in the median value of \Ho~(and of \omegam), and a remarkable reduction by a factor of more than 3, from $\sim\,$21\% to $\sim\,$6\%, for its uncertainty (from $\sim\,$63\% to $\sim\,$40\% for the uncertainty on \omegam).

In each lens galaxy cluster, the combination of the positions of several tens of spectroscopically confirmed multiple images and of one or more time-delays between the multiple images of a lensed QSO or SN will allow one to determine the lens Fermat potential differences with a $\sim\,$5\% uncertainty \citep[including both the statistical and systematic errors, as shown by][]{Grillo:2018, Grillo:2020, Acebron:2021}. The planned modeling of the extended surface brightness distributions and kinematic maps of some of multiple images will very likely reduce this uncertainty below 5\%. For each time-varying source lensed by a galaxy cluster, the longest time-delay between its multiple images will be measured with a $\sim\,$2\% error (as obtained so far for the known systems). This will result in a $\lesssim\,$6\% total uncertainty on the value of \Ho~estimated from a single lens cluster. The dataset that are already available for the first three lens clusters will already provide a combined $\approx$3\% uncertainty on \Ho, that will be reduced to $\lesssim\,$2\%, when a sample of $\sim\,$10 lens clusters will be completed, thanks also to the new data from the Rubin survey. 

\clearpage

\subsection{Cosmic Voids}

The largest discernible structures of the Universe make up the so-called cosmic web. It represents a network of compact nodes that are connected by filaments and walls of lower density \citep{Zeldovich1970}. The remaining space is taken up by {\it cosmic voids}, extended regions of very low matter content \citep[e.g.,][]{Zeldovich1982, Bertschinger1985, vdWeygaert1993}. The nodes are occupied by groups and clusters of galaxies, which makes them the most luminous and thus best identifiable individual structures at cosmological distances. The contrary is the case for voids, which host the least luminous galaxies in the cosmos and have only been discovered in the late 70's \citep{Gregory1978, Joeveer1978}. A systematic identification of voids not only requires a complete sampling of their boundaries, consisting of filaments and walls, but also the sensitivity to detect the faintest galaxies in their interiors. This has only recently become feasible with the advance of wide and deep redshift surveys that are able to reveal the three-dimensional structure of the cosmic web in great detail \citep[e.g., see][for some of the first void catalogs obtained from SDSS, VIPERS, BOSS, DES, 6dFGS, KiDS, and eBOSS]{Pan2012, Sutter2012a, Micheletti2014, Mao2017a, Sanchez2017, Achitouv2017, Brouwer2018, Hawken2020}.

Since then, void catalogs of ever-growing size have been compiled and analyzed to tackle unanswered questions in various fields of cosmology and astrophysics. For example, voids can been used to study environmental effects in the formation and evolution of galaxies \citep[e.g.,][]{Hoyle2005, Patiri2006, Kreckel2012, Ricciardelli2014b, Habouzit2020, Panchal2020}, to investigate the nature of gravity with the motivation to find modifications to the general theory of relativity \citep[e.g.,][]{Clampitt2013, Spolyar2013, Zivick2015, Cai2015, Barreira2015, Hamaus2015, Achitouv2016, Voivodic2017, Falck2018, Sahlen2018, Baker2018, Paillas2019, Davies2019, Perico2019, Alam2020, Contarini2021, Wilson2021}, or to reveal unknown properties of the standard model ingredients in cosmology, namely its initial conditions \citep{Chan2019}, dark energy \citep[e.g.,][]{Lee2009, Biswas2010, Lavaux2012, Sutter2012b, Bos2012, Hamaus2014c, Pisani2015a, Pollina2016, Verza2019}, dark matter \citep[e.g.,][]{Leclercq2015, Yang2015, Reed2015, Baldi2018}, and neutrinos \citep{Massara2015, Banerjee2016, Sahlen2019, Kreisch2019, Schuster2019, Zhang2020, Bayer2021, Kreisch2021}. It is the under-dense character of voids that makes them particularly sensitive to homogeneous or diffuse components of our Universe, such as dark energy and neutrinos. For example, dark energy dominates the matter-energy budget inside voids much earlier than in the cosmos as a whole. Thanks to their small mass, neutrinos can freely stream into the deep interiors of voids, while baryons and dark matter are mostly restricted to their boundaries due to gravitational interaction. Finally, screening mechanisms that efficiently hide possible deviations from general relativity in regions of high density or deep gravitational potential are not effective inside voids.

In order to encompass such a wide range of topics, various void-related observables have been considered. This includes cross-correlations with the CMB, which provide detections of the integrated Sachs–Wolfe effect \citep[ISW, e.g.,][]{Granett2008, Ilic2013, Cai2014, Planck2014, Nadathur2016, Kovacs2019, Kovacs2021} and of the Sunyaev–Zeldovich (SZ) effect \citep{Alonso2018}, or correlations with the distorted shapes of galaxies, revealing the matter content of voids via the gravitational lensing effect \citep[e.g.,][]{Melchior2014, Clampitt2015, Gruen2016, Sanchez2017, Cai2017, Brouwer2018, Fang2019, Vielzeuf2021, Jeffrey2021}. However, voids may also serve as cosmological probes themselves, because their dynamics are governed by the same physical laws that describe the evolution of the Universe as a whole. This enables us to predict their properties from first principles, and to compare these predictions with observations in order to constrain cosmological models.

\subsubsection{Basic idea and equations}
\label{sec:V1}

In this section, we discuss two of the most studied observables that have been investigated for cosmological applications with voids so far: the void size function and the void density profile (or void-galaxy cross-correlation function). These two observables are affected by the so-called \citet{Alcock1979} (AP) effect \citep[e.g.,][]{Ryden1995, Sutter2012b, Sutter2014b, Hamaus2014c, Hamaus2016, Mao2017b, Correa2019, Endo2020, Nadathur2020, Paillas2021} and by redshift-space distortions (RSD) \citep[e.g.,][]{Ryden1996, Padilla2005, Paz2013, Pisani2015b, Hamaus2015, Hamaus2017, Cai2016, Chuang2017, Hawken2017, Hawken2020, Achitouv2019, Aubert2020, Correa2021a, Correa2021b}, which themselves carry cosmologically relevant information.
For other methods that employ voids as cosmological probes, such as their pairwise clustering statistics on large scales \cite[e.g.,][]{Hamaus2014a, Hamaus2014c, Chan2014, Zhao2016, Chuang2017, Lares2017a, Voivodic2020}, the associated baryon acoustic oscillation (BAO) feature \citep[e.g.,][]{Kitaura2016, Liang2016, Chan2021}, the velocity statistics of voids \citep[e.g.,][]{Sutter2014c, Ruiz2015, Lambas2016, Ceccarelli2016, Wojtak2016, Lares2017b}, or marked tracer statistics that up-weight underdense regions \citep[e.g.,][]{Beisbart2000, Sheth2005, White2016, Philcox2020, Massara2021, Massara2022}, we refer the reader to the provided references. 

\subsubsubsection{Void size function}

\begin{figure}[t!]
	\centering
	\resizebox{\hsize}{!}{
		\includegraphics{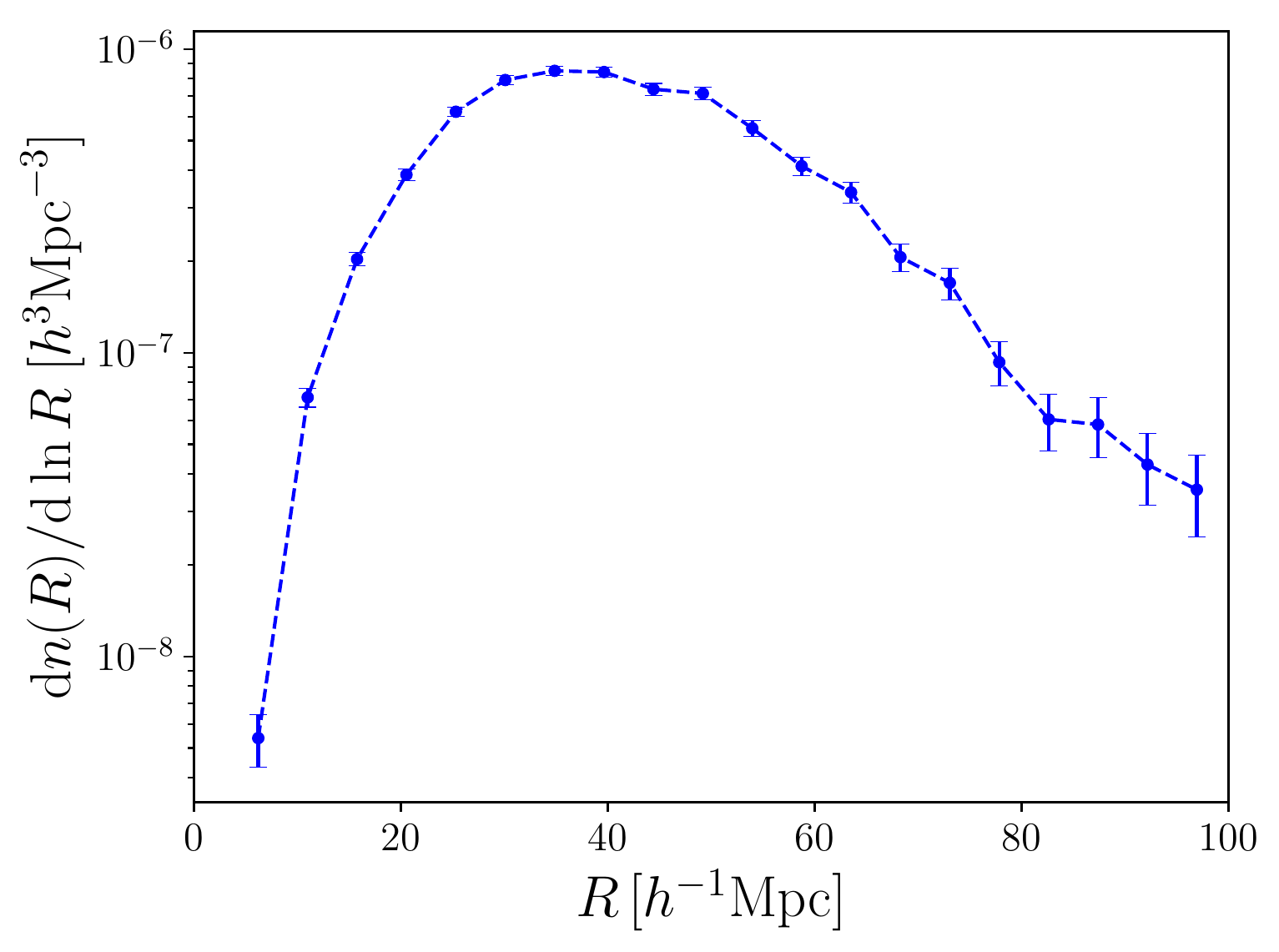}
		\includegraphics[trim=0 -1.5 0 0.5]{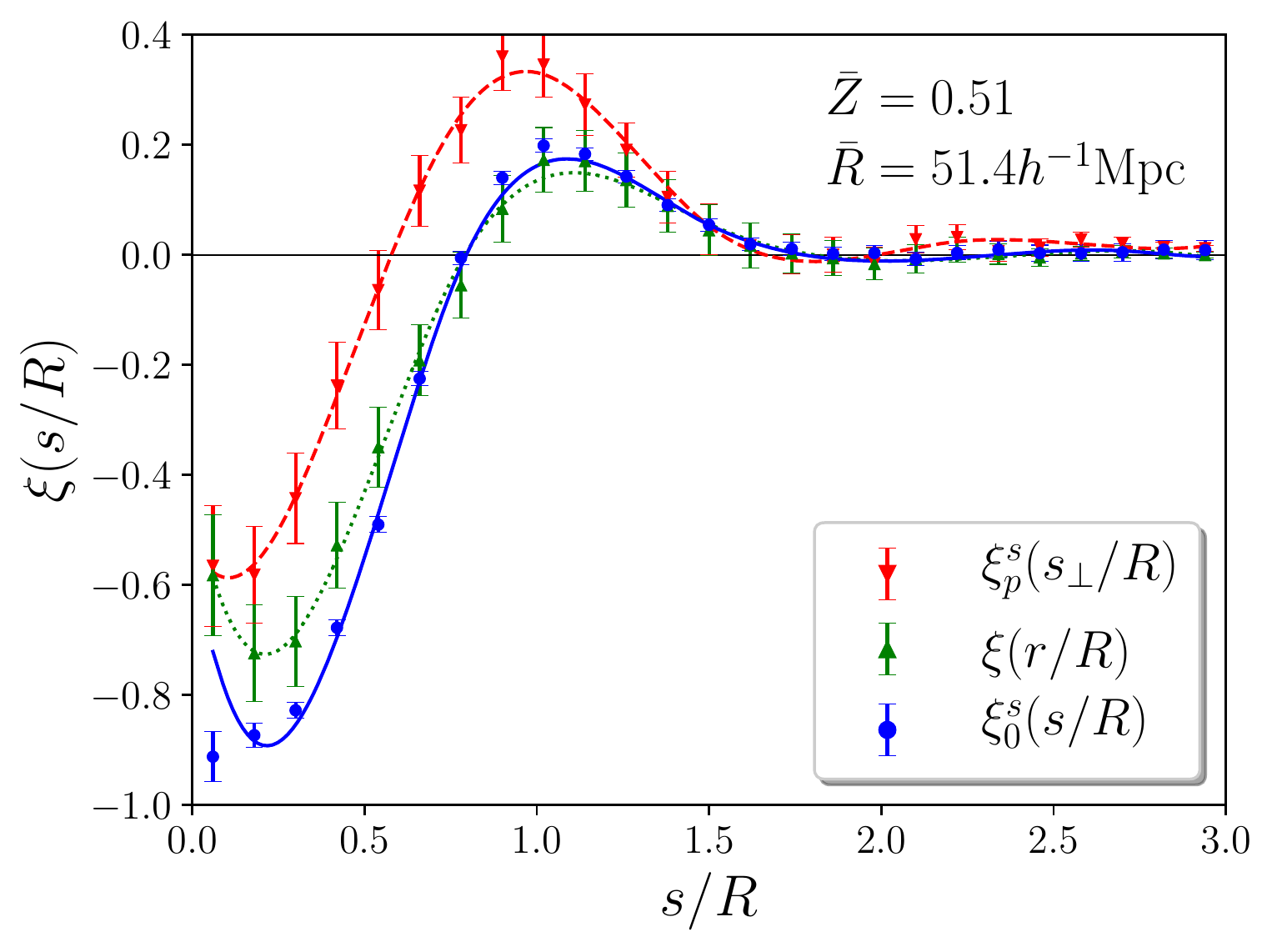}}
	\caption{Left: Void size function from the final BOSS data in the redshift range $0.20<z<0.75$. Right: Projected void density profile (void-galaxy cross-correlation function, red wedges) from the final BOSS data, and its real-space counterpart after deprojection (green triangles). The redshift-space monopole of the density profile (blue dots) is shown along with its best-fit model (blue solid line). The upturn towards the void center is due to residual noise in the deprojected profile, which is used in the model. Images reproduced with permission from~\citet{Hamaus2020}, copyright by IOP Publishing.}
	\label{fig:void_size}
\end{figure}

The void size function $\mathrm{d}n(R,z)/\mathrm{d}R$ specifies the number density of voids of a given size $R$ at redshift $z$. It is also known as void abundance. One can think of it in analogy to the cluster mass function $\mathrm{d}n(M,z)/\mathrm{d}M$, with the advantage of the void size being a directly observable quantity. In contrast, the cluster mass $M$ can in practice only be related to other observables, such as richness or X-ray luminosity. The void size function has already been measured in current data (see Fig.~\ref{fig:void_size}), but has not yet been used to extract cosmological constraints \citep[however, see][for constraints from extreme-value statistics of voids]{Sahlen2016}. The increase in expected void numbers from upcoming surveys in the next decade and the strong modeling activity performed on simulations will soon allow first applications to observational data \citep{Pisani2019}. Theoretical models for the void size function allow us to predict void numbers in the dark matter distribution from first principles \citep[e.g.,][]{Sheth2004, Furlanetto:2006jb,Platen2007,Paranjape2012, Jennings2013,Pisani2015a}. By accounting for tracer bias, it is possible to relate those predictions to observable voids in the tracer distribution \citep{Pollina2016,Ronconi2017,Ronconi2019,Contarini2019}, thereby providing estimates of expected void numbers in large-scale structure surveys. First and foremost, predicting void numbers is an important task, necessary to perform accurate forecasts for other probes relying on the statistics of voids. However, it turns out that the void size function is an extremely sensitive probe of cosmology in itself: by counting voids of different size in surveys one can obtain constraints on the dark energy EoS \citep{Pisani2015a, Verza2019}, the presence of massive neutrinos \citep{Sahlen2019,Kreisch2019,Schuster2019,Kreisch2021}, and modified gravity \citep{Clampitt2013,Lam2015,Cai2015,Zivick2015,Sahlen2016,Contarini2021}.

The most common setup to obtain predictions relies on the excursion-set formalism \citep{Bond1991}, applied to the hierarchical evolution of cosmic voids. It has first been developed by \citet{Sheth2004} and was later extended by \citet{Jennings2013}. Excursion-set theory provides predictions for void numbers based on spherical fluctuations in the initial (Lagrangian) density field. It calculates their conditional first-crossing distribution $f_{\ln\sigma}(\sigma)$ as a function of the root mean square matter fluctuations $\sigma$, smoothed on a scale $R_\mathrm{L}$. A fluctuation becomes a void when its Lagrangian density contrast $\delta^\mathrm{L}$, filtered on the scale $R_\mathrm{L}$, reaches the void formation threshold $\delta_\mathrm{v}^\mathrm{L}$ without crossing the collapse threshold $\delta_\mathrm{c}^\mathrm{L}$ on a scale larger than $R_\mathrm{L}$. The thresholds are determined via the nonlinear evolution of a spherically symmetric top-hat fluctuation \citep{Icke1984}, the moment of shell crossing conventionally defines the formation of a void \citep{Bertschinger1985,Blumenthal1992}. The void size function in Lagrangian space is then given by:
\begin{equation}\label{Vdn}
\frac{\mathrm{d} n_\mathrm{L}}{\mathrm{d} \ln R_\mathrm{L}} = \frac{f_{\ln \sigma}(\sigma)}{V(R_\mathrm{L})} \, \frac{\mathrm{d} \ln \sigma^{-1}}{\mathrm{d} \ln R_\mathrm{L}} \;\; ,
\end{equation}
where $V(R_\mathrm{L})=4\pi R_\mathrm{L}^3/3$, and the first-crossing distribution is \citep{Sheth2004}:
\begin{equation}
f_{\ln \sigma}(\sigma) = 2 \sum_{j=1}^{\infty} \, e^{-\frac{(j \pi x)^2}{2}} \, j \pi x^2 \, \sin{\left( j \pi \mathcal{D} \right)} \;\; ,
\end{equation}
with:
\begin{equation}
\mathcal{D} \equiv \frac{|\delta_\mathrm{v}^\mathrm{L}|}{\delta_\mathrm{c}^\mathrm{L} + |\delta_\mathrm{v}^\mathrm{L}|}\;, \qquad x \equiv \frac{\mathcal{D}}{|\delta_\mathrm{v}^\mathrm{L}|} \sigma \;\; .
\end{equation}
The label $\mathrm{L}$ indicates all quantities that are evaluated following linear theory in Lagrangian space. To ensure volume conservation between the linear and nonlinear density field, \citet{Jennings2013} impose:
\begin{equation}
V(R)\mathrm{d}n = V(R_\mathrm{L})\mathrm{d}n_\mathrm{L} \rvert_{R_\mathrm{L}(R)} \;\; .
\end{equation}
Together with the equality $\mathrm{d} \ln R=\mathrm{d} \ln R_\mathrm{L}$, which applies for the spherical top-hat model, one obtains the final expression for the so-called Vdn model, as extension from the original Sheth and van de Weygaert model: 
\begin{equation}
\frac{\mathrm{d} n}{\mathrm{d} \ln R} = \frac{f_{\ln \sigma}(\sigma)}{V(R)} \, \frac{\mathrm{d} \ln \sigma^{-1}}{\mathrm{d} \ln R} \;\; .
\end{equation}

However, in order to apply this model to data it is necessary to consider the complicating fact that in practice, voids are found in the distribution of tracers of the matter density field, that are typically galaxies. Moreover, the structures identified by a shape-agnostic void finding algorithm are not the idealistic spherically symmetric and isolated objects assumed in the theoretical model \citep[e.g.,][see Sect.~\ref{sec:V2}]{Platen2007,Neyrinck2008,Sutter2015}. To align the theory with observations, two important steps need to be taken into account. Firstly, the measured properties of real voids need to be linked with the idealistic top-hat model, such that their size and depth agree. For example, this can be achieved by identifying a sphere of radius $R$ around the void center, which yields a given density threshold $\delta_\mathrm{v}$. The spherical top-hat model suggests using $\delta_\mathrm{v}\simeq-0.8$ at the moment of shell crossing as a natural choice \citep{Bertschinger1985,Blumenthal1992}, but in principle the model keeps its validity with any other value \citep{Jennings2013,Verza2019}. Secondly, density fluctuations in the tracer distribution are biased with respect to the matter density field (see Sect.~\ref{sec:V4}). Therefore, a model for tracer bias needs to be incorporated in the theoretical formalism to predict the observable void size function \citep{Ronconi2019,Contarini2019,Contarini2021}.

\subsubsubsection{Void density profile}

Apart from their size, voids are characterized by their unique composition and geometry. While these properties may vary significantly from one void to another, they are more well-defined in an ensemble average sense. For example, in a statistically homogeneous and isotropic universe the average density profile of voids exhibits some universal characteristics: an extended under-dense core and a steep density run towards the void boundary \citep[e.g., see][and Fig.~\ref{fig:void_size}]{Ricciardelli2014a, Hamaus2014b}. The boundary itself features an over-dense ridge whose amplitude diminishes for increasingly large voids \citep[e.g.,][]{Sheth2004, Ceccarelli2013}. These characteristics can be parameterized by analytical fitting formulae for the isotropic void density profile. For example, one well-explored expression is given by:
\begin{equation}
	\delta(r) = \delta_\mathrm{c}\frac{1-(r/r_\mathrm{s})^\alpha}{1+(r/R)^\beta} \;\; , 
	\label{HSW}
\end{equation}
where $\delta\equiv\rho/\bar{\rho}-1$ is the density contrast with respect to the background density of the Universe $\bar{\rho}$~\citep{Hamaus2014b}. For voids with an effective radius $R$, it expresses the average density fluctuation as a function of comoving distance $r$ from the void center and contains four parameters: a scale radius $r_\mathrm{s}$ that determines where the density equals the background value, a central under-density $\delta_\mathrm{c}$, and two power-law indices $\alpha$ and $\beta$ that control its inner and outer slopes. It has further been shown that the latter two parameters linearly scale with $r_\mathrm{s}$, which can be exploited to reduce the dimensionality of the parameter space for the density profile to two. While the form of Eq.~\ref{HSW} has been motivated and tested by simulation studies \citep[e.g.,][]{Sutter2014a, Barreira2015, Pollina2017, Falck2018, Baker2018, Perico2019, Stopyra2021, Shim2021, Tavasoli2021}, it is also in good agreement with observations \citep[e.g.,][]{Sanchez2017, Chantavat2017, Pollina2019, Fang2019}. Typical values for the parameters in Eq.~\ref{HSW} are $r_\mathrm{s}\simeq R$, $\delta_\mathrm{c}\simeq-0.8$, $\alpha\simeq2$, and $\beta\simeq9$~\citep{Hamaus2014b}.

However, in redshift surveys the assumption of spherical symmetry is violated due to RSD. They arise as a consequence of the peculiar motions of galaxies on top of the Hubble flow, causing a Doppler shift in their emitted spectrum. This affects the distance-redshift relation, which only accounts for a Hubble redshift $z_h$. As a result, the comoving location $\mathbf{x}$ of a galaxy with observed redshift $z$ is given by:
\begin{equation}
	\mathbf{x}(z) = \mathbf{x}(z_h) + \frac{1+z_h}{H(z_h)}\mathbf{v}_\parallel \;\;,
	\label{x_rsd}
\end{equation}
where $\mathbf{v}_\parallel$ is the component of the galaxy velocity vector $\mathbf{v}$ along the line of sight, relative to the observer. The same argument applies to the location of a void center $\mathbf{X}$ at redshift $Z$ (we use capitals to designate void properties), i.e. for the separation vector~$\mathbf{s}$ between galaxy and void center in redshift space we obtain:
\begin{equation}
	 \mathbf{s} \equiv \mathbf{x}(z)-\mathbf{X}(Z) \simeq \mathbf{x}(z_h)-\mathbf{X}(Z_h) + \frac{1+z_h}{H(z_h)}\left(\mathbf{v}_\parallel-\mathbf{V}_\parallel\right) = \mathbf{r} + \frac{1+z_h}{H(z_h)}\mathbf{u}_\parallel \;\; ,
	\label{s(r)}
\end{equation}
where $\mathbf{r}$ is their comoving separation in real space and $\mathbf{u}_\parallel=\mathbf{v}_\parallel-\mathbf{V}_\parallel$ their relative velocity along the line of sight. Thus, a description of the mapping between real and redshift space requires a model for the dynamics of voids. It has been shown that the assumptions of average spherical symmetry and local mass conservation at linear order in the density contrast provide an accurate relation for the relative velocity field $\mathbf{u}$ \citep{Peebles1980,Hamaus2014b}:
\begin{equation}
	\mathbf{u}(\mathbf{r}) = -\frac{f(z_h)}{3}\frac{H(z_h)}{1+z_h}\Delta(r)\,\mathbf{r} \;\; ,
	\label{u(r)}
\end{equation}
where $f$ is the linear growth rate of density perturbations and $\Delta(r)$ the average density contrast within a radius $r\equiv|\mathbf{r}|$ from the void center:
\begin{equation}
	\Delta(r) = \frac{3}{r^3}\int_0^r\delta(r')r'^2\,\mathrm{d}r' \;\;. \label{Delta(r)}
\end{equation}
The vector $\mathbf{u}$ is directed along the radial direction $\mathbf{r}$ from the void center in real space, so the coordinate mapping from Eq.~\ref{s(r)} can now be written in terms of $\mathbf{r}$ and its component along the line of sight $\mathbf{r}_\parallel$:
\begin{equation}
	\mathbf{s} = \mathbf{r} - \frac{f(z_h)}{3}\Delta(r)\,\mathbf{r}_\parallel \;\; .
	\label{s(r)_lin}
\end{equation}
From this equation the coordinate transformation between real and redshift space is fully determined by the void density profile in real space, e.g. via the fitting formula of Eq.~\ref{HSW}. The linear growth rate $f$ only depends on the cosmological model, but within the realm of General Relativity it is well approximated by a power law of the matter content $\Omega_{\rm m}(z)$ at redshift $z$ with a growth index of $\gamma\simeq0.55$, $f(z)=\Omega_{\rm m}(z)^\gamma$ \citep{Lahav1991,Linder2005}.

\begin{figure}[t]
	\centering
	\resizebox{0.95\hsize}{!}{
		\includegraphics[trim= 0 0 0 40]{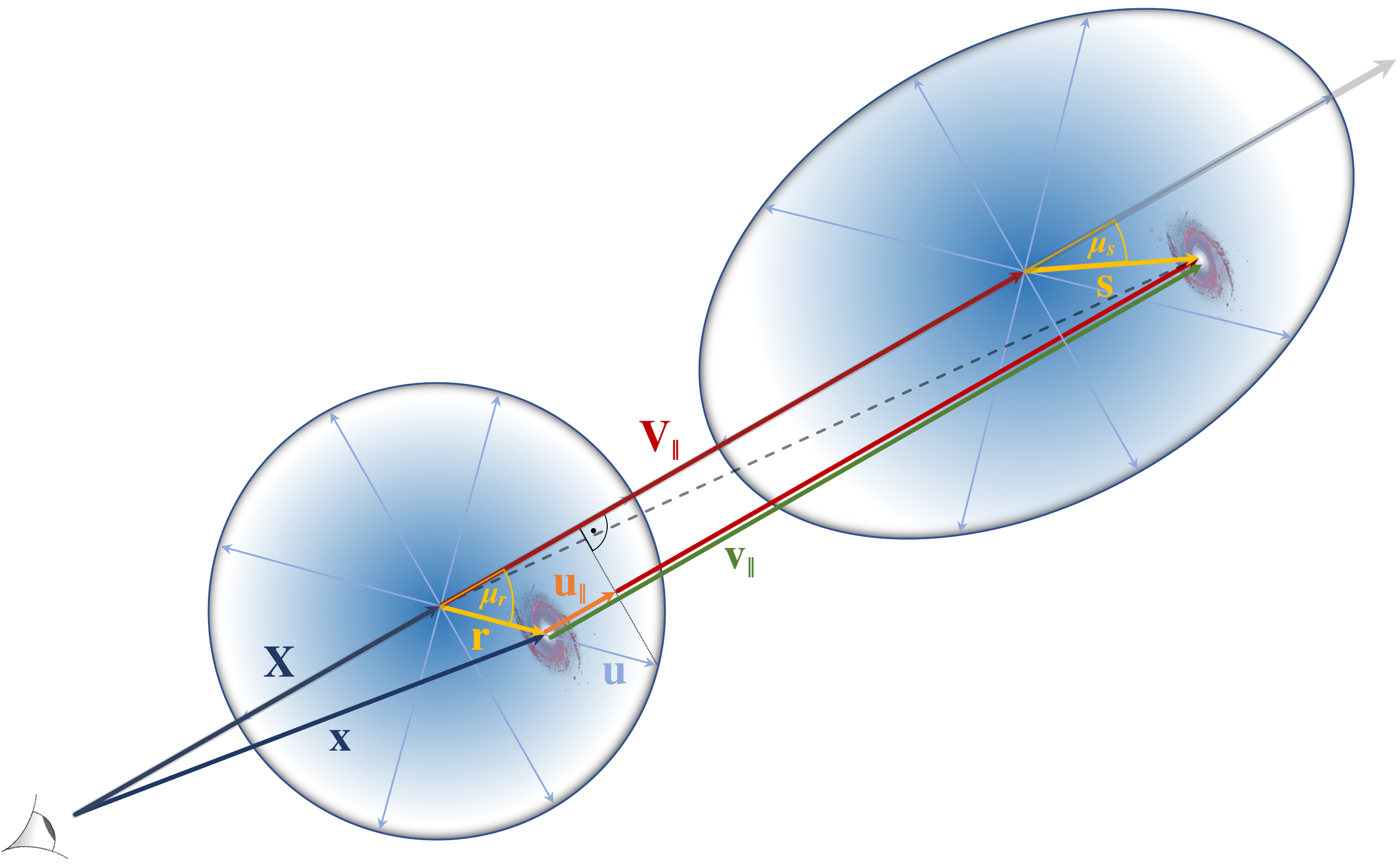}}
	\caption{Schematic representation of a void in real (left) and in redshift space (right). The separation vector~$\mathbf{r}$ between its center at $\mathbf{X}$ and a galaxy at $\mathbf{x}$ in real space is transformed via $\mathbf{s}=\mathbf{r}+\mathbf{u}_\parallel$ to redshift space, where $\mathbf{u}_\parallel=\mathbf{v}_\parallel-\mathbf{V}_\parallel$ is the relative line-of-sight velocity between them. For simplicity the illustration displays $\mu$ instead of $\arccos(\mu)$ to indicate angles to the line of sight and uses velocity displacements in units of $(1+z_h)/H(z_h)$. Image reproduced with permission from \citet{Hamaus2020}, copyright by IOP Publishing.}
	\label{fig:voidstretch}
\end{figure}

The coordinate mapping in Eq.~\ref{s(r)_lin} leads to an anisotropic distortion of voids along the line of sight, as illustrated schematically in Fig.~\ref{fig:voidstretch}. Therefore, an isotropic density profile is no longer sufficient to describe the average geometry and composition of voids. Instead, the corresponding observable quantity is the void-galaxy cross-correlation function $\xi^s(\mathbf{s})$ in redshift space, which not only depends on the magnitude $s$ of the separation vector, but also on the cosine of its angle to the line of sight $\mu_s=s_\parallel/s$. Because the number of galaxies around every void is conserved in the mapping from real to redshift space, the Jacobian $\partial\mathbf{s}/\partial\mathbf{r}$ relates $\delta(r)$ to $\xi^s(\mathbf{s})$ via:
\begin{equation}
	\int[1+b\delta(r)]\mathrm{d}^3r = \int[1+\xi^s(\mathbf{s})]\det\!\left(\frac{\partial\mathbf{s}}{\partial\mathbf{r}}\right)\mathrm{d}^3r \;\; . 
	\label{jacobian}
\end{equation}
Here we have additionally assumed a linear bias relation of the form $\xi(r)=b\delta(r)$ between galaxy and matter over-densities in real space, with a scale-independent bias parameter $b$ (see Sect.~\ref{sec:V4}). This assumption has been investigated with the help of $N$-body simulations \citep{Sutter2014a,Pollina2017,Contarini2019,Ronconi2019}, but also with galaxy-clustering and weak-lensing observations \citep{Pollina2019,Fang2019}, and was found to be remarkably accurate. Using Eq.~\ref{s(r)_lin} inside Eq.~\ref{jacobian} and an expansion to linear order in $\delta(r)$, one finally arrives at \citep{Cai2016,Hamaus2017}:
\begin{equation}
	\xi^s(\mathbf{s}) = b\delta(r) + \frac{f}{3}\Delta(r) + f\mu_r^2\left[\delta(r)-\Delta(r)\right] \;\; , 
	\label{xi_s}
\end{equation}
where $\mu_r=r_\parallel/r$. Given a density profile $\delta(r)$ and the mapping between $\mathbf{s}$ and $\mathbf{r}$ from Eq.~\ref{s(r)_lin}, one can now evaluate the void-galaxy cross-correlation function $\xi^s$ for any observed separation vector $\mathbf{s}$. Since $\mathbf{r}$ is unknown, one may initially evaluate $\Delta(r)$ at $r=s$ and then calculate $r$ via iteratively applying the following set of equations \citep{Hamaus2020}:
\begin{equation}
	r = \sqrt{r_\perp^2+r_\parallel^2}\;,\qquad
	r_\perp = s_\perp\;,\qquad r_\parallel = s_\parallel\left[1-\frac{f}{3}\Delta(r)\right]^{-1} \;\; ,
	\label{r_per_par}
\end{equation}
where $s_\perp$ is the perpendicular component of $\mathbf{s}$ to the line of sight, and hence unaffected by RSD. Equation~\ref{xi_s} can also be expanded in terms of Legendre polynomials, with monopole and quadrupole as the only non-vanishing multipoles at linear order \citep{Cai2016}.

It remains to determine the real-space density profile $\delta(r)$ to be used in the previous equations. Various approaches have been followed in the literature: they either make use analytic fitting formulae like Eq.~\ref{HSW} \citep{Paz2013,Hamaus2015,Hamaus2016,Correa2019}, calibrated measurements from simulations \citep{Achitouv2017,Nadathur2020}, or a deprojection technique to determine it from the observed data directly \citep{Pisani2014,Hawken2017,Hamaus2020}. The latter approach is based on the inverse Abel transform \citep{Abel1842,Bracewell1999}:
\begin{equation}
	\xi(r) = -\frac{1}{\pi}\int_r^\infty\frac{\mathrm{d}\xi^s_\mathrm{p}(s_\perp)}{\mathrm{d}s_\perp}\frac{\mathrm{d}s_\perp}{\sqrt{s_\perp^2-r^2}} \;\; ,
	\label{iAbel}
\end{equation}
exploiting the fact that the projected void-galaxy cross-correlation function in redshift space, $\xi^s_\mathrm{p}(s_\perp)=\int\xi^s(\mathbf{s})\,\mathrm{d}s_\parallel$, is insensitive to RSD, which only act along the line-of-sight component $s_\parallel$ of the separation vector $\mathbf{s}$ (see Fig.~\ref{fig:void_size}). Assuming linear bias, this provides the real-space density profile via $\delta(r)=\xi(r)/b$.

Moreover, it is possible to extend this dynamical model to the quasi-linear regime via the so-called Gaussian Streaming Model (GSM). Assuming the pairwise line-of-sight velocity $u_\parallel$ between void centers and galaxies to follow a Gaussian distribution, the void-galaxy cross-correlation function in redshift space is given by \citep[e.g.,][]{Paz2013,Hamaus2015,Cai2016}:
\begin{equation}
	1+\xi^s(\mathbf{s}) = \int\left[1+\xi(r)\right]\frac{1}{\sqrt{2\pi}\;\sigma_\parallel(r,\mu_r)}\exp\left\{-\frac{\left[u_\parallel-u(r)\mu_r\right]^2}{2\sigma_\parallel^2(r,\mu_r)}\right\}\mathrm{d}u_\parallel \;\; ,
	\label{GSM}
\end{equation}
which additionally requires the pairwise velocity dispersion along the line of sight $\sigma_\parallel(r,\mu_r)$ as a model ingredient. We refer to \citet{Hamaus2020} for a discussion on the advantages and disadvantages of the various modeling choices.

\subsubsection{Sample selection}
\label{sec:V2}

The observational identification of voids requires a distribution of tracers of the large-scale structure, as it is obtained via redshift surveys. Typically these tracers are galaxies with either spectroscopic or photometric redshift estimates, but other tracer types, such as galaxy clusters \citep{Pollina2019}, the Ly-$\alpha$ forest \citep{Stark2015,Krolewski2018,Porqueres2019}, or the 21cm emission from neutral Hydrogen \citep{White2017,Endo2020} have been considered for void finding as well. These observations commonly optimize the target selection based on their individual science cases, but voids can be extracted as a byproduct without additional expense. Therefore, the sample selection for voids usually derives from the target tracer selection, and is rarely optimized specifically for void detection \citep[however, see][for more details on the optimization of surveys for void detection]{vdWeygaert2011,Pisani2019}. 
Nevertheless, previous survey data has proven itself very valuable in providing void catalogs of high quality with significant sample sizes \citep[e.g.,][]{Sutter2012a,Mao2017a,Fang2019,Hamaus2020,Aubert2020,Nadathur2020}.

Various techniques for the identification of voids have been presented in the literature \citep[see][for an overview of different methods]{Colberg2008,Cautun2018}. They either consider a full distribution of tracers in 3D, or 2D projections along the line-of-sight direction.
The former approach is typically applied to spectroscopic, the latter to photometric data, although both techniques can be used in each case. 
Moreover, some void finders search for spherical domains with tracer densities below a given threshold, while others locate void boundaries of arbitrary geometry in a non-parametric fashion. The latter can be achieved with a so-called {\it watershed} algorithm \citep{Platen2007}, which requires the definition of a density field from the distribution of tracer particles. The density field itself can be estimated in various ways, for example via grid interpolation or adaptive methods, such as Delaunay or Voronoi tessellation. As a result, one obtains a nearly space-filling distribution of voids in the large-scale structure with individual properties, such as their size, shape, density, or center location, which can be considered as cosmological observables. Among the most popular software repositories that implements this is the public Void IDentification and Examination toolkit \texttt{VIDE}\footnote{\url{https://bitbucket.org/cosmicvoids/vide_public/wiki/Home}}~\citep{Sutter2015}. It is based on the code \texttt{ZOBOV} \citep{Neyrinck2008}, which performs a Voronoi tessellation and the watershed transform on a set of tracer particles. \texttt{VIDE} additionally handles the complexities arising from the survey geometry, which typically represents a masked light cone within a given redshift range. Voids intersecting with the boundary of the survey mask are usually excluded from the final void catalog. Furthermore, a cut on minimum void size based on the mean tracer separation is often used to mitigate the contamination from spurious voids that may arise via random density fluctuations (see Sect.~\ref{sec:V4}).

\subsubsection{Measurements}
\label{sec:V3}

The location of an astronomical object at cosmological distance is determined via its observed redshift $z$ and its position on the sky, expressed in angular coordinates $\vartheta$ (right ascension) and $\varphi$ (declination). In order to identify voids in the 3D distribution of tracers, we first need to perform a transformation to Cartesian coordinates $\mathbf{x}$ in comoving space:
\begin{equation}
	\mathbf{x}(z,\vartheta,\varphi) = (1+z) D_\mathrm{A}(z)\begin{pmatrix}\cos\vartheta\cos\varphi\\\sin\vartheta\cos\varphi\\\sin\varphi\end{pmatrix} \;\; ,
	\label{x_comoving}
\end{equation}
where $D_\mathrm{A}(z)$ is the angular diameter distance to a tracer at redshift $z$. It depends on the expansion history of the Universe via the Hubble function $H(z)$ and on the curvature of space via the parameter $\Omega_k$ as expressed in Eq.~\ref{eq:lumdist}. That equation can be also written as:
\begin{equation} D_\mathrm{A}(z) = \frac{c}{(1+z)H_0\sqrt{-\Omega_k}}\sin\left(\sqrt{-\Omega_k}\int_0^z\frac{H_0}{H(z')}\mathrm{d}z'\right) \;\;,
	\label{D_A(z)}
\end{equation}
where $c$ is the speed of light and $H_0\equiv H(z=0)$ the Hubble constant. Thus, in order to perform the coordinate transformation in Eq.~\ref{x_comoving} it is necessary to assume a particular cosmology. Within $\Lambda$CDM, for example, this requires values for the radiation, matter, and cosmological constant parameters $\Omega_\mathrm{r}$, \omegam, and \omegal, which determine the curvature parameter as $\Omega_k=1-\Omega_\mathrm{r}-\Omega_\mathrm{m}-\Omega_\Lambda$. The Hubble function is given by Eq.~\ref{eq:Hz1}.
Once the coordinate transformation is performed, voids can be identified in comoving space. It is then possible to ascribe a volume $V$ and an {\it effective radius} $R$ to every void. In particular, making use of a Voronoi tessellation, one can define these quantities via a sum over the cell volumes $\mathcal{V}_i$ of the individual tracer particles with index $i$ that belong to each void:
\begin{equation}
	V = \sum\nolimits_i\mathcal{V}_i\;,\qquad R = \left(\frac{3}{4\pi}V\right)^{1/3} \;\; .
	\label{Reff}
\end{equation}
Moreover, one can define a volume-weighted barycenter from the tracers at location $\mathbf{x}_i$, which serves as a good estimator for the geometric center of a void \citep[e.g.,][]{Sutter2012a,Cautun2016,Stopyra2021}:
\begin{equation}
	\mathbf{X} = \frac{\sum_i\mathbf{x}_i\mathcal{V}_i}{\sum_i\mathcal{V}_i} \;\; .
	\label{barycenter}
\end{equation}
Further properties, such as the inertia tensor with its eigenvalues and eigenvectors, the ellipticity, the minimum density, the density contrast, or the average density can be defined for each void based on its defining tracers \citep{Sutter2015}.

The separations between void centers and tracers in comoving space can be calculated via their differences in the angle on the sky $\delta\theta$ and in redshift $\delta z$ following Eq.~\ref{x_comoving}:
\begin{equation}
	s_\perp = (1+z)D_\mathrm{A}(z)\,\delta\theta\;,\qquad s_\parallel = \frac{c}{H(z)}\,\delta z \;\; . 
	\label{s_comoving}
\end{equation}
However, as both $D_\mathrm{A}(z)$ and $H(z)$ depend on the assumed cosmological model, so do the separations. It is therefore common practice to introduce two AP parameters $q_\perp$ and $q_\parallel$ that inherit the dependence on cosmology via:
\begin{equation}
	q_\perp = \frac{s_\perp^*}{s_\perp} = \frac{D_\mathrm{A}^*(z)}{D_\mathrm{A}(z)}\;,\qquad q_\parallel = \frac{s_\parallel^*}{s_\parallel} = \frac{H(z)}{H^*(z)} \;\; , 
	\label{AP}
\end{equation}
where the quantities with an asterisk are evaluated in the true underlying cosmology, which is unknown. In the special case where the assumed cosmology coincides with the true one, $q_\perp=q_\parallel=1$. In turn, measuring $q_\perp$ and $q_\parallel$ provides a measurement of $D_\mathrm{A}^*(z)$ and $H^*(z)$, respectively. However, without an absolute calibration scale the two parameters remain degenerate in the AP test. Only their ratio, known as the AP parameter:
\begin{equation}
	\varepsilon \equiv \frac{q_\perp}{q_\parallel} =  \frac{D_\mathrm{A}^*(z)H^*(z)}{D_\mathrm{A}(z)H(z)} \;\;
	\label{epsilon}
\end{equation}
can be determined, which provides a measurement of the product $D_\mathrm{A}^*(z)H^*(z)$ \citep{Sutter2012b,Hamaus2016}. Furthermore, the observed volume is proportional to $s_\parallel s^2_\perp$, which implies $R^*=q_\perp^{2/3}q_\parallel^{1/3}R$ for the true effective void radius \citep{Hamaus2020,Correa2021a}.

In practice, the AP test is applied to measurements of the void-galaxy cross-correlation function $\xi^s(s_\perp,s_\parallel)$ in redshift space. It is customary to use the \citet{Landy1993} estimator for this purpose:
\begin{equation}
	\hat{\xi}^s(s_\perp,s_\parallel) = \frac{\langle{\mathcal{D}_\mathrm{v} \mathcal{D}_\mathrm{g}}\rangle-\langle{\mathcal{D}_\mathrm{v} \mathcal{R}_\mathrm{g}}\rangle-\langle{\mathcal{R}_\mathrm{v} \mathcal{D}_\mathrm{g}}\rangle+\langle{\mathcal{R}_\mathrm{v} \mathcal{R}_\mathrm{g}}\rangle}{\langle{\mathcal{R}_\mathrm{v}\mathcal{R}_\mathrm{g}}\rangle}  \;\; ,
	\label{LS_estimator}
\end{equation}
where the angled brackets indicate normalized pair counts of void-center and galaxy positions in the data ($\mathcal{D}_\mathrm{v}$, $\mathcal{D}_\mathrm{g}$) and in random catalogs ($\mathcal{R}_\mathrm{v}$, $\mathcal{R}_\mathrm{g}$) without spatial correlations. The number of random objects has to be large enough to guarantee an unbiased estimate of $\xi^s$, it is typically set one to two orders of magnitude higher than the number of observed objects of each kind. From Eq.~\ref{LS_estimator} it is then straightforward to estimate the projected correlation function via line-of-sight integration:
\begin{equation}
	\hat{\xi}^s_\mathrm{p}(s_\perp) = \int\hat{\xi}^s(s_\perp,s_\parallel)\,\mathrm{d}s_\parallel \;\; .
	\label{xi_p}
\end{equation}
Application of the inverse Abel transform from Eq.~\ref{iAbel} then provides the real-space correlation function $\xi(r)$ (see Fig.~\ref{fig:void_size}), which is needed as a model ingredient for $\xi^s(\mathbf{s})$, as in Eqs.~\ref{xi_s} or \ref{GSM}. For example, assuming linear bias Eq.~\ref{xi_s} can be written as:
\begin{equation}
	\xi^s(s_\perp,s_\parallel) = \xi(r) + \frac{1}{3}\frac{f}{b}\overline{\xi}(r) + \frac{f}{b}\mu^2\left[\xi(r)-\overline{\xi}(r)\right] \;\; , 
	\label{xi_s_2}
\end{equation}
where $\overline{\xi}(r) = 3r^{-3}\!\int_0^r\xi(r')\,r'^2\,\mathrm{d}r'$. This can be compared to the measured $\hat{\xi}^s(s_\perp,s_\parallel)$ assuming a Gaussian likelihood:
\begin{equation}
	\mathcal{L}(\hat{\xi}^s|\mathbf{\Theta}) \propto \exp\left\{-\frac{1}{2}\sum\limits_{i,j}\left[\hat{\xi}^s(\mathbf{s}_i)-\xi^s(\mathbf{s}_i|\mathbf{\Theta})\right]\,\hat{\mathsf{C}}_{ij}^{-1}\left[\hat{\xi}^s(\mathbf{s}_j)-\xi^s(\mathbf{s}_j|\mathbf{\Theta})\right]\right\} \;\; ,
	\label{likelihood}
\end{equation}
with model parameter vector $\mathbf{\Theta}$ and covariance matrix:
\begin{equation}
	\hat{\mathsf{C}}_{ij} = \left\langle{\left[{\hat{\xi}^s(\mathbf{s}_i)-\langle\hat{\xi}^s(\mathbf{s}_i)\rangle}\right]\left[{\hat{\xi}^s(\mathbf{s}_j)-\langle\hat{\xi}^s(\mathbf{s}_j)\rangle}\right]}\right\rangle \;\; .
	\label{covariance}
\end{equation}
Here, angled brackets imply averages over an ensemble of observations. Because voids are spatially exclusive, they represent independent regions of the large-scale structure and the covariance matrix can be estimated via jackknife resampling of the observed sample of voids \citep[e.g.,][]{Paz2013,Hamaus2015,Cai2016,Correa2019}.

The likelihood can be used to determine the AP parameter $\varepsilon$, which is equivalent to a measurement of the product of angular diameter distance $D_\mathrm{A}(z)$ and Hubble expansion $H(z)$ at redshift $z$. Because these two quantities depend on the cosmological model via Eqs.~\ref{D_A(z)} and \ref{eq:Hz1}, measurements of $\varepsilon$ can be converted to constraints on the cosmological parameters that enter these two equations. A variation of $\varepsilon$ corresponds to a change in distance ratios along and perpendicular to the line of sight, which can be described as a geometric distortion of void shapes. However, voids are additionally affected by RSD due to the peculiar velocity flows in their immediate surroundings, as described in Sect.~\ref{sec:V1}. The magnitude of these velocities and hence the strength of dynamic distortions is controlled by the growth rate parameter $f$, which enters in the model Eq.~\ref{xi_s_2} via Eq.~\ref{u(r)}. In order to properly model the average shapes of voids, respectively the void-galaxy cross-correlation function in redshift space, geometric and dynamic distortions must be accounted for at the same time \citep{Hamaus2015,Hamaus2016}. Fortunately, the two types of distortions influence $\xi^s(s_\perp,s_\parallel)$ in fundamentally different ways, such that there is no significant degeneracy between the parameters $\varepsilon$ and $f/b$.

\subsubsection{Systematic effects}
\label{sec:V4}

The mass distribution on cosmological scales is predominantly constituted by invisible dark matter. Large-scale structure surveys merely allow us to infer this distribution via luminous tracers of the mass, but this inference is subject to bias, statistical noise, and other sources of error. As we rely on the spatial distribution of tracers for void identification, these complications necessarily propagate into the properties of voids as a source of systematic effects. The main known systematics are summarized in the list below.

\noindent
{\bf Clustering bias.}
The over-densities of tracers $\delta_\mathrm{t}$ generally differ from the fluctuations in the matter density field $\delta_\mathrm{m}$, a phenomenon referred to as {\it tracer bias} \citep{Desjacques2018}. At linear order in the perturbations this difference is quantified by a multiplicative constant $b$, denoted as linear bias, with $\delta_\mathrm{t}=b\delta_\mathrm{m}$ \citep{Kaiser1984}. For example, luminous red galaxies (LRGs) typically have $b>1$, because they populate relatively massive halos that form in the most over-dense environments \citep[e.g.,][]{GilMarin2015,Zhai2017}. Therefore, voids identified in the distribution of LRGs exhibit deeper interiors and higher compensation ridges compared to voids identified in the dark matter density field \citep{Sutter2014a,Pollina2017,Pollina2019}. As a consequence, basic void properties, such as their effective radius and density profile, depend on the bias of the tracer sample considered for their identification.

\noindent
{\bf Stochasticity.}
While the distribution of dark matter can be seen as a collisionless fluid, tracers of the mass consist of discrete objects, such as galaxies. Therefore, the density field of tracers $\delta_\mathrm{t}$ must be estimated from a finite number of objects per volume element, which is subject to discreteness noise, also referred to as {\it shot noise}. Typically, this shot noise is assumed to obey Poisson statistics, but corrections due to the finite extent of tracers and their nonlinear clustering appear \citep{Hamaus:2010im,2013PhRvD..88h3507B,Paech2017,Ginzburg2017,Friedrich2021}. Voids are necessarily affected by shot noise as well, if they are defined via tracer statistics. For example, even in a tracer distribution that is drawn from a homogeneous density field, chance fluctuations due to shot noise can result in spurious void detections \citep{Neyrinck2008}. Therefore, one may expect that not all voids identified in a real tracer distribution are genuine, but that there is a contamination of spurious voids depending on the sparsity of the considered tracer sample. With the help of simulations and mock catalogs the contamination fraction can be assessed, exploiting the fact that various void properties distinguish genuine from spurious voids. The use of machine learning methods is particularly effective to minimize the contamination by spurious voids \citep{Cousinou2019}.

\noindent
{\bf Nonlinear RSD.}
By design, large-scale structure surveys infer distances via the measured redshifts of sources, which are distorted due to their peculiar motion along the line of sight \citep{Kaiser1984}.  While just a decade ago peculiar velocities have been considered the strongest systematic effect to limit the extractable information from the void density profile \citep[e.g.,][]{Lavaux2012,Sutter2012b}, RSD models for voids have now reached a level of maturity that exploit peculiar velocities as an independent source of information. At the linear level, which is most relevant for voids, these RSDs can be modeled very accurately, as discussed in Sect.~\ref{sec:V1}, but their nonlinear regime is more complex and difficult to understand from first principles. A well-known example of an extreme type of nonlinear RSD is the so-called Finger-of-God (FoG) effect \citep{Jackson1972}. It arises around the most massive structures in the Universe observed in redshift space, galaxy clusters, and appears as an elongated feature along the line of sight. This apparent elongation is RSD caused by the virial motion of the cluster member galaxies. While the occurrence of FoGs inside voids is less likely, they can disrupt their over-dense boundaries. In turn, this can cause spurious mergers or segmentation of voids, preferentially along the line-of-sight direction, which results in an anisotropic selection effect \citep{Pisani2015b,Correa2021b}.

\noindent
{\bf Redshift error.} The measurement of redshift is error-prone itself. While this can be largely neglected for the high-resolution spectra obtained with spectroscopic redshift surveys, photometric surveys are subject to a relatively large photo-$z$ scatter that often amounts to a few percent uncertainty in redshift. Translated to a distance scale, this typically corresponds to several tens of Mpc and therefore strongly impacts the identification of voids whose extent is of the same order. 2D void finders have specifically been designed to reduce the impact of this error from photometric surveys \citep{Sanchez2017,Kovacs2019,Vielzeuf2021}. Another option is to rely on tracers with higher photo-$z$ accuracy, such as galaxy clusters, for the identification of voids \citep{Pollina2019}.

\noindent
{\bf Survey boundary.} Redshift surveys typically only observe a fraction of the full sky. In addition, objects in the foreground, such as stars or the plane of the Galaxy, have to be masked out. Together with the finite redshift range of the survey, this creates a complex geometry of the observed cosmological volume. Voids that intersect with a survey boundary are only partially observed and hence cannot be used for further analysis. This constraint concerns the largest voids most severely, as they are the most likely to extend beyond the edges of the survey. Thus, survey boundaries impact the detectable distribution of void sizes in a systematic way, which is not straightforward to predict \citep{Sutter2014}. To mitigate this effect, it is desirable to survey large contiguous fractions of the sky, and to discard voids that are too close to the survey boundary.

In reality the mentioned systematic effects do not occur in isolation, but impact the identification of voids jointly. It is therefore difficult to address them from purely theoretical grounds. As an alternative, various empirical approaches to handle systematics have been adopted in the literature. This can be realized in essentially two different ways: first, at the level of the data, such as performing a cleaning procedure to select voids based on their size and depth \citep[e.g.,][]{Contarini2019,Ronconi2019}, applying projections within redshift slices \citep[e.g.,][]{Sanchez2017}, or implementing a velocity reconstruction to control the void selection \citep{Nadathur2020}. Second, at the level of the model, which can be extended by additional nuisance parameters to allow some more flexibility \citep[e.g.,][]{Hamaus2020,Paillas2021}. Even though such extra parameters may not be uniquely associated with a given systematic effect, they can be marginalized over for the cosmological interpretation of the analysis.

\subsubsection{Main results and forecasts}
\label{sec:V5}

Voids have been considered for cosmological forecasts and constraints in various ways throughout the literature. Constraints from current data based on voids mainly rely on the void density profile, the theoretical modeling of the void size function has only recently reached maturity and will show its full power with larger samples of voids from the next generation of surveys. Therefore, here we focus on one of the most established applications to probe cosmology with voids: the AP test with the void-galaxy cross-correlation function $\xi^s(s_\perp,s_\parallel)$.

\begin{figure}[t!]
\centering
\resizebox{0.95\hsize}{!}{
\includegraphics[trim=10 0 0 0]{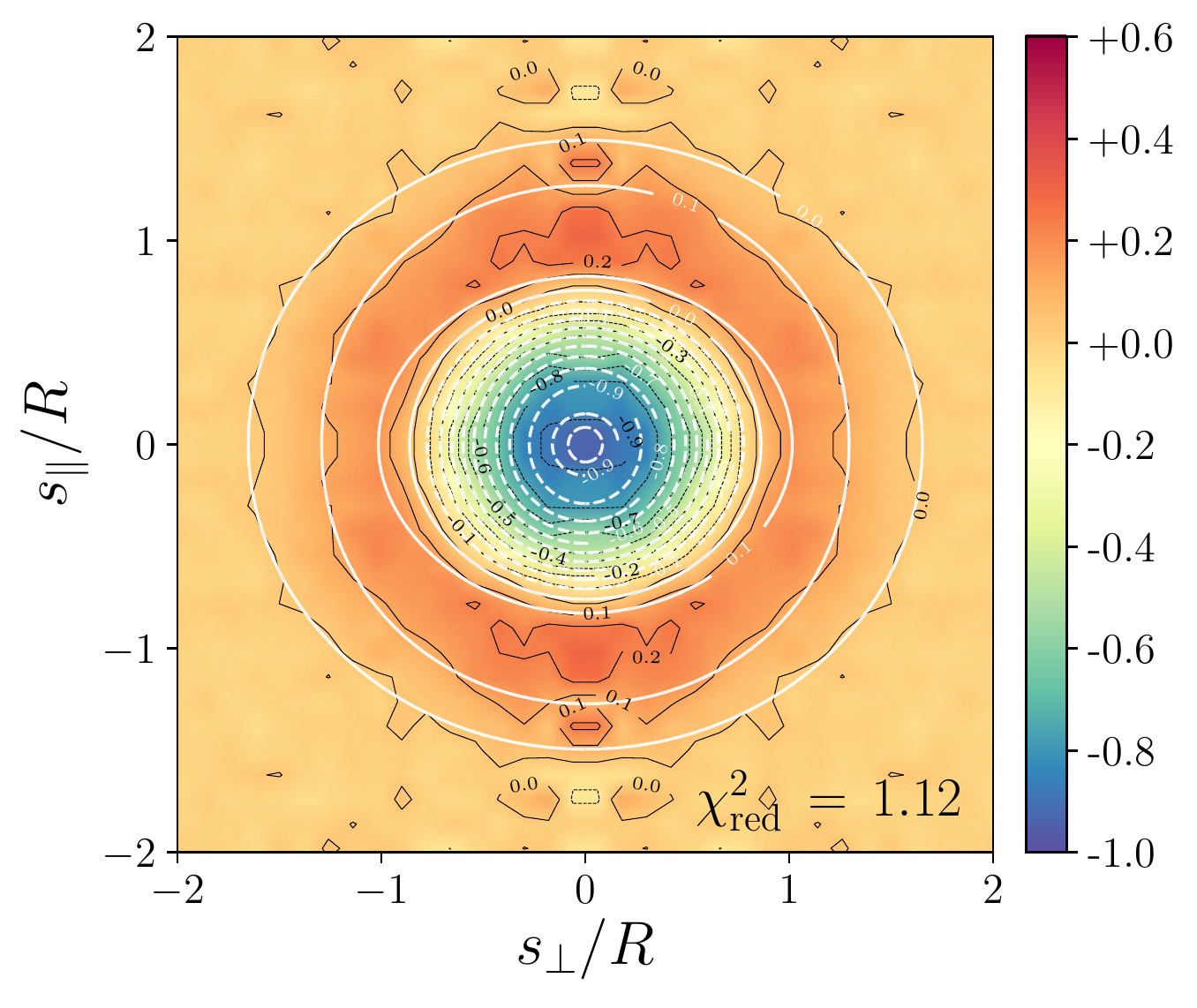}
\includegraphics[trim=0 -20 0 0]{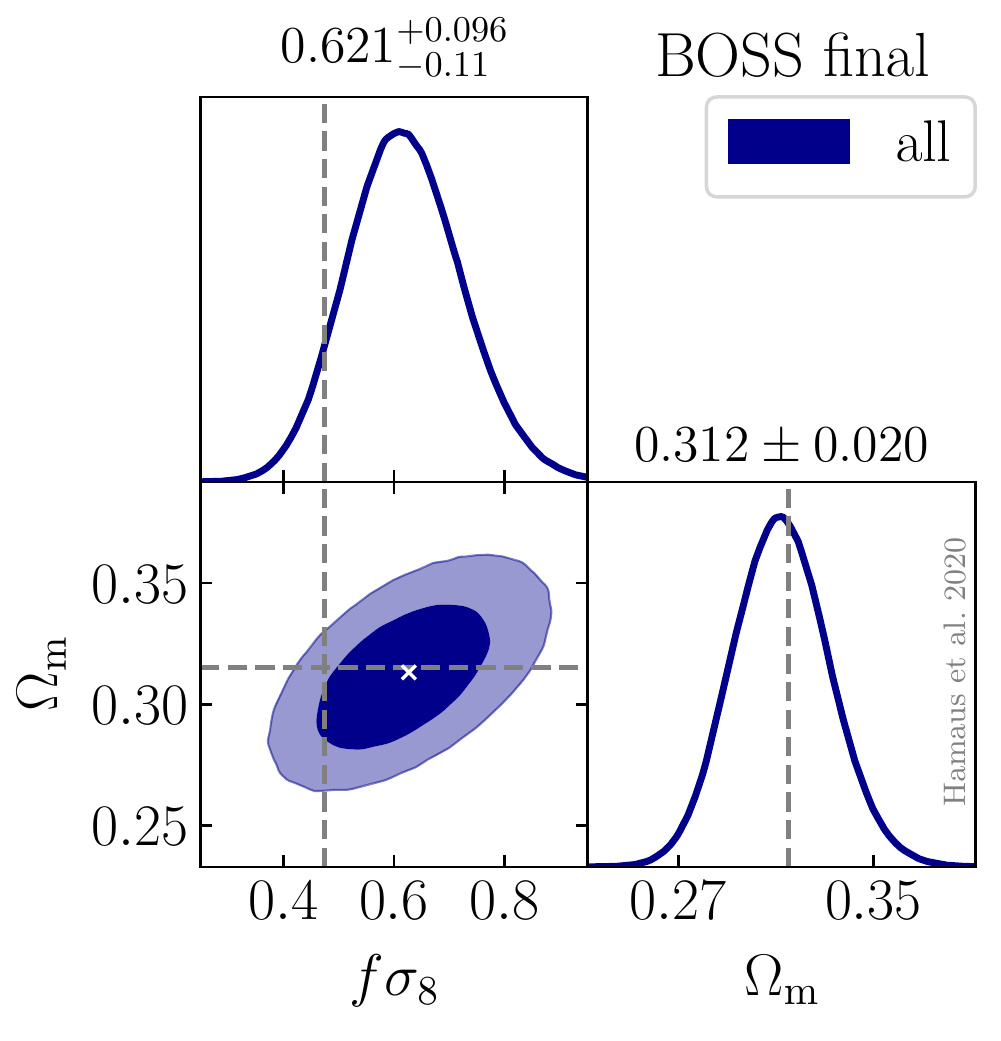}}
\caption{Left: Measurement of the void-galaxy cross-correlation function $\xi^s(s_\perp,s_\parallel)$ from voids in the final BOSS data (color scale with black contours) and the best-fit model (white contours). Right: Constraints on the parameters \omegam~and $f\sigma_8$ obtained from modeling the data in the left panel. A white cross indicates the best fit and dashed lines the mean parameter values obtained by the \citet{planck2018}. Images reproduced with permission from~\citet{Hamaus2020}, copyright by IOP Publishing.}\label{fig:xi2d_BOSS}
\end{figure}

Figure~\ref{fig:xi2d_BOSS} shows the results obtained by performing an AP test with voids identified in the final data release of the Baryon Oscillation Spectroscopic Survey~\citep[BOSS,][]{Dawson2013}. The left panel contains the measured $\hat{\xi}^s(s_\perp,s_\parallel)$ in bins of void-centric separations along and perpendicular to the line of sight, and the best-fit model indicated by white contour lines. Application of a MCMC sampler allows one to retrieve the posterior distribution of the model parameters $\varepsilon$ and $f/b$. Then, assuming a flat $\Lambda$CDM cosmology, $\varepsilon$ can be converted to \omegam, the only free parameter in the product $D_\mathrm{A}H$ within that model (since $\Omega_\mathrm{r}$ can be neglected and $\Omega_k=0$). Furthermore, with a measurement of the linear clustering amplitude of the tracer galaxies in BOSS, which is determined by the product of their bias $b$ and the amplitude of linear matter fluctuations $\sigma_8$, the ratio $f/b$ can be converted to the more commonly quoted combination $f\sigma_8$. Because $\sigma_8$ is defined in terms of $8\,h^{-1}\mathrm{Mpc}$, the posterior on $f\sigma_8$ should be marginalized over the Hubble constant $h$, which is often neglected \citep{Sanchez2020}.

The constraints on \omegam~and $f\sigma_8$ from the AP test with voids in the final BOSS data are shown in the right panel of Fig.~\ref{fig:xi2d_BOSS}. It demonstrates how competitive this relatively new method is, for example when compared to the more traditional approach that focuses on the pairwise clustering of galaxies \citep[e.g.,][]{Alam2017}. The latter is more challenging to model on small scales, due to the complex velocity statistics of galaxies in over-dense environments. However, on large scales it is imprinted with a characteristic scale of about $105\,h^{-1}$Mpc by the BAO feature that emerged during the radiation-dominated epoch of the early Universe, which can be used as a standard ruler to constrain $D_\mathrm{A}(z)$ and $H(z)$ individually. There are strong indications that the combination of such measurements with the AP test from voids can greatly improve the precision on cosmological parameters, thanks to their complementarity \citep[e.g.,][]{Nadathur2020,Paillas2021,Kreisch2021}. 

\begin{figure}[b!]
\centering
\resizebox{0.9\hsize}{!}{\includegraphics[trim=10 10 0 10]{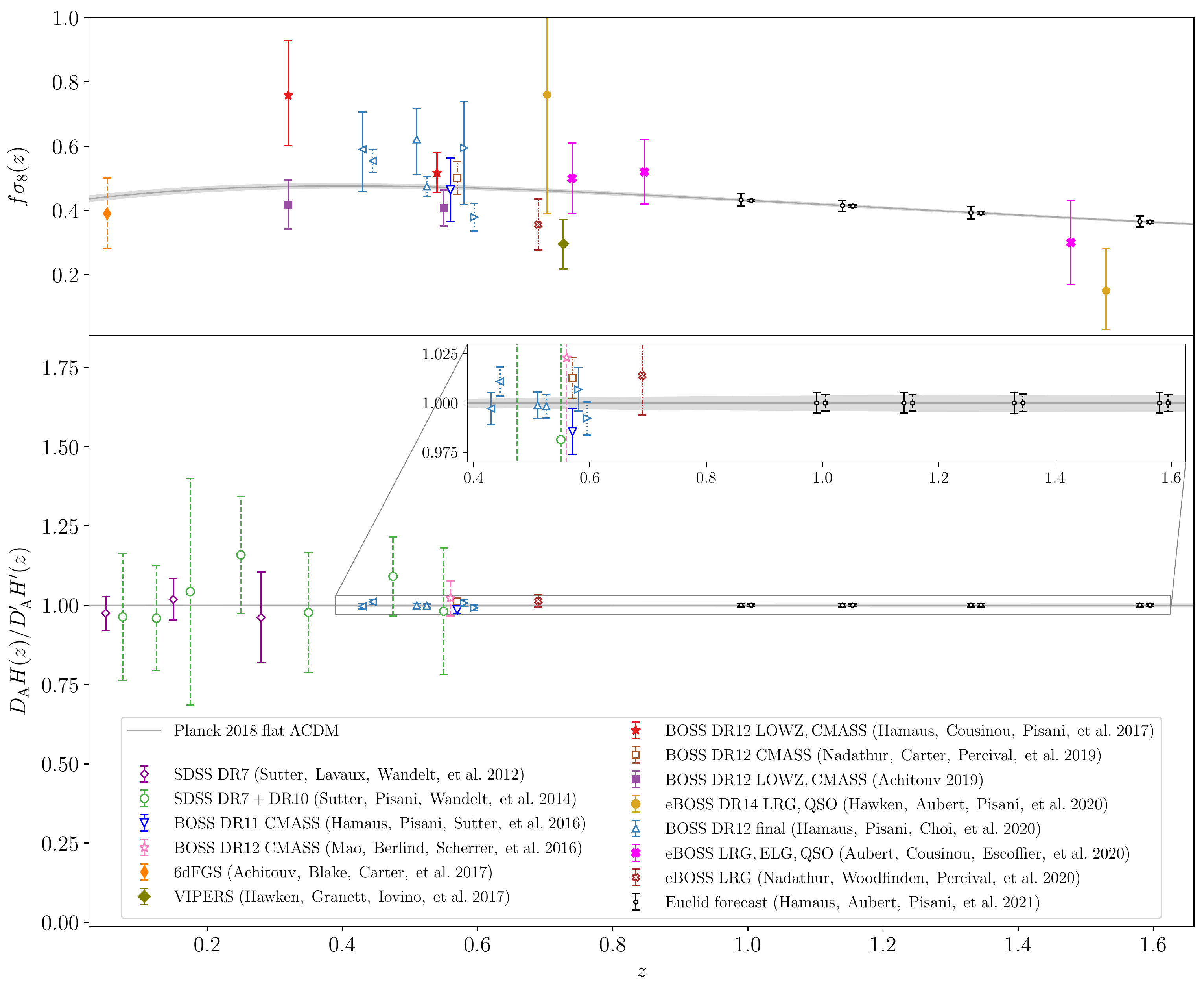}}
\caption{Comparison of constraints on growth via $f\sigma_8$ and geometry via $D_\mathrm{A} H$ ($68\%$ confidence intervals) obtained from cosmic voids in the literature, references are ordered chronologically in the figure legend. Gray lines with shaded error bands show the \citet{planck2018} baseline result as a reference, with corresponding values of $D_\mathrm{A}'H'$. Filled markers indicate growth rate measurements without consideration of the AP effect, while open markers include the AP test. The different line styles of error bars indicate various degrees of model assumptions made: model-independent (solid), calibrated on simulations (dashed), calibrated on mocks (dotted), calibrated on simulations and mocks (dash-dotted).}
	\label{fig:void_constraints}
\end{figure}

Figure~\ref{fig:void_constraints} summarizes the cosmological constraints that have been obtained from cosmic voids as a stand-alone probe in the literature. Despite being a young field of research, it has been blossoming with applications of increasingly accurate techniques applied to very different surveys, including SDSS \citep{Sutter2012b}, BOSS \citep{Sutter2014b,Hamaus2016,Mao2017b}, eBOSS \citep{Hawken2020,Aubert2020,Nadathur2020}, VIPERS \citep{Hawken2017}, and 6dFGS \citep{Achitouv2017}. All measurements of $D_\mathrm{A}H$ are based on the AP test, while some constraints on $f\sigma_8$ are only derived from dynamic distortions of void shapes and assume a fiducial cosmology with a fixed $D_\mathrm{A}H$. The method becomes particularly powerful towards higher redshift, where the observed volume and hence the available sample size of voids grows larger. Moreover, the product $D_\mathrm{A}(z)H(z)$ is an increasing function of redshift, so its measurement becomes more sensitive to changes in cosmological parameters at higher~$z$. These two trends will eventually be overcome by the declining amplitude of nonlinear fluctuations in the matter density field and the absence of observable tracers of the latter. However, upcoming surveys of the next generation, such as DESI \citep{DESI:2016}, Euclid \citep{Laureijs:2011gra}, PFS \citep{Takada:2014}, the Nancy Grace Roman Space Telescope \citep{Spergel:2015}, the Vera Rubin Observatory \citep{LSSTScience:2009jmu}, and SPHEREx \citep{Dore:2014} are expected to obtain void catalogs of unprecedented size, containing on the order of $10^5$ objects each~\citep{Pisani2019, Hamaus2021}. Compared to the current state of the art, this corresponds to an increase of about two orders of magnitude. Therefore, we expect the next generation of surveys to initiate an era of voids in the pursuit of precision cosmology.

\clearpage

\subsection{Neutral Hydrogen Intensity Mapping}

Traditionally, large-scale structure surveys aim to detect individual galaxies in three dimensions. This involves measuring the redshift of each galaxy as well as its angular position on the sky, and then creating a catalog and a corresponding 3D map. This procedure has been routinely used by optical galaxy surveys like SDSS and has led to constraints on dark energy, gravity, and the initial conditions of the Universe \citep[see, for example,][]{Beutler2012,Alam:2020sor,Mueller:2021tqa}. An alternative proposal is to map the large-scale structure of the Universe using the redshifted 21cm line from the spin flip transition in neutral hydrogen (\hinospace) with radio telescopes \citep{Battye:2004re,Chang:2007xk,Loeb:2008hg,Mao:2008ug,Peterson:2009ka,Seo:2009fq,Ansari:2011bv}. 

\subsubsection{Basic idea and equations}

\begin{figure}[b!]
\centering
\includegraphics[scale=0.6]{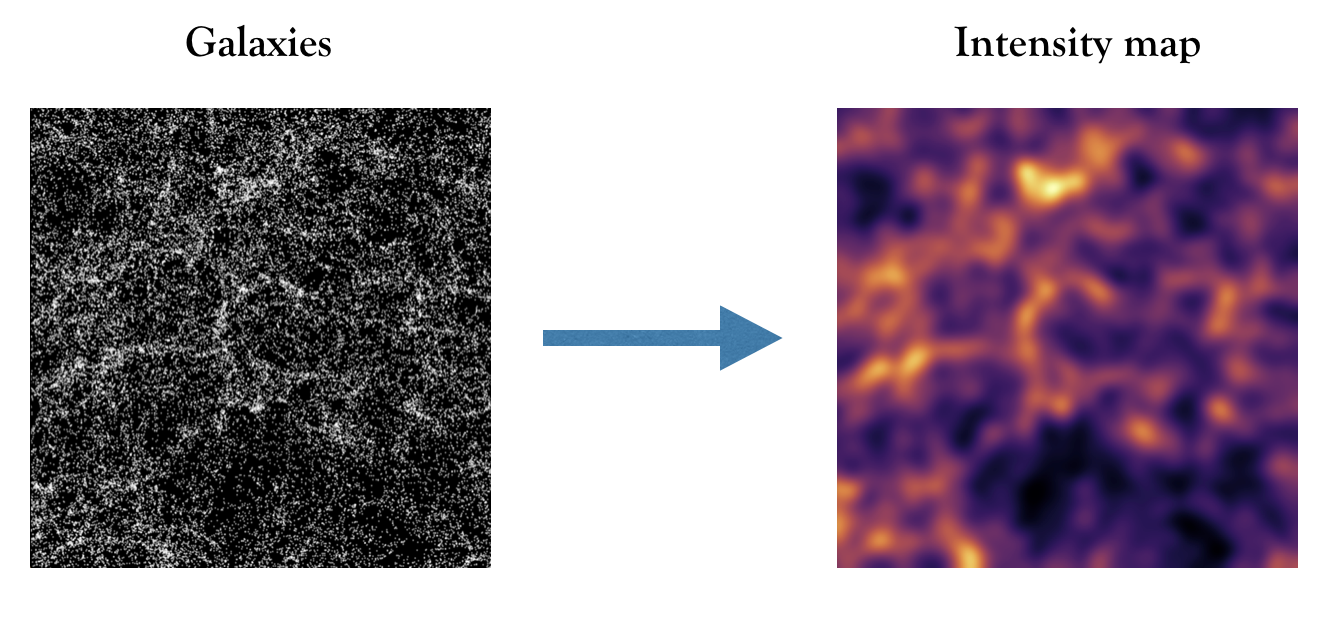}
\caption{From individual galaxies (left) to \hi intensity maps (right). Using the intensity mapping technique we can map the entire \hi flux from many galaxies together in large 3D pixels, and produce low angular resolution \hi brightness temperature maps that retain the large-scale statistical information. This figure was produced using the \texttt{MultiDark} simulations \citep{Klypin:2014kpa,Knebe:2017eei} and the methods in \cite{Cunnington:2020mnn}.}
\label{fig:IM-map}
\end{figure}

The \hi Intensity Mapping technique does not require the often difficult and expensive detection of individual galaxies. Instead, it maps the entire \hi flux coming from many galaxies together in large 3D pixels, across the sky and along time (see Fig.~\ref{fig:IM-map}).  With radio telescope arrays, the \hi intensity mapping method has the potential to provide the largest map of the Universe back to $\sim$1 billion years after the Big Bang. The data can then be used for precision cosmology and galaxy evolution studies \citep{Kovetz:2019uss, Ahmed:2019ocj}.

A number of \hi intensity mapping experiments are expected to launch in the next few years, with some of them already working with pathfinder data. These are the proposed MeerKLASS survey using the SKA Observatory's MeerKAT precursor \citep{Santos:2017qgq}, FAST \citep{Hu:2019okh}, BINGO \citep{Battye:2012tg, Wuensche:2018alk}, CHIME \citep{Bandura:2014gwa}, HIRAX \citep{Newburgh:2016mwi}, Tianlai \citep{Li:2020ast, Wu:2020jwm}, PUMA \citep{PUMA:2019jwd}, and CHORD \citep{Vanderlinde:2019tjt}. Existing experiments include \hi intensity mapping surveys performed with the Green Bank Telescope (GBT) \citep{Chang:2010jp, Switzer_2013, Switzer_2015, Masui:2012zc, Wolz:2021ofa} and Parkes \citep{Anderson:2017ert}. 

At cosmological distances, the 21cm line is redshifted to very low frequencies, which alleviates the danger of line confusion that often plagues other lines. There is a one-to-one correspondence of observing frequency, $\nu$, with redshift, $z$, given by:
\begin{equation}
    \nu = \frac{1420.4}{1+z} \, {\rm MHz}  \;\; .
\end{equation}
For this reason, there is no need of detecting and cataloging individual galaxies. Looking at some 3D region (voxel) on the sky, the radio telescope receives the total 21cm intensity from that region, as demonstrated in Fig.~\ref{fig:IM-map}.  This is a proxy  for the total amount of hydrogen in the voxel, which is then assumed to be a (biased) tracer for the total matter density. While the telescope beam can be quite large and erase the small-scale structure, the large-scale statistical information is retained. From Earth, the 21cm line is measurable up to very high redshifts, $z\sim50$, and could reach $z\sim200$ with a lunar instrument \citep{Furlanetto:2006jb}. This provides a unique opportunity for cosmology and astrophysics studies at high redshifts, where traditional galaxy surveys become shot noise limited.

It is important to consider the mode of operation of the telescope array. Purpose-built \hi intensity mapping experiments like CHIME and HIRAX are interferometers with elements that are closely packed together. Sparse arrays like MeerKAT and SKA-MID cannot provide enough short baselines to probe cosmological scales when used in interferometric mode. Instead, they need to operate in ``single-dish'' mode \citep{Battye:2012tg, Wang:2020lkn}, where the array is used as a collection of scanning auto-correlation dishes. This is necessary in order to map cosmological scales with sufficient sensitivity \citep{Bull:2014rha, Santos:2015gra, SKARedBook2018}. 

\begin{figure}[t!]
\centering
\includegraphics[scale=0.4]{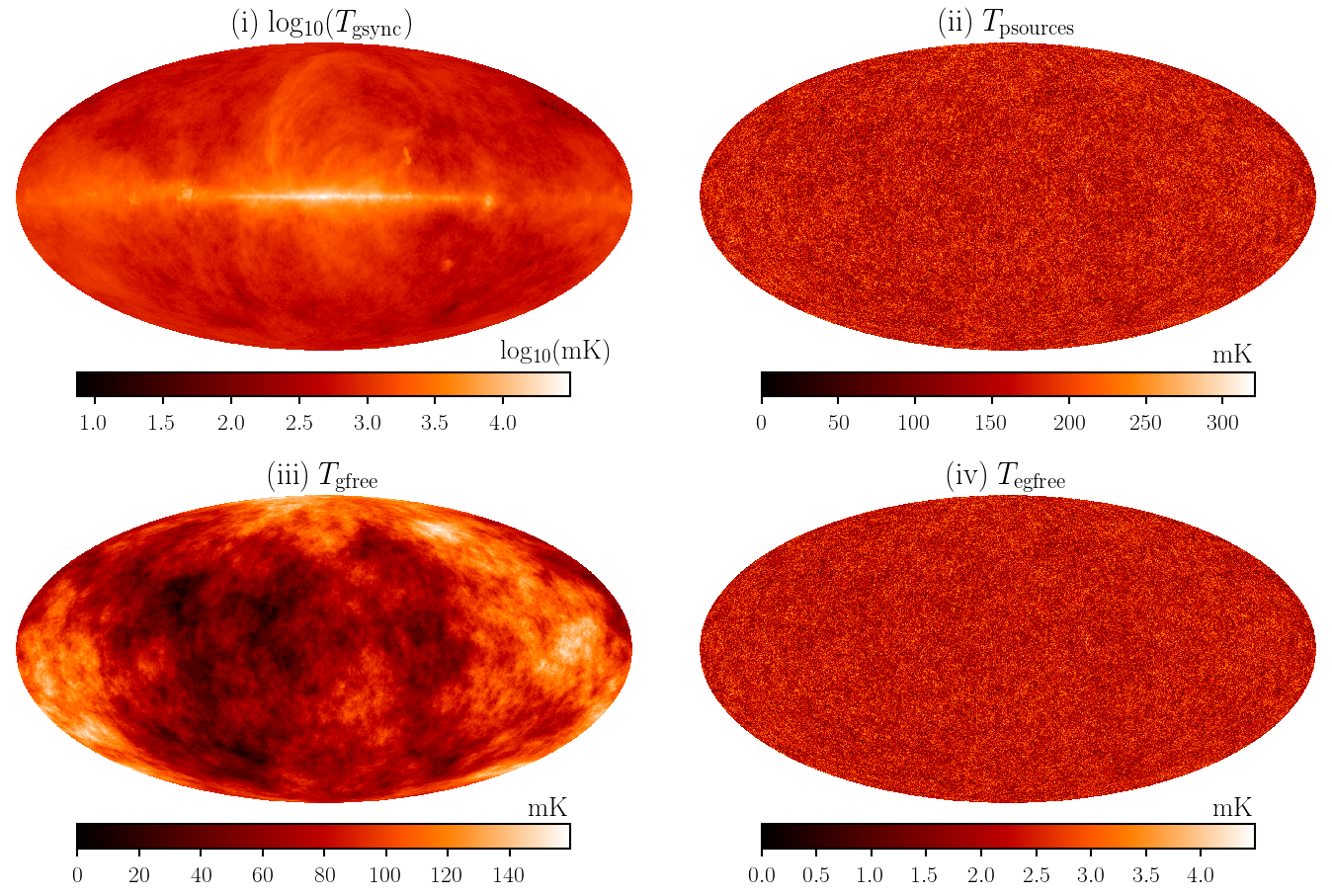}
\caption{Simulated full sky maps of different 21cm foreground components at a frequency of $1136$ MHz ($z = 0.25$). The frequency dependence of these foregrounds can be approximated by power laws with a running spectral index. Image reproduced with permission from \cite{Cunnington:2019lvb}, copyright by MNRAS.}
\label{fig:FG-maps}
\end{figure}

A major challenge for the \hi intensity mapping method is the presence of strong astrophysical emission, or foreground contamination: 21cm foregrounds such as galactic synchrotron \citep{Zheng:2016lul}, point sources, and free-free emission, are bright in the relevant frequency ranges and can be orders of magnitude stronger than the cosmological \hi signal (see Fig.~\ref{fig:FG-maps} and top panel of Fig.~\ref{fig:Cl-FG}). Hence, they have to be removed.

\subsubsubsection{Modeling the observed \hi signal}

We consider the 3D power spectrum as our main observable, and follow the formalism used in optical galaxy surveys analyses. Similarly to optical galaxies, redshift space distortions (RSD) introduce anisotropies in the observed \hi power spectrum. In order to account for this, we consider the power spectrum as a function of redshift $z$, $k$, and $\mu$, where $k$ is the amplitude of the wave vector and $\mu$ the cosine of the angle between the wave vector and the line-of-sight (LoS) component.
We model RSD by considering the Kaiser effect \citep{Kaiser:1987qv}, which is a large-scale effect dependent on the growth rate, $f$. To linear order, the anisotropic \hi power spectrum can be written as: 
\begin{equation}\label{eq:RSDPk}
	P_\hinospace(k, \mu) =  \left( \overline{T}_\hinospace b_\hinospace + \overline{T}_\hinospace f \mu^{2}\right)^{2} P_\text{M}(k) + P_\text{SN} \;\; ,
\end{equation}
where $P_\text{SN} = \overline{T}_\hinospace^2 (1/\overline{n})$ is the shot noise, $\overline{n}$ is the number density of objects, $P_\text{M}(k)$ is the underlying matter power spectrum, $b_\hinospace$ is the \hi bias, and $\overline{T}_\hinospace$ is the mean \hi brightness temperature \citep{Chang:2010jp, Battye:2012tg}:
\begin{equation}
\label{eq:TbarModelEq}
\overline{T}_\hinospace=44 \left(\frac{\Omega_{\mathrm{HI}}(z) h}{2.45 \times 10^{-4}}\right)  \frac{(1+z)^{2}}{H(z)/H_0} \mathrm{\mu K} \, .
\end{equation}
The $P_{\rm SN}$ contribution is expected to be subdominant (smaller than the thermal noise of the telescope) and is usually neglected \citep{Villaescusa-Navarro:2018vsg}. The \hi abundance and clustering properties have been studied using simulations and semi-analytical modeling \citep[see, e.g.,][]{Villaescusa-Navarro:2018vsg,Spinelli:2019smg}. In general, the clustering of \hi can be accurately described by perturbative methods \citep{Castorina:2019zho}, and maps can be constructed with N-body and approximate methods based on the halo model \citep{Alonso:2014sna, Villaescusa-Navarro:2014cma, Padmanabhan:2015wja, Padmanabhan:2016fgy, Carucci:2016yzq, Spinelli:2021emp, Avila:2021wih}. For interpreting current and forthcoming observations, there is a pressing need to work towards end-to-end simulations including observational effects  \citep{Spinelli:2021emp}.

\subsubsubsection{The telescope beam effect}

The telescope beam introduces one of the main instrumental effects in the case of single-dish intensity mapping experiments. We can model this effect using a damping term dependent on the physical smoothing scale of the beam \citep[see, e.g.,][]{Battye:2012tg,Villaescusa-Navarro:2016kbz}. Assuming the telescope beam can be modeled as a Gaussian, this is defined as $R_\text{beam} = \sigma_\theta r(z)$, where $\sigma_\theta =  \theta_\text{FWHM} / ( 2\sqrt{2\ln(2)} )$, $\theta_\text{FWHM} \sim \lambda/D_{\rm dish}$ is the full-width-half-maximum of the beam with diameter $D_{\rm dish}$ at observation wavelength
$\lambda = 21(1+z)$ cm, and $r(z)$ is the comoving distance to a redshift $z$. We emphasize that the angular resolution of single-dish surveys is very low, of the order $\sim$1 deg, while interferometers have much better angular resolution.

The Fourier transform of the telescope beam damping term is:
\begin{equation}
	\widetilde{B}_\perp(k,\mu) = \exp\left(\frac{-k^2 R_\text{beam}^2(1-\mu^2)}{2}\right) \;\; ,
\label{eq:gaussian-beam}
\end{equation}
and the power spectrum becomes:
\begin{equation}\label{RSDPk}
	P_\hinospace(k, \mu) = \widetilde{B}_\perp^2(k,\mu)
 \times \left[ \left( \overline{T}_\hinospace b_\hinospace + \overline{T}_\hinospace f \mu^{2}\right)^{2} P_\text{M}(k) + P_\text{SN} \right] \;\; .
\end{equation}
For surveys that are limited in frequency resolution, a similar effect will occur on the small radial scales. 
In cases where this might be relevant, a way to account for it is described in \cite{Blake:2019ddd}.   

\subsubsubsection{Thermal noise}

Instrumental noise is determined by the telescope configuration and survey strategy \citep[see, e.g.,][for detailed descriptions of representative cases]{Battye:2012tg, Bull:2014rha, Pourtsidou:2016dzn}. 
For a single-dish experiment, the pixel noise is assumed to be described by a Gaussian random field with spread given by:
\begin{equation}\label{NoiseEq}
    \sigma_{\mathrm{pix}}=\frac{T_{\mathrm{sys}}(\nu)}{\sqrt{\delta_\nu t_{\mathrm{total}}\left(\Omega_{\mathrm{pix}} / S_{\text {area }}\right) N_{\mathrm{dishes}}}} \;\; .
\end{equation}
Here, $T_{\rm sys}(\nu)$ is the system temperature (including receiver and sky components) at a given frequency, $S_{\rm area}$ the sky area, $\Omega_{\rm pix} = 1.33 \theta^2_{\rm FWHM}$, $N_{\rm dishes}$ the number of dishes (this can also include multiple feeds/beams per dish), $\delta_\nu$ the frequency channel bandwidth, and $t_{\rm obs}$ the total observation time. The combination $t_{\rm obs} (\Omega_{\rm pix} /S_{\rm area})$ represents the time spent at each pointing.
It follows that the noise power spectrum is: 
\begin{equation}
    P_\text{N} = \sigma_\text{pix}^2 V_\text{pix} \, , 
\end{equation}
where $V_\text{pix}$ is the voxel volume. Typical values for a cosmological survey using MeerKAT in single-dish mode are $T_{\rm sys} \sim 30 \, {\rm K}$, $N_{\rm dishes} = 64$, $S_{\rm area}= 5,000 \, {\rm deg}^2$ and $t_{\rm total} = 5,000 \, {\rm hrs}$.

For the simplest form of interferometer, a dual polarization array assuming uniform antennae distribution, the noise power spectrum is:
\begin{equation}
    P^{\mathrm{N}}=T_{\mathrm{sys}}^{2} r^{2} y_{\nu}\left(\frac{\lambda^{4}}{A_{\mathrm{e}}^{2}}\right) \frac{1}{2 n(u) t_{\mathrm{total}}}\left(\frac{S_{\mathrm{area}}}{\mathrm{FOV}}\right) \;\; .
\end{equation}
Here, $A_e$ is the effective beam area, ${\rm FOV} \approx \lambda / (D_{\rm dish})^2$, $r$ is the comoving distance to the observation redshift $z$, and $y_\nu = c(1+z)^2/(\nu_0H(z))$ with $\nu_0 = 1420$ MHz. The distribution function of the antennae $n(u)$ can be approximated as 
$n(u) \simeq N^2_f / 2\pi u^2_{\rm max}$ for the uniform case, where $N_f$ is the number of elements of the interferometer and $u_{\rm max} \simeq D_{\rm max}/\lambda$ with $D_{\rm max}$ the maximum baseline. Typical values for a compact instrument like HIRAX are $T_{\rm sys}=50 \, {\rm K}$, $N_f = 1024$, $D_{\rm dish} = 6 \, {\rm m}$, $D_{\rm max} = 250 \, {\rm m}$, $S_{\rm area}= 15,000 \, {\rm deg}^2$ and $t_{\rm total} = 10,000 \, {\rm hrs}$.

\subsubsection{Sample selection}

\hi intensity mapping maps the entire \hi flux coming from many galaxies together in large voxels. This means that we do not need to select individual galaxies. 
An advantage of the intensity mapping technique is that it is sensitive to all sources of \hi emission, regardless how faint. This is in contrast to traditional galaxy surveys, which are sensitive only above a flux cutoff. This makes \hi intensity mapping ideally suited to probe the global \hi content, a key quantity for galaxy formation and evolution studies. 

Another fundamental choice is the bandwidth of observation. For example, MeerKAT can perform cosmological observations using its L-band ($900-1420$ MHz) or UHF-band ($580-1000$ MHz) receivers. The former corresponds to a redshift range $0<z<0.58$, while the latter can probe $0.4<z<1.45$. Band 1 of SKA-MID corresponds to a very wide redshift range, $0.35<z<3$. Other examples are CHIME and HIRAX, with $0.8<z<2.5$. Depending on the bandwidth of observation as well as the sky area coverage and total observing time, these \hi intensity mapping surveys can measure Baryon Acoustic Oscillations and Redshift Space Distortions, and search for signatures of primordial non-Gaussianity. 

The selection of frequency bandwidth and patch of sky can also be tuned to try and mitigate known systematic effects: 
\begin{itemize}
    \item Human-made Radio Frequency Interference (RFI) is a major source of contamination \citep{Harper:2018ncl}. While methods for RFI flagging and removal do exist \citep{Offringa:2010kb, Akeret_2017}, it is important to perform observations in ``radio-quiet'' locations.
    \item Foreground contamination from Galactic synchrotron, free-free emission, and point sources, can be orders of magnitude larger than the \hi cosmological signal \citep[see, e.g.,][]{Oh:2003jy}. Different regions of the sky are contaminated by the various foregrounds differently, and regions of the sky that are particularly complex (for example the Galactic plane) should be avoided.
    \item At the time of writing, there has been no detection of the \hi auto-correlation signal due to residual foregrounds and other systematics \citep[see, e.g.,][]{Switzer_2013,Switzer_2015}. The only available detections come from cross-correlating \hi maps with spectroscopic optical galaxies \citep{Chang:2010jp, Masui:2012zc, Anderson:2017ert, Wolz:2021ofa}. While detecting the \hi auto-correlation is the primary aim, it is currently desirable that the chosen patch of sky overlaps with optical galaxy surveys.
\end{itemize}

\subsubsection{Measurements}
\label{sec:NHIM-measurements}

In this section, we summarize the main steps of a typical \hi intensity mapping data analysis procedure, based on the pioneering works by the GBT, Parkes, and MeerKLASS teams \citep{Chang:2010jp, Switzer_2013, Switzer_2015,Masui:2012zc, Anderson:2017ert, Wang:2020lkn}. 
In general, single-dish observations require a scanning strategy where the dishes are rapidly moving across the sky. The goal is to keep the instrument gains (which are set by the so-called $1/f$ noise \citep{Harper:2017gln}) constant, limited only by the thermal noise fluctuations while covering the relevant angular scales. The scanning strategy is also tuned in order to avoid false signals, for example ground spill and atmospheric emission. 

\begin{figure}[b!]
\centering
\includegraphics[scale=0.4]{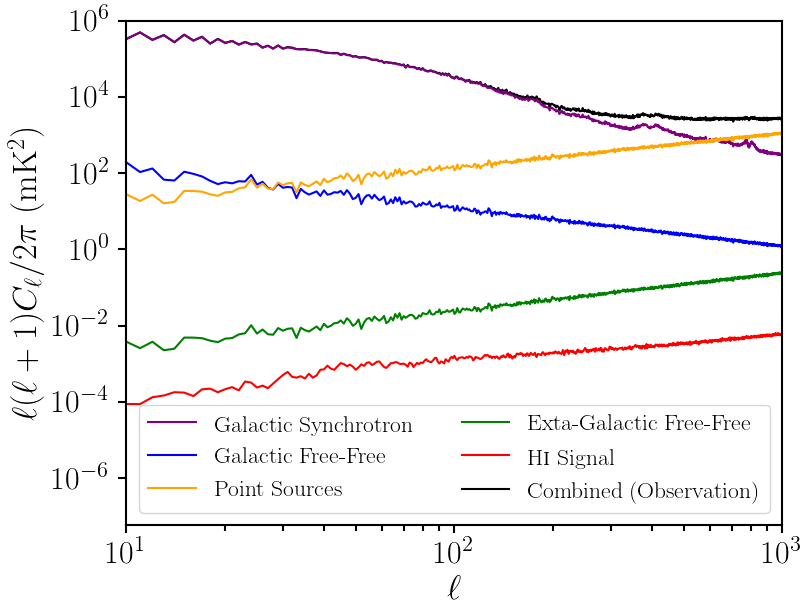}
\includegraphics[scale=0.4]{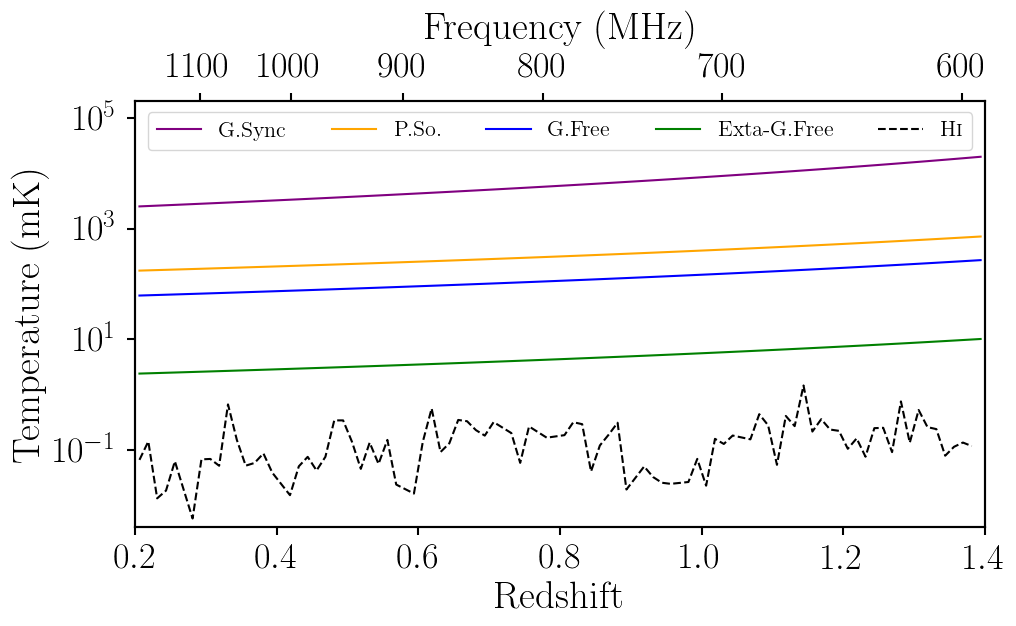}
\caption{Top: Angular power spectra for different simulated foregrounds, and the \hi cosmological signal. The black solid line represents the combined signal. All are at a frequency of 1136 MHz ($z = 0.25$). Bottom: Observed brightness temperatures along a chosen LoS through frequency (redshift). Images reproduced with permission from \cite{Cunnington:2019lvb}, copyright by MNRAS.}
\label{fig:Cl-FG}
\end{figure}

\noindent
{\bf From raw data to maps.} The raw data is stored in time-stream blocks. The first step of the data analysis is to mitigate RFI contamination. This is facilitated by the high spectral resolution of the data (e.g. 4096 channels across 200 MHz of bandwidth for the GBT).
Individual frequency channels are flagged and removed based on their variance. Any RFI in a block whose variance is not prominent enough to be flagged is identified as increased noise later on and down-weighted at the map-making stage. Some low-level RFI can be masked after map-making.
In addition, aliasing issues and high variance often result in removing channels within a few MHz of the band edges, as well as channels in the receiver's resonances. Before mapping, the data are re-binned (to $\sim$1 MHz bins). The time-stream data can be converted to sky maps with an inverse-noise weighted chi-squared minimization. This is a known method from CMB map-making \citep{Tegmark:1996qs}, and it produces the maximum likelihood (unbiased and optimal) estimate of the sky map assuming the noise is Gaussian. The algorithm also produces an inverse noise covariance matrix, useful for applying inverse-noise weights.

\noindent
{\bf Foreground subtraction.} 
Strong astrophysical foregrounds have to be separated from the cosmological \hi signal. Fortunately, these are expected to be spectrally smooth, following power-laws in frequency  \citep{Oh:2003jy, Santos:2004ju, Seo:2009fq}, and can be removed if the calibration of the instrument is well controlled. The top panel of Fig.~\ref{fig:Cl-FG} demonstrates the differences in amplitude of the various foregrounds compared to the cosmological \hi signal, while the bottom panel demonstrates the differences in spectral smoothness. 

Since cosmic \hi oscillates in a near-Gaussian fashion with frequency, in contrast to the slowly evolving foregrounds that are also orders of magnitude larger, the two can be separated \citep{Liu:2011hh,Wolz:2013wna,Shaw:2014khi, Alonso:2014dhk}. Blind component separation methods aim to identify a set of smooth functions (the dominant foreground components) and subtract them from the observed maps to uncover the cosmological \hi signal. 
There is a wealth of different foreground removal algorithms, including parameterized fitting, non-parametric fitting, and mode projection \citep[see][for a comprehensive review]{Liu:2019awk}. Principal Component Analysis (PCA) is a popular method that uses mode projection and exploits the fact that foregrounds are much larger in amplitude than the signal. PCA works by estimating the data frequency-frequency covariance\footnote{RFI can also be detected as frequency-frequency covariance in the foreground cleaning \citep{Switzer_2015}.} matrix and then performing an eigenvalue decomposition. The strongest modes in frequency (the foregrounds) can then be identified and projected out. An advantage of this ``blind'' approach is that it can take into account a distortion of the smoothness of the foregrounds by the instrument, as it works by determining which modes are dominant in the observed data. However, the price to pay is that inevitably a part of the cosmological \hi signal will also be removed \citep{Switzer_2015}. 
Other methods include Independent Component Analysis (ICA) \citep{Chapman:2012yj,Wolz:2015lwa} and Generalized Morphological Component Analysis (GMCA) \citep{Chapman:2012pn,Carucci:2020enz}. The former works by maximizing non-Gaussianity, and the latter is a sparsity-based algorithm that works with the spatial structure of the foregrounds in wavelet space. An example of a non-parametric fitting method is Gaussian Process Regression \citep{Mertens:2017gxw,Soares:2020zaq}. Other methods include the Generalized Needlet Internal Linear Combination (GNILC) \citep{Olivari:2017bfv} and Kernel PCA \citep{Irfan:2021bci}.
For recent comparisons of different foreground removal methods using real data and simulations, see \cite{Hothi:2020dgq,Cunnington:2020njn,Spinelli:2021emp}. 

\begin{figure}[t!]
\centering
\includegraphics[scale=0.65]{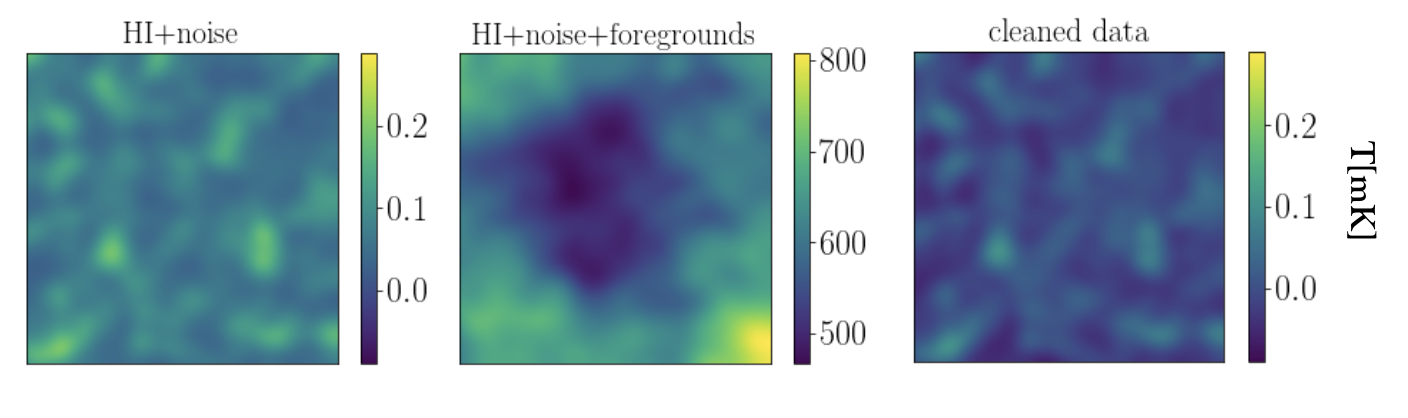}
\caption{Simulated \hi maps before and after foreground cleaning with PCA. From left to right: a map with simulated \hi signal with added thermal (instrumental) noise; the same map with added 21cm foregrounds; the ``cleaned'' map after performing foreground removal with PCA. This figure was produced using the publicly available code \texttt{gpr4im} \citep{Soares:2021ohm} and MeerKAT-like simulated data products at a frequency of $1136$ MHz ($z = 0.25$).}
\label{fig:IM-input-vs-cleaned}
\end{figure}

In Fig.~\ref{fig:IM-input-vs-cleaned} we have taken simulated \hi intensity maps, and added thermal (instrumental) noise and 21cm foregrounds. We have then performed a PCA foreground cleaning.
An important choice made by hand is how many principal components, $N_{\rm FG}$, to remove. In this case, we show results with $N_{\rm FG}=3$, which is expected to be near optimal for idealized simulated cases like the ones we have considered here \citep{Alonso:2014dhk, Wolz:2015lwa}. However, a much higher $N_{\rm FG} \sim 30$ has been required for real data analyses \citep{Masui:2012zc, Wolz:2021ofa}.

\subsubsubsection{Power Spectrum estimator} 

When performing an \hi intensity mapping survey, it is useful to separately analyze sub-datasets taken at different times (seasons) so that the thermal noise of the instrument is independent in each map. This way the \hi power spectrum can be constructed by cross-correlating (and then averaging over) different sub-datasets; this procedure has the advantage that the final power spectrum is free of the additive thermal noise bias \citep{Switzer_2013, Masui:2012zc}. The method can also suppress systematics like time-dependent RFI.

Intensity maps are over-temperatures measured as a discrete function of position, $\delta(\vec{x}_i)=T(\vec{x}_i)-\bar{T}$, where $\bar{T}$ is the mean temperature at each frequency slice. The total number of pixels, $N_{\rm pix}=N_x \cdot N_y \cdot N_z$, is defined by the angular grid and the number of frequency bins.
It follows that the Fourier transform of the temperature field is a function of wavevector $\vec{k}_\ell$. We can write:
\begin{equation}
    \tilde{\delta}(\vec{k}_\ell) 
    = \sum^{N_{\rm pix}}_{j=1} 
    \delta(\vec{x}_j)w(\vec{x}_j){\rm exp}(i\vec{k}_\ell \cdot \vec{x}_j) \;\; ,
\end{equation}
where $w(\vec{x}_j)$ is a weighting function normalized to unity.

\begin{figure}[t!]
\centering
\includegraphics[scale=0.4]{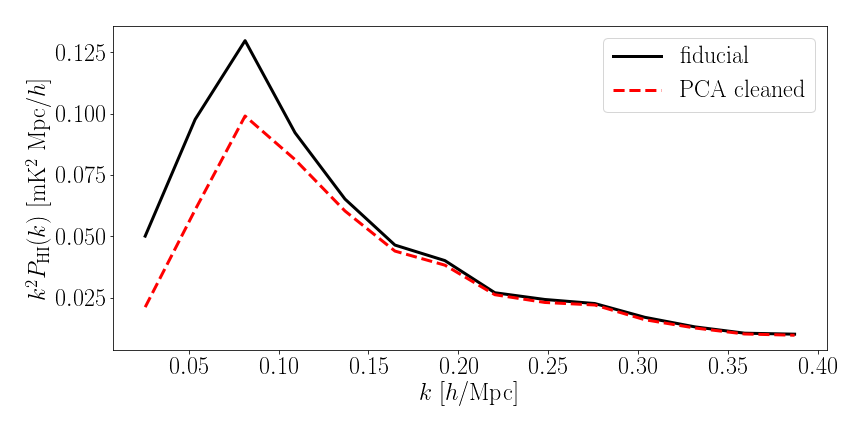}
\caption{Measured \hi power spectra demonstrating the \hi signal loss effect after foreground cleaning with PCA ($N_{\rm FG}=3$). This figure was produced using the publicly available code \texttt{IntensityTools} \citep{Cunnington:2019lvb,Blake:2019ddd,Soares:2020zaq} and MeerKAT-like simulated data products at $0.2 < z < 0.58$.}
\label{fig:Pk-input-vs-cleaned}
\end{figure}

Let us now introduce the inverse-noise weighted power spectrum estimator in the flat-sky approximation as used in the GBT and Parkes analyses\footnote{For a comprehensive description of all relevant observational effects and the derivation of general weighting schemes, as well as the publicly available pipeline, see \cite{Blake:2019ddd}.} \citep{Masui:2012zc, Wolz:2015lwa, Anderson:2017ert, Wolz:2021ofa}. For the cross-correlation of two sub-dataset maps $A$ and $B$, we have:
\begin{equation}
    \hat P^{AB}(\vec k_l)=\frac{V_{\rm cell} \mathrm{Re}\{ \tilde \delta^A(\vec k_l)\cdot
\tilde\delta^B(\vec k_l)^*\} 
}{\sum_{j=1}^{N_{\rm pix}} w^A(\vec x_j)\cdot  w^B(\vec x_j)} \;\; ,
\label{eq:PSest}
\end{equation}
with $V_{\rm cell} = V_s / N_{\rm pix}$, where $V_s = L_x \cdot L_y \cdot L_z$ is the comoving physical volume of the data cube. For \hi intensity maps, $w(\vec x_j)$ is given by the inverse noise map of each season. The estimator can be straightforwardly recast for the cross-correlation of intensity  maps with optical galaxies, denoted with subscript ``opt''. The total weighting factor is then $w(\vec x_j)=W(\vec x_j)w_{\rm opt}(\vec x_j)$, where $w_{\rm opt}(\vec x_j)$ is given by optimal weighting function $w_{\rm opt}(\vec x_i)=1/(1+W(\vec x_i)\times \bar N P_0)$, with $P_0=10^3 h^{-3}\rm{Mpc}^3$, and the selection function $W(\vec x_j)$ \citep{Feldman:1993ky}. The 1D power spectra, $\hat P(k)$, are determined by averaging all modes with $k = |\vec k|$ within the $k$ bin width. This is the well known power spectrum monopole. We can also compute higher order multipoles like the quadrupole and hexadecapole following the multipole expansion formalism \citep{Blake:2019ddd,Cunnington:2020mnn}.

In Fig.~\ref{fig:Pk-input-vs-cleaned} we show the measured power spectra from simulations with an input (fiducial) \hi signal, to which foregrounds are added and then removed using PCA. We can immediately see that the process of foreground cleaning results in large-scale \hi signal loss. Accounting for this effect is crucial in order to get unbiased \hi and cosmological constraints \citep{Masui:2012zc, Bernal:2019jdo,Cunnington:2020wdu,Soares:2020zaq}. More details on how this can be done will be presented in Sect.~\ref{sec:IM-results}, where we will also describe ways to estimate the statistical uncertainties in the measurements.

\subsubsection{Systematic effects}

The main known sources of systematic uncertainties affecting the \hi measurements are: foreground contamination, $1/f$ noise,  Radio Frequency Interference, calibration errors, and primary beam effects. We summarize these in the list below.

\noindent 
{\bf Foregrounds and polarization leakage.} 
We have already discussed that spectrally smooth foregrounds can be many orders of magnitude larger than the \hi cosmological signal (see \autoref{fig:Cl-FG}), and how their removal with methods like PCA or FastICA results in large-scale signal loss (\autoref{fig:Pk-input-vs-cleaned}). In addition, the interplay between polarized foregrounds and the instrument leads to polarization leakage, a non-smooth component that further complicates the cosmological analysis. For detailed studies on this subject in the context of \hi intensity mapping, see \cite{Shaw:2014khi, Alonso:2014sna, Alonso:2014dhk,  Carucci:2020enz,Cunnington:2020njn}. The auto-correlation of intensity maps is biased by residual foregrounds. However, these residuals and other survey-specific systematics are expected to drop out in cross-correlation with optical galaxy surveys, and that is why detections to date have only been achieved with cross-correlations \citep{Masui:2012zc, Anderson:2017ert, Wolz:2021ofa, Cunnington:2022uzo}. This means that significant advances in calibration, simulations, and data analysis techniques are needed for the 21cm foreground removal to work at the level required for precision cosmology. Working with pathfinder data \citep[see, e.g.,][]{Cunnington:2022uzo} will show whether this is possible with current and forthcoming instruments, or whether we need a future generation of purpose-built radio telescopes \citep{Ahmed:2019ocj}.

\noindent
{\bf $1/f$ noise.} 
This is a form of time-correlated noise component that manifests itself as gain fluctuations and leads to stripes in the \hi intensity maps. This noise can be mitigated with a fast enough scanning strategy, reduced by applying conservative PCA cleaning in the time-ordered data, and/or calibrated out. For detailed studies of this subject in the context of \hi intensity mapping, see \cite{Bigot-Sazy:2015jaa, Harper:2017gln,Li:2020bcr}.

\noindent 
{\bf RFI.} 
We have already mentioned Radio Frequency Interference (RFI), which can originate from terrestrial telecommunications as well as navigation satellites \citep{Harper:2018ncl}, and is a major problem for all radio observations. Even if the experiment employs RFI mitigation systems, it has been shown that RFI can still dominate thermal noise in several channels within the band \citep{Switzer_2013,Masui:2012zc,Wang:2020lkn}, resulting in significant signal loss ($\sim$11\% for the GBT). In Sect.~\ref{sec:NHIM-measurements} we discussed how RFI flagging and removal is performed, although this is likely not optimal considering the requirements of forthcoming \hi intensity mapping experiments. For example, missing frequency channels as a result of RFI flagging can compromise the performance of foreground removal methods \citep{Carucci:2020enz, Soares:2021ohm}. 

\noindent
{\bf Calibration.} 
Bandpass and flux calibration errors can have a large impact on the \hi signal recovery. A successful calibration process must calibrate the receiver gain fluctuations, account for the bandpass spectrum that multiplies the true sky signal, and calibrate the total power. 
The main calibration procedures are using periodic noise diodes as relative calibration references and tracking known astronomical sources, each of which has its own limitations and uncertainties \citep{2013PhDT.......570M, Newburgh:2014toa, Anderson:2017ert, Wang:2020lkn}. For the GBT observations used in \cite{Masui:2012zc}, uncertainties on the calibration of the reference flux scale and the measurements of calibration sources with respect to this reference, uncertainty of the measured fluxes, receiver non-linearity, beam shape irregularities and other variations led to a $9\%$ total calibration systematic error. This translates to systematic errors in the derived \hi constraints below the statistical errors for the GBT levels of thermal noise, but for future experiments aiming to perform high precision cosmological measurements calibration levels must be improved. 

\noindent
{\bf Primary beam effects.}
In the vast majority of \hi intensity mapping literature, the telescope beam effect is approximated by a perfect Gaussian smoothing, like in Eq.~\ref{eq:gaussian-beam}. But in reality, there are side-lobes in the beam profile; the primary beam can also distort the frequency structure of the foregrounds due its own dependence on frequency. A way to mitigate this and other issues related to the instrumental response is to convolve all maps to a common resolution, higher than the one of the largest beam in the frequency band \citep[see, e.g.,][]{Switzer_2015, Wolz:2015lwa}. However the way this convolution is done is based on the Gaussian beam model. Side-lobes further complicate foreground removal, and end-to-end simulations will be necessary in order to address this challenge \citep{Matshawule:2020fjz, Spinelli:2021emp}.

\subsubsection{Main results and forecasts}
\label{sec:IM-results}

Here we summarize the main data analysis results and forecasts. We begin by describing how the GBT measurements and power spectrum analyses have been performed. Then we present current constraints as well as forecasts on \hi parameters. We end this section by listing some of the cosmological forecasts that have been performed to demonstrate the ability of \hi intensity mapping surveys to constrain dark energy, gravity, and the initial conditions of the Universe. 

\subsubsubsection{Data analyses}

The most comprehensive \hi intensity mapping analyses to date have been performed using GBT observations, alone and in combination with optical galaxy surveys \citep{Chang:2010jp, Switzer_2013,Switzer_2015, Masui:2012zc, Wolz:2015lwa, Wolz:2021ofa}. In this section, we will describe how these analyses were performed, and present the detections and constraints they achieved. For a comprehensive description of the GBT pipeline and analysis software\footnote{\url{https://github.com/kiyo-masui/analysis_IM}} we refer the reader to \cite{2013PhDT.......570M}.

The GBT data we work with cover $100 \, {\rm deg}^2$ on the sky and a redshift range $0.6<z<1$. The data is contaminated by RFI and two telescope resonance frequencies. To suppress these effects, several frequency (redshift) channels were removed. The GBT also uses 4 Sections (sub-seasons) $\{A,B,C,D\}$ to suppress thermal noise bias as described in Sect.~\ref{sec:NHIM-measurements}. The noise is large and highly anisotropic towards the edges of the map, therefore 15 pixels per side are masked from the analysis. Foreground removal on the GBT data has been performed using PCA \citep{Switzer_2013, Masui:2012zc} and FastICA \citep{Wolz:2015lwa, Wolz:2021ofa}. 
The GBT beam can be approximated by a Gaussian with a frequency-dependent FWHM, and in order to mitigate some systematic effects as explained in section~\ref{sec:NHIM-measurements}, the maps are convolved to a common resolution of $0.44 \, {\rm deg}$.

\begin{figure}[t!]
\centering
\includegraphics[scale=0.7]{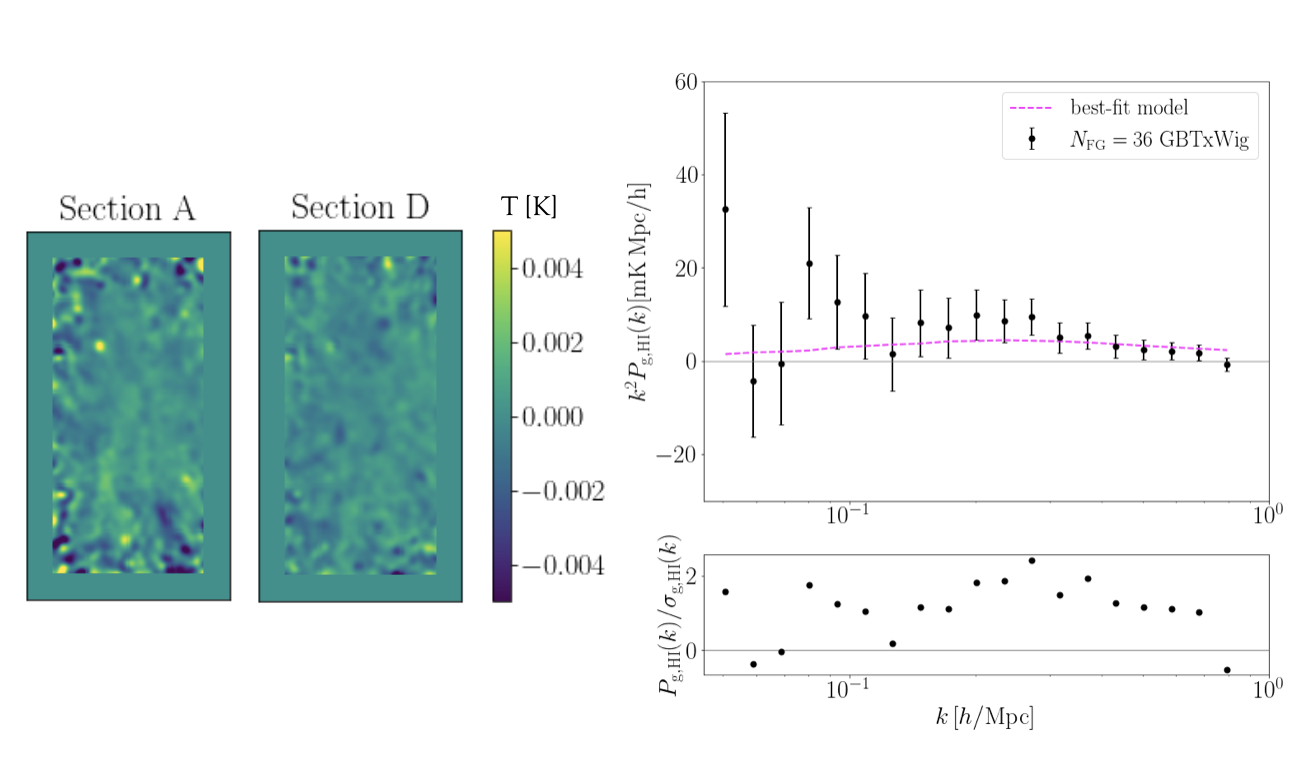}
\caption{Left: GBT data Sections A and D used in the analysis of \cite{Masui:2012zc, Wolz:2021ofa} with the masking choices detailed in the main text. The different Sections correspond to different data seasons and help mitigate thermal noise bias and other systematic effects. Right: GBT-WiggleZ cross-correlation power spectrum (top) and a null diagnostic test (bottom) using data from the analysis of \cite{Wolz:2021ofa}.}
\label{fig:IM-GBT-seasons}
\end{figure}

Here, we concentrate on the most recent analysis presented in \cite{Wolz:2021ofa}. In the left panel of Fig.~\ref{fig:IM-GBT-seasons} we show the (masked) Sections $A$ and $D$ after using $N_{\rm FG}=36$ in the FastICA foreground removal process. In principle, the cross-correlation of Sections (e.g. $A \times B$, $A \times D$, etc.) should be a proxy for the \hi auto power spectrum. 
For this, the GBT analysis using the estimator of Eq.~\ref{eq:PSest}. A correction is applied to the power spectrum estimate for the telescope beam effect using the discretized, Fourier-transform Gaussian beam of Eq.~\ref{eq:gaussian-beam}.

The GBT data are noise dominated. Therefore, for the cross-correlation of the different Sections we can estimate the measurement errors as:
\begin{equation}
    \sigma (\hat{P}^{AB}(k_i)) = P_{\rm noise}(k_i) / \sqrt{2 N(k_i)} \;\; ,
\end{equation}
with $N$ the number of independent measured modes in the $k$ bin, and the factor $1/\sqrt{2}$ accounts for the fact that the two maps are independent. There are various approaches for estimating $P_{\rm noise}$, such as using the power spectrum of each sub-dataset after the foreground removal as a proxy for the noise \citep[for more details, see, e.g.,][]{Wolz:2015lwa}. 

Despite the thermal noise bias mitigation, the \hi auto power spectrum result is an order of magnitude higher than what is expected from theory \citep{Switzer_2013}. This is because the data suffers from systematic effects and we have to resort to cross-correlations with optical galaxies to mitigate them and achieve a detection.  

The first detection of the cross-correlation between LSS and \hi intensity maps at $z \sim 1$ was reported in \citep{Chang:2010jp}, using data from the GBT and the DEEP2 galaxy survey.
A more significant detection using GBT intensity maps and overlapping WiggleZ galaxies was achieved in \cite{Masui:2012zc}, and again in \cite{Wolz:2021ofa} at the level of $\sim 5 \sigma$. The latter study also achieved $\sim 5 \sigma$ detections using the LRG and ELG samples from SDSS-eBOSS. In the right panel of Fig.~\ref{fig:IM-GBT-seasons} we show the measured GBT-WiggleZ cross-correlation power spectrum with $N_{\rm FG}=36$ used in FastICA for the foreground cleaning of the GBT \hi maps. We also show a null diagnostic test plotting the ratio of data and error. The error in the galaxy-\hi cross-correlation is estimated as:
\begin{equation}
    \sigma\left(\hat{P}_{\mathrm{g,HI}}\left(k_{i}\right)\right)=\sqrt{\frac{1}{2 \cdot N\left(k_{i}\right)}} \sqrt{\hat{P}_{\mathrm{g,HI}}\left(k_{i}\right)^{2}+\hat{P}_{\mathrm{g}}\left(k_{i}\right) \hat{P}^{A B}\left(k_{i}\right)} \;\; .
\end{equation}

An important component of the GBT data analysis is the use of the transfer function formalism to quantify and correct for the \hi signal loss due to foreground removal (see Fig.~\ref{fig:Pk-input-vs-cleaned}). We will give a brief description of how this works here, and we refer the interested reader to \cite{Switzer_2013, Wolz:2021ofa} for the details. Suppose we have a set of mock (simulated) \hi signal data, denoted by $m$, and our real (observed) data, $d$. Let us also denote cross-correlation of our real and mock data cubes by ``,''. If our real data was completely free of foreground contamination, we would simply have $P(d+m,m) = P(m)$.
This means that if we inject the mock into the data and then cross-correlate with the mock we would get back the power spectrum of the mock, i.e., $P(d,m)=0$ (the data corresponds to a different realization). But foreground effects will distort this picture and we wish to introduce a transfer function, $T(k)$, to compensate for the unavoidable signal loss. 
We will then have:
\begin{equation}
P(\mathrm{FG}(d+m),m) = P(m) \cdot T \;\; ,
\end{equation}
where $\mathrm{FG}(d+m)$ corresponds to foreground cleaning of the $(d+m)$ combined data cube, which takes into account the real data effects and systematics. 
The above formula defines the transfer function (in reality this is constructed in 2D, ($k_\parallel, k_\perp$)). 
The signal loss can reach very high levels ($\sim$50\%) depending on scale \citep{Wolz:2021ofa}, and therefore the transfer function correction is necessary in order to recover the true \hi power spectrum. Assuming the chosen fiducial cosmology (which is kept fixed in the GBT analyses) is correct, we can then proceed to perform a best-fit analysis for constraining \hi quantities. 

\subsubsubsection{\hi measurements and forecasts} 

In the post-reionization era, \hi intensity mapping provides an excellent probe of galaxy evolution.
We will first review the main findings of the GBT \citep{Masui:2012zc, Wolz:2021ofa} and Parkes \citep{Anderson:2017ert} cross-correlation analyses. The model for $P_{{\rm g},\hinospace}$ is given by:
\begin{equation}
    P_{{\rm g}, \hinospace}(k) = \bar{T}_\hinospace b_\hinospace b_{\rm g}r_{\hinospace,{\rm opt}} P_{\delta \delta}(k) \;\; ,
    \label{eq:model}
\end{equation} 
with $b_\hinospace$ the \hi bias, $b_{\rm g}$ the optical sample bias (WiggleZ, eBOSS ELGs, eBOSS LRGs), $r_{\hinospace,{\rm opt}}$ the galaxy-hydrogen correlation coefficient, and $P_{\delta \delta}(k)$ the nonlinear matter power spectrum including a linear RSD boost \citep[for more details and a discussion on the assumptions and limitations of this empirical model, see][]{Wolz:2021ofa}. The coefficient $r_{\hinospace,{\rm opt}}$ is dependent on the H\textsc{i} content of the galaxy sample.
The model is run through the same pipeline as the data to include weighting, beam, and window function effects. With the cosmology and optical bias values kept fixed, and using \autoref{eq:TbarModelEq}, we can fit the unknown pre-factor $\Omega_\hinospace b_\hinospace r_{\hinospace,{\rm opt}}$ to the data. 

Following this procedure, \cite{Masui:2012zc} measured the GBT maps cross-correlation with the WiggleZ 15hr and 1hr fields. Fitting data in the range of scales
$0.05 \, h{\rm Mpc}^{-1} < k < 0.8 \, h{\rm Mpc}^{-1}$, they found $10^3\Omega_\hinospace b_\hinospace r = 0.40 \pm 0.05$ for the combined,  $10^3\Omega_\hinospace b_\hinospace r = 0.46 \pm 0.08$ for the 15hr field and $10^3\Omega_\hinospace b_\hinospace r = 0.34 \pm 0.07$ for the 1hr field. 
For a more restrictive range of scales, their combined measurement was $10^3\Omega_\hinospace b_\hinospace r = 0.44 \pm 0.07$. 
The errors quoted are statistical, and \cite{Masui:2012zc} also estimated a $\pm 0.04$ systematic error.

\begin{figure}[t!]
\centering
\includegraphics[scale=0.55]{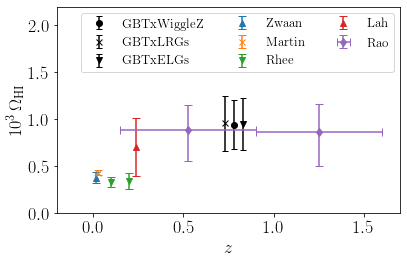}
\includegraphics[scale=0.55]{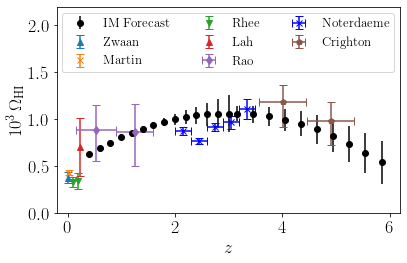}
\caption{Left: Estimates for $\Omega_\hinospace$ from \cite{Wolz:2021ofa} compared to other measurements in the literature (see \cite{Crighton:2015pza} and references therein). All estimates are at a central redshift $z=0.78$ but they have been staggered for clarity. Right: Forecasts for $\Omega_\hinospace$. This figure was produced using the publicly available code \texttt{IM-Fish} \citep{Pourtsidou:2016dzn} and SKA-like specifications \citep{SKARedBook2018}.}
\label{fig:IM-OmHI-constraints}
\end{figure}

With similar methodology and considering three different ranges of scales, \cite{Wolz:2021ofa} found  $\Omega_{\textrm{H\textsc{i}}} b_{\textrm{H\textsc{i}}} r_{\textrm{H\textsc{i}},{\rm Wig}} = [0.58 \pm 0.09 \, {\rm (stat) \pm 0.05 \, {\rm (sys)}}] \times 10^{-3}$ for GBT-WiggleZ, $\Omega_{\textrm{H\textsc{i}}} b_{\textrm{H\textsc{i}}} r_{\textrm{H\textsc{i}},{\rm ELG}} = [0.40 \pm 0.09 \, {\rm (stat) \pm 0.04 \, {\rm (sys)}}] \times 10^{-3}$ for GBT-ELG, and $\Omega_{\textrm{H\textsc{i}}} b_{\textrm{H\textsc{i}}} r_{\textrm{H\textsc{i}},{\rm LRG}} = [0.35 \pm 0.08 \, {\rm (stat) \pm 0.03 \, {\rm (sys)}}] \times 10^{-3}$ for GBT-LRG, at $z\simeq0.8$ and an effective scale $k_{\rm eff}=0.31 \, h/{\rm Mpc}$. Results were also reported at $k_{\rm eff}=0.24 \, h/{\rm Mpc}$ and $k_{\rm eff}=0.48 \, h/{\rm Mpc}$. The latter case corresponds to the same range of scales considered in \cite{Masui:2012zc}, who found $10^3\Omega_\hinospace b_\hinospace r_{\hinospace,{\rm Wig}} = 0.34 \pm 0.07$ for the same field. These results imply that red galaxies are more weakly correlated with \hi on the scales under consideration, suggesting that \hi is more associated with blue star-forming galaxies and tends to avoid red galaxies. 
This is in qualitative agreement with what was found in \cite{Anderson:2017ert}, at a lower redshift $z=0.08$ (it is also expected from galaxy evolution studies). \cite{Anderson:2017ert} cross-correlated Parkes \hi intensity maps with red and blue galaxies from the 2dF survey sample.
Making some further assumptions, \cite{Wolz:2021ofa} also derived constraints on $\Omega_{\rm HI}(z\simeq0.8)$, which are shown in the left panel of Fig.~\ref{fig:IM-OmHI-constraints}. With little information on \hi parameters beyond the local Universe, these are amongst the most precise $\Omega_{\rm HI}$ constraints in an under-explored redshift range. 

Forecasts using the proposed SKA-MID and SKA-LOW surveys are shown in the right panel of Fig.~\ref{fig:IM-OmHI-constraints}. These use the anisotropic \hi power spectrum (Eq.~\ref{eq:RSDPk}) to break the degeneracy between $\Omega_{\rm HI}$ and $b_{\rm HI}$ \citep{Wyithe:2008th, Masui:2010mp, Pourtsidou:2016dzn}. Similar measurements can be achieved with instruments like CHIME and HIRAX, the MIGHTEE survey \citep{ Paul:2020ank, Chen:2020uld}, and ASKAP \citep{Wolz:2017rlw}. 

\subsubsubsection{Cosmological forecasts}

The cosmological forecasts literature for \hi intensity mapping surveys is exhaustive. The main result is that assuming excellent calibration and mitigation of systematic and foreground contamination effects, \hi intensity mapping experiments can complement and compete with the largest and best Stage-IV optical galaxy surveys. Both single-dish and interferometric \hi intensity mapping surveys can probe dark energy, gravity, and the initial conditions of the Universe at a level comparable to optical surveys like Euclid \citep{Laureijs:2011gra,Blanchard:2019oqi} and VRO/LSST \citep{Mandelbaum:2018ouv}. Here, we summarize the main findings and caveats. Unless otherwise stated, we quote $1\sigma$ forecast marginal errors for the various  parameters, and give a few representative references for the interested reader. 

\begin{itemize}
    \item Large sky \hi intensity mapping surveys with radio telescopes like MeerKAT and SKA-MID (in single-dish mode), Tianlai, CHIME, HIRAX, PUMA, and SKA-LOW can use the \hi power spectrum to probe galaxy evolution and cosmology at a very wide redshift range ($0 < z < 6$). Using the CPL parameterization for the dark energy EoS (Eq.~\ref{eq:de_CPL}), the forecasts give $\sigma(w_0) \sim 0.05$, $\sigma(w_a) \sim 0.15$, with the fiducial values $(w_0,w_a)= (-1,0)$ for the $\Lambda$CDM model. Parameterizing the growth of structure as $f(z)=\Omega_m(z)^\gamma$ \citep{Lahav:1991, linder2003} the forecasts give $\sigma(\gamma) \sim 0.03$, with the fiducial value $\gamma = 0.55$ for GR. For more details, see e.g. \cite{Chang:2007xk, Masui:2010mp, Battye:2012tg, Hall:2012wd, Bull:2014rha, CosmicVisions21cm:2018rfq, Heneka:2018ins, Weltman:2018zrl, Liu:2019ygl}. The neutrino mass can also be constrained, $\sigma(M_\nu) \sim 0.3$ eV ($95\%$ CL) \citep{Villaescusa-Navarro:2015cca}. Most forecasts are very optimistic, assuming perfect instrument calibration and foreground removal. In addition, there are degeneracies between the \hi parameters and the cosmological parameters, for example \hi intensity mapping surveys without prior assumptions can constrain $\bar{T}_{\rm HI}f\sigma_8$ and not $f\sigma_8$ like optical galaxy surveys. Forecasts taking into account some of these caveats are presented in \cite{Padmanabhan:2018llf, Bernal:2019jdo, Camera:2019iwy, Soares:2020zaq}.
    
    \item The aforementioned surveys can probe ultra-large scales and constrain the primordial non-Gaussianity parameter, $f_{\rm NL}$, to a level $\sigma(f_{\rm NL}) \sim 1$ \citep{Camera:2013kpa, Alonso:2015uua, Karagiannis:2019jjx, Barreira:2021dpt}. However, foreground removal effects can lead to large degeneracies and biased estimates \citep{Cunnington:2020wdu}. These need to be controlled for \hi intensity mapping to reach its full potential. 
    
    \item Joint analyses of \hi intensity maps, optical galaxy surveys (galaxy clustering and cosmic shear), and CMB experiments, can be a powerful way to mitigate systematic effects and constrain \hi and cosmological parameters \citep{Wyithe:2008th, Masui:2010mp, Pourtsidou:2015mia, Villaescusa-Navarro:2015cca, Pourtsidou:2016dzn, SKARedBook2018, Viljoen:2020efi}. Using the multiple tracers method can suppress cosmic variance on large scales and provide the most precise measurements of primordial non-Gaussianity and general relativistic effects \citep[see, e.g.,][]{Alonso:2015sfa, Fonseca:2015laa, Fonseca:2016xvi,Witzemann:2018cdx}.  
    
    \item More futuristic prospects include \hi intensity mapping lensing \citep{Pourtsidou:2013hea, Jalilvand:2018ikk}, exploiting higher order statistics such as the bispectrum \citep{Karagiannis:2020dpq, Cunnington:2021czb}, and cross-correlations between gravitational wave detections and \hi intensity maps \citep{Scelfo:2021fqe}.
\end{itemize}

\clearpage

\subsection{Surface Brightness Fluctuations}

\cite{ts88} introduced the method of surface brightness fluctuations (SBF hereafter) as a way to obtain distances to stellar systems based on the discrete nature of star counts.
As refined in later papers \citep[e.g.,][]{tal90,jensen98,blake99a,cantiello05,mei05iv}, the SBF method uses the stochastic nature of star counts and luminosities to measure a quantity that is closely linked to the mean brightness of the red giant branch (RGB) star population in a galaxy or other stellar system. 

\subsubsection{Basic idea and equations}
\label{sbf.idea}

\begin{figure*}[p]
  \includegraphics[width=0.9\textwidth]{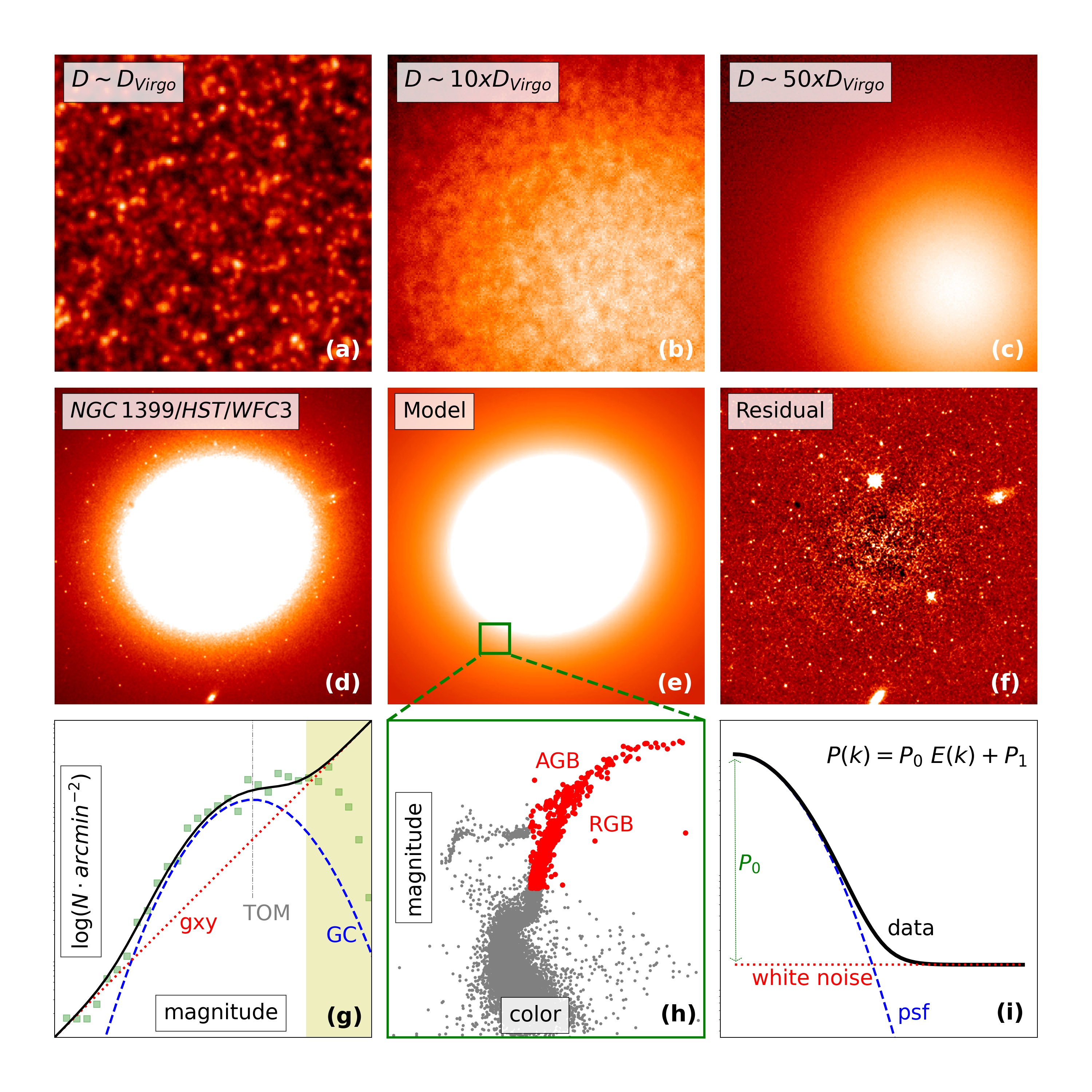}
  
  \caption{Illustration of SBF observations and measurements.  $(a)$~Simulation of the
    stellar population in a spheroidal galaxy at the distance of the Virgo cluster
    \citep[$D_{Virgo}\simeq16.5$ Mpc,][]{blake09} as observed with the E-ELT in $\sim$1 hour
    (Cantiello et al., 2021, in prep.). $(b)$~Same as in panel $(a)$, but for a galaxy ten
    times more distant. $(c)$~Same as in panel $(a)$, but for a galaxy fifty times more
    distant. Stars, which appear marginally resolved in panel $(a)$, blend together into a
    smooth brightness profile at larger distances. $(d)$~Near-infrared image of NGC\,1399
    from the HST WFC3 camera. $(e)$~Model of NGC\,1399's surface brightness distribution
    derived from the WFC3/IR image.  $(f)$~Residual frame, obtained from the galaxy image
    $(d)$ minus the model $(e)$.  $(g)$~Typical luminosity function analysis for estimating
    the ``residual variance'' $P_r$ due to contaminating sources: green squares show the
    data, the blue curve and red line show the fits to the globular cluster and background
    galaxy luminosity functions, respectively, and the solid black line is the combined
    model luminosity function \citep[data and fits are from][]{cantiello11}.
    The vertical gray dashed line indicates the GCLF turnover magnitude and the shaded area
    shows the magnitude interval where the detection is incomplete. $(h)$~Color-magnitude
    diagram of an old stellar population \citep[data for the MW globular cluster NGC\,1851
      from][]{piotto02}; the RGB/AGB population is highlighted with red dots.
    $(f)$~A schematic illustration of the SBF power spectrum analysis. Images reproduced with permission from \cite{blake09}, \cite{cantiello11}, and \cite{piotto02}, copyright by Astrophysical Journal and Astronomy \& Astrophysics}
  \label{sbf.fig1_sbf}
\end{figure*}

Qualitatively, the idea behind the method is simple, as illustrated in Fig.~\ref{sbf.fig1_sbf} (panels $a$-$c$). Stars that can be individually identified in nearby stellar systems gradually blend into a smooth brightness distribution as the distance increases, but the discrete nature of the stars can be discerned through statistical fluctuations in the integrated flux of the stars per resolution element.
These fluctuations are lower relative to the mean surface brightness (i.e., the galaxy appears smoother) at larger distances.

Observationally, {\it SBF is the ratio of the intrinsic variance (correcting for the blurring by the point spread function, PSF) of the stellar light distribution of a region of a galaxy to the mean surface brightness within the same region.}
In the nearby Universe, galaxy surface brightness is independent of distance, but the variance per unit solid angle decreases as distance squared. The ratio of the variance to the mean has units of flux and constitutes the SBF observable.
Although it may be harder to visualize than other standard candles, such as supernovae or Cepheids, it is as rigorously defined, and scales in the same way with distance. Physically, the SBF is related to the ratio of the first and second moments of the stellar luminosity function within the region analyzed.

For example, consider a galaxy that projects a stellar population of $n_i$ stars of flux $f_i$ (where $i=1, ..., N$ covers the entire flux interval, i.e. all evolutionary phases, of the stellar population) on a particular pixel $k$ in an image.
Along an isophote (the locus of pixels of equal surface brightness within the galaxy) there are many, say $M$, independent {\it realizations} of the population [$n_i$, $f_i$]. Each pixel can be considered a realization. Ignoring, for the moment, the PSF blurring, these realizations obey Poisson statistics, and the first two moments of the stellar intensity distribution can be written as:
\begin{itemize}
    \item $(N\times M)^{-1}\times \sum_{k=1}^{M} \sum_{i=1}^{N} (n_{k,i} \times f_{k,i}) $
    is the average surface brightness per realization (or pixel);
    \item $(N\times M)^{-1}\times \sum_{k=1}^{M} \sum_{i=1}^{N} (n_{k,i} \times f_{k,i}^2)$
      is the mean-squared flux of the realizations.
\end{itemize}

\noindent
The index $k$ runs over the pixel realizations, and $i$ runs over the stellar luminosity function bins. If we assume that the same form of the underlying luminosity function applies to all the pixels in the region being analyzed, then the mean flux per stellar bin is independent of the pixel: $f_{k,i} = f_i$.  
The mean SBF flux, which is defined by \citet{ts88} as the ratio of the second to the first moment of the flux along the isophote, then becomes:
\begin{equation}
    \bar{f} = \frac{\,\sum_{i} \sum_{k} n_{k,i} \times f_{i} ^ 2} 
               {\,\sum_{i} \sum_{k} n_{k,i} \times f_{i}} \, = \,
              \frac{\,\sum_{i} n_{i} \,f_{i} ^ 2} 
               {\,\sum_{i} n_{i} \,f_{i}\,} \;\; .            
    \label{sbf.def}
\end{equation}
\noindent
Thus, \fbar\ is the flux-weighted mean stellar flux in the region of the isophote being analyzed.  The corresponding luminosity is \Lbar, which is equal to the ratio of the first two moments of the stellar luminosity function and can be readily calculated from stellar population models.
Because of the squared weighting in the numerator, the SBF signal is dominated by the brightest stars in the population.

In practice, the SBF measurement is done over finite regions of the galaxy. It is not necessary for the surface brightness to be constant, but the stellar luminosity function should not vary significantly over the region.  One deals with the varying surface brightness by subtracting a smooth model for the light distribution and measuring the fluctuations in the residual image. In this case, the numerator of \fbar\ becomes the variance with respect to the mean surface brightness. For a fully rigorous statistical treatment of SBF statistics, see \citet{cervino08}.

The SBF apparent magnitude is defined as $\mbar=-2.5\,\log(\bar{f})+ m_\textrm{ZP}$, where $m_\textrm{ZP}$ is the magnitude zero-point of the system. Although \mbar\ can be measured for any galaxy, this does not mean that a useful distance can be derived. One also needs a reliable calibration of \Mbar, the absolute magnitude that gives the correct distance modulus $(m{-}M)$ for a galaxy with the measured \mbar.
\Mbar\ depends only on the photometric bandpass and the stellar population in the galaxy. Thus, unlike other galaxy-based distance indicators, SBF does not depend on the mass, effective radius, dynamics, or environment of the galaxy, although these properties may influence the stellar population.

The measured SBF magnitude \mbar\ and a proper calibration of \Mbar\ are presently used to determine accurate distances moduli to galaxies within $D\sim150$ Mpc, enabling robust constraints on the Hubble parameter in the Local Volume via the Hubble-Lema\^{i}tre law: 
\begin{equation}
H_0=\langle v/D \rangle \;\; ,
\end{equation}
where $v$ is the  flow-corrected recessional velocity of the target galaxies.

Of course, not all stellar populations are created equal. In particular, galaxies that have undergone recent star formation have poorly calibrated values of \Mbar, which causes systematic uncertainty in their SBF distance. For this and other reasons, elliptical galaxies are the preferred targets for SBF studies. Until recently, challenges with data depth and quality have prevented precise distance measurements for significant samples of galaxies reaching into the Hubble flow, but datasets and observing strategies have improved \citep{cantiello18b, blake21, jensen21}, making SBF a powerful cosmological probe with a bright future.

\subsubsection{Sample Selection}
\label{sbf.sample}

Measuring SBF magnitudes requires careful modeling and subtraction of the galaxy light distribution. As described later (Sect.~\ref{sbf.measure}), any small-scale residual features with structure on the scale of the PSF will complicate the analysis.  Such features may be associated with dust, bars, shells, or other irregularities. In severe cases, the SBF signal may be entirely overwhelmed.  As a result, the smoothest, most featureless galaxies are the prime targets for the SBF method; that is to say, \mbar\ is easiest to measure in giant ellipticals and other early-type galaxies with substantial bulge components.

Of course, the stellar fluctuations must be sufficiently bright in order to detect the SBF signal; for distances in the Hubble flow, this requires the reddest optical bands or observing in the near-IR. The contamination from dust is also reduced at these wavelengths. However, beyond $\sim2$ \micron, the uncertainties in the calibration become too large for precise distances.  Below we discuss these issues of sample and bandpass selection in more detail.

\noindent 
{\bf Choosing the galaxies.} In addition to simplifying the galaxy subtraction for the \mbar\ measurement, early-type galaxies tend to be dominated by old stellar populations (Fig.~\ref{sbf.fig1_sbf}, panels $e$ and $h$), which simplifies the \Mbar\ calibration. This can be seen from empirical plots of the \Mbar\ versus color relations \citep[e.g.,][]{blake09,blake10,jensen15,cantiello18b,carlsten19} for galaxies at a common distance.
The empirical relations show that for red early-type galaxies, the correlations between the absolute SBF magnitude in red or near-IR bands and broad-baseline optical color have lower scatter than for fainter, bluer galaxies. In selected passbands (see below), the small intrinsic scatter in the \Mbar-color calibration relations for red galaxies in principle allows distance precision as low as $\sim3$\%. In practice, the errors are larger because of measurement uncertainties, as discussed in Sect.~\ref{sbf.measure}.

Consistent with observations, stellar population models predict less scatter in \Mbar\ at a given color for metallicities similar to those found in massive galaxies \citep[e.g.,][]{bva01, mei05v}.  At the blue end, galaxies may have low metallicities, younger ages, or a combination of both.  At these blue colors, the SBF is more affected by age than metallicity; thus two galaxies with similar optical colors may have significantly different SBF magnitudes if they have had different star formation histories, as discussed by \citet{greco21}. As a result, the observed scatter in the SBF calibration can be quite large at the blue end, and it becomes difficult to measure individual distances with a precision better than $\sim10\%$ due to the calibration effects alone.

Thus, for both measurement and calibration reasons, the ideal target galaxies for the SBF method are red early-type galaxies with no recent star formation and little or no dust. In spite of this, SBF measurements cover practically the entire mass range of galaxies \citep[e.g.,][]{blake09,jensen21,vdk18,carlsten19} and a wide range of morphologies, including the bulges of spirals \citep{tonry00} and ultra-diffuse galaxies \citep{blake18}. As long as there is a clean area of the galaxy without recent star formation, it is possible to derive an SBF distance. 

Another consideration in defining an observational sample is that the SBF must be detected to high signal-to-noise (S/N), including the effect of correcting for contaminating sources.
This puts a practical limit on the distance to which the SBF measurements can be made. Of course, for cosmologically interesting measurements, the galaxies must be distant enough to be in the Hubble flow (i.e., $d\gtrsim50$ Mpc). The depth requirement and the related distance limit depend sensitively on the bandpass, which we discuss next.

\noindent
{\bf Choosing the bandpass.}
The most common photometric bands used for SBF distance measurements in recent years have been $i,I,z,J,H$, and $K$, spanning the wavelength range from $\sim0.8$ to $\sim2.2$ \micron\ \citep{tonry01,jensen03,jensen21,mei07,blake09,blake10,cantiello07,cantiello18b,biscardi08}. At shorter wavelengths, the SBF signal is much fainter, and the slope of the \Mbar\ relation with color tends to be steeper because of increased sensitivity to stellar population effects \citep[e.g.][]{worthey93,bva01,cantiello03}. Of course, dust is also a bigger problem in bluer bands.

The intrinsic scatter of \Mbar\ as a function of color is as low as $\sim0.05$ mag for red galaxies in passbands near 1\,\micron\ \citep{blake09,blake21}. The intrinsic dispersion is less well constrained at longer wavelengths, but appears to be closer to 0.1 mag in the $H$ and $K$ bands \citep{jensen03,jensen15}, likely because of the increased stochastic effect of small numbers of luminous red asymptotic giant branch stars (AGB), the properties of which depend sensitively on population age \citep{raimondo05,raimondo09}.

Another issue that is worse in bluer bands (like $B$ or $V$) is the contamination of the SBF signal by globular clusters host in the galaxy, point-like sources that produce extra variance in the image. In bands where the SBF is fainter, the globular clusters must be identified and removed to fainter magnitudes. Even in the $I$ band, for elliptical galaxies with typical globular cluster frequencies, sources must be detected and masked to $\lesssim0.3$ mag of the peak, or ``turnover,'' of the globular cluster luminosity function (GCLF), in order to decrease the contamination to the $\sim20$\% level \citep{blake95}, which reduces the uncertainty in the correction to $\sim5$\% \citep{tal90}.  

In contrast, with the much brighter fluctuations in the $K$ band, it is only necessary to reach within $\sim2$ mag of the GCLF turnover to reduce the contamination to the same level \citep{jensen98}. Thus, the stellar population scatter in \Mbar\ is a much bigger issue than globular cluster contamination for SBF measurements near 2\,\micron.

Currently, the most efficient instrumental system available for SBF distances is the F110W (broad~$J$) filter of the WFC3/IR on the Hubble Space Telescope (HST). Using this setup, it is possible to measure distances for early-type galaxies out to 80 Mpc with a median statistical uncertainty of 4\% \citep{blake21,jensen21} in only one HST orbit. One of the reasons this system works well is that the near-IR sky background is much fainter from space. For future ground-based surveys, such as the one provided by the Vera Rubin Observatory, we expect that the $y$ band, which covers the red end of the optical spectrum similar to F110W but extends less far into the near-IR, may prove the best choice for SBF measurements.  At present, however, we lack data in this band for testing.  The following section describes SBF measurements for the preferred targets of early-type galaxies in bands suitable for accurate distance determination.

\subsubsection{Measurements}

SBF distance determination consists of two parts: measuring the fully corrected apparent SBF magnitude \mbar\, and determining the best value for the absolute \Mbar\ from a calibration based on distance-independent stellar population properties, typically broadband color.

\subsubsubsection{Measuring SBF magnitudes}
\label{sbf.measure}

In the absence of any atmospheric and instrumental blurring and external sources of fluctuation, the SBF signal of a stellar system would simply be the statistical variance due to the varying numbers and luminosities of the stars in each pixel, normalized by the local mean flux. In reality, PSF blurring creates a correlation between adjacent pixels; therefore, the SBF signal is measured in Fourier space by determining the amplitude of the component on the scale of the PSF in the image power spectrum.  If the large-scale light distribution of the galaxy is well-subtracted, then the power spectrum will consist mainly of a white noise component and a component convolved with the PSF. There may be additional power at lower wavenumbers (larger scales) due to imperfect galaxy subtraction, or at higher wavenumbers (smaller scales) due to correction of geometric distortion of the image \citep{mei05iv,cantiello05}, but these wavenumbers can be omitted from the analysis.

The detailed process of measuring SBF magnitudes is described in numerous papers with some variations based on the bandpass and other properties of the data \citep{blake99a,blake09,cantiello05,cantiello07,jensen98,jensen21,mei05iv,mei05v}. See these papers for details on putting the method into practice. Here, we highlight the main steps of the procedure in order of execution, and the products of each step.

\begin{enumerate}[label=\roman*)]
\item {\it Galaxy model} (Fig.~\ref{sbf.fig1_sbf}, panel $(e)$): a smooth isophotal model of the galaxy surface brightness after sky subtraction; the resulting {\it model frame} corresponds to the first moment of the light distribution.

\item {\it Residual frame} (Fig.~\ref{sbf.fig1_sbf}, panel $(f)$): difference image obtained by subtracting the galaxy surface brightness model (and a low-order fit to the background) from the original sky-subtracted image.

\item {\it Mask frame}: mask made by identifying all sources of non-SBF variance (dust, globular clusters, foreground stars, background galaxies, bright satellite galaxies, tidal features, bars, etc.) down to a specified S/N threshold and masking them out.

\item {\it Fluctuation frame}: the masked residual frame normalized by the square root of the model frame, used to measure the normalized stellar fluctuations; also contains contaminating fluctuations from unexcised sources fainter than the detection limit, plus white noise resulting from photon counting statistics and detector read noise.

\item {\it Power spectrum frame}: 2-D Fourier power spectrum of the fluctuation frame, used to derive the SBF amplitude after azimuthal averaging (Fig.~\ref{sbf.fig1_sbf}, panel $(i)$). Because the stellar fluctuations are convolved with the PSF of the image, in the Fourier domain they are multiplied to the Fourier transform of the PSF (convolved with the window function of the mask, see below).\\
Once an accurate PSF template is created from stars in the field and normalized, the fluctuation amplitude is obtained as the constant $P_0$ in Eq.~\ref{pk.eq} below. This is obtained by fitting the azimuthally averaged power spectrum of the fluctuation frame, $P(k)$, with the expectation power spectrum $E(k)$. Here $E(k)$ is a convolution of the PSF power spectrum and the mask function with which fluctuation frame was multiplied. In addition, the power spectrum includes a constant white-noise component; thus, the full power spectrum is modeled as:
\begin{equation}
  P(k)=P_0{\,\times\,}E(k)+P_1 \;\; .
  \label{pk.eq}
\end{equation}

\item {\it Correction for background fluctuations} (Fig.~\ref{sbf.fig1_sbf}, panel $(g)$): globular clusters and background galaxies that are too faint for direct detection will remain in the image after masking, and their flux will contribute to the $P_0$ component of the power spectrum. To correct for this contamination, we calculate the ``residual power'' $P_r$ from contaminating sources by extrapolating a fit to the combined GCLF and background galaxy luminosity function. The ability to detect and remove the globular clusters is often the limiting factor in how far SBF distances can be measured. Contamination due to background galaxies is normally much less for giant ellipticals.

\item {\it SBF magnitude}: Using the measured fluctuation amplitude $P_0$ and the estimated contribution from contaminating sources $P_r$, the stellar fluctuation signal is $P_f=P_0-P_r$. This corresponds to \fbar\ in Eq. \ref{sbf.def}. Thus, converting to the SBF magnitude: $\mbar=-2.5\log(P_f)+m_\mathrm{ZP}$, where $m_\mathrm{ZP}$ is the appropriate photometric zero-point magnitude.

\end{enumerate}

\subsubsubsection{SBF Calibration}

To obtain a distance from the measured \mbar, one must adopt an absolute SBF magnitude \Mbar\ for the stellar population. This can be done using either an empirical calibration or theoretical predictions from stellar population synthesis models. With some exceptions \citep[e.g.,][]{biscardi08}, the vast majority of published SBF distances rely on empirical calibrations \citep{tonry01, blake01b,blake09, cantiello18b, jensen03,jensen21}.

The ground-based SBF survey by Tonry and collaborators \citep{tonry97,tonry01,blake99} measured $I$-band SBF magnitudes \mIbar\ and \VI\ colors for 300 galaxies out to about 40 Mpc and derived the first high-quality empirical SBF calibration. To do this, \citet{tonry97} plotted \mIbar\ as a function of \VI\ for nearby groups and clusters, determining a single linear slope for the color dependence of \mIbar.
The zero-point of the calibration was then determined from SBF measurements in the bulges of six spiral galaxies that also had distances measured from Cepheids \citep{tonry00}. This was revised slightly by \citet{blake02} using a recalibrated set of Cepheid distances from \citet{freedman01}. The resulting linear calibration fully specified \MIbar\ as a function of \VI. The intrinsic scatter about this relation was estimated to be of order 0.05 mag, although it was fairly uncertain because the median statistical error on the distances was roughly four times larger.

The same basic approach, with some variations, has been used to derive empirical \mbar-color calibrations for the SBF method in $V$ \citep{bva01}, $K$ \citep{jensen03}, ACS/F850LP \citep{mei07,blake09}, WFC3/F110W \citep{jensen15}, and $i$ \citep{cantiello18b}. Higher-order polynomials were used for the ACS and WFC3 calibrations, while \citet{cantiello18b} presented calibrations that combined two color indices, rather than just one as in  previous cases.

In general, the empirical approach works well and the resulting calibrations agree with theoretical predictions within the uncertainties. The weak point remains the distance zero-point, which is tied to the Cepheid distance scale via measurements in spiral galaxies, which are not ideal targets for the SBF method. However, alternative zero-point calibrations based on the tip of the red giant branch (TRGB) have also been presented \citep{mould09,blake21}, and these agree well with the Cepheid-based calibration. Fully theoretical calibrations of \Mbar\ versus color do not rely on other distance indicators, and thus do not carry systematic uncertainties from Cepheids or other primary distance indicators. However, the often {poor agreement} among different sets of models shows that theoretical calibrations still carry substantial systematic uncertainties, especially in the near-IR bands \citep[e.g.,][]{jensen15}, which are observationally most promising for future SBF studies.

As a final remark, we note that since the empirical \Mbar\ calibrations are parameterized by photometric color, precise measurements of the galaxy colors are required for high-quality distance estimates.  Thus, great care must be dedicated to observational details such as photometric calibration, flat-fielding, sky subtraction, etc.

\subsubsubsection{Statistical uncertainties}

Before moving on to systematic effects, we summarize the statistical uncertainties in SBF distance measurements.  These can be grouped into three categories: {\it i)} random errors in the photometric calibration, {\it ii)} errors in the measurement of the fluctuations themselves, including the corrections for background contamination, and {\it iii)} random uncertainty in \Mbar\ resulting from stellar population effects and errors in the galaxy color estimate.

The first category includes effects such as flat-fielding, background estimation, uncertainty in the galactic extinction, and uncertainties in the photometric zero-point, that are not specific to the SBF method. These effects are typically at the 1\% level, but care must be taken to account for them in a consistent way, as they may contribute to various parts of the SBF measurement process. For example, extinction uncertainty affects both \mbar\ and the color estimate used for determining \Mbar.

Factors contributing to statistical uncertainties in the SBF measurement include: the accuracy of the galaxy surface brightness model, the fit to the image power spectrum to determine $P_0$, the extrapolation of the luminosity function fit to estimate the $P_r$ term from contaminating sources (the error is typically 20-25\% of the $P_r$ correction itself), and the match of the PSF template to the data being analyzed. These errors can be minimized by optimizing the observing (including instrument, exposure time, and bandpass) and data analysis strategies. As shown in several works \citep[e.g.,][]{blake09,blake10,cantiello18b,jensen21}, the total statistical uncertainty on \mbar\ can be kept as low as 0.04 to 0.05 mag.

Concerning random errors in \Mbar, if the images have high S/N and are in the same bandpasses used for the SBF calibration so that no photometric transformation is needed, then the error in \Mbar\ due to the color uncertainty can be kept to the $\sim0.01$ mag level. In this case, the random error in \Mbar\ is dominated by intrinsic scatter in the calibration due to stellar population effects.  In the $I,z,$ and $J$ bands, this scatter is estimated to be 0.05 to 0.06 mag \citep[e.g.,][]{tonry97,blake09,cantiello18}. 

In summary, measurement uncertainties in well-designed SBF observations can be reduced to the $\sim\,$0.05 mag level. If these observations are of red galaxies in a well-suited bandpass near 1\,\micron, the intrinsic scatter in the calibration relation will be at a similar level. Combining these two sources of error gives a total statistical error as low as $\sim0.07$ mag, or about 3.3\% in distance per galaxy, although $\sim\,$4\% is more typical for the median statistical error in well-designed SBF distance samples \citep[e.g.,][]{blake21,jensen21}. 

\subsubsection{Systematic effects}
\label{sbf.systematic}

The dominant systematic uncertainty affecting all SBF distances is the zero-point of the absolute \Mbar\ calibration. This zero-point was determined by comparing ground-based $I$-band SBF magnitudes for the bulges of spiral galaxies with measured Cepheid distances \citep{tonry97,tonry00,ajhar01,blake02}. In most cases, the SBF zero-points in other bands have been set by tying the measurements to the $I$-band SBF distances \citep[e.g.,][]{mei07,blake09,jensen15,cantiello18b}.

The most recent analysis of the systematic uncertainty in SBF distances was by \citet{blake21}, who revised the zero~point to account for the improved LMC distance determined to $\sim\,$1\% precision by \citet{pietrzynski19}. They concluded that the zero-point uncertainty in the Cepheid-calibrated SBF distances in the WFC3/F110W band (the most useful for constraining \Ho) is 0.09 mag, or 4.2\% in distance. This is larger than the typical HST SBF measurement error.

Since the SBF method works best for early-type galaxies with old stellar populations, and these do not contain the young Cepheid stars (see Sect.~\ref{sbf.sample} above), it is worth exploring other means for calibrating SBF. The TRGB method is ideal for measuring distances of early-type galaxies and obtaining an independent calibration of SBF.  Like Cepheids, it is possible to calibrate the TRGB method with geometric distances from Gaia \citep[e.g.,][]{soltis21}, but unlike with Cepheids, the stellar population underlying both SBF and the TRGB is the same, \ie, old low-mass stars.

A first attempt to calibrate SBF with TRGB \citep{mould09} used a sample of 16 galaxies within 10 Mpc dominated by relatively blue dwarf galaxies and found negligible change with respect to the Cepheid calibration, but the distance uncertainties were large and the colors did not extend to the range occupied by massive red ellipticals, the preferred SBF targets at large distances.
More recently, \citet{blake21} rederived the SBF zero-point using the few TRGB distances available for massive early-type galaxies. They concluded that the mean offset between the Cepheid and TRGB calibrations of SBF was $0.01\pm0.10$ mag. Because these two calibrations were independent and consistent, they could be combined to improve the precision on the SBF zero-point; this reduces the systematic error in the SBF distances to just over 3\%.

Another potential systematic effect comes from SBF $k$-corrections for galaxies at larger distances. These must be estimated from stellar population models. Based on the model calculations by \citet{liu00}, \citet{jensen21} estimated the SBF $k$-corrections in F110W to be less than 0.01~mag at 100 Mpc, the limit of their sample. Thus, $k$-corrections are not currently a significant problem, but they could become more important for future studies. We come back to this issue in Sect.~\ref{sbf.forecasts}.

In conclusion, the systematic uncertainty on SBF distances is slightly larger than 4\% when based solely on Cepheids. Combining the best current Cepheid and TRGB calibrations for SBF, the systematic error in distance drops to about 3\%. Ultimately, the TRGB method should provide much better precision because it can be used in the same type of galaxies, giant ellipticals, which are best for SBF measurements, while Cepheids only occur in galaxies that are inherently problematic for the SBF method. \citet{blake21} estimated that with a sample of $\sim\,$15 giant ellipticals having both high-quality SBF and TRGB distances, it would be possible to reduce the systematic uncertainty in the SBF zero-point to the 2\% level, including the uncertainty in the TRGB absolute magnitude calibration, which should soon approach the 1\% level, thanks to Gaia. Such an overlapping sample of SBF and TRGB distances to giant ellipticals becomes feasible
with the advent of JWST.

\subsubsection{Main results and forecasts}

\subsubsubsection{Main results}
\label{sbf.mainresults}

The SBF method has been in use for several decades.  About 600 independent SBF distances (for $\sim\,$400 distinct galaxies) have been measured from the Local group to $\sim\,$130~Mpc. Samples with at least 20 galaxies include: \cite{tonry01,jensen03,jensen21,mieske05,mieske06,blake09,cantiello18b,cohen18,carlsten19}.
Soon, there will be another $\gtrsim200$ from the Next Generation Virgo Survey (Cantiello et al., 2022, in prep.). Although the method is capable of high precision, the quality of published SBF distances is quite heterogeneous, with errors typically 4-5\% for HST measurements, 10\% for ground-based data on giant ellipticals, and up to 30\% for some dwarfs galaxies.

SBF distances have been used to map the velocity field of the local Universe \citep{tonry00} and constrain the cosmic mass density \citep{blake99}. Although the ground-based samples used for these studies were variable in quality and only extended to 40~Mpc, the results agree well with modern analyses. More recently, SBF measurements have been used to probe the structure of nearby clusters \citep{mei07,blake09,cantiello18b}, estimate supermassive black hole masses \citep{ehtVI,nguyen20,liepold20}, investigate satellite galaxy systems \citep{cohen18,carlsten19}, confirm the lack of dark matter in some ultra-diffuse galaxies \citep{vdk18,blake18}, and measure the most precise distance to the host galaxy of the binary neutron star merger event GW170817 \citep{cantiello18}. 
SBF has also been used for various determinations of the Hubble constant, \Ho \citep{tonry00,blake99,blake02,jensen01,biscardi08}.
Here we focus on two recent \Ho~studies.

\citet{khetan21}, presented a recalibration of the peak magnitudes of 24 local SNe Ia using a heterogeneous sample of ground and space-based SBF distances from the literature. Adopting a hierarchical Bayesian approach, the authors then extended the calibration to a sample of 96 SNe\,Ia at redshifts $0.02< z< 0.08$ and derived  $H_0 = 70.5\pm2.4\,(\mathit{stat}) \pm3.4\,(\mathit{sys})$ \Hunit.
Note that in this case, SBF is used as an intermediate rung in the distance ladder, between Cepheids and SNe\,Ia, rather than constraining \Ho~directly. When updated for consistency with the improved LMC distance from \citet{pietrzynski19}, the result becomes  $H_0 = 71.2\pm2.4\pm3.4$ \Hunit.

\begin{figure*}[t!]
  \includegraphics[width=0.95\textwidth]{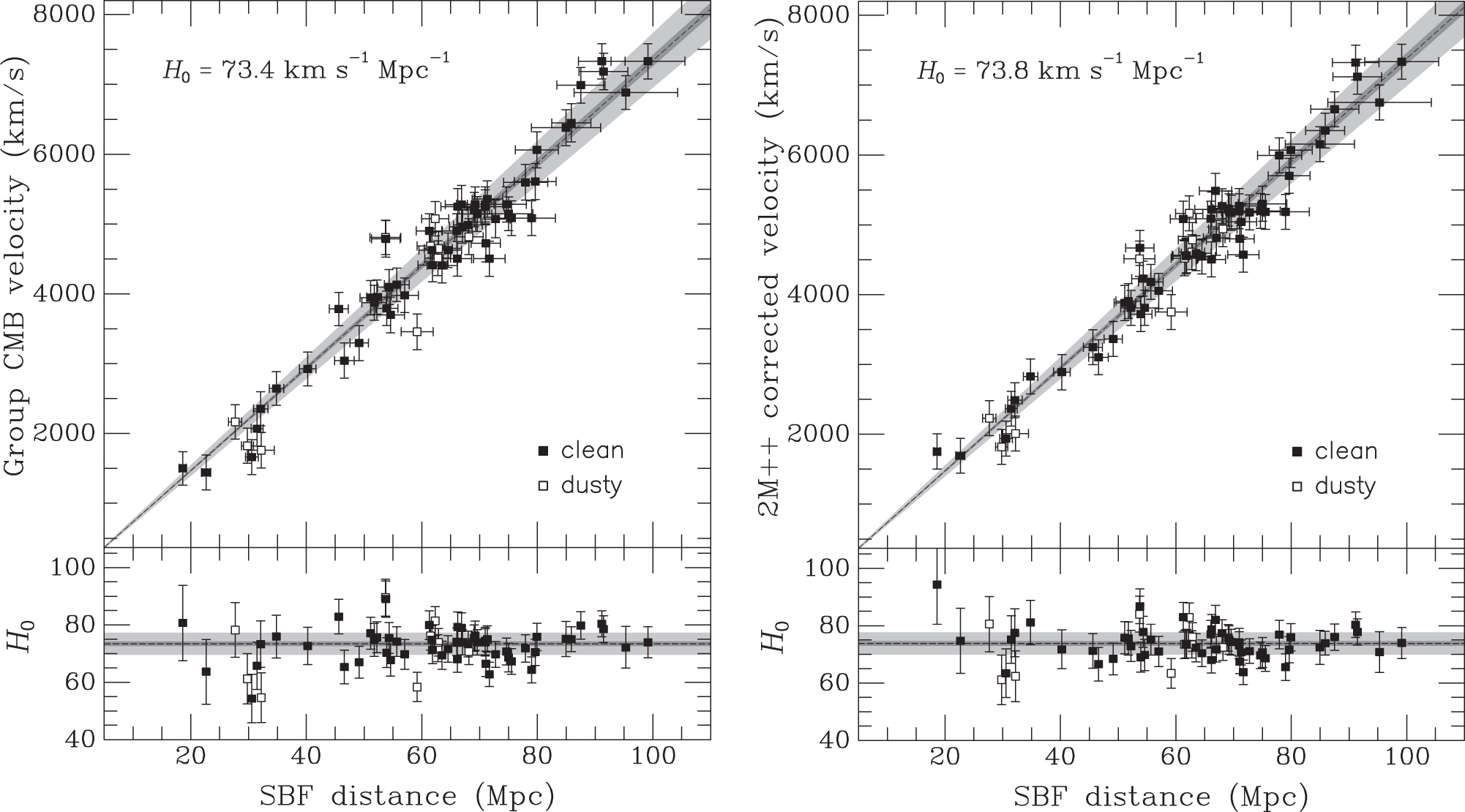}
  \caption{Hubble diagrams and residuals from \citet{blake21} based on Cepheid-calibrated
    WFC3/IR SBF distances tabulated by  \citet{jensen21}.
  The velocities are group-averaged values in the cosmic microwave background rest frame
  without correction for peculiar motions (left) and corrected using the 2M++ flow model
  derived from the redshift-space density field analysis of \citet{carrick15}.
  Solid symbols indicate ``clean''  galaxies, for which no dust or 
  spiral structure is evident; open symbols are for galaxies with obvious dust and/or spiral
  structure. The best-fit Hubble constants are indicated, and the statistical and
  systematic error ranges are shown in dark and light gray, respectively.
  The reduced $\chi^2$ improves from 0.97 for the fit on the left to 0.89 for the fit
  using the flow-model, but the value of \Ho~depends on the overall velocity scale factor,
  and the study adopted the model-independent version for this.
  \Ho~would increase by 0.3\% for the TRGB-based SBF calibration. Image reproduced with permission from \cite{blake21}, copyright by AstroPhysical Journal.%
    }
  \label{fig:sbfH0}
\end{figure*}

\citet[][]{blake21}, using the homogeneous sample of 63 SBF distances measured by \citet{jensen21} for bright, mainly early-type, galaxies out to 100 Mpc observed with the F110W filter of HST's WFC3/IR, derived $H_0 = 73.3 \pm0.7\pm2.4$ \ksm. The systematic (second) error mainly represents the SBF zero-point uncertainty after combining the Cepheid and TRGB calibrations. Because peculiar velocities can have an important impact over this distance range, \citet{blake21} tested four different treatments of the galaxy velocities, including two large-scale flow models, and included this effect in the systematic error estimate. Figure~\ref{fig:sbfH0} shows example Hubble diagrams from the study. The observed scatter in the Hubble diagram is consistent with the combined uncertainties from the SBF distances and the corrected recessional velocities.

The \Ho~result by \citet{blake21} agrees well with most other local measurements and with \citet{khetan21} to within $1\sigma$.  It disagrees by more than $2\sigma$ with the value of \Ho~based on the cosmic microwave background, assuming the standard $\Lambda$CDM model \citep{planck20}, reinforcing the tension. More WFC3/IR SBF distances are being obtained by ongoing HST programs; these will improve the constraints on the velocity model and further reduce the uncertainties on~\Ho.

\subsubsubsection{Forecasts}
\label{sbf.forecasts}

The outlook for SBF is bright for several reasons: the next generation of wide-field survey telescopes will produce imaging data suitable for SBF measurements; JWST and the AO-assisted ELT facilities will allow the method to be pushed to unprecedented distances; and new samples of TRGB distances, tied to Gaia parallaxes, will drastically reduce the systematic uncertainty in the SBF zero-point calibration.
Sec~\ref{sbf.systematic} already discussed the expected zero-point improvement from the TRGB calibration. Here we comment on the other two anticipated opportunities for SBF studies.

\noindent
{\bf Wide-field surveys.}
Forthcoming large sky surveys, such as the Vera Rubin Observatory \citep{LSSTScience:2009jmu} and the Euclid Wide Survey \citep{Laureijs:2011gra}, will produce breakthroughs in many fields of astronomy, including excellent opportunities to use SBF to map the spatial distribution of galaxies in the low-redshift Universe. The detailed simulations by \citet{greco21} indicate that Rubin will produce an unprecedented dataset for SBF studies. The multi-band $ugrizy$ Rubin dataset, with typical seeing of 0.7\arcsec, and final $5\sigma$ point source depth of $i_{5\sigma}\sim26.8$ mag, will make it possible to measure SBF distances with 10\% accuracy out to at least 70 Mpc, twice as far as the limit of the ground-based SBF survey of \citet{tonry01}.  

The Euclid satellite\footnote{https://sci.esa.int/web/euclid} has one-fourth the collecting area of HST but, compared to Rubin, it has the advantage of near-IR coverage and a sharp ($\sim$0.2\arcsec), stable PSF. Taking as reference the Euclid/NISP H band, with a predicted $5\sigma$ point source depth of $H_{5\sigma}\sim24$ mag,  Euclid should enable SBF distances for all suitable galaxies out to at least half the distance as Rubin ($\sim30-40$ Mpc), and possibly more. Another future wide-field mission of enormous interest for SBF is the Nancy Grace Roman Space Telescope\footnote{https://roman.gsfc.nasa.gov/} \citep{Spergel:2015}. It will have the same aperture as HST and similar resolution, but $\sim100$ times the field of view and better IR sensitivity. With a $5\sigma$ point-source depth of 28 mag in 1 hr in the $J$ and $H$ bands, Roman will deliver phenomenal survey depth and breadth, making it the ultimate machine for producing SBF distances. More detailed simulations are needed to
quantitatively refine the expectations for both Euclid and Roman.

\noindent
{\bf Going deeper.}
JWST \citep{Gardner:2006} and the forthcoming 30--40m class of Extremely Large Telescopes will have near-IR imaging capabilities far exceeding that of HST.  As discussed above, JWST should greatly improve the SBF zero-point calibration by enabling much more extensive direct comparisons of SBF and Gaia-calibrated TRGB distances in giant ellipticals. This will significantly reduce the systematic uncertainty in \Ho, making SBF competitive with SNe Ia in this area.

In addition, these new facilities will make it feasible to go far beyond the previous limit of 100-150 Mpc achieved with HST \citep{jensen01,jensen21,biscardi08}.  With its sharper (FWHM$\sim$0.1\arcsec), better sampled PSF in the near-IR and $\sim7$ times the collecting area of HST, JWST should enable SBF distance measurements out to $\sim\,$300 Mpc. As always, the limiting factors will be contamination from globular clusters, and a newly significant consideration will be the quality of the $k$-corrections derived from stellar population models (Sect.~\ref{sbf.systematic}). Further work is needed on this issue.

The ELTs hold even greater promise for pushing SBF to unprecedented depths, potentially out to $z\sim\,$0.1, and perhaps even directly probing cosmic acceleration and dark energy as a complement to SNe Ia. However, this depends critically on the ability to measure precise and reliable SBF magnitudes using adaptive optics (AO). Although some studies have been
made of this topic \citep{sbf4elt,jensen12}, quantitative demonstrations of AO-assisted SBF measurements are lacking.  Further work, using actual AO data, is much needed, and again $k$-corrections will be an important ingredient in deriving accurate calibrations. We appeal to the stellar population modelers of the world to dedicate some effort to this important problem. 

\clearpage

\subsection{Stellar Ages}
\label{SA}

The expansion rate of the Universe determines the look-back time. This opens up the possibility to use time (or age) measurements to constrain the background parameters of the cosmological model. The cosmic chronometers method (see Sect.~\ref{sec:CC}) uses relative ages to determine $H(z)$, but absolute ages can also be used in a complementary way. In fact, historically, absolute ages were  used  already in the 50's and more extensively in the 90's (see below) to impose competitive (then) constraint on the cosmological model. While the age of any old astrophysical object could in principle serve the purpose to constrain the age of the Universe, historically stellar ages have been a promising avenue as they can be determined with precision and accuracy to date much superior than for any other type objects.

\subsubsection{Basic idea and equations}

\begin{figure*}[t]
 \begin{centering}
\includegraphics[width=0.32\textwidth]{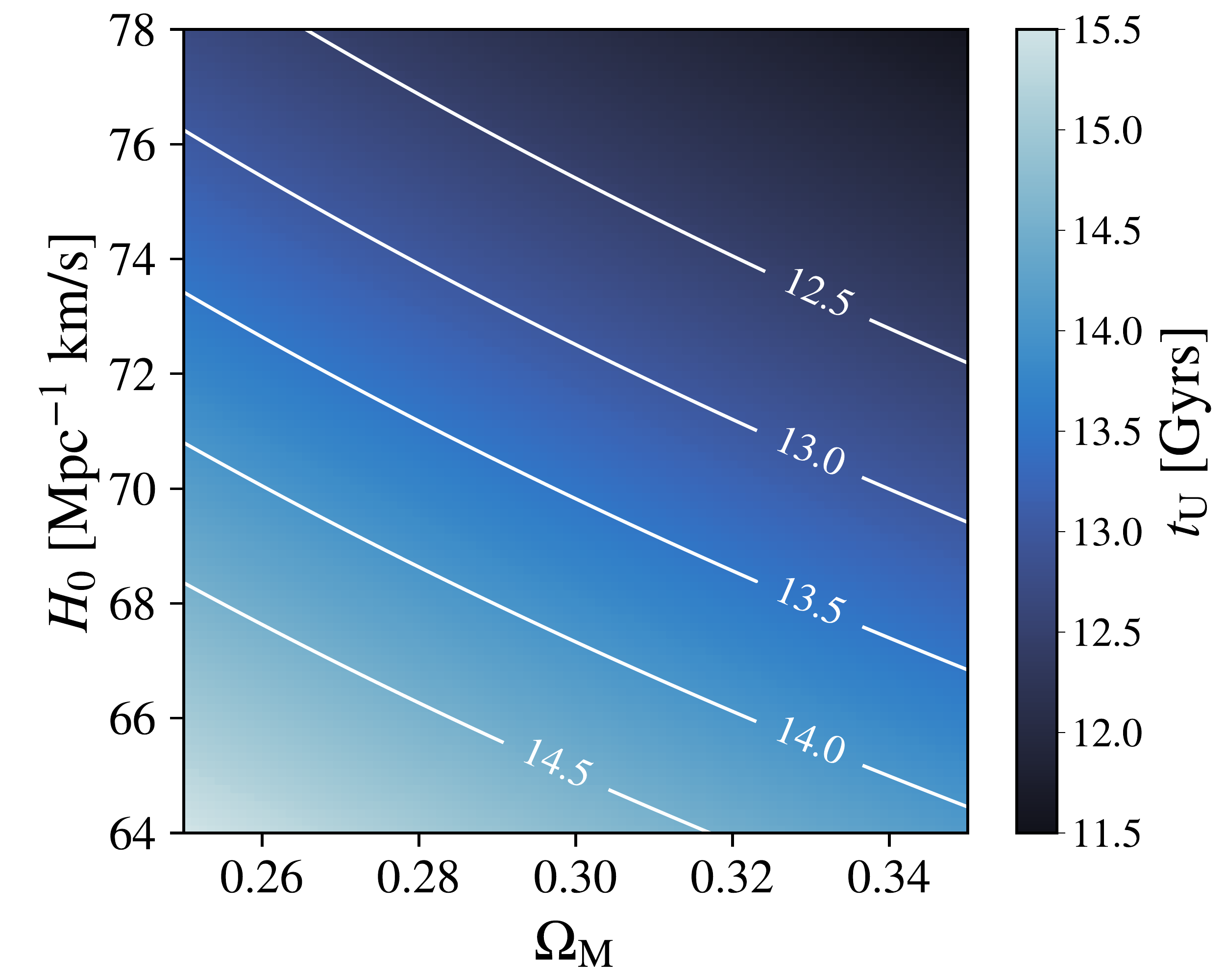}
\includegraphics[width=0.32\textwidth]{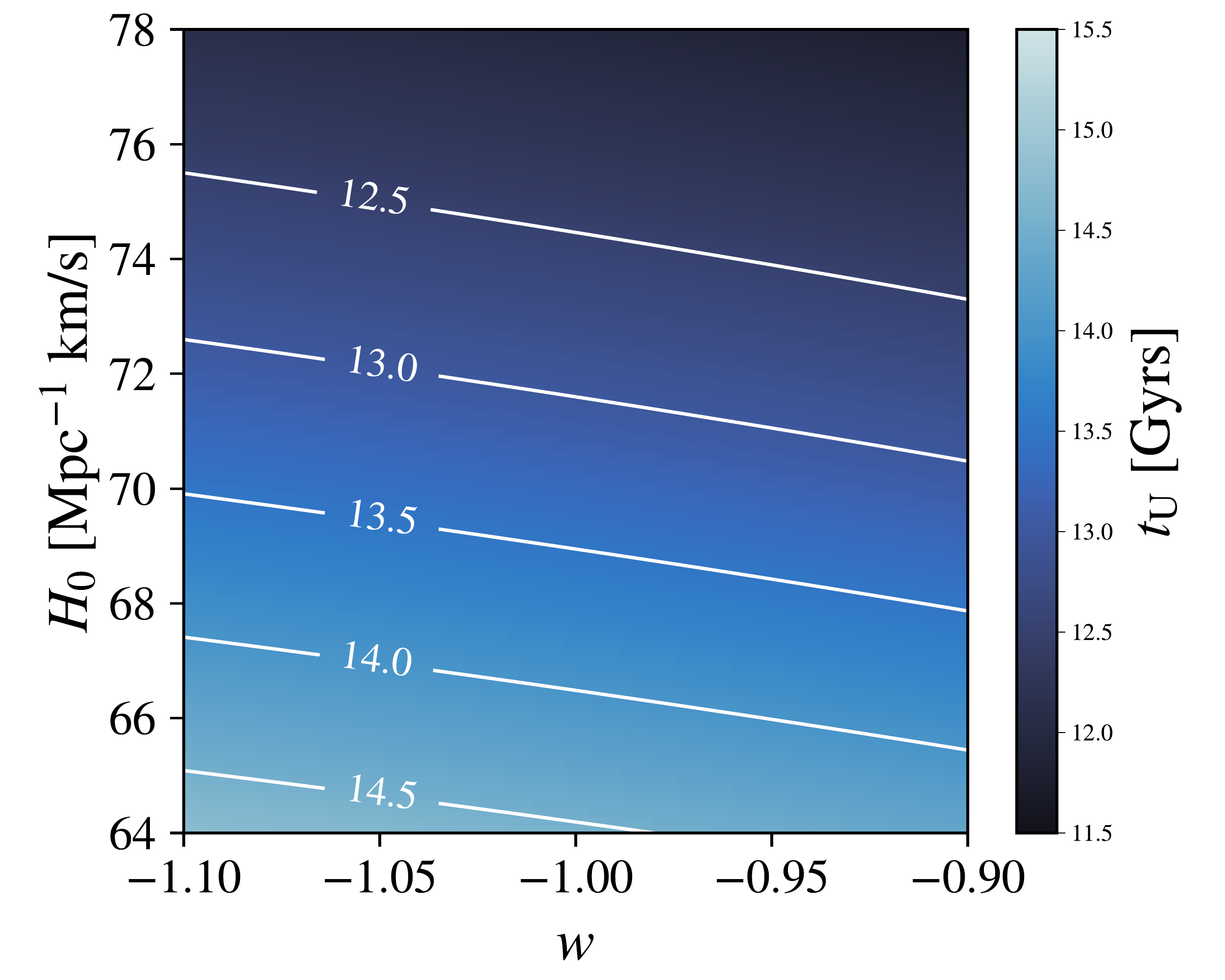}
\includegraphics[width=0.32\textwidth]{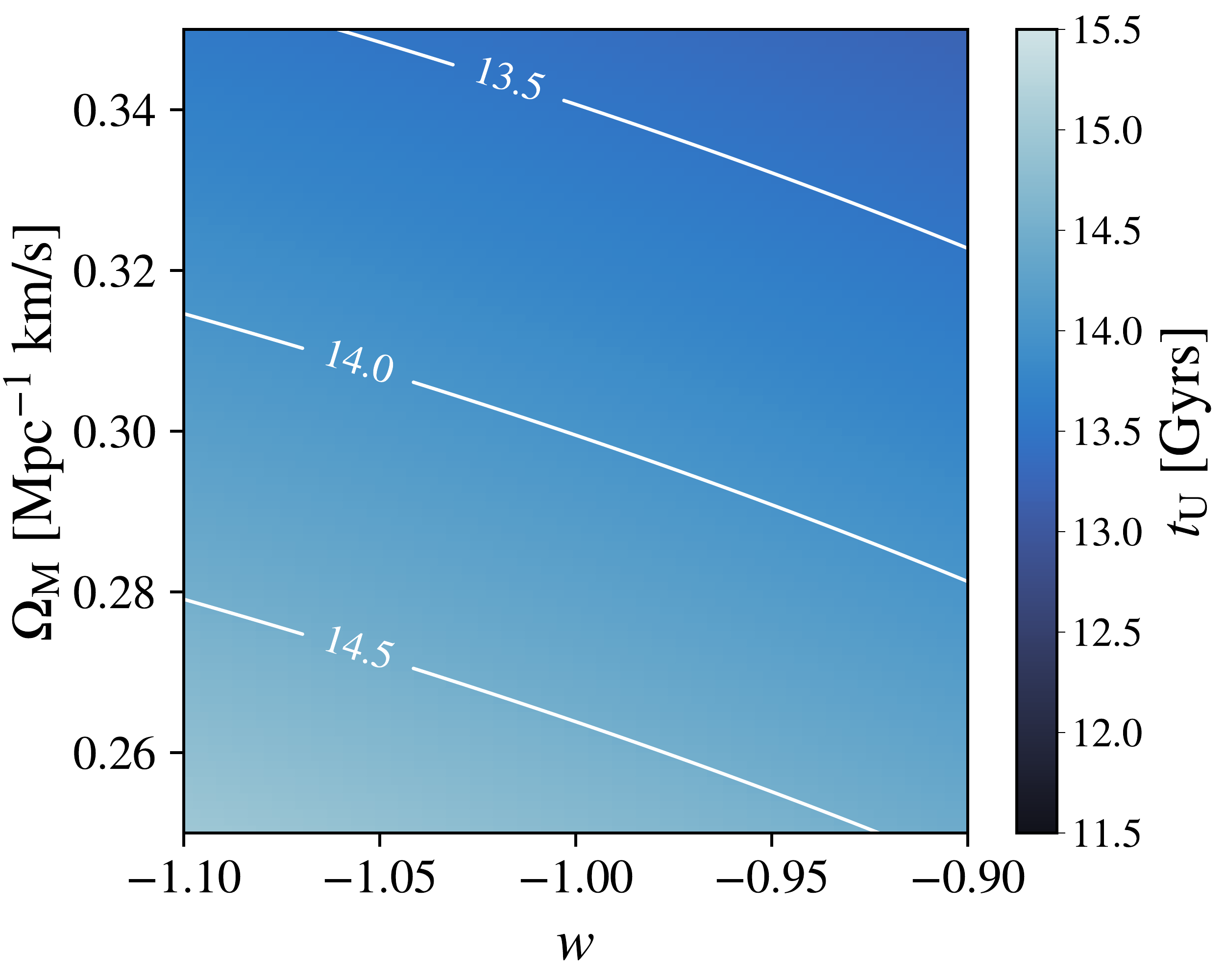}
\caption{Age of the Universe (in Gyr) as function of \Ho~and \omegam~for $w=-1$ (left panel), \Ho~and $w$ for \omegam=0.3138 (central panel), and \omegam~and $w$ for $h=0.6736$ (right panel). When a parameter is not varied, it is fixed to Planck18 $\Lambda$CDM best-fit value \citep{planck2018}. White lines mark contours with constant value of $t_{\rm U}$. Image reproduced with permission from \cite{Triangles}, copyright by APS.}  
\label{fig:tUpars}
\end{centering}
\end{figure*}

The look-back time $t$ as function of redshift is given by:
\begin{equation}
    t(z) =\frac{977.8}{H_0}\int_0^z \frac{{\rm d}z^\prime }{(1+z^\prime)E(z')}\,{\rm Gyr} \;\; ,
    \label{eq:tz}
\end{equation}
with $E(z)\equiv H(z)/H_0$ and $H(z)$ in \Hunit. Following Eq.~\eqref{eq:tz}, the age of the Universe is $t_{\rm U}\equiv t(\infty)$. We show the dependence of $t_{\rm U}$ on \Ho, \omegam~and a constant EoS parameter $w$ for dark energy in a $w$CDM model in Fig.~\ref{fig:tUpars}. It is evident that the strongest dependence is on \Ho, while \omegam~and $w$ have less influence. 

The integral in Eq.~\eqref{eq:tz} is dominated by contributions from redshifts below few tens, decreasing as $z$ grows. Therefore, any exotic pre-recombination physics does not significantly affect the age of the Universe. On the other hand,  $E(z)$ is bound to be very close to that of a CMB-calibrated $\Lambda$CDM model at $z\lesssim 2.4$, as shown in \cite{Triangles}. Hence, a precise and robust determination of $t_{\rm U}$ which does not significantly rely on a cosmological model, in combination with BAO and SNe Ia, may weigh in on proposed solutions to the \Ho~tension. If an independent (and model-agnostic) determination of $t_{\rm U}$  were to coincide with {\it Planck}'s inferred value assuming $\Lambda$CDM, alternative models involving exotic physics relevant only in the early Universe would need to invoke additional modifications also of the late-Universe expansion history to reproduce all observations as their prediction for $t_{\rm U}$ would be too low.

The color-magnitude diagram (CMD) of co-eval stellar populations in the Milky Way, or any other nearby galaxies where this is observationally possible, can be used to infer the age of its oldest stars. The age can also be estimated  for individual stars if their metallicity and the distance are  known. For resolved stellar populations, however, an independent measurement of the distance is not strictly necessary as the full morphology of the color-magnitude diagram can, in principle, provide a determination of the absolute age. There is extensive literature on the dating of stellar populations; reviews can be found in, e.g., \cite{Catelan,Soderblom,Bolte+}. In this section, we will focus on the most recent developments in the field. 
Ages of stars can also be computed via nucleo-cosmochronology \citep[see, e.g.,][]{Christlieb2016}, which consists in measuring global abundances of radioactive elements like Uranium and Thorium to estimate the age of the parent star. Another method is to use the cooling luminosity function of white dwarfs \citep[see, e.g.,][and references therein for the current status of these methods]{Catelan}; while useful, they are not still at the accuracy level of stellar ages measured via the observed color-magnitude diagram, and we will not discuss them further. We will instead focus on the use of the color-magnitude diagram on Globular Clusters (GCs) as new developments are providing stellar ages at the few \% level accuracy.

The first quantitative attempt to compute the age of the globular cluster M3 was made by Haselgrove and Hoyle more than 60 years ago \citep{Hoyle}. In this work, stellar models were computed on the early Cambridge mainframe computer and its results compared ``by eye'' to the observed color-magnitude diagram. A few stellar phases were computed by solving the equations of stellar structure; this output was compared to observations. Their estimated age for M3 is only 50\% off from its current value.\footnote{Their low age estimate is due to the use of an incorrect distance to M3, since the stellar model used deviated just $\sim$10\% from current models' prediction of  the effective temperature and gravity of stars, with their same, correct assumptions~\cite{Bolte+}.} This was the first true attempt to use computer models to fit resolved stellar populations and thus obtain cosmological parameters: the age of the Universe in this case. Previous estimates of the ages of GCs involved just analytic calculations, which significantly impacted the accuracy of the results, given the complexity of the stellar structure equations (see e.g.,~\cite{Sandage}).

\begin{figure}[t!]
\centering
\includegraphics[scale=0.35]{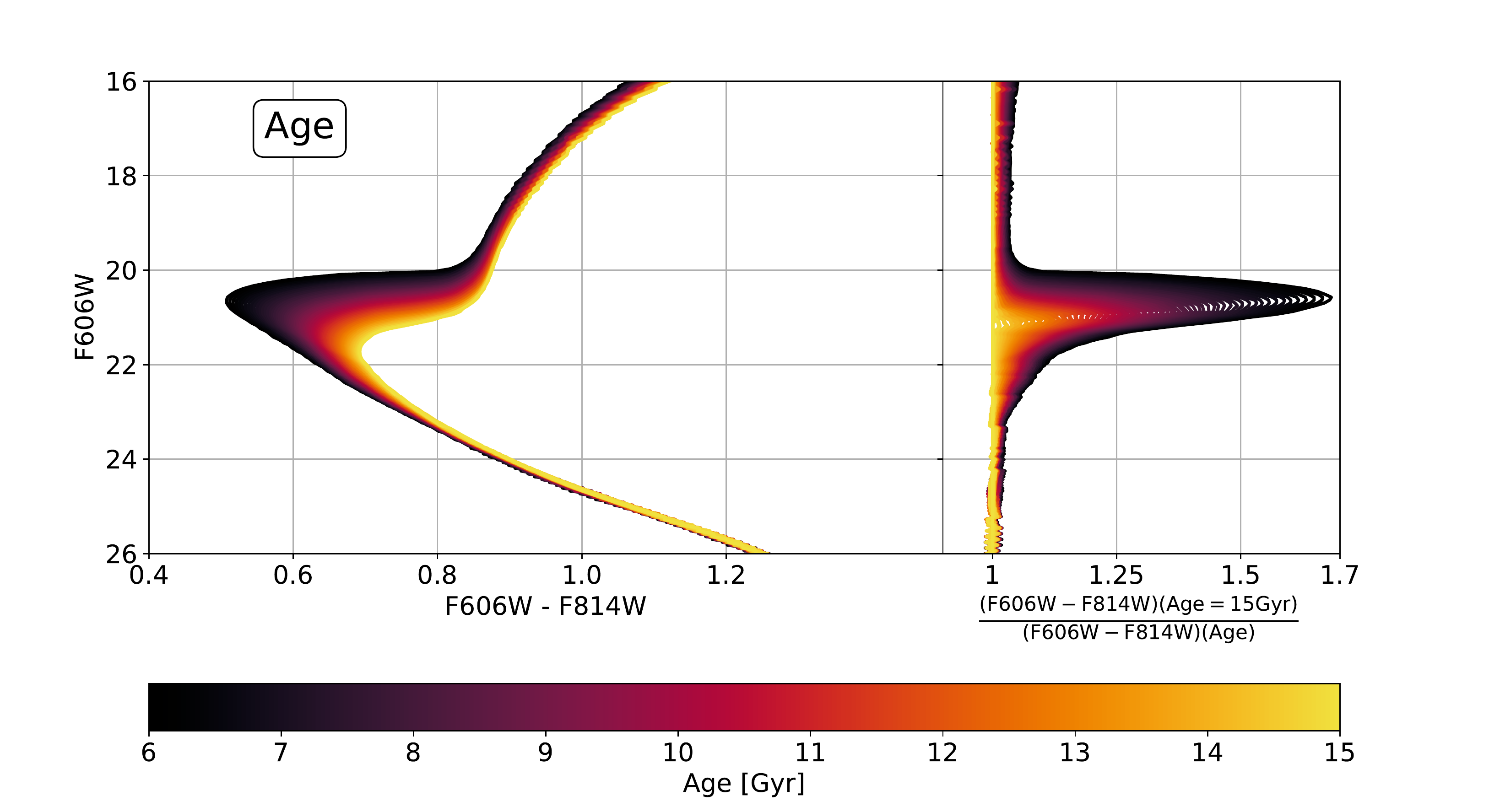} 
\includegraphics[scale=0.35]{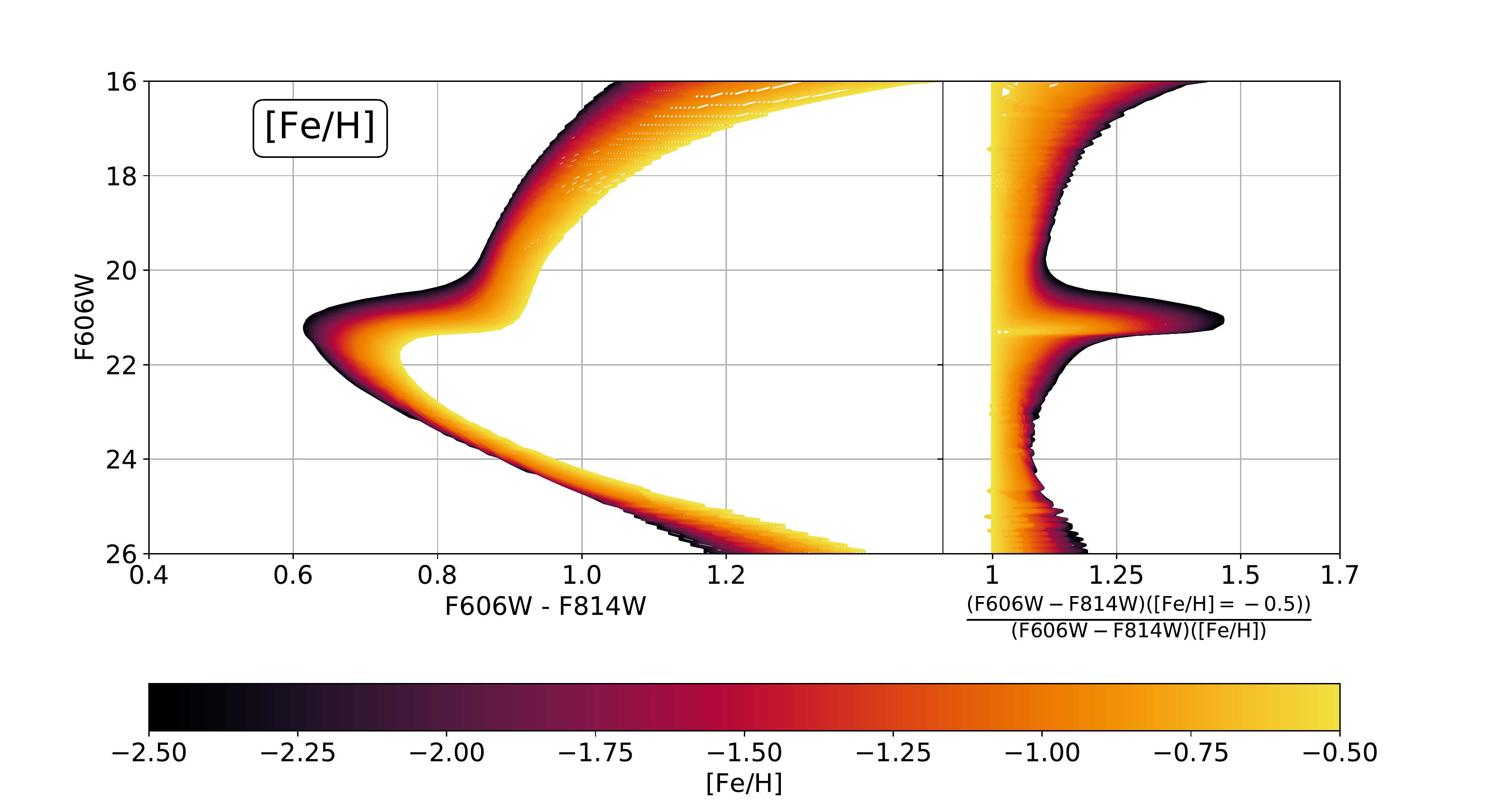}
\includegraphics[scale=0.35]{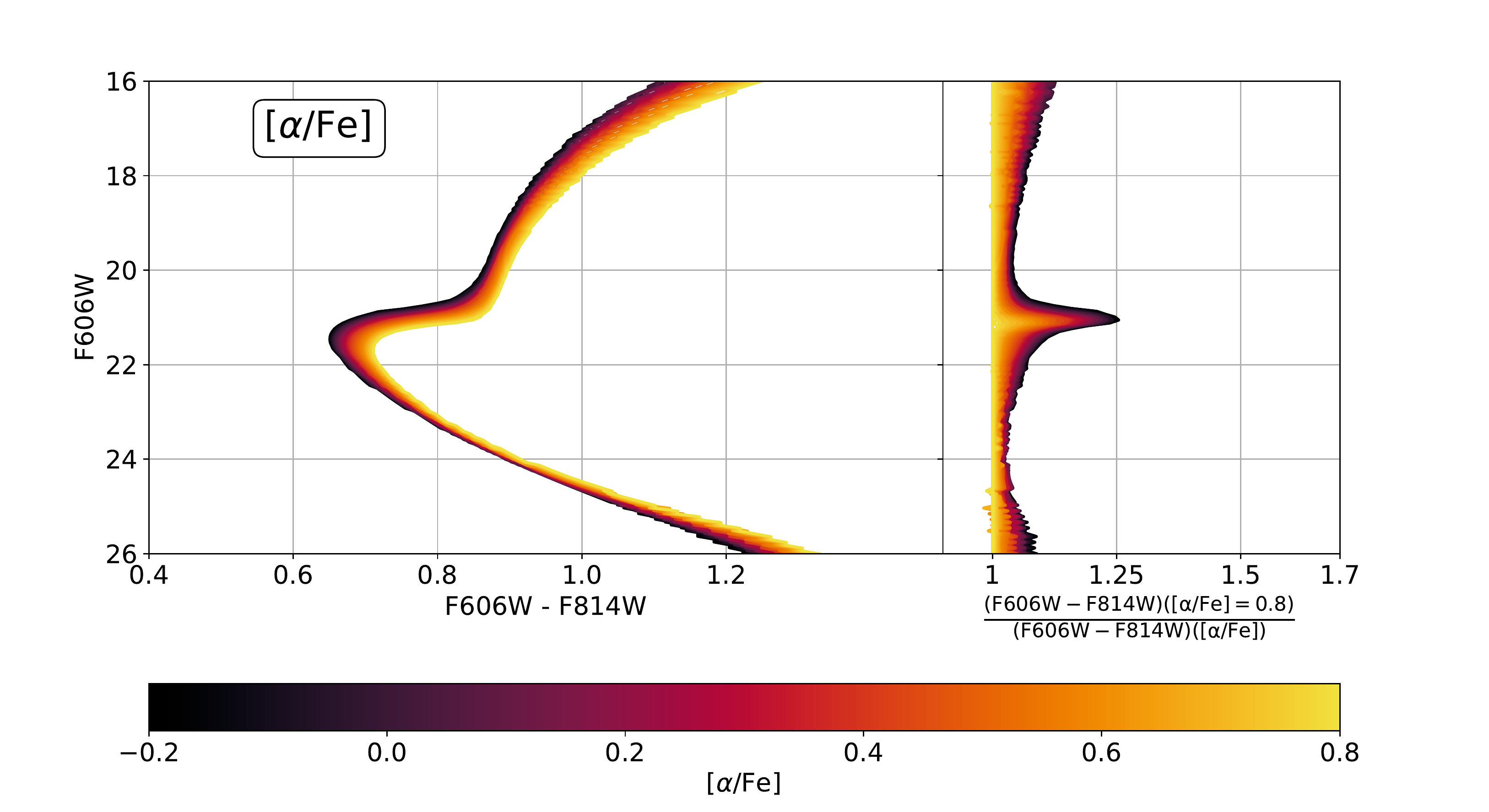} 
\caption{Dependence of the stellar isochrone on variations of age, metallicity and [$\alpha$/Fe] of the GC with all other parameters fixed. Right panels show the relative difference in color. Image reproduced with permission from \cite{Valcin1}, copyright by IOP Publishing.}
\label{diff1}
\end{figure}

Historically, the age of the oldest stellar populations in the Milky Way has been measured using the luminosity of the Main-Sequence Turn-Off Point(MSTOP) in the color-magnitude diagram of GCs. 
In this way, however, the full richness of information contained in the whole color magnitude diagram is discarded, and only one point kept. While it is true that the MSTOP contains significant information about the age of the stellar population, other parts of the CMD diagram do as well, especially around the sub-giant branch and the main sequence below the MSTOP; this is crucial to break degeneracies with distance and metallicity content (see Fig.~\ref{diff1}).

Globular clusters are (almost, more on this below) single stellar populations of stars \citep[see, e.g.,][]{Bolte+}. It has long been recognized that they are among the most metal poor ($\sim$1\% of the solar metallicity) stellar systems in the Milky Way, and exhibit color-magnitude diagrams characteristic of old ($>$ 10 Gyr) stellar populations \citep{OMalley,Catelan,Bolte+}. 

Of great interest is the fact that determination of stellar ages in the '90s provided one of the first hints that the dominant cosmological model at the time (an Einstein-de-Sitter Universe) needed revision \citep[see, e.g.,][]{ostriker,JimenezGC,Spinrad}. Old stellar populations were determined to be older than $1/H_0$, the age of the Universe in that model \citep[see, e.g.][]{JimenezGC}. Of course, the age of stellar objects at $z=0$ is just a lower limit to the age of the Universe and, by itself, does not constrain the cosmological model, as changes in \Ho~and \omegam~can accommodate an Einstein-de-Sitter Universe. 

In the past, in order to break this degeneracy, a determination of the stellar ages of the oldest galaxies at $z \gg 0$ proved crucial. This was first achieved by \cite{Dunlop}. It is revealing to see Fig.~18 in \cite{Spinrad}, which shows the exclusion of the  Einstein-de-Sitter Universe once the ages of GCs are taken into account.  
This philosophy has been further developed in the cosmic chronometer method, with the first  cosmological-model-independent determination of the redshift evolution of the Hubble parameter, $H(z)$ (see Sect.~\ref{sec:CC} and references therein).

The determination of the absolute age of a GC inferred using only the MSTOP luminosity is degenerate with other properties of the GC. As already shown in the pioneering work of \cite{Hoyle}, the distance uncertainty to the GC entails the largest contribution to the error budget: a given \% level of relative uncertainty in the distance determination involves roughly the same level of uncertainty in the inference of the age. Other sources of uncertainty are:  the metallicity content, the Helium fraction, the dust absorption  \citep{Bolte+}, and theoretical systematics regarding the physics and modeling of stellar evolution. 

However, there is more information enclosed in the full-color magnitude diagram of a GC than that enclosed in its MSTOP.
As first pointed out in~\cite{JimenezPadoanLF,PadoanJimenezLF}, the full color-magnitude diagram has features that allow for a joint fit of the distance scale and the age (see Fig.~\ref{diff1} for a visual rendering of this). On the one hand, Fig.~2 in \cite{JimenezPadoanGC} shows how the different portions of the color-magnitude diagram constrain the corresponding physical quantities. Figure~1 in \cite{PadoanJimenezLF} and Figure~3 in \cite{JimenezPadoanGC} show how the luminosity function is not a pure power-law, but has features that contain information about the different physical parameters of the GC. This technique enabled the estimation of the ages of the GCs M68 \citep{JimenezPadoanLF}, M5 and M55 \citep{JimenezPadoanGC}. Moreover, in principle, exploiting the morphology of the horizontal branch makes it possible to determine the ages of GCs independently of the distance \citep{JimenezGC96}.

Further, on the observational front, the gathering of Hubble Space Telescope (HST) photometry for a significant sample of galactic GCs has been a game changer. HST has provided very accurate photometry with a very compact point spread function, thus easing the problems of crowding when attempting to extract the color-magnitude diagram for a GC and making it much easier to control contamination from foreground and background field stars. 

For these reasons, a precise and robust determination of the age of a GC requires a global fit of all these quantities from the full color-magnitude diagram  of the cluster. In order to exploit this information, and due to degeneracies among GC parameters,  a suitable statistical approach is needed. Bayesian techniques, which have recently become the workhorse of  cosmological parameter inference, are of particular interest.  
In the perspective of possibly using the estimated age of the oldest stellar populations in a cosmological context as a route to constrain the age of the Universe, it is of value to adopt Bayesian techniques in this context too.
  
There are only a few recent attempts at using Bayesian techniques to fit GCs' color-magnitude diagrams, albeit only using some of their features \citep[see, e.g.,][]{BayesianGC}. Other attempts to use Bayesian techniques to age-date individual stars from the GAIA catalog can be found in \cite{Lund}. A limitation of  the methodology presented in \cite{BayesianGC} is the large number of parameters needed in the likelihood. Actually, for a GC of $N _{\rm stars}$ there are, in principle, $4 \times N_{\rm stars} + 5$ model's parameters (effectively $3 \times N_{\rm stars} + 5$), where the variables for each star are: initial stellar mass, photometry, ratio of secondary to primary initial stellar masses \citep[fixed to 0 in][]{BayesianGC}, and cluster membership indicator. In addition, there are 5 (4) additional GC variables, namely: age, metallicity \citep[fixed in the analysis of][]{BayesianGC}, distance modulus, absorption, and Helium fraction. For a cluster of 10,000 or more stars, the computational cost of this approach is very high. To overcome this issue, \cite{BayesianGC} randomly selected a sub-sample of 3,000 stars, half above and half below the MSTOP of the cluster,  ``to ensure a reasonable sample of stars on the sub-giant and red-giant branches''. Another difficulty arises from the fact that  the cluster membership indicator variable can take only the value of 0 or 1 (i.e., whether a star belongs to the cluster or not). This creates a sample of two populations referred  to as a {\it finite mixture distributions} \citep{BayesianGC}. 

Recently, a Bayesian analysis of the GC CMD using all features in it has been carried out by \cite{Valcin1,Valcin2}. This has resulted in the join determination of ages, metallicities and distances for 68 GCs observed by the HST/ACS project. The main advantage of the \cite{Valcin1,Valcin2} approach is that by using all features in the CMD diagram it is possible not only to obtain ages with smaller uncertainties, but also remove some of the systematic uncertainties \citep{Valcin2}. 

\subsubsection{Sample selection}

To obtain a lower limit to the age of the Universe one needs to select the objects hosting the oldest stars. This in itself is an obvious circular argument as we will only know which stars are the oldest after having measured their age. The most useful approach is to select those GCs with the lowest metallicity, as they will likely be the first formed in the Universe. In reality, since the Milky Way only contains a couple of hundred of GCs, the most natural approach would be to just compute the age for all of them and then select.

This procedure can be also applied to other stellar clusters, like open clusters, but these ones always tend to be significantly younger than GCs.

To measure the ages of stars in GCs the sample selection is fairly straightforward. One selects the stars that belong to the globular cluster. The best procedure to do this is to plot individual stars in the color-magnitude diagram to identify the locus of cluster members. While there are technicalities involved in computing photometry in crowded fields and how to identify cluster members, care need to be taken but these are issues well under-control. 
Indeed, we may already have all data needed as almost all known globular clusters in the Milky Way are known. it would be useful to obtained resolved stellar populations of GCs in other galaxies, like Andromeda; this is something that JWST may achieve in the near future. The most important revolution will come from using full sky surveys to measure ages of stars systematically.
For now, suitable observations for a representative sub-sample of 68 GC are available \citep{Valcin1,Valcin2}.
\subsubsection{Measurements}

Accurate photometry is the main requirement for obtaining color-magnitude diagrams of GCs. In addition, it would be desirable to obtain as much spectroscopy as possible from the resolved stars, as this would help reduce the reliance on Bayesian priors on metallicity. 

Of course, good data require outstanding analysis tools. Simply fitting the MSTOP does not do justice to the data, as this discards precious information on other parameters besides the age of the GC. The recent use of fully Bayesian techniques \citep[like, e.g., in][]{Valcin1} shows that there is more information in the CMD. Future uses of likelihood-free inference can further extract all information from the CMD.

\subsubsection{Systematic effects}

Systematics are the main source of uncertainties when obtaining the absolute age of GCs; note, however, that relative ages are less prone to systematic uncertainties. Systematic uncertainties are dominated by uncertainties in stellar evolution and distance determination to the GC. The best study of systematic uncertainties in the age determination of GCs is the work by Chaboyer and collaborators \citep[see, e.g.,][]{OMalley}. 
They are mostly arising from three sources:  uncertainties in  {\it i)}  nuclear reaction rates,  {\it ii)}in the modeling of convection in the outer layers of low mass stars, and  {\it iiii)} in the estimation of the distance to the GC. These are the same systematic uncertainties that affect the age determination obtained using only the MSTOP, but are ameliorated when using the full morphology of the color-magnitude diagram \citep[see][]{Valcin2}. In particular, both distance and uncertainties in convection of the star's outer layers can be significantly reduced when using the full CMD. In addition, independent measurements obtained by the Gaia space mission will drastically reduce the uncertainty on distance. It is worth noting that the uncertainties concerning stellar nuclear rates could be greatly reduced by producing better theoretical computations \citep[see also][]{Boylan}.

Another source of (systematic) error is the uncertainty in $z_{\rm form}$ to infer the  the age of the Universe from the age of the star; this was addressed in detail in \cite{Jimenez2019}. The determination of $z_{\rm form}$ will be improved dramatically by the upcoming observations from JWST \citep{Gardner:2006} which will conclusively map the mass function of objects at $10<z<20$.

\begin{figure}[b!]
\centering
\includegraphics[scale=0.6]{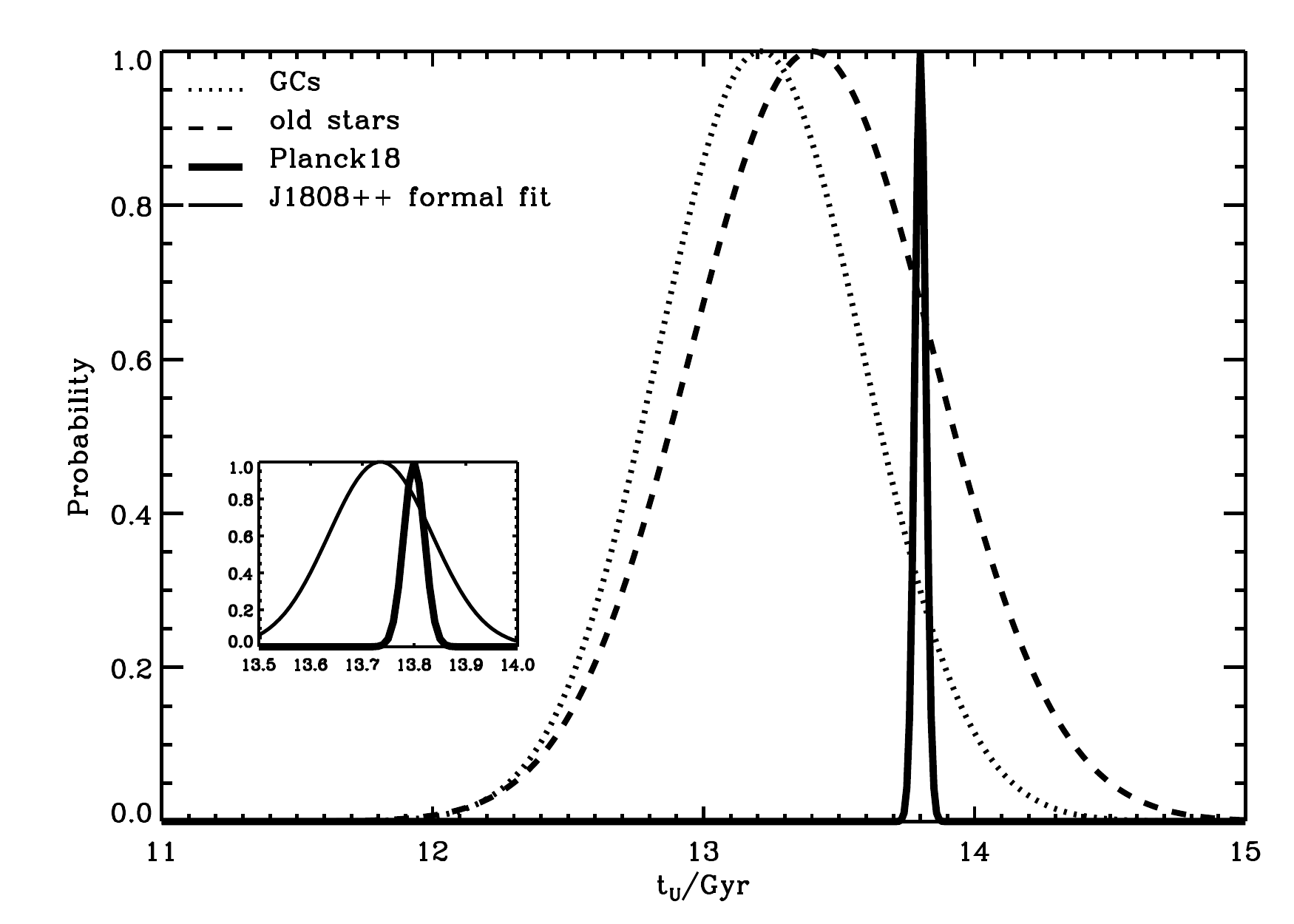} 
\caption{Probability distribution for the age of the Universe obtained using stellar ages (thin set of lines) and derived by Planck18 from the CMB assuming the $\Lambda$CDM model \citep[thick solid,][]{planck2018}. All determinations are in good agreement.  Just as an example of what kind of accuracy could be obtained if systematic uncertainties were all under control, the inset shows the age of the Universe  for the formal determination and formal  uncertainty of J18082002-5104378, which is fully compatible with Planck18. The formal GCs ages for 69 ACS clusters from \cite{OMalley} would look similar to the J18082002-5104378 line. Image reproduced with permission from \cite{Jimenez2019}, copyright by IOP Publishing.}
\label{fig:agesMSTOP}
\end{figure}

\subsubsection{Main results and forecasts}

\begin{figure}[t!]
\centering
\includegraphics[scale=0.5]{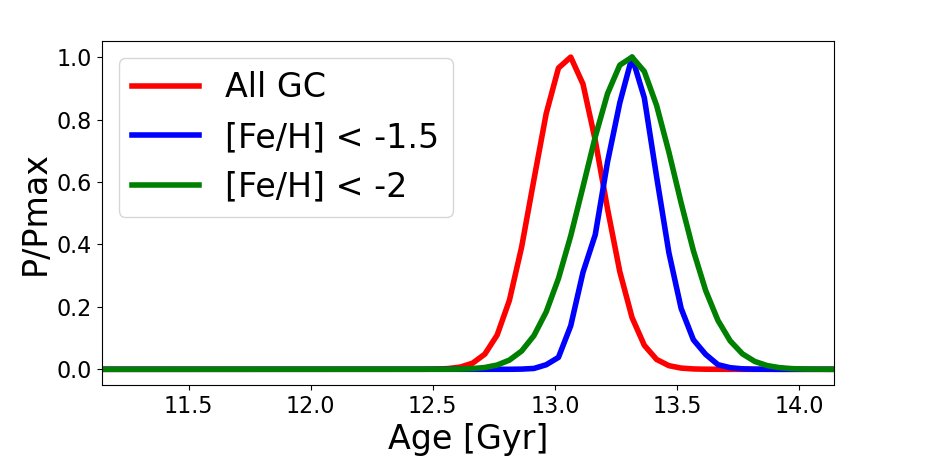} 
\caption{Age distribution for globular clusters using the Bayesian method of~\cite{Valcin1} when using the full CMD with different metallicity cuts. The behavior is consistent with the expected age-metallicity relation. Only the statistical uncertainty is displayed. An additional uncertainty of 0.25 Gyr~\citep{Valcin2} at 68\% confidence level needs to be added to account for the systematic uncertainty. Image reproduced with permission from \cite{Valcin1}, copyright by IOP Publishing.}
\label{fig:agesBayesian}
\end{figure}

\begin{figure}[b!]
\centering
\includegraphics[scale=0.35]{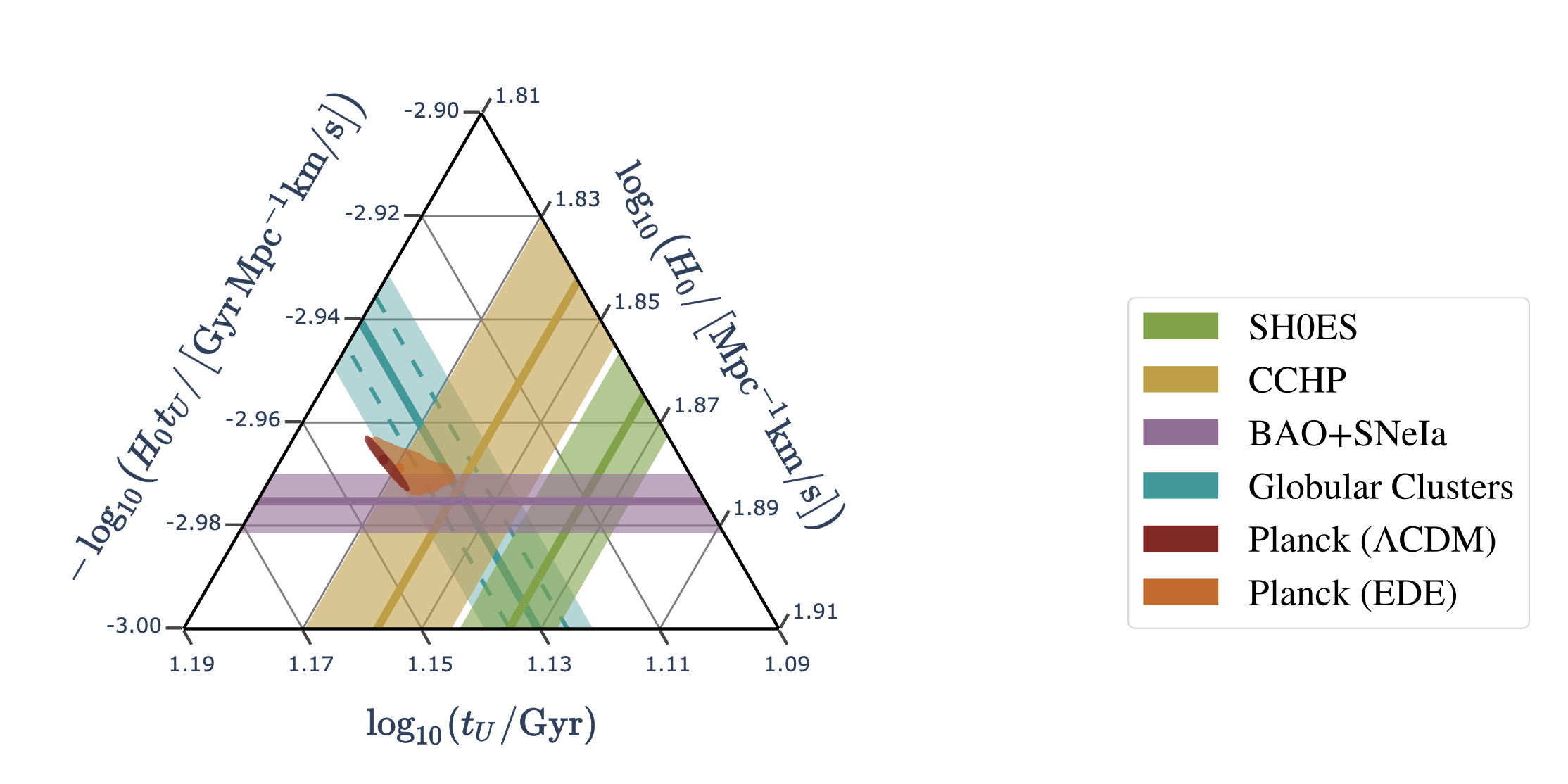} 
\caption{Triad plot ``new cosmic triangle'':
68\% confidence level marginalized constraints on the new cosmic triangle:  the triad corresponding to the age of the Universe and the Hubble constant (upper left) is shown. Note that all points in the figure sum up to 0, while the ticks in the axes determine the direction of equal values for each axis. Note that absolute ages of GCs are consistent with the model-dependent Planck18 value. Image reproduced with permission from \cite{Triangles}, copyright by Physical Review D.}
\label{fig:agesTriangle}
\end{figure}

The most recent determinations of the ages of GCs using the MSTOP and full CMD Bayesian method are shown in Fig.~\ref{fig:agesMSTOP} and \ref{fig:agesBayesian}.
In Fig.~\ref{fig:agesMSTOP}, taken from \cite{Jimenez2019} ages determinations from the literature were used, including the ages of individual stars,and CMB-derived age. 
In Fig.~\ref{fig:agesBayesian} the age of the Universe is computed using the method of \cite{Valcin1}.
Despite the very different observables and approaches, there is good agreement among all the age determinations. 

Stellar ages have proven to be also extremely useful to unveil the nature of the Hubble tension, as done in \cite{sunny,krishnan2021}. 
A summary of how the absolute age $t_U$ determination can weigh in on the current ``debate'' on the expansion rate of the Universe is shown in Fig.~\ref{fig:agesTriangle}, and further elaborated in \cite{Triangles}. 
There are two independent physical quantities ($H_0$, $t_U$), but three quantities are measured independently: $t_U$ from absolute ages, $H_0$ from cosmic distance ladder, and, in a $\Lambda$CDM model, $H_0t_U$ from standard rules and candles (BAO and SNe Ia), and a combination of constraints on $t_U$ and $H_0$ from CMB. This is an over-constrained system which can be be represented on a triad plot (see \cite{Triangles}, referred to as ``new cosmic triangle'') such that $\log t_U+log H_0\equiv log (H0 t_U)$, Fig.~\ref{fig:agesTriangle}. 
While BAO, SNe Ia and CMB inferences depend on the cosmological model, the stellar ages ones (and the distance ladder ones) are cosmology-independent. 
It is interesting to note that GC ages, CMB, and BAO+SNe determinations agree, indicating that a $\Lambda$CDM-like expansion history is a good fit to the data, in the redshift window heavily weighed by these data.  

The future for more accurate GC ages lies in two fronts: reducing the systematic uncertainty in stellar modeling by constructing improved models and reducing the width of priors used for metallicity and distances by resorting to additional, complementary observations.
Direct distances from Gaia \citep{Gaia:2016} are particularly promising.
Especially useful will be the final Gaia data release which will provide \% or sub-\% direct (parallax based) distances to GCs. This will tremendously narrow the adopted distance prior range. Another important ingredient will be the direct spectroscopic determination of chemical abundances in individual stars in GCs, specially below the MSTOP. The JWST telescope \citep{Gardner:2006} will enable enormous progress in these two directions. If these two priors are constrained at the \% level from direct observations, then the only remaining systematic uncertainty will be that from constructing the stellar models.
As shown in \cite{Valcin2}, when using the full CMD, the dominant uncertainty left in stellar models is the one due to nuclear reaction rates which  can in principle  be improved by a combination of laboratory and theory efforts.

\clearpage

\subsection{Secular Redshift Drift}
\label{sec:RD}

Any non-empty universe will exhibit an accelerating or decelerating Hubble expansion, which can be observed as a secular redshift drift.  \citet{sandage1962} first proposed observing this effect in the optical spectra of galaxies to measure the cosmic deceleration. \citet{loeb1998} later suggested using the neutral hydrogen Lyman $\alpha$ forest of absorption lines toward quasars, and this concept has been developed as a key science case for large optical telescopes \citep[e.g.,][]{corasaniti2007,liske2008}. 
Large radio telescopes may likewise probe the redshift drift using neutral hydrogen via the 21cm emission line from galaxy surveys or using \hi 21cm absorption toward quasars \citep[e.g.,][]{darling2012,yu2014,kloeckner2015}. Measurements require exquisite, repeatable, long-term wavelength calibration that will most likely rely on a stable local oscillator in both the optical and radio wavelength regimes.

The secular redshift drift is a means to directly observe the cosmic acceleration that does not rely on models, standard candles, standard rulers, or the cosmological distance ladder.  It is capable of directly testing standard dark energy cosmology and can be used as a probe of cosmological inhomogeneities and thus test the FLRW paradigm and general anisotropic models \citep[e.g.,][]{quartin2010}.
However, the signal is so small (of order \Ho$\Delta t$, where $\Delta t$ is the duration of observation) that it is unlikely to provide competitive constraints on cosmological parameters in an era of precision cosmology. For example, \citet{alves2019} predict that a 40m-class optical telescope Ly$\alpha$ forest program combined with an \hi 21cm emission line survey and \hi 21cm absorption line monitoring can provide independent constraints on \Ho, \omegam, and $w_0$ of order 19\%, 7\%, and 13\%, respectively, in a flat $w$CDM model (marginalized 1$\sigma$ uncertainties).

Nevertheless, a measurement of $\dot{z}$ is a model-independent indication of the presence of dark energy \citep{heinesen2021}, and offers a means to directly determine the cosmic expansion history.
It also offers some improvement on cosmological priors when combined with more traditional measurements \citep{alves2019}, and notably tends to break parameter degeneracies in traditional comsological probes \citep{martins2021}.

In the following, we describe the expected secular redshift drift, its dependence on cosmological parameters, measurement methods including sample selection and systematic effects, and forecasts of the measurement precision and the resulting constraints on cosmological parameters.  

\subsubsection{Basic idea and equations}

The observed secular redshift drift, the rate of change of redshift in the current epoch $t_0$, is to first order the difference between the Hubble expansion of a coasting universe at redshift $z$ and the true Hubble expansion at that redshift \citep[e.g.,][]{loeb1998}:
\begin{equation}
\frac{d z}{d t_0} \equiv \dot{z} = (1+z)\, H_0 - H(z) \;\; .
\end{equation}
The derivation of this relationship relies only on the null interval obeying $c\,dt = a(t)\,dr$ and the definitions $1+z = a(t_0)/a(t_e)$ and $H = \dot{a}/a$:  for 
redshifts measured at times $t_0$ and $t_0+\Delta t_0$, the redshift change is 
\begin{equation}
    \Delta z = \frac{a(t_0+\Delta t)}{a(t_e+\Delta t_e)} - \frac{a(t_0)}{a(t_e)} 
             \simeq \left[\frac{a(t_0)}{a(t_e)}\, \frac{\dot{a}(t_0)}{a(t_0)}  -  \frac{\dot{a}(t_e)}{a(t_e)}\right] \Delta t_0 
\end{equation}
for $\Delta t_0 \ll t_0$.
The redshift drift can be recast in terms of an observed acceleration:
\begin{equation}
\frac{d v}{d t_0} = \frac{c\, \dot{z}}{1+z} = c H_0\left(1 - \frac{E(z)}{1+z}\right) \;\; ,
\end{equation}
where $E(z)$ is the unitless rescaled Hubble parameter (Eq.~\ref{eq:Ez}) that depends on the contents and curvature of the universe. 
Measurements of the secular redshift drift thus encode the Hubble constant, the matter density, the curvature, and the dark energy density and its equation of state. \citet{alves2019} show that the redshift drift is most sensitive to \Ho~and \omegam~(or \omegal) in a canonical flat $\Lambda$CDM cosmology.  In $w$CDM or $w_0 w_a$CDM models, the effect is less sensitive to $w_0$ and least sensitive to $w_a$ (but these broad statements vary somewhat as a function of redshift and the span of redshifts explored by a given probe).

\begin{figure}[t]
  \begin{center}
    \includegraphics[scale=0.42,trim=20 50 0 20,clip=false]{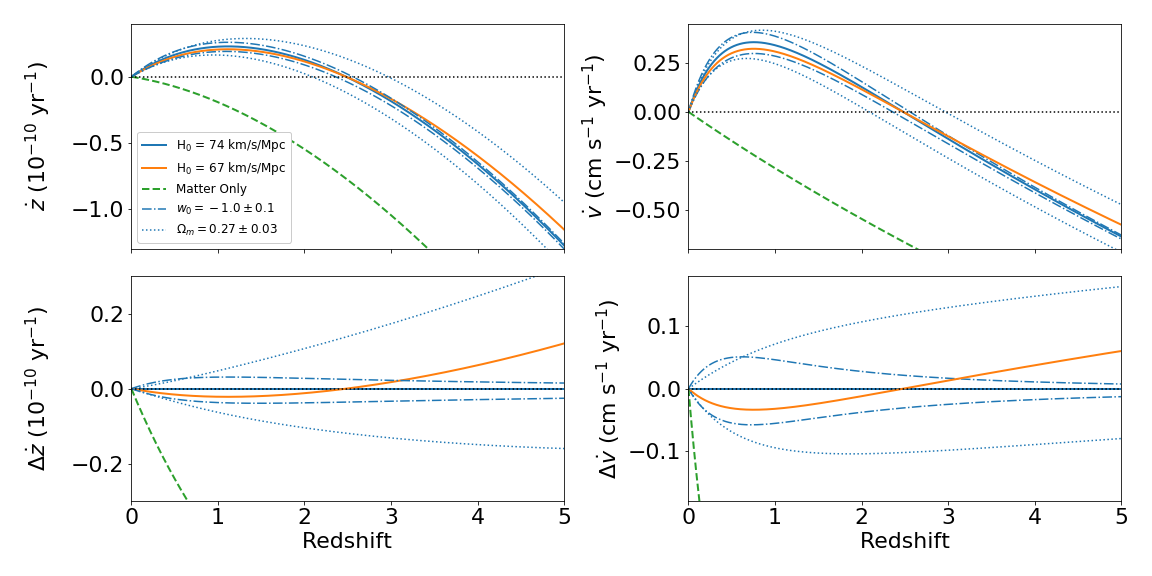}
  \end{center}
  \caption{ Secular redshift drift (left) and apparent acceleration (right) versus redshift.  All loci assume
    \Ho=74 \Hunit, \omegam=0.27, \omegal=0.73, and $w_0 = -1$ unless
    otherwise indicated. The top row shows the full signal, and the bottom row shows the difference between several models and the fiducial cosmology.}
  \label{fig:theory}
\end{figure}

Figure~\ref{fig:theory} shows sample tracks of $\dot{z}$ and $\dot{v}$
versus redshift for a few cosmologies as well as their differences from a fiducial model. There are a few noteworthy features of the redshift drift:  {\it i)} the redshifts of the peak $\dot{z}$, the peak acceleration, and the null between acceleration and deceleration are all independent of \Ho, but {\it ii)} the amplitude of the peaks (and the amplitude of the curves generally) scale with \Ho. The redshifts of the peaks and the null depend sensitively on the energy densities, including the curvature, but are somewhat insensitive to $w_0$ and $w_a$ when these are close to the canonical values.
For example, the $\dot{z} = 0$ redshift varies by roughly $z=2.5 \mp 0.5$ for \omegam$= 0.27 \pm 0.03$ in a flat $\Lambda$CDM cosmology.

Measurements of $\dot{z}$ at a variety of redshifts can thus probe epochs of acceleration caused by dark energy ($z \lesssim 2.5$) as well as epochs of deceleration caused by matter ($z \gtrsim 2.5$). This measurement is challenging because the size of the acceleration is small: it reaches a peak value of roughly 0.4 cm~s$^{-1}$~yr$^{-1}$ at $z\simeq 0.76$. The peak in $\dot{z}$ is $\sim$2$\times 10^{-11}$~yr$^{-1}$ (or $\sim H_0/3$) at $z\simeq 1.1$, as shown in Fig.~\ref{fig:theory}.  
Provided one can achieve adequate precision and measurement stability over years to decades, nearly any redshift indicator can be used to measure the secular redshift drift, including spectral lines (emission or absorption) and spectral edges or continuum breaks.

Since the most accessible measurements rely on spectral line centroiding, high signal-to-noise observations of many narrow lines are required, and narrow lines tend to be absorption lines (we exclude astrophysical masers from consideration). The technique therefore favors reasonable optical depth (but unsaturated) absorption lines toward bright optical or radio continuum sources. The Lyman $\alpha$ forest provides a high-$N$, high-$\sigma$ per line regime while radio absorption lines provide (for now) low-$N$, low-$\sigma$ measurements. The two regimes are likely to be competitive in the long-run, although the Ly$\alpha$ forest method will likely be less susceptible to and be better able to average out gravitational accelerations caused by the local environment and large scale structure (see Sect.~\ref{subsubsecRD:systematics}).

\hi 21cm radio emission from galaxies has also been proposed as a means to measure $\dot{z}$ using large galaxy surveys \citep{kloeckner2015}.  This approach relies on large samples, $\sim$10$^7$ galaxies per measurement, in order to overcome the large line width that samples the full rotation curve of galaxies, the large expected internal accelerations, and the acceleration caused by large-scale structures (as will be discussed in Sect.~\ref{subsubsecRD:systematics}).  

\subsubsection{Sample selection}
\label{subsubsecRD:sample}

There exist three main methods for detecting the secular redshift drift:
(1) Lyman $\alpha$ forest absorption lines toward bright quasars,
(2) \hi 21cm emission from galaxies, and
(3) \hi 21cm absorption toward radio sources.
There are additional methods beyond these that have not been as well developed such as molecular absorption lines toward bright (sub)mm continuum sources. It is also certain that additional clever ideas will arise \citep[see especially][]{kim2015} as the notion of directly measuring the cosmic acceleration gains traction and becomes more realistic with new facilities.

Bright quasars are needed to maximize signal-to-noise in high-resolution spectra of the Ly$\alpha$ forest. Quasars must also be redshifted to place Ly$\alpha$ redward of the atmospheric UV cutoff. To maximize spectral coverage per observation, optimal quasars would have $z\simeq5$ and be as bright as possible. The number of monitored quasars does not need to be large because the large-N statistics arise from the hundreds of absorption lines seen along each sight-line \citep[e.g.,][]{liske2008}.

\noindent
\hi 21cm emission line surveys rely on areal coverage and redshift selection. Redshift selection for a fiducial Square Kilometer Array (SKA) survey is flux-limited, and the ability to measure the redshift drift is limited by the number of detected galaxies and their signal-to-noise. Typically, $\sim$10$^7$ galaxies need to be observed within a redshift bin, and \citet{kloeckner2015} predict that $\dot{z}$ can be measured up to $z\sim1$.

\noindent  
At present, there are only $\sim$140 \hi 21cm absorption line systems known, which is a consequence of limited surveys, limited bandwidths, radio frequency interference (RFI), and flux sensitivity (absorption systems are generally only detected toward Jy-level continuum radio sources at $\sim$1~GHz). As areal coverage and sensitivity of surveys increase with SKA prototypes, the Five-hundred-meter Aperture Spherical radio Telescope (FAST), Canadian Hydrogen Intensity Mapping Experiment (CHIME), and ultimately the full SKA, the number of known systems is expected to increase by more than an order of magnitude.

Most planned or current surveys expect to detect at least hundreds of new \hi 21cm absorption line systems. For example, the ASKAP FLASH survey expects to detect several hundred new 21cm absorption line systems at $z\lesssim1$ \citep{sadler2020}. \citet{jiao2020} describe a commensal FAST survey that is predicted to detect roughly 800, 1900, and 2600 \hi 21cm absorption systems with $z<0.37$ in 1, 5, and 10-year surveys, whereas \citet{zhang2021} predict more than 1500 absorbers would be detected at $z<0.37$ by FAST.
CHIME, however, will survey the northern sky continuously and is predicted to detect $\sim$10$^5$ absorption lines in $0.8 < z < 2.5$  \citep{yu2014}.

\subsubsection{Measurements}

\begin{figure}[t]
  \begin{center}
    \includegraphics[scale=0.42,trim=20 50 0 20,clip=false]{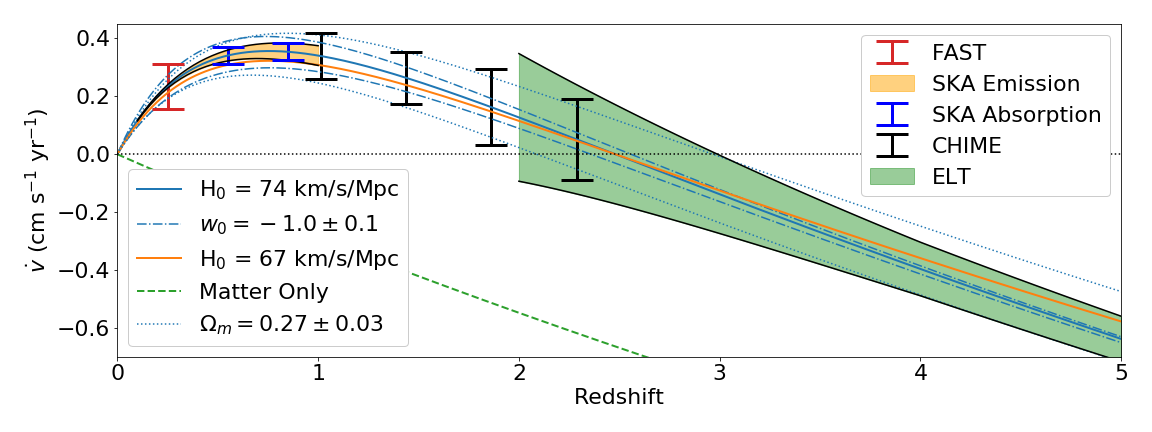}
  \end{center}
  \caption{ Forecast acceleration measurements versus redshift for the SKA using \hi 21cm emission from galaxies \citep{kloeckner2015}, CHIME using \hi 21cm absorption \citep{yu2014}, \hi 21cm absorption using FAST and the SKA estimated from projected detections (see text), and an ELT program that monitors the Ly$\alpha$ forest \citep{liske2008}. The cosmological tracks follow Figure \ref{fig:theory}. The shaded loci and error bars indicate 1$\sigma$ uncertainties.}
  \label{fig:measurements}
\end{figure}

Here we focus on the expected precision obtained by redshift drift measurements (forecasts for cosmological parameters based on the following predicted measurements are described in Sect.~\ref{subsubsecRD:forecasts}). Figure~\ref{fig:measurements} depicts the following predictions:
\begin{enumerate}
    \item Following \citet{liske2008} Eq.~16, we predict measurements based on a generic 42~m ELT. In the figure, we assume a two-epoch Ly$\alpha$ forest monitoring program of 10 quasars with S/N of 3000 spanning 20 years. Such a program is expected to reach acceleration uncertainties of 0.22--0.08 cm~s$^{-1}$~yr$^{-1}$ over redshifts $z=$2--5. It may be possible to improve upon this prediction using absorption lines beyond Ly$\alpha$, such as other Lyman series lines or metal lines that arise from higher column density clouds \citep{liske2008}. Moreover, \citet{cooke2020} presented a ``Ly$\alpha$ cell'' calibration technique that uses relative accelerations of metal and Ly$\alpha$ forest lines to provide a larger lever arm on the signal and to allow internal wavelength calibration of spectra.  Finally, \citet{eikenberry2019a,eikenberry2019b} proposed a dedicated non-ELT facility comprising many small telescopes that could reduce the detection time to 5 years.  

    \item The full SKA, following \citet{kloeckner2015} \citep[see also][]{martins2016}, is predicted to use 21cm emission from galaxies to measure $\dot{z}$ with 1--10\% uncertainty over redshifts $z=$0.1--1.0 in two epochs spanning 12 years. Galaxy-scale emission line profiles are broad (100's of km~s$^{-1}$, modulo inclination), which translates into a factor of $\sim$1000 in sample size needed to roughly match absorption line centroiding, all else equal.  We suggest that emission line edges and object-by-object cross-correlation may improve the expected performance of this technique but that the sensitivity of this technique to $\dot{z}$ needs to be modeled in detail using observed 21 cm emission line profiles.

    \item Provided the expected populations of \hi 21cm absorption line systems are detected by FAST and SKA precursors (as discussed in Sect.~\ref{subsubsecRD:sample}), we can modify the \citet{darling2012} predictions to make new estimates of the redshift drift measurement. A 20-year FAST monitoring program of 1000 absorption lines at $z<0.37$ will obtain acceleration precision of roughly $\pm 0.08$~cm~s$^{-1}$~yr$^{-1}$.  
    Likewise, a 10-year SKA program observing two redshift bins at $z=0.55$ and $z=0.85$ with 500 lines each can reach rms acceleration noise of $\sim$0.08~cm~s$^{-1}$~yr$^{-1}$, which is similar to the expectation for 21cm emission.  
\end{enumerate}

\citet{yu2014} predict that CHIME can reach 0.08--0.14~cm~s$^{-1}$~yr$^{-1}$ uncertainties spanning the range $z=0.8$--2.5 in a 10-year survey. The key differences between CHIME and FAST or SKA programs are the 100-fold higher number of expected absorption line systems and the daily observation of every system over 10 years. If absorption line systems are detected at the predicted rate, this suggests that CHIME will be competitive with two- or  few-epoch surveys of $\sim$10$^3$ systems that require much larger collecting areas.

Figure~\ref{fig:measurements} shows the measurement forecasts and illustrates how the signal can be detected but cannot generally discriminate between cosmologies that are consistent with current paradigms.  They can, however definitively and directly demonstrate the influence of dark energy on the cosmic expansion without use of standard distance indicators or models.

\subsubsection{Systematic effects}
\label{subsubsecRD:systematics}

Systematic effects include the ability to obtain stable and repeatable wavelength or frequency calibration, the relative angular motions of absorbing gas with respect to illumination sources, illumination source variability in size, flux, and spectral properties, motion of the observer, peculiar velocity and accelerations, and gravitational accelerations internal to and between monitored objects. Observations are made from a very non-inertial reference frame that reflects multiple accelerations and rotations, although these will be well-determined in the near future to better precision than is needed for the $\dot{z}$ measurement.

The requisite calibration stability relies on a local oscillator, and current radio facilities already support this level of precision \citep[e.g.,][]{cooke2020}. Optical spectroscopy will require stable references such as laser combs and actively-controlled high-precision spectrographs \citep[e.g.,][]{eikenberry2019a,eikenberry2019b}.  

Gravitational accelerations within galaxies, between galaxies, and within galaxy clusters are of order 1 cm~s$^{-1}$~yr$^{-1}$. For example, the barycenter acceleration due to its orbit within the Galaxy is $\sim$0.7~cm~s$^{-1}$~yr$^{-1}$ \citep[e.g.,][]{titov2011,charlot2020,klioner2021}, which is larger than the peak cosmological acceleration.
The $\dot{z}$ signal, however, has a well-defined sign at low and high redshifts (away from the null value), while gravitational accelerations will be randomly distributed and null-centered. The net effect of peculiar accelerations will therefore be added noise, which may drive up sample sizes, integration times, and program duration. Gravitational accelerations will be largest for 21cm emission and absorption lines.

\hi 21cm absorption lines can be intrinsic to the host of the illumination source or intervening between the illumination and the observer, but are generally going to have column densities associated with damped Ly$\alpha$ systems and therefore associated with galaxies rather than intergalactic clouds. Peculiar accelerations are of larger concern in these systems than in the Ly$\alpha$ forest \citep{cooke2020}, particularly in light of the comparatively smaller number of clouds that will be used for the measurements, except for CHIME (if the expected absorption line population is realized).

\citet{loeb1998} and \citet{liske2008} explored the impact of peculiar acceleration on the Ly$\alpha$ forest and found that it is significantly smaller than the cosmological signal.
\citet{cooke2020} used hydrodynamical
simulations to calculate peculiar
accelerations in the Ly$\alpha$ forest and
in gas in galaxies and founds that 
the Ly$\alpha$ forest peculiar accelerations are much smaller than the 
cosmological signal 
except at the $\dot{z}$ zero-crossing region.  Gas in galaxies, however, 
shows accelerations of the same order
of magnitude up to 2 dex higher than 
the redshift drift, which supports the 
concern about systematic effects in 
21 cm measurements.

\subsubsection{Main results and forecasts}\label{subsubsecRD:forecasts}

\begin{figure}[t]
  \begin{center}
    \includegraphics[scale=0.42,trim=20 -102 0 20,clip=false]{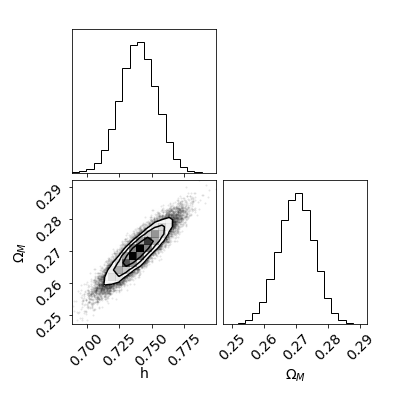}
    \includegraphics[scale=0.42,trim=20 50 0 20,clip=false]{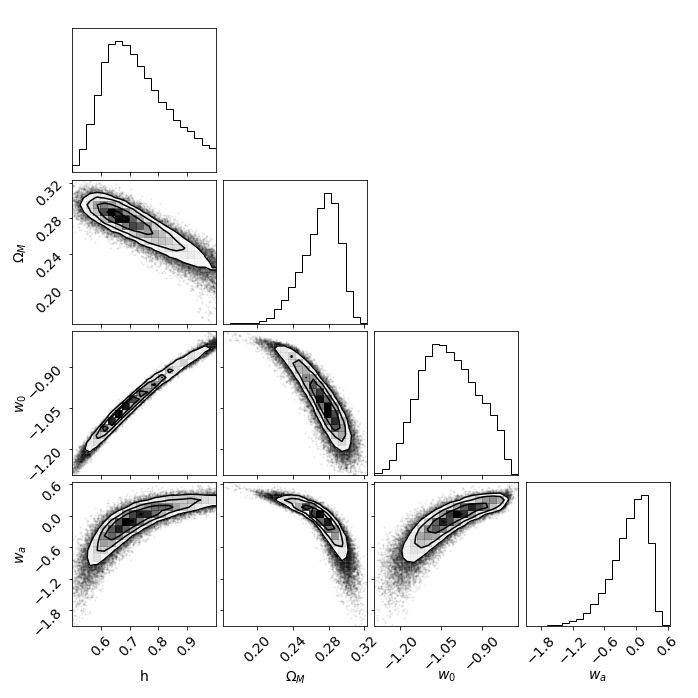}
  \end{center}
  \caption{ Forecast cosmological parameter constraints using the combined secular redshift drift measurements presented in
    Figure \ref{fig:measurements} for a flat $\Lambda$CDM model (left) and a flat CPL model (right).}
   \label{fig:forecasts}
\end{figure}

Secular redshift drift measurements on their own will not compete with other ``precision cosmology'' probes in terms of per cent-level constraints on cosmological parameters. However, the method does offer a model-independent method to directly detect the cosmic acceleration that does not rely on standard candles, standard rulers, or the cosmic distance ladder, and therefore has completely different systematics from canonical cosmological probes.  It is also a powerful probe of isotropy and the general FLRW model \citep{quartin2010}.

Using the combined data and uncertainties for all methods shown in Fig.~\ref{fig:measurements}, we run an MCMC analysis to forecast constraints on the parameters of three different cosmological models:
(1) a flat $\Lambda$CDM model (with two free parameters, \Ho and \omegam),  
(2) a geometrically unconstrained $\Lambda$CDM model (where also \omegal is free to vary), and
(3) a flat $w_0 w_a$CDM model (with four free parameters, namely \Ho, \omegam, $w_0$, and $w_a$).
The fiducial parameter values used for the forecasts are \Ho=74 \Hunit,
\omegam=0.27, \omegal=0.73, $w_0 = -1$, and $w_a = 0$.  
The less constrained models show the largest uncertainties in large part due to strong correlations between parameters.
The best-constrained cosmology is the flat $\Lambda$CDM model, which provides uncertainties on \Ho~and \omegam~of $\pm2$\%. The unconstrained $\Lambda$CDM model has uncertainties in \Ho, \omegam, and \omegal~of $\sim$40\% and are highly degenerate. Finally, the flat $w_0 w_a$CDM model shows a mixed picture with uncertainties of 17\% in  \Ho, 8\% in \omegam, $\pm0.1$ in $w_0$, and $\pm0.3$ in $w_a$, with strong correlation between all parameters.  In analyses comparing various redshift drift measurement methods, \citet{martins2021} and \citet{esteves2021} show that there is no ``best'' method and caution that the choice of measurement should be tailored to specific science goals (e.g., constraining \omegam\ versus the dark energy equation of state).

The correlation between parameters measured by the secular redshift drift suggests that this method would benefit from joint analyses with other
cosmological probes (non-standard and otherwise, see Sect.~\ref{sec:synergy}). For example, \citet{alves2019} combine the expected ELT, SKA 21cm emission, and CHIME measurements to make
individual and joint forecasts for flat $\Lambda$CDM, $w$CDM, and $w_0 w_a$CDM cosmologies, both with and without priors.
When current or future expected priors are included, cosmological parameter constraints of $\sim$1\% can be obtained.
Moreover, \citet{martins2021} show that the redshift drift can break parameter degeneracies in traditional cosmological probes.

The larger impact of the secular redshift drift measurement is its ability to unambiguously and directly identify the influence of dark energy on the Hubble expansion.  This statement applies individually for any of the measurement methods described above, including the Ly$\alpha$ forest technique that would only measure deceleration:  the amplitude of $\dot{z}$ changes dramatically in the absence of dark energy.  {\it Any} method that can measure a non-zero cosmic acceleration can differentiate between cosmologies with and without dark energy (as shown in Figs.~\ref{fig:theory} and \ref{fig:measurements}).

\clearpage

\subsection{Clustering of Standard Candles}
\label{sec:CSC}   

For over a decade after the seminal 1998 papers~\citep{perlmutter1998,perlmutter1999,riess1998} SNe Ia have been one of the most important observables in cosmology. Their prominence as a probe of the background cosmology has more recently been shadowed by the large increase in the available data of both the CMB and of BAO in large-scale structure. There are, however, two reasons why supernovae could return to the forefront of cosmology. First, the Vera Rubin Observatory Legacy Survey of Space and Time \citep[LSST,][]{LSSTScience:2009jmu} should increase the available number of events by at least two orders of magnitude. Second, supernovae are also able to probe cosmology beyond the background level. 

There have been two approaches to extract information on linear perturbation parameters from supernovae. First, they can be used as probes of gravitational lensing. They can in fact be used both in the weak \citep{Quartin:2013moa,Castro:2014oja,Scovacricchi:2016ylt,Macaulay:2016uwy,DES:2020kbf} and strong lensing regimes \citep{Zumalacarregui:2017qqd,Grillo:2018ume,Grillo:2020yvj}. The main observable is the induced change in their scatter at a given redshift. The second approach is to measure the correlations between supernova magnitudes induced by the peculiar velocity field. This field can be computed to good precision in linear perturbation theory and is correlated to the density contrast \citep{hui2006}. Measurements of these correlations have been more recently explored in detail in a number of papers \citep{Castro:2015rrx,Howlett:2017asw,Garcia:2019ita,Amendola:2019lvy,Graziani:2020kkr}. Interestingly, such correlations can also be probed with good precision by upcoming standard siren data (see also Section~\ref{sec:ss}), as discussed in~\cite{Palmese:2020kxn,Alfradique:2022tox}.

Here, we review this latter approach and the forecasts performed for future survey. The advantages of peculiar velocity measurements is that they are well described by linear perturbation theory and  both velocity and density tracers have different degeneracies with the linear bias, making them very complementary. 

\subsubsection{Basic idea and equations}

The first measurement of peculiar velocity correlation in real supernova data was carried out by~\cite{Gordon:2007zw} using 271 SNe and the MLCS2k2 light-curve fitting method, reaching a $3.6\sigma$ detection. \citet{Castro:2015rrx} proposed a more thorough methodology to extract peculiar velocity information from supernova data. Combining the SN velocity and SN lensing observables in the JLA supernova catalog \citep{Betoule:2014frx}, a joint measurement of $\sigma_8$ and the growth rate index $\gamma$ was obtained from SN data alone. This included marginalization over 8 nuisance parameters for both light-curve fitting (using SALT2), lensing and peculiar velocities and other 4 cosmological parameters. It was also shown by \citet{Castro:2015rrx} that SN lensing and velocities constraints were very complementary, with degeneracy directions differing by $60^\circ$ in the $\sigma_8,\,\gamma$ plane. It was shown that both SN lensing and velocities were also very complementary to the CMB growth of structure constraints. 

A measurement of $f \sigma_8$ at low redshifts, where the dependence on cosmology is weak, was also obtained with SN velocities by~\cite{Huterer:2016uyq,Boruah:2019icj}. The former used the Supercal SN catalog and 6dFGS data; the latter used the A2 (Second Amendment) SN catalog combined with 2MTF and SFI++ data, and also included velocity estimates based on the Tully-Fisher method. \cite{Qin:2019axr} combined the density and velocity measurements (using the Fundamental Plane relation instead of supernovae) and discussed how to recover the momentum power spectrum (see below for a discussion on the momentum).

A summary of current constraints are listed in Table~\ref{tab:CSC-current}. Note that none of these current measurements employ the full Clustering of Standard Candles method, as described below. In particular, only \cite{Qin:2019axr} combined velocity and density power spectrum measurements, and in that case the cross-spectrum was not analyzed and velocities were not estimated using standard candles.

\renewcommand{\arraystretch}{1.4}
\begin{table}
    \centering
    \setlength{\tabcolsep}{4.1pt}
    \begin{tabular}{|c c c c|}
    \hline
    data & remarks &  measurement  & reference  \\
    \hline
    271 SN (custom cat.) & no $P_{\delta\delta}$, $P_{\delta v}$  & $\sigma_8 = 0.79\pm0.22$  & \citet{Gordon:2007zw} \\       
    JLA SN & includes SN lensing, no $P_{\delta\delta}$, $P_{\delta v}$  & \makecell*{$\sigma_8 = 0.65^{+0.23}_{-0.37}$, $\gamma = 1.38^{+1.7}_{-0.65}$ \\ $\sigma_8 = 0.40^{+0.21}_{-0.23}$, $\gamma$ fixed}  & \citet{Castro:2015rrx} \\
    Supercal SN + 6dFGS & no $P_{\delta\delta}$, $P_{\delta v}$  & $f\sigma_8 (z=0.02) = 0.428^{+0.048}_{-0.045}$  & \citet{Huterer:2016uyq} \\    
    2MTF, 6dFGSv & $v$ from FP, no $P_{\delta v}$  & $f\sigma_8 (z=0.03) = 0.404^{+0.082}_{-0.081}$  & \cite{Qin:2019axr} \\
    A2 SN, 2MTF, SFI++ & $v$ from SN and TF, no $P_{\delta\delta}$, $P_{\delta v}$  & $f\sigma_8 (z=0.028) = 0.400\pm0.017$  & \cite{Boruah:2019icj} \\    
    \hline
    \end{tabular}
    \caption{\label{tab:CSC-current} Current measurements using techniques similar to the Clustering of Standard Candles but which do not employ the full methodology. FP and TF stands for the fundamental plane and Tully-Fisher methods.} 
\end{table}
\renewcommand{\arraystretch}{1.0}

Measurements of the velocity power spectrum (which can be more precisely measured with standard candles) can be combined to great gain with measurements of the density power spectrum and the density-velocity cross-spectrum. This was first proposed by \citet{Howlett:2017asw} (henceforth H17), which performed Fisher Matrix forecasts for measuring $f\sigma_8$ combining density and velocity spectra. The former measured with galaxies, the latter with SN. Similar forecasts were also performed by \cite{Palmese:2020kxn,Alfradique:2022tox} combining future standard siren and galaxy survey data. 

\begin{figure}[t!]
    \centering
    \includegraphics[width=.6\columnwidth]{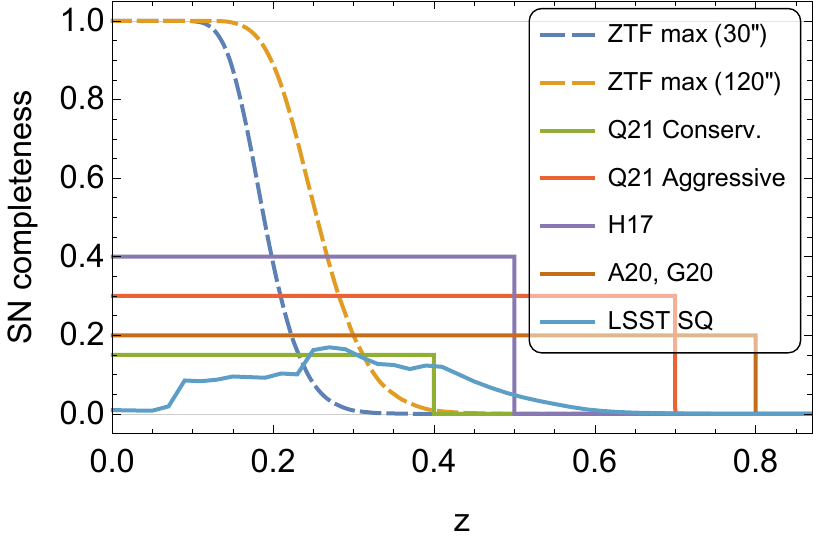}
    \caption{\label{fig:completeness}  
    Comparison of assumed SN completeness in forecasts. In dashed lines represent the maximum ZTF theoretical completeness using the limiting magnitude in the deepest filter for the standard 30' exposure time and also for a possible 120' exposure. The horizontal solid lines represent assumptions made different works. For Rubin we also show results from the survey strategy as of 2019, which was obtained after applying the proposed photometric quality cuts for a 5-year survey (LSST SQ, for Status Quo), which greatly reduces the completeness. 
    }
\end{figure}

The above promising results prompted a study of the capabilities of Rubin to perform measurements of the velocity power spectrum with SNe. \citet{Garcia:2019ita} investigated the constraints that could be achieved with Rubin using the official survey strategy under investigation by the collaboration at the time. As was known, that strategy was not optimal for SN science, and the inferred SN completeness using the SNANA code \citep{Kessler:2009ys} and the proposed quality cuts was very low both at low $z$ and for $z>0.5$. Figure~\ref{fig:completeness} illustrates this result (dubbed LSST Status Quo, or LSST SQ in short), as well as the assumed completeness by a few other recent works. Nevertheless, even without further refinements, this was already enough to achieve very interesting velocity measurements with Rubin. It was also shown in that paper that the velocity constraints in the $\sigma_8,\, \gamma$ plane exhibit moderate non-Gaussianity (they are banana-shaped, instead of ellipsoidal), and thus the Fisher Matrix forecasts on the errors were not very accurate. Since the combination of velocity and density spectra makes for much tighter constraints than using velocity alone, the Fisher Matrix results for the combined cases is expected be a good approximation of the full likelihood results.

\citet{Garcia:2019ita} also investigated how to improve the observing strategy, and found that the same observing time provides the similar cosmological information whether one observes a larger area, or a smaller area during more years. In fact, it was shown that even with optimistic Rubin SN numbers, the SN velocity spectrum is still observed far from the Cosmic Variance regime, and for a broad range of SN number densities $n_{\rm s}$ the uncertainties still scale as $n_{\rm s}^{-1/2}$, which is the same power with which uncertainties generally scale with the survey area. This means that in terms of SN clustering, the most important feature is to have a high cadence in order to achieve higher SN completeness. \cite{LSSTDarkEnergyScience:2021ryz} recently revisited the impact of different survey strategies on SN velocity measurements.

Figure~\ref{fig:completeness} also illustrates that the Zwicky Transient Facility (ZTF) would in principle be capable of observing a catalog of SN with high completeness for $z<0.3$. In fact, recently a first measurement of the clustering of both core collapse and type Ia SN was performed by ZTF \citep{Tsaprazi2021}.

The combined measurements of velocity and density spectra were also studied in \cite{Amendola:2019lvy} (henceforth A21) where a model-independent methodology was proposed to extract competitive constraints in $E(z)$ without almost any assumption regarding the cosmological model in any stage of the analysis. It was shown, using SN only both as density and velocity tracers, that it was possible to achieve 5–13\% (9–40\%) measurements in redshift bins of $\Delta z = 0.1$ up to at least $z = 0.6$. These results included marginalization over a large number of bias parameters, which were allowed to vary freely in both $z$ and $k$. It was also discussed that using SN one cannot however measure \Ho~with this method. Moreover the constrains on $E(z)$ blow up in the limit $z \rightarrow 0$.

\cite{Quartin:2021dmr} (henceforth Q21) recently proposed further to analyze galaxy and supernova data in a more exhaustive way by using SN both as density and velocity tracers. This combines the complementarity of the velocity measurements with the benefits of a multi-tracer analysis~\citep[see, e.g.][]{Seljak:2009,McDonald:2009,Abramo:2012,Abramo:2013}. Here, instead of different galaxy populations, the multiple tracers are galaxies and supernovae. This leads to 6 different power spectra (3 auto and 3 cross spectra), and thus it was dubbed the $6\times2$pt method. This was shown to increase the precision with respect to the $3\times2$pt methods studied in both H17 and~\cite{Amendola:2019lvy}, at no cost in terms of extra data being needed. This extra precision was achieved not only in the cosmological parameters, but also in the bias parameters, making this approach more robust to uncertainties in the galaxy bias.

One should note that in general velocity tracers inhabit galaxies. This means that we can only observe the velocity fields where there are galaxies. This means that we observe a mass-weighted velocity field, also referred to as the momentum field~(\cite{Howlett:2019bky}): $\boldsymbol{p}(\mathbf{r}) \equiv \boldsymbol{v}(\mathbf{r}) (1+\delta_g(\mathbf{r}))$. At larger scales both momentum and velocity field coincide, but already at scales of $\sim$0.1 $h$/Mpc the former picks up non-linearities arising from quadratic terms. Nevertheless, these can be modeled using perturbation theory in a straightforward manner, so we will neglect them here for simplicity.

Let us denote with $\delta_m$ the density contrast of matter, and with $\delta_T = b_T \delta_m$ the density contrast of a tracer field of sources (subscript $T$) that are or can be standardized, e.g. SNe Ia, where $b_T$ is the bias, in general dependent in an unknown way on space and time. In the linear regime and in Fourier space, we know that, due to the continuity equation, the peculiar velocity field $v$ and the matter density contrast of a tracer field  are related by:
\begin{equation}
    \boldsymbol{v}_T = i H\beta \frac{\boldsymbol{k}}{k^{2}(1+z)}\,\delta_{T} \;\; ,   
\end{equation}
where $\beta=f/b_T$, and $f=d\log\delta_m/d\log a$ being the growth rate. The only component of the velocity field that is observable is, however,  the longitudinal velocity $v_{T\parallel}=\boldsymbol{v}\cdot\boldsymbol{r}/r$ \citep[although see][]{Hotinli:2018yyc}, so the relation becomes:
\begin{equation}
    v_{\parallel}=i\frac{H}{k(1+z)}\beta\frac{\boldsymbol{k}\cdot\boldsymbol{r}}{kr}\delta_{T}
    =i\frac{H\mu}{k(1+z)}\beta\delta_{T} \;\; ,
    \label{eq:cont}
\end{equation}
where $\mu=\cos\theta_{\boldsymbol{k},\boldsymbol{r}}$ is the angle between $\boldsymbol{k}$ and the line of sight $\boldsymbol{r}$. From this expression we see that, if we can measure both $\delta_T$ and $v_{\parallel}$, we have access to the combination of $H\beta$, assuming that we also know $k,\mu$ (and of course the redshift $z$). So in order to measure $H(z)$ we need to measure $\beta$: this can be estimated through the redshift distortion of the galaxy power spectrum.
However, we also need to be able to convert the raw data of redshift and angles into $k$ and $\mu$. To solve this problem, we will make use of the fact that $k,\mu$ depend on the observables (redshift and angles) through the angular diameter distance $D_{\rm A}$ and through $H(z)$ itself. We assume the Etherington relation between the luminosity distance $D_{\rm L}$ and the angular diameter distances is valid, so that $D_{\rm L}=D_{\rm A}(1+z)^2$. We also assume that $D_{\rm L}$ is measured directly from the standard candles, while \Ho~is given by local measurements, so that we know the combination $H_0 D_{\rm L}$. Although we could include the error on the estimation of $H_0 D_{\rm L}$ in our formalism, we will show at the end that it is way below the other uncertainties, so we may neglect it.

A peculiar velocity $v$ (in units of $c$) induces a change in the luminosity distance $D_{\rm L}$  given by~\citep{2006PhRvD..73l3526H}:
\begin{equation}
    \frac{\delta {D_{\rm L}}}{D_{\rm L}}=v\left[2-\frac{d\log D_{\rm L}}{d\log(1+z)}\right] \;\; .
\end{equation}
Since $m=M+25+5\log D_{\rm L}$, a small change in $D_{\rm L}$ induces a change $\delta m$ in the apparent magnitude given by:
\begin{equation}
    \frac{\delta  D_{\rm L}}{D_{\rm L}}=\frac{\log10}{5}\delta m \;\;,
\end{equation}
so that finally the radial peculiar velocity of a standard candle is obtained from the scatter $\delta m$ of its apparent magnitude as:
\begin{equation}
    v=\frac{\log 10}{5}  \delta m \left[2-\frac{d\log D_{\rm L}}{d\log(1+z)}\right]^{-1} \;\;. 
    \label{eq:magn-vel}
\end{equation}

Let us now consider three Gaussian fields in Fourier space with zero mean: the density contrast $\delta_{\rm s}$ of the standard candles, their peculiar velocity field $v_{\rm s}$, and the galaxy density contrast $\delta_{\rm g}$. A fraction of the supernovae could be hosted by one of the galaxies in the sample, but we expect this fraction to be small. We consider the same growth rate index $f$ for every tracer field, which equates to assuming universal gravity. We also introduce the linear bias for each species, $b_{\rm g,s}=\delta_{\rm g,s}/\delta_{\rm tot}$, where $\delta_{\rm tot}$ is the underlying total matter density contrast. The functions $b_{\rm g,s}$ are in general arbitrary functions of space and time. Following \cite{Quartin:2021dmr} we can write the six observed power spectra as:
\begin{align}
     P_{\rm gg}(k,\mu,z) &= \Upsilon \big[1+ \beta_{\rm g} \mu^{2}\big]^2 \,b_{\rm g}^{2} \,S_{\rm g}^2\, D_+^2 P_{\textrm{mm}}(k) + \frac{1}{n_{\rm g}} \;\; , \label{eq:pgg} \\
     P_{\rm ss}(k,\mu,z) &= \Upsilon \big[1+ \beta_{\rm s} \mu^{2}\big]^2 \,b_{\rm s}^{2}\,S_{\rm s}^2 \, D_+^2 P_{\textrm{mm}}(k) + \frac{1}{n_{\rm s}} \;\; , \label{eq:pss} \\
     P_{\rm gs}(k,\mu,z) &= \Upsilon \big[1+ \beta_{\rm g} \mu^{2}\big]\big[1+ \beta_{\rm s} \mu^{2}\big] \,b_{\rm g} \,b_{\rm s}\,S_{\rm g} \,S_{\rm s} \, D_+^2 P_{\textrm{mm}}(k) + \frac{n_{\rm gs}}{n_{\rm g}n_{\rm s}} \;\; , \label{eq:pgs} \\
     P_{\rm gv}(k,\mu,z) &= \Upsilon \frac{H\mu}{k(1+z)} \!\big[1 + \beta_{\rm g}\mu^{2}\big] b_{\rm g}\, S_{\rm g}\, S_{\rm v}  \,f  D_+^2 P_{\textrm{mm}}(k) \;\; , \label{eq:pgv} \\
     P_{\rm sv}(k,\mu,z) &= \Upsilon \frac{H\mu}{k(1+z)} \!\big[1 + \beta_{\rm s}\mu^{2}\big] b_{\rm s}\, S_{\rm s}\, S_{\rm v} \, f  D_+^2 P_{\textrm{mm}}(k) \;\; , \label{eq:psv} \\
     P_{\rm vv}(k,\mu, z) &= \Upsilon \left[\frac{H\mu}{k(1+z)}\right]^2 S_{\rm v}^2 \,f^{2} \,D_+^2 P_{\textrm{mm}}(k)+ \frac{\sigma^2_{v, {\rm eff}}}{n_{\rm s}} \;\; ,
    \label{eq:pvv}
\end{align}
where $\beta_i \equiv f/b_i$,  $\mu \equiv \hat{k} \cdot \hat{r}$, $S_{\rm g,v,s}$ are damping terms  and $P_{\textrm{mm}}$ is the matter power spectrum at $z=0$.  

All observed spectra are multiplied by a volume-correcting factor $\Upsilon$ \citep{1996MNRAS.282..877B,Seo:2003pu}, 
where:
\begin{equation}
     \Upsilon = \frac{H D_{L,r}^2}{H_{r}D_{\rm L}^2} \;\;,
\end{equation}
because we need first to choose a reference cosmology, e.g. $\Lambda$CDM (subscript $r$), and then correct for any other cosmology. For the same reason, the AP effect \citep{Alcock:1979mp}, which introduces corrections to $k$ and $\mu$ that depend on $H,D$ \citep[see, e.g.,][]{2000ApJ...528...30M,Amendola:2004be}, has also been taken into account by replacing all $k,\mu$'s in the rhs of Eqs. \eqref{eq:pgg}-\eqref{eq:pvv} with the AP-corrected $k,\mu$'s.

The non-linear smoothing factors $S_{v,g,s}$ (important only at small scales) can be taken following \cite{2014MNRAS.445.4267K,Howlett:2017asw} to be:
\begin{equation}\label{eq:nonlinear-smooth}
    S_{\rm v,g,s}=\exp\left[-\frac{1}{4}(k\mu\sigma_{\rm v,g,s})^{2}\right] \;\; .
\end{equation}
In this expression, $\sigma_{v,g,s}$ are assumed to be independent of redshift. Q21 set as fiducial values $\sigma_{\rm g}=\sigma_{\rm s} = 4.24 \;{\rm Mpc}/h$ and $\sigma_{\rm v}= 8.5$ Mpc$/h$. H17 used very similar values. These fiducial values nevertheless have little impact in the forecasts. Finally, the noise term in the velocity correlation is given by~\citep{2006PhRvD..73l3526H,Davis:2010jq}:
\begin{equation}
    \sigma_{v,{\rm eff}}^{2}\equiv\left[\frac{\log10}{5}\sigma_{\rm int}\right]^2\!\left[2-\frac{d\log D_{\rm L}}{d\log(1+z)}\right]^{-2}\!\!\!+\frac{\sigma_{v{\rm ,nonlin}}^{2}}{c^{2}} \;\; ,
    \label{eq:new-sv-1}
\end{equation}
where $\sigma_{\rm int}$ is the intrinsic variance of the source's magnitude.

The $6\times2$pt results in a $3\times3$ matrix of correlation:
   \begin{equation}
    \mathbf{C} = \left(\begin{array}{ccc}
        \!P_{\rm gg} &  \!P_{\rm gs}  & \!P_{\rm gv} \\
        \!P_{\rm gs} &  \!P_{\rm ss}  & \!P_{\rm sv} \\
        \!P_{\rm gv} &  \!P_{\rm sv}  & \!P_{\rm vv}
    \end{array}\right) \;\; .
\end{equation}
The probability distribution of our random variables, i.e. $x_{a}=\sqrt{V}\{\delta_g,\delta_s,v_s\}$, is assumed Gaussian with zero mean and covariance matrix given by $\mathbf{C}$. The Fisher matrix associated to the unknown parameters is thus \citep{Abramo:2019ejj}:
\begin{equation}
    F_{\alpha\beta} \,=\, VV_{k}\bar{F}_{\alpha\beta} \;\; ,
\end{equation}
where $V_{k}=(2\pi)^{-3}2\pi k^{2}\Delta_{k}$ is a volume element in Fourier space and $\bar{F}_{\alpha\beta}$ is:
\begin{equation}
    \bar{F}_{\alpha\beta}=\frac{1}{2}\int_{-1}^{+1}d\mu\frac{\partial C_{ab}}{\partial\theta_{\alpha}}C_{ad}^{-1}\frac{\partial C_{cd}}{\partial\theta_{\beta}}C_{bc}^{-1} \;\; ,
\end{equation}
where the integrand is evaluated at the fiducial value and $\theta_\alpha$ are the cosmological parameters we want to estimate. For a $z$-shell of volume $V(z)$ and for $\Delta_{k}\approx2\pi/V^{1/3}$, we have:
\begin{equation}
   VV_{k} = \frac{k^{2}V^{2/3}}{2\pi} \;\; .
\end{equation}
The $k$-cells were chosen in A21 and Q21 with equal $\Delta_{k}=2\pi/V(z)^{1/3}$ between $k_{\rm min}(z)$ and $k_{\rm max}$, and $k_{\rm min}=2\pi/V(z)^{1/3}$ following \citep{Garcia:2019ita}. A21 and Q21 assumed $k_{\rm max} = 0.1~h/$Mpc, whereas H17 assumed $k_{\rm max} = 0.2~h/$Mpc (see Table~\ref{tab:sn-surveys}). As discussed in A21, the latter value is responsible for substantial increases in precision.

\begin{table}
    \centering
    \setlength{\tabcolsep}{2.2pt}
    \begin{tabular}{|c|cccc|cccc|cccc|}
    \hline 
    & \multicolumn{4}{c}{{Q21 Conservative}} & \multicolumn{4}{|c}{\rm{Q21 Aggressive}} & \multicolumn{4}{|c|}{\rm{H17 All Rubin SNe}} \\
    $z_{\rm bin}$ & V & $10^{3}\cdot n_{\rm s}$ &  $b_{\rm g}$ & $k_{\rm max}$ & V & $10^{3}\cdot n_{\rm s}$ & $b_{\rm g}$ & $k_{\rm max}$ & V & $10^{3}\cdot n_{\rm s}$ & $b_{\rm g}$ & $k_{\rm max}$ \\
    & $[\text{Gpc}/h]^{3}$ & $[h/\text{Mpc}]^3$ &  & $h$/Mpc & $[\text{Gpc}/h]^{3}$ & $[h/\text{Mpc}]^3$ & & $h$/Mpc & $[\text{Gpc}/h]^{3}$ & $[h/\text{Mpc}]^3$ & & $h$/Mpc \\
    \hline
    0.05 & 0.019 & 0.048 & 1.38 & 0.1 & 0.046 & 0.096 & 1.38 & 0.1 & 0.046 & 0.143 & 1.38 & 0.2 \\  
    0.15 & 0.123 & 0.052 & 1.45 & 0.1 & 0.296 & 0.105 & 1.45 & 0.1 & 0.296 & 0.157 & 1.45 & 0.2 \\  
    0.25 & 0.303 & 0.057 & 1.53 & 0.1 & 0.727 & 0.114 & 1.53 & 0.1 & 0.727 & 0.172 & 1.53 & 0.2 \\  
    0.35 & 0.531 & 0.061 & 2.04 & 0.1 & 1.27  & 0.122 & 2.04 & 0.1 & 1.27  & 0.186 & 1.61 & 0.2 \\  
    0.45 &  --   &  --   &  --  & --  & 1.88  & 0.131 & 2.15 & 0.1 & 1.88  & 0.200 & 1.69 & 0.2 \\  
    0.55 &  --   &  --   &  --  & --  & 2.51  & 0.139 & 2.26 & 0.1 &  --   &  --   &  --  & --  \\  
    0.65 &  --   &  --   &  --  & --  & 3.13  & 0.148 & 2.37 & 0.1 &  --   &  --   &  --  & --  \\  
    \hline
    \end{tabular}
    \caption{\label{tab:sn-surveys}  Survey specifications for the proposed forecast scenarios in H17 and Q21. The assumed SN completeness in each case is represented in Figure~\ref{fig:completeness}. The supernova bias is assumed to be $1.0/D_+(z)$; the galaxy bias is assumed to be $1.34/D_+(z)$ for $z\le0.3$ (mostly BGs), $1.7/D_+(z)$ for $z>0.3$ (mostly LRGs).  The $z$ bins have $\Delta z=0.1$ and are centered on $z_{\rm bin}$. 
    }
\end{table}

\subsubsection{Measurements and sample selection}

The equations above and the results below assume spectroscopic measurements of both galaxies and supernovae. If one has to rely on photometric data only, the corresponding photo-$z$ errors will degrade the clustering measurements along the line-of-sight, resulting in larger effective non-linear smoothing factors $S_{v,g,s}$. For supernovae, the absence of spectroscopic follow-ups will result in contamination from core collapse supernovae, which could be a source of bias as discussed below.

The need for galaxy spectra does not substantially decrease the final precision of the method as due to cosmic variance the information saturates for relatively low number densities, which should be reached with surveys like DESI \citep{DESI:2016} and 4MOST \citep{Dejong:2019}. For instance, in the Q21 Conservative case, only half a million galaxies with spectra would be required. This will only pose a real challenge in the cases where one tries to push to higher redshifts ($z \gtrsim 0.5$), since the absolute number of galaxies needed to get close to the cosmic-variance limit in each redshift bin increases roughly with $z^2$.

\subsubsection{Systematic effects}

The sources of systematic effects in the $6 \times 2$pt method are the same as for any supernovae and large scale structure survey. Here we limit ourselves to the list of the most important ones, referring  to the literature for details.

On the supernovae side, one has to expect various sources of systematic uncertainties. For instance, one can incorrectly classify core collapse supernovae or other transient phenomena as SNe Ia. This is specially problematic if SNe lack spectra, although there is an on-going effort to improve photometric classification techniques \citep[see, e.g.,][]{Lochner:2016hbn,COIN:2018niv,Villar:2020epn}. Without further improvements in photometric classification, extra dispersion would need to be included in the SN distances to avoid biases, which \cite{VargasdosSantos:2019ovq} showed that could lead to an effective reduction on the number of SN by up to two thirds.

Secondly, the standardization of SNe Ia might be more complicate of what usually assumed, with dependencies on environment, host mass, metallicities, etc., that are still not perfectly accounted and corrected for. Gravitational lensing of the sources is another possible form of bias, although the overall effect is expected to be negligible. The smoothing factors $S_{g,s,v}$ that we introduced in the previous section might also deviate from the simple parameterization we adopted, perhaps with a redshift dependence. If the SNe Ia redshifts are evaluated through photometric methods, there are of course additional sources of uncertainties, which could however be modeled by larger, and redshift dependent, smoothing factors.

On the large-scale structure side, one should of course consider carefully other effects. First, finite surveys induce window-function distortions on the power spectrum shape that have to be taken into account, although on the forthcoming large surveys this problem is probably under control. Secondly, the redshift bins cannot really be taken as independent, and some correction is also expected \citep[see, e.g.,][]{2017MNRAS.470..688B}. Moreover, magnitude lensing is also affecting the clustering \citep[see, e.g.,][]{2016PhRvD..94d3007C}. 

Perhaps the most problematic systematics is however the impact of non-linearities. The assumption of linearity enters in fact our calculation in several ways: in the $P(k)$ shape, in the Kaiser redshift correction, in the velocity-density contrast relation, and in the overall Gaussian assumption. The non-linearity is actually in principle accounted for by the smoothing factors $S_{\rm v,g,s}$, but of course these functions are calibrated only through $\Lambda$CDM simulations and might differ sensibly in alternative cosmologies. 
Already at $k=0.1\,h$/Mpc one-loop corrections become relevant, specially when one allows conservative priors for all nuisance parameters involved, as discussed in~\cite{Amendola:2022vte}. Nevertheless, inclusion of all these parameters may allow the extension to higher values of $k_{\rm max}$, especially at higher redshifts.  For instance for
the analysis of BOSS data~\cite{Chudaykin:2020ghx} and \cite{Ivanov:2019pdj} employed $k_{\rm max} = 0.20\,h$/Mpc and $k_{\rm max} = 0.25\,h$/Mpc, respectively. It remains to be investigated how to best generalize the clustering of standard candles to include one loop corrections and to which scales it can be relied upon.

\begin{figure}
    \centering
    \includegraphics[width=.44\columnwidth]{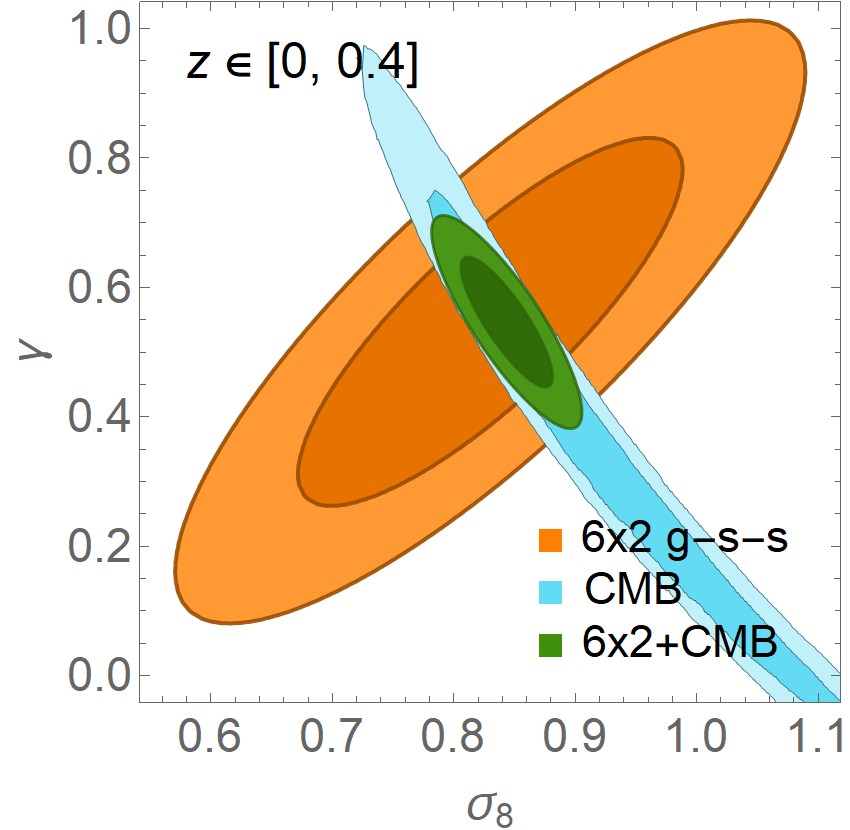}
    \includegraphics[width=.44\columnwidth]{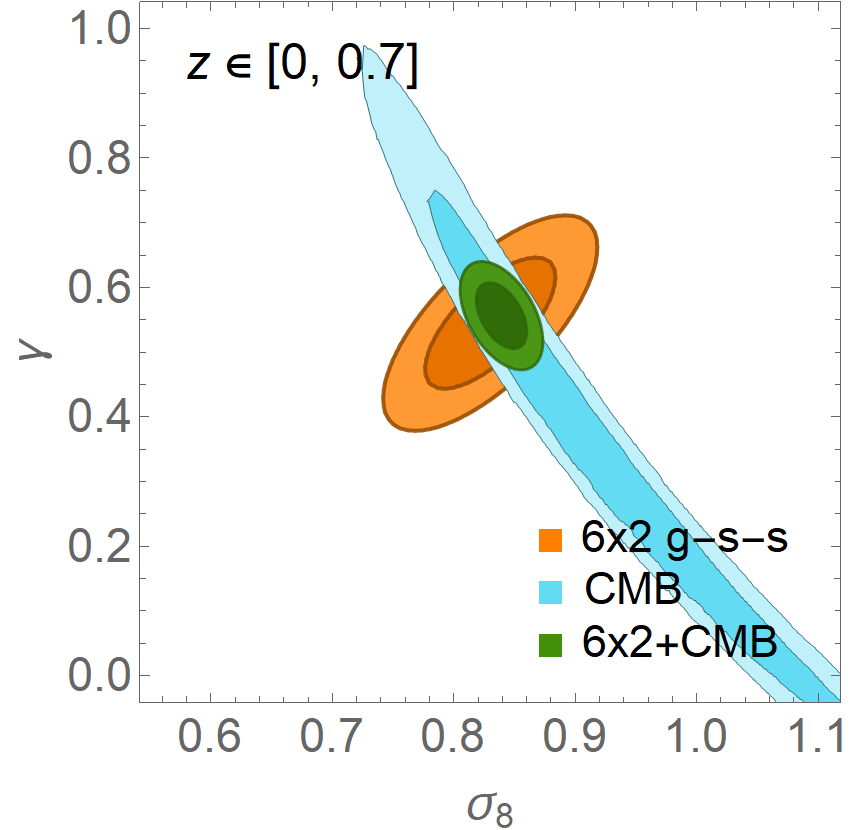}
    \caption{\label{fig:6x2-2D-plot} $1$ and $2\sigma$ marginalized forecasts in $\{\sigma_8,\, \gamma\}$ in from the $6\times2$pt method for the Q21 Conservative (left) and Q21 Aggressive forecasts (right panel). Also shown are the CMB-only and joint constraints. As can be seen, the $6\times2$pt and CMB constraints are very complementary. Image adapted with permission from \cite{Quartin:2021dmr}.}
\end{figure}

\subsubsection{Main results and forecasts}

The results obtained in H17 for the $3\times2$pt case using galaxies and supernovae are summarized in Tab.~\ref{tab:H17fsigma8} for the case dubbed All Rubin SNe, which is described in detail in Tab.~\ref{tab:sn-surveys}. Here we just recast the H17 results in wider redshift bins. As can be seen, constraints in $f\sigma_8$ between $3$ and $10\%$ can be achieved in that case.

Forecasts of the $6\times2$pt were performed in Q21 allowing a cosmological model with 5 parameters: $\{\sigma_8$, $\gamma$, \omegam, $\Omega_{k0}$, $h\}$ and making use of 3 global nuisance parameters describing the non-linear smoothing factors in Eq.~\eqref{eq:nonlinear-smooth} and allowing each bias parameter to be free in each redshift bin. The final, marginalized, constraints in each parameter is given in Tab.~\ref{tab:6x2-errors}, and the 2-D contours in $\{\sigma_8, \,\gamma\}$ are depicted in Fig.~\ref{fig:6x2-2D-plot}. As discussed in Q21, neglecting the AP corrections or assuming flatness has little impact on the $\sigma_8$ and $\gamma$ constraints (the other parameters are affected to a higher degree). This figure also illustrates the CMB contours, which were extracted from \cite{Mantz:2014paa}. We point the reader to Q21 for more details.

The $6\times2$pt method can also be combined with the traditional Hubble diagram distance measurements with standard candles. This synergy has been investigated by~\cite{Alfradique:2022tox}, where it was shown that although the improvements to $\{\sigma_8, \,\gamma\}$ are negligible, this combination yields large gains for $\Omega_{k0}$, and should be able to constrain it to less than 2\% using either Rubin SN or third generation standard siren measurements. In the latter case, $h$ could also be measured with over an order magnitude increased precision.

\begin{table}
    \centering
    \setlength{\tabcolsep}{4.1pt}
    \begin{tabular}{|c c c c c c|}
    \hline
    $z_{\rm bin}$     & 0.05 & 0.15 & 0.25 & 0.35 & 0.45   \\
    H17 All Rubin SNe, $\sigma(\ln f\sigma_8)$   & 0.10  & 0.058   & 0.041  & 0.033  & 0.028 \\
    \hline
    \end{tabular}
    \caption{\label{tab:H17fsigma8} Relative 1$\sigma$ errors in $f\sigma_8$ using the $3\times2$pt $g$--$s$ method.  Adapted from \cite{Howlett:2017asw}.
    }
\end{table}

\begin{table}
    \centering
    \setlength{\tabcolsep}{4.1pt}
    \begin{tabular}{|c c c c c c c c|}
    \hline
    $1\sigma$ uncertainties in:  & $\sigma_8$ & $\gamma$ & $h$ & \omegam & $\Omega_{k0}$ & $\langle \ln b_g \rangle$ & $\langle \ln b_s \rangle$  \\
    \hline
    Q21 Conservative   & 0.10  & 0.19  & 0.039 & 0.015  & 0.26  & 0.14  & 0.15 \\
    Q21 Aggressive     & 0.036 & 0.067 & 0.013 & 0.0047 & 0.074 & 0.050 & 0.052\\
    \hline
    \end{tabular}
    \caption{\label{tab:6x2-errors} Fully marginalized absolute forecast uncertainties in each cosmological parameter using the $6\times2$pt method.  The (relative) bias uncertainties are the average over all redshift bins, but their redshift dependence is small, only around $\sim$10\%.  Adapted from \cite{Quartin:2021dmr}.
    }
\end{table}

\begin{figure}
    \centering
    \includegraphics[width=.7\columnwidth]{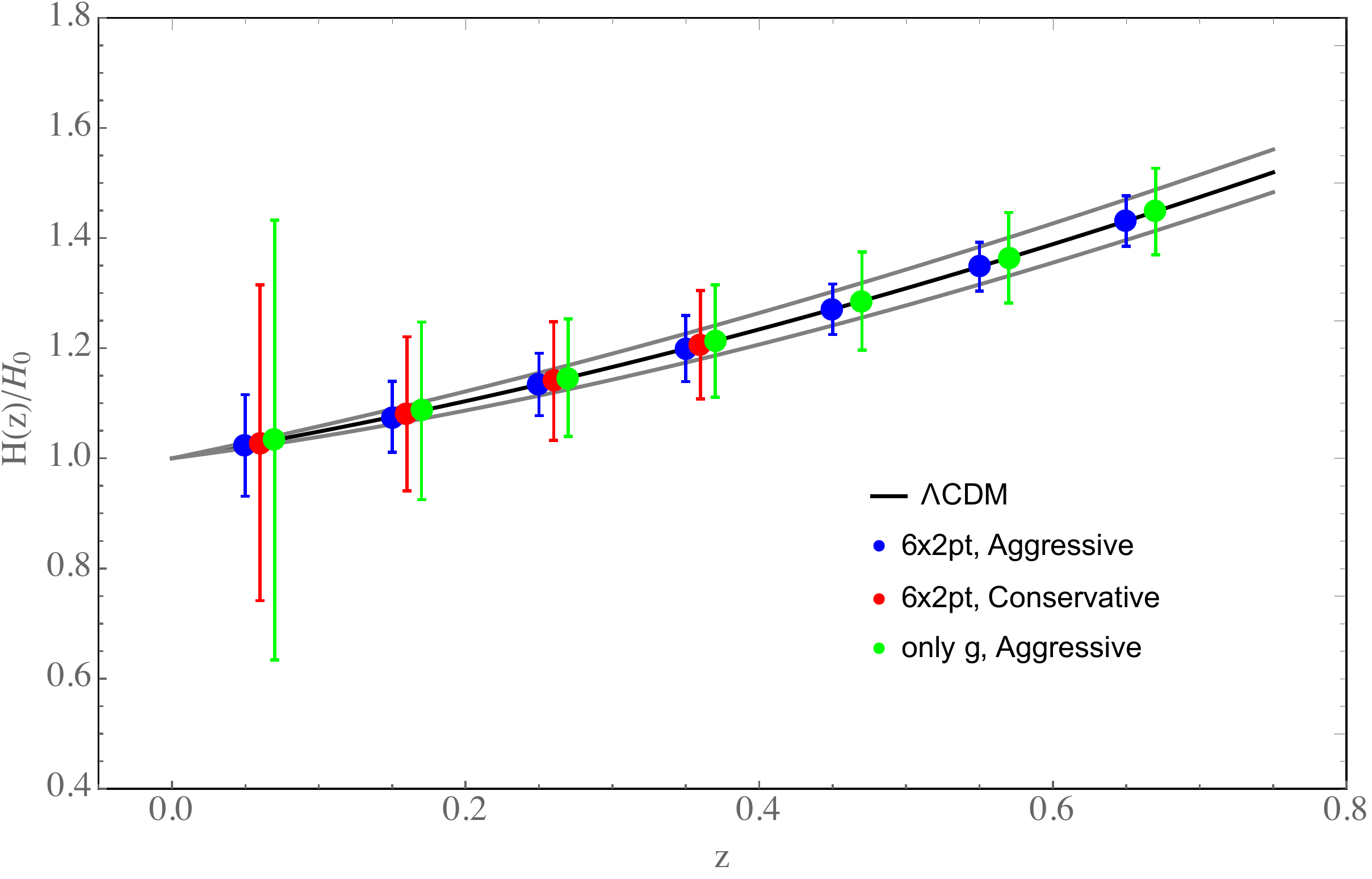}
    \caption{\label{fig:h-plot} Errors in $H$ with the model-independent approach compared to using only galaxy clustering (red and green errorbars are slightly displaced for clarity). The two grey continuous lines represent $H(z)$ for $w=-0.9$ (top) and $w=-1.1$ (bottom), as a convenient graphical reference.}
\end{figure}

Finally, using the methodology discussed in \cite{Amendola:2019lvy}, one can employ the $6\times2$pt method also in another way, namely, to produce forecasts without assuming a parameterization of $H(z),P(k,z)$ and $\beta_{\rm g,s}(k,z)$. This is obtained by employing the data directly in every $k,z$-bin, avoiding therefore the need for assuming a specific cosmological model. Q21 showed that one can obtain uncertainties on $E(z)$ around 3-4\% in the farthest bin of the Aggressive survey, as shown in Fig.~\ref{fig:h-plot}. 

\clearpage

\section{Synergies and complementarities between cosmological probes}
\label{sec:synergy}

In Sect.~\ref{sec:probes}, we extensively discussed all the characteristic and peculiarities of the new emerging cosmological probes, individually.
At the end of this review, it is useful to explore also the improvement that can be achieved from the synergical complementarity of the various probes when, potentially, they are combined together; this will allow us to assess if, and how much, they complete each other, and what we could learn from studying them jointly. 

\begin{figure}[b!]
\centering
\includegraphics[width=\textwidth]{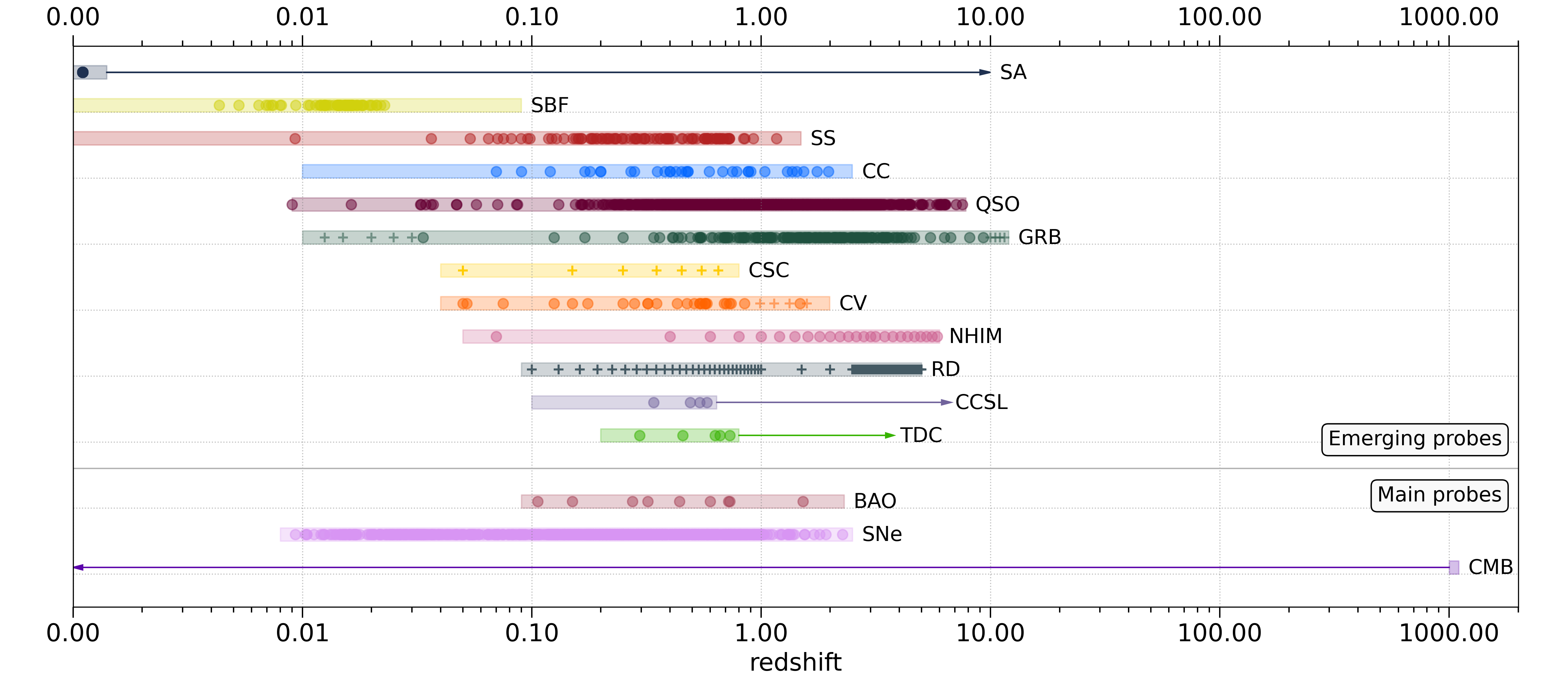}
\caption{Redshift distribution of the various emerging cosmological probes considered in this review. From top to bottom are shown stellar ages (SA), surface brightness fluctuations (SBF), standard sirens (SS), cosmic chronometers (CC), quasars (QSO), gamma-ray burst (GRB), clustering of standard candles (CSC), cosmic voids (CV), neutral hydrogen intensity mapping (NHIM), secular redshift drift (RD), cosmography with cluster strong lensing (CCSL), and time delay cosmography (TDC). The horizontal bands show, for each probe, the expected redshift range considering both current and future measurements, where the circle dots represent current measurements described in the review and cross signs the forecasts. The arrows indicate when a probe carries integrated information from a larger redshift, as in the case of stellar ages, mapping all the expansion history since their formation, or TDC and CCSL, carrying the information not only of the lenses (dotted points) but also of the sources. In the lower part of the figure is shown, for comparison, the redshift distribution of the main cosmological probes, namely baryon acoustic oscillations (BAO), supernovae (SNe), and cosmic microwave background (CMB).}
\label{fig:z_probes}
\end{figure}

The first important point to look at is the redshift range specifically mapped by each probe, as presented in Fig.~\ref{fig:z_probes}. The horizontal bands show the redshift range of the various methods as discussed in the corresponding sections, either currently covered or expected to be covered with future surveys. The dotted points represent current measurements, while the crosses indicate future forecasts. In some cases, a cosmological probe includes an integrated information from a higher redshift, as in the case of TDC and CCSL, being the measurements of sources at a much larger distance than the lenses, or of SA, providing information also of the entire expansion history up to the formation redshift of the star considered; in the plot, we display that information with arrows. The various methods have been ordered from top to bottom as a function of the spanned range.
In the bottom part of the plot are shown, for comparison, the main cosmological probes, namely CMB, BAO and SNe. The first point that it is interesting to notice is how the new emerging cosmological probes richly complement the main probes covering different ranges of cosmic times, from the very local ones ($z<0.1$ for SA, SBF, and SS), and extending to very high redshifts (up to $z\sim 10-12$ for QSO, GRB, NHIM, and RD). They allow us to span almost 13.4 Gyr of cosmic time, a significantly larger range than the one reachable by current probes. It is also relevant how a significant fraction of these methods overlap with the range of BAO and SNe ($0.1<z<2$ for CC, CSC, CV, CCSL, and TDC), providing a crucially wider compilation of late-Universe probes that can result decisive in breaking the dichotomy between late- and early-Universe results, and in validating the results obtained from standard probes.

\renewcommand{\arraystretch}{1.5}
\begin{table}[p]
\centering
\begin{tabular}{|p{0.115\textwidth}|p{0.12\textwidth}|p{0.35\textwidth}|p{0.33\textwidth}|}
\hline \hline
{\bf Probe} & {\bf Measure} & \multicolumn{1}{l|}{{\bf Strength}} & \multicolumn{1}{l|}{{\bf Weakness}} \\
\hline
cosmic chronometers & $H(z)$ & independent of cosmological model assumptions; differential approach mitigates several potential biases & need to be calibrated on SPS \mbox{models}; CC sample has to be accurately selected to minimize contamination 
\\
quasars with $L_{\rm X}-L_{\rm UV}$ & $D_{\rm L}(z)$ & detectable over a wide redshift range; cosmology-independent estimate of $D_{\rm L}$ & large scatter of the Hubble diagram when compared to SNe Ia; small statistics at redshifts $z\geq 4$ 
\\
gamma-ray bursts & $D_{\rm L}(z)$ & detectable over a wide $z$ range; poorly affected by dust or gas absorption; no evidence of evolutionary effects; independent measure of $D_{\rm L}$; correlations can be self-calibrated using expected large datasets in the near future & used correlations cannot be currently calibrated on nearby events (very low number / peculiar events); GRB prompt gamma-ray emission physics still to be fully understood \\
standard sirens/GWs & $D_{\rm L}(z)$, \Ho & no calibration other than GR; independent measurement of \Ho & need deep and complete galaxy catalogs and/or astrophysical calibration of compact binary population for dark sirens, quick search and follow-up for bright sirens
\\
time-delay cosmography & $D_{\Delta t}$, $D_{\rm A}$ & cosmology-independent direct measurement of $D_{\Delta t}$, an absolute angular diameter distance product, to measure $H_0$ & breaking lensing degeneracies requires high-precision ancillary data products (i.e. dynamic measurements)
\\
cosmography with cluster strong lensing & $D_{\Delta t}$, $D_{\rm A}(z_{\rm d},z_{\rm s})$ & one-step measurement, sensitive to both $H_0$ and Universe geometry; less prone to inherent lensing systematics & complexity of lens modeling, time-consuming spectroscopic and monitoring observational campaigns 
\\
cosmic voids & $D_{\rm A}H(z)$, $f\sigma_8(z)$ & cosmology independent; pure \mbox{geometry}; linear dynamics; orthogonal to other probes & need large contiguous survey volumes; accurate tracer redshifts for RSD \& AP test 
\\
HI intensity mapping & $P_{\rm HI}(k,z)$, $P_{\rm HI,g}(k,z)$ & galaxy evolution and cosmology, large volumes, high redshifts  & foreground contamination, RFI, instrumental systematics 
\\
surface brightness fluctuations & \Ho & can be empirically (Cepheids or TRGB) and theoretically (SPS) calibrated; small internal scatter (band-dependent); Hubble flow reached in a single observation (no temporal monitoring required) & can be affected by dust or young stars, if any; practical limit with space-based resolution is around $\sim$200 Mpc; precise colors needed for calibration 
\\
stellar ages & $t_{\rm U}$, \Ho & no cosmological assumption; direct and complementary local probe & need further assessment to reduce systematics involved in stellar ages determination; need larger sample to increase the accuracy
\\
redshift drift & $H(z)$ & cosmology-independent; direct test of dark energy & small signal; very long timeline; stability of the measurement 
\\
clustering of standard candles & $E(z)$, $f\sigma_8(z)$ (or $\sigma_8$, $\gamma$) & data  will be available naturally from SNe and redshift surveys; model-independent measurement of $E(z)$; can be combined with traditional Hubble diagram measurements & SNe systematics may bias distance measurements, which are used by the method; non-linearities in clustering must be kept under control.
\\
\hline \hline
\end{tabular}
\caption{Summary table of the emerging cosmological probes, highlighting for each one what observable they are constraining, comparing their strengths and weaknesses.}
\label{tab:probes}
\end{table}
\renewcommand{\arraystretch}{1.}

\renewcommand{\arraystretch}{1.5}
\begin{longtable}{|p{0.115\textwidth}|p{0.26\textwidth}|p{0.26\textwidth}|p{0.26\textwidth}|}
\caption{Current maturity and constraining power of the emerging cosmological probes, with expected timescales for development.}
\label{tab:probes2}\\
\hline \hline
{\bf Probe} & {\bf Timescales} & {\bf Power} & {\bf Maturity} \\
\hline
\endfirsthead
\multicolumn{4}{c}%
{{\bfseries \tablename\ \thetable{}} -- {\it continued from previous page}} \\
\hline
{\bf Probe} & {\bf Timescales} & {\bf Power} & {\bf Maturity} \\
\hline
\endhead
cosmic chronometers & No dedicated survey available, need to rely on legacy data from ongoing and future planned surveys; current sample of $O(10^6)$ CC can increase by $\sim 10^6$ CC up to $z\sim0.7$ (SDSS-IV), and by $\sim10^3$ up to $z\sim1.5$ (Euclid + ATLAS). & current precision is $\simeq$ 8\% on \Ho\ and 20\% on \omegam\ for a flat $\Lambda$CDM model, increasing to $\simeq$ 4\% on \Ho\ with future analyses (in an open wCDM model). & Relatively young technique, to reach full maturity a more stringent assessment of reliable SPS models is crucial, since this is the main source of systematics effects. \\
quasars with $L_{\rm X}-L_{\rm UV}$  & Increased data from current $\sim 10^3$ to $\sim10^4$ quasars is expected in a few years with ongoing (e.g.~\textit{eROSITA}, 4MOST) and planned surveys (e.g.~Euclid, LSST).  & Current precision is $\simeq5\%$ on \omegam\ (flat $\Lambda$CDM) and $\simeq15\%$ on $w_0$ (flat $w_0w_a$CDM) once combined with Type Ia SNe. & Relatively young technique. Improvement both in the sample selection and data analysis is necessary. \\
gamma-ray bursts & Significant improvement in the size of the sample of GRBs with measured redshift and accurate spectral and timing parameters expected starting from 2023 (Chinese-French SVOM satellite). Further substantial step forward would be granted by next generation GRB missions currently being proposed for the next decade (e.g., THESEUS, Gamow Explorer, Hi-Z Gundam). &  Current precision is  $\simeq 7\%$ on \omegam\ (for a flat $\Lambda$CDM) and $\simeq 13\%$ on $w_0$ (in a flat CPL model); the expected Figure of Merit from future analyses will be $FoM= 1.9$ when using 772 GRBs, increasing by a factor 2.5 when using a sample of 1500 GRBs. & Cosmological exploitation of GRBs well established since more than one decade with already mature methods in sample selection, techniques for measuring cosmological parameters, calibration methods. Substantial progress expected from the continuing theoretical progress on GRB emission processes and the expected increase and improvement of the samples.\\
standard sirens/GWs & Bright sirens: $\sigma H_0\!\sim\!15\%/\sqrt{N}$ with $N$ events, $\sim10$/year well-localized GW events with EM counterparts expected when current detectors reach design sensitivity in $\sim2025$. Dark and spectral sirens: larger errors \& converge more slowly, but use more events and probe $H(z)$ to higher $z$. 
& Current precision is $\sim14\%$ on $H_0$, with weak constraints on $H(z)$ out to $z \sim 1$. Expect to reach $\lesssim2\%$ on $H_0$ within a decade. & Bright siren approach is straightforward, but so far only has one event. Dark and spectral siren approaches are still being developed to gain control over systematic uncertainties, but maturing quickly.  \\
time-delay \mbox{cosmography} & Significant expansion of samples within years expected based on recent quadruply lensed quasar discoveries. Further improvements expected from JWST and ground-based 8-m class kinematics measurements. Hundreds of lensed SNe and lensed quasars time delays with LSST.
& Current precision on $H_0$ is $\sim$2-8\% depending on the assumptions on the mass density profile. Future constraints with increased sample size and kinematic constraints will allow a $\sim1\%$ percent $H_0$ measurement without relying on mass profile assumptions. & Relatively mature field with more than two decades of work and progress being made. Demand on follow up and the involvement of a diverse set of observations limited progress. Accelerations of the analyses and moving towards large samples of lenses are underway. \\
cosmography \mbox{with cluster} strong lensing & Currently, only five lens galaxy clusters with multiple images of time-varying sources (3 QSOs and 2 SNe) and measured time-delays are known. LSST will discover a few tens of new QSOs and SNe multiply imaged by galaxy clusters and will measure their time-delays. The latter can require time consuming monitor campaigns. & From the first study on a single lens cluster, the values of \Ho, \omegam, and $w_0$ were measured with, respectively, $\sim$6\%, $\sim$40\%, and $\sim$30\% (including both statistical and systematic 1$\sigma$) uncertainties. The data already available for three lens clusters will likely provide a combined $\sim$3\% uncertainty on \Ho, that will be reduced to $\lesssim$2\% with a future sample of ten lens clusters. & Relatively new method, based on solid theory and with a decade of preparatory work on cluster strong lensing with extensive photometric and spectroscopic data. Progress is being made on improving the modeling of lenses and sources and on enlarging the sample size. \\
cosmic voids & $10^3$--$10^4$ voids available in existing data (BOSS, DES, eBOSS), $10^5$--$10^6$ voids expected in ongoing/planned surveys (DESI, Euclid, PFS, Roman, Rubin, SPHEREx). & Currently up to $\sim5\%$ precision on \omegam\ and $f\sigma_8$ (flat $\Lambda$CDM); future surveys should reach $<1\%$ constraints on \omegam\ and $f\sigma_8$, and $\sim10\%$ precision on $w$ (flat $w$CDM). & RSD \& AP test with void density profile at high maturity level. Other statistics, such as the void size function and void lensing, will be exploited in the near future. \\
neutral hydrogen intensity mapping & Currently available data from GBT, MeerKAT, CHIME. Forthcoming and proposed surveys (within $\sim 10$ years) include HIRAX, MeerKLASS/SKA, PUMA, FAST, Tianlai, CHORD. & Current detections in cross-correlation with optical galaxy samples: $\!\!\sim \! 20\%$ precision on $\Omega_{\textrm{H\textsc{i}}}b_{\textrm{H\textsc{i}}}r$. Future surveys should constrain HI and cosmological parameters to $\sim \! 1\%$, complementing Stage-IV optical galaxy surveys.  & Relatively new technique, no auto-correlation detection yet. Improvement in instrumental calibration, systematics characterisation and mitigation, and detailed understanding of foreground removal effects is necessary. \\
surface brightness fluctuations & In about a decade observations from deep and wide optical and near-IR photometric surveys (mainly Rubin and Euclid) will provide accurate SBF measurements for $\sim10^4$ (and likely more) galaxies. & Current precision on single distances, for well-identified targets, can reach $4\%$: thousands of accurate measurements will guarantee constraining $H_0$ at $<1\%$ precision & SBF with AO-assisted facilities not yet fully tested; data from ELTs may allow pushing the technique to a factor of a few larger distances compared with the present 100-150 Mpc limit, where different cosmological scenarios have a measurable impact relative to the predictable SBF accuracy.\\
stellar ages & Data are already available. GAIA and JWST will provide outstanding data to determine distances and metallicities. & Can constrain age of the universe to sub-percentage accuracy fully independent of the cosmological model. & Arguably the most mature probe. It was the first indication of $\Lambda$ in cosmology: the so-called age crisis problem. Biggest limitation is systematics from stellar models, which can be mitigated by better modeling. Distances to old stars are already determined with good accuracy.\\
redshift drift & At least a decade of observation is required after large HI absorption samples are constructed and after ELT and/or SKA development. & Expect $\sim2\%$ precision on $H_0$ and \omegam\  (flat $\Lambda$CDM). & This is a young and unproven technique, that will have substantial challenges in systematics and calibration and will require new instrumentation. \\
\mbox{clustering of} standard candles & For SN, precision possible with Rubin detections and 4MOST spectra; for SS precision only possible with third generation GW detectors + substantial follow-up resources. & Precision for Rubin SN and galaxies with spectra (5 yrs): $f \sigma_8(z)$ at $<1\%$ ($z<0.5$); $\sigma_8$ at $5-10\%$, $\gamma$ at $12-30\%$; $H_0$ at < 5\%.
& Young technique; potential systematics to be investigated. Effects of non-linearities must be kept under control, similar to (already mature) full-shape $P(k)$ measurements.
\\
\hline \hline
\end{longtable}
\renewcommand{\arraystretch}{1.}

Beyond a different redshift distribution, each probe has its own strengths and weaknesses: in Tab.~\ref{tab:probes} we summarize them, presenting also which quantity they are primarily constraining. From the table, it is evident their wide diversity. In the first place, we have that SBF and stellar ages, as also highlighted in Fig.~\ref{fig:z_probes}, are mostly analyzing very local samples, and as a consequence will be in particular relevant in constraining local cosmological parameters. The advantage is that they require no cosmology-dependent calibrations, being either based on the direct estimate of the stellar age, or on calibration on other observables, with a very small scatter. They represent ideal methods to obtain complementary local estimate of the Hubble constant \Ho~and the age of the Universe $t_U$. Similarly, standard sirens provide a direct and cosmology-independent estimate of the luminosity distance of sources detected through GWs that can, when a counterpart is identified directly (bright sirens) or statistically (dark sirens), lead to another direct and independent measurement of \Ho; moreover, as described in Sect.~\ref{sec:ss}, future observing runs and GW observatory will allow also a measurement of the Hubble parameter $H(z)$ with an accuracy comparable to, and competitive with, other methods.

Moving at higher redshifts, it is better to keep into account also the redshift dependence of the cosmological components to better comprehend the strengths of the various methods. The dark energy component, in particular, dominates at smaller redshifts, $z\lesssim 0.5$, while at larger redshift the contribution of the matter, or of an evolving dark energy component, starts to become more significant. From this point of view, GRB and QSO represent optimal expansions to the Hubble diagram with respect to SNe, being able to measure the luminosity distance up to $z\sim8-12$, providing ideal samples to test possible deviations from a standard $\Lambda$CDM model. 

On the other hand, the strength of CC, RD and CSC is to provide multiple cosmology-independent estimate of $H(z)$ (or $E(z)$), that can constrain the expansion history of the Universe up to $z\sim2$ with minimal assumptions, not needing to assume a specific background model (as done, e.g., for SNe or BAO).
\begin{figure}[b!]
\centering
\includegraphics[width=0.95\textwidth]{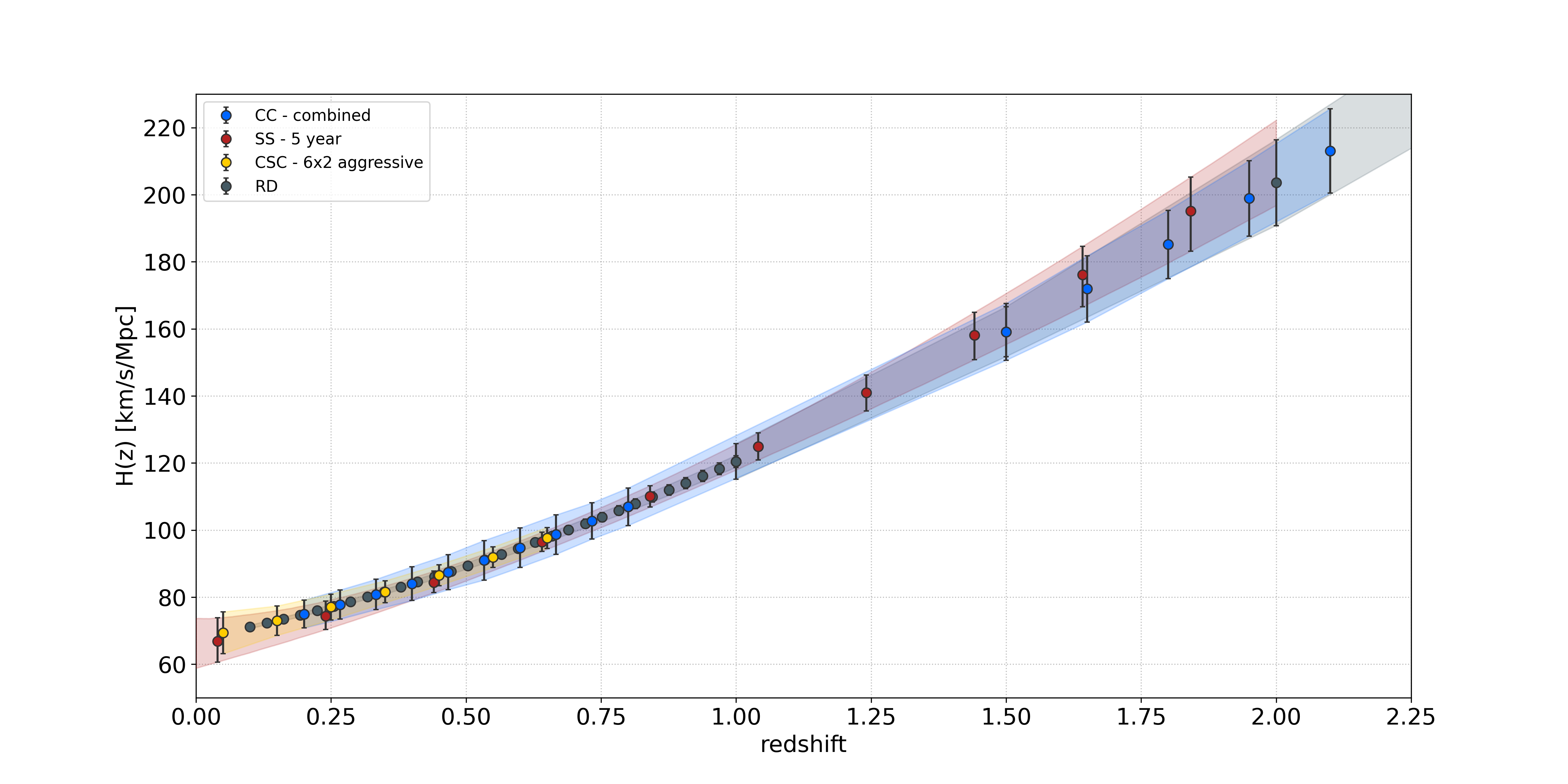}
\caption{Forecasts of the Hubble parameter $H(z)$ with future measurements from cosmic chronometers (CC), standard sirens (SS), clustering of standard candles (CSC), and redshift drift (RD), as discussed in the corresponding sections. The colored dotted points show the forecasts, with bands helping the visualization.}
\label{fig:Hzforecasts}
\end{figure}
Fig.~\ref{fig:Hzforecasts} shows a comparison of the forecasts of future $H(z)$ measurements that can be obtained with these cosmological probes, as presented in Sect.~\ref{sec:CC}, \ref{sec:ss}, \ref{sec:RD}, and \ref{sec:CSC}. It is interesting to notice that in the future in the range $0<z<1$ we will have several method that will provide a percent or sub-percent accuracy measurement of the Hubble parameter in a cosmology-independent way. This will be crucial, because having multiple independent probes will allow us to check for consistency and keep systematic errors under control. At the same time, such an accuracy over a wide redshift range will provide an ideal dataset to really test a vast range of cosmological models and constrain the components of our Universe. For example, the combination of the different strengths of the various methods will play a fundamental role also to test deviations from a standard FLRW metric \citep[see,e.g.,][]{Rasanen:2014mca,cao2019}, to measure cosmological parameters in non-standard models \citep{dagostino2019,dagostino2020}, and to explore trends in redshift and angular direction possibly providing hints to address the Hubble tension \citep[e.g.,][]{krishnan2020,dainotti2022b}.
To complement the information of Tab.~\ref{tab:probes}, in Tab.~\ref{tab:probes2} we also summarize for each probe the foreseen timescale for a development of the method, highlighting the current or future survey expected to provide a significant improvement in terms of statistic or methodological advance and the expected time frame, the constraining power of each probe in the cosmological parameter they are mainly measuring, and their current maturity.

\begin{figure}[t!]
\centering
\includegraphics[width=0.75\textwidth]{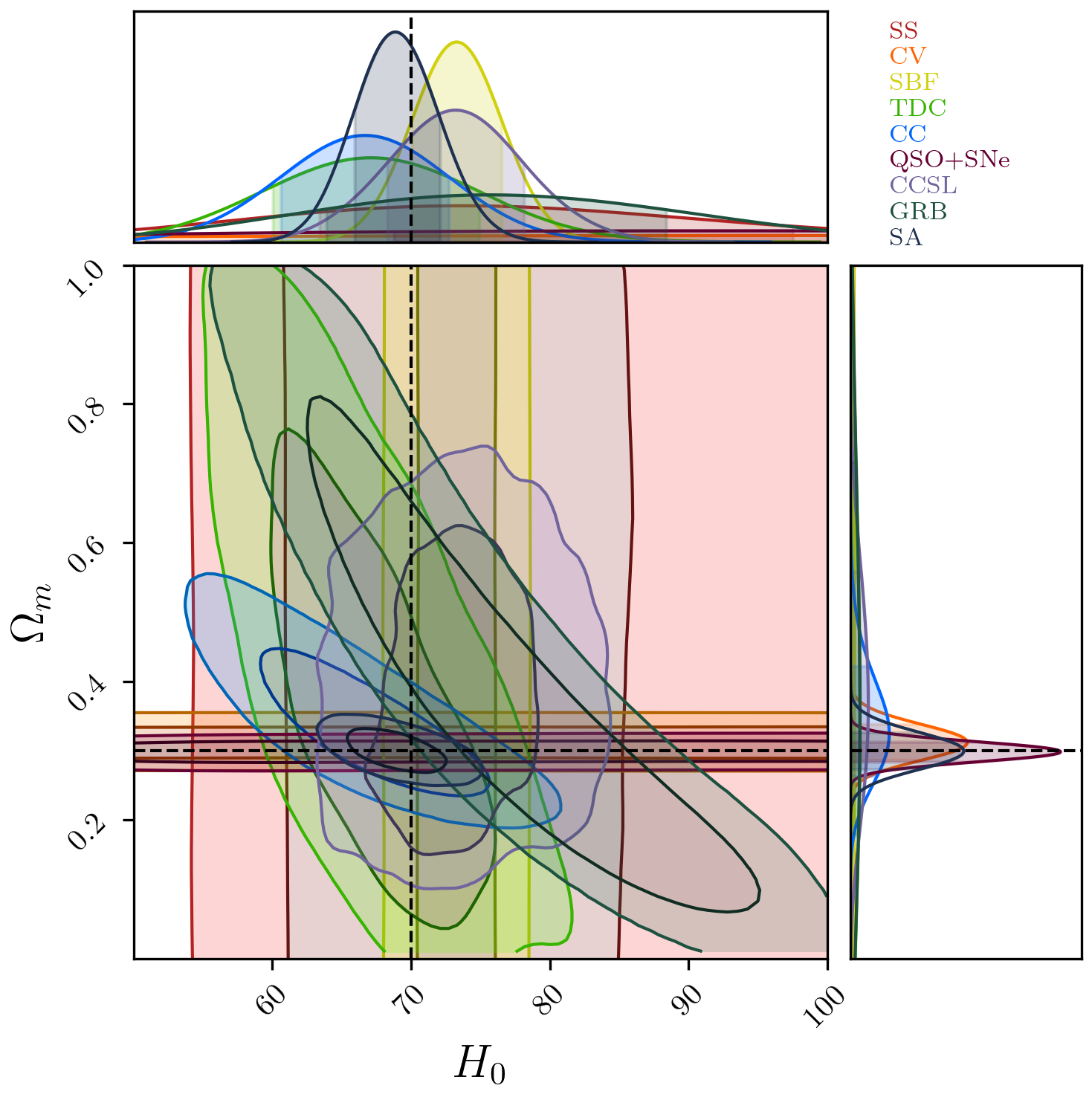}
\caption{Current constraints on cosmological parameters from the various cosmological probes covered in this review, namely cosmic chronometers (CC, this paper), quasars \citep[QSO,][]{lusso2020}, standard sirens \citep[SS][]{2017PhRvL.119p1101A,2017Natur.551...85A}, time-delay cosmography \citep[TDC,][]{Birrer:tdcosmoiv}, surface brightness fluctuations \citep[SBF,][]{blake21}, cosmic voids \citep[CV,][]{Hamaus2020}, cosmography with cluster strong lensing \citep[CCSL, SN Refsdal case][]{Grillo:2020}, gamma-ray bursts \citep[GRB ``Amati'' relation][an updated sample of 212 objects]{Amati19}, and stellar ages 
\citep[SA,][]{Jimenez2019}. The figure shows the contour plot in the \Ho-\omegam~plane for a flat $\Lambda$CDM cosmology, with their marginalized projection; the darker and lighter contours show the 68\% and 95\% confidence levels, respectively. In the case of QSO, as discussed in Sect.~\ref{sec:QSO}, also information from SNe Ia have been added to normalize the Hubble diagram; for SA, a Gaussian prior \omegam=$0.3\pm0.02$ is assumed \citep{Jimenez2019}. The dashed lines indicate, for illustrative purposes, the values \Ho=70 \Hunit and \omegam=0.3.}
\label{fig:combined_constraints}
\end{figure}

To conclude, in Fig.~\ref{fig:combined_constraints} are shown, on a common plane, the current constraints achieved from the various cosmological probes discussed in this review. Since, as discussed above, different probes are sensitive to different parameters, we decided to explore a parameter space that maximizes the number of probes available, in particular the constraints that can be obtained in a flat $\Lambda$CDM cosmology where the parameters free to vary are the Hubble constant \Ho~and \omegam. In this plane, it is possible to fully explore the complementarity and synergy between the various emerging cosmological probes. First of all, we notice that, as also discussed previously, there are methods providing a constraint only on one of the two parameters. This is the case of CV and QSO, and of SBF, SS, and partially SA, that, as discussed in the previous sections, cannot measure \Ho~  and \omegam, respectively. It is useful to underline here that, in the case of QSO, we have included in the constraints also data from SNe, as discussed in Sect.~\ref{sec:QSO}, and that in the case of SA we have assumed a Gaussian prior on \omegam=$0.30\pm0.02$ as  in \cite{Jimenez2019}. On the other hand, there are probes that are sensitive to both \Ho~and \omegam, namely CC, TDC, CCSL, shown considering the SN Refsdal case \citep[see Sect.~\ref{sec:CCSL} and][]{Grillo:2020}, and GRB, here exploited with the ``Amati relation'' approach (see Sect.~\ref{sec:GRB}). 

Two points are worth underlining here. The first one is that all probes, despite different accuracies, are converging on a common part of the \Ho-\omegam~plane. Given the extreme diversity between the methods considered, this is very relevant because it builds up the possibility of combining different probes to improve the accuracy on the estimated parameters. Such combinations are, at the moment, beyond the scope of this review, because it requires one to carefully address all possible systematics and covariances between the various probes, but Fig.~\ref{fig:combined_constraints} appears extremely promising. The second one is that the various probes present also a significant degree of orthogonality, due to the different sensitivities discussed above. This has been proven to be extremely important in the past, where the extreme accuracy reached by the main probes was mainly based on the orthogonality between the constraints from SNe, BAO and CMB \citep[see, e.g.,][]{scolnic2018}. Finding a similar level of complementarity also between the new emerging probes represents a good omen toward the use of these new methods in modern cosmology, to better constrain cosmological parameters, provide additional evidences to help solve current tensions, keep under control systematic effects of both the main and the new probes, and, potentially, discover new physics.

\clearpage

\section{Summary and conclusions}
\label{sec:conclusions}

In this article, we have reviewed the new emerging cosmological probes that are contributing (and are expected to contribute in the near future) to modern cosmology. In particular, we have discussed cosmic chronometers, quasars, gamma-ray bursts, gravitational waves used as standard sirens, time-delay cosmography, cosmography with cluster strong lensing, cosmic voids, neutral hydrogen intensity mapping, surface brightness fluctuations, the ages of the oldest stellar objects, secular redshift drift, and clustering of standard candles. We presented, for each cosmological probe, the main equations involved in the method, how a sample can be selected and the method applied, reviewed the main results and expected forecasts, and discussed the systematics involved, showing also possible paths on how to mitigate or minimize those.

These emerging cosmological probes represent a valuable resource for the next years, since they could allow us to go beyond the main cosmological probes currently exploited (SNe, BAO, CMB, weak lensing). In particular, they will provide crucial additional information to check for possible systematics in current analyses, increase the number of independent measurements of cosmological probes, and give new hints to address the current tensions in cosmology, possibly strengthening the need for new physics (see, e.g. Fig.~\ref{fig:combined_constraints} and Tables ~\ref{tab:probes} and \ref{tab:probes2}). As also shown in Fig.~\ref{fig:Hzforecasts}, these probes will represent also an important dataset in the future to obtain constraints on the expansion history of the Universe at the percent precision independently of assumptions on a particular cosmological model, being ideal complementary probes to the excellent results we are obtaining from the other main probes. 
The exploitation of new and complementary cosmological probes will be fundamental also in view of the new surveys and missions that are currently undergoing or planned, such as SDSS BOSS Data Release 16 \citep{boss16}, DESI \citep{DESI:2016}, Gaia \citep{Gaia:2016}, JWST \citep{Gardner:2006}, Euclid \citep{Laureijs:2011gra}, PFS \citep{Takada:2014}, the Nancy Grace Roman Space Telescope \citep{Spergel:2015}, the LSST on Vera Rubin Observatory \citep{LSSTScience:2009jmu}, the next LIGO-Virgo-Kagra observing runs \citep{2015CQGra..32g4001L,2015CQGra..32b4001A,2020arXiv200505574A} and future GW experiments like Cosmic Explorer \citep{Reitze:2019} and the Einstein Telescope \citep{Punturo:2010}, the MIGHTEE survey \citep{ Paul:2020ank, Chen:2020uld}, ASKAP \citep{Wolz:2017rlw}, SPHEREx \citep{Dore:2014}, and the ATLAS mission \citep{Atlasmission}.

As a final note, we acknowledge that other alternative probes are also available and could provide valuable information in the future, including fast radio bursts \citep[FRB,][]{jaroszynski2019, wucknitz2021}, HII galaxies \citep{terlevich2015}, black hole shadows \citep{tsupko2020,vagnozzi2020,perlick2022,renzi2022},  Type II SNe \citep{dejaeger2020} and SNe Ia lensing \citep{Quartin:2013moa,Castro:2014oja,Scovacricchi:2016ylt,Zumalacarregui:2017qqd}.
While we are currently not including those in this review because they are not at the same level of maturity of the discussed probes, we are looking forward to them for possible applications in the future.


\section*{Acknowledgments}

MM and MC acknowledge support from MIUR, PRIN 2017 (grant 20179ZF5KS). MM and AC acknowledge the grants ASI n.I/023/12/0 and ASI n.2018-23-HH.0. AC acknowledges the support from grant PRIN MIUR 2017 - 20173ML3WW\_001. CG, MM, and PR acknowledge financial support through grants PRIN-MIUR 2017WSCC32, 2020SKSTHZ. EL acknowledges the support of grant ID: 45780 Fondazione Cassa di Risparmio Firenze. APi acknowledges the support from NASA ROSES grant 12-EUCLID12-0004, and NASA grant 15-WFIRST15-0008 to the Nancy Grace Roman Space Telescope Science Investigation Team ``Cosmology with the High Latitude Survey''. Funding for the work of RJ and LV was partially provided by  project PGC2018-098866- B-I00 MCIN/AEI/10.13039/501100011033 y FEDER ``Una manera de hacer Europa'', and the  ``Center of Excellence Maria de Maeztu 2020-2023'' award to the ICCUB (CEX2019- 000918-M funded by MCIN/AEI/10.13039/501100011033). APo is a UK Research and Innovation Future Leaders Fellow [grant number MR/S016066/1]. APo is grateful to Dida Markovic and Marta Spinelli for detailed feedback, and to Chris Blake, Steven Cunnington and Paula Soares for their suggestions and assistance. APo warmly thanks the Cosmic Dawn Center, University of Copenhagen, for hospitality while part of this work was in progress. LV acknowledges support from the  European Union Horizon 2020 research and innovation program ERC (BePreSySe, grant agreement 725327). JD acknowledges support from NASA grant 20-ADAP20-0222. NH acknowledges support from the Excellence Cluster ORIGINS, which is funded by the Deutsche Forschungsgemeinschaft (DFG, German Research Foundation) under Germany's Excellence Strategy -- EXC-2094 -- 390783311. LA acknowledges financial contribution from the agreement ASI-INAF n.2017-14-H.O.
MQ is supported by the Brazilian research agencies CNPq, FAPERJ and CAPES. LAm acknowledges support from DFG project  456622116.  LAm and MQ acknowledge  financial support by the Coordenação de Aperfeiçoamento de Pessoal de Nível Superior - Brasil (CAPES) - Finance Code 001, and  support from the CAPES-DAAD bilateral project  ``Data Analysis and Model Testing in the Era of Precision Cosmology''. We thank N. Borghi, F. Cogato, and A. Huchet for their useful contribution and discussion.

\bibliographystyle{spbasic}
\bibliography{bib}

\begin{thebibliography}{1061}
\providecommand{\natexlab}[1]{#1}
\providecommand{\url}[1]{{#1}}
\providecommand{\urlprefix}{URL }
\expandafter\ifx\csname urlstyle\endcsname\relax
  \providecommand{\doi}[1]{DOI~\discretionary{}{}{}#1}\else
  \providecommand{\doi}{DOI~\discretionary{}{}{}\begingroup
  \urlstyle{rm}\Url}\fi
\providecommand{\eprint}[2][]{\url{#2}}

\bibitem[{Abbott et~al.(2017)}]{Abbott2017gamma}
Abbott B., et~al. (2017) {Gravitational Waves and Gamma-Rays from a Binary
  Neutron Star Merger: GW170817 and GRB 170817A}. \emph{The Astrophysical
  Journal} 848(2):L13, \doi{10.3847/2041-8213/aa920c}

\bibitem[{{Abbott} et~al.(2017{\natexlab{a}}){Abbott}, {Abbott}, {Abbott},
  {Acernese}, {Ackley}, {Adams}, {Adams}, {Addesso}, {Adhikari}, {Adya}, and
  et~al.}]{2017Natur.551...85A}
{Abbott} B.~P., {Abbott} R., {Abbott} T.~D., {Acernese} F., {Ackley} K.,
  {Adams} C., {Adams} T., {Addesso} P., et~al. (2017{\natexlab{a}}) {A
  gravitational-wave standard siren measurement of the Hubble constant}.
  \emph{\nat} 551(7678):85--88, \doi{10.1038/nature24471}, \eprint{1710.05835}

\bibitem[{{Abbott} et~al.(2017{\natexlab{b}}){Abbott}, {Abbott}, {Abbott},
  {Acernese}, {Ackley}, {Adams}, {Adams}, {Addesso}, {Adhikari}, {Adya}, and
  et~al.}]{2017PhRvL.119n1101A}
{Abbott} B.~P., {Abbott} R., {Abbott} T.~D., {Acernese} F., {Ackley} K.,
  {Adams} C., {Adams} T., {Addesso} P., et~al. (2017{\natexlab{b}}) {GW170814:
  A Three-Detector Observation of Gravitational Waves from a Binary Black Hole
  Coalescence}. \emph{\prl} 119(14):141101,
  \doi{10.1103/PhysRevLett.119.141101}, \eprint{1709.09660}

\bibitem[{{Abbott} et~al.(2017{\natexlab{c}}){Abbott}, {Abbott}, {Abbott},
  {Acernese}, {Ackley}, {Adams}, {Adams}, {Addesso}, {Adhikari}, {Adya}, and
  et~al.}]{2017PhRvL.119p1101A}
{Abbott} B.~P., {Abbott} R., {Abbott} T.~D., {Acernese} F., {Ackley} K.,
  {Adams} C., {Adams} T., {Addesso} P., et~al. (2017{\natexlab{c}}) {GW170817:
  Observation of Gravitational Waves from a Binary Neutron Star Inspiral}.
  \emph{\prl} 119(16):161101, \doi{10.1103/PhysRevLett.119.161101},
  \eprint{1710.05832}

\bibitem[{{Abbott} et~al.(2018){Abbott}, {Abbott}, {Abbott}, {Abernathy},
  {Acernese}, {Ackley}, {Adams}, {Adams}, {Addesso}, {Adhikari}, and
  et~al.}]{2018LRR....21....3A}
{Abbott} B.~P., {Abbott} R., {Abbott} T.~D., {Abernathy} M.~R., {Acernese} F.,
  {Ackley} K., {Adams} C., {Adams} T., et~al. (2018) {Prospects for observing
  and localizing gravitational-wave transients with Advanced LIGO, Advanced
  Virgo and KAGRA}. \emph{Living Reviews in Relativity} 21(1):3,
  \doi{10.1007/s41114-018-0012-9}, \eprint{1304.0670}

\bibitem[{{Abbott} et~al.(2019{\natexlab{a}}){Abbott}, {Abbott}, {Abbott},
  {Acernese}, {Ackley}, {Adams}, {Adams}, {Addesso}, {Adhikari}, {Adya}, and
  et~al.}]{2019PhRvX...9a1001A}
{Abbott} B.~P., {Abbott} R., {Abbott} T.~D., {Acernese} F., {Ackley} K.,
  {Adams} C., {Adams} T., {Addesso} P., et~al. (2019{\natexlab{a}}) {Properties
  of the Binary Neutron Star Merger GW170817}. \emph{Physical Review X}
  9(1):011001, \doi{10.1103/PhysRevX.9.011001}, \eprint{1805.11579}

\bibitem[{{Abbott} et~al.(2019{\natexlab{b}}){Abbott}, {Abbott}, {Abbott},
  {Acernese}, {Ackley}, {Adams}, {Adams}, {Addesso}, {Adhikari}, {Adya}, and
  et~al.}]{2019PhRvL.123a1102A}
{Abbott} B.~P., {Abbott} R., {Abbott} T.~D., {Acernese} F., {Ackley} K.,
  {Adams} C., {Adams} T., {Addesso} P., et~al. (2019{\natexlab{b}}) {Tests of
  General Relativity with GW170817}. \emph{\prl} 123(1):011102,
  \doi{10.1103/PhysRevLett.123.011102}, \eprint{1811.00364}

\bibitem[{{Abbott} et~al.(2021{\natexlab{a}}){Abbott}, {Abbott}, {Abbott},
  {Abraham}, {Acernese}, {Ackley}, {Adams}, {Adhikari}, {Adya}, {Affeldt}, and
  et~al.}]{2021ApJ...909..218A}
{Abbott} B.~P., {Abbott} R., {Abbott} T.~D., {Abraham} S., {Acernese} F.,
  {Ackley} K., {Adams} C., {Adhikari} R.~X., et~al. (2021{\natexlab{a}}) {A
  Gravitational-wave Measurement of the Hubble Constant Following the Second
  Observing Run of Advanced LIGO and Virgo}. \emph{\apj} 909(2):218,
  \doi{10.3847/1538-4357/abdcb7}, \eprint{1908.06060}

\bibitem[{{Abbott} et~al.(2020){Abbott}, {Abbott}, {Abraham}, {Acernese},
  {Ackley}, {Adams}, {Adhikari}, {Adya}, {Affeldt}, {Agathos}, and
  et~al.}]{2020ApJ...896L..44A}
{Abbott} R., {Abbott} T.~D., {Abraham} S., {Acernese} F., {Ackley} K., {Adams}
  C., {Adhikari} R.~X., {Adya} V.~B., et~al. (2020) {GW190814: Gravitational
  Waves from the Coalescence of a 23 Solar Mass Black Hole with a 2.6 Solar
  Mass Compact Object}. \emph{\apjl} 896(2):L44,
  \doi{10.3847/2041-8213/ab960f}, \eprint{2006.12611}

\bibitem[{{Abbott} et~al.(2021{\natexlab{b}}){Abbott}, {Abbott}, {Abraham},
  {Acernese}, {Ackley}, {Adams}, {Adams}, {Adhikari}, {Adya}, {Affeldt}, and
  et~al.}]{2021ApJ...913L...7A}
{Abbott} R., {Abbott} T.~D., {Abraham} S., {Acernese} F., {Ackley} K., {Adams}
  A., {Adams} C., {Adhikari} R.~X., et~al. (2021{\natexlab{b}}) {Population
  Properties of Compact Objects from the Second LIGO-Virgo Gravitational-Wave
  Transient Catalog}. \emph{\apjl} 913(1):L7, \doi{10.3847/2041-8213/abe949},
  \eprint{2010.14533}

\bibitem[{Abel(1842)}]{Abel1842}
Abel N.~H. (1842) Oeuvres Completes. SEd. L. Sylow and S. Lie, New York:
  Johnson Reprint Corp.

\bibitem[{Abramo(2011)}]{Abramo:2012}
Abramo L.~R. (2011) The full fisher matrix for galaxy surveys. \emph{MNRAS 420,
  3, pp 2032 (2012)} \urlprefix\url{http://arxiv.org/abs/1108.5449},
  \eprint{1108.5449}

\bibitem[{Abramo and Amendola(2019)}]{Abramo:2019ejj}
Abramo L.~R., Amendola L. (2019) {Fisher matrix for multiple tracers: model
  independent constraints on the redshift distortion parameter}. \emph{JCAP}
  1906(06):030, \doi{10.1088/1475-7516/2019/06/030}, \eprint{1904.00673}

\bibitem[{Abramo and Leonard(2013)}]{Abramo:2013}
Abramo L.~R., Leonard K.~E. (2013) {Why multi-tracer surveys beat cosmic
  variance}. \emph{Mon Not Roy Astron Soc} 432:318, \doi{10.1093/mnras/stt465},
  \eprint{1302.5444}

\bibitem[{{Acebron} et~al.(2017){Acebron}, {Jullo}, {Limousin}, {Tilquin},
  {Giocoli}, {Jauzac}, {Mahler}, and {Richard}}]{Acebron:2017}
{Acebron} A., {Jullo} E., {Limousin} M., {Tilquin} A., {Giocoli} C., {Jauzac}
  M., {Mahler} G., {Richard} J. (2017) {Hubble Frontier Fields: systematic
  errors in strong lensing models of galaxy clusters - implications for
  cosmography}. \emph{\mnras} 470(2):1809--1825, \doi{10.1093/mnras/stx1330},
  \eprint{1704.05380}

\bibitem[{{Acebron} et~al.(2021){Acebron}, {Grillo}, {Bergamini}, {Mercurio},
  {Rosati}, {Bartosch Caminha}, {Tozzi}, {Brammer}, {Meneghetti}, {Morelli},
  {Nonino}, and {Vanzella}}]{Acebron:2021}
{Acebron} A., {Grillo} C., {Bergamini} P., {Mercurio} A., {Rosati} P.,
  {Bartosch Caminha} G., {Tozzi} P., {Brammer} G.~B., et~al. (2021) {VLT/MUSE
  observations of SDSS J1029+2623: towards a high-precision strong lensing
  model}. \emph{arXiv e-prints} arXiv:2111.05871, \eprint{2111.05871}

\bibitem[{{Acernese} et~al.(2015){Acernese}, {Agathos}, {Agatsuma}, {Aisa},
  {Allemandou}, {Allocca}, {Amarni}, {Astone}, {Balestri}, {Ballardin}, and
  et~al.}]{2015CQGra..32b4001A}
{Acernese} F., {Agathos} M., {Agatsuma} K., {Aisa} D., {Allemandou} N.,
  {Allocca} A., {Amarni} J., {Astone} P., et~al. (2015) {Advanced Virgo: a
  second-generation interferometric gravitational wave detector}.
  \emph{Classical and Quantum Gravity} 32(2):024001,
  \doi{10.1088/0264-9381/32/2/024001}, \eprint{1408.3978}

\bibitem[{{Achitouv}(2016)}]{Achitouv2016}
{Achitouv} I. (2016) {Testing the imprint of nonstandard cosmologies on void
  profiles using Monte Carlo random walks}. \emph{\prd} 94(10):103524,
  \doi{10.1103/PhysRevD.94.103524}, \eprint{1609.01284}

\bibitem[{{Achitouv}(2019)}]{Achitouv2019}
{Achitouv} I. (2019) {New constraints on the linear growth rate using cosmic
  voids in the SDSS DR12 datasets}. \emph{\prd} 100(12):123513,
  \doi{10.1103/PhysRevD.100.123513}, \eprint{1903.05645}

\bibitem[{{Achitouv} et~al.(2017){Achitouv}, {Blake}, {Carter}, {Koda}, and
  {Beutler}}]{Achitouv2017}
{Achitouv} I., {Blake} C., {Carter} P., {Koda} J., {Beutler} F. (2017)
  {Consistency of the growth rate in different environments with the 6-degree
  Field Galaxy Survey: Measurement of the void-galaxy and galaxy-galaxy
  correlation functions}. \emph{\prd} 95(8):083502,
  \doi{10.1103/PhysRevD.95.083502}, \eprint{1606.03092}

\bibitem[{{Agnello} et~al.(2015){Agnello}, {Kelly}, {Treu}, and
  {Marshall}}]{Agnello15}
{Agnello} A., {Kelly} B.~C., {Treu} T., {Marshall} P.~J. (2015) {Data mining
  for gravitationally lensed quasars}. \emph{\mnras} 448(2):1446--1462,
  \doi{10.1093/mnras/stv037}, \eprint{1410.4565}

\bibitem[{{Agnello} et~al.(2016){Agnello}, {Sonnenfeld}, {Suyu}, {Treu},
  {Fassnacht}, {Mason}, {Brada{\v{c}}}, and {Auger}}]{Agnello:2016}
{Agnello} A., {Sonnenfeld} A., {Suyu} S.~H., {Treu} T., {Fassnacht} C.~D.,
  {Mason} C., {Brada{\v{c}}} M., {Auger} M.~W. (2016) {Spectroscopy and
  high-resolution imaging of the gravitational lens SDSS J1206+4332}.
  \emph{\mnras} 458(4):3830--3838, \doi{10.1093/mnras/stw529}

\bibitem[{{Agnello} et~al.(2018){Agnello}, {Lin}, {Kuropatkin}, {Buckley-Geer},
  {Anguita}, {Schechter}, {Morishita}, {Motta}, {Rojas}, {Treu}, {Amara},
  {Auger}, {Courbin}, {Fassnacht}, {Frieman}, {More}, {Marshall}, {McMahon},
  {Meylan}, {Suyu}, {Glazebrook}, {Morgan}, {Nord}, {Abbott}, {Abdalla},
  {Annis}, {Bechtol}, {Benoit-L{\'e}vy}, {Bertin}, {Bernstein}, {Brooks},
  {Burke}, {Rosell}, {Carretero}, {Cunha}, {D'Andrea}, {da Costa}, {Desai},
  {Drlica-Wagner}, {Eifler}, {Flaugher}, {Garc{\'\i}a-Bellido}, {Gaztanaga},
  {Gerdes}, {Gruen}, {Gruendl}, {Gschwend}, {Gutierrez}, {Honscheid}, {James},
  {Kuehn}, {Lahav}, {Lima}, {Maia}, {March}, {Menanteau}, {Miquel}, {Ogando},
  {Plazas}, {Sanchez}, {Scarpine}, {Schindler}, {Schubnell}, {Sevilla-Noarbe},
  {Smith}, {Soares-Santos}, {Sobreira}, {Suchyta}, {Swanson}, {Tarle},
  {Tucker}, and {Wechsler}}]{Agnello:2018}
{Agnello} A., {Lin} H., {Kuropatkin} N., {Buckley-Geer} E., {Anguita} T.,
  {Schechter} P.~L., {Morishita} T., {Motta} V., et~al. (2018) {DES meets Gaia:
  discovery of strongly lensed quasars from a multiplet search}. \emph{\mnras}
  479(4):4345--4354, \doi{10.1093/mnras/sty1419}, \eprint{1711.03971}

\bibitem[{{Ahmed} et~al.(2019){Ahmed}, {Alonso}, {Amin}, {Ansari}, {Arena},
  {Bandura}, {Beardsley}, {Bull}, {Castorina}, {Chang}, {Dav{\'e}}, {Dillon},
  {van Engelen}, {Ewall-Wice}, {Ferraro}, and et~al.}]{Ahmed:2019ocj}
{Ahmed} Z., {Alonso} D., {Amin} M.~A., {Ansari} R., {Arena} E.~J., {Bandura}
  K., {Beardsley} A., {Bull} P., et~al. (2019) {Research and Development for HI
  Intensity Mapping}. \emph{arXiv e-prints} arXiv:1907.13090,
  \eprint{1907.13090}

\bibitem[{{Ahumada} et~al.(2020){Ahumada}, {Prieto}, {Almeida}, {Anders},
  {Anderson}, {Andrews}, {Anguiano}, {Arcodia}, {Armengaud}, {Aubert}, {Avila},
  {Avila-Reese}, {Badenes}, {Balland}, and et~al.}]{boss16}
{Ahumada} R., {Prieto} C.~A., {Almeida} A., {Anders} F., {Anderson} S.~F.,
  {Andrews} B.~H., {Anguiano} B., {Arcodia} R., et~al. (2020) {The 16th Data
  Release of the Sloan Digital Sky Surveys: First Release from the APOGEE-2
  Southern Survey and Full Release of eBOSS Spectra}. \emph{\apjs} 249(1):3,
  \doi{10.3847/1538-4365/ab929e}, \eprint{1912.02905}

\bibitem[{{Aiola} et~al.(2020){Aiola}, {Calabrese}, {Maurin}, {Naess},
  {Schmitt}, {Abitbol}, {Addison}, {Ade}, {Alonso}, {Amiri}, {Amodeo},
  {Angile}, {Austermann}, {Baildon}, {Battaglia}, and
  https://www.overleaf.com/4448358379csmrynjykryg}]{Aiola:2020}
{Aiola} S., {Calabrese} E., {Maurin} L., {Naess} S., {Schmitt} B.~L., {Abitbol}
  M.~H., {Addison} G.~E., {Ade} P. A.~R., et~al. (2020) {The Atacama Cosmology
  Telescope: DR4 maps and cosmological parameters}. \emph{\jcap} 2020(12):047,
  \doi{10.1088/1475-7516/2020/12/047}, \eprint{2007.07288}

\bibitem[{{Ajhar} et~al.(2001){Ajhar}, {Tonry}, {Blakeslee}, {Riess}, and
  {Schmidt}}]{ajhar01}
{Ajhar} E.~A., {Tonry} J.~L., {Blakeslee} J.~P., {Riess} A.~G., {Schmidt} B.~P.
  (2001) {Reconciliation of the Surface Brightness Fluctuation and Type Ia
  Supernova Distance Scales}. \emph{\apj} 559(2):584--591,
  \doi{10.1086/322342}, \eprint{astro-ph/0105366}

\bibitem[{Akeret et~al.(2017)Akeret, Seehars, Chang, Monstein, Amara, and
  Refregier}]{Akeret_2017}
Akeret J., Seehars S., Chang C., Monstein C., Amara A., Refregier A. (2017)
  Hide \& seek: End-to-end packages to simulate and process radio survey data.
  \emph{Astronomy and Computing} 18:8–17, \doi{10.1016/j.ascom.2016.11.001},
  \urlprefix\url{http://dx.doi.org/10.1016/j.ascom.2016.11.001}

\bibitem[{{Akutsu} et~al.(2020){Akutsu}, {Ando}, {Arai}, {Arai}, {Araki},
  {Araya}, {Aritomi}, {Aso}, {Bae}, {Bae}, and et~al.}]{2020arXiv200505574A}
{Akutsu} T., {Ando} M., {Arai} K., {Arai} Y., {Araki} S., {Araya} A., {Aritomi}
  N., {Aso} Y., et~al. (2020) {Overview of KAGRA: Detector design and
  construction history}. \emph{arXiv e-prints} arXiv:2005.05574,
  \eprint{2005.05574}

\bibitem[{{Alam} et~al.(2017){Alam}, {Ata}, {Bailey}, {Beutler}, {Bizyaev},
  {Blazek}, {Bolton}, {Brownstein}, {Burden}, {Chuang}, {Comparat}, {Cuesta},
  {Dawson}, {Eisenstein}, and et~al.}]{Alam2017}
{Alam} S., {Ata} M., {Bailey} S., {Beutler} F., {Bizyaev} D., {Blazek} J.~A.,
  {Bolton} A.~S., {Brownstein} J.~R., et~al. (2017) {The clustering of galaxies
  in the completed SDSS-III Baryon Oscillation Spectroscopic Survey:
  cosmological analysis of the DR12 galaxy sample}. \emph{\mnras}
  470(3):2617--2652, \doi{10.1093/mnras/stx721}, \eprint{1607.03155}

\bibitem[{{Alam} et~al.(2020){Alam}, {Aviles}, {Bean}, {Cai}, {Cautun},
  {Cervantes-Cota}, {Cuesta-Lazaro}, {Chandrachani Devi}, {Eggemeier},
  {Fromenteau}, {Gonzalez-Morales}, {Halenka}, {He}, {Hellwing},
  {Hernandez-Aguayo}, {Ishak}, {Koyama}, {Li}, {de la Macorra}, {Menesses
  Rizo}, {Miller}, {Mueller}, {Niz}, {Ntelis}, {Rodriguez Otero}, {Sabiu},
  {Slepian}, {Stark}, {Valenzuela}, {Valogiannis}, {Vargas-Magana}, {Winther},
  {Zarrouk}, {Zhao}, and {Zheng}}]{Alam2020}
{Alam} S., {Aviles} A., {Bean} R., {Cai} Y.-C., {Cautun} M., {Cervantes-Cota}
  J.~L., {Cuesta-Lazaro} C., {Chandrachani Devi} N., et~al. (2020) {Testing the
  theory of gravity with DESI: estimators, predictions and simulation
  requirements}. \emph{arXiv e-prints} arXiv:2011.05771, \eprint{2011.05771}

\bibitem[{Alam et~al.(2021)}]{Alam:2020sor}
Alam S., et~al. (2021) {Completed SDSS-IV extended Baryon Oscillation
  Spectroscopic Survey: Cosmological implications from two decades of
  spectroscopic surveys at the Apache Point Observatory}. \emph{Phys Rev D}
  103(8):083533, \doi{10.1103/PhysRevD.103.083533}, \eprint{2007.08991}

\bibitem[{{Albrecht} et~al.(2006){Albrecht}, {Bernstein}, {Cahn}, {Freedman},
  {Hewitt}, {Hu}, {Huth}, {Kamionkowski}, {Kolb}, {Knox}, {Mather}, {Staggs},
  and {Suntzeff}}]{albrecht2006}
{Albrecht} A., {Bernstein} G., {Cahn} R., {Freedman} W.~L., {Hewitt} J., {Hu}
  W., {Huth} J., {Kamionkowski} M., et~al. (2006) {Report of the Dark Energy
  Task Force}. \emph{arXiv e-prints} astro-ph/0609591,
  \eprint{astro-ph/0609591}

\bibitem[{{Alcock} and {Paczynski}(1979)}]{Alcock1979}
{Alcock} C., {Paczynski} B. (1979) {An evolution free test for non-zero
  cosmological constant}. \emph{\nat} 281:358, \doi{10.1038/281358a0}

\bibitem[{Alcock and Paczynski(1979)}]{Alcock:1979mp}
Alcock C., Paczynski B. (1979) {An evolution free test for non-zero
  cosmological constant}. \emph{Nature} 281:358--359, \doi{10.1038/281358a0}

\bibitem[{{Alfradique} et~al.(2022){Alfradique}, {Quartin}, {Amendola},
  {Castro}, and {Toubiana}}]{Alfradique:2022tox}
{Alfradique} V., {Quartin} M., {Amendola} L., {Castro} T., {Toubiana} A. (2022)
  {The lure of sirens: joint distance and velocity measurements with third
  generation detectors}. \emph{arXiv e-prints} arXiv:2205.14034,
  \eprint{2205.14034}

\bibitem[{{Aljaf} et~al.(2021){Aljaf}, {Gregoris}, and
  {Khurshudyan}}]{aljaf2021}
{Aljaf} M., {Gregoris} D., {Khurshudyan} M. (2021) {Constraints on interacting
  dark energy models through cosmic chronometers and Gaussian process}.
  \emph{European Physical Journal C} 81(6):544,
  \doi{10.1140/epjc/s10052-021-09306-2}, \eprint{2005.01891}

\bibitem[{{Allen} et~al.(2011){Allen}, {Evrard}, and {Mantz}}]{allen2011}
{Allen} S.~W., {Evrard} A.~E., {Mantz} A.~B. (2011) {Cosmological Parameters
  from Observations of Galaxy Clusters}. \emph{\araa} 49(1):409--470,
  \doi{10.1146/annurev-astro-081710-102514}, \eprint{1103.4829}

\bibitem[{Alonso and Ferreira(2015)}]{Alonso:2015sfa}
Alonso D., Ferreira P.~G. (2015) {Constraining ultralarge-scale cosmology with
  multiple tracers in optical and radio surveys}. \emph{Phys Rev D}
  92(6):063525, \doi{10.1103/PhysRevD.92.063525}, \eprint{1507.03550}

\bibitem[{Alonso et~al.(2014)Alonso, Ferreira, and Santos}]{Alonso:2014sna}
Alonso D., Ferreira P.~G., Santos M.~G. (2014) {Fast simulations for intensity
  mapping experiments}. \emph{Mon Not Roy Astron Soc} 444(4):3183--3197,
  \doi{10.1093/mnras/stu1666}, \eprint{1405.1751}

\bibitem[{Alonso et~al.(2015{\natexlab{a}})Alonso, Bull, Ferreira, Maartens,
  and Santos}]{Alonso:2015uua}
Alonso D., Bull P., Ferreira P.~G., Maartens R., Santos M. (2015{\natexlab{a}})
  {Ultra large-scale cosmology in next-generation experiments with single
  tracers}. \emph{Astrophys J} 814(2):145, \doi{10.1088/0004-637X/814/2/145},
  \eprint{1505.07596}

\bibitem[{Alonso et~al.(2015{\natexlab{b}})Alonso, Bull, Ferreira, and
  Santos}]{Alonso:2014dhk}
Alonso D., Bull P., Ferreira P.~G., Santos M.~G. (2015{\natexlab{b}}) {Blind
  foreground subtraction for intensity mapping experiments}. \emph{Mon Not Roy
  Astron Soc} 447:400, \doi{10.1093/mnras/stu2474}, \eprint{1409.8667}

\bibitem[{{Alonso} et~al.(2018){Alonso}, {Hill}, {Hlo{\v{z}}ek}, and
  {Spergel}}]{Alonso2018}
{Alonso} D., {Hill} J.~C., {Hlo{\v{z}}ek} R., {Spergel} D.~N. (2018)
  {Measurement of the thermal Sunyaev-Zel'dovich effect around cosmic voids}.
  \emph{\prd} 97(6):063514, \doi{10.1103/PhysRevD.97.063514},
  \eprint{1709.01489}

\bibitem[{{Alves} et~al.(2019){Alves}, {Leite}, {Martins}, {Matos}, and
  {Silva}}]{alves2019}
{Alves} C.~S., {Leite} A.~C.~O., {Martins} C.~J.~A.~P., {Matos} J.~G.~B.,
  {Silva} T.~A. (2019) {Forecasts of redshift drift constraints on cosmological
  parameters}. \emph{\mnras} 488(3):3607--3624, \doi{10.1093/mnras/stz1934},
  \eprint{1907.05151}

\bibitem[{{Amati}(2006)}]{Amati06}
{Amati} L. (2006) {The E$_{p,i}$-E$_{iso}$ correlation in gamma-ray bursts:
  updated observational status, re-analysis and main implications}.
  \emph{\mnras} 372(1):233--245, \doi{10.1111/j.1365-2966.2006.10840.x},
  \eprint{astro-ph/0601553}

\bibitem[{{Amati} and {Della Valle}(2013)}]{Amati13b}
{Amati} L., {Della Valle} M. (2013) {Measuring Cosmological Parameters with
  Gamma Ray Bursts}. \emph{International Journal of Modern Physics D}
  22(14):1330028, \doi{10.1142/S0218271813300280}, \eprint{1310.3141}

\bibitem[{{Amati} et~al.(2002){Amati}, {Frontera}, {Tavani}, {in't Zand},
  {Antonelli}, {Costa}, {Feroci}, {Guidorzi}, {Heise}, {Masetti}, {Montanari},
  {Nicastro}, {Palazzi}, {Pian}, {Piro}, and {Soffitta}}]{Amati02}
{Amati} L., {Frontera} F., {Tavani} M., {in't Zand} J.~J.~M., {Antonelli} A.,
  {Costa} E., {Feroci} M., {Guidorzi} C., et~al. (2002) {Intrinsic spectra and
  energetics of BeppoSAX Gamma-Ray Bursts with known redshifts}. \emph{\aap}
  390:81--89, \doi{10.1051/0004-6361:20020722}, \eprint{astro-ph/0205230}

\bibitem[{{Amati} et~al.(2008){Amati}, {Guidorzi}, {Frontera}, {Della Valle},
  {Finelli}, {Landi}, and {Montanari}}]{Amati08}
{Amati} L., {Guidorzi} C., {Frontera} F., {Della Valle} M., {Finelli} F.,
  {Landi} R., {Montanari} E. (2008) {Measuring the cosmological parameters with
  the E$_{p,i}$-E$_{iso}$ correlation of gamma-ray bursts}. \emph{\mnras}
  391(2):577--584, \doi{10.1111/j.1365-2966.2008.13943.x}, \eprint{0805.0377}

\bibitem[{{Amati} et~al.(2009){Amati}, {Frontera}, and {Guidorzi}}]{Amati09}
{Amati} L., {Frontera} F., {Guidorzi} C. (2009) {Extremely energetic Fermi
  gamma-ray bursts obey spectral energy correlations}. \emph{\aap}
  508(1):173--180, \doi{10.1051/0004-6361/200912788}, \eprint{0907.0384}

\bibitem[{{Amati} et~al.(2018){Amati}, {O'Brien}, {G{\"o}tz}, {Bozzo},
  {Tenzer}, {Frontera}, {Ghirlanda}, {Labanti}, {Osborne}, {Stratta}, {Tanvir},
  {Willingale}, {Attina}, {Campana}, {Castro-Tirado}, {Contini}, {Fuschino},
  {Gomboc}, {Hudec}, {Orleanski}, {Renotte}, {Rodic}, {Bagoly}, {Blain},
  {Callanan}, {Covino}, {Ferrara}, {Le Floch}, {Marisaldi}, {Mereghetti},
  {Rosati}, {Vacchi}, {D'Avanzo}, {Giommi}, {Piranomonte}, {Piro}, {Reglero},
  {Rossi}, {Santangelo}, {Salvaterra}, {Tagliaferri}, {Vergani}, {Vinciguerra},
  {Briggs}, {Campolongo}, {Ciolfi}, {Connaughton}, {Cordier}, {Morelli},
  {Orlandini}, {Adami}, {Argan}, {Atteia}, {Auricchio}, {Balazs}, {Baldazzi},
  {Basa}, {Basak}, {Bellutti}, {Bernardini}, {Bertuccio}, {Braga}, {Branchesi},
  {Brandt}, {Brocato}, {Budtz-Jorgensen}, {Bulgarelli}, {Burderi}, {Camp},
  {Capozziello}, {Caruana}, {Casella}, {Cenko}, {Chardonnet}, {Ciardi},
  {Colafrancesco}, {Dainotti}, {D'Elia}, {De Martino}, {De Pasquale}, {Del
  Monte}, {Della Valle}, {Drago}, {Evangelista}, {Feroci}, {Finelli},
  {Fiorini}, {Fynbo}, {Gal-Yam}, {Gendre}, {Ghisellini}, {Grado}, {Guidorzi},
  {Hafizi}, {Hanlon}, {Hjorth}, {Izzo}, {Kiss}, {Kumar}, {Kuvvetli}, {Lavagna},
  {Li}, {Longo}, {Lyutikov}, {Maio}, {Maiorano}, {Malcovati}, {Malesani},
  {Margutti}, {Martin-Carrillo}, {Masetti}, {McBreen}, {Mignani}, {Morgante},
  {Mundell}, {Nargaard-Nielsen}, {Nicastro}, {Palazzi}, {Paltani}, {Panessa},
  {Pareschi}, {Pe'er}, {Penacchioni}, {Pian}, {Piedipalumbo}, {Piran}, {Rauw},
  {Razzano}, {Read}, {Rezzolla}, {Romano}, {Ruffini}, {Savaglio}, {Sguera},
  {Schady}, {Skidmore}, {Song}, {Stanway}, {Starling}, {Topinka}, {Troja}, {van
  Putten}, {Vanzella}, {Vercellone}, {Wilson-Hodge}, {Yonetoku}, {Zampa},
  {Zampa}, {Zhang}, {Zhang}, {Zhang}, {Zhang}, {Antonelli}, {Bianco}, {Boci},
  {Boer}, {Botticella}, {Boulade}, {Butler}, {Campana}, {Capitanio}, {Celotti},
  {Chen}, {Colpi}, {Comastri}, {Cuby}, {Dadina}, {De Luca}, {Dong}, {Ettori},
  {Gandhi}, {Geza}, {Greiner}, {Guiriec}, {Harms}, {Hernanz}, {Hornstrup},
  {Hutchinson}, {Israel}, {Jonker}, {Kaneko}, {Kawai}, {Wiersema}, {Korpela},
  {Lebrun}, {Lu}, {MacFadyen}, {Malaguti}, {Maraschi}, {Melandri}, {Modjaz},
  {Morris}, {Omodei}, {Paizis}, {P{\'a}ta}, {Petrosian}, {Rachevski}, {Rhoads},
  {Ryde}, {Sabau-Graziati}, {Shigehiro}, {Sims}, {Soomin}, {Sz{\'e}csi},
  {Urata}, {Uslenghi}, {Valenziano}, {Vianello}, {Vojtech}, {Watson}, and
  {Zicha}}]{Amati18}
{Amati} L., {O'Brien} P., {G{\"o}tz} D., {Bozzo} E., {Tenzer} C., {Frontera}
  F., {Ghirlanda} G., {Labanti} C., et~al. (2018) {The THESEUS space mission
  concept: science case, design and expected performances}. \emph{Advances in
  Space Research} 62(1):191--244, \doi{10.1016/j.asr.2018.03.010},
  \eprint{1710.04638}

\bibitem[{{Amati} et~al.(2019){Amati}, {D'Agostino}, {Luongo}, {Muccino}, and
  {Tantalo}}]{Amati19}
{Amati} L., {D'Agostino} R., {Luongo} O., {Muccino} M., {Tantalo} M. (2019)
  {Addressing the circularity problem in the E$_{p}$-E$_{iso}$ correlation of
  gamma-ray bursts}. \emph{\mnras} 486(1):L46--L51,
  \doi{10.1093/mnrasl/slz056}, \eprint{1811.08934}

\bibitem[{Amendola and Quartin(2021)}]{Amendola:2019lvy}
Amendola L., Quartin M. (2021) {Measuring the Hubble function with standard
  candle clustering}. \emph{Mon Not Roy Astron Soc} 504(3):3884--3889,
  \doi{10.1093/mnras/stab887}, \eprint{1912.10255}

\bibitem[{Amendola et~al.(2005)Amendola, Quercellini, and
  Giallongo}]{Amendola:2004be}
Amendola L., Quercellini C., Giallongo E. (2005) {Constraints on perfect fluid
  and scalar field dark energy models from future redshift surveys}. \emph{Mon
  Not Roy Astron Soc} 357:429--439, \doi{10.1111/j.1365-2966.2004.08558.x},
  \eprint{astro-ph/0404599}

\bibitem[{Amendola et~al.(2018)Amendola, Kunz, Saltas, and
  Sawicki}]{Amendola:2017orw}
Amendola L., Kunz M., Saltas I.~D., Sawicki I. (2018) {The fate of large-scale
  structure in modified gravity after GW170817 and GRB170817A}. \emph{Phys Rev
  Lett} 120(13):131101, \doi{10.1103/PhysRevLett.120.131101},
  \eprint{1711.04825}

\bibitem[{{Amendola} et~al.(2022){Amendola}, {Pietroni}, and
  {Quartin}}]{Amendola:2022vte}
{Amendola} L., {Pietroni} M., {Quartin} M. (2022) {Fisher matrix for the
  one-loop galaxy power spectrum: measuring expansion and growth rates without
  assuming a cosmological model}. \emph{arXiv e-prints} arXiv:2205.00569,
  \eprint{2205.00569}

\bibitem[{{Amon} et~al.(2021){Amon}, {Gruen}, {Troxel}, {MacCrann}, {Dodelson},
  {Choi}, {Doux}, {Secco}, {Samuroff}, {Krause}, {Cordero}, {Myles}, and
  et~al.}]{Amon2021}
{Amon} A., {Gruen} D., {Troxel} M.~A., {MacCrann} N., {Dodelson} S., {Choi} A.,
  {Doux} C., {Secco} L.~F., et~al. (2021) {Dark Energy Survey Year 3 Results:
  Cosmology from Cosmic Shear and Robustness to Data Calibration}. \emph{arXiv
  e-prints} arXiv:2105.13543, \eprint{2105.13543}

\bibitem[{Anderson et~al.(2018)}]{Anderson:2017ert}
Anderson C., et~al. (2018) {Low-amplitude clustering in low-redshift 21-cm
  intensity maps cross-correlated with 2dF galaxy densities}. \emph{\mnras}
  476(3):3382--3392, \doi{10.1093/mnras/sty346}, \eprint{1710.00424}

\bibitem[{{Angora} et~al.(2020){Angora}, {Rosati}, {Brescia}, {Mercurio},
  {Grillo}, {Caminha}, {Meneghetti}, {Nonino}, {Vanzella}, {Bergamini},
  {Biviano}, and {Lombardi}}]{angora:2020}
{Angora} G., {Rosati} P., {Brescia} M., {Mercurio} A., {Grillo} C., {Caminha}
  G., {Meneghetti} M., {Nonino} M., et~al. (2020) {The search for galaxy
  cluster members with deep learning of panchromatic HST imaging and extensive
  spectroscopy}. \emph{\aap} 643:A177, \doi{10.1051/0004-6361/202039083},
  \eprint{2009.08224}

\bibitem[{{Annunziatella} et~al.(2017){Annunziatella}, {Bonamigo}, {Grillo},
  {Mercurio}, {Rosati}, {Caminha}, {Biviano}, {Girardi}, {Gobat}, {Lombardi},
  and {Munari}}]{Annunziatella:2017}
{Annunziatella} M., {Bonamigo} M., {Grillo} C., {Mercurio} A., {Rosati} P.,
  {Caminha} G., {Biviano} A., {Girardi} M., et~al. (2017) {Mass Profile
  Decomposition of the Frontier Fields Cluster MACS J0416-2403: Insights on the
  Dark-matter Inner Profile}. \emph{\apj} 851(2):81,
  \doi{10.3847/1538-4357/aa9845}, \eprint{1711.02109}

\bibitem[{Ansari et~al.(2012)Ansari, Campagne, Colom, Goff, Magneville, Martin,
  Moniez, Rich, and Yeche}]{Ansari:2011bv}
Ansari R., Campagne J.~E., Colom P., Goff J. M.~L., Magneville C., Martin
  J.~M., Moniez M., Rich J., et~al. (2012) {21 cm observation of LSS at
  z\textasciitilde{}1 Instrument sensitivity and foreground subtraction}.
  \emph{Astron Astrophys} 540:A129, \doi{10.1051/0004-6361/201117837},
  \eprint{1108.1474}

\bibitem[{{Arjona} and {Nesseris}(2021)}]{arjona2021}
{Arjona} R., {Nesseris} S. (2021) {Novel null tests for the spatial curvature
  and homogeneity of the Universe and their machine learning reconstructions}.
  \emph{\prd} 103(10):103539, \doi{10.1103/PhysRevD.103.103539},
  \eprint{2103.06789}

\bibitem[{{Arnouts} et~al.(2013){Arnouts}, {Le Floc'h}, {Chevallard},
  {Johnson}, {Ilbert}, {Treyer}, {Aussel}, {Capak}, {Sanders}, {Scoville},
  {McCracken}, {Milliard}, {Pozzetti}, and {Salvato}}]{arnouts2013}
{Arnouts} S., {Le Floc'h} E., {Chevallard} J., {Johnson} B.~D., {Ilbert} O.,
  {Treyer} M., {Aussel} H., {Capak} P., et~al. (2013) {Encoding of the infrared
  excess in the NUVrK color diagram for star-forming galaxies}. \emph{\aap}
  558:A67, \doi{10.1051/0004-6361/201321768}, \eprint{1309.0008}

\bibitem[{{Asgari} et~al.(2020){Asgari}, {Tr{\"o}ster}, {Heymans},
  {Hildebrandt}, {van den Busch}, {Wright}, {Choi}, {Erben}, {Joachimi},
  {Joudaki}, {Kannawadi}, {Kuijken}, {Lin}, {Schneider}, and
  {Zuntz}}]{Asgari2020}
{Asgari} M., {Tr{\"o}ster} T., {Heymans} C., {Hildebrandt} H., {van den Busch}
  J.~L., {Wright} A.~H., {Choi} A., {Erben} T., et~al. (2020) {KiDS+VIKING-450
  and DES-Y1 combined: Mitigating baryon feedback uncertainty with COSEBIs}.
  \emph{\aap} 634:A127, \doi{10.1051/0004-6361/201936512}, \eprint{1910.05336}

\bibitem[{{Asgari} et~al.(2021){Asgari}, {Lin}, {Joachimi}, {Giblin},
  {Heymans}, {Hildebrandt}, {Kannawadi}, {St{\"o}lzner}, {Tr{\"o}ster}, {van
  den Busch}, {Wright}, {Bilicki}, {Blake}, {de Jong}, {Dvornik}, {Erben},
  {Getman}, {Hoekstra}, {K{\"o}hlinger}, {Kuijken}, {Miller}, {Radovich},
  {Schneider}, {Shan}, and {Valentijn}}]{Asgari2021}
{Asgari} M., {Lin} C.-A., {Joachimi} B., {Giblin} B., {Heymans} C.,
  {Hildebrandt} H., {Kannawadi} A., {St{\"o}lzner} B., et~al. (2021) {KiDS-1000
  cosmology: Cosmic shear constraints and comparison between two point
  statistics}. \emph{\aap} 645:A104, \doi{10.1051/0004-6361/202039070},
  \eprint{2007.15633}

\bibitem[{{Ashton} et~al.(2021){Ashton}, {Ackley}, {Hernandez}, and
  {Piotrzkowski}}]{Ashton:2020kyr}
{Ashton} G., {Ackley} K., {Hernandez} I.~M., {Piotrzkowski} B. (2021) {Current
  observations are insufficient to confidently associate the binary black hole
  merger GW190521 with AGN J124942.3 + 344929}. \emph{Classical and Quantum
  Gravity} 38(23):235004, \doi{10.1088/1361-6382/ac33bb}, \eprint{2009.12346}

\bibitem[{{Astier} et~al.(2006){Astier}, {Guy}, {Regnault}, {Pain}, {Aubourg},
  {Balam}, {Basa}, {Carlberg}, {Fabbro}, {Fouchez}, {Hook}, {Howell}, {Lafoux},
  {Neill}, {Palanque-Delabrouille}, {Perrett}, and et~al.}]{astier2006}
{Astier} P., {Guy} J., {Regnault} N., {Pain} R., {Aubourg} E., {Balam} D.,
  {Basa} S., {Carlberg} R.~G., et~al. (2006) {The Supernova Legacy Survey:
  measurement of {\ensuremath{\Omega}}$_{M}$,
  {\ensuremath{\Omega}}$_{{\ensuremath{\Lambda}}}$ and w from the first year
  data set}. \emph{\aap} 447(1):31--48, \doi{10.1051/0004-6361:20054185},
  \eprint{astro-ph/0510447}

\bibitem[{{Aubert} et~al.(2020){Aubert}, {Cousinou}, {Escoffier}, {Hawken},
  {Nadathur}, {Alam}, {Bautista}, {Burtin}, {de Mattia}, {Gil-Mar{\'\i}n},
  {Hou}, {Jullo}, {Neveux}, {Rossi}, {Smith}, {Tamone}, and {Vargas
  Maga{\~n}a}}]{Aubert2020}
{Aubert} M., {Cousinou} M.-C., {Escoffier} S., {Hawken} A.~J., {Nadathur} S.,
  {Alam} S., {Bautista} J., {Burtin} E., et~al. (2020) {The Completed SDSS-IV
  Extended Baryon Oscillation Spectroscopic Survey: Growth rate of structure
  measurement from cosmic voids}. \emph{arXiv e-prints} arXiv:2007.09013,
  \eprint{2007.09013}

\bibitem[{{Auger} et~al.(2009){Auger}, {Treu}, {Bolton}, {Gavazzi}, {Koopmans},
  {Marshall}, {Bundy}, and {Moustakas}}]{Auger:2009}
{Auger} M.~W., {Treu} T., {Bolton} A.~S., {Gavazzi} R., {Koopmans} L.~V.~E.,
  {Marshall} P.~J., {Bundy} K., {Moustakas} L.~A. (2009) {The Sloan Lens ACS
  Survey. IX. Colors, Lensing, and Stellar Masses of Early-Type Galaxies}.
  \emph{\apj} 705(2):1099--1115, \doi{10.1088/0004-637X/705/2/1099},
  \eprint{0911.2471}

\bibitem[{{Avila} et~al.(2022){Avila}, {Vos-Gin{\'e}s}, {Cunnington},
  {Stevens}, {Yepes}, {Knebe}, and {Chuang}}]{Avila:2021wih}
{Avila} S., {Vos-Gin{\'e}s} B., {Cunnington} S., {Stevens} A. R.~H., {Yepes}
  G., {Knebe} A., {Chuang} C.-H. (2022) {H I IM correlation function from UNIT
  simulations: BAO and observationally induced anisotropy}. \emph{\mnras}
  510(1):292--308, \doi{10.1093/mnras/stab3406}, \eprint{2105.10454}

\bibitem[{{Avni} and {Tananbaum}(1982)}]{avnitananbaum82}
{Avni} Y., {Tananbaum} H. (1982) {On the cosmological evolution of the X-ray
  emission from quasars}. \emph{\apjl} 262:L17--L21, \doi{10.1086/183903}

\bibitem[{{Ayuso} et~al.(2021){Ayuso}, {Lazkoz}, and {Salzano}}]{ayuso2021}
{Ayuso} I., {Lazkoz} R., {Salzano} V. (2021) {Observational constraints on
  cosmological solutions of f (Q ) theories}. \emph{\prd} 103(6):063505,
  \doi{10.1103/PhysRevD.103.063505}, \eprint{2012.00046}

\bibitem[{{Azzam}(2012)}]{Azzam2012}
{Azzam} W.~J. (2012) {Dependence of the GRB Lag-Luminosity Relation on Redshift
  in the Source Frame}. \emph{International Journal of Astronomy and
  Astrophysics} 2(1):1--5, \doi{10.4236/ijaa.2012.21001}, \eprint{1203.6530}

\bibitem[{{Bailoni} et~al.(2017){Bailoni}, {Spurio Mancini}, and
  {Amendola}}]{2017MNRAS.470..688B}
{Bailoni} A., {Spurio Mancini} A., {Amendola} L. (2017) {Improving Fisher
  matrix forecasts for galaxy surveys: window function, bin cross-correlation
  and bin redshift uncertainty}. \emph{\mnras} 470(1):688--705,
  \doi{10.1093/mnras/stx1209}, \eprint{1608.00458}

\bibitem[{{Baker} et~al.(2018){Baker}, {Clampitt}, {Jain}, and
  {Trodden}}]{Baker2018}
{Baker} T., {Clampitt} J., {Jain} B., {Trodden} M. (2018) {Void lensing as a
  test of gravity}. \emph{\prd} 98(2):023511, \doi{10.1103/PhysRevD.98.023511},
  \eprint{1803.07533}

\bibitem[{Baker et~al.(2017)}]{Baker2017}
Baker T., et~al. (2017) {Strong Constraints on Cosmological Gravity from
  GW170817 and GRB 170817A}. \emph{Physical Review Letters} 119(25):251301,
  \doi{10.1103/PhysRevLett.119.251301}

\bibitem[{{Baldauf} et~al.(2013){Baldauf}, {Seljak}, {Smith}, {Hamaus}, and
  {Desjacques}}]{2013PhRvD..88h3507B}
{Baldauf} T., {Seljak} U., {Smith} R.~E., {Hamaus} N., {Desjacques} V. (2013)
  {Halo stochasticity from exclusion and nonlinear clustering}. \emph{Phys Rev
  D} 88(8):083507, \doi{10.1103/PhysRevD.88.083507}, \eprint{1305.2917}

\bibitem[{{Baldi} and {Villaescusa-Navarro}(2018)}]{Baldi2018}
{Baldi} M., {Villaescusa-Navarro} F. (2018) {Cosmic degeneracies - II.
  Structure formation in joint simulations of warm dark matter and f(R)
  gravity}. \emph{\mnras} 473:3226--3240, \doi{10.1093/mnras/stx2594}

\bibitem[{{Baldry} et~al.(2004){Baldry}, {Glazebrook}, {Brinkmann},
  {Ivezi{\'c}}, {Lupton}, {Nichol}, and {Szalay}}]{baldry2004}
{Baldry} I.~K., {Glazebrook} K., {Brinkmann} J., {Ivezi{\'c}} {\v{Z}}.,
  {Lupton} R.~H., {Nichol} R.~C., {Szalay} A.~S. (2004) {Quantifying the
  Bimodal Color-Magnitude Distribution of Galaxies}. \emph{\apj}
  600(2):681--694, \doi{10.1086/380092}, \eprint{astro-ph/0309710}

\bibitem[{{Baldry} et~al.(2006){Baldry}, {Balogh}, {Bower}, {Glazebrook},
  {Nichol}, {Bamford}, and {Budavari}}]{baldry2006}
{Baldry} I.~K., {Balogh} M.~L., {Bower} R.~G., {Glazebrook} K., {Nichol} R.~C.,
  {Bamford} S.~P., {Budavari} T. (2006) {Galaxy bimodality versus stellar mass
  and environment}. \emph{\mnras} 373(2):469--483,
  \doi{10.1111/j.1365-2966.2006.11081.x}, \eprint{astro-ph/0607648}

\bibitem[{{Baldry} et~al.(2008){Baldry}, {Glazebrook}, and
  {Driver}}]{baldry2008}
{Baldry} I.~K., {Glazebrook} K., {Driver} S.~P. (2008) {On the galaxy stellar
  mass function, the mass-metallicity relation and the implied baryonic mass
  function}. \emph{\mnras} 388(3):945--959,
  \doi{10.1111/j.1365-2966.2008.13348.x}, \eprint{0804.2892}

\bibitem[{{Balestra} et~al.(2016){Balestra}, {Mercurio}, {Sartoris}, {Girardi},
  {Grillo}, {Nonino}, {Rosati}, {Biviano}, {Ettori}, {Forman}, {Jones},
  {Koekemoer}, {Medezinski}, {Merten}, and et~al.}]{Balestra:2016}
{Balestra} I., {Mercurio} A., {Sartoris} B., {Girardi} M., {Grillo} C.,
  {Nonino} M., {Rosati} P., {Biviano} A., et~al. (2016) {CLASH-VLT: Dissecting
  the Frontier Fields Galaxy Cluster MACS J0416.1-2403 with
  {\ensuremath{\sim}}800 Spectra of Member Galaxies}. \emph{\apjs} 224(2):33,
  \doi{10.3847/0067-0049/224/2/33}, \eprint{1511.02522}

\bibitem[{{Ballinger} et~al.(1996){Ballinger}, {Peacock}, and
  {Heavens}}]{1996MNRAS.282..877B}
{Ballinger} W.~E., {Peacock} J.~A., {Heavens} A.~F. (1996) {Measuring the
  cosmological constant with redshift surveys}. \emph{\mnras} 282:877,
  \doi{10.1093/mnras/282.3.877}, \eprint{astro-ph/9605017}

\bibitem[{{Balogh} et~al.(1999){Balogh}, {Morris}, {Yee}, {Carlberg}, and
  {Ellingson}}]{balogh1999}
{Balogh} M.~L., {Morris} S.~L., {Yee} H.~K.~C., {Carlberg} R.~G., {Ellingson}
  E. (1999) {Differential Galaxy Evolution in Cluster and Field Galaxies at
  z\~{}0.3}. \emph{\apj} 527:54--79, \doi{10.1086/308056},
  \eprint{astro-ph/9906470}

\bibitem[{{Band} et~al.(1993){Band}, {Matteson}, {Ford}, {Schaefer}, {Palmer},
  {Teegarden}, {Cline}, {Briggs}, {Paciesas}, {Pendleton}, {Fishman},
  {Kouveliotou}, {Meegan}, {Wilson}, and {Lestrade}}]{Band93}
{Band} D., {Matteson} J., {Ford} L., {Schaefer} B., {Palmer} D., {Teegarden}
  B., {Cline} T., {Briggs} M., et~al. (1993) {BATSE Observations of Gamma-Ray
  Burst Spectra. I. Spectral Diversity}. \emph{\apj} 413:281,
  \doi{10.1086/172995}

\bibitem[{{Band} and {Preece}(2005)}]{Band05}
{Band} D.~L., {Preece} R.~D. (2005) {Testing the Gamma-Ray Burst Energy
  Relationships}. \emph{\apj} 627(1):319--323, \doi{10.1086/430402},
  \eprint{astro-ph/0501559}

\bibitem[{Bandura et~al.(2014)}]{Bandura:2014gwa}
Bandura K., et~al. (2014) {Canadian Hydrogen Intensity Mapping Experiment
  (CHIME) Pathfinder}. \emph{Proc SPIE Int Soc Opt Eng} 9145:22,
  \doi{10.1117/12.2054950}, \eprint{1406.2288}

\bibitem[{{Banerjee} and {Dalal}(2016)}]{Banerjee2016}
{Banerjee} A., {Dalal} N. (2016) {Simulating nonlinear cosmological structure
  formation with massive neutrinos}. \emph{\jcap} 11:015,
  \doi{10.1088/1475-7516/2016/11/015}, \eprint{1606.06167}

\bibitem[{{Bargiacchi} et~al.(2021{\natexlab{a}}){Bargiacchi}, {Benetti},
  {Capozziello}, {Lusso}, {Risaliti}, and {Signorini}}]{bargiacchi2021arXiv}
{Bargiacchi} G., {Benetti} M., {Capozziello} S., {Lusso} E., {Risaliti} G.,
  {Signorini} M. (2021{\natexlab{a}}) {Quasar cosmology: dark energy evolution
  and spatial curvature}. \emph{arXiv e-prints} arXiv:2111.02420,
  \eprint{2111.02420}

\bibitem[{{Bargiacchi} et~al.(2021{\natexlab{b}}){Bargiacchi}, {Risaliti},
  {Benetti}, {Capozziello}, {Lusso}, {Saccardi}, and
  {Signorini}}]{bargiacchi2021}
{Bargiacchi} G., {Risaliti} G., {Benetti} M., {Capozziello} S., {Lusso} E.,
  {Saccardi} A., {Signorini} M. (2021{\natexlab{b}}) {Cosmography by
  orthogonalized logarithmic polynomials}. \emph{arXiv e-prints}
  arXiv:2101.08278, \eprint{2101.08278}

\bibitem[{Barreira(2021)}]{Barreira:2021dpt}
Barreira A. (2021) {The local PNG bias of neutral Hydrogen, ${\rm H_I}$}.
  \eprint{2112.03253}

\bibitem[{{Barreira} et~al.(2015){Barreira}, {Cautun}, {Li}, {Baugh}, and
  {Pascoli}}]{Barreira2015}
{Barreira} A., {Cautun} M., {Li} B., {Baugh} C.~M., {Pascoli} S. (2015) {Weak
  lensing by voids in modified lensing potentials}. \emph{\jcap} 2015(8):028,
  \doi{10.1088/1475-7516/2015/08/028}, \eprint{1505.05809}

\bibitem[{{Bartelmann} and {Schneider}(2001)}]{bartelmann2001}
{Bartelmann} M., {Schneider} P. (2001) {Weak gravitational lensing}.
  \emph{\physrep} 340(4-5):291--472, \doi{10.1016/S0370-1573(00)00082-X},
  \eprint{astro-ph/9912508}

\bibitem[{{Basak} and {Rao}(2013)}]{Basak13}
{Basak} R., {Rao} A.~R. (2013) {Pulse-wise Amati correlation in Fermi gamma-ray
  bursts}. \emph{\mnras} 436(4):3082--3088, \doi{10.1093/mnras/stt1790},
  \eprint{1309.5233}

\bibitem[{{Basilakos} and {Nesseris}(2017)}]{basilakos2017}
{Basilakos} S., {Nesseris} S. (2017) {Conjoined constraints on modified gravity
  from the expansion history and cosmic growth}. \emph{\prd} 96(6):063517,
  \doi{10.1103/PhysRevD.96.063517}, \eprint{1705.08797}

\bibitem[{Battye et~al.(2004)Battye, Davies, and Weller}]{Battye:2004re}
Battye R.~A., Davies R.~D., Weller J. (2004) {Neutral hydrogen surveys for high
  redshift galaxy clusters and proto-clusters}. \emph{Mon Not Roy Astron Soc}
  355:1339--1347, \doi{10.1111/j.1365-2966.2004.08416.x},
  \eprint{astro-ph/0401340}

\bibitem[{Battye et~al.(2013)Battye, Browne, Dickinson, Heron, Maffei, and
  Pourtsidou}]{Battye:2012tg}
Battye R.~A., Browne I. W.~A., Dickinson C., Heron G., Maffei B., Pourtsidou A.
  (2013) {HI intensity mapping : a single dish approach}. \emph{Mon Not Roy
  Astron Soc} 434:1239--1256, \doi{10.1093/mnras/stt1082}, \eprint{1209.0343}

\bibitem[{{Bautista} et~al.(2021){Bautista}, {Paviot}, {Vargas Maga{\~n}a}, {de
  la Torre}, {Fromenteau}, {Gil-Mar{\'\i}n}, {Ross}, {Burtin}, {Dawson}, {Hou},
  {Kneib}, {de Mattia}, {Percival}, {Rossi}, {Tojeiro}, {Zhao}, {Zhao}, {Alam},
  {Brownstein}, {Chapman}, {Choi}, {Chuang}, {Escoffier}, {de la Macorra}, {du
  Mas des Bourboux}, {Mohammad}, {Moon}, {M{\"u}ller}, {Nadathur}, {Newman},
  {Schneider}, {Seo}, and {Wang}}]{bautista2021}
{Bautista} J.~E., {Paviot} R., {Vargas Maga{\~n}a} M., {de la Torre} S.,
  {Fromenteau} S., {Gil-Mar{\'\i}n} H., {Ross} A.~J., {Burtin} E., et~al.
  (2021) {The completed SDSS-IV extended Baryon Oscillation Spectroscopic
  Survey: measurement of the BAO and growth rate of structure of the luminous
  red galaxy sample from the anisotropic correlation function between redshifts
  0.6 and 1}. \emph{\mnras} 500(1):736--762, \doi{10.1093/mnras/staa2800},
  \eprint{2007.08993}

\bibitem[{{Bayer} et~al.(2021){Bayer}, {Villaescusa-Navarro}, {Massara}, {Liu},
  {Spergel}, {Verde}, {Wandelt}, {Viel}, and {Ho}}]{Bayer2021}
{Bayer} A.~E., {Villaescusa-Navarro} F., {Massara} E., {Liu} J., {Spergel}
  D.~N., {Verde} L., {Wandelt} B.~D., {Viel} M., et~al. (2021) {Detecting
  Neutrino Mass by Combining Matter Clustering, Halos, and Voids}. \emph{\apj}
  919(1):24, \doi{10.3847/1538-4357/ac0e91}, \eprint{2102.05049}

\bibitem[{{Becker} et~al.(1995){Becker}, {White}, and {Helfand}}]{becker1995}
{Becker} R.~H., {White} R.~L., {Helfand} D.~J. (1995) {The FIRST Survey: Faint
  Images of the Radio Sky at Twenty Centimeters}. \emph{\apj} 450:559,
  \doi{10.1086/176166}

\bibitem[{{Beisbart} and {Kerscher}(2000)}]{Beisbart2000}
{Beisbart} C., {Kerscher} M. (2000) {Luminosity- and Morphology-dependent
  Clustering of Galaxies}. \emph{\apj} 545(1):6--25, \doi{10.1086/317788},
  \eprint{astro-ph/0003358}

\bibitem[{{Belli} et~al.(2017){Belli}, {Genzel}, {F{\"o}rster Schreiber},
  {Wisnioski}, {Wilman}, {Wuyts}, {Mendel}, {Beifiori}, {Bender}, {Brammer},
  {Burkert}, {Chan}, {Davies}, {Davies}, {Fabricius}, {Fossati}, {Galametz},
  {Lang}, {Lutz}, {Momcheva}, {Nelson}, {Saglia}, {Tacconi}, {Tadaki},
  {{\"U}bler}, and {van Dokkum}}]{belli2017}
{Belli} S., {Genzel} R., {F{\"o}rster Schreiber} N.~M., {Wisnioski} E.,
  {Wilman} D.~J., {Wuyts} S., {Mendel} J.~T., {Beifiori} A., et~al. (2017)
  {KMOS$^{3D}$ Reveals Low-level Star Formation Activity in Massive Quiescent
  Galaxies at 0.7 < z < 2.7}. \emph{\apjl} 841(1):L6,
  \doi{10.3847/2041-8213/aa70e5}, \eprint{1703.07778}

\bibitem[{{Belli} et~al.(2019){Belli}, {Newman}, and {Ellis}}]{belli2019}
{Belli} S., {Newman} A.~B., {Ellis} R.~S. (2019) {MOSFIRE Spectroscopy of
  Quiescent Galaxies at 1.5 \&lt; z \&lt; 2.5. II. Star Formation Histories and
  Galaxy Quenching}. \emph{\apj} 874(1):17, \doi{10.3847/1538-4357/ab07af},
  \eprint{1810.00008}

\bibitem[{{Benetti} and {Capozziello}(2019)}]{benetti2019}
{Benetti} M., {Capozziello} S. (2019) {Connecting early and late epochs by
  f(z)CDM cosmography}. \emph{\jcap} 2019(12):008,
  \doi{10.1088/1475-7516/2019/12/008}, \eprint{1910.09975}

\bibitem[{{Benetti} et~al.(2021){Benetti}, {Lobato Graef}, and
  {Vagnozzi}}]{benetti2021}
{Benetti} M., {Lobato Graef} L., {Vagnozzi} S. (2021) {Primordial gravitational
  waves from NANOGrav: a broken power-law approach}. \emph{arXiv e-prints}
  arXiv:2111.04758, \eprint{2111.04758}

\bibitem[{{Bennett} et~al.(2003){Bennett}, {Halpern}, {Hinshaw}, {Jarosik},
  {Kogut}, {Limon}, {Meyer}, {Page}, {Spergel}, {Tucker}, {Wollack}, {Wright},
  {Barnes}, {Greason}, {Hill}, {Komatsu}, {Nolta}, {Odegard}, {Peiris},
  {Verde}, and {Weiland}}]{WMAP}
{Bennett} C.~L., {Halpern} M., {Hinshaw} G., {Jarosik} N., {Kogut} A., {Limon}
  M., {Meyer} S.~S., {Page} L., et~al. (2003) {First-Year Wilkinson Microwave
  Anisotropy Probe (WMAP) Observations: Preliminary Maps and Basic Results}.
  \emph{\apjs} 148(1):1--27, \doi{10.1086/377253}, \eprint{astro-ph/0302207}

\bibitem[{{Bera} et~al.(2020){Bera}, {Rana}, {More}, and
  {Bose}}]{2020ApJ...902...79B}
{Bera} S., {Rana} D., {More} S., {Bose} S. (2020) {Incompleteness Matters Not:
  Inference of H$_{0}$ from Binary Black Hole-Galaxy Cross-correlations}.
  \emph{\apj} 902(1):79, \doi{10.3847/1538-4357/abb4e0}, \eprint{2007.04271}

\bibitem[{{Bergamini} et~al.(2021{\natexlab{a}}){Bergamini}, {Agnello}, and
  {Caminha}}]{Bergamini:2021b}
{Bergamini} P., {Agnello} A., {Caminha} G.~B. (2021{\natexlab{a}}) {Cluster
  strong lensing with hierarchical inference. Formalism, functional tests, and
  public code release}. \emph{\aap} 648:A123,
  \doi{10.1051/0004-6361/201937138}, \eprint{2008.11728}

\bibitem[{{Bergamini} et~al.(2021{\natexlab{b}}){Bergamini}, {Rosati},
  {Vanzella}, {Caminha}, {Grillo}, {Mercurio}, {Meneghetti}, {Angora},
  {Calura}, {Nonino}, and {Tozzi}}]{bergamini:2021}
{Bergamini} P., {Rosati} P., {Vanzella} E., {Caminha} G.~B., {Grillo} C.,
  {Mercurio} A., {Meneghetti} M., {Angora} G., et~al. (2021{\natexlab{b}}) {A
  new high-precision strong lensing model of the galaxy cluster MACS
  J0416.1-2403. Robust characterization of the cluster mass distribution from
  VLT/MUSE deep observations}. \emph{\aap} 645:A140,
  \doi{10.1051/0004-6361/202039564}, \eprint{2010.00027}

\bibitem[{{Bernal} et~al.(2016){Bernal}, {Verde}, and {Riess}}]{Bernal2016}
{Bernal} J.~L., {Verde} L., {Riess} A.~G. (2016) {The trouble with H$_{0}$}.
  \emph{\jcap} 2016(10):019, \doi{10.1088/1475-7516/2016/10/019},
  \eprint{1607.05617}

\bibitem[{Bernal et~al.(2019)Bernal, Breysse, Gil-Mar\'\i{}n, and
  Kovetz}]{Bernal:2019jdo}
Bernal J.~L., Breysse P.~C., Gil-Mar\'\i{}n H., Kovetz E.~D. (2019)
  {User\textquoteright{}s guide to extracting cosmological information from
  line-intensity maps}. \emph{Phys Rev D} 100(12):123522,
  \doi{10.1103/PhysRevD.100.123522}, \eprint{1907.10067}

\bibitem[{{Bernal} et~al.(2021){Bernal}, {Verde}, {Jimenez}, {Kamionkowski},
  {Valcin}, and {Wandelt}}]{Triangles}
{Bernal} J.~L., {Verde} L., {Jimenez} R., {Kamionkowski} M., {Valcin} D.,
  {Wandelt} B.~D. (2021) {Trouble beyond H$_{0}$ and the new cosmic triangles}.
  \emph{\prd} 103(10):103533, \doi{10.1103/PhysRevD.103.103533},
  \eprint{2102.05066}

\bibitem[{{Bernardini} et~al.(2012){Bernardini}, {Margutti}, {Zaninoni}, and
  {Chincarini}}]{Bernardini12}
{Bernardini} M.~G., {Margutti} R., {Zaninoni} E., {Chincarini} G. (2012) {A
  universal scaling for short and long gamma-ray bursts: E$_{X,iso}$ -
  E$_{{\ensuremath{\gamma}},iso}$ - E$_{pk}$}. \emph{\mnras} 425(2):1199--1204,
  \doi{10.1111/j.1365-2966.2012.21487.x}, \eprint{1203.1060}

\bibitem[{{Bertrand} et~al.(2019){Bertrand}, {Cordier}, {Jianyan}, {Wei}, and
  {SVOM Collaboration}}]{Cordier19}
{Bertrand}, {Cordier}, {Jianyan}, {Wei}, {SVOM Collaboration} (2019) {The SVOM
  mission.} \emph{\memsai} 90:242

\bibitem[{{Bertschinger}(1985)}]{Bertschinger1985}
{Bertschinger} E. (1985) {The self-similar evolution of holes in an Einstein-de
  Sitter universe}. \emph{\apjs} 58:1--37, \doi{10.1086/191027}

\bibitem[{{Bessa} et~al.(2021){Bessa}, {Campista}, and {Bernui}}]{bessa2021}
{Bessa} P., {Campista} M., {Bernui} A. (2021) {Observational constraints on
  Starobinsky $f(R)$ cosmology from cosmic expansion and structure growth
  data}. \emph{arXiv e-prints} arXiv:2112.00822, \eprint{2112.00822}

\bibitem[{Betoule et~al.(2014)}]{Betoule:2014frx}
Betoule M., et~al. (2014) {Improved cosmological constraints from a joint
  analysis of the SDSS-II and SNLS supernova samples}. \emph{AstronAstrophys}
  568:A22, \doi{10.1051/0004-6361/201423413}, \eprint{1401.4064}

\bibitem[{Beutler et~al.(2012)Beutler, Blake et~al.}]{Beutler2012}
Beutler F., Blake M. C.and~Colless, et~al. (2012) {The 6dF Galaxy Survey: z=0
  measurements of the growth rate and $\sigma$8}. \emph{Mon Not Roy Astron Soc}
  423(4):3430--3444, \doi{10.1111/j.1365-2966.2012.21136.x}

\bibitem[{Bigot-Sazy et~al.(2015)Bigot-Sazy, Dickinson, Battye, Browne, Ma,
  Maffei, Noviello, Remazeilles, and Wilkinson}]{Bigot-Sazy:2015jaa}
Bigot-Sazy M.~A., Dickinson C., Battye R.~A., Browne I. W.~A., Ma Y.~Z., Maffei
  B., Noviello F., Remazeilles M., et~al. (2015) {Simulations for single-dish
  intensity mapping experiments}. \emph{Mon Not Roy Astron Soc}
  454(3):3240--3253, \doi{10.1093/mnras/stv2153}, \eprint{1507.04561}

\bibitem[{{Bilicki} and {Seikel}(2012)}]{bilicki2012}
{Bilicki} M., {Seikel} M. (2012) {We do not live in the R$_{h}$ = ct universe}.
  \emph{\mnras} 425(3):1664--1668, \doi{10.1111/j.1365-2966.2012.21575.x},
  \eprint{1206.5130}

\bibitem[{{Binney} and {Mamon}(1982)}]{Binney:1982}
{Binney} J., {Mamon} G.~A. (1982) {M/L and velocity anisotropy from
  observations of spherical galaxies, of must M 87 have a massive black hole ?}
  \emph{\mnras} 200:361--375, \doi{10.1093/mnras/200.2.361}

\bibitem[{{Binney} and {Tremaine}(2008)}]{BinneyTremaine:2008}
{Binney} J., {Tremaine} S. (2008) {Galactic Dynamics: Second Edition}.
  Princeton University Press

\bibitem[{{Birrer}(2021)}]{Birrer:2021curvedarcs}
{Birrer} S. (2021) {Gravitational lensing formalism in a curved arc basis: A
  continuous description of observables and degeneracies from the weak to the
  strong lensing regime}. \emph{arXiv e-prints} arXiv:2104.09522,
  \eprint{2104.09522}

\bibitem[{{Birrer} and {Amara}(2018)}]{Birrer:lenstronomy}
{Birrer} S., {Amara} A. (2018) {lenstronomy: Multi-purpose gravitational lens
  modelling software package}. \emph{Physics of the Dark Universe} 22:189--201,
  \doi{10.1016/j.dark.2018.11.002}, \eprint{1803.09746}

\bibitem[{{Birrer} and {Treu}(2021)}]{BirrerTreu:2021}
{Birrer} S., {Treu} T. (2021) {TDCOSMO. V. Strategies for precise and accurate
  measurements of the Hubble constant with strong lensing}. \emph{\aap}
  649:A61, \doi{10.1051/0004-6361/202039179}, \eprint{2008.06157}

\bibitem[{{Birrer} et~al.(2015){Birrer}, {Amara}, and
  {Refregier}}]{Birrer:2015}
{Birrer} S., {Amara} A., {Refregier} A. (2015) {Gravitational Lens Modeling
  with Basis Sets}. \emph{\apj} 813(2):102, \doi{10.1088/0004-637X/813/2/102},
  \eprint{1504.07629}

\bibitem[{{Birrer} et~al.(2016){Birrer}, {Amara}, and
  {Refregier}}]{Birrer:2016}
{Birrer} S., {Amara} A., {Refregier} A. (2016) {The mass-sheet degeneracy and
  time-delay cosmography: analysis of the strong lens RXJ1131-1231}.
  \emph{\jcap} 2016(8):020, \doi{10.1088/1475-7516/2016/08/020},
  \eprint{1511.03662}

\bibitem[{{Birrer} et~al.(2017){Birrer}, {Welschen}, {Amara}, and
  {Refregier}}]{Birrer:2017los}
{Birrer} S., {Welschen} C., {Amara} A., {Refregier} A. (2017) {Line-of-sight
  effects in strong lensing: putting theory into practice}. \emph{\jcap}
  2017(4):049, \doi{10.1088/1475-7516/2017/04/049}, \eprint{1610.01599}

\bibitem[{{Birrer} et~al.(2019){Birrer}, {Treu}, {Rusu}, {Bonvin}, {Fassnacht},
  {Chan}, {Agnello}, {Shajib}, {Chen}, {Auger}, {Courbin}, {Hilbert}, {Sluse},
  {Suyu}, {Wong}, {Marshall}, {Lemaux}, and {Meylan}}]{Birrer:2019}
{Birrer} S., {Treu} T., {Rusu} C.~E., {Bonvin} V., {Fassnacht} C.~D., {Chan}
  J.~H.~H., {Agnello} A., {Shajib} A.~J., et~al. (2019) {H0LiCOW - IX.
  Cosmographic analysis of the doubly imaged quasar SDSS 1206+4332 and a new
  measurement of the Hubble constant}. \emph{\mnras} 484(4):4726--4753,
  \doi{10.1093/mnras/stz200}, \eprint{1809.01274}

\bibitem[{{Birrer} et~al.(2020){Birrer}, {Shajib}, {Galan}, {Millon}, {Treu},
  {Agnello}, {Auger}, {Chen}, {Christensen}, {Collett}, {Courbin}, {Fassnacht},
  {Koopmans}, {Marshall}, {Park}, {Rusu}, {Sluse}, {Spiniello}, {Suyu},
  {Wagner-Carena}, {Wong}, {Barnab{\`e}}, {Bolton}, {Czoske}, {Ding},
  {Frieman}, and {Van de Vyvere}}]{Birrer:tdcosmoiv}
{Birrer} S., {Shajib} A.~J., {Galan} A., {Millon} M., {Treu} T., {Agnello} A.,
  {Auger} M., {Chen} G.~C.~F., et~al. (2020) {TDCOSMO. IV. Hierarchical
  time-delay cosmography {\textendash} joint inference of the Hubble constant
  and galaxy density profiles}. \emph{\aap} 643:A165,
  \doi{10.1051/0004-6361/202038861}, \eprint{2007.02941}

\bibitem[{{Birrer} et~al.(2021){Birrer}, {Dhawan}, and
  {Shajib}}]{Birrer:2021glSNe}
{Birrer} S., {Dhawan} S., {Shajib} A.~J. (2021) {The Hubble constant from
  strongly lensed supernovae with standardizable magnifications}. \emph{arXiv
  e-prints} arXiv:2107.12385, \eprint{2107.12385}

\bibitem[{{Biscardi} et~al.(2008){Biscardi}, {Raimondo}, {Cantiello}, and
  {Brocato}}]{biscardi08}
{Biscardi} I., {Raimondo} G., {Cantiello} M., {Brocato} E. (2008) {Optical
  Surface Brightness Fluctuations of Shell Galaxies toward 100 Mpc}.
  \emph{\apj} 678(1):168--178, \doi{10.1086/587126}, \eprint{0802.2509}

\bibitem[{{Bisogni} et~al.(2017){Bisogni}, {Risaliti}, and {Lusso}}]{brl2017}
{Bisogni} S., {Risaliti} G., {Lusso} E. (2017) {A Hubble Diagram for Quasars}.
  \emph{Frontiers in Astronomy and Space Sciences} 4:68,
  \doi{10.3389/fspas.2017.00068}, \eprint{1712.07515}

\bibitem[{{Bisogni} et~al.(2021){Bisogni}, {Lusso}, {Civano}, {Nardini},
  {Risaliti}, {Elvis}, and {Fabbiano}}]{bisogni2021}
{Bisogni} S., {Lusso} E., {Civano} F., {Nardini} E., {Risaliti} G., {Elvis} M.,
  {Fabbiano} G. (2021) {The Chandra view of the relation between X-ray and UV
  emission in quasars}. \emph{\aap} 655:A109,
  \doi{10.1051/0004-6361/202140852}, \eprint{2109.03252}

\bibitem[{{Biswas} et~al.(2010){Biswas}, {Alizadeh}, and
  {Wandelt}}]{Biswas2010}
{Biswas} R., {Alizadeh} E., {Wandelt} B.~D. (2010) {Voids as a precision probe
  of dark energy}. \emph{\prd} 82(2):023002, \doi{10.1103/PhysRevD.82.023002},
  \eprint{1002.0014}

\bibitem[{Blake(2019)}]{Blake:2019ddd}
Blake C. (2019) {Power spectrum modelling of galaxy and radio intensity maps
  including observational effects}. \emph{Mon Not Roy Astron Soc}
  489(1):153--167, \doi{10.1093/mnras/stz2145}, \eprint{1902.07439}

\bibitem[{{Blakeslee} and {Cantiello}(2018)}]{blake18}
{Blakeslee} J.~P., {Cantiello} M. (2018) {Independent Analysis of the Distance
  to NGC{\,}1052-DF2}. \emph{Research Notes of the American Astronomical
  Society} 2(3):146, \doi{10.3847/2515-5172/aad90e}, \eprint{1808.02176}

\bibitem[{{Blakeslee} and {Tonry}(1995)}]{blake95}
{Blakeslee} J.~P., {Tonry} J.~L. (1995) {Measurements of Globular Cluster
  Specific Frequencies and Luminosity Function Widths in Coma}. \emph{\apj}
  442:579, \doi{10.1086/175461}

\bibitem[{{Blakeslee} et~al.(1999{\natexlab{a}}){Blakeslee}, {Ajhar}, and
  {Tonry}}]{blake99a}
{Blakeslee} J.~P., {Ajhar} E.~A., {Tonry} J.~L. (1999{\natexlab{a}}) {Distances
  from Surface Brightness Fluctuations}. In: {Heck} A., {Caputo} F. (eds)
  Post-Hipparcos Cosmic Candles, Astrophysics and Space Science Library, vol
  237, p 181, \doi{10.1007/978-94-011-4734-7\_11}, \eprint{astro-ph/9807124}

\bibitem[{{Blakeslee} et~al.(1999{\natexlab{b}}){Blakeslee}, {Davis}, {Tonry},
  {Dressler}, and {Ajhar}}]{blake99}
{Blakeslee} J.~P., {Davis} M., {Tonry} J.~L., {Dressler} A., {Ajhar} E.~A.
  (1999{\natexlab{b}}) {A First Comparison of the Surface Brightness
  Fluctuation Survey Distances with the Galaxy Density Field: Implications for
  H$_{0}$ and {\ensuremath{\Omega}}}. \emph{\apjl} 527(2):L73--L76,
  \doi{10.1086/312404}, \eprint{astro-ph/9910340}

\bibitem[{{Blakeslee} et~al.(2001{\natexlab{a}}){Blakeslee}, {Lucey}, {Barris},
  {Hudson}, and {Tonry}}]{blake01b}
{Blakeslee} J.~P., {Lucey} J.~R., {Barris} B.~J., {Hudson} M.~J., {Tonry} J.~L.
  (2001{\natexlab{a}}) {A synthesis of data from fundamental plane and surface
  brightness fluctuation surveys}. \emph{\mnras} 327(3):1004--1020,
  \doi{10.1046/j.1365-8711.2001.04800.x}, \eprint{astro-ph/0108194}

\bibitem[{{Blakeslee} et~al.(2001{\natexlab{b}}){Blakeslee}, {Vazdekis}, and
  {Ajhar}}]{bva01}
{Blakeslee} J.~P., {Vazdekis} A., {Ajhar} E.~A. (2001{\natexlab{b}}) {Stellar
  populations and surface brightness fluctuations: new observations and
  models}. \emph{\mnras} 320:193

\bibitem[{{Blakeslee} et~al.(2002){Blakeslee}, {Lucey}, {Tonry}, {Hudson},
  {Narayanan}, and {Barris}}]{blake02}
{Blakeslee} J.~P., {Lucey} J.~R., {Tonry} J.~L., {Hudson} M.~J., {Narayanan}
  V.~K., {Barris} B.~J. (2002) {Early-type galaxy distances from the
  Fundamental Plane and surface brightness fluctuations}. \emph{\mnras}
  330(2):443--457, \doi{10.1046/j.1365-8711.2002.05080.x},
  \eprint{astro-ph/0111183}

\bibitem[{{Blakeslee} et~al.(2009){Blakeslee}, {Jord{\'a}n}, {Mei},
  {C{\^o}t{\'e}}, {Ferrarese}, {Infante}, {Peng}, {Tonry}, and
  {West}}]{blake09}
{Blakeslee} J.~P., {Jord{\'a}n} A., {Mei} S., {C{\^o}t{\'e}} P., {Ferrarese}
  L., {Infante} L., {Peng} E.~W., {Tonry} J.~L., et~al. (2009) {The ACS Fornax
  Cluster Survey. V. Measurement and Recalibration of Surface Brightness
  Fluctuations and a Precise Value of the Fornax-Virgo Relative Distance}.
  \emph{\apj} 694:556--572, \doi{10.1088/0004-637X/694/1/556},
  \eprint{0901.1138}

\bibitem[{{Blakeslee} et~al.(2010){Blakeslee}, {Cantiello}, {Mei},
  {C{\^o}t{\'e}}, {Barber DeGraaff}, {Ferrarese}, {Jord{\'a}n}, {Peng},
  {Tonry}, and {Worthey}}]{blake10}
{Blakeslee} J.~P., {Cantiello} M., {Mei} S., {C{\^o}t{\'e}} P., {Barber
  DeGraaff} R., {Ferrarese} L., {Jord{\'a}n} A., {Peng} E.~W., et~al. (2010)
  {Surface Brightness Fluctuations in the Hubble Space Telescope ACS/WFC F814W
  Bandpass and an Update on Galaxy Distances}. \emph{\apj} 724(1):657--668,
  \doi{10.1088/0004-637X/724/1/657}, \eprint{1009.3270}

\bibitem[{Blakeslee et~al.(2021)Blakeslee, Jensen, Ma, Milne, and
  Greene}]{blake21}
Blakeslee J.~P., Jensen J.~B., Ma C.-P., Milne P.~A., Greene J.~E. (2021) The
  hubble constant from infrared surface brightness fluctuation distances.
  \emph{The Astrophysical Journal} 911(1):65, \doi{10.3847/1538-4357/abe86a},
  \urlprefix\url{https://doi.org/10.3847/1538-4357/abe86a}

\bibitem[{Blanchard et~al.(2020)}]{Blanchard:2019oqi}
Blanchard A., et~al. (2020) {Euclid preparation: VII. Forecast validation for
  Euclid cosmological probes}. \emph{Astron Astrophys} 642:A191,
  \doi{10.1051/0004-6361/202038071}, \eprint{1910.09273}

\bibitem[{{Blandford} and {Narayan}(1986)}]{Blandford:1986}
{Blandford} R., {Narayan} R. (1986) {Fermat's Principle, Caustics, and the
  Classification of Gravitational Lens Images}. \emph{\apj} 310:568,
  \doi{10.1086/164709}

\bibitem[{{Blum} et~al.(2020){Blum}, {Castorina}, and
  {Simonovi{\'c}}}]{Blum:2020}
{Blum} K., {Castorina} E., {Simonovi{\'c}} M. (2020) {Could Quasar Lensing Time
  Delays Hint to a Core Component in Halos, Instead of H$_{0}$ Tension?}
  \emph{\apjl} 892(2):L27, \doi{10.3847/2041-8213/ab8012}, \eprint{2001.07182}

\bibitem[{{Blumenthal} et~al.(1992){Blumenthal}, {da Costa}, {Goldwirth},
  {Lecar}, and {Piran}}]{Blumenthal1992}
{Blumenthal} G.~R., {da Costa} L.~N., {Goldwirth} D.~S., {Lecar} M., {Piran} T.
  (1992) {The Largest Possible Voids}. \emph{\apj} 388:234,
  \doi{10.1086/171147}

\bibitem[{{Bolton} et~al.(2008){Bolton}, {Burles}, {Koopmans}, {Treu},
  {Gavazzi}, {Moustakas}, {Wayth}, and {Schlegel}}]{Bolton:2008}
{Bolton} A.~S., {Burles} S., {Koopmans} L. V.~E., {Treu} T., {Gavazzi} R.,
  {Moustakas} L.~A., {Wayth} R., {Schlegel} D.~J. (2008) {The Sloan Lens ACS
  Survey. V. The Full ACS Strong-Lens Sample}. \emph{\apj} 682(2):964--984,
  \doi{10.1086/589327}, \eprint{0805.1931}

\bibitem[{{Bonamigo} et~al.(2017){Bonamigo}, {Grillo}, {Ettori}, {Caminha},
  {Rosati}, {Mercurio}, {Annunziatella}, {Balestra}, and
  {Lombardi}}]{Bonamigo:2017}
{Bonamigo} M., {Grillo} C., {Ettori} S., {Caminha} G.~B., {Rosati} P.,
  {Mercurio} A., {Annunziatella} M., {Balestra} I., et~al. (2017) {Joining
  X-Ray to Lensing: An Accurate Combined Analysis of MACS J0416.1-2403}.
  \emph{\apj} 842(2):132, \doi{10.3847/1538-4357/aa75cc}, \eprint{1705.10322}

\bibitem[{{Bonamigo} et~al.(2018){Bonamigo}, {Grillo}, {Ettori}, {Caminha},
  {Rosati}, {Mercurio}, {Munari}, {Annunziatella}, {Balestra}, and
  {Lombardi}}]{Bonamigo:2018}
{Bonamigo} M., {Grillo} C., {Ettori} S., {Caminha} G.~B., {Rosati} P.,
  {Mercurio} A., {Munari} E., {Annunziatella} M., et~al. (2018) {Dissection of
  the Collisional and Collisionless Mass Components in a Mini Sample of CLASH
  and HFF Massive Galaxy Clusters at z {\ensuremath{\approx}} 0.4}. \emph{\apj}
  864(1):98, \doi{10.3847/1538-4357/aad4a7}, \eprint{1807.10286}

\bibitem[{{Bond} et~al.(1991){Bond}, {Cole}, {Efstathiou}, and
  {Kaiser}}]{Bond1991}
{Bond} J.~R., {Cole} S., {Efstathiou} G., {Kaiser} N. (1991) {Excursion Set
  Mass Functions for Hierarchical Gaussian Fluctuations}. \emph{\apj} 379:440,
  \doi{10.1086/170520}

\bibitem[{{Bonilla} et~al.(2021){Bonilla}, {Kumar}, and {Nunes}}]{bonilla2021}
{Bonilla} A., {Kumar} S., {Nunes} R.~C. (2021) {Measurements of H$_{0}$ and
  reconstruction of the dark energy properties from a model-independent joint
  analysis}. \emph{European Physical Journal C} 81(2):127,
  \doi{10.1140/epjc/s10052-021-08925-z}, \eprint{2011.07140}

\bibitem[{{Bonvin} et~al.(2017){Bonvin}, {Courbin}, {Suyu}, {Marshall}, {Rusu},
  {Sluse}, {Tewes}, {Wong}, {Collett}, {Fassnacht}, {Treu}, {Auger}, {Hilbert},
  {Koopmans}, {Meylan}, {Rumbaugh}, {Sonnenfeld}, and
  {Spiniello}}]{Bonvin:2017}
{Bonvin} V., {Courbin} F., {Suyu} S.~H., {Marshall} P.~J., {Rusu} C.~E.,
  {Sluse} D., {Tewes} M., {Wong} K.~C., et~al. (2017) {H0LiCOW - V. New
  COSMOGRAIL time delays of HE 0435-1223: H$_{0}$ to 3.8 per cent precision
  from strong lensing in a flat {\ensuremath{\Lambda}}CDM model}. \emph{\mnras}
  465(4):4914--4930, \doi{10.1093/mnras/stw3006}, \eprint{1607.01790}

\bibitem[{{Bora} and {Desai}(2021)}]{bora2021}
{Bora} K., {Desai} S. (2021) {A test of cosmic distance duality relation using
  SPT-SZ galaxy clusters, Type Ia supernovae, and cosmic chronometers}.
  \emph{\jcap} 2021(6):052, \doi{10.1088/1475-7516/2021/06/052},
  \eprint{2104.00974}

\bibitem[{Boran et~al.(2018)Boran, Desai, Kahya, and Woodard}]{Boran:2017rdn}
Boran S., Desai S., Kahya E.~O., Woodard R.~P. (2018) {GW170817 Falsifies Dark
  Matter Emulators}. \emph{Phys Rev D} 97(4):041501,
  \doi{10.1103/PhysRevD.97.041501}, \eprint{1710.06168}

\bibitem[{{Borghi} et~al.(2021{\natexlab{a}}){Borghi}, {Moresco}, and
  {Cimatti}}]{borghi2021b}
{Borghi} N., {Moresco} M., {Cimatti} A. (2021{\natexlab{a}}) {Towards a Better
  Understanding of Cosmic Chronometers: A new measurement of $H(z)$ at
  $z\sim0.7$}. \emph{arXiv e-prints} arXiv:2110.04304, \eprint{2110.04304}

\bibitem[{{Borghi} et~al.(2021{\natexlab{b}}){Borghi}, {Moresco}, {Cimatti},
  {Huchet}, {Quai}, and {Pozzetti}}]{borghi2021}
{Borghi} N., {Moresco} M., {Cimatti} A., {Huchet} A., {Quai} S., {Pozzetti} L.
  (2021{\natexlab{b}}) {Towards a Better Understanding of Cosmic Chronometers:
  Stellar Population Properties of Passive Galaxies at Intermediate Redshift}.
  \emph{arXiv e-prints} arXiv:2106.14894, \eprint{2106.14894}

\bibitem[{{Borhanian} et~al.(2020){Borhanian}, {Dhani}, {Gupta}, {Arun}, and
  {Sathyaprakash}}]{2020arXiv200702883B}
{Borhanian} S., {Dhani} A., {Gupta} A., {Arun} K.~G., {Sathyaprakash} B.~S.
  (2020) {Dark Sirens to Resolve the Hubble-Lema{\^\i}tre Tension}. \emph{arXiv
  e-prints} arXiv:2007.02883, \eprint{2007.02883}

\bibitem[{{Boruah} et~al.(2020){Boruah}, {Hudson}, and
  {Lavaux}}]{Boruah:2019icj}
{Boruah} S.~S., {Hudson} M.~J., {Lavaux} G. (2020) {Cosmic flows in the nearby
  Universe: new peculiar velocities from SNe and cosmological constraints}.
  \emph{\mnras} 498(2):2703--2718, \doi{10.1093/mnras/staa2485},
  \eprint{1912.09383}

\bibitem[{{Bos} et~al.(2012){Bos}, {van de Weygaert}, {Dolag}, and
  {Pettorino}}]{Bos2012}
{Bos} E.~G.~P., {van de Weygaert} R., {Dolag} K., {Pettorino} V. (2012) {The
  darkness that shaped the void: dark energy and cosmic voids}. \emph{\mnras}
  426(1):440--461, \doi{10.1111/j.1365-2966.2012.21478.x}, \eprint{1205.4238}

\bibitem[{{Bosnjak} et~al.(2008){Bosnjak}, {Celotti}, {Longo}, and
  {Barbiellini}}]{Bosnjak08}
{Bosnjak} Z., {Celotti} A., {Longo} F., {Barbiellini} G. (2008)
  {Energetics-spectral correlations versus the BATSE gamma-ray bursts
  population}. \emph{\mnras} 384(2):599--604,
  \doi{10.1111/j.1365-2966.2007.12672.x}

\bibitem[{Boylan-Kolchin and Weisz(2021)}]{Boylan}
Boylan-Kolchin M., Weisz D.~R. (2021) Uncertain times: the redshift–time
  relation from cosmology and stars. \emph{Monthly Notices of the Royal
  Astronomical Society} 505(2):2764–2783, \doi{10.1093/mnras/stab1521},
  \urlprefix\url{http://dx.doi.org/10.1093/mnras/stab1521}

\bibitem[{Bracewell(1999)}]{Bracewell1999}
Bracewell R. (1999) The Fourier Transform and Its Applications, 3rd ed.
  McGraw-Hill

\bibitem[{{Brouwer} et~al.(2018){Brouwer}, {Demchenko}, {Harnois-D{\'e}raps},
  {Bilicki}, {Heymans}, {Hoekstra}, {Kuijken}, {Alpaslan}, {Brough}, {Cai},
  {Costa-Duarte}, {Dvornik}, {Erben}, {Hildebrandt}, {Holwerda}, {Schneider},
  {Sif{\'o}n}, and {van Uitert}}]{Brouwer2018}
{Brouwer} M.~M., {Demchenko} V., {Harnois-D{\'e}raps} J., {Bilicki} M.,
  {Heymans} C., {Hoekstra} H., {Kuijken} K., {Alpaslan} M., et~al. (2018)
  {Studying galaxy troughs and ridges using weak gravitational lensing with the
  Kilo-Degree Survey}. \emph{\mnras} 481(4):5189--5209,
  \doi{10.1093/mnras/sty2589}, \eprint{1805.00562}

\bibitem[{{Bruzual} and {Charlot}(2003)}]{bruzual2003}
{Bruzual} G., {Charlot} S. (2003) {Stellar population synthesis at the
  resolution of 2003}. \emph{\mnras} 344:1000--1028,
  \doi{10.1046/j.1365-8711.2003.06897.x}, \eprint{astro-ph/0309134}

\bibitem[{{Bruzual A.}(1983)}]{bruzual1983}
{Bruzual A.} G. (1983) {Spectral evolution of galaxies. I - Early-type
  systems}. \emph{\apj} 273:105--127, \doi{10.1086/161352}

\bibitem[{{Buckley-Geer} et~al.(2020){Buckley-Geer}, {Lin}, {Rusu}, {Poh},
  {Palmese}, {Agnello}, {Christensen}, {Frieman}, {Shajib}, {Treu}, {Collett},
  {Birrer}, {Anguita}, {Fassnacht}, {Meylan}, {Mukherjee}, et~al., and {DES
  Collaboration}}]{Buckley-Geer:2020}
{Buckley-Geer} E.~J., {Lin} H., {Rusu} C.~E., {Poh} J., {Palmese} A., {Agnello}
  A., {Christensen} L., {Frieman} J., et~al. (2020) {STRIDES: Spectroscopic and
  photometric characterization of the environment and effects of mass along the
  line of sight to the gravitational lenses DES J0408-5354 and WGD 2038-4008}.
  \emph{\mnras} 498(3):3241--3274, \doi{10.1093/mnras/staa2563},
  \eprint{2003.12117}

\bibitem[{Bull et~al.(2015)Bull, Ferreira, Patel, and Santos}]{Bull:2014rha}
Bull P., Ferreira P.~G., Patel P., Santos M.~G. (2015) {Late-time cosmology
  with 21cm intensity mapping experiments}. \emph{Astrophys J} 803(1):21,
  \doi{10.1088/0004-637X/803/1/21}, \eprint{1405.1452}

\bibitem[{{Ca{\~n}ameras} et~al.(2020){Ca{\~n}ameras}, {Schuldt}, {Suyu},
  {Taubenberger}, {Meinhardt}, {Leal-Taix{\'e}}, {Lemon}, {Rojas}, and
  {Savary}}]{Canameras2020}
{Ca{\~n}ameras} R., {Schuldt} S., {Suyu} S.~H., {Taubenberger} S., {Meinhardt}
  T., {Leal-Taix{\'e}} L., {Lemon} C., {Rojas} K., et~al. (2020) {HOLISMOKES.
  II. Identifying galaxy-scale strong gravitational lenses in Pan-STARRS using
  convolutional neural networks}. \emph{\aap} 644:A163,
  \doi{10.1051/0004-6361/202038219}, \eprint{2004.13048}

\bibitem[{{Cai} et~al.(2014){Cai}, {Neyrinck}, {Szapudi}, {Cole}, and
  {Frenk}}]{Cai2014}
{Cai} Y.-C., {Neyrinck} M.~C., {Szapudi} I., {Cole} S., {Frenk} C.~S. (2014) {A
  Possible Cold Imprint of Voids on the Microwave Background Radiation}.
  \emph{\apj} 786(2):110, \doi{10.1088/0004-637X/786/2/110}, \eprint{1301.6136}

\bibitem[{{Cai} et~al.(2015){Cai}, {Padilla}, and {Li}}]{Cai2015}
{Cai} Y.-C., {Padilla} N., {Li} B. (2015) {Testing gravity using cosmic voids}.
  \emph{\mnras} 451:1036--1055, \doi{10.1093/mnras/stv777}, \eprint{1410.1510}

\bibitem[{{Cai} et~al.(2016){Cai}, {Taylor}, {Peacock}, and
  {Padilla}}]{Cai2016}
{Cai} Y.-C., {Taylor} A., {Peacock} J.~A., {Padilla} N. (2016) {Redshift-space
  distortions around voids}. \emph{\mnras} 462:2465--2477,
  \doi{10.1093/mnras/stw1809}, \eprint{1603.05184}

\bibitem[{{Cai} et~al.(2017){Cai}, {Neyrinck}, {Mao}, {Peacock}, {Szapudi}, and
  {Berlind}}]{Cai2017}
{Cai} Y.-C., {Neyrinck} M., {Mao} Q., {Peacock} J.~A., {Szapudi} I., {Berlind}
  A.~A. (2017) {The lensing and temperature imprints of voids on the cosmic
  microwave background}. \emph{\mnras} 466(3):3364--3375,
  \doi{10.1093/mnras/stw3299}, \eprint{1609.00301}

\bibitem[{{Calder{\'o}n Bustillo} et~al.(2021){Calder{\'o}n Bustillo}, {Leong},
  {Dietrich}, and {Lasky}}]{2021ApJ...912L..10C}
{Calder{\'o}n Bustillo} J., {Leong} S. H.~W., {Dietrich} T., {Lasky} P.~D.
  (2021) {Mapping the Universe Expansion: Enabling Percent-level Measurements
  of the Hubble Constant with a Single Binary Neutron-star Merger Detection}.
  \emph{\apjl} 912(1):L10, \doi{10.3847/2041-8213/abf502}, \eprint{2006.11525}

\bibitem[{Camera and Padmanabhan(2020)}]{Camera:2019iwy}
Camera S., Padmanabhan H. (2020) {Beyond \ensuremath{\Lambda}CDM with H i
  intensity mapping: robustness of cosmological constraints in the presence of
  astrophysics}. \emph{Mon Not Roy Astron Soc} 496(4):4115--4126,
  \doi{10.1093/mnras/staa1663}, \eprint{1910.00022}

\bibitem[{Camera et~al.(2013)Camera, Santos, Ferreira, and
  Ferramacho}]{Camera:2013kpa}
Camera S., Santos M.~G., Ferreira P.~G., Ferramacho L. (2013) {Cosmology on
  Ultra-Large Scales with HI Intensity Mapping: Limits on Primordial
  non-Gaussianity}. \emph{Phys Rev Lett} 111:171302,
  \doi{10.1103/PhysRevLett.111.171302}, \eprint{1305.6928}

\bibitem[{{Caminha} et~al.(2016){Caminha}, {Grillo}, {Rosati}, {Balestra},
  {Karman}, {Lombardi}, {Mercurio}, {Nonino}, {Tozzi}, {Zitrin}, {Biviano},
  {Girardi}, {Koekemoer}, and et~al.}]{Caminha:2016}
{Caminha} G.~B., {Grillo} C., {Rosati} P., {Balestra} I., {Karman} W.,
  {Lombardi} M., {Mercurio} A., {Nonino} M., et~al. (2016) {CLASH-VLT: A highly
  precise strong lensing model of the galaxy cluster RXC J2248.7-4431 (Abell
  S1063) and prospects for cosmography}. \emph{\aap} 587:A80,
  \doi{10.1051/0004-6361/201527670}, \eprint{1512.04555}

\bibitem[{{Caminha} et~al.(2017{\natexlab{a}}){Caminha}, {Grillo}, {Rosati},
  {Balestra}, {Mercurio}, {Vanzella}, {Biviano}, {Caputi}, {Delgado-Correal},
  {Karman}, {Lombardi}, {Meneghetti}, {Sartoris}, and {Tozzi}}]{Caminha:2017a}
{Caminha} G.~B., {Grillo} C., {Rosati} P., {Balestra} I., {Mercurio} A.,
  {Vanzella} E., {Biviano} A., {Caputi} K.~I., et~al. (2017{\natexlab{a}}) {A
  refined mass distribution of the cluster MACS J0416.1-2403 from a new large
  set of spectroscopic multiply lensed sources}. \emph{\aap} 600:A90,
  \doi{10.1051/0004-6361/201629297}, \eprint{1607.03462}

\bibitem[{{Caminha} et~al.(2017{\natexlab{b}}){Caminha}, {Grillo}, {Rosati},
  {Meneghetti}, {Mercurio}, {Ettori}, {Balestra}, {Biviano}, {Umetsu},
  {Vanzella}, {Annunziatella}, {Bonamigo}, {Delgado-Correal}, {Girardi},
  {Lombardi}, {Nonino}, {Sartoris}, {Tozzi}, {Bartelmann}, {Bradley}, {Caputi},
  {Coe}, {Ford}, {Fritz}, {Gobat}, {Postman}, {Seitz}, and
  {Zitrin}}]{Caminha:2017b}
{Caminha} G.~B., {Grillo} C., {Rosati} P., {Meneghetti} M., {Mercurio} A.,
  {Ettori} S., {Balestra} I., {Biviano} A., et~al. (2017{\natexlab{b}}) {Mass
  distribution in the core of MACS J1206. Robust modeling from an exceptionally
  large sample of central multiple images}. \emph{\aap} 607:A93,
  \doi{10.1051/0004-6361/201731498}, \eprint{1707.00690}

\bibitem[{{Caminha} et~al.(2019){Caminha}, {Rosati}, {Grillo}, {Rosani},
  {Caputi}, {Meneghetti}, {Mercurio}, {Balestra}, {Bergamini}, {Biviano},
  {Nonino}, {Umetsu}, {Vanzella}, {Annunziatella}, {Broadhurst},
  {Delgado-Correal}, {Demarco}, {Koekemoer}, {Lombardi}, {Maier}, {Verdugo},
  and {Zitrin}}]{Caminha:2019}
{Caminha} G.~B., {Rosati} P., {Grillo} C., {Rosani} G., {Caputi} K.~I.,
  {Meneghetti} M., {Mercurio} A., {Balestra} I., et~al. (2019) {Strong lensing
  models of eight CLASH clusters from extensive spectroscopy: Accurate total
  mass reconstructions in the cores}. \emph{\aap} 632:A36,
  \doi{10.1051/0004-6361/201935454}, \eprint{1903.05103}

\bibitem[{{Caminha} et~al.(2021){Caminha}, {Suyu}, {Grillo}, and
  {Rosati}}]{Caminha:2021}
{Caminha} G.~B., {Suyu} S.~H., {Grillo} C., {Rosati} P. (2021) {Galaxy cluster
  strong lensing cosmography: cosmological constraints from a sample of regular
  galaxy clusters}. \emph{arXiv e-prints} arXiv:2110.06232, \eprint{2110.06232}

\bibitem[{{Cantiello} et~al.(2003){Cantiello}, {Raimondo}, {Brocato}, and
  {Capaccioli}}]{cantiello03}
{Cantiello} M., {Raimondo} G., {Brocato} E., {Capaccioli} M. (2003) {New
  Optical and Near-Infrared Surface Brightness Fluctuation Models: A Primary
  Distance Indicator Ranging from Globular Clusters to Distant Galaxies?}
  \emph{The Astrophysical Journal} 125:2783, \doi{10.1086/375322}

\bibitem[{{Cantiello} et~al.(2005){Cantiello}, {Blakeslee}, {Raimondo}, {Mei},
  {Brocato}, and {Capaccioli}}]{cantiello05}
{Cantiello} M., {Blakeslee} J.~P., {Raimondo} G., {Mei} S., {Brocato} E.,
  {Capaccioli} M. (2005) {Detection of Radial Surface Brightness Fluctuations
  and Color Gradients in Elliptical Galaxies with the Advanced Camera for
  Surveys}. \emph{\apj} 634:239, \doi{10.1086/491694},
  \eprint{astro-ph/0507699}

\bibitem[{{Cantiello} et~al.(2007){Cantiello}, {Raimondo}, {Blakeslee},
  {Brocato}, and {Capaccioli}}]{cantiello07}
{Cantiello} M., {Raimondo} G., {Blakeslee} J.~P., {Brocato} E., {Capaccioli} M.
  (2007) {Detection of Surface Brightness Fluctuations in Elliptical Galaxies
  Imaged with the Advanced Camera for Surveys: B- and I-Band Measurements}.
  \emph{\apj} 662:940--951, \doi{10.1086/517984},
  \eprint{arXiv:astro-ph/0703539}

\bibitem[{{Cantiello} et~al.(2011){Cantiello}, {Biscardi}, {Brocato}, and
  {Raimondo}}]{cantiello11}
{Cantiello} M., {Biscardi} I., {Brocato} E., {Raimondo} G. (2011) {VLT optical
  BVR observations of two bright supernova Ia hosts in the Virgo cluster.
  Surface brightness fluctuation analysis}. \emph{\aap} 532:A154,
  \doi{10.1051/0004-6361/201116667}

\bibitem[{{Cantiello} et~al.(2018{\natexlab{a}}){Cantiello}, {Blakeslee},
  {Ferrarese}, {C{\^o}t{\'e}}, {Roediger}, {Raimondo}, {Peng}, {Gwyn},
  {Durrell}, and {Cuillandre}}]{cantiello18b}
{Cantiello} M., {Blakeslee} J.~P., {Ferrarese} L., {C{\^o}t{\'e}} P.,
  {Roediger} J.~C., {Raimondo} G., {Peng} E.~W., {Gwyn} S., et~al.
  (2018{\natexlab{a}}) {The Next Generation Virgo Cluster Survey (NGVS). XVIII.
  Measurement and Calibration of Surface Brightness Fluctuation Distances for
  Bright Galaxies in Virgo (and Beyond)}. \emph{\apj} 856(2):126,
  \doi{10.3847/1538-4357/aab043}, \eprint{1802.05526}

\bibitem[{{Cantiello} et~al.(2018{\natexlab{b}}){Cantiello}, {Jensen},
  {Blakeslee}, {Berger}, {Levan}, {Tanvir}, {Raimondo}, {Brocato}, {Alexander},
  {Blanchard}, {Branchesi}, {Cano}, {Chornock}, {Covino}, {Cowperthwaite},
  {D'Avanzo}, {Eftekhari}, {Fong}, {Fruchter}, {Grado}, {Hjorth}, {Holz},
  {Lyman}, {Mandel}, {Margutti}, {Nicholl}, {Villar}, and
  {Williams}}]{cantiello18}
{Cantiello} M., {Jensen} J.~B., {Blakeslee} J.~P., {Berger} E., {Levan} A.~J.,
  {Tanvir} N.~R., {Raimondo} G., {Brocato} E., et~al. (2018{\natexlab{b}}) {A
  Precise Distance to the Host Galaxy of the Binary Neutron Star Merger
  GW170817 Using Surface Brightness Fluctuations}. \emph{\apjl} 854(2):L31,
  \doi{10.3847/2041-8213/aaad64}, \eprint{1801.06080}

\bibitem[{{Cao} and {Ratra}(2022)}]{Cao22}
{Cao} S., {Ratra} B. (2022) {Using lower redshift, non-CMB, data to constrain
  the Hubble constant and other cosmological parameters}. \emph{\mnras}
  513(4):5686--5700, \doi{10.1093/mnras/stac1184}, \eprint{2203.10825}

\bibitem[{{Cao} et~al.(2019){Cao}, {Qi}, {Cao}, {Biesiada}, {Li}, {Pan}, and
  {Zhu}}]{cao2019}
{Cao} S., {Qi} J., {Cao} Z., {Biesiada} M., {Li} J., {Pan} Y., {Zhu} Z.-H.
  (2019) {Direct test of the FLRW metric from strongly lensed gravitational
  wave observations}. \emph{Scientific Reports} 9:11608,
  \doi{10.1038/s41598-019-47616-4}, \eprint{1910.10365}

\bibitem[{{Cao} et~al.(2021){Cao}, {Ryan}, {Khadka}, and {Ratra}}]{Cao21}
{Cao} S., {Ryan} J., {Khadka} N., {Ratra} B. (2021) {Cosmological constraints
  from higher redshift gamma-ray burst, H II starburst galaxy, and quasar (and
  other) data}. \emph{\mnras} 501(1):1520--1538, \doi{10.1093/mnras/staa3748},
  \eprint{2009.12953}

\bibitem[{{Capozziello} and {Ruchika}(2019)}]{capozziello2019}
{Capozziello} S., {Ruchika} A.~A. Sen (2019) {Model-independent constraints on
  dark energy evolution from low-redshift observations}. \emph{\mnras}
  484(4):4484--4494, \doi{10.1093/mnras/stz176}, \eprint{1806.03943}

\bibitem[{{Capozziello} et~al.(2018){Capozziello}, {D'Agostino}, and
  {Luongo}}]{capozziello2018}
{Capozziello} S., {D'Agostino} R., {Luongo} O. (2018) {Cosmographic analysis
  with Chebyshev polynomials}. \emph{\mnras} 476(3):3924--3938,
  \doi{10.1093/mnras/sty422}, \eprint{1712.04380}

\bibitem[{{Capozziello} et~al.(2019){Capozziello}, {D'Agostino}, and
  {Luongo}}]{capozziello2019b}
{Capozziello} S., {D'Agostino} R., {Luongo} O. (2019) {Extended gravity
  cosmography}. \emph{International Journal of Modern Physics D}
  28(10):1930016, \doi{10.1142/S0218271819300167}, \eprint{1904.01427}

\bibitem[{{Cappellari}(2017)}]{cappellari2017}
{Cappellari} M. (2017) {Improving the full spectrum fitting method: accurate
  convolution with Gauss-Hermite functions}. \emph{\mnras} 466(1):798--811,
  \doi{10.1093/mnras/stw3020}, \eprint{1607.08538}

\bibitem[{{Caputi} et~al.(2012){Caputi}, {Dunlop}, {McLure}, {Huang}, {Fazio},
  {Ashby}, {Castellano}, {Fontana}, {Cirasuolo}, and et~al.}]{caputi2012}
{Caputi} K.~I., {Dunlop} J.~S., {McLure} R.~J., {Huang} J.~S., {Fazio} G.~G.,
  {Ashby} M.~L.~N., {Castellano} M., {Fontana} A., et~al. (2012) {The Nature of
  Extremely Red H - [4.5] > 4 Galaxies Revealed with SEDS and CANDELS}.
  \emph{\apjl} 750(1):L20, \doi{10.1088/2041-8205/750/1/L20},
  \eprint{1202.0496}

\bibitem[{{Cardiel}(2010)}]{cardiel2010}
{Cardiel} N. (2010) {indexf: Line-strength Indices in Fully Calibrated FITS
  Spectra}. \eprint{1010.046}

\bibitem[{{Cardona} et~al.(2016){Cardona}, {Durrer}, {Kunz}, and
  {Montanari}}]{2016PhRvD..94d3007C}
{Cardona} W., {Durrer} R., {Kunz} M., {Montanari} F. (2016) {Lensing
  convergence and the neutrino mass scale in galaxy redshift surveys}.
  \emph{\prd} 94(4):043007, \doi{10.1103/PhysRevD.94.043007},
  \eprint{1603.06481}

\bibitem[{{Cardone} et~al.(2009){Cardone}, {Capozziello}, and
  {Dainotti}}]{Cardone09}
{Cardone} V.~F., {Capozziello} S., {Dainotti} M.~G. (2009) {An updated
  gamma-ray bursts Hubble diagram}. \emph{\mnras} 400(208):775--790,
  \doi{10.1111/j.1365-2966.2009.15456.x}, \eprint{0901.3194}

\bibitem[{{Carlsten} et~al.(2019){Carlsten}, {Beaton}, {Greco}, and
  {Greene}}]{carlsten19}
{Carlsten} S.~G., {Beaton} R.~L., {Greco} J.~P., {Greene} J.~E. (2019) {Using
  Surface Brightness Fluctuations to Study Nearby Satellite Galaxy Systems:
  Calibration and Methodology}. \emph{\apj} 879(1):13,
  \doi{10.3847/1538-4357/ab22c1}, \eprint{1901.07575}

\bibitem[{{Carlstrom} et~al.(2011){Carlstrom}, {Ade}, {Aird}, {Benson},
  {Bleem}, {Busetti}, {Chang}, {Chauvin}, {Cho}, {Crawford}, {Crites}, {Dobbs},
  {Halverson}, {Heimsath}, and et~al.}]{SPT}
{Carlstrom} J.~E., {Ade} P.~A.~R., {Aird} K.~A., {Benson} B.~A., {Bleem} L.~E.,
  {Busetti} S., {Chang} C.~L., {Chauvin} E., et~al. (2011) {The 10 Meter South
  Pole Telescope}. \emph{\pasp} 123(903):568, \doi{10.1086/659879},
  \eprint{0907.4445}

\bibitem[{{Carnall} et~al.(2018){Carnall}, {McLure}, {Dunlop}, and
  {Dav{\'e}}}]{carnall2018}
{Carnall} A.~C., {McLure} R.~J., {Dunlop} J.~S., {Dav{\'e}} R. (2018)
  {Inferring the star formation histories of massive quiescent galaxies with
  BAGPIPES: evidence for multiple quenching mechanisms}. \emph{\mnras}
  480(4):4379--4401, \doi{10.1093/mnras/sty2169}, \eprint{1712.04452}

\bibitem[{{Carnall} et~al.(2019){Carnall}, {McLure}, {Dunlop}, {Cullen},
  {McLeod}, {Wild}, {Johnson}, {Appleby}, {Dav{\'e}}, {Amorin}, {Bolzonella},
  {Castellano}, {Cimatti}, {Cucciati}, {Gargiulo}, {Garilli}, {Marchi},
  {Pentericci}, {Pozzetti}, {Schreiber}, {Talia}, and {Zamorani}}]{carnall2019}
{Carnall} A.~C., {McLure} R.~J., {Dunlop} J.~S., {Cullen} F., {McLeod} D.~J.,
  {Wild} V., {Johnson} B.~D., {Appleby} S., et~al. (2019) {The VANDELS survey:
  the star-formation histories of massive quiescent galaxies at 1.0 \&lt; z
  \&lt; 1.3}. \emph{\mnras} 490(1):417--439, \doi{10.1093/mnras/stz2544},
  \eprint{1903.11082}

\bibitem[{{Carrick} et~al.(2015){Carrick}, {Turnbull}, {Lavaux}, and
  {Hudson}}]{carrick15}
{Carrick} J., {Turnbull} S.~J., {Lavaux} G., {Hudson} M.~J. (2015)
  {Cosmological parameters from the comparison of peculiar velocities with
  predictions from the 2M++ density field}. \emph{\mnras} 450(1):317--332,
  \doi{10.1093/mnras/stv547}, \eprint{1504.04627}

\bibitem[{Carucci et~al.(2017)Carucci, Villaescusa-Navarro, and
  Viel}]{Carucci:2016yzq}
Carucci I.~P., Villaescusa-Navarro F., Viel M. (2017) {The cross-correlation
  between 21 cm intensity mapping maps and the Ly\ensuremath{\alpha} forest in
  the post-reionization era}. \emph{JCAP} 04:001,
  \doi{10.1088/1475-7516/2017/04/001}, \eprint{1611.07527}

\bibitem[{Carucci et~al.(2020)Carucci, Irfan, and Bobin}]{Carucci:2020enz}
Carucci I.~P., Irfan M.~O., Bobin J. (2020) {Recovery of 21 cm intensity maps
  with sparse component separation}. \emph{Mon Not Roy Astron Soc}
  499(1):304--319, \doi{10.1093/mnras/staa2854}, \eprint{2006.05996}

\bibitem[{Castorina and White(2019)}]{Castorina:2019zho}
Castorina E., White M. (2019) {Measuring the growth of structure with intensity
  mapping surveys}. \emph{JCAP} 06:025, \doi{10.1088/1475-7516/2019/06/025},
  \eprint{1902.07147}

\bibitem[{{Castro} and {Quartin}(2014)}]{Castro:2014oja}
{Castro} T., {Quartin} M. (2014) {First measurement of {$\sigma$}$_{8}$ using
  supernova magnitudes only}. \emph{\mnras} 443:L6--L10,
  \doi{10.1093/mnrasl/slu071}, \eprint{1403.0293}

\bibitem[{Castro et~al.(2016)Castro, Quartin, and
  Benitez-Herrera}]{Castro:2015rrx}
Castro T., Quartin M., Benitez-Herrera S. (2016) {Turning noise into signal:
  learning from the scatter in the Hubble diagram}. \emph{Phys Dark Univ}
  13:66--76, \doi{10.1016/j.dark.2016.04.006}, \eprint{1511.08695}

\bibitem[{{Castro-Rodr{\'\i}guez} and
  {L{\'o}pez-Corredoira}(2012)}]{castro2012}
{Castro-Rodr{\'\i}guez} N., {L{\'o}pez-Corredoira} M. (2012) {The age of
  extremely red and massive galaxies at very high redshift}. \emph{\aap}
  537:A31, \doi{10.1051/0004-6361/201117418}, \eprint{1111.2726}

\bibitem[{{Catelan}(2018)}]{Catelan}
{Catelan} M. (2018) {The ages of (the oldest) stars}. In: {Chiappini} C.,
  {Minchev} I., {Starkenburg} E., {Valentini} M. (eds) Rediscovering Our
  Galaxy, vol 334, pp 11--20, \doi{10.1017/S1743921318000868},
  \eprint{1709.08656}

\bibitem[{{Cautun} et~al.(2016){Cautun}, {Cai}, and {Frenk}}]{Cautun2016}
{Cautun} M., {Cai} Y.-C., {Frenk} C.~S. (2016) {The view from the boundary: a
  new void stacking method}. \emph{\mnras} 457(3):2540--2553,
  \doi{10.1093/mnras/stw154}, \eprint{1509.00010}

\bibitem[{{Cautun} et~al.(2018){Cautun}, {Paillas}, {Cai}, {Bose}, {Armijo},
  {Li}, and {Padilla}}]{Cautun2018}
{Cautun} M., {Paillas} E., {Cai} Y.-C., {Bose} S., {Armijo} J., {Li} B.,
  {Padilla} N. (2018) {The Santiago-Harvard-Edinburgh-Durham void comparison -
  I. SHEDding light on chameleon gravity tests}. \emph{\mnras}
  476(3):3195--3217, \doi{10.1093/mnras/sty463}, \eprint{1710.01730}

\bibitem[{{Ceccarelli} et~al.(2013){Ceccarelli}, {Paz}, {Lares}, {Padilla}, and
  {Lambas}}]{Ceccarelli2013}
{Ceccarelli} L., {Paz} D., {Lares} M., {Padilla} N., {Lambas} D.~G. (2013)
  {Clues on void evolution - I. Large-scale galaxy distributions around voids}.
  \emph{\mnras} 434(2):1435--1442, \doi{10.1093/mnras/stt1097},
  \eprint{1306.5798}

\bibitem[{{Ceccarelli} et~al.(2016){Ceccarelli}, {Ruiz}, {Lares}, {Paz},
  {Maldonado}, {Luparello}, and {Garcia Lambas}}]{Ceccarelli2016}
{Ceccarelli} L., {Ruiz} A.~N., {Lares} M., {Paz} D.~J., {Maldonado} V.~E.,
  {Luparello} H.~E., {Garcia Lambas} D. (2016) {The sparkling Universe: a
  scenario for cosmic void motions}. \emph{\mnras} 461(4):4013--4021,
  \doi{10.1093/mnras/stw1524}, \eprint{1511.06741}

\bibitem[{{Cervi{\~n}o} et~al.(2008){Cervi{\~n}o}, {Luridiana}, and
  {Jamet}}]{cervino08}
{Cervi{\~n}o} M., {Luridiana} V., {Jamet} L. (2008) {On surface brightness
  fluctuations: probabilistic and statistical bases. I. Stellar population and
  theoretical surface brightness fluctuations}. \emph{\aap} 491(3):693--701,
  \doi{10.1051/0004-6361:20077515}, \eprint{0809.4491}

\bibitem[{{Chabrier}(2003)}]{chabrier2003}
{Chabrier} G. (2003) {Galactic Stellar and Substellar Initial Mass Function}.
  \emph{\pasp} 115(809):763--795, \doi{10.1086/376392},
  \eprint{astro-ph/0304382}

\bibitem[{{Chan} and {Hamaus}(2021)}]{Chan2021}
{Chan} K.~C., {Hamaus} N. (2021) {Volume statistics as a probe of large-scale
  structure}. \emph{\prd} 103(4):043502, \doi{10.1103/PhysRevD.103.043502},
  \eprint{2010.13955}

\bibitem[{{Chan} et~al.(2014){Chan}, {Hamaus}, and {Desjacques}}]{Chan2014}
{Chan} K.~C., {Hamaus} N., {Desjacques} V. (2014) {Large-scale clustering of
  cosmic voids}. \emph{\prd} 90(10):103521, \doi{10.1103/PhysRevD.90.103521},
  \eprint{1409.3849}

\bibitem[{{Chan} et~al.(2019){Chan}, {Hamaus}, and {Biagetti}}]{Chan2019}
{Chan} K.~C., {Hamaus} N., {Biagetti} M. (2019) {Constraint of void bias on
  primordial non-Gaussianity}. \emph{\prd} 99(12):121304,
  \doi{10.1103/PhysRevD.99.121304}, \eprint{1812.04024}

\bibitem[{Chang et~al.(2008)Chang, Pen, Peterson, and McDonald}]{Chang:2007xk}
Chang T.-C., Pen U.-L., Peterson J.~B., McDonald P. (2008) {Baryon Acoustic
  Oscillation Intensity Mapping as a Test of Dark Energy}. \emph{Phys Rev Lett}
  100:091303, \doi{10.1103/PhysRevLett.100.091303}, \eprint{0709.3672}

\bibitem[{Chang et~al.(2010)Chang, Pen, Bandura, and Peterson}]{Chang:2010jp}
Chang T.-C., Pen U.-L., Bandura K., Peterson J.~B. (2010) {Hydrogen 21-cm
  Intensity Mapping at redshift 0.8}. \emph{Nature} 466:463--465,
  \doi{10.1038/nature09187}, \eprint{1007.3709}

\bibitem[{{Chantavat} et~al.(2017){Chantavat}, {Sawangwit}, and
  {Wandelt}}]{Chantavat2017}
{Chantavat} T., {Sawangwit} U., {Wandelt} B.~D. (2017) {Void Profile from
  Planck Lensing Potential Map}. \emph{\apj} 836(2):156,
  \doi{10.3847/1538-4357/836/2/156}, \eprint{1702.01009}

\bibitem[{Chapman et~al.(2012)Chapman, Abdalla, Harker, Jelic, Labropoulos,
  Zaroubi, Brentjens, de~Bruyn, and Koopmans}]{Chapman:2012yj}
Chapman E., Abdalla F.~B., Harker G., Jelic V., Labropoulos P., Zaroubi S.,
  Brentjens M.~A., de~Bruyn A.~G., et~al. (2012) {Foreground Removal using
  FastICA: A Showcase of LOFAR-EoR}. \emph{Mon Not Roy Astron Soc}
  423:2518--2532, \doi{10.1111/j.1365-2966.2012.21065.x}, \eprint{1201.2190}

\bibitem[{Chapman et~al.(2013)}]{Chapman:2012pn}
Chapman E., et~al. (2013) {The Scale of the Problem : Recovering Images of
  Reionization with GMCA}. \emph{Mon Not Roy Astron Soc} 429:165,
  \doi{10.1093/mnras/sts333}, \eprint{1209.4769}

\bibitem[{{Charlot} et~al.(2020){Charlot}, {Jacobs}, {Gordon}, {Lambert}, {de
  Witt}, {B{\"o}hm}, {Fey}, {Heinkelmann}, {Skurikhina}, {Titov}, {Arias},
  {Bolotin}, {Bourda}, {Ma}, {Malkin}, {Nothnagel}, {Mayer}, {MacMillan},
  {Nilsson}, and {Gaume}}]{charlot2020}
{Charlot} P., {Jacobs} C.~S., {Gordon} D., {Lambert} S., {de Witt} A.,
  {B{\"o}hm} J., {Fey} A.~L., {Heinkelmann} R., et~al. (2020) {The third
  realization of the International Celestial Reference Frame by very long
  baseline interferometry}. \emph{\aap} 644:A159,
  \doi{10.1051/0004-6361/202038368}, \eprint{2010.13625}

\bibitem[{{Chatterjee} et~al.(2021){Chatterjee}, {Hegade K.~R.}, {Holder},
  {Holz}, {Perkins}, {Yagi}, and {Yunes}}]{2021PhRvD.104h3528C}
{Chatterjee} D., {Hegade K.~R.} A., {Holder} G., {Holz} D.~E., {Perkins} S.,
  {Yagi} K., {Yunes} N. (2021) {Cosmology with Love: Measuring the Hubble
  constant using neutron star universal relations}. \emph{\prd} 104(8):083528,
  \doi{10.1103/PhysRevD.104.083528}, \eprint{2106.06589}

\bibitem[{{Chen} et~al.(2016{\natexlab{a}}){Chen}, {Suyu}, {Wong}, {Fassnacht},
  {Chiueh}, {Halkola}, {Hu}, {Auger}, {Koopmans}, {Lagattuta}, {McKean}, and
  {Vegetti}}]{Chen:2016}
{Chen} G. C.~F., {Suyu} S.~H., {Wong} K.~C., {Fassnacht} C.~D., {Chiueh} T.,
  {Halkola} A., {Hu} I.~S., {Auger} M.~W., et~al. (2016{\natexlab{a}}) {SHARP -
  III. First use of adaptive-optics imaging to constrain cosmology with
  gravitational lens time delays}. \emph{\mnras} 462(4):3457--3475,
  \doi{10.1093/mnras/stw991}, \eprint{1601.01321}

\bibitem[{{Chen} et~al.(2019){Chen}, {Fassnacht}, {Suyu}, {Rusu}, {Chan},
  {Wong}, {Auger}, {Hilbert}, {Bonvin}, {Birrer}, {Millon}, {Koopmans},
  {Lagattuta}, {McKean}, {Vegetti}, {Courbin}, {Ding}, {Halkola}, {Jee},
  {Shajib}, {Sluse}, {Sonnenfeld}, and {Treu}}]{Chen:2019}
{Chen} G. C.~F., {Fassnacht} C.~D., {Suyu} S.~H., {Rusu} C.~E., {Chan} J.
  H.~H., {Wong} K.~C., {Auger} M.~W., {Hilbert} S., et~al. (2019) {A SHARP view
  of H0LiCOW: H$_{0}$ from three time-delay gravitational lens systems with
  adaptive optics imaging}. \emph{\mnras} 490(2):1743--1773,
  \doi{10.1093/mnras/stz2547}, \eprint{1907.02533}

\bibitem[{{Chen} et~al.(2021){Chen}, {Fassnacht}, {Suyu}, {Koopmans},
  {Lagattuta}, {McKean}, {Auger}, {Vegetti}, and {Treu}}]{Chen:2021}
{Chen} G. C.~F., {Fassnacht} C.~D., {Suyu} S.~H., {Koopmans} L. V.~E.,
  {Lagattuta} D.~J., {McKean} J.~P., {Auger} M.~W., {Vegetti} S., et~al. (2021)
  {SHARP VIII: J0924+0219 lens mass distribution and time-delay prediction
  through adaptive-optics imaging}. \emph{arXiv e-prints} arXiv:2107.10304,
  \eprint{2107.10304}

\bibitem[{{Chen} and {Holz}(2016)}]{2016arXiv161201471C}
{Chen} H.-Y., {Holz} D.~E. (2016) {Finding the One: Identifying the Host
  Galaxies of Gravitational-Wave Sources}. \emph{arXiv e-prints}
  arXiv:1612.01471, \eprint{1612.01471}

\bibitem[{{Chen} et~al.(2018){Chen}, {Fishbach}, and
  {Holz}}]{2018Natur.562..545C}
{Chen} H.-Y., {Fishbach} M., {Holz} D.~E. (2018) {A two per cent Hubble
  constant measurement from standard sirens within five years}. \emph{\nat}
  562(7728):545--547, \doi{10.1038/s41586-018-0606-0}, \eprint{1712.06531}

\bibitem[{{Chen} et~al.(2016{\natexlab{b}}){Chen}, {Zhou}, and {Fu}}]{chen2016}
{Chen} Z., {Zhou} B., {Fu} X. (2016{\natexlab{b}}) {Testing the
  Distance-Duality Relation from Hubble, Galaxy Clusters and Type Ia Supernovae
  Data with Model Independent Methods}. \emph{International Journal of
  Theoretical Physics} 55(2):1229--1240, \doi{10.1007/s10773-015-2765-1}

\bibitem[{Chen et~al.(2021)Chen, Wolz, Spinelli, and Murray}]{Chen:2020uld}
Chen Z., Wolz L., Spinelli M., Murray S.~G. (2021) {Extracting H i astrophysics
  from interferometric intensity mapping}. \emph{Mon Not Roy Astron Soc}
  502(4):5259--5276, \doi{10.1093/mnras/stab386}, \eprint{2010.07985}

\bibitem[{{Chernoff} and {Finn}(1993)}]{1993ApJ...411L...5C}
{Chernoff} D.~F., {Finn} L.~S. (1993) {Gravitational Radiation, Inspiraling
  Binaries, and Cosmology}. \emph{\apjl} 411:L5, \doi{10.1086/186898},
  \eprint{gr-qc/9304020}

\bibitem[{{Chevallard} and {Charlot}(2016)}]{chevallard2016}
{Chevallard} J., {Charlot} S. (2016) {Modelling and interpreting spectral
  energy distributions of galaxies with BEAGLE}. \emph{\mnras}
  462(2):1415--1443, \doi{10.1093/mnras/stw1756}, \eprint{1603.03037}

\bibitem[{{Chevallier} and {Polarski}(2001)}]{chevallier2001}
{Chevallier} M., {Polarski} D. (2001) {Accelerating Universes with Scaling Dark
  Matter}. \emph{International Journal of Modern Physics D} 10(2):213--223,
  \doi{10.1142/S0218271801000822}, \eprint{gr-qc/0009008}

\bibitem[{{Chiriv{\`\i}} et~al.(2018){Chiriv{\`\i}}, {Suyu}, {Grillo},
  {Halkola}, {Balestra}, {Caminha}, {Mercurio}, and {Rosati}}]{Chirivi:2018}
{Chiriv{\`\i}} G., {Suyu} S.~H., {Grillo} C., {Halkola} A., {Balestra} I.,
  {Caminha} G.~B., {Mercurio} A., {Rosati} P. (2018) {MACS J0416.1-2403: Impact
  of line-of-sight structures on strong gravitational lensing modelling of
  galaxy clusters}. \emph{\aap} 614:A8, \doi{10.1051/0004-6361/201731433},
  \eprint{1706.07815}

\bibitem[{{Choi} et~al.(2014){Choi}, {Conroy}, {Moustakas}, {Graves}, {Holden},
  {Brodwin}, {Brown}, and {van Dokkum}}]{choi2014}
{Choi} J., {Conroy} C., {Moustakas} J., {Graves} G.~J., {Holden} B.~P.,
  {Brodwin} M., {Brown} M.~J.~I., {van Dokkum} P.~G. (2014) {The Assembly
  Histories of Quiescent Galaxies since z = 0.7 from Absorption Line
  Spectroscopy}. \emph{\apj} 792:95, \doi{10.1088/0004-637X/792/2/95},
  \eprint{1403.4932}

\bibitem[{{Christlieb}(2016)}]{Christlieb2016}
{Christlieb} N. (2016) {Age determination of metal-poor halo stars using
  nucleochronometry}. \emph{Astronomische Nachrichten} 337(8-9):931,
  \doi{10.1002/asna.201612401}

\bibitem[{{Chuang} et~al.(2017){Chuang}, {Kitaura}, {Liang}, {Font-Ribera},
  {Zhao}, {McDonald}, and {Tao}}]{Chuang2017}
{Chuang} C.-H., {Kitaura} F.-S., {Liang} Y., {Font-Ribera} A., {Zhao} C.,
  {McDonald} P., {Tao} C. (2017) {Linear redshift space distortions for cosmic
  voids based on galaxies in redshift space}. \emph{\prd} 95:063528,
  \doi{10.1103/PhysRevD.95.063528}, \eprint{1605.05352}

\bibitem[{Chudaykin et~al.(2021)Chudaykin, Dolgikh, and
  Ivanov}]{Chudaykin:2020ghx}
Chudaykin A., Dolgikh K., Ivanov M.~M. (2021) {Constraints on the curvature of
  the Universe and dynamical dark energy from the Full-shape and BAO data}.
  \emph{Phys Rev D} 103(2):023507, \doi{10.1103/PhysRevD.103.023507},
  \eprint{2009.10106}

\bibitem[{{Cid Fernandes} et~al.(2005){Cid Fernandes}, {Mateus}, {Sodr{\'e}},
  {Stasi{\'n}ska}, and {Gomes}}]{cidfernandes2005}
{Cid Fernandes} R., {Mateus} A., {Sodr{\'e}} L., {Stasi{\'n}ska} G., {Gomes}
  J.~M. (2005) {Semi-empirical analysis of Sloan Digital Sky Survey galaxies -
  I. Spectral synthesis method}. \emph{\mnras} 358(2):363--378,
  \doi{10.1111/j.1365-2966.2005.08752.x}, \eprint{astro-ph/0412481}

\bibitem[{{Cimatti} et~al.(2004){Cimatti}, {Daddi}, {Renzini}, {Cassata},
  {Vanzella}, {Pozzetti}, {Cristiani}, {Fontana}, {Rodighiero}, {Mignoli}, and
  {Zamorani}}]{cimatti2004}
{Cimatti} A., {Daddi} E., {Renzini} A., {Cassata} P., {Vanzella} E., {Pozzetti}
  L., {Cristiani} S., {Fontana} A., et~al. (2004) {Old galaxies in the young
  Universe}. \emph{\nat} 430(6996):184--187, \doi{10.1038/nature02668},
  \eprint{astro-ph/0407131}

\bibitem[{{Citro} et~al.(2016){Citro}, {Pozzetti}, {Moresco}, and
  {Cimatti}}]{citro2016}
{Citro} A., {Pozzetti} L., {Moresco} M., {Cimatti} A. (2016) {Inferring the
  star-formation histories of the most massive and passive early-type galaxies
  at z \&lt; 0.3}. \emph{\aap} 592:A19, \doi{10.1051/0004-6361/201527772},
  \eprint{1604.07826}

\bibitem[{{Citro} et~al.(2017){Citro}, {Pozzetti}, {Quai}, {Moresco},
  {Vallini}, and {Cimatti}}]{citro2017}
{Citro} A., {Pozzetti} L., {Quai} S., {Moresco} M., {Vallini} L., {Cimatti} A.
  (2017) {A methodology to select galaxies just after the quenching of star
  formation}. \emph{\mnras} 469:3108--3124, \doi{10.1093/mnras/stx932},
  \eprint{1704.05462}

\bibitem[{{Clampitt} and {Jain}(2015)}]{Clampitt2015}
{Clampitt} J., {Jain} B. (2015) {Lensing measurements of the mass distribution
  in SDSS voids}. \emph{\mnras} 454:3357--3365, \doi{10.1093/mnras/stv2215},
  \eprint{1404.1834}

\bibitem[{{Clampitt} et~al.(2013){Clampitt}, {Cai}, and {Li}}]{Clampitt2013}
{Clampitt} J., {Cai} Y.-C., {Li} B. (2013) {Voids in modified gravity:
  excursion set predictions}. \emph{\mnras} 431(1):749--766,
  \doi{10.1093/mnras/stt219}, \eprint{1212.2216}

\bibitem[{{Coe} et~al.(2019){Coe}, {Salmon}, {Brada{\v{c}}}, {Bradley},
  {Sharon}, {Zitrin}, {Acebron}, {Cerny}, {Cibirka}, {Strait}, and
  et~al.}]{Coe:2019}
{Coe} D., {Salmon} B., {Brada{\v{c}}} M., {Bradley} L.~D., {Sharon} K.,
  {Zitrin} A., {Acebron} A., {Cerny} C., et~al. (2019) {RELICS: Reionization
  Lensing Cluster Survey}. \emph{\apj} 884(1):85,
  \doi{10.3847/1538-4357/ab412b}, \eprint{1903.02002}

\bibitem[{{Cohen} et~al.(2018){Cohen}, {van Dokkum}, {Danieli}, {Romanowsky},
  {Abraham}, {Merritt}, {Zhang}, {Mowla}, {Kruijssen}, {Conroy}, and
  {Wasserman}}]{cohen18}
{Cohen} Y., {van Dokkum} P., {Danieli} S., {Romanowsky} A.~J., {Abraham} R.,
  {Merritt} A., {Zhang} J., {Mowla} L., et~al. (2018) {The Dragonfly Nearby
  Galaxies Survey. V. HST/ACS Observations of 23 Low Surface Brightness Objects
  in the Fields of NGC 1052, NGC 1084, M96, and NGC 4258}. \emph{\apj}
  868(2):96, \doi{10.3847/1538-4357/aae7c8}, \eprint{1807.06016}

\bibitem[{{Colberg} et~al.(2008){Colberg}, {Pearce}, {Foster}, {Platen},
  {Brunino}, {Neyrinck}, {Basilakos}, {Fairall}, {Feldman}, {Gottl{\"o}ber},
  {Hahn}, {Hoyle}, {M{\"u}ller}, {Nelson}, {Plionis}, {Porciani}, {Shandarin},
  {Vogeley}, and {van de Weygaert}}]{Colberg2008}
{Colberg} J.~M., {Pearce} F., {Foster} C., {Platen} E., {Brunino} R.,
  {Neyrinck} M., {Basilakos} S., {Fairall} A., et~al. (2008) {The
  Aspen-Amsterdam void finder comparison project}. \emph{\mnras}
  387(2):933--944, \doi{10.1111/j.1365-2966.2008.13307.x}, \eprint{0803.0918}

\bibitem[{{Cole} et~al.(2005){Cole}, {Percival}, {Peacock}, {Norberg}, {Baugh},
  {Frenk}, {Baldry}, {Bland-Hawthorn}, {Bridges}, {Cannon}, {Colless},
  {Collins}, {Couch}, {Cross}, {Dalton}, {Eke}, {De Propris}, {Driver},
  {Efstathiou}, {Ellis}, {Glazebrook}, {Jackson}, {Jenkins}, {Lahav}, {Lewis},
  {Lumsden}, {Maddox}, {Madgwick}, {Peterson}, {Sutherland}, and
  {Taylor}}]{cole2005}
{Cole} S., {Percival} W.~J., {Peacock} J.~A., {Norberg} P., {Baugh} C.~M.,
  {Frenk} C.~S., {Baldry} I., {Bland-Hawthorn} J., et~al. (2005) {The 2dF
  Galaxy Redshift Survey: power-spectrum analysis of the final data set and
  cosmological implications}. \emph{\mnras} 362(2):505--534,
  \doi{10.1111/j.1365-2966.2005.09318.x}, \eprint{astro-ph/0501174}

\bibitem[{{Coles} et~al.(2014){Coles}, {Read}, and {Saha}}]{Coles:2014}
{Coles} J.~P., {Read} J.~I., {Saha} P. (2014) {Gravitational lens recovery with
  GLASS: measuring the mass profile and shape of a lens}. \emph{\mnras}
  445(3):2181--2197, \doi{10.1093/mnras/stu1781}, \eprint{1401.7990}

\bibitem[{{Colg{\'a}in} and {Sheikh-Jabbari}(2021)}]{colgain2021b}
{Colg{\'a}in} E.~{\'O}., {Sheikh-Jabbari} M.~M. (2021) {Elucidating
  cosmological model dependence with $H_0$}. \emph{arXiv e-prints}
  arXiv:2101.08565, \eprint{2101.08565}

\bibitem[{{Colg{\'a}in} and {Yavartanoo}(2019)}]{colgain2019}
{Colg{\'a}in} E.~{\'O}., {Yavartanoo} H. (2019) {Testing the Swampland: $H_0$
  tension}. \emph{arXiv e-prints} arXiv:1905.02555, \eprint{1905.02555}

\bibitem[{{Colg{\'a}in} et~al.(2021){Colg{\'a}in}, {Sheikh-Jabbari}, and
  {Yin}}]{colgain2021a}
{Colg{\'a}in} E.~{\'O}., {Sheikh-Jabbari} M.~M., {Yin} L. (2021) {Can dark
  energy be dynamical?} \emph{arXiv e-prints} arXiv:2104.01930,
  \eprint{2104.01930}

\bibitem[{{Colg{\'a}in} et~al.(2022{\natexlab{a}}){Colg{\'a}in},
  {Sheikh-Jabbari}, {Solomon}, {Bargiacchi}, {Capozziello}, {Dainotti}, and
  {Stojkovic}}]{colgain2022b}
{Colg{\'a}in} E.~{\'O}., {Sheikh-Jabbari} M.~M., {Solomon} R., {Bargiacchi} G.,
  {Capozziello} S., {Dainotti} M.~G., {Stojkovic} D. (2022{\natexlab{a}})
  {Revealing Intrinsic Flat $\Lambda$CDM Biases with Standardizable Candles}.
  \emph{arXiv e-prints} arXiv:2203.10558, \eprint{2203.10558}

\bibitem[{{Colg{\'a}in} et~al.(2022{\natexlab{b}}){Colg{\'a}in},
  {Sheikh-Jabbari}, {Solomon}, {Dainotti}, and {Stojkovic}}]{colgain2022a}
{Colg{\'a}in} E.~{\'O}., {Sheikh-Jabbari} M.~M., {Solomon} R., {Dainotti}
  M.~G., {Stojkovic} D. (2022{\natexlab{b}}) {Putting Flat $\Lambda$CDM In The
  (Redshift) Bin}. \emph{arXiv e-prints} arXiv:2206.11447, \eprint{2206.11447}

\bibitem[{{Collett}(2015)}]{Collett:2015}
{Collett} T.~E. (2015) {The Population of Galaxy-Galaxy Strong Lenses in
  Forthcoming Optical Imaging Surveys}. \emph{\apj} 811(1):20,
  \doi{10.1088/0004-637X/811/1/20}, \eprint{1507.02657}

\bibitem[{{Comparat} et~al.(2017){Comparat}, {Maraston}, {Goddard},
  {Gonzalez-Perez}, {Lian}, {Meneses-Goytia}, {Thomas}, {Brownstein},
  {Tojeiro}, {Finoguenov}, {Merloni}, {Prada}, {Salvato}, {Zhu}, {Zou}, and
  {Brinkmann}}]{comparat2017}
{Comparat} J., {Maraston} C., {Goddard} D., {Gonzalez-Perez} V., {Lian} J.,
  {Meneses-Goytia} S., {Thomas} D., {Brownstein} J.~R., et~al. (2017) {Stellar
  population properties for 2 million galaxies from SDSS DR14 and DEEP2 DR4
  from full spectral fitting}. \emph{arXiv e-prints} arXiv:1711.06575,
  \eprint{1711.06575}

\bibitem[{{Comparat} et~al.(2020){Comparat}, {Merloni}, {Dwelly}, {Salvato},
  {Schwope}, {Coffey}, {Wolf}, {Arcodia}, {Liu}, {Buchner}, {Nandra},
  {Georgakakis}, {Clerc}, {Brusa}, {Brownstein}, {Schneider}, {Pan}, and
  {Bizyaev}}]{comparat2020}
{Comparat} J., {Merloni} A., {Dwelly} T., {Salvato} M., {Schwope} A., {Coffey}
  D., {Wolf} J., {Arcodia} R., et~al. (2020) {The final SDSS-IV/SPIDERS X-ray
  point source spectroscopic catalogue}. \emph{\aap} 636:A97,
  \doi{10.1051/0004-6361/201937272}, \eprint{1912.03068}

\bibitem[{{Conroy} and {Gunn}(2010)}]{conroy2010}
{Conroy} C., {Gunn} J.~E. (2010) {The Propagation of Uncertainties in Stellar
  Population Synthesis Modeling. III. Model Calibration, Comparison, and
  Evaluation}. \emph{\apj} 712(2):833--857, \doi{10.1088/0004-637X/712/2/833},
  \eprint{0911.3151}

\bibitem[{{Conroy} and {van Dokkum}(2012)}]{conroy2012}
{Conroy} C., {van Dokkum} P. (2012) {Counting Low-mass Stars in Integrated
  Light}. \emph{\apj} 747(1):69, \doi{10.1088/0004-637X/747/1/69},
  \eprint{1109.0007}

\bibitem[{{Conroy} et~al.(2009){Conroy}, {Gunn}, and {White}}]{conroy2009}
{Conroy} C., {Gunn} J.~E., {White} M. (2009) {The Propagation of Uncertainties
  in Stellar Population Synthesis Modeling. I. The Relevance of Uncertain
  Aspects of Stellar Evolution and the Initial Mass Function to the Derived
  Physical Properties of Galaxies}. \emph{\apj} 699(1):486--506,
  \doi{10.1088/0004-637X/699/1/486}, \eprint{0809.4261}

\bibitem[{{Conroy} et~al.(2014){Conroy}, {Graves}, and {van
  Dokkum}}]{conroy2014}
{Conroy} C., {Graves} G.~J., {van Dokkum} P.~G. (2014) {Early-type Galaxy
  Archeology: Ages, Abundance Ratios, and Effective Temperatures from
  Full-spectrum Fitting}. \emph{\apj} 780(1):33,
  \doi{10.1088/0004-637X/780/1/33}, \eprint{1303.6629}

\bibitem[{{Contarini} et~al.(2019){Contarini}, {Ronconi}, {Marulli},
  {Moscardini}, {Veropalumbo}, and {Baldi}}]{Contarini2019}
{Contarini} S., {Ronconi} T., {Marulli} F., {Moscardini} L., {Veropalumbo} A.,
  {Baldi} M. (2019) {Cosmological exploitation of the size function of cosmic
  voids identified in the distribution of biased tracers}. \emph{\mnras}
  488(3):3526--3540, \doi{10.1093/mnras/stz1989}, \eprint{1904.01022}

\bibitem[{{Contarini} et~al.(2021){Contarini}, {Marulli}, {Moscardini},
  {Veropalumbo}, {Giocoli}, and {Baldi}}]{Contarini2021}
{Contarini} S., {Marulli} F., {Moscardini} L., {Veropalumbo} A., {Giocoli} C.,
  {Baldi} M. (2021) {Cosmic voids in modified gravity models with massive
  neutrinos}. \emph{\mnras} 504(4):5021--5038, \doi{10.1093/mnras/stab1112},
  \eprint{2009.03309}

\bibitem[{{Cooke}(2020)}]{cooke2020}
{Cooke} R. (2020) {The ACCELERATION programme: I. Cosmology with the redshift
  drift}. \emph{\mnras} 492(2):2044--2057, \doi{10.1093/mnras/stz3465},
  \eprint{1912.04983}

\bibitem[{{Corasaniti} et~al.(2007){Corasaniti}, {Huterer}, and
  {Melchiorri}}]{corasaniti2007}
{Corasaniti} P.-S., {Huterer} D., {Melchiorri} A.~r. (2007) {Exploring the dark
  energy redshift desert with the Sandage-Loeb test}. \emph{\prd} 75(6):062001,
  \doi{10.1103/PhysRevD.75.062001}, \eprint{astro-ph/0701433}

\bibitem[{{Correa} et~al.(2019){Correa}, {Paz}, {Padilla}, {Ruiz}, {Angulo},
  and {S{\'a}nchez}}]{Correa2019}
{Correa} C.~M., {Paz} D.~J., {Padilla} N.~D., {Ruiz} A.~N., {Angulo} R.~E.,
  {S{\'a}nchez} A.~G. (2019) {Non-fiducial cosmological test from geometrical
  and dynamical distortions around voids}. \emph{\mnras} 485(4):5761--5772,
  \doi{10.1093/mnras/stz821}, \eprint{1811.12251}

\bibitem[{{Correa} et~al.(2021){Correa}, {Paz}, {S{\'a}nchez}, {Ruiz},
  {Padilla}, and {Angulo}}]{Correa2021a}
{Correa} C.~M., {Paz} D.~J., {S{\'a}nchez} A.~G., {Ruiz} A.~N., {Padilla}
  N.~D., {Angulo} R.~E. (2021) {Redshift-space effects in voids and their
  impact on cosmological tests. Part I: the void size function}. \emph{\mnras}
  500(1):911--925, \doi{10.1093/mnras/staa3252}, \eprint{2007.12064}

\bibitem[{{Correa} et~al.(2022){Correa}, {Paz}, {Padilla}, {S{\'a}nchez},
  {Ruiz}, and {Angulo}}]{Correa2021b}
{Correa} C.~M., {Paz} D.~J., {Padilla} N.~D., {S{\'a}nchez} A.~G., {Ruiz}
  A.~N., {Angulo} R.~E. (2022) {Redshift-space effects in voids and their
  impact on cosmological tests - II. The void-galaxy cross-correlation
  function}. \emph{\mnras} 509(2):1871--1884, \doi{10.1093/mnras/stab3070},
  \eprint{2107.01314}

\bibitem[{{Cosmic Visions 21 cm Collaboration} et~al.(2018){Cosmic Visions 21
  cm Collaboration}, {Ansari}, {Arena}, {Bandura}, {Bull}, {Castorina},
  {Chang}, {Chen}, {Connor}, {Foreman}, {Frisch}, and
  et~al.}]{CosmicVisions21cm:2018rfq}
{Cosmic Visions 21 cm Collaboration}, {Ansari} R., {Arena} E.~J., {Bandura} K.,
  {Bull} P., {Castorina} E., {Chang} T.-C., {Chen} S.-F., et~al. (2018)
  {Inflation and Early Dark Energy with a Stage II Hydrogen Intensity Mapping
  Experiment}. \emph{arXiv e-prints} arXiv:1810.09572, \eprint{1810.09572}

\bibitem[{{Courbin} et~al.(2011){Courbin}, {Chantry}, {Revaz}, {Sluse},
  {Faure}, {Tewes}, {Eulaers}, {Koleva}, {Asfandiyarov}, {Dye}, {Magain}, {van
  Winckel}, {Coles}, {Saha}, {Ibrahimov}, and {Meylan}}]{Courbin2011}
{Courbin} F., {Chantry} V., {Revaz} Y., {Sluse} D., {Faure} C., {Tewes} M.,
  {Eulaers} E., {Koleva} M., et~al. (2011) {COSMOGRAIL: the COSmological
  MOnitoring of GRAvItational Lenses. IX. Time delays, lens dynamics and
  baryonic fraction in HE 0435-1223}. \emph{\aap} 536:A53,
  \doi{10.1051/0004-6361/201015709}, \eprint{1009.1473}

\bibitem[{{Courbin} et~al.(2018{\natexlab{a}}){Courbin}, {Bonvin},
  {Buckley-Geer}, {Fassnacht}, {Frieman}, {Lin}, {Marshall}, {Suyu}, {Treu},
  {Anguita}, {Motta}, and et~al.}]{Courbin2018}
{Courbin} F., {Bonvin} V., {Buckley-Geer} E., {Fassnacht} C.~D., {Frieman} J.,
  {Lin} H., {Marshall} P.~J., {Suyu} S.~H., et~al. (2018{\natexlab{a}})
  {COSMOGRAIL: the COSmological MOnitoring of GRAvItational Lenses. XVI. Time
  delays for the quadruply imaged quasar DES J0408-5354 with high-cadence
  photometric monitoring}. \emph{\aap} 609:A71,
  \doi{10.1051/0004-6361/201731461}

\bibitem[{{Courbin} et~al.(2018{\natexlab{b}}){Courbin}, {Bonvin},
  {Buckley-Geer}, {Fassnacht}, {Frieman}, {Lin}, {Marshall}, {Suyu}, {Treu},
  {Anguita}, {Motta}, and et~al.}]{Courbin:2018}
{Courbin} F., {Bonvin} V., {Buckley-Geer} E., {Fassnacht} C.~D., {Frieman} J.,
  {Lin} H., {Marshall} P.~J., {Suyu} S.~H., et~al. (2018{\natexlab{b}})
  {COSMOGRAIL: the COSmological MOnitoring of GRAvItational Lenses. XVI. Time
  delays for the quadruply imaged quasar DES J0408-5354 with high-cadence
  photometric monitoring}. \emph{\aap} 609:A71,
  \doi{10.1051/0004-6361/201731461}

\bibitem[{{Cousinou} et~al.(2019){Cousinou}, {Pisani}, {Tilquin}, {Hamaus},
  {Hawken}, and {Escoffier}}]{Cousinou2019}
{Cousinou} M.~C., {Pisani} A., {Tilquin} A., {Hamaus} N., {Hawken} A.~J.,
  {Escoffier} S. (2019) {Multivariate analysis of cosmic void characteristics}.
  \emph{Astronomy and Computing} 27:53, \doi{10.1016/j.ascom.2019.03.001},
  \eprint{1805.07181}

\bibitem[{Creminelli and Vernizzi(2017)}]{Creminelli2017}
Creminelli P., Vernizzi F. (2017) {Dark Energy after GW170817 and GRB170817A}.
  \emph{Physical Review Letters} 119(25):251302,
  \doi{10.1103/PhysRevLett.119.251302}

\bibitem[{Crighton et~al.(2015)}]{Crighton:2015pza}
Crighton N.~H., et~al. (2015) {The neutral hydrogen cosmological mass density
  at z = 5}. \emph{Mon Not Roy Astron Soc} 452(1):217--234,
  \doi{10.1093/mnras/stv1182}, \eprint{1506.02037}

\bibitem[{{Crisostomi} and {Koyama}(2018)}]{Crisostomi:2017pjs}
{Crisostomi} M., {Koyama} K. (2018) {Self-accelerating universe in
  scalar-tensor theories after GW170817}. \emph{\prd} 97(8):084004,
  \doi{10.1103/PhysRevD.97.084004}, \eprint{1712.06556}

\bibitem[{Cunnington et~al.(2019)Cunnington, Wolz, Pourtsidou, and
  Bacon}]{Cunnington:2019lvb}
Cunnington S., Wolz L., Pourtsidou A., Bacon D. (2019) {Impact of foregrounds
  on HI intensity mapping cross-correlations with optical surveys}. \emph{Mon
  Not Roy Astron Soc} 488(4):5452--5472, \doi{10.1093/mnras/stz1916},
  \eprint{1904.01479}

\bibitem[{Cunnington et~al.(2020{\natexlab{a}})Cunnington, Camera, and
  Pourtsidou}]{Cunnington:2020wdu}
Cunnington S., Camera S., Pourtsidou A. (2020{\natexlab{a}}) {The degeneracy
  between primordial non-Gaussianity and foregrounds in 21 cm intensity mapping
  experiments}. \emph{Mon Not Roy Astron Soc} 499(3):4054--4067,
  \doi{10.1093/mnras/staa2986}, \eprint{2007.12126}

\bibitem[{Cunnington et~al.(2020{\natexlab{b}})Cunnington, Pourtsidou, Soares,
  Blake, and Bacon}]{Cunnington:2020mnn}
Cunnington S., Pourtsidou A., Soares P.~S., Blake C., Bacon D.
  (2020{\natexlab{b}}) {Multipole expansion for H i intensity mapping
  experiments: simulations and modelling}. \emph{Mon Not Roy Astron Soc}
  496(1):415--433, \doi{10.1093/mnras/staa1524}, \eprint{2002.05626}

\bibitem[{Cunnington et~al.(2021{\natexlab{a}})Cunnington, Irfan, Carucci,
  Pourtsidou, and Bobin}]{Cunnington:2020njn}
Cunnington S., Irfan M.~O., Carucci I.~P., Pourtsidou A., Bobin J.
  (2021{\natexlab{a}}) {21-cm foregrounds and polarization leakage: cleaning
  and mitigation strategies}. \emph{Mon Not Roy Astron Soc} 504(1):208--227,
  \doi{10.1093/mnras/stab856}, \eprint{2010.02907}

\bibitem[{Cunnington et~al.(2021{\natexlab{b}})Cunnington, Watkinson, and
  Pourtsidou}]{Cunnington:2021czb}
Cunnington S., Watkinson C., Pourtsidou A. (2021{\natexlab{b}}) {The H\,i
  intensity mapping bispectrum including observational effects}. \emph{Mon Not
  Roy Astron Soc} 507(2):1623--1639, \doi{10.1093/mnras/stab2200},
  \eprint{2102.11153}

\bibitem[{Cunnington et~al.(2022)}]{Cunnington:2022uzo}
Cunnington S., et~al. (2022) {HI intensity mapping with MeerKAT: power spectrum
  detection in cross-correlation with WiggleZ galaxies}. \eprint{2206.01579}

\bibitem[{{Cutri} et~al.(2003){Cutri}, {Skrutskie}, {van Dyk}, {Beichman},
  {Carpenter}, {Chester}, {Cambresy}, {Evans}, {Fowler}, {Gizis}, {Howard},
  {Huchra}, {Jarrett}, {Kopan}, {Kirkpatrick}, {Light}, {Marsh}, {McCallon},
  {Schneider}, {Stiening}, {Sykes}, {Weinberg}, {Wheaton}, {Wheelock}, and
  {Zacarias}}]{cutri2003}
{Cutri} R.~M., {Skrutskie} M.~F., {van Dyk} S., {Beichman} C.~A., {Carpenter}
  J.~M., {Chester} T., {Cambresy} L., {Evans} T., et~al. (2003) {VizieR Online
  Data Catalog: 2MASS All-Sky Catalog of Point Sources (Cutri+ 2003)}.
  \emph{VizieR Online Data Catalog} II/246

\bibitem[{{Daddi} et~al.(2004){Daddi}, {Cimatti}, {Renzini}, {Vernet},
  {Conselice}, {Pozzetti}, {Mignoli}, {Tozzi}, {Broadhurst}, {di Serego
  Alighieri}, {Fontana}, {Nonino}, {Rosati}, and {Zamorani}}]{daddi2004}
{Daddi} E., {Cimatti} A., {Renzini} A., {Vernet} J., {Conselice} C., {Pozzetti}
  L., {Mignoli} M., {Tozzi} P., et~al. (2004) {Near-Infrared Bright Galaxies at
  z\raisebox{-0.5ex}\textasciitilde=2. Entering the Spheroid Formation Epoch?}
  \emph{\apjl} 600(2):L127--L130, \doi{10.1086/381020},
  \eprint{astro-ph/0308456}

\bibitem[{{Daddi} et~al.(2005){Daddi}, {Renzini}, {Pirzkal}, {Cimatti},
  {Malhotra}, {Stiavelli}, {Xu}, {Pasquali}, {Rhoads}, {Brusa}, {di Serego
  Alighieri}, {Ferguson}, {Koekemoer}, {Moustakas}, {Panagia}, and
  {Windhorst}}]{daddi2005}
{Daddi} E., {Renzini} A., {Pirzkal} N., {Cimatti} A., {Malhotra} S.,
  {Stiavelli} M., {Xu} C., {Pasquali} A., et~al. (2005) {Passively Evolving
  Early-Type Galaxies at 1.4 \&lt;\raisebox{-0.5ex}\textasciitilde z
  \&lt;\raisebox{-0.5ex}\textasciitilde 2.5 in the Hubble Ultra Deep Field}.
  \emph{\apj} 626(2):680--697, \doi{10.1086/430104}, \eprint{astro-ph/0503102}

\bibitem[{{D'Agostino} and {Nunes}(2019)}]{dagostino2019}
{D'Agostino} R., {Nunes} R.~C. (2019) {Probing observational bounds on
  scalar-tensor theories from standard sirens}. \emph{\prd} 100(4):044041,
  \doi{10.1103/PhysRevD.100.044041}, \eprint{1907.05516}

\bibitem[{{D'Agostino} and {Nunes}(2020)}]{dagostino2020}
{D'Agostino} R., {Nunes} R.~C. (2020) {Measurements of H$_{0}$ in modified
  gravity theories: The role of lensed quasars in the late-time Universe}.
  \emph{\prd} 101(10):103505, \doi{10.1103/PhysRevD.101.103505},
  \eprint{2002.06381}

\bibitem[{{Dahle} et~al.(2013){Dahle}, {Gladders}, {Sharon}, {Bayliss},
  {Wuyts}, {Abramson}, {Koester}, {Groeneboom}, {Brinckmann}, {Kristensen},
  {Lindholmer}, {Nielsen}, {Krogager}, and {Fynbo}}]{sdss2222}
{Dahle} H., {Gladders} M.~D., {Sharon} K., {Bayliss} M.~B., {Wuyts} E.,
  {Abramson} L.~E., {Koester} B.~P., {Groeneboom} N., et~al. (2013) {SDSS
  J2222+2745: A Gravitationally Lensed Sextuple Quasar with a Maximum Image
  Separation of 15.''1 Discovered in the Sloan Giant Arcs Survey}. \emph{\apj}
  773(2):146, \doi{10.1088/0004-637X/773/2/146}, \eprint{1211.1091}

\bibitem[{{Dahle} et~al.(2015){Dahle}, {Gladders}, {Sharon}, {Bayliss}, and
  {Rigby}}]{Dahle2015}
{Dahle} H., {Gladders} M.~D., {Sharon} K., {Bayliss} M.~B., {Rigby} J.~R.
  (2015) {Time Delay Measurements for the Cluster-lensed Sextuple Quasar SDSS
  J2222+2745}. \emph{\apj} 813(1):67, \doi{10.1088/0004-637X/813/1/67},
  \eprint{1505.06187}

\bibitem[{{Dainotti} and {Amati}(2018)}]{Dainotti18}
{Dainotti} M.~G., {Amati} L. (2018) {Gamma-ray Burst Prompt Correlations:
  Selection and Instrumental Effects}. \emph{\pasp} 130(987):051001,
  \doi{10.1088/1538-3873/aaa8d7}, \eprint{1704.00844}

\bibitem[{{Dainotti} et~al.(2008){Dainotti}, {Cardone}, and
  {Capozziello}}]{Dainotti08}
{Dainotti} M.~G., {Cardone} V.~F., {Capozziello} S. (2008) {A time-luminosity
  correlation for {\ensuremath{\gamma}}-ray bursts in the X-rays}.
  \emph{\mnras} 391(1):L79--L83, \doi{10.1111/j.1745-3933.2008.00560.x},
  \eprint{0809.1389}

\bibitem[{{Dainotti} et~al.(2020){Dainotti}, {Livermore}, {Kann}, {Li},
  {Oates}, {Yi}, {Zhang}, {Gendre}, {Cenko}, and {Fraija}}]{Dainotti20}
{Dainotti} M.~G., {Livermore} S., {Kann} D.~A., {Li} L., {Oates} S., {Yi} S.,
  {Zhang} B., {Gendre} B., et~al. (2020) {The Optical Luminosity-Time
  Correlation for More than 100 Gamma-Ray Burst Afterglows}. \emph{\apjl}
  905(2):L26, \doi{10.3847/2041-8213/abcda9}, \eprint{2011.14493}

\bibitem[{{Dainotti} et~al.(2021){Dainotti}, {De Simone}, {Schiavone},
  {Montani}, {Rinaldi}, and {Lambiase}}]{Dainotti2021}
{Dainotti} M.~G., {De Simone} B., {Schiavone} T., {Montani} G., {Rinaldi} E.,
  {Lambiase} G. (2021) {On the Hubble Constant Tension in the SNe Ia Pantheon
  Sample}. \emph{\apj} 912(2):150, \doi{10.3847/1538-4357/abeb73},
  \eprint{2103.02117}

\bibitem[{{Dainotti} et~al.(2022{\natexlab{a}}){Dainotti}, {Bargiacchi},
  {Lenart}, {Capozziello}, {{\'O} Colg{\'a}in}, {Solomon}, {Stojkovic}, and
  {Sheikh-Jabbari}}]{dainotti2022}
{Dainotti} M.~G., {Bargiacchi} G., {Lenart} A.~{\L}., {Capozziello} S., {{\'O}
  Colg{\'a}in} E., {Solomon} R., {Stojkovic} D., {Sheikh-Jabbari} M.~M.
  (2022{\natexlab{a}}) {Quasar Standardization: Overcoming Selection Biases and
  Redshift Evolution}. \emph{\apj} 931(2):106, \doi{10.3847/1538-4357/ac6593},
  \eprint{2203.12914}

\bibitem[{{Dainotti} et~al.(2022{\natexlab{b}}){Dainotti}, {De Simone},
  {Schiavone}, {Montani}, {Rinaldi}, {Lambiase}, {Bogdan}, and
  {Ugale}}]{dainotti2022b}
{Dainotti} M.~G., {De Simone} B.~D., {Schiavone} T., {Montani} G., {Rinaldi}
  E., {Lambiase} G., {Bogdan} M., {Ugale} S. (2022{\natexlab{b}}) {On the
  Evolution of the Hubble Constant with the SNe Ia Pantheon Sample and Baryon
  Acoustic Oscillations: A Feasibility Study for GRB-Cosmology in 2030}.
  \emph{Galaxies} 10(1):24, \doi{10.3390/galaxies10010024}, \eprint{2201.09848}

\bibitem[{{Dalal} et~al.(2006){Dalal}, {Holz}, {Hughes}, and
  {Jain}}]{2006PhRvD..74f3006D}
{Dalal} N., {Holz} D.~E., {Hughes} S.~A., {Jain} B. (2006) {Short GRB and
  binary black hole standard sirens as a probe of dark energy}. \emph{\prd}
  74(6):063006, \doi{10.1103/PhysRevD.74.063006}, \eprint{astro-ph/0601275}

\bibitem[{{D{\'a}lya} et~al.(2018){D{\'a}lya}, {Galg{\'o}czi}, {Dobos}, {Frei},
  {Heng}, {Macas}, {Messenger}, {Raffai}, and {de Souza}}]{2018MNRAS.479.2374D}
{D{\'a}lya} G., {Galg{\'o}czi} G., {Dobos} L., {Frei} Z., {Heng} I.~S., {Macas}
  R., {Messenger} C., {Raffai} P., et~al. (2018) {GLADE: A galaxy catalogue for
  multimessenger searches in the advanced gravitational-wave detector era}.
  \emph{\mnras} 479(2):2374--2381, \doi{10.1093/mnras/sty1703},
  \eprint{1804.05709}

\bibitem[{{Dark Energy Survey Collaboration} et~al.(2016){Dark Energy Survey
  Collaboration}, {Abbott}, {Abdalla}, {Aleksi{\'c}}, {Allam}, {Amara},
  {Bacon}, {Balbinot}, {Banerji}, {Bechtol}, {Benoit-L{\'e}vy}, {Bernstein},
  {Bertin}, {Blazek}, {Bonnett}, {Bridle}, {Brooks}, {Brunner}, and
  et~al.}]{Abbott:2016ktf}
{Dark Energy Survey Collaboration}, {Abbott} T., {Abdalla} F.~B., {Aleksi{\'c}}
  J., {Allam} S., {Amara} A., {Bacon} D., {Balbinot} E., et~al. (2016) {The
  Dark Energy Survey: more than dark energy - an overview}. \emph{\mnras}
  460(2):1270--1299, \doi{10.1093/mnras/stw641}, \eprint{1601.00329}

\bibitem[{{Darling}(2012)}]{darling2012}
{Darling} J. (2012) {Toward a Direct Measurement of the Cosmic Acceleration}.
  \emph{ApJL} 761(2):L26, \doi{10.1088/2041-8205/761/2/L26}, \eprint{1211.4585}

\bibitem[{{Davidzon} et~al.(2017){Davidzon}, {Ilbert}, {Laigle}, {Coupon},
  {McCracken}, {Delvecchio}, {Masters}, {Capak}, {Hsieh}, {Le F{\`e}vre},
  {Tresse}, {Bethermin}, {Chang}, {Faisst}, {Le Floc'h}, {Steinhardt}, {Toft},
  {Aussel}, {Dubois}, {Hasinger}, {Salvato}, {Sanders}, {Scoville}, and
  {Silverman}}]{davidzon2017}
{Davidzon} I., {Ilbert} O., {Laigle} C., {Coupon} J., {McCracken} H.~J.,
  {Delvecchio} I., {Masters} D., {Capak} P., et~al. (2017) {The COSMOS2015
  galaxy stellar mass function . Thirteen billion years of stellar mass
  assembly in ten snapshots}. \emph{\aap} 605:A70,
  \doi{10.1051/0004-6361/201730419}, \eprint{1701.02734}

\bibitem[{{Davies} et~al.(2019){Davies}, {Cautun}, and {Li}}]{Davies2019}
{Davies} C.~T., {Cautun} M., {Li} B. (2019) {Cosmological test of gravity using
  weak lensing voids}. \emph{\mnras} 490(4):4907--4917,
  \doi{10.1093/mnras/stz2933}, \eprint{1907.06657}

\bibitem[{Davis et~al.(2011)}]{Davis:2010jq}
Davis T.~M., et~al. (2011) {The Effect of Peculiar Velocities on Supernova
  Cosmology}. \emph{Astrophys J} 741:67, \doi{10.1088/0004-637X/741/1/67},
  \eprint{1012.2912}

\bibitem[{{Dawson} et~al.(2013){Dawson}, {Schlegel}, {Ahn}, {Anderson},
  {Aubourg}, {Bailey}, {Barkhouser}, {Bautista}, {Beifiori}, {Berlind},
  {Bhardwaj}, {Bizyaev}, {Blake}, and et~al.}]{Dawson2013}
{Dawson} K.~S., {Schlegel} D.~J., {Ahn} C.~P., {Anderson} S.~F., {Aubourg}
  {\'E}., {Bailey} S., {Barkhouser} R.~H., {Bautista} J.~E., et~al. (2013) {The
  Baryon Oscillation Spectroscopic Survey of SDSS-III}. \emph{\aj} 145(1):10,
  \doi{10.1088/0004-6256/145/1/10}, \eprint{1208.0022}

\bibitem[{{de Jaeger} et~al.(2020){de Jaeger}, {Stahl}, {Zheng}, {Filippenko},
  {Riess}, and {Galbany}}]{dejaeger2020}
{de Jaeger} T., {Stahl} B.~E., {Zheng} W., {Filippenko} A.~V., {Riess} A.~G.,
  {Galbany} L. (2020) {A measurement of the Hubble constant from Type II
  supernovae}. \emph{\mnras} 496(3):3402--3411, \doi{10.1093/mnras/staa1801},
  \eprint{2006.03412}

\bibitem[{{de Jong} et~al.(2019){de Jong}, {Agertz}, {Berbel}, {Aird},
  {Alexander}, {Amarsi}, {Anders}, {Andrae}, {Ansarinejad}, {Ansorge},
  {Antilogus}, {Anwand-Heerwart}, {Arentsen}, {Arnadottir}, {Asplund}, {Auger},
  {Azais}, and et~al.}]{Dejong:2019}
{de Jong} R.~S., {Agertz} O., {Berbel} A.~A., {Aird} J., {Alexander} D.~A.,
  {Amarsi} A., {Anders} F., {Andrae} R., et~al. (2019) {4MOST: Project overview
  and information for the First Call for Proposals}. \emph{The Messenger}
  175:3--11, \doi{10.18727/0722-6691/5117}, \eprint{1903.02464}

\bibitem[{De~Paolis et~al.(2020)De~Paolis, Nucita, Strafella, Licchelli, and
  Ingrosso}]{DePaolis:2020onl}
De~Paolis F., Nucita A.~A., Strafella F., Licchelli D., Ingrosso G. (2020) {A
  Quasar microlensing event towards J1249+3449?} \emph{Mon Not Roy Astron Soc}
  499(1):L87--L90, \doi{10.1093/mnrasl/slaa140}, \eprint{2008.02692}

\bibitem[{{Del Pozzo}(2012)}]{2012PhRvD..86d3011D}
{Del Pozzo} W. (2012) {Inference of cosmological parameters from gravitational
  waves: Applications to second generation interferometers}. \emph{\prd}
  86(4):043011, \doi{10.1103/PhysRevD.86.043011}, \eprint{1108.1317}

\bibitem[{{Del Pozzo} et~al.(2017){Del Pozzo}, {Li}, and
  {Messenger}}]{2017PhRvD..95d3502D}
{Del Pozzo} W., {Li} T. G.~F., {Messenger} C. (2017) {Cosmological inference
  using only gravitational wave observations of binary neutron stars}.
  \emph{\prd} 95(4):043502, \doi{10.1103/PhysRevD.95.043502},
  \eprint{1506.06590}

\bibitem[{{Delubac} et~al.(2015){Delubac}, {Bautista}, {Busca}, {Rich},
  {Kirkby}, {Bailey}, {Font-Ribera}, {Slosar}, {Lee}, {Pieri}, and
  et~al.}]{delubac2015}
{Delubac} T., {Bautista} J.~E., {Busca} N.~G., {Rich} J., {Kirkby} D., {Bailey}
  S., {Font-Ribera} A., {Slosar} A., et~al. (2015) {Baryon acoustic
  oscillations in the Ly{\ensuremath{\alpha}} forest of BOSS DR11 quasars}.
  \emph{\aap} 574:A59, \doi{10.1051/0004-6361/201423969}, \eprint{1404.1801}

\bibitem[{{Demianski} et~al.(2017){Demianski}, {Piedipalumbo}, {Sawant}, and
  {Amati}}]{demianski17}
{Demianski} M., {Piedipalumbo} E., {Sawant} D., {Amati} L. (2017) {Cosmology
  with gamma-ray bursts. I. The Hubble diagram through the calibrated
  E$_{p,I}$-E$_{iso}$ correlation}. \emph{\aap} 598:A112,
  \doi{10.1051/0004-6361/201628909}, \eprint{1610.00854}

\bibitem[{{Demianski} et~al.(2020){Demianski}, {Lusso}, {Paolillo},
  {Piedipalumbo}, and {Risaliti}}]{demianski2020}
{Demianski} M., {Lusso} E., {Paolillo} M., {Piedipalumbo} E., {Risaliti} G.
  (2020) {Investigating dark energy equation of state with high redshift Hubble
  diagram}. \emph{Frontiers in Astronomy and Space Sciences} 7:69,
  \doi{10.3389/fspas.2020.521056}, \eprint{2010.05289}

\bibitem[{{Demianski} et~al.(2021){Demianski}, {Piedipalumbo}, {Sawant}, and
  {Amati}}]{demianski21}
{Demianski} M., {Piedipalumbo} E., {Sawant} D., {Amati} L. (2021) {Prospects of
  high redshift constraints on dark energy models with the E$_{p, i}$ -
  E$_{iso}$ correlation in long gamma ray bursts}. \emph{\mnras}
  506(1):903--918, \doi{10.1093/mnras/stab1669}

\bibitem[{{DES Collaboration} et~al.(2021){DES Collaboration}, {Abbott},
  {Aguena}, {Alarcon}, {Allam}, {Alves}, {Amon}, {Andrade-Oliveira}, {Annis},
  {Avila}, {Bacon}, {Baxter}, {Bechtol}, {Becker}, {Bernstein}, {Bhargava},
  {Birrer}, and et~al.}]{DES2021}
{DES Collaboration}, {Abbott} T.~M.~C., {Aguena} M., {Alarcon} A., {Allam} S.,
  {Alves} O., {Amon} A., {Andrade-Oliveira} F., et~al. (2021) {Dark Energy
  Survey Year 3 Results: Cosmological Constraints from Galaxy Clustering and
  Weak Lensing}. \emph{arXiv e-prints} arXiv:2105.13549, \eprint{2105.13549}

\bibitem[{{Deshmukh} et~al.(2018){Deshmukh}, {Caputi}, {Ashby}, {Cowley},
  {McCracken}, {Fynbo}, {Le F{\`e}vre}, {Milvang-Jensen}, and
  {Ilbert}}]{deshmukh2018}
{Deshmukh} S., {Caputi} K.~I., {Ashby} M.~L.~N., {Cowley} W.~I., {McCracken}
  H.~J., {Fynbo} J.~P.~U., {Le F{\`e}vre} O., {Milvang-Jensen} B., et~al.
  (2018) {The Spitzer Matching Survey of the UltraVISTA Ultra-deep Stripes
  (SMUVS): The Evolution of Dusty and Nondusty Galaxies with Stellar Mass at z
  = 2-6}. \emph{\apj} 864(2):166, \doi{10.3847/1538-4357/aad9f5},
  \eprint{1712.03905}

\bibitem[{{DESI Collaboration} et~al.(2016){DESI Collaboration}, {Aghamousa},
  {Aguilar}, {Ahlen}, {Alam}, {Allen}, {Allende Prieto}, {Annis}, {Bailey},
  {Balland}, {Ballester}, {Baltay}, {Beaufore}, {Bebek}, and
  et~al.}]{DESI:2016}
{DESI Collaboration}, {Aghamousa} A., {Aguilar} J., {Ahlen} S., {Alam} S.,
  {Allen} L.~E., {Allende Prieto} C., {Annis} J., et~al. (2016) {The DESI
  Experiment Part I: Science,Targeting, and Survey Design}. \emph{arXiv
  e-prints} arXiv:1611.00036, \eprint{1611.00036}

\bibitem[{{Desjacques} et~al.(2018){Desjacques}, {Jeong}, and
  {Schmidt}}]{Desjacques2018}
{Desjacques} V., {Jeong} D., {Schmidt} F. (2018) {Large-scale galaxy bias}.
  \emph{\physrep} 733:1--193, \doi{10.1016/j.physrep.2017.12.002},
  \eprint{1611.09787}

\bibitem[{{Di Valentino} et~al.(2016){Di Valentino}, {Melchiorri}, and
  {Silk}}]{DiValentino2016}
{Di Valentino} E., {Melchiorri} A., {Silk} J. (2016) {Reconciling Planck with
  the local value of H$_{0}$ in extended parameter space}. \emph{Physics
  Letters B} 761:242--246, \doi{10.1016/j.physletb.2016.08.043},
  \eprint{1606.00634}

\bibitem[{{Di Valentino} et~al.(2020){Di Valentino}, {Gariazzo}, {Mena}, and
  {Vagnozzi}}]{divalentino2020}
{Di Valentino} E., {Gariazzo} S., {Mena} O., {Vagnozzi} S. (2020) {Soundness of
  dark energy properties}. \emph{\jcap} 2020(7):045,
  \doi{10.1088/1475-7516/2020/07/045}, \eprint{2005.02062}

\bibitem[{{Di Valentino} et~al.(2021){Di Valentino}, {Mena}, {Pan},
  {Visinelli}, {Yang}, {Melchiorri}, {Mota}, {Riess}, and
  {Silk}}]{DiValentino2021}
{Di Valentino} E., {Mena} O., {Pan} S., {Visinelli} L., {Yang} W., {Melchiorri}
  A., {Mota} D.~F., {Riess} A.~G., et~al. (2021) {In the Realm of the Hubble
  tension $-$ a Review of Solutions}. \emph{arXiv e-prints} arXiv:2103.01183,
  \eprint{2103.01183}

\bibitem[{{D{\'\i}az-Garc{\'\i}a} et~al.(2019){D{\'\i}az-Garc{\'\i}a},
  {Cenarro}, {L{\'o}pez-Sanjuan}, {Ferreras}, {Cervi{\~n}o},
  {Fern{\'a}ndez-Soto}, {Gonz{\'a}lez Delgado}, {M{\'a}rquez}, and
  et~al.}]{diaz2019}
{D{\'\i}az-Garc{\'\i}a} L.~A., {Cenarro} A.~J., {L{\'o}pez-Sanjuan} C.,
  {Ferreras} I., {Cervi{\~n}o} M., {Fern{\'a}ndez-Soto} A., {Gonz{\'a}lez
  Delgado} R.~M., {M{\'a}rquez} I., et~al. (2019) {Stellar populations of
  galaxies in the ALHAMBRA survey up to z {\ensuremath{\sim}} 1. II. Stellar
  content of quiescent galaxies within the dust-corrected stellar mass-colour
  and the UVJ colour-colour diagrams}. \emph{\aap} 631:A156,
  \doi{10.1051/0004-6361/201832788}, \eprint{1711.10590}

\bibitem[{{Ding} et~al.(2019){Ding}, {Biesiada}, {Zheng}, {Liao}, {Li}, and
  {Zhu}}]{2019JCAP...04..033D}
{Ding} X., {Biesiada} M., {Zheng} X., {Liao} K., {Li} Z., {Zhu} Z.-H. (2019)
  {Cosmological inference from standard sirens without redshift measurements}.
  \emph{\jcap} 2019(4):033, \doi{10.1088/1475-7516/2019/04/033},
  \eprint{1801.05073}

\bibitem[{{Dobie} et~al.(2020){Dobie}, {Kaplan}, {Hotokezaka}, {Murphy},
  {Deller}, {Hallinan}, and {Nissanke}}]{2020MNRAS.494.2449D}
{Dobie} D., {Kaplan} D.~L., {Hotokezaka} K., {Murphy} T., {Deller} A.,
  {Hallinan} G., {Nissanke} S. (2020) {Constraining properties of neutron star
  merger outflows with radio observations}. \emph{\mnras} 494(2):2449--2464,
  \doi{10.1093/mnras/staa789}, \eprint{1910.13662}

\bibitem[{{Dor{\'e}} et~al.(2014){Dor{\'e}}, {Bock}, {Ashby}, {Capak},
  {Cooray}, {de Putter}, {Eifler}, {Flagey}, {Gong}, {Habib}, {Heitmann},
  {Hirata}, {Jeong}, {Katti}, {Korngut}, {Krause}, {Lee}, {Masters},
  {Mauskopf}, {Melnick}, {Mennesson}, {Nguyen}, {{\"O}berg}, {Pullen},
  {Raccanelli}, {Smith}, {Song}, {Tolls}, {Unwin}, {Venumadhav}, {Viero},
  {Werner}, and {Zemcov}}]{Dore:2014}
{Dor{\'e}} O., {Bock} J., {Ashby} M., {Capak} P., {Cooray} A., {de Putter} R.,
  {Eifler} T., {Flagey} N., et~al. (2014) {Cosmology with the SPHEREX All-Sky
  Spectral Survey}. \emph{arXiv e-prints} arXiv:1412.4872, \eprint{1412.4872}

\bibitem[{{Dunlop} et~al.(1996){Dunlop}, {Peacock}, {Spinrad}, {Dey},
  {Jimenez}, {Stern}, and {Windhorst}}]{Dunlop}
{Dunlop} J., {Peacock} J., {Spinrad} H., {Dey} A., {Jimenez} R., {Stern} D.,
  {Windhorst} R. (1996) {A 3.5-Gyr-old galaxy at redshift 1.55}. \emph{\nat}
  381(6583):581--584, \doi{10.1038/381581a0}

\bibitem[{{Edelson} et~al.(2015){Edelson}, {Gelbord}, {Horne}, {McHardy},
  {Peterson}, {Ar{\'e}valo}, {Breeveld}, {De Rosa}, {Evans}, {Goad}, {Kriss},
  {Brandt}, {Gehrels}, and et~al.}]{edelson2015}
{Edelson} R., {Gelbord} J.~M., {Horne} K., {McHardy} I.~M., {Peterson} B.~M.,
  {Ar{\'e}valo} P., {Breeveld} A.~A., {De Rosa} G., et~al. (2015) {Space
  Telescope and Optical Reverberation Mapping Project. II. Swift and HST
  Reverberation Mapping of the Accretion Disk of NGC 5548}. \emph{\apj}
  806(1):129, \doi{10.1088/0004-637X/806/1/129}, \eprint{1501.05951}

\bibitem[{{Efstathiou}(2020)}]{Efstathiou2020}
{Efstathiou} G. (2020) {A Lockdown Perspective on the Hubble Tension (with
  comments from the SH0ES team)}. \emph{arXiv e-prints} arXiv:2007.10716,
  \eprint{2007.10716}

\bibitem[{{Efstathiou}(2021)}]{Efstathiou2021}
{Efstathiou} G. (2021) {To H0 or not to H0?} \emph{arXiv e-prints}
  arXiv:2103.08723, \eprint{2103.08723}

\bibitem[{{Eikenberry} et~al.(2019{\natexlab{a}}){Eikenberry}, {Gonzalez},
  {Darling}, {Liske}, {Slepian}, {Mueller}, {Conklin}, {Fulda}, {Mendes de
  Oliveira}, {Bentz}, {Jeram}, {Dong}, {Townsend}, {Izuti Nakazono}, {Quimby},
  and {Welsh}}]{eikenberry2019b}
{Eikenberry} S., {Gonzalez} A., {Darling} J., {Liske} J., {Slepian} Z.,
  {Mueller} G., {Conklin} J., {Fulda} P., et~al. (2019{\natexlab{a}}) {The
  Cosmic Accelerometer}. In: BAAS, vol~51, p 137, \eprint{1907.08271}

\bibitem[{{Eikenberry} et~al.(2019{\natexlab{b}}){Eikenberry}, {Gonzalez},
  {Darling}, {Slepian}, {Mueller}, {Conklin}, {Fulda}, {Jeram}, {Dong}, and
  {Townsend}}]{eikenberry2019a}
{Eikenberry} S., {Gonzalez} A., {Darling} J., {Slepian} Z., {Mueller} G.,
  {Conklin} J., {Fulda} P., {Jeram} S., et~al. (2019{\natexlab{b}}) {A Direct
  Measure of Cosmic Acceleration}. \emph{BAAS} 51(3):283

\bibitem[{{Eisenstein} et~al.(2005){Eisenstein}, {Zehavi}, {Hogg},
  {Scoccimarro}, {Blanton}, {Nichol}, {Scranton}, {Seo}, {Tegmark}, {Zheng},
  {Anderson}, {Annis}, {Bahcall}, {Brinkmann}, and et~al.}]{eisenstein2005}
{Eisenstein} D.~J., {Zehavi} I., {Hogg} D.~W., {Scoccimarro} R., {Blanton}
  M.~R., {Nichol} R.~C., {Scranton} R., {Seo} H.-J., et~al. (2005) {Detection
  of the Baryon Acoustic Peak in the Large-Scale Correlation Function of SDSS
  Luminous Red Galaxies}. \emph{\apj} 633(2):560--574, \doi{10.1086/466512},
  \eprint{astro-ph/0501171}

\bibitem[{{Endo} et~al.(2020){Endo}, {Tashiro}, and {Nishizawa}}]{Endo2020}
{Endo} T., {Tashiro} H., {Nishizawa} A.~J. (2020) {The Alcock Paczynski test
  with voids in 21 cm intensity field}. \emph{\mnras} 499(1):587--596,
  \doi{10.1093/mnras/staa2822}, \eprint{2002.00348}

\bibitem[{{Esteves} et~al.(2021){Esteves}, {Martins}, {Pereira}, and
  {Alves}}]{esteves2021}
{Esteves} J., {Martins} C.~J.~A.~P., {Pereira} B.~G., {Alves} C.~S. (2021)
  {Cosmological impact of redshift drift measurements}. \emph{\mnras}
  508(1):L53--L57, \doi{10.1093/mnrasl/slab102}, \eprint{2108.10739}

\bibitem[{{Estrada-Carpenter} et~al.(2019){Estrada-Carpenter}, {Papovich},
  {Momcheva}, {Brammer}, {Long}, {Quadri}, {Bridge}, {Dickinson}, {Ferguson},
  {Finkelstein}, {Giavalisco}, {Gosmeyer}, {Lotz}, {Salmon}, {Skelton},
  {Trump}, and {Weiner}}]{estrada2019}
{Estrada-Carpenter} V., {Papovich} C., {Momcheva} I., {Brammer} G., {Long} J.,
  {Quadri} R.~F., {Bridge} J., {Dickinson} M., et~al. (2019) {CLEAR. I. Ages
  and Metallicities of Quiescent Galaxies at 1.0 \&lt; z \&lt; 1.8 Derived from
  Deep Hubble Space Telescope Grism Data}. \emph{\apj} 870(2):133,
  \doi{10.3847/1538-4357/aaf22e}, \eprint{1810.02824}

\bibitem[{{Eulaers} et~al.(2013){Eulaers}, {Tewes}, {Magain}, {Courbin},
  {Asfandiyarov}, {Ehgamberdiev}, {Rathna Kumar}, {Stalin}, {Prabhu}, {Meylan},
  and {Van Winckel}}]{Eulaers2013}
{Eulaers} E., {Tewes} M., {Magain} P., {Courbin} F., {Asfandiyarov} I.,
  {Ehgamberdiev} S., {Rathna Kumar} S., {Stalin} C.~S., et~al. (2013)
  {COSMOGRAIL: the COSmological MOnitoring of GRAvItational Lenses. XII. Time
  delays of the doubly lensed quasars SDSS J1206+4332 and HS 2209+1914}.
  \emph{\aap} 553:A121, \doi{10.1051/0004-6361/201321140}, \eprint{1304.4474}

\bibitem[{{Event Horizon Telescope Collaboration} et~al.(2019){Event Horizon
  Telescope Collaboration}, {Akiyama}, {Alberdi}, {Alef}, {Asada}, {Azulay},
  {Baczko}, {Ball}, {Balokovi{\'c}}, {Barrett}, {Bintley}, {Blackburn},
  {Boland}, {Bouman}, {Bower}, and et~al.}]{ehtVI}
{Event Horizon Telescope Collaboration}, {Akiyama} K., {Alberdi} A., {Alef} W.,
  {Asada} K., {Azulay} R., {Baczko} A.-K., {Ball} D., et~al. (2019) {First M87
  Event Horizon Telescope Results. VI. The Shadow and Mass of the Central Black
  Hole}. \emph{\apjl} 875(1):L6, \doi{10.3847/2041-8213/ab1141},
  \eprint{1906.11243}

\bibitem[{{Ezquiaga} and {Holz}(2021)}]{2021ApJ...909L..23E}
{Ezquiaga} J.~M., {Holz} D.~E. (2021) {Jumping the Gap: Searching for LIGO's
  Biggest Black Holes}. \emph{\apjl} 909(2):L23, \doi{10.3847/2041-8213/abe638}

\bibitem[{{Ezquiaga} and {Holz}(2022)}]{2022arXiv220208240M}
{Ezquiaga} J.~M., {Holz} D.~E. (2022) {Spectral sirens: cosmology from the full
  mass distribution of compact binaries}. \emph{arXiv e-prints}
  arXiv:2202.08240, \eprint{2202.08240}

\bibitem[{Ezquiaga and Zumalac{\'{a}}rregui(2017)}]{Ezquiaga2017}
Ezquiaga J.~M., Zumalac{\'{a}}rregui M. (2017) {Dark Energy After GW170817:
  Dead Ends and the Road Ahead}. \emph{Physical Review Letters} 119(25):251304,
  \doi{10.1103/PhysRevLett.119.251304}

\bibitem[{{Falck} et~al.(2018){Falck}, {Koyama}, {Zhao}, and
  {Cautun}}]{Falck2018}
{Falck} B., {Koyama} K., {Zhao} G.-B., {Cautun} M. (2018) {Using voids to
  unscreen modified gravity}. \emph{\mnras} 475(3):3262--3272,
  \doi{10.1093/mnras/stx3288}, \eprint{1704.08942}

\bibitem[{{Falco} et~al.(1985){Falco}, {Gorenstein}, and
  {Shapiro}}]{Falco:1985}
{Falco} E.~E., {Gorenstein} M.~V., {Shapiro} I.~I. (1985) {On model-dependent
  bounds on H 0 from gravitational images : application to Q 0957+561 A, B.}
  \emph{\apj} 289:L1--L4, \doi{10.1086/184422}

\bibitem[{{Fana Dirirsa} et~al.(2019){Fana Dirirsa}, {Razzaque}, {Piron},
  {Arimoto}, {Axelsson}, {Kocevski}, {Longo}, {Ohno}, and {Zhu}}]{Fana19}
{Fana Dirirsa} F., {Razzaque} S., {Piron} F., {Arimoto} M., {Axelsson} M.,
  {Kocevski} D., {Longo} F., {Ohno} M., et~al. (2019) {Spectral Analysis of
  Fermi-LAT Gamma-Ray Bursts with Known Redshift and their Potential Use as
  Cosmological Standard Candles}. \emph{\apj} 887(1):13,
  \doi{10.3847/1538-4357/ab4e11}, \eprint{1910.07009}

\bibitem[{{Fang} et~al.(2018){Fang}, {Faber}, {Koo}, {Rodr{\'\i}guez-Puebla},
  {Guo}, {Barro}, {Behroozi}, {Brammer}, {Chen}, {Dekel}, and
  et~al.}]{fang2018}
{Fang} J.~J., {Faber} S.~M., {Koo} D.~C., {Rodr{\'\i}guez-Puebla} A., {Guo} Y.,
  {Barro} G., {Behroozi} P., {Brammer} G., et~al. (2018) {Demographics of
  Star-forming Galaxies since z {\ensuremath{\sim}} 2.5. I. The UVJ Diagram in
  CANDELS}. \emph{\apj} 858(2):100, \doi{10.3847/1538-4357/aabcba},
  \eprint{1710.05489}

\bibitem[{{Fang} et~al.(2019){Fang}, {Hamaus}, {Jain}, {Pandey}, {Pollina},
  {S{\'a}nchez}, {Kov{\'a}cs}, {Chang}, {Carretero}, {Castander}, {Choi},
  {Crocce}, {DeRose}, {Fosalba}, {Gatti}, {Gazta{\~n}aga}, et~al., and {DES
  Collaboration}}]{Fang2019}
{Fang} Y., {Hamaus} N., {Jain} B., {Pandey} S., {Pollina} G., {S{\'a}nchez} C.,
  {Kov{\'a}cs} A., {Chang} C., et~al. (2019) {Dark Energy Survey year 1
  results: the relationship between mass and light around cosmic voids}.
  \emph{\mnras} 490(3):3573--3587, \doi{10.1093/mnras/stz2805},
  \eprint{1909.01386}

\bibitem[{{Farmer} et~al.(2019){Farmer}, {Renzo}, {de Mink}, {Marchant}, and
  {Justham}}]{2019ApJ...887...53F}
{Farmer} R., {Renzo} M., {de Mink} S.~E., {Marchant} P., {Justham} S. (2019)
  {Mind the Gap: The Location of the Lower Edge of the Pair-instability
  Supernova Black Hole Mass Gap}. \emph{\apj} 887(1):53,
  \doi{10.3847/1538-4357/ab518b}, \eprint{1910.12874}

\bibitem[{{Farr} et~al.(2019){Farr}, {Fishbach}, {Ye}, and
  {Holz}}]{2019ApJ...883L..42F}
{Farr} W.~M., {Fishbach} M., {Ye} J., {Holz} D.~E. (2019) {A Future
  Percent-level Measurement of the Hubble Expansion at Redshift 0.8 with
  Advanced LIGO}. \emph{\apjl} 883(2):L42, \doi{10.3847/2041-8213/ab4284},
  \eprint{1908.09084}

\bibitem[{{Fassnacht} and {Lubin}(2002)}]{Fassnacht2002}
{Fassnacht} C.~D., {Lubin} L.~M. (2002) {The Gravitational Lens-Galaxy Group
  Connection. I. Discovery of a Group Coincident with CLASS B0712+472}.
  \emph{\aj} 123(2):627--636, \doi{10.1086/338648}, \eprint{astro-ph/0111205}

\bibitem[{{Fassnacht} et~al.(2002){Fassnacht}, {Xanthopoulos}, {Koopmans}, and
  {Rusin}}]{Fassnacht:2002}
{Fassnacht} C.~D., {Xanthopoulos} E., {Koopmans} L.~V.~E., {Rusin} D. (2002) {A
  Determination of H$_{0}$ with the CLASS Gravitational Lens B1608+656. III. A
  Significant Improvement in the Precision of the Time Delay Measurements}.
  \emph{\apj} 581(2):823--835, \doi{10.1086/344368}, \eprint{astro-ph/0208420}

\bibitem[{Feldman et~al.(1994)Feldman, Kaiser, and Peacock}]{Feldman:1993ky}
Feldman H.~A., Kaiser N., Peacock J.~A. (1994) {Power spectrum analysis of
  three-dimensional redshift surveys}. \emph{Astrophys J} 426:23--37,
  \doi{10.1086/174036}, \eprint{astro-ph/9304022}

\bibitem[{{Fenimore} and {Ramirez-Ruiz}(2000)}]{Fenimore2000}
{Fenimore} E.~E., {Ramirez-Ruiz} E. (2000) {Redshifts For 220 BATSE Gamma-Ray
  Bursts Determined by Variability and the Cosmological Consequences}.
  \emph{arXiv e-prints} astro-ph/0004176, \eprint{astro-ph/0004176}

\bibitem[{{Ferreras} et~al.(1999){Ferreras}, {Charlot}, and
  {Silk}}]{ferreras1999}
{Ferreras} I., {Charlot} S., {Silk} J. (1999) {The Age and Metallicity Range of
  Early-Type Galaxies in Clusters}. \emph{\apj} 521(1):81--89,
  \doi{10.1086/307513}, \eprint{astro-ph/9803235}

\bibitem[{{Finke} et~al.(2021){Finke}, {Foffa}, {Iacovelli}, {Maggiore}, and
  {Mancarella}}]{2021arXiv210112660F}
{Finke} A., {Foffa} S., {Iacovelli} F., {Maggiore} M., {Mancarella} M. (2021)
  {Cosmology with LIGO/Virgo dark sirens: Hubble parameter and modified
  gravitational wave propagation}. \emph{arXiv e-prints} arXiv:2101.12660,
  \eprint{2101.12660}

\bibitem[{{Firmani} et~al.(2006){Firmani}, {Ghisellini}, {Avila-Reese}, and
  {Ghirlanda}}]{Firmani06}
{Firmani} C., {Ghisellini} G., {Avila-Reese} V., {Ghirlanda} G. (2006)
  {Discovery of a tight correlation among the prompt emission properties of
  long gamma-ray bursts}. \emph{\mnras} 370(1):185--197,
  \doi{10.1111/j.1365-2966.2006.10445.x}, \eprint{astro-ph/0605073}

\bibitem[{{Fishbach} and {Holz}(2017)}]{2017ApJ...851L..25F}
{Fishbach} M., {Holz} D.~E. (2017) {Where Are LIGO{\textquoteright}s Big Black
  Holes?} \emph{\apjl} 851(2):L25, \doi{10.3847/2041-8213/aa9bf6},
  \eprint{1709.08584}

\bibitem[{{Fishbach} et~al.(2019){Fishbach}, {Gray}, {Maga{\~n}a Hernandez},
  {Qi}, {Sur}, {Acernese}, {Aiello}, {Allocca}, {Aloy}, {Amato}, and
  et~al.}]{2019ApJ...871L..13F}
{Fishbach} M., {Gray} R., {Maga{\~n}a Hernandez} I., {Qi} H., {Sur} A.,
  {Acernese} F., {Aiello} L., {Allocca} A., et~al. (2019) {A Standard Siren
  Measurement of the Hubble Constant from GW170817 without the Electromagnetic
  Counterpart}. \emph{\apjl} 871(1):L13, \doi{10.3847/2041-8213/aaf96e},
  \eprint{1807.05667}

\bibitem[{{Fishbach} et~al.(2021){Fishbach}, {Doctor}, {Callister}, {Edelman},
  {Ye}, {Essick}, {Farr}, {Farr}, and {Holz}}]{2021ApJ...912...98F}
{Fishbach} M., {Doctor} Z., {Callister} T., {Edelman} B., {Ye} J., {Essick} R.,
  {Farr} W.~M., {Farr} B., et~al. (2021) {When Are LIGO/Virgo's Big Black Hole
  Mergers?} \emph{\apj} 912(2):98, \doi{10.3847/1538-4357/abee11},
  \eprint{2101.07699}

\bibitem[{{Fitzpatrick}(1999)}]{F99}
{Fitzpatrick} E.~L. (1999) {Correcting for the Effects of Interstellar
  Extinction}. \emph{The Publications of the Astronomical Society of the
  Pacific} 111:63--75, \doi{10.1086/316293}, \eprint{astro-ph/9809387}

\bibitem[{{Fixsen}(2009)}]{fixsen2009}
{Fixsen} D.~J. (2009) {The Temperature of the Cosmic Microwave Background}.
  \emph{\apj} 707(2):916--920, \doi{10.1088/0004-637X/707/2/916},
  \eprint{0911.1955}

\bibitem[{{Fleury} et~al.(2021{\natexlab{a}}){Fleury}, {Larena}, and
  {Uzan}}]{Fleury:2021metric}
{Fleury} P., {Larena} J., {Uzan} J.-P. (2021{\natexlab{a}}) {Gravitational
  lenses in arbitrary space-times}. \emph{Classical and Quantum Gravity}
  38(8):085002, \doi{10.1088/1361-6382/abea2d}, \eprint{2011.04440}

\bibitem[{{Fleury} et~al.(2021{\natexlab{b}}){Fleury}, {Larena}, and
  {Uzan}}]{Fleury:2021los}
{Fleury} P., {Larena} J., {Uzan} J.-P. (2021{\natexlab{b}}) {Line-of-sight
  effects in strong gravitational lensing}. \emph{arXiv e-prints}
  arXiv:2104.08883, \eprint{2104.08883}

\bibitem[{{Fohlmeister} et~al.(2013){Fohlmeister}, {Kochanek}, {Falco},
  {Wambsganss}, {Oguri}, and {Dai}}]{Fohlmeister2013}
{Fohlmeister} J., {Kochanek} C.~S., {Falco} E.~E., {Wambsganss} J., {Oguri} M.,
  {Dai} X. (2013) {A Two-year Time Delay for the Lensed Quasar SDSS
  J1029+2623}. \emph{\apj} 764(2):186, \doi{10.1088/0004-637X/764/2/186},
  \eprint{1207.5776}

\bibitem[{Fonseca et~al.(2015)Fonseca, Camera, Santos, and
  Maartens}]{Fonseca:2015laa}
Fonseca J., Camera S., Santos M., Maartens R. (2015) {Hunting down
  horizon-scale effects with multi-wavelength surveys}. \emph{Astrophys J}
  812(2):L22, \doi{10.1088/2041-8205/812/2/L22}, \eprint{1507.04605}

\bibitem[{Fonseca et~al.(2017)Fonseca, Maartens, and Santos}]{Fonseca:2016xvi}
Fonseca J., Maartens R., Santos M.~G. (2017) {Probing the primordial Universe
  with MeerKAT and DES}. \emph{Mon Not Roy Astron Soc} 466(3):2780--2786,
  \doi{10.1093/mnras/stw3248}, \eprint{1611.01322}

\bibitem[{{Font-Ribera} et~al.(2014){Font-Ribera}, {Kirkby}, {Busca},
  {Miralda-Escud{\'e}}, {Ross}, {Slosar}, {Rich}, {Aubourg}, {Bailey},
  {Bhardwaj}, {Bautista}, and et~al.}]{font-ribera2014}
{Font-Ribera} A., {Kirkby} D., {Busca} N., {Miralda-Escud{\'e}} J., {Ross}
  N.~P., {Slosar} A., {Rich} J., {Aubourg} {\'E}., et~al. (2014) {Quasar-Lyman
  {\ensuremath{\alpha}} forest cross-correlation from BOSS DR11: Baryon
  Acoustic Oscillations}. \emph{\jcap} 2014(5):027,
  \doi{10.1088/1475-7516/2014/05/027}, \eprint{1311.1767}

\bibitem[{{Fontana} et~al.(2006){Fontana}, {Salimbeni}, {Grazian}, {Giallongo},
  {Pentericci}, {Nonino}, {Fontanot}, {Menci}, {Monaco}, {Cristiani},
  {Vanzella}, {de Santis}, and {Gallozzi}}]{fontana2006}
{Fontana} A., {Salimbeni} S., {Grazian} A., {Giallongo} E., {Pentericci} L.,
  {Nonino} M., {Fontanot} F., {Menci} N., et~al. (2006) {The Galaxy mass
  function up to z =4 in the GOODS-MUSIC sample: into the epoch of formation of
  massive galaxies}. \emph{\aap} 459(3):745--757,
  \doi{10.1051/0004-6361:20065475}, \eprint{astro-ph/0609068}

\bibitem[{{Foreman-Mackey} et~al.(2013){Foreman-Mackey}, {Hogg}, {Lang}, and
  {Goodman}}]{foreman2013}
{Foreman-Mackey} D., {Hogg} D.~W., {Lang} D., {Goodman} J. (2013) {emcee: The
  MCMC Hammer}. \emph{\pasp} 125(925):306, \doi{10.1086/670067},
  \eprint{1202.3665}

\bibitem[{{Fowler} and {Hoyle}(1964)}]{1964ApJS....9..201F}
{Fowler} W.~A., {Hoyle} F. (1964) {Neutrino Processes and Pair Formation in
  Massive Stars and Supernovae.} \emph{\apjs} 9:201, \doi{10.1086/190103}

\bibitem[{{Foxley-Marrable} et~al.(2018){Foxley-Marrable}, {Collett},
  {Vernardos}, {Goldstein}, and {Bacon}}]{Foxley-Marrable:2018}
{Foxley-Marrable} M., {Collett} T.~E., {Vernardos} G., {Goldstein} D.~A.,
  {Bacon} D. (2018) {The impact of microlensing on the standardization of
  strongly lensed Type Ia supernovae}. \emph{\mnras} 478(4):5081--5090,
  \doi{10.1093/mnras/sty1346}, \eprint{1802.07738}

\bibitem[{{Franx} and {van Dokkum}(1996)}]{franx1996}
{Franx} M., {van Dokkum} P.~G. (1996) {Measuring the Evolution of the M\%TL
  ratio from the Fundamental Plane in CL 0024+16 at Z =0.39}. In: {Bender} R.,
  {Davies} R.~L. (eds) New Light on Galaxy Evolution, vol 171, p 233,
  \eprint{astro-ph/9603029}

\bibitem[{{Franx} et~al.(2003){Franx}, {Labb{\'e}}, {Rudnick}, {van Dokkum},
  {Daddi}, {F{\"o}rster Schreiber}, {Moorwood}, {Rix}, {R{\"o}ttgering}, {van
  der Wel}, {van der Werf}, and {van Starkenburg}}]{franx2003}
{Franx} M., {Labb{\'e}} I., {Rudnick} G., {van Dokkum} P.~G., {Daddi} E.,
  {F{\"o}rster Schreiber} N.~M., {Moorwood} A., {Rix} H.-W., et~al. (2003) {A
  Significant Population of Red, Near-Infrared-selected High-Redshift
  Galaxies}. \emph{\apjl} 587(2):L79--L82, \doi{10.1086/375155},
  \eprint{astro-ph/0303163}

\bibitem[{{Franzetti} et~al.(2007){Franzetti}, {Scodeggio}, {Garilli},
  {Vergani}, {Maccagni}, {Guzzo}, {Tresse}, {Ilbert}, {Lamareille}, {Contini},
  {Le F{\`e}vre}, {Zamorani}, {Brinchmann}, {Charlot}, {Bottini}, {Le Brun},
  and et~al.}]{franzetti2007}
{Franzetti} P., {Scodeggio} M., {Garilli} B., {Vergani} D., {Maccagni} D.,
  {Guzzo} L., {Tresse} L., {Ilbert} O., et~al. (2007) {The VIMOS-VLT deep
  survey. Color bimodality and the mix of galaxy populations up to z$\sim$2}.
  \emph{\aap} 465(3):711--723, \doi{10.1051/0004-6361:20065942},
  \eprint{astro-ph/0607075}

\bibitem[{{Freedman} et~al.(2001){Freedman}, {Madore}, {Gibson}, {Ferrarese},
  {Kelson}, {Sakai}, {Mould}, {Kennicutt}, {Ford}, {Graham}, {Huchra},
  {Hughes}, {Illingworth}, {Macri}, and {Stetson}}]{freedman01}
{Freedman} W.~L., {Madore} B.~F., {Gibson} B.~K., {Ferrarese} L., {Kelson}
  D.~D., {Sakai} S., {Mould} J.~R., {Kennicutt} J. Robert~C., et~al. (2001)
  {Final Results from the Hubble Space Telescope Key Project to Measure the
  Hubble Constant}. \emph{\apj} 553(1):47--72, \doi{10.1086/320638},
  \eprint{astro-ph/0012376}

\bibitem[{{Friedrich} et~al.(2021){Friedrich}, {Halder}, {Boyle}, {Uhlemann},
  {Britt}, {Codis}, {Gruen}, and {Hahn}}]{Friedrich2021}
{Friedrich} O., {Halder} A., {Boyle} A., {Uhlemann} C., {Britt} D., {Codis} S.,
  {Gruen} D., {Hahn} C. (2021) {The PDF perspective on the tracer-matter
  connection: Lagrangian bias and non-Poissonian shot noise}. \emph{arXiv
  e-prints} arXiv:2107.02300, \eprint{2107.02300}

\bibitem[{{Frontera} et~al.(2012){Frontera}, {Amati}, {Guidorzi}, {Landi}, and
  {in't Zand}}]{Frontera12}
{Frontera} F., {Amati} L., {Guidorzi} C., {Landi} R., {in't Zand} J. (2012)
  {Broadband Time-resolved E $_{ p, i }$-L $_{iso}$ Correlation in Gamma-Ray
  Bursts}. \emph{\apj} 754(2):138, \doi{10.1088/0004-637X/754/2/138},
  \eprint{1206.5626}

\bibitem[{Fryer et~al.(2001)Fryer, Woosley, and Heger}]{Fryer:2000my}
Fryer C.~L., Woosley S.~E., Heger A. (2001) {Pair instability supernovae,
  gravity waves, and gamma-ray transients}. \emph{Astrophys J} 550:372--382,
  \doi{10.1086/319719}, \eprint{astro-ph/0007176}

\bibitem[{Furlanetto et~al.(2006)Furlanetto, Oh, and
  Briggs}]{Furlanetto:2006jb}
Furlanetto S., Oh S.~P., Briggs F. (2006) {Cosmology at Low Frequencies: The 21
  cm Transition and the High-Redshift Universe}. \emph{Phys Rept} 433:181--301,
  \doi{10.1016/j.physrep.2006.08.002}, \eprint{astro-ph/0608032}

\bibitem[{{Gaia Collaboration} et~al.(2016){Gaia Collaboration}, {Prusti}, {de
  Bruijne}, {Brown}, {Vallenari}, {Babusiaux}, {Bailer-Jones}, {Bastian},
  {Biermann}, {Evans}, {Eyer}, {Jansen}, {Jordi}, {Klioner}, {Lammers},
  {Lindegren}, {Luri}, and et~al.}]{Gaia:2016}
{Gaia Collaboration}, {Prusti} T., {de Bruijne} J.~H.~J., {Brown} A.~G.~A.,
  {Vallenari} A., {Babusiaux} C., {Bailer-Jones} C.~A.~L., {Bastian} U., et~al.
  (2016) {The Gaia mission}. \emph{\aap} 595:A1,
  \doi{10.1051/0004-6361/201629272}, \eprint{1609.04153}

\bibitem[{{Gaia Collaboration} et~al.(2021){Gaia Collaboration}, {Klioner},
  {Mignard}, {Lindegren}, {Bastian}, {McMillan}, {Hern{\'a}ndez}, {Hobbs},
  {Ramos-Lerate}, {Biermann}, {Bombrun}, {de Torres}, {Gerlach}, {Geyer},
  {Hilger}, {Lammers}, {Steidelm{\"u}ller}, and et~al.}]{klioner2021}
{Gaia Collaboration}, {Klioner} S.~A., {Mignard} F., {Lindegren} L., {Bastian}
  U., {McMillan} P.~J., {Hern{\'a}ndez} J., {Hobbs} D., et~al. (2021) {Gaia
  Early Data Release 3. Acceleration of the Solar System from Gaia astrometry}.
  \emph{\aap} 649:A9, \doi{10.1051/0004-6361/202039734}, \eprint{2012.02036}

\bibitem[{{Gallazzi} et~al.(2005){Gallazzi}, {Charlot}, {Brinchmann}, {White},
  and {Tremonti}}]{gallazzi2005}
{Gallazzi} A., {Charlot} S., {Brinchmann} J., {White} S.~D.~M., {Tremonti}
  C.~A. (2005) {The ages and metallicities of galaxies in the local universe}.
  \emph{\mnras} 362:41--58, \doi{10.1111/j.1365-2966.2005.09321.x},
  \eprint{astro-ph/0506539}

\bibitem[{{Gallazzi} et~al.(2014){Gallazzi}, {Bell}, {Zibetti}, {Brinchmann},
  and {Kelson}}]{gallazzi2014}
{Gallazzi} A., {Bell} E.~F., {Zibetti} S., {Brinchmann} J., {Kelson} D.~D.
  (2014) {Charting the Evolution of the Ages and Metallicities of Massive
  Galaxies since z = 0.7}. \emph{\apj} 788(1):72,
  \doi{10.1088/0004-637X/788/1/72}, \eprint{1404.5624}

\bibitem[{Garcia et~al.(2020)Garcia, Quartin, and Siffert}]{Garcia:2019ita}
Garcia K., Quartin M., Siffert B.~B. (2020) {On the amount of peculiar velocity
  field information in supernovae from LSST and beyond}. \emph{Phys Dark Univ}
  29:100519, \doi{10.1016/j.dark.2020.100519}, \eprint{1905.00746}

\bibitem[{{Gardner} et~al.(2006){Gardner}, {Mather}, {Clampin}, {Doyon},
  {Greenhouse}, {Hammel}, {Hutchings}, {Jakobsen}, {Lilly}, {Long}, {Lunine},
  {McCaughrean}, {Mountain}, {Nella}, {Rieke}, {Rieke}, {Rix}, {Smith},
  {Sonneborn}, {Stiavelli}, {Stockman}, {Windhorst}, and
  {Wright}}]{Gardner:2006}
{Gardner} J.~P., {Mather} J.~C., {Clampin} M., {Doyon} R., {Greenhouse} M.~A.,
  {Hammel} H.~B., {Hutchings} J.~B., {Jakobsen} P., et~al. (2006) {The James
  Webb Space Telescope}. \emph{\ssr} 123(4):485--606,
  \doi{10.1007/s11214-006-8315-7}, \eprint{astro-ph/0606175}

\bibitem[{{Gavazzi} et~al.(2002){Gavazzi}, {Bonfanti}, {Sanvito}, {Boselli},
  and {Scodeggio}}]{gavazzi2002}
{Gavazzi} G., {Bonfanti} C., {Sanvito} G., {Boselli} A., {Scodeggio} M. (2002)
  {Spectrophotometry of Galaxies in the Virgo Cluster. I. The Star Formation
  History}. \emph{\apj} 576(1):135--151, \doi{10.1086/341730}

\bibitem[{{Gehrels} et~al.(2009){Gehrels}, {Ramirez-Ruiz}, and
  {Fox}}]{Gehrels09}
{Gehrels} N., {Ramirez-Ruiz} E., {Fox} D.~B. (2009) {Gamma-Ray Bursts in the
  Swift Era}. \emph{\araa} 47(1):567--617,
  \doi{10.1146/annurev.astro.46.060407.145147}, \eprint{0909.1531}

\bibitem[{{Ghirlanda} et~al.(2004){Ghirlanda}, {Ghisellini}, and
  {Lazzati}}]{Ghirlanda04}
{Ghirlanda} G., {Ghisellini} G., {Lazzati} D. (2004) {The Collimation-corrected
  Gamma-Ray Burst Energies Correlate with the Peak Energy of Their
  {\ensuremath{\nu}}F$_{{\ensuremath{\nu}}}$ Spectrum}. \emph{\apj}
  616(1):331--338, \doi{10.1086/424913}, \eprint{astro-ph/0405602}

\bibitem[{{Ghirlanda} et~al.(2005){Ghirlanda}, {Ghisellini}, and
  {Firmani}}]{Ghirlanda05}
{Ghirlanda} G., {Ghisellini} G., {Firmani} C. (2005) {Probing the existence of
  the E$_{peak}$-E$_{iso}$ correlation in long gamma ray bursts}. \emph{\mnras}
  361(1):L10--L14, \doi{10.1111/j.1745-3933.2005.00053.x},
  \eprint{astro-ph/0502186}

\bibitem[{{Ghirlanda} et~al.(2008){Ghirlanda}, {Nava}, {Ghisellini}, {Firmani},
  and {Cabrera}}]{Ghirlanda08}
{Ghirlanda} G., {Nava} L., {Ghisellini} G., {Firmani} C., {Cabrera} J.~I.
  (2008) {The E$_{peak}$-E$_{iso}$ plane of long gamma-ray bursts and selection
  effects}. \emph{\mnras} 387(1):319--330,
  \doi{10.1111/j.1365-2966.2008.13232.x}, \eprint{0804.1675}

\bibitem[{{Ghirlanda} et~al.(2010){Ghirlanda}, {Nava}, and
  {Ghisellini}}]{Ghirlanda10}
{Ghirlanda} G., {Nava} L., {Ghisellini} G. (2010) {Spectral-luminosity relation
  within individual Fermi gamma rays bursts}. \emph{\aap} 511:A43,
  \doi{10.1051/0004-6361/200913134}, \eprint{0908.2807}

\bibitem[{{Gil-Mar{\'\i}n} et~al.(2015){Gil-Mar{\'\i}n}, {Nore{\~n}a}, {Verde},
  {Percival}, {Wagner}, {Manera}, and {Schneider}}]{GilMarin2015}
{Gil-Mar{\'\i}n} H., {Nore{\~n}a} J., {Verde} L., {Percival} W.~J., {Wagner}
  C., {Manera} M., {Schneider} D.~P. (2015) {The power spectrum and bispectrum
  of SDSS DR11 BOSS galaxies - I. Bias and gravity}. \emph{\mnras}
  451(1):539--580, \doi{10.1093/mnras/stv961}, \eprint{1407.5668}

\bibitem[{{Gilmore} and {Natarajan}(2009)}]{2009MNRAS.396..354G}
{Gilmore} J., {Natarajan} P. (2009) {Cosmography with cluster strong lensing}.
  \emph{\mnras} 396(1):354--364, \doi{10.1111/j.1365-2966.2009.14612.x},
  \eprint{astro-ph/0605245}

\bibitem[{{Ginzburg} et~al.(2017){Ginzburg}, {Desjacques}, and
  {Chan}}]{Ginzburg2017}
{Ginzburg} D., {Desjacques} V., {Chan} K.~C. (2017) {Shot noise and biased
  tracers: A new look at the halo model}. \emph{\prd} 96(8):083528,
  \doi{10.1103/PhysRevD.96.083528}, \eprint{1706.08738}

\bibitem[{{Girelli} et~al.(2019){Girelli}, {Bolzonella}, and
  {Cimatti}}]{girelli19}
{Girelli} G., {Bolzonella} M., {Cimatti} A. (2019) {Massive and old quiescent
  galaxies at high redshift}. \emph{\aap} 632:A80,
  \doi{10.1051/0004-6361/201834547}, \eprint{1910.07544}

\bibitem[{{Goldstein} et~al.(2018){Goldstein}, {Nugent}, {Kasen}, and
  {Collett}}]{Goldstein:2018}
{Goldstein} D.~A., {Nugent} P.~E., {Kasen} D.~N., {Collett} T.~E. (2018)
  {Precise Time Delays from Strongly Gravitationally Lensed Type Ia Supernovae
  with Chromatically Microlensed Images}. \emph{\apj} 855(1):22,
  \doi{10.3847/1538-4357/aaa975}, \eprint{1708.00003}

\bibitem[{{Goldstein} et~al.(2019){Goldstein}, {Nugent}, and
  {Goobar}}]{Goldstein:2019}
{Goldstein} D.~A., {Nugent} P.~E., {Goobar} A. (2019) {Rates and Properties of
  Supernovae Strongly Gravitationally Lensed by Elliptical Galaxies in
  Time-domain Imaging Surveys}. \emph{\apjs} 243(1):6,
  \doi{10.3847/1538-4365/ab1fe0}, \eprint{1809.10147}

\bibitem[{{G{\'o}mez-Valent} and {Amendola}(2018)}]{gomezvalent2018}
{G{\'o}mez-Valent} A., {Amendola} L. (2018) {H$_{0}$ from cosmic chronometers
  and Type Ia supernovae, with Gaussian Processes and the novel Weighted
  Polynomial Regression method}. \emph{\jcap} 2018(4):051,
  \doi{10.1088/1475-7516/2018/04/051}, \eprint{1802.01505}

\bibitem[{{Gonzalez} et~al.(2021){Gonzalez}, {Benetti}, {von Marttens}, and
  {Alcaniz}}]{gonzalez2021}
{Gonzalez} J.~E., {Benetti} M., {von Marttens} R., {Alcaniz} J. (2021) {Testing
  the consistency between cosmological data: the impact of spatial curvature
  and the dark energy EoS}. \emph{arXiv e-prints} arXiv:2104.13455,
  \eprint{2104.13455}

\bibitem[{{Goobar} et~al.(2017){Goobar}, {Amanullah}, {Kulkarni}, {Nugent},
  {Johansson}, {Steidel}, {Law}, {M{\"o}rtsell}, {Quimby}, {Blagorodnova}, and
  et~al.}]{Goobar:2017}
{Goobar} A., {Amanullah} R., {Kulkarni} S.~R., {Nugent} P.~E., {Johansson} J.,
  {Steidel} C., {Law} D., {M{\"o}rtsell} E., et~al. (2017) {iPTF16geu: A
  multiply imaged, gravitationally lensed type Ia supernova}. \emph{Science}
  356(6335):291--295, \doi{10.1126/science.aal2729}, \eprint{1611.00014}

\bibitem[{Gordon et~al.(2007)Gordon, Land, and Slosar}]{Gordon:2007zw}
Gordon C., Land K., Slosar A. (2007) {Cosmological Constraints from Type Ia
  Supernovae Peculiar Velocity Measurements}. \emph{Phys Rev Lett} 99:081301,
  \doi{10.1103/PhysRevLett.99.081301}, \eprint{0705.1718}

\bibitem[{{Gorenstein} et~al.(1988){Gorenstein}, {Falco}, and
  {Shapiro}}]{Gorenstein:1988}
{Gorenstein} M.~V., {Falco} E.~E., {Shapiro} I.~I. (1988) {Degeneracies in
  Parameter Estimates for Models of Gravitational Lens Systems}. \emph{\apj}
  327:693, \doi{10.1086/166226}

\bibitem[{{Gouliermis} et~al.(2005){Gouliermis}, {Brandner}, {Butler}, and
  {Hippler}}]{sbf4elt}
{Gouliermis} D., {Brandner} W., {Butler} D., {Hippler} S. (2005) {Surface
  Brightness Fluctuations: A Case for Extremely Large Telescopes}. In:
  {Brandner} W., {Kasper} M.~E. (eds) Science with Adaptive Optics, p 334,
  \doi{10.1007/10828557\_57}, \eprint{astro-ph/0311602}

\bibitem[{{Graham} et~al.(2020){Graham}, {Ford}, {McKernan}, {Ross}, {Stern},
  {Burdge}, {Coughlin}, {Djorgovski}, {Drake}, {Duev}, and
  et~al.}]{2020PhRvL.124y1102G}
{Graham} M.~J., {Ford} K.~E.~S., {McKernan} B., {Ross} N.~P., {Stern} D.,
  {Burdge} K., {Coughlin} M., {Djorgovski} S.~G., et~al. (2020) {Candidate
  Electromagnetic Counterpart to the Binary Black Hole Merger
  Gravitational-Wave Event S190521g$^{*}$}. \emph{\prl} 124(25):251102,
  \doi{10.1103/PhysRevLett.124.251102}, \eprint{2006.14122}

\bibitem[{{Granata} et~al.(2021){Granata}, {Mercurio}, {Grillo}, {Tortorelli},
  {Bergamini}, {Meneghetti}, {Rosati}, {Bartosch Caminha}, and
  {Nonino}}]{Granata:2021}
{Granata} G., {Mercurio} A., {Grillo} C., {Tortorelli} L., {Bergamini} P.,
  {Meneghetti} M., {Rosati} P., {Bartosch Caminha} G., et~al. (2021) {Improved
  strong lensing modelling of galaxy clusters using the Fundamental Plane: the
  case of Abell S1063}. \emph{arXiv e-prints} arXiv:2107.09079,
  \eprint{2107.09079}

\bibitem[{{Granett} et~al.(2008){Granett}, {Neyrinck}, and
  {Szapudi}}]{Granett2008}
{Granett} B.~R., {Neyrinck} M.~C., {Szapudi} I. (2008) {An Imprint of
  Superstructures on the Microwave Background due to the Integrated Sachs-Wolfe
  Effect}. \emph{\apjl} 683:L99, \doi{10.1086/591670}, \eprint{0805.3695}

\bibitem[{{Gray} et~al.(2020){Gray}, {Hernandez}, {Qi}, {Sur}, {Brady}, {Chen},
  {Farr}, {Fishbach}, {Gair}, {Ghosh}, and et~al.}]{2020PhRvD.101l2001G}
{Gray} R., {Hernandez} I.~M., {Qi} H., {Sur} A., {Brady} P.~R., {Chen} H.-Y.,
  {Farr} W.~M., {Fishbach} M., et~al. (2020) {Cosmological inference using
  gravitational wave standard sirens: A mock data analysis}. \emph{\prd}
  101(12):122001, \doi{10.1103/PhysRevD.101.122001}, \eprint{1908.06050}

\bibitem[{{Gray} et~al.(2021){Gray}, {Messenger}, and
  {Veitch}}]{2021arXiv211104629G}
{Gray} R., {Messenger} C., {Veitch} J. (2021) {A Pixelated Approach to Galaxy
  Catalogue Incompleteness: Improving the Dark Siren Measurement of the Hubble
  Constant}. \emph{arXiv e-prints} arXiv:2111.04629, \eprint{2111.04629}

\bibitem[{{Graziani} et~al.(2020){Graziani}, {Rigault}, {Regnault}, {Gris},
  {M{\"o}ller}, {Antilogus}, {Astier}, {Betoule}, {Bongard}, {Briday},
  {Cohen-Tanugi}, {Copin}, {Courtois}, {Fouchez}, {Gangler}, {Guinet},
  {Hawken}, {Kim}, {L{\'e}get}, {Neveu}, {Ntelis}, {Rosnet}, and
  {Nuss}}]{Graziani:2020kkr}
{Graziani} R., {Rigault} M., {Regnault} N., {Gris} P., {M{\"o}ller} A.,
  {Antilogus} P., {Astier} P., {Betoule} M., et~al. (2020) {Peculiar velocity
  cosmology with type Ia supernovae}. \emph{arXiv e-prints} arXiv:2001.09095,
  \eprint{2001.09095}

\bibitem[{{Greco} et~al.(2021){Greco}, {van Dokkum}, {Danieli}, {Carlsten}, and
  {Conroy}}]{greco21}
{Greco} J.~P., {van Dokkum} P., {Danieli} S., {Carlsten} S.~G., {Conroy} C.
  (2021) {Measuring Distances to Low-luminosity Galaxies Using Surface
  Brightness Fluctuations}. \emph{\apj} 908(1):24,
  \doi{10.3847/1538-4357/abd030}, \eprint{2004.07273}

\bibitem[{{Greene} et~al.(2013){Greene}, {Suyu}, {Treu}, {Hilbert}, {Auger},
  {Collett}, {Marshall}, {Fassnacht}, {Blandford}, {Brada{\v{c}}}, and
  {Koopmans}}]{Greene:2013}
{Greene} Z.~S., {Suyu} S.~H., {Treu} T., {Hilbert} S., {Auger} M.~W., {Collett}
  T.~E., {Marshall} P.~J., {Fassnacht} C.~D., et~al. (2013) {Improving the
  Precision of Time-delay Cosmography with Observations of Galaxies along the
  Line of Sight}. \emph{\apj} 768(1):39, \doi{10.1088/0004-637X/768/1/39},
  \eprint{1303.3588}

\bibitem[{{Gregory} and {Thompson}(1978)}]{Gregory1978}
{Gregory} S.~A., {Thompson} L.~A. (1978) {The Coma/A1367 supercluster and its
  environs.} \emph{\apj} 222:784--799, \doi{10.1086/156198}

\bibitem[{{Grillo} et~al.(2015){Grillo}, {Suyu}, {Rosati}, {Mercurio},
  {Balestra}, {Munari}, {Nonino}, {Caminha}, {Lombardi}, {De Lucia}, {Borgani},
  {Gobat}, {Biviano}, {Girardi}, {Umetsu}, {Coe}, {Koekemoer}, {Postman},
  {Zitrin}, {Halkola}, {Broadhurst}, {Sartoris}, {Presotto}, {Annunziatella},
  {Maier}, {Fritz}, {Vanzella}, and {Frye}}]{Grillo:2015}
{Grillo} C., {Suyu} S.~H., {Rosati} P., {Mercurio} A., {Balestra} I., {Munari}
  E., {Nonino} M., {Caminha} G.~B., et~al. (2015) {CLASH-VLT: Insights on the
  Mass Substructures in the Frontier Fields Cluster MACS J0416.1-2403 through
  Accurate Strong Lens Modeling}. \emph{\apj} 800(1):38,
  \doi{10.1088/0004-637X/800/1/38}, \eprint{1407.7866}

\bibitem[{{Grillo} et~al.(2016){Grillo}, {Karman}, {Suyu}, {Rosati},
  {Balestra}, {Mercurio}, {Lombardi}, {Treu}, {Caminha}, {Halkola}, {Rodney},
  {Gavazzi}, and {Caputi}}]{Grillo:2016}
{Grillo} C., {Karman} W., {Suyu} S.~H., {Rosati} P., {Balestra} I., {Mercurio}
  A., {Lombardi} M., {Treu} T., et~al. (2016) {The Story of Supernova
  {\textquotedblleft}Refsdal{\textquotedblright} Told by Muse}. \emph{\apj}
  822(2):78, \doi{10.3847/0004-637X/822/2/78}, \eprint{1511.04093}

\bibitem[{{Grillo} et~al.(2018){Grillo}, {Rosati}, {Suyu}, {Balestra},
  {Caminha}, {Halkola}, {Kelly}, {Lombardi}, {Mercurio}, {Rodney}, and
  {Treu}}]{Grillo:2018}
{Grillo} C., {Rosati} P., {Suyu} S.~H., {Balestra} I., {Caminha} G.~B.,
  {Halkola} A., {Kelly} P.~L., {Lombardi} M., et~al. (2018) {Measuring the
  Value of the Hubble Constant {\textquotedblleft}{\`a} la
  Refsdal{\textquotedblright}}. \emph{\apj} 860(2):94,
  \doi{10.3847/1538-4357/aac2c9}, \eprint{1802.01584}

\bibitem[{{Grillo} et~al.(2020){Grillo}, {Rosati}, {Suyu}, {Caminha},
  {Mercurio}, and {Halkola}}]{Grillo:2020}
{Grillo} C., {Rosati} P., {Suyu} S.~H., {Caminha} G.~B., {Mercurio} A.,
  {Halkola} A. (2020) {On the Accuracy of Time-delay Cosmography in the
  Frontier Fields Cluster MACS J1149.5+2223 with Supernova Refsdal}.
  \emph{\apj} 898(1):87, \doi{10.3847/1538-4357/ab9a4c}, \eprint{2001.02232}

\bibitem[{Grillo et~al.(2020)Grillo, Rosati, Suyu, Caminha, Mercurio, and
  Halkola}]{Grillo:2020yvj}
Grillo C., Rosati P., Suyu S.~H., Caminha G.~B., Mercurio A., Halkola A. (2020)
  {On the accuracy of time-delay cosmography in the Frontier Fields Cluster
  MACS J1149.5+2223 with supernova Refsdal}. \emph{Astrophys J} 898(1):87,
  \doi{10.3847/1538-4357/ab9a4c}, \eprint{2001.02232}

\bibitem[{Grillo et~al.(2018)}]{Grillo:2018ume}
Grillo C., et~al. (2018) {Measuring the Value of the Hubble Constant ''\`a la
  Refsdal''}. \emph{Astrophys J} 860(2):94, \doi{10.3847/1538-4357/aac2c9},
  \eprint{1802.01584}

\bibitem[{{Grogin} and {Narayan}(1996)}]{Grogin:1996}
{Grogin} N.~A., {Narayan} R. (1996) {A New Model of the Gravitational Lens
  0957+561 and a Limit on the Hubble Constant}. \emph{\apj} 464:92,
  \doi{10.1086/177302}

\bibitem[{{Gruen} et~al.(2016){Gruen}, {Friedrich}, {Amara}, {Bacon},
  {Bonnett}, {Hartley}, {Jain}, {Jarvis}, {Kacprzak}, {Krause}, {Mana}, {Rozo},
  {Rykoff}, and et~al.}]{Gruen2016}
{Gruen} D., {Friedrich} O., {Amara} A., {Bacon} D., {Bonnett} C., {Hartley} W.,
  {Jain} B., {Jarvis} M., et~al. (2016) {Weak lensing by galaxy troughs in DES
  Science Verification data}. \emph{\mnras} 455(3):3367--3380,
  \doi{10.1093/mnras/stv2506}, \eprint{1507.05090}

\bibitem[{{Grupe} et~al.(2010){Grupe}, {Komossa}, {Leighly}, and
  {Page}}]{grupe2010}
{Grupe} D., {Komossa} S., {Leighly} K.~M., {Page} K.~L. (2010) {The
  Simultaneous Optical-to-X-Ray Spectral Energy Distribution of Soft X-Ray
  Selected Active Galactic Nuclei Observed by Swift}. \emph{\apjs}
  187(1):64--106, \doi{10.1088/0067-0049/187/1/64}, \eprint{1001.3140}

\bibitem[{{Guy} et~al.(2010){Guy}, {Sullivan}, {Conley}, {Regnault}, {Astier},
  {Balland}, {Basa}, {Carlberg}, {Fouchez}, {Hardin}, {Hook}, {Howell}, {Pain},
  {Palanque-Delabrouille}, {Perrett}, {Pritchet}, {Rich}, {Ruhlmann-Kleider},
  {Balam}, {Baumont}, {Ellis}, {Fabbro}, {Fakhouri}, {Fourmanoit},
  {Gonz{\'a}lez-Gait{\'a}n}, {Graham}, {Hsiao}, {Kronborg}, {Lidman}, {Mourao},
  {Perlmutter}, {Ripoche}, {Suzuki}, and {Walker}}]{guy2010}
{Guy} J., {Sullivan} M., {Conley} A., {Regnault} N., {Astier} P., {Balland} C.,
  {Basa} S., {Carlberg} R.~G., et~al. (2010) {The Supernova Legacy Survey
  3-year sample: Type Ia supernovae photometric distances and cosmological
  constraints}. \emph{\aap} 523:A7, \doi{10.1051/0004-6361/201014468},
  \eprint{1010.4743}

\bibitem[{{Habouzit} et~al.(2020){Habouzit}, {Pisani}, {Goulding}, {Dubois},
  {Somerville}, and {Greene}}]{Habouzit2020}
{Habouzit} M., {Pisani} A., {Goulding} A., {Dubois} Y., {Somerville} R.~S.,
  {Greene} J.~E. (2020) {Properties of simulated galaxies and supermassive
  black holes in cosmic voids}. \emph{\mnras} 493(1):899--921,
  \doi{10.1093/mnras/staa219}, \eprint{1912.06662}

\bibitem[{Hall et~al.(2013)Hall, Bonvin, and Challinor}]{Hall:2012wd}
Hall A., Bonvin C., Challinor A. (2013) {Testing General Relativity with 21-cm
  intensity mapping}. \emph{Phys Rev D} 87(6):064026,
  \doi{10.1103/PhysRevD.87.064026}, \eprint{1212.0728}

\bibitem[{Hamaus et~al.(2010)Hamaus, Seljak, Desjacques, Smith, and
  Baldauf}]{Hamaus:2010im}
Hamaus N., Seljak U., Desjacques V., Smith R.~E., Baldauf T. (2010) {Minimizing
  the Stochasticity of Halos in Large-Scale Structure Surveys}. \emph{Phys Rev}
  D82:043515, \doi{10.1103/PhysRevD.82.043515}, \eprint{1004.5377}

\bibitem[{{Hamaus} et~al.(2014{\natexlab{a}}){Hamaus}, {Sutter}, {Lavaux}, and
  {Wandelt}}]{Hamaus2014c}
{Hamaus} N., {Sutter} P.~M., {Lavaux} G., {Wandelt} B.~D. (2014{\natexlab{a}})
  {Testing cosmic geometry without dynamic distortions using voids}.
  \emph{\jcap} 12:013, \doi{10.1088/1475-7516/2014/12/013}, \eprint{1409.3580}

\bibitem[{{Hamaus} et~al.(2014{\natexlab{b}}){Hamaus}, {Sutter}, and
  {Wandelt}}]{Hamaus2014b}
{Hamaus} N., {Sutter} P.~M., {Wandelt} B.~D. (2014{\natexlab{b}}) {Universal
  Density Profile for Cosmic Voids}. \emph{Physical Review Letters}
  112(25):251302, \doi{10.1103/PhysRevLett.112.251302}, \eprint{1403.5499}

\bibitem[{{Hamaus} et~al.(2014{\natexlab{c}}){Hamaus}, {Wandelt}, {Sutter},
  {Lavaux}, and {Warren}}]{Hamaus2014a}
{Hamaus} N., {Wandelt} B.~D., {Sutter} P.~M., {Lavaux} G., {Warren} M.~S.
  (2014{\natexlab{c}}) {Cosmology with Void-Galaxy Correlations}.
  \emph{Physical Review Letters} 112(4):041304,
  \doi{10.1103/PhysRevLett.112.041304}, \eprint{1307.2571}

\bibitem[{{Hamaus} et~al.(2015){Hamaus}, {Sutter}, {Lavaux}, and
  {Wandelt}}]{Hamaus2015}
{Hamaus} N., {Sutter} P.~M., {Lavaux} G., {Wandelt} B.~D. (2015) {Probing
  cosmology and gravity with redshift-space distortions around voids}.
  \emph{\jcap} 11:036, \doi{10.1088/1475-7516/2015/11/036}, \eprint{1507.04363}

\bibitem[{{Hamaus} et~al.(2016){Hamaus}, {Pisani}, {Sutter}, {Lavaux},
  {Escoffier}, {Wandelt}, and {Weller}}]{Hamaus2016}
{Hamaus} N., {Pisani} A., {Sutter} P.~M., {Lavaux} G., {Escoffier} S.,
  {Wandelt} B.~D., {Weller} J. (2016) {Constraints on Cosmology and Gravity
  from the Dynamics of Voids}. \emph{Physical Review Letters} 117(9):091302,
  \doi{10.1103/PhysRevLett.117.091302}, \eprint{1602.01784}

\bibitem[{{Hamaus} et~al.(2017){Hamaus}, {Cousinou}, {Pisani}, {Aubert},
  {Escoffier}, and {Weller}}]{Hamaus2017}
{Hamaus} N., {Cousinou} M.-C., {Pisani} A., {Aubert} M., {Escoffier} S.,
  {Weller} J. (2017) {Multipole analysis of redshift-space distortions around
  cosmic voids}. \emph{\jcap} 7:014, \doi{10.1088/1475-7516/2017/07/014},
  \eprint{1705.05328}

\bibitem[{{Hamaus} et~al.(2020){Hamaus}, {Pisani}, {Choi}, {Lavaux}, {Wandelt},
  and {Weller}}]{Hamaus2020}
{Hamaus} N., {Pisani} A., {Choi} J.-A., {Lavaux} G., {Wandelt} B.~D., {Weller}
  J. (2020) {Precision cosmology with voids in the final BOSS data}.
  \emph{\jcap} 2020(12):023, \doi{10.1088/1475-7516/2020/12/023},
  \eprint{2007.07895}

\bibitem[{{Hamaus} et~al.(2021){Hamaus}, {Aubert}, {Pisani}, {Contarini},
  {Verza}, {Cousinou}, {Escoffier}, {Hawken}, {Lavaux}, {Pollina}, {Wandelt},
  {Weller}, {Bonici}, {Carbone}, {Guzzo}, {Kovacs}, {Marulli}, {Massara},
  {Moscardini}, and et~al.}]{Hamaus2021}
{Hamaus} N., {Aubert} M., {Pisani} A., {Contarini} S., {Verza} G., {Cousinou}
  M.~C., {Escoffier} S., {Hawken} A., et~al. (2021) {Euclid: Forecasts from
  redshift-space distortions and the Alcock-Paczynski test with cosmic voids}.
  \emph{arXiv e-prints} arXiv:2108.10347, \eprint{2108.10347}

\bibitem[{{Haridasu} et~al.(2018){Haridasu}, {Lukovi{\'c}}, {Moresco}, and
  {Vittorio}}]{haridasu2018}
{Haridasu} B.~S., {Lukovi{\'c}} V.~V., {Moresco} M., {Vittorio} N. (2018) {An
  improved model-independent assessment of the late-time cosmic expansion}.
  \emph{\jcap} 2018(10):015, \doi{10.1088/1475-7516/2018/10/015},
  \eprint{1805.03595}

\bibitem[{Harper and Dickinson(2018)}]{Harper:2018ncl}
Harper S., Dickinson C. (2018) {Potential impact of global navigation satellite
  services on total power HI intensity mapping surveys}. \emph{Mon Not Roy
  Astron Soc} 479(2):2024--2036, \doi{10.1093/mnras/sty1495},
  \eprint{1803.06314}

\bibitem[{Harper et~al.(2018)Harper, Dickinson, Battye, Roychowdhury, Browne,
  Ma, Olivari, and Chen}]{Harper:2017gln}
Harper S., Dickinson C., Battye R., Roychowdhury S., Browne I., Ma Y.-Z.,
  Olivari L., Chen T. (2018) {Impact of Simulated 1/f Noise for HI Intensity
  Mapping Experiments}. \emph{Mon Not Roy Astron Soc} 478(2):2416--2437,
  \doi{10.1093/mnras/sty1238}, \eprint{1711.07843}

\bibitem[{{Haselgrove} and {Hoyle}(1956)}]{Hoyle}
{Haselgrove} C.~B., {Hoyle} F. (1956) {A preliminary determination of the age
  of type II stars}. \emph{\mnras} 116:527, \doi{10.1093/mnras/116.5.527}

\bibitem[{{Hawken} et~al.(2017){Hawken}, {Granett}, {Iovino}, {Guzzo},
  {Peacock}, {de la Torre}, {Garilli}, {Bolzonella}, {Scodeggio}, {Abbas},
  {Adami}, and et~al.}]{Hawken2017}
{Hawken} A.~J., {Granett} B.~R., {Iovino} A., {Guzzo} L., {Peacock} J.~A., {de
  la Torre} S., {Garilli} B., {Bolzonella} M., et~al. (2017) {The VIMOS Public
  Extragalactic Redshift Survey. Measuring the growth rate of structure around
  cosmic voids}. \emph{\aap} 607:A54, \doi{10.1051/0004-6361/201629678},
  \eprint{1611.07046}

\bibitem[{{Hawken} et~al.(2020){Hawken}, {Aubert}, {Pisani}, {Cousinou},
  {Escoffier}, {Nadathur}, {Rossi}, and {Schneider}}]{Hawken2020}
{Hawken} A.~J., {Aubert} M., {Pisani} A., {Cousinou} M.-C., {Escoffier} S.,
  {Nadathur} S., {Rossi} G., {Schneider} D.~P. (2020) {Constraints on the
  growth of structure around cosmic voids in eBOSS DR14}. \emph{\jcap}
  2020(6):012, \doi{10.1088/1475-7516/2020/06/012}, \eprint{1909.04394}

\bibitem[{{Heavens} et~al.(2004{\natexlab{a}}){Heavens}, {Panter}, {Jimenez},
  and {Dunlop}}]{Heavens2004}
{Heavens} A., {Panter} B., {Jimenez} R., {Dunlop} J. (2004{\natexlab{a}}) {The
  star-formation history of the Universe from the stellar populations of nearby
  galaxies}. \emph{\nat} 428(6983):625--627, \doi{10.1038/nature02474},
  \eprint{astro-ph/0403293}

\bibitem[{{Heavens} et~al.(2004{\natexlab{b}}){Heavens}, {Panter}, {Jimenez},
  and {Dunlop}}]{MOPEDII}
{Heavens} A., {Panter} B., {Jimenez} R., {Dunlop} J. (2004{\natexlab{b}}) {The
  star-formation history of the Universe from the stellar populations of nearby
  galaxies}. \emph{\nat} 428(6983):625--627, \doi{10.1038/nature02474},
  \eprint{astro-ph/0403293}

\bibitem[{{Heavens} et~al.(2000){Heavens}, {Jimenez}, and {Lahav}}]{MOPEDI}
{Heavens} A.~F., {Jimenez} R., {Lahav} O. (2000) {Massive lossless data
  compression and multiple parameter estimation from galaxy spectra}.
  \emph{\mnras} 317(4):965--972, \doi{10.1046/j.1365-8711.2000.03692.x},
  \eprint{astro-ph/9911102}

\bibitem[{Heger and Woosley(2002)}]{Heger:2001cd}
Heger A., Woosley S.~E. (2002) {The nucleosynthetic signature of population
  III}. \emph{Astrophys J} 567:532--543, \doi{10.1086/338487},
  \eprint{astro-ph/0107037}

\bibitem[{{Heinesen}(2021)}]{heinesen2021}
{Heinesen} A. (2021) {Redshift drift as a model independent probe of dark
  energy}. \emph{\prd} 103(8):L081302, \doi{10.1103/PhysRevD.103.L081302},
  \eprint{2102.03774}

\bibitem[{Heneka and Amendola(2018)}]{Heneka:2018ins}
Heneka C., Amendola L. (2018) {General Modified Gravity With 21cm Intensity
  Mapping: Simulations and Forecast}. \emph{JCAP} 10:004,
  \doi{10.1088/1475-7516/2018/10/004}, \eprint{1805.03629}

\bibitem[{{Hennawi} et~al.(2007){Hennawi}, {Dalal}, {Bode}, and
  {Ostriker}}]{Hennawi:2007}
{Hennawi} J.~F., {Dalal} N., {Bode} P., {Ostriker} J.~P. (2007) {Characterizing
  the Cluster Lens Population}. \emph{\apj} 654(2):714--730,
  \doi{10.1086/497362}, \eprint{astro-ph/0506171}

\bibitem[{{Hennawi} et~al.(2008){Hennawi}, {Gladders}, {Oguri}, {Dalal},
  {Koester}, {Natarajan}, {Strauss}, {Inada}, {Kayo}, {Lin}, {Lampeitl},
  {Annis}, {Bahcall}, and {Schneider}}]{Hennawi08}
{Hennawi} J.~F., {Gladders} M.~D., {Oguri} M., {Dalal} N., {Koester} B.,
  {Natarajan} P., {Strauss} M.~A., {Inada} N., et~al. (2008) {A New Survey for
  Giant Arcs}. \emph{\aj} 135(2):664--681, \doi{10.1088/0004-6256/135/2/664},
  \eprint{astro-ph/0610061}

\bibitem[{{Heymans} et~al.(2013){Heymans}, {Grocutt}, {Heavens}, {Kilbinger},
  {Kitching}, {Simpson}, {Benjamin}, {Erben}, {Hildebrandt}, {Hoekstra},
  {Mellier}, {Miller}, {Van Waerbeke}, {Brown}, {Coupon}, {Fu},
  {Harnois-D{\'e}raps}, {Hudson}, {Kuijken}, {Rowe}, {Schrabback}, {Semboloni},
  {Vafaei}, and {Velander}}]{Heymans2013}
{Heymans} C., {Grocutt} E., {Heavens} A., {Kilbinger} M., {Kitching} T.~D.,
  {Simpson} F., {Benjamin} J., {Erben} T., et~al. (2013) {CFHTLenS tomographic
  weak lensing cosmological parameter constraints: Mitigating the impact of
  intrinsic galaxy alignments}. \emph{\mnras} 432(3):2433--2453,
  \doi{10.1093/mnras/stt601}, \eprint{1303.1808}

\bibitem[{{Heymans} et~al.(2021){Heymans}, {Tr{\"o}ster}, {Asgari}, {Blake},
  {Hildebrandt}, {Joachimi}, {Kuijken}, {Lin}, {S{\'a}nchez}, {van den Busch},
  {Wright}, {Amon}, {Bilicki}, and et~al.}]{Heymans2021}
{Heymans} C., {Tr{\"o}ster} T., {Asgari} M., {Blake} C., {Hildebrandt} H.,
  {Joachimi} B., {Kuijken} K., {Lin} C.-A., et~al. (2021) {KiDS-1000 Cosmology:
  Multi-probe weak gravitational lensing and spectroscopic galaxy clustering
  constraints}. \emph{\aap} 646:A140, \doi{10.1051/0004-6361/202039063},
  \eprint{2007.15632}

\bibitem[{{Hildebrandt} et~al.(2017){Hildebrandt}, {Viola}, {Heymans},
  {Joudaki}, {Kuijken}, {Blake}, {Erben}, {Joachimi}, {Klaes}, {Miller},
  {Morrison}, {Nakajima}, {Verdoes Kleijn}, {Amon}, {Choi}, {Covone}, and
  et~al.}]{Hildebrandt2017}
{Hildebrandt} H., {Viola} M., {Heymans} C., {Joudaki} S., {Kuijken} K., {Blake}
  C., {Erben} T., {Joachimi} B., et~al. (2017) {KiDS-450: cosmological
  parameter constraints from tomographic weak gravitational lensing}.
  \emph{\mnras} 465(2):1454--1498, \doi{10.1093/mnras/stw2805},
  \eprint{1606.05338}

\bibitem[{{Hirata} et~al.(2010){Hirata}, {Holz}, and
  {Cutler}}]{2010PhRvD..81l4046H}
{Hirata} C.~M., {Holz} D.~E., {Cutler} C. (2010) {Reducing the weak lensing
  noise for the gravitational wave Hubble diagram using the non-Gaussianity of
  the magnification distribution}. \emph{\prd} 81(12):124046,
  \doi{10.1103/PhysRevD.81.124046}, \eprint{1004.3988}

\bibitem[{{Hojjati} et~al.(2013){Hojjati}, {Kim}, and {Linder}}]{Hojjati:2013}
{Hojjati} A., {Kim} A.~G., {Linder} E.~V. (2013) {Robust strong lensing time
  delay estimation}. \emph{\prd} 87(12):123512,
  \doi{10.1103/PhysRevD.87.123512}, \eprint{1304.0309}

\bibitem[{{Holanda} et~al.(2013){Holanda}, {Carvalho}, and
  {Alcaniz}}]{holanda2013}
{Holanda} R.~F.~L., {Carvalho} J.~C., {Alcaniz} J.~S. (2013) {Model-independent
  constraints on the cosmic opacity}. \emph{\jcap} 2013(4):027,
  \doi{10.1088/1475-7516/2013/04/027}, \eprint{1207.1694}

\bibitem[{{Holz} and {Hughes}(2005)}]{2005ApJ...629...15H}
{Holz} D.~E., {Hughes} S.~A. (2005) {Using Gravitational-Wave Standard Sirens}.
  \emph{\apj} 629(1):15--22, \doi{10.1086/431341}, \eprint{astro-ph/0504616}

\bibitem[{Hothi et~al.(2020)}]{Hothi:2020dgq}
Hothi I., et~al. (2020) {Comparing foreground removal techniques for recovery
  of the LOFAR-EoR 21 cm power spectrum}. \emph{Mon Not Roy Astron Soc}
  500(2):2264--2277, \doi{10.1093/mnras/staa3446}, \eprint{2011.01284}

\bibitem[{Hotinli et~al.(2019)Hotinli, Meyers, Dalal, Jaffe, Johnson, Mertens,
  Münchmeyer, Smith, and van Engelen}]{Hotinli:2018yyc}
Hotinli S.~C., Meyers J., Dalal N., Jaffe A.~H., Johnson M.~C., Mertens J.~B.,
  Münchmeyer M., Smith K.~M., et~al. (2019) {Transverse Velocities with the
  Moving Lens Effect}. \emph{Phys Rev Lett} 123(6):061301,
  \doi{10.1103/PhysRevLett.123.061301}, \eprint{1812.03167}

\bibitem[{{Hou} et~al.(2021){Hou}, {S{\'a}nchez}, {Ross}, {Smith}, {Neveux},
  {Bautista}, {Burtin}, {Zhao}, {Scoccimarro}, {Dawson}, {de Mattia}, and
  et~al.}]{hou2021}
{Hou} J., {S{\'a}nchez} A.~G., {Ross} A.~J., {Smith} A., {Neveux} R.,
  {Bautista} J., {Burtin} E., {Zhao} C., et~al. (2021) {The completed SDSS-IV
  extended Baryon Oscillation Spectroscopic Survey: BAO and RSD measurements
  from anisotropic clustering analysis of the quasar sample in configuration
  space between redshift 0.8 and 2.2}. \emph{\mnras} 500(1):1201--1221,
  \doi{10.1093/mnras/staa3234}, \eprint{2007.08998}

\bibitem[{Howlett(2019)}]{Howlett:2019bky}
Howlett C. (2019) {The redshift-space momentum power spectrum \textendash{} I.
  Optimal estimation from peculiar velocity surveys}. \emph{Mon Not Roy Astron
  Soc} 487(4):5209--5234, \doi{10.1093/mnras/stz1403}, \eprint{1906.02875}

\bibitem[{Howlett et~al.(2017)Howlett, Robotham, Lagos, and
  Kim}]{Howlett:2017asw}
Howlett C., Robotham A. S.~G., Lagos C. D.~P., Kim A.~G. (2017) {Measuring the
  growth rate of structure with Type IA Supernovae from LSST}. \emph{Astrophys
  J} 847(2):128, \doi{10.3847/1538-4357/aa88c8}, \eprint{1708.08236}

\bibitem[{{Hoyle} et~al.(2005){Hoyle}, {Rojas}, {Vogeley}, and
  {Brinkmann}}]{Hoyle2005}
{Hoyle} F., {Rojas} R.~R., {Vogeley} M.~S., {Brinkmann} J. (2005) {The
  Luminosity Function of Void Galaxies in the Sloan Digital Sky Survey}.
  \emph{\apj} 620(2):618--628, \doi{10.1086/427176}, \eprint{astro-ph/0309728}

\bibitem[{{Hu} et~al.(2021){Hu}, {Wang}, and {Dai}}]{Hu21}
{Hu} J.~P., {Wang} F.~Y., {Dai} Z.~G. (2021) {Measuring cosmological parameters
  with a luminosity-time correlation of gamma-ray bursts}. \emph{\mnras}
  507(1):730--742, \doi{10.1093/mnras/stab2180}, \eprint{2107.12718}

\bibitem[{Hu et~al.(2020)Hu, Wang, Wu, Wang, Zhang, and Chen}]{Hu:2019okh}
Hu W., Wang X., Wu F., Wang Y., Zhang P., Chen X. (2020) {Forecast for FAST:
  from Galaxies Survey to Intensity Mapping}. \emph{Mon Not Roy Astron Soc}
  493(4):5854--5870, \doi{10.1093/mnras/staa650}, \eprint{1909.10946}

\bibitem[{{Huang} et~al.(2020){Huang}, {Storfer}, {Ravi}, {Pilon}, {Domingo},
  {Schlegel}, {Bailey}, {Dey}, {Gupta}, {Herrera}, {Juneau}, {Landriau},
  {Lang}, {Meisner}, {Moustakas}, {Myers}, {Schlafly}, {Valdes}, {Weaver},
  {Yang}, and {Y{\`e}che}}]{Huang2020}
{Huang} X., {Storfer} C., {Ravi} V., {Pilon} A., {Domingo} M., {Schlegel}
  D.~J., {Bailey} S., {Dey} A., et~al. (2020) {Finding Strong Gravitational
  Lenses in the DESI DECam Legacy Survey}. \emph{\apj} 894(1):78,
  \doi{10.3847/1538-4357/ab7ffb}, \eprint{1906.00970}

\bibitem[{{Huang} et~al.(2021){Huang}, {Haster}, {Vitale}, {Varma}, {Foucart},
  and {Biscoveanu}}]{2021PhRvD.103h3001H}
{Huang} Y., {Haster} C.-J., {Vitale} S., {Varma} V., {Foucart} F., {Biscoveanu}
  S. (2021) {Statistical and systematic uncertainties in extracting the source
  properties of neutron star-black hole binaries with gravitational waves}.
  \emph{\prd} 103(8):083001, \doi{10.1103/PhysRevD.103.083001},
  \eprint{2005.11850}

\bibitem[{{Huber} et~al.(2021){Huber}, {Suyu}, {Noebauer}, {Chan}, {Kromer},
  {Sim}, {Sluse}, and {Taubenberger}}]{Huber:2021}
{Huber} S., {Suyu} S.~H., {Noebauer} U.~M., {Chan} J.~H.~H., {Kromer} M., {Sim}
  S.~A., {Sluse} D., {Taubenberger} S. (2021) {HOLISMOKES. III. Achromatic
  phase of strongly lensed Type Ia supernovae}. \emph{\aap} 646:A110,
  \doi{10.1051/0004-6361/202039218}, \eprint{2008.10393}

\bibitem[{{Hui} and {Greene}(2006{\natexlab{a}})}]{hui2006}
{Hui} L., {Greene} P.~B. (2006{\natexlab{a}}) {Correlated fluctuations in
  luminosity distance and the importance of peculiar motion in supernova
  surveys}. \emph{\prd} 73(12):123526, \doi{10.1103/PhysRevD.73.123526},
  \eprint{astro-ph/0512159}

\bibitem[{{Hui} and {Greene}(2006{\natexlab{b}})}]{2006PhRvD..73l3526H}
{Hui} L., {Greene} P.~B. (2006{\natexlab{b}}) {Correlated fluctuations in
  luminosity distance and the importance of peculiar motion in supernova
  surveys}. \emph{\prd} 73(12):123526, \doi{10.1103/PhysRevD.73.123526},
  \eprint{astro-ph/0512159}

\bibitem[{{Huterer} and {Shafer}(2018)}]{huterer2018}
{Huterer} D., {Shafer} D.~L. (2018) {Dark energy two decades after:
  observables, probes, consistency tests}. \emph{Reports on Progress in
  Physics} 81(1):016901, \doi{10.1088/1361-6633/aa997e}, \eprint{1709.01091}

\bibitem[{Huterer et~al.(2017)Huterer, Shafer, Scolnic, and
  Schmidt}]{Huterer:2016uyq}
Huterer D., Shafer D., Scolnic D., Schmidt F. (2017) {Testing $\Lambda$CDM at
  the lowest redshifts with SN Ia and galaxy velocities}. \emph{JCAP}
  1705(05):015, \doi{10.1088/1475-7516/2017/05/015}, \eprint{1611.09862}

\bibitem[{{Icke}(1984)}]{Icke1984}
{Icke} V. (1984) {Voids and filaments}. \emph{\mnras} 206:1P--3P,
  \doi{10.1093/mnras/206.1.1P}

\bibitem[{{Ilbert} et~al.(2006){Ilbert}, {Lauger}, {Tresse}, {Buat}, {Arnouts},
  {Le F{\`e}vre}, {Burgarella}, {Zucca}, {Bardelli}, {Zamorani}, {Bottini},
  {Garilli}, {Le Brun}, {Maccagni}, {Picat}, {Scaramella}, {Scodeggio}, and
  et~al.}]{ilbert2006}
{Ilbert} O., {Lauger} S., {Tresse} L., {Buat} V., {Arnouts} S., {Le F{\`e}vre}
  O., {Burgarella} D., {Zucca} E., et~al. (2006) {The VIMOS-VLT Deep Survey.
  Galaxy luminosity function per morphological type up to z = 1.2}. \emph{\aap}
  453(3):809--815, \doi{10.1051/0004-6361:20053632}, \eprint{astro-ph/0604010}

\bibitem[{{Ilbert} et~al.(2009){Ilbert}, {Capak}, {Salvato}, {Aussel},
  {McCracken}, {Sanders}, {Scoville}, {Kartaltepe}, {Arnouts}, {Le Floc'h},
  {Mobasher}, {Taniguchi}, {Lamareille}, {Leauthaud}, {Sasaki}, {Thompson},
  {Zamojski}, {Zamorani}, {Bardelli}, and et~al.}]{ilbert2009}
{Ilbert} O., {Capak} P., {Salvato} M., {Aussel} H., {McCracken} H.~J.,
  {Sanders} D.~B., {Scoville} N., {Kartaltepe} J., et~al. (2009) {Cosmos
  Photometric Redshifts with 30-Bands for 2-deg$^{2}$}. \emph{\apj}
  690(2):1236--1249, \doi{10.1088/0004-637X/690/2/1236}, \eprint{0809.2101}

\bibitem[{{Ilbert} et~al.(2010){Ilbert}, {Salvato}, {Le Floc'h}, {Aussel},
  {Capak}, {McCracken}, {Mobasher}, {Kartaltepe}, {Scoville}, {Sanders},
  {Arnouts}, {Bundy}, {Cassata}, {Kneib}, {Koekemoer}, {Le F{\`e}vre}, {Lilly},
  {Surace}, {Taniguchi}, {Tasca}, {Thompson}, {Tresse}, {Zamojski}, {Zamorani},
  and {Zucca}}]{ilbert2010}
{Ilbert} O., {Salvato} M., {Le Floc'h} E., {Aussel} H., {Capak} P., {McCracken}
  H.~J., {Mobasher} B., {Kartaltepe} J., et~al. (2010) {Galaxy Stellar Mass
  Assembly Between 0.2 < z < 2 from the S-COSMOS Survey}. \emph{\apj}
  709(2):644--663, \doi{10.1088/0004-637X/709/2/644}, \eprint{0903.0102}

\bibitem[{{Ilbert} et~al.(2013){Ilbert}, {McCracken}, {Le F{\`e}vre}, {Capak},
  {Dunlop}, {Karim}, {Renzini}, {Caputi}, {Boissier}, {Arnouts}, {Aussel},
  {Comparat}, {Guo}, {Hudelot}, {Kartaltepe}, {Kneib}, {Krogager}, {Le Floc'h},
  {Lilly}, {Mellier}, and et~al.}]{ilbert2013}
{Ilbert} O., {McCracken} H.~J., {Le F{\`e}vre} O., {Capak} P., {Dunlop} J.,
  {Karim} A., {Renzini} M.~A., {Caputi} K., et~al. (2013) {Mass assembly in
  quiescent and star-forming galaxies since $z\simeq4$ from UltraVISTA}.
  \emph{\aap} 556:A55, \doi{10.1051/0004-6361/201321100}, \eprint{1301.3157}

\bibitem[{{Ili{\'c}} et~al.(2013){Ili{\'c}}, {Langer}, and
  {Douspis}}]{Ilic2013}
{Ili{\'c}} S., {Langer} M., {Douspis} M. (2013) {Detecting the integrated
  Sachs-Wolfe effect with stacked voids}. \emph{\aap} 556:A51,
  \doi{10.1051/0004-6361/201321150}, \eprint{1301.5849}

\bibitem[{{Inada} et~al.(2003){Inada}, {Oguri}, {Pindor}, {Hennawi}, {Chiu},
  {Zheng}, {Ichikawa}, {Gregg}, {Becker}, {Suto}, {Strauss}, {Turner}, and
  et~al.}]{sdss1004}
{Inada} N., {Oguri} M., {Pindor} B., {Hennawi} J.~F., {Chiu} K., {Zheng} W.,
  {Ichikawa} S.-I., {Gregg} M.~D., et~al. (2003) {A gravitationally lensed
  quasar with quadruple images separated by 14.62arcseconds}. \emph{\nat}
  426(6968):810--812, \doi{10.1038/nature02153}, \eprint{astro-ph/0312427}

\bibitem[{{Inada} et~al.(2006){Inada}, {Oguri}, {Morokuma}, {Doi}, {Yasuda},
  {Becker}, {Richards}, {Kochanek}, {Kayo}, {Konishi}, {Utsunomiya}, {Shin},
  {Strauss}, {Sheldon}, {York}, {Hennawi}, {Schneider}, {Dai}, and
  {Fukugita}}]{sdss1029}
{Inada} N., {Oguri} M., {Morokuma} T., {Doi} M., {Yasuda} N., {Becker} R.~H.,
  {Richards} G.~T., {Kochanek} C.~S., et~al. (2006) {SDSS J1029+2623: A
  Gravitationally Lensed Quasar with an Image Separation of 22.5''}.
  \emph{\apjl} 653(2):L97--L100, \doi{10.1086/510671},
  \eprint{astro-ph/0611275}

\bibitem[{Irfan and Bull(2021)}]{Irfan:2021bci}
Irfan M.~O., Bull P. (2021) {Cleaning foregrounds from single-dish 21\,cm
  intensity maps with Kernel principal component analysis}. \emph{Mon Not Roy
  Astron Soc} 508(3):3551--3568, \doi{10.1093/mnras/stab2855},
  \eprint{2107.02267}

\bibitem[{Ishida et~al.(2019)}]{COIN:2018niv}
Ishida E. E.~O., et~al. (2019) {Optimizing spectroscopic follow-up strategies
  for supernova photometric classification with active learning}. \emph{Mon Not
  Roy Astron Soc} 483(1):2--18, \doi{10.1093/mnras/sty3015},
  \eprint{1804.03765}

\bibitem[{Ivanov et~al.(2020)Ivanov, Simonovi\'c, and
  Zaldarriaga}]{Ivanov:2019pdj}
Ivanov M.~M., Simonovi\'c M., Zaldarriaga M. (2020) {Cosmological Parameters
  from the BOSS Galaxy Power Spectrum}. \emph{JCAP} 05:042,
  \doi{10.1088/1475-7516/2020/05/042}, \eprint{1909.05277}

\bibitem[{{Izzo} et~al.(2015){Izzo}, {Muccino}, {Zaninoni}, {Amati}, and {Della
  Valle}}]{Izzo15}
{Izzo} L., {Muccino} M., {Zaninoni} E., {Amati} L., {Della Valle} M. (2015)
  {New measurements of {\ensuremath{\Omega}}$_{m}$ from gamma-ray bursts}.
  \emph{\aap} 582:A115, \doi{10.1051/0004-6361/201526461}, \eprint{1508.05898}

\bibitem[{{J{\~o}eveer} et~al.(1978){J{\~o}eveer}, {Einasto}, and
  {Tago}}]{Joeveer1978}
{J{\~o}eveer} M., {Einasto} J., {Tago} E. (1978) {Spatial distribution of
  galaxies and of clusters of galaxies in the southern galactic hemisphere}.
  \emph{\mnras} 185:357--370, \doi{10.1093/mnras/185.2.357}

\bibitem[{{Jackson}(1972)}]{Jackson1972}
{Jackson} J.~C. (1972) {A critique of Rees's theory of primordial gravitational
  radiation}. \emph{\mnras} 156:1P, \doi{10.1093/mnras/156.1.1P},
  \eprint{0810.3908}

\bibitem[{Jalilvand et~al.(2019)Jalilvand, Majerotto, Durrer, and
  Kunz}]{Jalilvand:2018ikk}
Jalilvand M., Majerotto E., Durrer R., Kunz M. (2019) {Intensity mapping of the
  21 cm emission: lensing}. \emph{JCAP} 01:020,
  \doi{10.1088/1475-7516/2019/01/020}, \eprint{1807.01351}

\bibitem[{{Jaroszynski}(2019)}]{jaroszynski2019}
{Jaroszynski} M. (2019) {Fast radio bursts and cosmological tests}.
  \emph{\mnras} 484(2):1637--1644, \doi{10.1093/mnras/sty3529},
  \eprint{1812.11936}

\bibitem[{{Jee} et~al.(2015){Jee}, {Komatsu}, and {Suyu}}]{Jee:2015}
{Jee} I., {Komatsu} E., {Suyu} S.~H. (2015) {Measuring angular diameter
  distances of strong gravitational lenses}. \emph{\jcap} 2015(11):033,
  \doi{10.1088/1475-7516/2015/11/033}, \eprint{1410.7770}

\bibitem[{{Jeffrey} et~al.(2021){Jeffrey}, {Gatti}, {Chang}, {Whiteway},
  {Demirbozan}, {Kovacs}, {Pollina}, {Bacon}, {Hamaus}, {Kacprzak}, {Lahav},
  {Lanusse}, et~al., and {DES Collaboration}}]{Jeffrey2021}
{Jeffrey} N., {Gatti} M., {Chang} C., {Whiteway} L., {Demirbozan} U., {Kovacs}
  A., {Pollina} G., {Bacon} D., et~al. (2021) {Dark Energy Survey Year 3
  results: Curved-sky weak lensing mass map reconstruction}. \emph{\mnras}
  505(3):4626--4645, \doi{10.1093/mnras/stab1495}, \eprint{2105.13539}

\bibitem[{{Jennings} et~al.(2013){Jennings}, {Li}, and {Hu}}]{Jennings2013}
{Jennings} E., {Li} Y., {Hu} W. (2013) {The abundance of voids and the
  excursion set formalism}. \emph{\mnras} 434:2167--2181,
  \doi{10.1093/mnras/stt1169}, \eprint{1304.6087}

\bibitem[{{Jensen}(2012)}]{jensen12}
{Jensen} J.~B. (2012) {Surface Brightness Fluctuation PSF Fitting Techniques
  with Natural Guide Star Adaptive Optics}. In: American Astronomical Society
  Meeting Abstracts \#220, American Astronomical Society Meeting Abstracts, vol
  220, p 332.01

\bibitem[{{Jensen} et~al.(1998){Jensen}, {Tonry}, and {Luppino}}]{jensen98}
{Jensen} J.~B., {Tonry} J.~L., {Luppino} G.~A. (1998) {Measuring Distances
  Using Infrared Surface Brightness Fluctuations}. \emph{\apj} 505(1):111--128,
  \doi{10.1086/306163}, \eprint{astro-ph/9804169}

\bibitem[{{Jensen} et~al.(2001){Jensen}, {Tonry}, {Thompson}, {Ajhar}, {Lauer},
  {Rieke}, {Postman}, and {Liu}}]{jensen01}
{Jensen} J.~B., {Tonry} J.~L., {Thompson} R.~I., {Ajhar} E.~A., {Lauer} T.~R.,
  {Rieke} M.~J., {Postman} M., {Liu} M.~C. (2001) {The Infrared Surface
  Brightness Fluctuation Hubble Constant}. \emph{\apj} 550:503--521,
  \doi{10.1086/319819}

\bibitem[{{Jensen} et~al.(2003){Jensen}, {Tonry}, {Barris}, {Thompson}, {Liu},
  {Rieke}, {Ajhar}, and {Blakeslee}}]{jensen03}
{Jensen} J.~B., {Tonry} J.~L., {Barris} B.~J., {Thompson} R.~I., {Liu} M.~C.,
  {Rieke} M.~J., {Ajhar} E.~A., {Blakeslee} J.~P. (2003) {Measuring Distances
  and Probing the Unresolved Stellar Populations of Galaxies Using Infrared
  Surface Brightness Fluctuations}. \emph{\apj} 583:712--726,
  \doi{10.1086/345430}

\bibitem[{{Jensen} et~al.(2015){Jensen}, {Blakeslee}, {Gibson}, {Lee},
  {Cantiello}, {Raimondo}, {Boyer}, and {Cho}}]{jensen15}
{Jensen} J.~B., {Blakeslee} J.~P., {Gibson} Z., {Lee} H.-c., {Cantiello} M.,
  {Raimondo} G., {Boyer} N., {Cho} H. (2015) {Measuring Infrared Surface
  Brightness Fluctuation Distances with HST WFC3: Calibration and Advice}.
  \emph{\apj} 808:91, \doi{10.1088/0004-637X/808/1/91}, \eprint{1505.00400}

\bibitem[{{Jensen} et~al.(2021){Jensen}, {Blakeslee}, {Ma}, {Milne}, {Brown},
  {Cantiello}, {Garnavich}, {Greene}, {Lucey}, {Phan}, {Tully}, and
  {Wood}}]{jensen21}
{Jensen} J.~B., {Blakeslee} J.~P., {Ma} C.-P., {Milne} P.~A., {Brown} P.~J.,
  {Cantiello} M., {Garnavich} P.~M., {Greene} J.~E., et~al. (2021) {Infrared
  Surface Brightness Fluctuation Distances for MASSIVE and Type Ia Supernova
  Host Galaxies}. \emph{arXiv e-prints} arXiv:2105.08299, \eprint{2105.08299}

\bibitem[{{Jiao} et~al.(2020){Jiao}, {Zhang}, {Zhang}, {Yu}, {Zhu}, and
  {Li}}]{jiao2020}
{Jiao} K., {Zhang} J.-C., {Zhang} T.-J., {Yu} H.-R., {Zhu} M., {Li} D. (2020)
  {Toward a direct measurement of the cosmic acceleration: roadmap and forecast
  on FAST}. \emph{\jcap} 2020(1):054, \doi{10.1088/1475-7516/2020/01/054},
  \eprint{1905.01184}

\bibitem[{Jimenez and Loeb(2002)}]{Jimenez_2002}
Jimenez R., Loeb A. (2002) {Constraining cosmological parameters based on
  relative galaxy ages}. \emph{Astrophys J} 573:37--42, \doi{10.1086/340549},
  \eprint{astro-ph/0106145}

\bibitem[{{Jimenez} and {Loeb}(2002)}]{jimenez2002}
{Jimenez} R., {Loeb} A. (2002) {Constraining Cosmological Parameters Based on
  Relative Galaxy Ages}. \emph{\apj} 573:37--42, \doi{10.1086/340549},
  \eprint{astro-ph/0106145}

\bibitem[{{Jimenez} and {Padoan}(1996)}]{JimenezPadoanLF}
{Jimenez} R., {Padoan} P. (1996) {A New Self-consistency Check on the Ages of
  Globular Clusters}. \emph{\apjl} 463:L17, \doi{10.1086/310053}

\bibitem[{{Jimenez} and {Padoan}(1998)}]{JimenezPadoanGC}
{Jimenez} R., {Padoan} P. (1998) {The Ages and Distances of Globular Clusters
  with the Luminosity Function Method: The Case of M5 and M55}. \emph{\apj}
  498(2):704--709, \doi{10.1086/305593}, \eprint{astro-ph/9701141}

\bibitem[{{Jimenez} et~al.(1996{\natexlab{a}}){Jimenez}, {Thejll}, {Jorgensen},
  {MacDonald}, and {Pagel}}]{JimenezGC}
{Jimenez} R., {Thejll} P., {Jorgensen} U.~G., {MacDonald} J., {Pagel} B.
  (1996{\natexlab{a}}) {Ages of globular clusters: a new approach}.
  \emph{\mnras} 282(3):926--942, \doi{10.1093/mnras/282.3.926},
  \eprint{astro-ph/9602132}

\bibitem[{{Jimenez} et~al.(1996{\natexlab{b}}){Jimenez}, {Thejll}, {Jorgensen},
  {MacDonald}, and {Pagel}}]{JimenezGC96}
{Jimenez} R., {Thejll} P., {Jorgensen} U.~G., {MacDonald} J., {Pagel} B.
  (1996{\natexlab{b}}) {Ages of globular clusters: a new approach}.
  \emph{\mnras} 282(3):926--942, \doi{10.1093/mnras/282.3.926},
  \eprint{astro-ph/9602132}

\bibitem[{{Jimenez} et~al.(2003){Jimenez}, {Verde}, {Treu}, and
  {Stern}}]{jimenez2003}
{Jimenez} R., {Verde} L., {Treu} T., {Stern} D. (2003) {Constraints on the
  Equation of State of Dark Energy and the Hubble Constant from Stellar Ages
  and the Cosmic Microwave Background}. \emph{\apj} 593:622--629,
  \doi{10.1086/376595}, \eprint{astro-ph/0302560}

\bibitem[{{Jimenez} et~al.(2004){Jimenez}, {MacDonald}, {Dunlop}, {Padoan}, and
  {Peacock}}]{jimenez2004}
{Jimenez} R., {MacDonald} J., {Dunlop} J.~S., {Padoan} P., {Peacock} J.~A.
  (2004) {Synthetic stellar populations: single stellar populations, stellar
  interior models and primordial protogalaxies}. \emph{\mnras} 349(1):240--254,
  \doi{10.1111/j.1365-2966.2004.07492.x}, \eprint{astro-ph/0402271}

\bibitem[{{Jimenez} et~al.(2019){Jimenez}, {Cimatti}, {Verde}, {Moresco}, and
  {Wandelt}}]{Jimenez2019}
{Jimenez} R., {Cimatti} A., {Verde} L., {Moresco} M., {Wandelt} B. (2019) {The
  local and distant Universe: stellar ages and H$_{0}$}. \emph{\jcap}
  2019(3):043, \doi{10.1088/1475-7516/2019/03/043}, \eprint{1902.07081}

\bibitem[{{Joudaki} et~al.(2017){Joudaki}, {Blake}, {Heymans}, {Choi},
  {Harnois-Deraps}, {Hildebrandt}, {Joachimi}, {Johnson}, {Mead}, {Parkinson},
  {Viola}, and {van Waerbeke}}]{Joudaki2017}
{Joudaki} S., {Blake} C., {Heymans} C., {Choi} A., {Harnois-Deraps} J.,
  {Hildebrandt} H., {Joachimi} B., {Johnson} A., et~al. (2017) {CFHTLenS
  revisited: assessing concordance with Planck including astrophysical
  systematics}. \emph{\mnras} 465(2):2033--2052, \doi{10.1093/mnras/stw2665},
  \eprint{1601.05786}

\bibitem[{{Joudaki} et~al.(2020){Joudaki}, {Hildebrandt}, {Traykova},
  {Chisari}, {Heymans}, {Kannawadi}, {Kuijken}, {Wright}, {Asgari}, {Erben},
  {Hoekstra}, {Joachimi}, {Miller}, {Tr{\"o}ster}, and {van den
  Busch}}]{Joudaki2020}
{Joudaki} S., {Hildebrandt} H., {Traykova} D., {Chisari} N.~E., {Heymans} C.,
  {Kannawadi} A., {Kuijken} K., {Wright} A.~H., et~al. (2020) {KiDS+VIKING-450
  and DES-Y1 combined: Cosmology with cosmic shear}. \emph{\aap} 638:L1,
  \doi{10.1051/0004-6361/201936154}, \eprint{1906.09262}

\bibitem[{{Jullo} and {Kneib}(2009)}]{jullokneib:2009}
{Jullo} E., {Kneib} J.~P. (2009) {Multiscale cluster lens mass mapping - I.
  Strong lensing modelling}. \emph{\mnras} 395(3):1319--1332,
  \doi{10.1111/j.1365-2966.2009.14654.x}, \eprint{0901.3792}

\bibitem[{{Jullo} et~al.(2007){Jullo}, {Kneib}, {Limousin},
  {El{\'\i}asd{\'o}ttir}, {Marshall}, and {Verdugo}}]{jullo:2007}
{Jullo} E., {Kneib} J.~P., {Limousin} M., {El{\'\i}asd{\'o}ttir} {\'A}.,
  {Marshall} P.~J., {Verdugo} T. (2007) {A Bayesian approach to strong lensing
  modelling of galaxy clusters}. \emph{New Journal of Physics} 9(12):447,
  \doi{10.1088/1367-2630/9/12/447}, \eprint{0706.0048}

\bibitem[{{Jullo} et~al.(2010){Jullo}, {Natarajan}, {Kneib}, {D'Aloisio},
  {Limousin}, {Richard}, and {Schimd}}]{Jullo:2010}
{Jullo} E., {Natarajan} P., {Kneib} J.~P., {D'Aloisio} A., {Limousin} M.,
  {Richard} J., {Schimd} C. (2010) {Cosmological constraints from strong
  gravitational lensing in clusters of galaxies.} \emph{Science} 329:924--927,
  \doi{10.1126/science.1185759}, \eprint{1008.4802}

\bibitem[{{Kaiser}(1984)}]{Kaiser1984}
{Kaiser} N. (1984) {On the spatial correlations of Abell clusters.}
  \emph{\apjl} 284:L9--L12, \doi{10.1086/184341}

\bibitem[{Kaiser(1987)}]{Kaiser:1987qv}
Kaiser N. (1987) {Clustering in real space and in redshift space}. \emph{Mon
  Not Roy Astron Soc} 227:1--27

\bibitem[{Karagiannis et~al.(2020)Karagiannis, Slosar, and
  Liguori}]{Karagiannis:2019jjx}
Karagiannis D., Slosar A., Liguori M. (2020) {Forecasts on Primordial
  non-Gaussianity from 21 cm Intensity Mapping experiments}. \emph{JCAP}
  11:052, \doi{10.1088/1475-7516/2020/11/052}, \eprint{1911.03964}

\bibitem[{Karagiannis et~al.(2021)Karagiannis, Fonseca, Maartens, and
  Camera}]{Karagiannis:2020dpq}
Karagiannis D., Fonseca J., Maartens R., Camera S. (2021) {Probing primordial
  non-Gaussianity with the power spectrum and bispectrum of future 21 cm
  intensity maps}. \emph{Phys Dark Univ} 32:100821,
  \doi{10.1016/j.dark.2021.100821}, \eprint{2010.07034}

\bibitem[{{Kelly}(2007)}]{kelly2007}
{Kelly} B.~C. (2007) {Some Aspects of Measurement Error in Linear Regression of
  Astronomical Data}. \emph{\apj} 665(2):1489--1506, \doi{10.1086/519947},
  \eprint{0705.2774}

\bibitem[{{Kelly} et~al.(2015){Kelly}, {Rodney}, {Treu}, {Foley}, {Brammer},
  {Schmidt}, {Zitrin}, {Sonnenfeld}, {Strolger}, {Graur}, {Filippenko}, {Jha},
  {Riess}, {Bradac}, {Weiner}, {Scolnic}, {Malkan}, {von der Linden}, {Trenti},
  {Hjorth}, {Gavazzi}, {Fontana}, {Merten}, {McCully}, {Jones}, {Postman},
  {Dressler}, {Patel}, {Cenko}, {Graham}, and {Tucker}}]{Kelly:2015}
{Kelly} P.~L., {Rodney} S.~A., {Treu} T., {Foley} R.~J., {Brammer} G.,
  {Schmidt} K.~B., {Zitrin} A., {Sonnenfeld} A., et~al. (2015) {Multiple images
  of a highly magnified supernova formed by an early-type cluster galaxy lens}.
  \emph{Science} 347(6226):1123--1126, \doi{10.1126/science.aaa3350},
  \eprint{1411.6009}

\bibitem[{{Kennicutt}(1998)}]{kennicutt1998}
{Kennicutt} J. Robert~C. (1998) {Star Formation in Galaxies Along the Hubble
  Sequence}. \emph{\araa} 36:189--232, \doi{10.1146/annurev.astro.36.1.189},
  \eprint{astro-ph/9807187}

\bibitem[{Kessler et~al.(2009)Kessler, Becker, Cinabro, Vanderplas, Frieman
  et~al.}]{Kessler:2009ys}
Kessler R., Becker A., Cinabro D., Vanderplas J., Frieman J.~A., et~al. (2009)
  {First-year Sloan Digital Sky Survey-II (SDSS-II) Supernova Results: Hubble
  Diagram and Cosmological Parameters}. \emph{AstrophysJSuppl} 185:32--84,
  \doi{10.1088/0067-0049/185/1/32}, \eprint{0908.4274}

\bibitem[{{Khadka} and {Ratra}(2020{\natexlab{a}})}]{Khadka20}
{Khadka} N., {Ratra} B. (2020{\natexlab{a}}) {Constraints on cosmological
  parameters from gamma-ray burst peak photon energy and bolometric fluence
  measurements and other data}. \emph{\mnras} 499(1):391--403,
  \doi{10.1093/mnras/staa2779}, \eprint{2007.13907}

\bibitem[{{Khadka} and {Ratra}(2020{\natexlab{b}})}]{kr2020a}
{Khadka} N., {Ratra} B. (2020{\natexlab{b}}) {Quasar X-ray and UV flux, baryon
  acoustic oscillation, and Hubble parameter measurement constraints on
  cosmological model parameters}. \emph{\mnras} 492(3):4456--4468,
  \doi{10.1093/mnras/staa101}, \eprint{1909.01400}

\bibitem[{{Khadka} and {Ratra}(2020{\natexlab{c}})}]{kr2020b}
{Khadka} N., {Ratra} B. (2020{\natexlab{c}}) {Using quasar X-ray and UV flux
  measurements to constrain cosmological model parameters}. \emph{\mnras}
  497(1):263--278, \doi{10.1093/mnras/staa1855}, \eprint{2004.09979}

\bibitem[{{Khadka} and {Ratra}(2021)}]{kr2021}
{Khadka} N., {Ratra} B. (2021) {Determining the range of validity of quasar
  X-ray and UV flux measurements for constraining cosmological model
  parameters}. \emph{\mnras} 502(4):6140--6156, \doi{10.1093/mnras/stab486},
  \eprint{2012.09291}

\bibitem[{{Khadka} et~al.(2021){Khadka}, {Luongo}, {Muccino}, and
  {Ratra}}]{Khadka21}
{Khadka} N., {Luongo} O., {Muccino} M., {Ratra} B. (2021) {Do gamma-ray burst
  measurements provide a useful test of cosmological models?} \emph{\jcap}
  2021(9):042, \doi{10.1088/1475-7516/2021/09/042}, \eprint{2105.12692}

\bibitem[{{Khetan} et~al.(2021){Khetan}, {Izzo}, {Branchesi}, {Wojtak},
  {Cantiello}, {Murugeshan}, {Agnello}, {Cappellaro}, {Della Valle}, {Gall},
  {Hjorth}, {Benetti}, {Brocato}, {Burke}, {Hiramatsu}, {Howell}, {Tomasella},
  and {Valenti}}]{khetan21}
{Khetan} N., {Izzo} L., {Branchesi} M., {Wojtak} R., {Cantiello} M.,
  {Murugeshan} C., {Agnello} A., {Cappellaro} E., et~al. (2021) {A new
  measurement of the Hubble constant using Type Ia supernovae calibrated with
  surface brightness fluctuations}. \emph{\aap} 647:A72,
  \doi{10.1051/0004-6361/202039196}, \eprint{2008.07754}

\bibitem[{Kim et~al.(2015)Kim, Linder, Edelstein, and Erskine}]{kim2015}
Kim A.~G., Linder E.~V., Edelstein J., Erskine D. (2015) Giving cosmic redshift
  drift a whirl. \emph{Astroparticle Physics} 62:195--205,
  \doi{https://doi.org/10.1016/j.astropartphys.2014.09.004},
  \urlprefix\url{https://www.sciencedirect.com/science/article/pii/S0927650514001339}

\bibitem[{{Kitaura} et~al.(2016){Kitaura}, {Chuang}, {Liang}, {Zhao}, {Tao},
  {Rodr{\'{\i}}guez-Torres}, {Eisenstein}, {Gil-Mar{\'{\i}}n}, {Kneib},
  {McBride}, {Percival}, {Ross}, {S{\'a}nchez}, {Tinker}, {Tojeiro},
  {Vargas-Magana}, and {Zhao}}]{Kitaura2016}
{Kitaura} F.-S., {Chuang} C.-H., {Liang} Y., {Zhao} C., {Tao} C.,
  {Rodr{\'{\i}}guez-Torres} S., {Eisenstein} D.~J., {Gil-Mar{\'{\i}}n} H.,
  et~al. (2016) {Signatures of the Primordial Universe from Its Emptiness:
  Measurement of Baryon Acoustic Oscillations from Minima of the Density
  Field}. \emph{Physical Review Letters} 116(17):171301,
  \doi{10.1103/PhysRevLett.116.171301}, \eprint{1511.04405}

\bibitem[{{Kloeckner} et~al.(2015){Kloeckner}, {Obreschkow}, {Martins},
  {Raccanelli}, {Champion}, {Roy}, {Lobanov}, {Wagner}, and
  {Keller}}]{kloeckner2015}
{Kloeckner} H.~R., {Obreschkow} D., {Martins} C., {Raccanelli} A., {Champion}
  D., {Roy} A.~L., {Lobanov} A., {Wagner} J., et~al. (2015) {Real time
  cosmology - A direct measure of the expansion rate of the Universe with the
  SKA}. In: Advancing Astrophysics with the Square Kilometre Array (AASKA14),
  p~27, \eprint{1501.03822}

\bibitem[{Klypin et~al.(2016)Klypin, Yepes, Gottlober, Prada, and
  Hess}]{Klypin:2014kpa}
Klypin A., Yepes G., Gottlober S., Prada F., Hess S. (2016) {MultiDark
  simulations: the story of dark matter halo concentrations and density
  profiles}. \emph{Mon Not Roy Astron Soc} 457(4):4340--4359,
  \doi{10.1093/mnras/stw248}, \eprint{1411.4001}

\bibitem[{Knebe et~al.(2018)}]{Knebe:2017eei}
Knebe A., et~al. (2018) {MultiDark-Galaxies: data release and first results}.
  \emph{Mon Not Roy Astron Soc} 474(4):5206--5231, \doi{10.1093/mnras/stx2662},
  \eprint{1710.08150}

\bibitem[{{Kneib} et~al.(1996){Kneib}, {Ellis}, {Smail}, {Couch}, and
  {Sharples}}]{kneib:1996}
{Kneib} J.~P., {Ellis} R.~S., {Smail} I., {Couch} W.~J., {Sharples} R.~M.
  (1996) {Hubble Space Telescope Observations of the Lensing Cluster Abell
  2218}. \emph{\apj} 471:643, \doi{10.1086/177995}, \eprint{astro-ph/9511015}

\bibitem[{{Kochanek}(2002)}]{Kochanek:2002}
{Kochanek} C.~S. (2002) {What Do Gravitational Lens Time Delays Measure?}
  \emph{\apj} 578(1):25--32, \doi{10.1086/342476}, \eprint{astro-ph/0205319}

\bibitem[{{Kochanek}(2020)}]{Kochanek:2020}
{Kochanek} C.~S. (2020) {Overconstrained gravitational lens models and the
  Hubble constant}. \emph{\mnras} 493(2):1725--1735,
  \doi{10.1093/mnras/staa344}, \eprint{1911.05083}

\bibitem[{{Kochanek}(2021)}]{Kochanek:2021}
{Kochanek} C.~S. (2021) {Overconstrained models of time delay lenses redux: how
  the angular tail wags the radial dog}. \emph{\mnras} 501(4):5021--5028,
  \doi{10.1093/mnras/staa4033}, \eprint{2003.08395}

\bibitem[{{Koda} et~al.(2014){Koda}, {Blake}, {Davis}, {Magoulas}, {Springob},
  {Scrimgeour}, {Johnson}, {Poole}, and {Staveley-Smith}}]{2014MNRAS.445.4267K}
{Koda} J., {Blake} C., {Davis} T., {Magoulas} C., {Springob} C.~M.,
  {Scrimgeour} M., {Johnson} A., {Poole} G.~B., et~al. (2014) {Are peculiar
  velocity surveys competitive as a cosmological probe?} \emph{\mnras}
  445(4):4267--4286, \doi{10.1093/mnras/stu1610}, \eprint{1312.1022}

\bibitem[{{Kodama} et~al.(2008){Kodama}, {Yonetoku}, {Murakami}, {Tanabe},
  {Tsutsui}, and {Nakamura}}]{kodama08}
{Kodama} Y., {Yonetoku} D., {Murakami} T., {Tanabe} S., {Tsutsui} R.,
  {Nakamura} T. (2008) {Gamma-ray bursts in 1.8 < z < 5.6 suggest that the time
  variation of the dark energy is small}. \emph{\mnras} 391(1):L1--L4,
  \doi{10.1111/j.1745-3933.2008.00508.x}, \eprint{0802.3428}

\bibitem[{{Koksbang}(2021)}]{koksbang2021}
{Koksbang} S.~M. (2021) {Searching for Signals of Inhomogeneity Using Multiple
  Probes of the Cosmic Expansion Rate H (z )}. \emph{\prl} 126(23):231101,
  \doi{10.1103/PhysRevLett.126.231101}, \eprint{2105.11880}

\bibitem[{{Kolatt} and {Bartelmann}(1998)}]{Kolatt:1998}
{Kolatt} T.~S., {Bartelmann} M. (1998) {Gravitational lensing of type IA
  supernovae by galaxy clusters}. \emph{\mnras} 296(3):763--772,
  \doi{10.1046/j.1365-8711.1998.01466.x}, \eprint{astro-ph/9708120}

\bibitem[{{Koleva} et~al.(2009){Koleva}, {Prugniel}, {Bouchard}, and
  {Wu}}]{koleva2009}
{Koleva} M., {Prugniel} P., {Bouchard} A., {Wu} Y. (2009) {ULySS: a full
  spectrum fitting package}. \emph{\aap} 501(3):1269--1279,
  \doi{10.1051/0004-6361/200811467}, \eprint{0903.2979}

\bibitem[{{Koopmans} et~al.(2003){Koopmans}, {Treu}, {Fassnacht}, {Blandford},
  and {Surpi}}]{Koopmans:2003}
{Koopmans} L.~V.~E., {Treu} T., {Fassnacht} C.~D., {Blandford} R.~D., {Surpi}
  G. (2003) {The Hubble Constant from the Gravitational Lens B1608+656}.
  \emph{\apj} 599(1):70--85, \doi{10.1086/379226}, \eprint{astro-ph/0306216}

\bibitem[{{Kouveliotou} et~al.(1993){Kouveliotou}, {Meegan}, {Fishman}, {Bhat},
  {Briggs}, {Koshut}, {Paciesas}, and {Pendleton}}]{kouveliotou93}
{Kouveliotou} C., {Meegan} C.~A., {Fishman} G.~J., {Bhat} N.~P., {Briggs}
  M.~S., {Koshut} T.~M., {Paciesas} W.~S., {Pendleton} G.~N. (1993)
  {Identification of Two Classes of Gamma-Ray Bursts}. \emph{\apjl} 413:L101,
  \doi{10.1086/186969}

\bibitem[{{Kov{\'a}cs} et~al.(2019){Kov{\'a}cs}, {S{\'a}nchez},
  {Garc{\'\i}a-Bellido}, {Elvin-Poole}, {Hamaus}, {Miranda}, {Nadathur},
  {Abbott}, {Abdalla}, {Annis}, {Avila}, et~al., and {DES
  Collaboration}}]{Kovacs2019}
{Kov{\'a}cs} A., {S{\'a}nchez} C., {Garc{\'\i}a-Bellido} J., {Elvin-Poole} J.,
  {Hamaus} N., {Miranda} V., {Nadathur} S., {Abbott} T., et~al. (2019) {More
  out of less: an excess integrated Sachs-Wolfe signal from supervoids mapped
  out by the Dark Energy Survey}. \emph{\mnras} 484(4):5267--5277,
  \doi{10.1093/mnras/stz341}, \eprint{1811.07812}

\bibitem[{{Kov{\'a}cs} et~al.(2021){Kov{\'a}cs}, {Beck}, {Smith}, {R{\'a}cz},
  {Csabai}, and {Szapudi}}]{Kovacs2021}
{Kov{\'a}cs} A., {Beck} R., {Smith} A., {R{\'a}cz} G., {Csabai} I., {Szapudi}
  I. (2021) {Evidence for a high-z ISW signal from supervoids in the
  distribution of eBOSS quasars}. \emph{arXiv e-prints} arXiv:2107.13038,
  \eprint{2107.13038}

\bibitem[{Kovetz et~al.(2020)}]{Kovetz:2019uss}
Kovetz E.~D., et~al. (2020) {Astrophysics and Cosmology with Line-Intensity
  Mapping}. \emph{Bull Am Astron Soc} 51(3):101, \eprint{1903.04496}

\bibitem[{{Kreckel} et~al.(2012){Kreckel}, {Platen}, {Arag{\'o}n-Calvo}, {van
  Gorkom}, {van de Weygaert}, {van der Hulst}, and {Beygu}}]{Kreckel2012}
{Kreckel} K., {Platen} E., {Arag{\'o}n-Calvo} M.~A., {van Gorkom} J.~H., {van
  de Weygaert} R., {van der Hulst} J.~M., {Beygu} B. (2012) {The Void Galaxy
  Survey: Optical Properties and H I Morphology and Kinematics}. \emph{AJ}
  144(1):16, \doi{10.1088/0004-6256/144/1/16}, \eprint{1204.5185}

\bibitem[{{Kreisch} et~al.(2019){Kreisch}, {Pisani}, {Carbone}, {Liu},
  {Hawken}, {Massara}, {Spergel}, and {Wandelt}}]{Kreisch2019}
{Kreisch} C.~D., {Pisani} A., {Carbone} C., {Liu} J., {Hawken} A.~J., {Massara}
  E., {Spergel} D.~N., {Wandelt} B.~D. (2019) {Massive Neutrinos Leave
  Fingerprints on Cosmic Voids}. \emph{\mnras} p 1877,
  \doi{10.1093/mnras/stz1944}, \eprint{1808.07464}

\bibitem[{{Kreisch} et~al.(2021){Kreisch}, {Pisani}, {Villaescusa-Navarro},
  {Spergel}, {Wandelt}, {Hamaus}, and {Bayer}}]{Kreisch2021}
{Kreisch} C.~D., {Pisani} A., {Villaescusa-Navarro} F., {Spergel} D.~N.,
  {Wandelt} B.~D., {Hamaus} N., {Bayer} A.~E. (2021) {The GIGANTES dataset:
  precision cosmology from voids in the machine learning era}. \emph{arXiv
  e-prints} arXiv:2107.02304, \eprint{2107.02304}

\bibitem[{{Kriek} et~al.(2019){Kriek}, {Price}, {Conroy}, {Suess}, {Mowla},
  {Pasha}, {Bezanson}, {van Dokkum}, and {Barro}}]{kriek2019}
{Kriek} M., {Price} S.~H., {Conroy} C., {Suess} K.~A., {Mowla} L., {Pasha} I.,
  {Bezanson} R., {van Dokkum} P., et~al. (2019) {Stellar Metallicities and
  Elemental Abundance Ratios of $z\sim1.4$ Massive Quiescent Galaxies}.
  \emph{\apjl} 880(2):L31, \doi{10.3847/2041-8213/ab2e75}, \eprint{1907.04327}

\bibitem[{{Krishnan} et~al.(2020){Krishnan}, {Colg{\'a}in}, {Ruchika},
  {Sheikh-Jabbari}, and {Yang}}]{krishnan2020}
{Krishnan} C., {Colg{\'a}in} E.~{\'O}., {Ruchika} A.~A. Sen, {Sheikh-Jabbari}
  M.~M., {Yang} T. (2020) {Is there an early Universe solution to Hubble
  tension?} \emph{\prd} 102(10):103525, \doi{10.1103/PhysRevD.102.103525},
  \eprint{2002.06044}

\bibitem[{{Krishnan} et~al.(2021){Krishnan}, {{\'O} Colg{\'a}in},
  {Sheikh-Jabbari}, and {Yang}}]{krishnan2021}
{Krishnan} C., {{\'O} Colg{\'a}in} E., {Sheikh-Jabbari} M.~M., {Yang} T. (2021)
  {Running Hubble tension and a H0 diagnostic}. \emph{\prd} 103(10):103509,
  \doi{10.1103/PhysRevD.103.103509}, \eprint{2011.02858}

\bibitem[{{Krolewski} et~al.(2018){Krolewski}, {Lee}, {White}, {Hennawi},
  {Schlegel}, {Nugent}, {Luki{\'c}}, {Stark}, {Koekemoer}, {Le F{\`e}vre},
  {Lemaux}, {Maier}, {Rich}, {Salvato}, and {Tasca}}]{Krolewski2018}
{Krolewski} A., {Lee} K.-G., {White} M., {Hennawi} J.~F., {Schlegel} D.~J.,
  {Nugent} P.~E., {Luki{\'c}} Z., {Stark} C.~W., et~al. (2018) {Detection of z
  {\ensuremath{\sim}} 2.3 Cosmic Voids from 3D Ly{\ensuremath{\alpha}} Forest
  Tomography in the COSMOS Field}. \emph{\apj} 861(1):60,
  \doi{10.3847/1538-4357/aac829}, \eprint{1710.02612}

\bibitem[{{Krone-Martins} et~al.(2018){Krone-Martins}, {Delchambre}, {Wertz},
  {Ducourant}, {Mignard}, {Teixeira}, {Kl{\"u}ter}, {Le Campion}, {Galluccio},
  {Surdej}, {Bastian}, {Wambsganss}, {Graham}, {Djorgovski}, and
  {Slezak}}]{Krone-Martins:2018}
{Krone-Martins} A., {Delchambre} L., {Wertz} O., {Ducourant} C., {Mignard} F.,
  {Teixeira} R., {Kl{\"u}ter} J., {Le Campion} J.~F., et~al. (2018) {Gaia GraL:
  Gaia DR2 gravitational lens systems. I. New quadruply imaged quasar
  candidates around known quasars}. \emph{\aap} 616:L11,
  \doi{10.1051/0004-6361/201833337}, \eprint{1804.11051}

\bibitem[{{Kroupa}(2001)}]{kroupa2001}
{Kroupa} P. (2001) {On the variation of the initial mass function}.
  \emph{\mnras} 322(2):231--246, \doi{10.1046/j.1365-8711.2001.04022.x},
  \eprint{astro-ph/0009005}

\bibitem[{{Kumar} and {Zhang}(2015)}]{Kumar15}
{Kumar} P., {Zhang} B. (2015) {The physics of gamma-ray bursts \& relativistic
  jets}. \emph{\physrep} 561:1--109, \doi{10.1016/j.physrep.2014.09.008},
  \eprint{1410.0679}

\bibitem[{{La Franca} et~al.(2014){La Franca}, {Bianchi}, {Ponti}, {Branchini},
  and {Matt}}]{lafranca2014}
{La Franca} F., {Bianchi} S., {Ponti} G., {Branchini} E., {Matt} G. (2014) {A
  New Cosmological Distance Measure Using Active Galactic Nucleus X-Ray
  Variability}. \emph{\apjl} 787(1):L12, \doi{10.1088/2041-8205/787/1/L12},
  \eprint{1404.2607}

\bibitem[{{Lagattuta} et~al.(2017){Lagattuta}, {Richard}, {Cl{\'e}ment},
  {Mahler}, {Patr{\'\i}cio}, {Pell{\'o}}, {Soucail}, {Schmidt}, {Wisotzki},
  {Martinez}, and {Bina}}]{Lagattuta:2017}
{Lagattuta} D.~J., {Richard} J., {Cl{\'e}ment} B., {Mahler} G., {Patr{\'\i}cio}
  V., {Pell{\'o}} R., {Soucail} G., {Schmidt} K.~B., et~al. (2017) {Lens
  modelling Abell 370: crowning the final frontier field with MUSE}.
  \emph{\mnras} 469(4):3946--3964, \doi{10.1093/mnras/stx1079},
  \eprint{1611.01513}

\bibitem[{{Lagattuta} et~al.(2019){Lagattuta}, {Richard}, {Bauer},
  {Cl{\'e}ment}, {Mahler}, {Soucail}, {Carton}, {Kneib}, {Laporte}, {Martinez},
  {Patr{\'\i}cio}, {Payne}, {Pell{\'o}}, {Schmidt}, and {de la
  Vieuville}}]{Lagattuta:2019}
{Lagattuta} D.~J., {Richard} J., {Bauer} F.~E., {Cl{\'e}ment} B., {Mahler} G.,
  {Soucail} G., {Carton} D., {Kneib} J.-P., et~al. (2019) {Probing 3D structure
  with a large MUSE mosaic: extending the mass model of Frontier Field Abell
  370}. \emph{\mnras} 485(3):3738--3760, \doi{10.1093/mnras/stz620},
  \eprint{1904.02158}

\bibitem[{{Lagos} et~al.(2019){Lagos}, {Fishbach}, {Landry}, and
  {Holz}}]{2019PhRvD..99h3504L}
{Lagos} M., {Fishbach} M., {Landry} P., {Holz} D.~E. (2019) {Standard sirens
  with a running Planck mass}. \emph{\prd} 99(8):083504,
  \doi{10.1103/PhysRevD.99.083504}, \eprint{1901.03321}

\bibitem[{{Lahav} et~al.(1991{\natexlab{a}}){Lahav}, {Lilje}, {Primack}, and
  {Rees}}]{Lahav1991}
{Lahav} O., {Lilje} P.~B., {Primack} J.~R., {Rees} M.~J. (1991{\natexlab{a}})
  {Dynamical effects of the cosmological constant.} \emph{\mnras} 251:128--136,
  \doi{10.1093/mnras/251.1.128}

\bibitem[{{Lahav} et~al.(1991{\natexlab{b}}){Lahav}, {Lilje}, {Primack}, and
  {Rees}}]{Lahav:1991}
{Lahav} O., {Lilje} P.~B., {Primack} J.~R., {Rees} M.~J. (1991{\natexlab{b}})
  {Dynamical effects of the cosmological constant}. \emph{\mnras} 251:128--136

\bibitem[{{Lam} et~al.(2015){Lam}, {Clampitt}, {Cai}, and {Li}}]{Lam2015}
{Lam} T.~Y., {Clampitt} J., {Cai} Y.-C., {Li} B. (2015) {Voids in modified
  gravity reloaded: Eulerian void assignment}. \emph{\mnras} 450(3):3319--3330,
  \doi{10.1093/mnras/stv797}, \eprint{1408.5338}

\bibitem[{{Lambas} et~al.(2016){Lambas}, {Lares}, {Ceccarelli}, {Ruiz}, {Paz},
  {Maldonado}, and {Luparello}}]{Lambas2016}
{Lambas} D.~G., {Lares} M., {Ceccarelli} L., {Ruiz} A.~N., {Paz} D.~J.,
  {Maldonado} V.~E., {Luparello} H.~E. (2016) {The sparkling Universe: the
  coherent motions of cosmic voids}. \emph{\mnras} 455(1):L99--L103,
  \doi{10.1093/mnrasl/slv151}, \eprint{1510.00712}

\bibitem[{{Landy} and {Szalay}(1993)}]{Landy1993}
{Landy} S.~D., {Szalay} A.~S. (1993) {Bias and variance of angular correlation
  functions}. \emph{\apj} 412:64--71, \doi{10.1086/172900}

\bibitem[{{Lares} et~al.(2017{\natexlab{a}}){Lares}, {Luparello}, {Maldonado},
  {Ruiz}, {Paz}, {Ceccarelli}, and {Garcia Lambas}}]{Lares2017b}
{Lares} M., {Luparello} H.~E., {Maldonado} V., {Ruiz} A.~N., {Paz} D.~J.,
  {Ceccarelli} L., {Garcia Lambas} D. (2017{\natexlab{a}}) {Voids and
  superstructures: correlations and induced large-scale velocity flows}.
  \emph{\mnras} 470(1):85--94, \doi{10.1093/mnras/stx1227}, \eprint{1705.06541}

\bibitem[{{Lares} et~al.(2017{\natexlab{b}}){Lares}, {Ruiz}, {Luparello},
  {Ceccarelli}, {Garcia Lambas}, and {Paz}}]{Lares2017a}
{Lares} M., {Ruiz} A.~N., {Luparello} H.~E., {Ceccarelli} L., {Garcia Lambas}
  D., {Paz} D.~J. (2017{\natexlab{b}}) {The sparkling Universe: clustering of
  voids and void clumps}. \emph{\mnras} 468(4):4822--4830,
  \doi{10.1093/mnras/stx825}, \eprint{1703.10428}

\bibitem[{{Laureijs} et~al.(2011){Laureijs}, {Amiaux}, {Arduini},
  {Augu{\`e}res}, {Brinchmann}, {Cole}, {Cropper}, {Dabin}, {Duvet}, {Ealet},
  {Garilli}, {Gondoin}, {Guzzo}, {Hoar}, {Hoekstra}, and
  https://www.overleaf.com/4448358379csmrynjykryg}]{Laureijs:2011gra}
{Laureijs} R., {Amiaux} J., {Arduini} S., {Augu{\`e}res} J.~L., {Brinchmann}
  J., {Cole} R., {Cropper} M., {Dabin} C., et~al. (2011) {Euclid Definition
  Study Report}. \emph{arXiv e-prints} arXiv:1110.3193, \eprint{1110.3193}

\bibitem[{{Lavaux} and {Wandelt}(2012)}]{Lavaux2012}
{Lavaux} G., {Wandelt} B.~D. (2012) {Precision Cosmography with Stacked Voids}.
  \emph{\apj} 754:109, \doi{10.1088/0004-637X/754/2/109}

\bibitem[{{Lawrence} et~al.(2007){Lawrence}, {Warren}, {Almaini}, {Edge},
  {Hambly}, {Jameson}, {Lucas}, {Casali}, {Adamson}, {Dye}, {Emerson},
  {Foucaud}, {Hewett}, {Hirst}, {Hodgkin}, {Irwin}, {Lodieu}, {McMahon},
  {Simpson}, {Smail}, {Mortlock}, and {Folger}}]{lawrence2007}
{Lawrence} A., {Warren} S.~J., {Almaini} O., {Edge} A.~C., {Hambly} N.~C.,
  {Jameson} R.~F., {Lucas} P., {Casali} M., et~al. (2007) {The UKIRT Infrared
  Deep Sky Survey (UKIDSS)}. \emph{\mnras} 379(4):1599--1617,
  \doi{10.1111/j.1365-2966.2007.12040.x}, \eprint{astro-ph/0604426}

\bibitem[{{Le Borgne} et~al.(2006){Le Borgne}, {Abraham}, {Daniel}, {McCarthy},
  {Glazebrook}, {Savaglio}, {Crampton}, {Juneau}, {Carlberg}, {Chen}, {Marzke},
  {Roth}, {J{\o}rgensen}, and {Murowinski}}]{leborgne2006}
{Le Borgne} D., {Abraham} R., {Daniel} K., {McCarthy} P.~J., {Glazebrook} K.,
  {Savaglio} S., {Crampton} D., {Juneau} S., et~al. (2006) {Gemini Deep Deep
  Survey. VI. Massive H{\ensuremath{\delta}}-strong Galaxies at z
  \raisebox{-0.5ex}\textasciitilde= 1}. \emph{\apj} 642(1):48--62,
  \doi{10.1086/500005}, \eprint{astro-ph/0503401}

\bibitem[{{Le Borgne} et~al.(2003){Le Borgne}, {Bruzual}, {Pell{\'o}},
  {Lan{\c{c}}on}, {Rocca-Volmerange}, {Sanahuja}, {Schaerer}, {Soubiran}, and
  {V{\'\i}lchez-G{\'o}mez}}]{leborgne2003}
{Le Borgne} J.~F., {Bruzual} G., {Pell{\'o}} R., {Lan{\c{c}}on} A.,
  {Rocca-Volmerange} B., {Sanahuja} B., {Schaerer} D., {Soubiran} C., et~al.
  (2003) {STELIB: A library of stellar spectra at R
  \raisebox{-0.5ex}\textasciitilde 2000}. \emph{\aap} 402:433--442,
  \doi{10.1051/0004-6361:20030243}, \eprint{astro-ph/0302334}

\bibitem[{{Leandro} et~al.(2021){Leandro}, {Marra}, and
  {Sturani}}]{2021arXiv210907537L}
{Leandro} H., {Marra} V., {Sturani} R. (2021) {Measuring the Hubble constant
  with black sirens}. \emph{arXiv e-prints} arXiv:2109.07537,
  \eprint{2109.07537}

\bibitem[{{Leclercq} et~al.(2015){Leclercq}, {Jasche}, {Sutter}, {Hamaus}, and
  {Wandelt}}]{Leclercq2015}
{Leclercq} F., {Jasche} J., {Sutter} P.~M., {Hamaus} N., {Wandelt} B. (2015)
  {Dark matter voids in the SDSS galaxy survey}. \emph{\jcap} 3:047,
  \doi{10.1088/1475-7516/2015/03/047}, \eprint{1410.0355}

\bibitem[{{Lee} and {Park}(2009)}]{Lee2009}
{Lee} J., {Park} D. (2009) {Constraining the Dark Energy Equation of State with
  Cosmic Voids}. \emph{\apjl} 696(1):L10--L12,
  \doi{10.1088/0004-637X/696/1/L10}, \eprint{0704.0881}

\bibitem[{{Lee} et~al.(1993){Lee}, {Freedman}, and {Madore}}]{lee1993}
{Lee} M.~G., {Freedman} W.~L., {Madore} B.~F. (1993) {The Tip of the Red Giant
  Branch as a Distance Indicator for Resolved Galaxies}. \emph{\apj} 417:553,
  \doi{10.1086/173334}

\bibitem[{{Leizerovich} et~al.(2021){Leizerovich}, {Kraiselburd}, {Landau}, and
  {Sc{\'o}ccola}}]{2021arXiv211201492L}
{Leizerovich} M., {Kraiselburd} L., {Landau} S.~J., {Sc{\'o}ccola} C.~G. (2021)
  {Testing f(R) gravity models with quasar X-ray and UV fluxes}. \emph{arXiv
  e-prints} arXiv:2112.01492, \eprint{2112.01492}

\bibitem[{{Leja} et~al.(2019){Leja}, {Tacchella}, and {Conroy}}]{leja2019}
{Leja} J., {Tacchella} S., {Conroy} C. (2019) {Beyond UVJ: More Efficient
  Selection of Quiescent Galaxies with Ultraviolet/Mid-infrared Fluxes}.
  \emph{\apjl} 880(1):L9, \doi{10.3847/2041-8213/ab2f8c}, \eprint{1907.02970}

\bibitem[{{Lemon} et~al.(2019){Lemon}, {Auger}, and {McMahon}}]{Lemon:2019}
{Lemon} C.~A., {Auger} M.~W., {McMahon} R.~G. (2019) {Gravitationally lensed
  quasars in Gaia - III. 22 new lensed quasars from Gaia data release 2}.
  \emph{\mnras} 483(3):4242--4258, \doi{10.1093/mnras/sty3366},
  \eprint{1810.04480}

\bibitem[{Li et~al.(2020)}]{Li:2020ast}
Li J., et~al. (2020) {The Tianlai Cylinder Pathfinder array: System functions
  and basic performance analysis}. \emph{Sci China Phys Mech Astron}
  63(12):129862, \doi{10.1007/s11433-020-1594-8}, \eprint{2006.05605}

\bibitem[{{Li} et~al.(2021){Li}, {Keeley}, {Shafieloo}, {Zheng}, {Cao},
  {Biesiada}, and {Zhu}}]{2021MNRAS.507..919L}
{Li} X., {Keeley} R.~E., {Shafieloo} A., {Zheng} X., {Cao} S., {Biesiada} M.,
  {Zhu} Z.-H. (2021) {Hubble diagram at higher redshifts: model independent
  calibration of quasars}. \emph{\mnras} 507(1):919--926,
  \doi{10.1093/mnras/stab2154}, \eprint{2103.16032}

\bibitem[{Li et~al.(2021)Li, Santos, Grainge, Harper, and Wang}]{Li:2020bcr}
Li Y., Santos M.~G., Grainge K., Harper S., Wang J. (2021) {HI intensity
  mapping with MeerKAT: 1/f noise analysis}. \emph{Mon Not Roy Astron Soc}
  501(3):4344--4358, \doi{10.1093/mnras/staa3856}, \eprint{2007.01767}

\bibitem[{{Liang} and {Zhang}(2005)}]{Liang05}
{Liang} E., {Zhang} B. (2005) {Model-independent Multivariable Gamma-Ray Burst
  Luminosity Indicator and Its Possible Cosmological Implications}. \emph{\apj}
  633(2):611--623, \doi{10.1086/491594}, \eprint{astro-ph/0504404}

\bibitem[{{Liang} et~al.(2008){Liang}, {Xiao}, {Liu}, and {Zhang}}]{Liang08}
{Liang} N., {Xiao} W.~K., {Liu} Y., {Zhang} S.~N. (2008) {A
  Cosmology-Independent Calibration of Gamma-Ray Burst Luminosity Relations and
  the Hubble Diagram}. \emph{\apj} 685(1):354--360, \doi{10.1086/590903},
  \eprint{0802.4262}

\bibitem[{{Liang} et~al.(2016){Liang}, {Zhao}, {Chuang}, {Kitaura}, and
  {Tao}}]{Liang2016}
{Liang} Y., {Zhao} C., {Chuang} C.-H., {Kitaura} F.-S., {Tao} C. (2016)
  {Measuring baryon acoustic oscillations from the clustering of voids}.
  \emph{\mnras} 459:4020--4028, \doi{10.1093/mnras/stw884}, \eprint{1511.04391}

\bibitem[{{Liepold} et~al.(2020){Liepold}, {Quenneville}, {Ma}, {Walsh},
  {McConnell}, {Greene}, and {Blakeslee}}]{liepold20}
{Liepold} C.~M., {Quenneville} M.~E., {Ma} C.-P., {Walsh} J.~L., {McConnell}
  N.~J., {Greene} J.~E., {Blakeslee} J.~P. (2020) {The MASSIVE Survey. XV. A
  Stellar Dynamical Mass Measurement of the Supermassive Black Hole in Massive
  Elliptical Galaxy NGC 1453}. \emph{\apj} 891(1):4,
  \doi{10.3847/1538-4357/ab6f71}, \eprint{2001.08753}

\bibitem[{{LIGO Scientific Collaboration} et~al.(2015){LIGO Scientific
  Collaboration}, {Aasi}, {Abbott}, {Abbott}, {Abbott}, {Abernathy}, {Ackley},
  {Adams}, {Adams}, {Addesso}, and et~al.}]{2015CQGra..32g4001L}
{LIGO Scientific Collaboration}, {Aasi} J., {Abbott} B.~P., {Abbott} R.,
  {Abbott} T., {Abernathy} M.~R., {Ackley} K., {Adams} C., et~al. (2015)
  {Advanced LIGO}. \emph{Classical and Quantum Gravity} 32(7):074001,
  \doi{10.1088/0264-9381/32/7/074001}, \eprint{1411.4547}

\bibitem[{{Lin} et~al.(2015){Lin}, {Li}, {Wang}, and {Chang}}]{Lin15}
{Lin} H.-N., {Li} X., {Wang} S., {Chang} Z. (2015) {Are long gamma-ray bursts
  standard candles?} \emph{\mnras} 453(1):128--132,
  \doi{10.1093/mnras/stv1624}, \eprint{1504.07026}

\bibitem[{{Lin} et~al.(2016{\natexlab{a}}){Lin}, {Li}, and {Chang}}]{Lin16a}
{Lin} H.-N., {Li} X., {Chang} Z. (2016{\natexlab{a}}) {Effect of gamma-ray
  burst (GRB) spectra on the empirical luminosity correlations and the GRB
  Hubble diagram}. \emph{\mnras} 459(3):2501--2512, \doi{10.1093/mnras/stw817},
  \eprint{1604.02285}

\bibitem[{{Lin} et~al.(2016{\natexlab{b}}){Lin}, {Li}, and {Chang}}]{Lin16b}
{Lin} H.-N., {Li} X., {Chang} Z. (2016{\natexlab{b}}) {Model-independent
  distance calibration of high-redshift gamma-ray bursts and constrain on the
  {\ensuremath{\Lambda}}CDM model}. \emph{\mnras} 455(2):2131--2138,
  \doi{10.1093/mnras/stv2471}, \eprint{1507.06662}

\bibitem[{{Lin} et~al.(2020){Lin}, {Mack}, and {Hou}}]{lin2020}
{Lin} W., {Mack} K.~J., {Hou} L. (2020) {Investigating the Hubble Constant
  Tension: Two Numbers in the Standard Cosmological Model}. \emph{\apjl}
  904(2):L22, \doi{10.3847/2041-8213/abc894}, \eprint{1910.02978}

\bibitem[{{Lin} et~al.(2021){Lin}, {Chen}, and {Mack}}]{lin2021}
{Lin} W., {Chen} X., {Mack} K.~J. (2021) {Early-Universe-Physics Insensitive
  and Uncalibrated Cosmic Standards: Constraints on $\Omega_{\rm{m}}$ and
  Implications for the Hubble Tension}. \emph{arXiv e-prints} arXiv:2102.05701,
  \eprint{2102.05701}

\bibitem[{{Linder}(2003)}]{linder2003}
{Linder} E.~V. (2003) {Exploring the Expansion History of the Universe}.
  \emph{Phys\ Rev\ Lett} 90(9):091301, \doi{10.1103/PhysRevLett.90.091301},
  \eprint{astro-ph/0208512}

\bibitem[{{Linder}(2005)}]{Linder2005}
{Linder} E.~V. (2005) {Cosmic growth history and expansion history}.
  \emph{\prd} 72(4):043529, \doi{10.1103/PhysRevD.72.043529},
  \eprint{astro-ph/0507263}

\bibitem[{{Linder}(2017)}]{linder2017}
{Linder} E.~V. (2017) {Cosmic growth and expansion conjoined}.
  \emph{Astroparticle Physics} 86:41--45,
  \doi{10.1016/j.astropartphys.2016.11.002}, \eprint{1610.05321}

\bibitem[{{Liske} et~al.(2008){Liske}, {Grazian}, {Vanzella}, {Dessauges},
  {Viel}, {Pasquini}, {Haehnelt}, {Cristiani}, {Pepe}, {Avila}, {Bonifacio},
  {Bouchy}, {Dekker}, {Delabre}, {D'Odorico}, {D'Odorico}, {Levshakov},
  {Lovis}, {Mayor}, {Molaro}, {Moscardini}, {Murphy}, {Queloz}, {Shaver},
  {Udry}, {Wiklind}, and {Zucker}}]{liske2008}
{Liske} J., {Grazian} A., {Vanzella} E., {Dessauges} M., {Viel} M., {Pasquini}
  L., {Haehnelt} M., {Cristiani} S., et~al. (2008) {Cosmic dynamics in the era
  of Extremely Large Telescopes}. \emph{MNRAS} 386(3):1192--1218,
  \doi{10.1111/j.1365-2966.2008.13090.x}, \eprint{0802.1532}

\bibitem[{Liu and Shaw(2020)}]{Liu:2019awk}
Liu A., Shaw J.~R. (2020) {Data Analysis for Precision 21 cm Cosmology}.
  \emph{Publ Astron Soc Pac} 132(1012):062001, \doi{10.1088/1538-3873/ab5bfd},
  \eprint{1907.08211}

\bibitem[{Liu and Tegmark(2011)}]{Liu:2011hh}
Liu A., Tegmark M. (2011) {A Method for 21cm Power Spectrum Estimation in the
  Presence of Foregrounds}. \emph{Phys Rev D} 83:103006,
  \doi{10.1103/PhysRevD.83.103006}, \eprint{1103.0281}

\bibitem[{{Liu} et~al.(2000){Liu}, {Charlot}, and {Graham}}]{liu00}
{Liu} M.~C., {Charlot} S., {Graham} J.~R. (2000) {Theoretical Predictions for
  Surface Brightness Fluctuations and Implications for Stellar Populations of
  Elliptical Galaxies}. \emph{\apj} 543(2):644--668, \doi{10.1086/317147},
  \eprint{astro-ph/0004367}

\bibitem[{Liu et~al.(2020)Liu, Heneka, and Amendola}]{Liu:2019ygl}
Liu X.-W., Heneka C., Amendola L. (2020) {Constraining coupled quintessence
  with the 21cm signal}. \emph{JCAP} 05(05):038,
  \doi{10.1088/1475-7516/2020/05/038}, \eprint{1910.02763}

\bibitem[{{Lloyd} et~al.(2000){Lloyd}, {Petrosian}, and {Mallozzi}}]{Lloyd00}
{Lloyd} N.~M., {Petrosian} V., {Mallozzi} R.~S. (2000) {Cosmological versus
  Intrinsic: The Correlation between Intensity and the Peak of the
  {\ensuremath{\nu}}F$_{{\ensuremath{\nu}}}$ Spectrum of Gamma-Ray Bursts}.
  \emph{\apj} 534(1):227--238, \doi{10.1086/308742}, \eprint{astro-ph/9908191}

\bibitem[{Lochner et~al.(2016)Lochner, McEwen, Peiris, Lahav, and
  Winter}]{Lochner:2016hbn}
Lochner M., McEwen J.~D., Peiris H.~V., Lahav O., Winter M.~K. (2016)
  {Photometric Supernova Classification With Machine Learning}. \emph{Astrophys
  J Suppl} 225(2):31, \doi{10.3847/0067-0049/225/2/31}, \eprint{1603.00882}

\bibitem[{{Lochner} et~al.(2021){Lochner}, {Scolnic}, {Almoubayyed}, {Anguita},
  {Awan}, {Gawiser}, {Gontcho}, {Gris}, {Huber}, {Jha}, {Jones}, {Kim},
  {Mandelbaum}, {Marshall}, {Petrushevska}, {Regnault}, {Setzer}, {Suyu},
  {Yoachim}, {Biswas}, {Blaineau}, {Hook}, {Moniez}, {Neilsen}, {Peiris},
  {Rothchild}, and {Stubbs}}]{LSSTDarkEnergyScience:2021ryz}
{Lochner} M., {Scolnic} D., {Almoubayyed} H., {Anguita} T., {Awan} H.,
  {Gawiser} E., {Gontcho} S. G.~A., {Gris} P., et~al. (2021) {The Impact of
  Observing Strategy on Cosmological Constraints with LSST}. \emph{arXiv
  e-prints} arXiv:2104.05676, \eprint{2104.05676}

\bibitem[{{Loeb}(1998)}]{loeb1998}
{Loeb} A. (1998) {Direct Measurement of Cosmological Parameters from the Cosmic
  Deceleration of Extragalactic Objects}. \emph{ApJL} 499(2):L111--L114,
  \doi{10.1086/311375}, \eprint{astro-ph/9802122}

\bibitem[{Loeb and Wyithe(2008)}]{Loeb:2008hg}
Loeb A., Wyithe S. (2008) {Precise Measurement of the Cosmological Power
  Spectrum With a Dedicated 21cm Survey After Reionization}. \emph{Phys Rev
  Lett} 100:161301, \doi{10.1103/PhysRevLett.100.161301}, \eprint{0801.1677}

\bibitem[{{L{\'o}pez-Corredoira} et~al.(2016){L{\'o}pez-Corredoira}, {Melia},
  {Lusso}, and {Risaliti}}]{corr2016}
{L{\'o}pez-Corredoira} M., {Melia} F., {Lusso} E., {Risaliti} G. (2016)
  {Cosmological test with the QSO Hubble diagram}. \emph{International Journal
  of Modern Physics D} 25(5):1650060, \doi{10.1142/S0218271816500607},
  \eprint{1602.06743}

\bibitem[{{Lotz} et~al.(2014){Lotz}, {Mountain}, {Grogin}, {Koekemoer},
  {Capak}, {Mack}, {Coe}, {Barker}, {Adler}, {Avila}, {Anderson}, {Casertano},
  {Christian}, {Gonzaga}, {Ferguson}, and et~al.}]{Lotz:2014}
{Lotz} J., {Mountain} M., {Grogin} N.~A., {Koekemoer} A.~M., {Capak} P.~L.,
  {Mack} J., {Coe} D.~A., {Barker} E.~A., et~al. (2014) {The HST Frontier
  Fields}. In: American Astronomical Society Meeting Abstracts \#223, American
  Astronomical Society Meeting Abstracts, vol 223, p 254.01

\bibitem[{{LSST Science Collaboration} et~al.(2009){LSST Science
  Collaboration}, {Abell}, {Allison}, {Anderson}, {Andrew}, {Angel}, {Armus},
  {Arnett}, {Asztalos}, {Axelrod}, {Bailey}, {Ballantyne}, {Bankert},
  {Barkhouse}, {Barr}, {Barrientos}, {Barth}, {Bartlett}, and
  et~al.}]{LSSTScience:2009jmu}
{LSST Science Collaboration}, {Abell} P.~A., {Allison} J., {Anderson} S.~F.,
  {Andrew} J.~R., {Angel} J. R.~P., {Armus} L., {Arnett} D., et~al. (2009)
  {LSST Science Book, Version 2.0}. \emph{arXiv e-prints} arXiv:0912.0201,
  \eprint{0912.0201}

\bibitem[{{Lu} et~al.(2012){Lu}, {Wei}, {Liang}, {Zhang}, {L{\"u}}, {L{\"u}},
  {Lei}, and {Zhang}}]{Lu12}
{Lu} R.-J., {Wei} J.-J., {Liang} E.-W., {Zhang} B.-B., {L{\"u}} H.-J., {L{\"u}}
  L.-Z., {Lei} W.-H., {Zhang} B. (2012) {A Comprehensive Analysis of Fermi
  Gamma-Ray Burst Data. II. E $_{p}$ Evolution Patterns and Implications for
  the Observed Spectrum-Luminosity Relations}. \emph{\apj} 756(2):112,
  \doi{10.1088/0004-637X/756/2/112}, \eprint{1204.0714}

\bibitem[{{Lusso}(2019)}]{lusso2019a}
{Lusso} E. (2019) {The nonlinear X-ray/ultraviolet relation in active galactic
  nuclei: Contribution of instrumental effects on the X-ray variability}.
  \emph{Astronomische Nachrichten} 340(4):267--272,
  \doi{10.1002/asna.201913608}, \eprint{1812.03179}

\bibitem[{{Lusso}(2020)}]{lusso2020f}
{Lusso} E. (2020) {Cosmology with quasars: predictions for eROSITA from a
  quasar Hubble diagram}. \emph{Frontiers in Astronomy and Space Sciences} 7:8,
  \doi{10.3389/fspas.2020.00008}, \eprint{2002.02464}

\bibitem[{{Lusso} and {Risaliti}(2016)}]{lr16}
{Lusso} E., {Risaliti} G. (2016) {The Tight Relation between X-Ray and
  Ultraviolet Luminosity of Quasars}. \emph{\apj} 819:154,
  \doi{10.3847/0004-637X/819/2/154}, \eprint{1602.01090}

\bibitem[{{Lusso} and {Risaliti}(2017)}]{lr17}
{Lusso} E., {Risaliti} G. (2017) {Quasars as standard candles. I. The physical
  relation between disc and coronal emission}. \emph{\aap} 602:A79,
  \doi{10.1051/0004-6361/201630079}, \eprint{1703.05299}

\bibitem[{{Lusso} et~al.(2015){Lusso}, {Worseck}, {Hennawi}, {Prochaska},
  {Vignali}, {Stern}, and {O'Meara}}]{lusso2015}
{Lusso} E., {Worseck} G., {Hennawi} J.~F., {Prochaska} J.~X., {Vignali} C.,
  {Stern} J., {O'Meara} J.~M. (2015) {The first ultraviolet quasar-stacked
  spectrum at z {$\simeq$} 2.4 from WFC3}. \emph{\mnras} 449:4204--4220,
  \doi{10.1093/mnras/stv516}, \eprint{1503.02075}

\bibitem[{{Lusso} et~al.(2019{\natexlab{a}}){Lusso}, {Piedipalumbo},
  {Risaliti}, {Paolillo}, {Bisogni}, {Nardini}, and {Amati}}]{lusso19b}
{Lusso} E., {Piedipalumbo} E., {Risaliti} G., {Paolillo} M., {Bisogni} S.,
  {Nardini} E., {Amati} L. (2019{\natexlab{a}}) {Tension with the flat
  {\ensuremath{\Lambda}}CDM model from a high-redshift Hubble diagram of
  supernovae, quasars, and gamma-ray bursts}. \emph{\aap} 628:L4,
  \doi{10.1051/0004-6361/201936223}, \eprint{1907.07692}

\bibitem[{{Lusso} et~al.(2019{\natexlab{b}}){Lusso}, {Piedipalumbo},
  {Risaliti}, {Paolillo}, {Bisogni}, {Nardini}, and {Amati}}]{lusso2019b}
{Lusso} E., {Piedipalumbo} E., {Risaliti} G., {Paolillo} M., {Bisogni} S.,
  {Nardini} E., {Amati} L. (2019{\natexlab{b}}) {Tension with the flat
  {\ensuremath{\Lambda}}CDM model from a high-redshift Hubble diagram of
  supernovae, quasars, and gamma-ray bursts}. \emph{\aap} 628:L4,
  \doi{10.1051/0004-6361/201936223}, \eprint{1907.07692}

\bibitem[{{Lusso} et~al.(2020){Lusso}, {Risaliti}, {Nardini}, {Bargiacchi},
  {Benetti}, {Bisogni}, {Capozziello}, {Civano}, {Eggleston}, {Elvis},
  {Fabbiano}, {Gilli}, {Marconi}, {Paolillo}, {Piedipalumbo}, {Salvestrini},
  {Signorini}, and {Vignali}}]{lusso2020}
{Lusso} E., {Risaliti} G., {Nardini} E., {Bargiacchi} G., {Benetti} M.,
  {Bisogni} S., {Capozziello} S., {Civano} F., et~al. (2020) {Quasars as
  standard candles. III. Validation of a new sample for cosmological studies}.
  \emph{\aap} 642:A150, \doi{10.1051/0004-6361/202038899}, \eprint{2008.08586}

\bibitem[{Macaulay et~al.(2017)Macaulay, Davis, Scovacricchi, Bacon, Collett,
  and Nichol}]{Macaulay:2016uwy}
Macaulay E., Davis T.~M., Scovacricchi D., Bacon D., Collett T.~E., Nichol
  R.~C. (2017) {The effects of velocities and lensing on moments of the Hubble
  diagram}. \emph{Mon Not Roy Astron Soc} 467(1):259--272,
  \doi{10.1093/mnras/stw3339}, \eprint{1607.03966}

\bibitem[{Macaulay et~al.(2020)}]{DES:2020kbf}
Macaulay E., et~al. (2020) {Weak Lensing of Type Ia Supernovae from the Dark
  Energy Survey}. \emph{Mon Not Roy Astron Soc} 496(3):4051--4059,
  \doi{10.1093/mnras/staa1852}, \eprint{2007.07956}

\bibitem[{{MacCrann} et~al.(2015){MacCrann}, {Zuntz}, {Bridle}, {Jain}, and
  {Becker}}]{MacCrann2015}
{MacCrann} N., {Zuntz} J., {Bridle} S., {Jain} B., {Becker} M.~R. (2015)
  {Cosmic discordance: are Planck CMB and CFHTLenS weak lensing measurements
  out of tune?} \emph{\mnras} 451(3):2877--2888, \doi{10.1093/mnras/stv1154},
  \eprint{1408.4742}

\bibitem[{{MacLeod} and {Hogan}(2008)}]{2008PhRvD..77d3512M}
{MacLeod} C.~L., {Hogan} C.~J. (2008) {Precision of Hubble constant derived
  using black hole binary absolute distances and statistical redshift
  information}. \emph{\prd} 77(4):043512, \doi{10.1103/PhysRevD.77.043512},
  \eprint{0712.0618}

\bibitem[{{Magira} et~al.(2000){Magira}, {Jing}, and
  {Suto}}]{2000ApJ...528...30M}
{Magira} H., {Jing} Y.~P., {Suto} Y. (2000) {Cosmological Redshift-Space
  Distortion on Clustering of High-Redshift Objects: Correction for Nonlinear
  Effects in the Power Spectrum and Tests with N-Body Simulations}. \emph{\apj}
  528(1):30--50, \doi{10.1086/308170}, \eprint{astro-ph/9907438}

\bibitem[{{Magris C.} et~al.(2003){Magris C.}, {Binette}, and {Bruzual
  A.}}]{magris2003}
{Magris C.} G., {Binette} L., {Bruzual A.} G. (2003) {ADEMIS: A Library of
  Evolutionary Models for Emission-Line Galaxies. I. Dust-free Models}.
  \emph{\apjs} 149(2):313--326, \doi{10.1086/378975}, \eprint{astro-ph/0311233}

\bibitem[{{Mancarella} et~al.(2021){Mancarella}, {Genoud-Prachex}, and
  {Maggiore}}]{2021arXiv211205728M}
{Mancarella} M., {Genoud-Prachex} E., {Maggiore} M. (2021) {Cosmology and
  modified gravitational wave propagation from binary black hole population
  models}. \emph{arXiv e-prints} arXiv:2112.05728, \eprint{2112.05728}

\bibitem[{{Mandel} et~al.(2019){Mandel}, {Farr}, and
  {Gair}}]{2019MNRAS.486.1086M}
{Mandel} I., {Farr} W.~M., {Gair} J.~R. (2019) {Extracting distribution
  parameters from multiple uncertain observations with selection biases}.
  \emph{\mnras} 486(1):1086--1093, \doi{10.1093/mnras/stz896},
  \eprint{1809.02063}

\bibitem[{Mantz et~al.(2015)}]{Mantz:2014paa}
Mantz A.~B., et~al. (2015) {Weighing the giants IV. Cosmology and neutrino
  mass}. \emph{Mon Not Roy Astron Soc} 446:2205--2225,
  \doi{10.1093/mnras/stu2096}, \eprint{1407.4516}

\bibitem[{{Mao} et~al.(2017{\natexlab{a}}){Mao}, {Berlind}, {Scherrer},
  {Neyrinck}, {Scoccimarro}, {Tinker}, {McBride}, and {Schneider}}]{Mao2017b}
{Mao} Q., {Berlind} A.~A., {Scherrer} R.~J., {Neyrinck} M.~C., {Scoccimarro}
  R., {Tinker} J.~L., {McBride} C.~K., {Schneider} D.~P. (2017{\natexlab{a}})
  {Cosmic Voids in the SDSS DR12 BOSS Galaxy Sample: The Alcock-Paczynski
  Test}. \emph{\apj} 835:160, \doi{10.3847/1538-4357/835/2/160},
  \eprint{1602.06306}

\bibitem[{{Mao} et~al.(2017{\natexlab{b}}){Mao}, {Berlind}, {Scherrer},
  {Neyrinck}, {Scoccimarro}, {Tinker}, {McBride}, {Schneider}, {Pan},
  {Bizyaev}, {Malanushenko}, and {Malanushenko}}]{Mao2017a}
{Mao} Q., {Berlind} A.~A., {Scherrer} R.~J., {Neyrinck} M.~C., {Scoccimarro}
  R., {Tinker} J.~L., {McBride} C.~K., {Schneider} D.~P., et~al.
  (2017{\natexlab{b}}) {A Cosmic Void Catalog of SDSS DR12 BOSS Galaxies}.
  \emph{\apj} 835(2):161, \doi{10.3847/1538-4357/835/2/161},
  \eprint{1602.02771}

\bibitem[{Mao et~al.(2008)Mao, Tegmark, McQuinn, Zaldarriaga, and
  Zahn}]{Mao:2008ug}
Mao Y., Tegmark M., McQuinn M., Zaldarriaga M., Zahn O. (2008) {How accurately
  can 21 cm tomography constrain cosmology?} \emph{Phys Rev D} 78:023529,
  \doi{10.1103/PhysRevD.78.023529}, \eprint{0802.1710}

\bibitem[{{Maraston} and {Str{\"o}mb{\"a}ck}(2011)}]{maraston2011}
{Maraston} C., {Str{\"o}mb{\"a}ck} G. (2011) {Stellar population models at high
  spectral resolution}. \emph{\mnras} 418(4):2785--2811,
  \doi{10.1111/j.1365-2966.2011.19738.x}, \eprint{1109.0543}

\bibitem[{{Mar{\'\i}n-Franch} et~al.(2009){Mar{\'\i}n-Franch}, {Aparicio},
  {Piotto}, {Rosenberg}, {Chaboyer}, {Sarajedini}, {Siegel}, {Anderson},
  {Bedin}, {Dotter}, {Hempel}, {King}, {Majewski}, {Milone}, {Paust}, and
  {Reid}}]{marinfranch2009}
{Mar{\'\i}n-Franch} A., {Aparicio} A., {Piotto} G., {Rosenberg} A., {Chaboyer}
  B., {Sarajedini} A., {Siegel} M., {Anderson} J., et~al. (2009) {The ACS
  Survey of Galactic Globular Clusters. VII. Relative Ages}. \emph{\apj}
  694(2):1498--1516, \doi{10.1088/0004-637X/694/2/1498}, \eprint{0812.4541}

\bibitem[{{Martin} et~al.(2005){Martin}, {Fanson}, {Schiminovich}, {Morrissey},
  {Friedman}, {Barlow}, {Conrow}, {Grange}, {Jelinsky}, {Milliard}, {Siegmund},
  {Bianchi}, {Byun}, {Donas}, {Forster}, {Heckman}, {Lee}, {Madore}, {Malina},
  {Neff}, {Rich}, {Small}, {Surber}, {Szalay}, {Welsh}, and
  {Wyder}}]{martin2005}
{Martin} D.~C., {Fanson} J., {Schiminovich} D., {Morrissey} P., {Friedman}
  P.~G., {Barlow} T.~A., {Conrow} T., {Grange} R., et~al. (2005) {The Galaxy
  Evolution Explorer: A Space Ultraviolet Survey Mission}. \emph{\apjl}
  619(1):L1--L6, \doi{10.1086/426387}, \eprint{astro-ph/0411302}

\bibitem[{{Martins} et~al.(2016){Martins}, {Martinelli}, {Calabrese}, and
  {Ramos}}]{martins2016}
{Martins} C.~J.~A.~P., {Martinelli} M., {Calabrese} E., {Ramos} M.~P.~L.~P.
  (2016) {Real-time cosmography with redshift derivatives}. \emph{\prd}
  94(4):043001, \doi{10.1103/PhysRevD.94.043001}, \eprint{1606.07261}

\bibitem[{{Martins} et~al.(2021){Martins}, {Alves}, {Esteves}, {Lapel}, and
  {Pereira}}]{martins2021}
{Martins} C.~J.~A.~P., {Alves} C.~S., {Esteves} J., {Lapel} A., {Pereira} B.~G.
  (2021) {Closing the cosmological loop with the redshift drift}. \emph{arXiv
  e-prints} arXiv:2110.12242, \eprint{2110.12242}

\bibitem[{{Martone} et~al.(2017){Martone}, {Izzo}, {Della Valle}, {Amati},
  {Longo}, and {G{\"o}tz}}]{Martone17}
{Martone} R., {Izzo} L., {Della Valle} M., {Amati} L., {Longo} G., {G{\"o}tz}
  D. (2017) {False outliers of the E$_{p,i}$ - E$_{iso}$ correlation?}
  \emph{\aap} 608:A52, \doi{10.1051/0004-6361/201730704}, \eprint{1708.03873}

\bibitem[{{Massara} et~al.(2015){Massara}, {Villaescusa-Navarro}, {Viel}, and
  {Sutter}}]{Massara2015}
{Massara} E., {Villaescusa-Navarro} F., {Viel} M., {Sutter} P.~M. (2015) {Voids
  in massive neutrino cosmologies}. \emph{\jcap} 11:018,
  \doi{10.1088/1475-7516/2015/11/018}, \eprint{1506.03088}

\bibitem[{{Massara} et~al.(2021){Massara}, {Villaescusa-Navarro}, {Ho},
  {Dalal}, and {Spergel}}]{Massara2021}
{Massara} E., {Villaescusa-Navarro} F., {Ho} S., {Dalal} N., {Spergel} D.~N.
  (2021) {Using the Marked Power Spectrum to Detect the Signature of Neutrinos
  in Large-Scale Structure}. \emph{\prl} 126(1):011301,
  \doi{10.1103/PhysRevLett.126.011301}, \eprint{2001.11024}

\bibitem[{{Massara} et~al.(2022){Massara}, {Villaescusa-Navarro}, {Hahn},
  {Abidi}, {Eickenberg}, {Ho}, {Lemos}, {Moradinezhad Dizgah}, and
  {R{\'e}galdo-Saint Blancard}}]{Massara2022}
{Massara} E., {Villaescusa-Navarro} F., {Hahn} C., {Abidi} M.~M., {Eickenberg}
  M., {Ho} S., {Lemos} P., {Moradinezhad Dizgah} A., et~al. (2022)
  {Cosmological Information in the Marked Power Spectrum of the Galaxy Field}.
  \emph{arXiv e-prints} arXiv:2206.01709, \eprint{2206.01709}

\bibitem[{{Mastrogiovanni} et~al.(2021){Mastrogiovanni}, {Leyde},
  {Karathanasis}, {Chassande-Mottin}, {Steer}, {Gair}, {Ghosh}, {Gray},
  {Mukherjee}, and {Rinaldi}}]{2021PhRvD.104f2009M}
{Mastrogiovanni} S., {Leyde} K., {Karathanasis} C., {Chassande-Mottin} E.,
  {Steer} D.~A., {Gair} J., {Ghosh} A., {Gray} R., et~al. (2021) {On the
  importance of source population models for gravitational-wave cosmology}.
  \emph{\prd} 104(6):062009, \doi{10.1103/PhysRevD.104.062009},
  \eprint{2103.14663}

\bibitem[{{Masui}(2013)}]{2013PhDT.......570M}
{Masui} K.~W. (2013) {Advancing precision cosmology with 21 cm intensity
  mapping}. PhD thesis, University of Toronto (Canada)

\bibitem[{Masui et~al.(2010)Masui, McDonald, and Pen}]{Masui:2010mp}
Masui K.~W., McDonald P., Pen U.-L. (2010) {Near term measurements with 21 cm
  intensity mapping: neutral hydrogen fraction and BAO at z\ensuremath{<}2}.
  \emph{Phys Rev D} 81:103527, \doi{10.1103/PhysRevD.81.103527},
  \eprint{1001.4811}

\bibitem[{Masui et~al.(2013)}]{Masui:2012zc}
Masui K.~W., et~al. (2013) {Measurement of 21 cm brightness fluctuations at z ~
  0.8 in cross-correlation}. \emph{Astrophys J} 763:L20,
  \doi{10.1088/2041-8205/763/1/L20}, \eprint{1208.0331}

\bibitem[{{Matshawule} et~al.(2021){Matshawule}, {Spinelli}, {Santos}, and
  {Ngobese}}]{Matshawule:2020fjz}
{Matshawule} S.~D., {Spinelli} M., {Santos} M.~G., {Ngobese} S. (2021) {H I
  intensity mapping with MeerKAT: primary beam effects on foreground cleaning}.
  \emph{\mnras} 506(4):5075--5092, \doi{10.1093/mnras/stab1688},
  \eprint{2011.10815}

\bibitem[{{Mawatari} et~al.(2016){Mawatari}, {Yamada}, {Fazio}, {Huang}, and
  {Ashby}}]{mawatari2016}
{Mawatari} K., {Yamada} T., {Fazio} G.~G., {Huang} J.-S., {Ashby} M. L.~N.
  (2016) {Possible identification of massive and evolved galaxies at z
  {\ensuremath{\gtrsim}} 5}. \emph{\pasj} 68(3):46, \doi{10.1093/pasj/psw041},
  \eprint{1603.08394}

\bibitem[{{McCully} et~al.(2014){McCully}, {Keeton}, {Wong}, and
  {Zabludoff}}]{McCully:2014}
{McCully} C., {Keeton} C.~R., {Wong} K.~C., {Zabludoff} A.~I. (2014) {A new
  hybrid framework to efficiently model lines of sight to gravitational
  lenses}. \emph{\mnras} 443(4):3631--3642, \doi{10.1093/mnras/stu1316},
  \eprint{1401.0197}

\bibitem[{{McCully} et~al.(2017){McCully}, {Keeton}, {Wong}, and
  {Zabludoff}}]{McCully:2017}
{McCully} C., {Keeton} C.~R., {Wong} K.~C., {Zabludoff} A.~I. (2017)
  {Quantifying Environmental and Line-of-sight Effects in Models of Strong
  Gravitational Lens Systems}. \emph{\apj} 836(1):141,
  \doi{10.3847/1538-4357/836/1/141}, \eprint{1601.05417}

\bibitem[{{McDermid} et~al.(2015){McDermid}, {Alatalo}, {Blitz}, {Bournaud},
  {Bureau}, {Cappellari}, {Crocker}, {Davies}, {Davis}, {de Zeeuw}, {Duc},
  {Emsellem}, {Khochfar}, {Krajnovi{\'c}}, {Kuntschner}, {Morganti}, {Naab},
  {Oosterloo}, {Sarzi}, {Scott}, {Serra}, {Weijmans}, and
  {Young}}]{mcdermid2015}
{McDermid} R.~M., {Alatalo} K., {Blitz} L., {Bournaud} F., {Bureau} M.,
  {Cappellari} M., {Crocker} A.~F., {Davies} R.~L., et~al. (2015) {The
  ATLAS$^{3D}$ Project - XXX. Star formation histories and stellar population
  scaling relations of early-type galaxies}. \emph{\mnras} 448(4):3484--3513,
  \doi{10.1093/mnras/stv105}, \eprint{1501.03723}

\bibitem[{McDonald and Seljak(2009)}]{McDonald:2009}
McDonald P., Seljak U. (2009) How to measure redshift-space distortions without
  sample variance. \emph{JCAP} 0910:007,
  \urlprefix\url{http://arxiv.org/abs/0810.0323}, \eprint{0810.0323}

\bibitem[{{Medvedev} et~al.(2020){Medvedev}, {Sazonov}, {Gilfanov}, {Burenin},
  {Khorunzhev}, {Meshcheryakov}, {Sunyaev}, {Bikmaev}, and
  {Irtuganov}}]{medvedev2020}
{Medvedev} P., {Sazonov} S., {Gilfanov} M., {Burenin} R., {Khorunzhev} G.,
  {Meshcheryakov} A., {Sunyaev} R., {Bikmaev} I., et~al. (2020) {SRG/eROSITA
  uncovers the most X-ray luminous quasar at z>6}. \emph{arXiv e-prints}
  arXiv:2007.04735, \eprint{2007.04735}

\bibitem[{{Mei} et~al.(2005{\natexlab{a}}){Mei}, {Blakeslee}, {Tonry}, {Jord{\'
  a}n}, {Peng}, {C{\^ o}t{\' e}}, {Ferrarese}, {Merritt}, {Milosavljevi{\' c}},
  and {West}}]{mei05iv}
{Mei} S., {Blakeslee} J.~P., {Tonry} J.~L., {Jord{\' a}n} A., {Peng} E.~W.,
  {C{\^ o}t{\' e}} P., {Ferrarese} L., {Merritt} D., et~al.
  (2005{\natexlab{a}}) {The ACS Virgo Cluster Survey. IV. Data Reduction
  Procedures for Surface Brightness Fluctuation Measurements with the Advanced
  Camera for Surveys}. \emph{\apjs} 156:113--125, \doi{10.1086/426544}

\bibitem[{{Mei} et~al.(2005{\natexlab{b}}){Mei}, {Blakeslee}, {Tonry},
  {Jord{\'a}n}, {Peng}, {C{\^o}t{\'e}}, {Ferrarese}, {West}, {Merritt}, and
  {Milosavljevi{\'c}}}]{mei05v}
{Mei} S., {Blakeslee} J.~P., {Tonry} J.~L., {Jord{\'a}n} A., {Peng} E.~W.,
  {C{\^o}t{\'e}} P., {Ferrarese} L., {West} M.~J., et~al. (2005{\natexlab{b}})
  {The Advanced Camera for Surveys Virgo Cluster Survey. V. Surface Brightness
  Fluctuation Calibration for Giant and Dwarf Early-Type Galaxies}. \emph{\apj}
  625(1):121--129, \doi{10.1086/429554}

\bibitem[{{Mei} et~al.(2007){Mei}, {Blakeslee}, {C{\^o}t{\'e}}, {Tonry},
  {West}, {Ferrarese}, {Jord{\'a}n}, {Peng}, {Anthony}, and {Merritt}}]{mei07}
{Mei} S., {Blakeslee} J.~P., {C{\^o}t{\'e}} P., {Tonry} J.~L., {West} M.~J.,
  {Ferrarese} L., {Jord{\'a}n} A., {Peng} E.~W., et~al. (2007) {The ACS Virgo
  Cluster Survey. XIII. SBF Distance Catalog and the Three-dimensional
  Structure of the Virgo Cluster}. \emph{\apj} 655(1):144--162,
  \doi{10.1086/509598}, \eprint{astro-ph/0702510}

\bibitem[{{Melchior} et~al.(2014){Melchior}, {Sutter}, {Sheldon}, {Krause}, and
  {Wandelt}}]{Melchior2014}
{Melchior} P., {Sutter} P.~M., {Sheldon} E.~S., {Krause} E., {Wandelt} B.~D.
  (2014) {First measurement of gravitational lensing by cosmic voids in SDSS}.
  \emph{\mnras} 440:2922--2927, \doi{10.1093/mnras/stu456}, \eprint{1309.2045}

\bibitem[{{Melia}(2019)}]{melia2019}
{Melia} F. (2019) {Cosmological test using the Hubble diagram of high-z
  quasars}. \emph{\mnras} 489(1):517--523, \doi{10.1093/mnras/stz2120},
  \eprint{1907.13127}

\bibitem[{{Meneghetti} et~al.(2007){Meneghetti}, {Argazzi}, {Pace},
  {Moscardini}, {Dolag}, {Bartelmann}, {Li}, and {Oguri}}]{Meneghetti:2007}
{Meneghetti} M., {Argazzi} R., {Pace} F., {Moscardini} L., {Dolag} K.,
  {Bartelmann} M., {Li} G., {Oguri} M. (2007) {Arc sensitivity to cluster
  ellipticity, asymmetries, and substructures}. \emph{\aap} 461(1):25--38,
  \doi{10.1051/0004-6361:20065722}, \eprint{astro-ph/0606006}

\bibitem[{{Meneghetti} et~al.(2010){Meneghetti}, {Fedeli}, {Pace},
  {Gottl{\"o}ber}, and {Yepes}}]{Meneghetti:2010a}
{Meneghetti} M., {Fedeli} C., {Pace} F., {Gottl{\"o}ber} S., {Yepes} G. (2010)
  {Strong lensing in the MARENOSTRUM UNIVERSE. I. Biases in the cluster lens
  population}. \emph{\aap} 519:A90, \doi{10.1051/0004-6361/201014098},
  \eprint{1003.4544}

\bibitem[{{Menzel} et~al.(2016){Menzel}, {Merloni}, {Georgakakis}, {Salvato},
  {Aubourg}, {Brandt}, {Brusa}, {Buchner}, {Dwelly}, {Nandra}, {P{\^a}ris},
  {Petitjean}, and {Schwope}}]{menzel2016}
{Menzel} M.-L., {Merloni} A., {Georgakakis} A., {Salvato} M., {Aubourg} E.,
  {Brandt} W.~N., {Brusa} M., {Buchner} J., et~al. (2016) {A spectroscopic
  survey of X-ray-selected AGNs in the northern XMM-XXL field}. \emph{\mnras}
  457:110--132, \doi{10.1093/mnras/stv2749}, \eprint{1511.07870}

\bibitem[{{Merlin} et~al.(2018){Merlin}, {Fontana}, {Castellano}, {Santini},
  {Torelli}, {Boutsia}, {Wang}, {Grazian}, {Pentericci}, {Schreiber}, {Ciesla},
  {McLure}, {Derriere}, {Dunlop}, and {Elbaz}}]{merlin2018}
{Merlin} E., {Fontana} A., {Castellano} M., {Santini} P., {Torelli} M.,
  {Boutsia} K., {Wang} T., {Grazian} A., et~al. (2018) {Chasing passive
  galaxies in the early Universe: a critical analysis in CANDELS GOODS-South}.
  \emph{\mnras} 473(2):2098--2123, \doi{10.1093/mnras/stx2385},
  \eprint{1709.00429}

\bibitem[{{Merlin} et~al.(2019){Merlin}, {Fortuni}, {Torelli}, {Santini},
  {Castellano}, {Fontana}, {Grazian}, {Pentericci}, {Pilo}, and
  {Schmidt}}]{merlin2019}
{Merlin} E., {Fortuni} F., {Torelli} M., {Santini} P., {Castellano} M.,
  {Fontana} A., {Grazian} A., {Pentericci} L., et~al. (2019) {Red and dead
  CANDELS: massive passive galaxies at the dawn of the Universe}. \emph{\mnras}
  490(3):3309--3328, \doi{10.1093/mnras/stz2615}, \eprint{1909.07996}

\bibitem[{{Merloni} et~al.(2012){Merloni}, {Predehl}, {Becker},
  {B{\"o}hringer}, {Boller}, {Brunner}, {Brusa}, {Dennerl}, {Freyberg},
  {Friedrich}, {Georgakakis}, {Haberl}, {Hasinger}, {Meidinger}, {Mohr},
  {Nandra}, {Rau}, {Reiprich}, {Robrade}, {Salvato}, {Santangelo}, {Sasaki},
  {Schwope}, {Wilms}, and {German eROSITA Consortium}}]{2012arXiv1209.3114M}
{Merloni} A., {Predehl} P., {Becker} W., {B{\"o}hringer} H., {Boller} T.,
  {Brunner} H., {Brusa} M., {Dennerl} K., et~al. (2012) {eROSITA Science Book:
  Mapping the Structure of the Energetic Universe}. \emph{arXiv e-prints}
  arXiv:1209.3114, \eprint{1209.3114}

\bibitem[{{Merloni} et~al.(2014){Merloni}, {Bongiorno}, {Brusa}, {Iwasawa},
  {Mainieri}, {Magnelli}, {Salvato}, {Berta}, {Cappelluti}, {Comastri},
  {Fiore}, {Gilli}, {Koekemoer}, {Le Floc'h}, {Lusso}, {Lutz}, {Miyaji},
  {Pozzi}, {Riguccini}, {Rosario}, {Silverman}, {Symeonidis}, {Treister},
  {Vignali}, and {Zamorani}}]{merloni2014}
{Merloni} A., {Bongiorno} A., {Brusa} M., {Iwasawa} K., {Mainieri} V.,
  {Magnelli} B., {Salvato} M., {Berta} S., et~al. (2014) {The incidence of
  obscuration in active galactic nuclei}. \emph{\mnras} 437(4):3550--3567,
  \doi{10.1093/mnras/stt2149}, \eprint{1311.1305}

\bibitem[{Mertens et~al.(2018)Mertens, Ghosh, and Koopmans}]{Mertens:2017gxw}
Mertens F.~G., Ghosh A., Koopmans L. V.~E. (2018) {Statistical 21-cm Signal
  Separation via Gaussian Process Regression Analysis}. \emph{Mon Not Roy
  Astron Soc} 478(3):3640--3652, \doi{10.1093/mnras/sty1207},
  \eprint{1711.10834}

\bibitem[{{Messenger} and {Read}(2012)}]{2012PhRvL.108i1101M}
{Messenger} C., {Read} J. (2012) {Measuring a Cosmological Distance-Redshift
  Relationship Using Only Gravitational Wave Observations of Binary Neutron
  Star Coalescences}. \emph{\prl} 108(9):091101,
  \doi{10.1103/PhysRevLett.108.091101}, \eprint{1107.5725}

\bibitem[{{M{\'e}sz{\'a}ros}(2002)}]{Meszaros02}
{M{\'e}sz{\'a}ros} P. (2002) {Theories of Gamma-Ray Bursts}. \emph{\araa}
  40:137--169, \doi{10.1146/annurev.astro.40.060401.093821},
  \eprint{astro-ph/0111170}

\bibitem[{{Metcalf} et~al.(2019){Metcalf}, {Meneghetti}, {Avestruz},
  {Bellagamba}, {Bom}, {Bertin}, {Cabanac}, {Courbin}, {Davies},
  {Decenci{\`e}re}, {Flamary}, {Gavazzi}, {Geiger}, {Hartley},
  {Huertas-Company}, {Jackson}, {Jacobs}, {Jullo}, {Kneib}, {Koopmans},
  {Lanusse}, {Li}, {Ma}, {Makler}, {Li}, {Lightman}, {Petrillo}, {Serjeant},
  {Sch{\"a}fer}, {Sonnenfeld}, {Tagore}, {Tortora}, {Tuccillo},
  {Valent{\'\i}n}, {Velasco-Forero}, {Verdoes Kleijn}, and
  {Vernardos}}]{Metcalf2019}
{Metcalf} R.~B., {Meneghetti} M., {Avestruz} C., {Bellagamba} F., {Bom} C.~R.,
  {Bertin} E., {Cabanac} R., {Courbin} F., et~al. (2019) {The strong
  gravitational lens finding challenge}. \emph{\aap} 625:A119,
  \doi{10.1051/0004-6361/201832797}, \eprint{1802.03609}

\bibitem[{{Micheletti} et~al.(2014){Micheletti}, {Iovino}, {Hawken}, {Granett},
  {Bolzonella}, {Cappi}, {Guzzo}, {Abbas}, {Adami}, {Arnouts}, {Bel},
  {Bottini}, {Branchini}, {Coupon}, {Cucciati}, {Davidzon}, {De Lucia}, {de la
  Torre}, {Fritz}, {Franzetti}, {Fumana}, {Garilli}, {Ilbert}, {Krywult}, {Le
  Brun}, {Le F{\`e}vre}, and et~al.}]{Micheletti2014}
{Micheletti} D., {Iovino} A., {Hawken} A.~J., {Granett} B.~R., {Bolzonella} M.,
  {Cappi} A., {Guzzo} L., {Abbas} U., et~al. (2014) {The VIMOS Public
  Extragalactic Redshift Survey. Searching for cosmic voids}. \emph{\aap}
  570:A106, \doi{10.1051/0004-6361/201424107}, \eprint{1407.2969}

\bibitem[{{Mieske} et~al.(2005){Mieske}, {Hilker}, and {Infante}}]{mieske05}
{Mieske} S., {Hilker} M., {Infante} L. (2005) {The distance to Hydra and
  Centaurus from surface brightness fluctuations: Consequences for the Great
  Attractor model}. \emph{\aap} 438(1):103--119,
  \doi{10.1051/0004-6361:20041583}, \eprint{astro-ph/0503647}

\bibitem[{{Mieske} et~al.(2006){Mieske}, {Hilker}, and {Infante}}]{mieske06}
{Mieske} S., {Hilker} M., {Infante} L. (2006) {An I-band calibration of surface
  brightness fluctuation measurements at blue colours}. \emph{\aap}
  458(3):1013--1023, \doi{10.1051/0004-6361:20054685}

\bibitem[{{Mignoli} et~al.(2009){Mignoli}, {Zamorani}, {Scodeggio}, {Cimatti},
  {Halliday}, {Lilly}, {Pozzetti}, {Vergani}, {Carollo}, {Contini}, {Le
  F{\'e}vre}, and et~al.}]{mignoli2009}
{Mignoli} M., {Zamorani} G., {Scodeggio} M., {Cimatti} A., {Halliday} C.,
  {Lilly} S.~J., {Pozzetti} L., {Vergani} D., et~al. (2009) {The zCOSMOS
  redshift survey: the three-dimensional classification cube and bimodality in
  galaxy physical properties}. \emph{\aap} 493(1):39--49,
  \doi{10.1051/0004-6361:200810520}, \eprint{0810.2245}

\bibitem[{{Millon} et~al.(2020{\natexlab{a}}){Millon}, {Courbin}, {Bonvin},
  {Buckley-Geer}, {Fassnacht}, {Frieman}, {Marshall}, {Suyu}, {Treu},
  {Anguita}, {Motta}, {Agnello}, {Chan}, {Chao}, {Chijani}, {Gilman},
  {Gilmore}, {Lemon}, {Lucey}, {Melo}, {Paic}, {Rojas}, {Sluse}, {Williams},
  {Hempel}, {Kim}, {Lachaume}, and {Rabus}}]{millon2020c}
{Millon} M., {Courbin} F., {Bonvin} V., {Buckley-Geer} E., {Fassnacht} C.~D.,
  {Frieman} J., {Marshall} P.~J., {Suyu} S.~H., et~al. (2020{\natexlab{a}})
  {TDCOSMO. II. Six new time delays in lensed quasars from high-cadence
  monitoring at the MPIA 2.2 m telescope}. \emph{\aap} 642:A193,
  \doi{10.1051/0004-6361/202038698}, \eprint{2006.10066}

\bibitem[{{Millon} et~al.(2020{\natexlab{b}}){Millon}, {Galan}, {Courbin},
  {Treu}, {Suyu}, {Ding}, {Birrer}, {Chen}, {Shajib}, {Sluse}, {Wong},
  {Agnello}, {Auger}, {Buckley-Geer}, {Chan}, {Collett}, {Fassnacht},
  {Hilbert}, {Koopmans}, {Motta}, {Mukherjee}, {Rusu}, {Sonnenfeld},
  {Spiniello}, and {Van de Vyvere}}]{Millon2020a}
{Millon} M., {Galan} A., {Courbin} F., {Treu} T., {Suyu} S.~H., {Ding} X.,
  {Birrer} S., {Chen} G.~C.~F., et~al. (2020{\natexlab{b}}) {TDCOSMO. I. An
  exploration of systematic uncertainties in the inference of H$_{0}$ from
  time-delay cosmography}. \emph{\aap} 639:A101,
  \doi{10.1051/0004-6361/201937351}, \eprint{1912.08027}

\bibitem[{{Mingo} et~al.(2016){Mingo}, {Watson}, {Rosen}, {Hardcastle}, {Ruiz},
  {Blain}, {Carrera}, {Mateos}, {Pineau}, and {Stewart}}]{mingo2016}
{Mingo} B., {Watson} M.~G., {Rosen} S.~R., {Hardcastle} M.~J., {Ruiz} A.,
  {Blain} A., {Carrera} F.~J., {Mateos} S., et~al. (2016) {The MIXR sample: AGN
  activity versus star formation across the cross-correlation of WISE, 3XMM,
  and FIRST/NVSS}. \emph{\mnras} 462:2631--2667, \doi{10.1093/mnras/stw1826},
  \eprint{1607.06471}

\bibitem[{{Momcheva} et~al.(2006){Momcheva}, {Williams}, {Keeton}, and
  {Zabludoff}}]{Momcheva2006}
{Momcheva} I., {Williams} K., {Keeton} C., {Zabludoff} A. (2006) {A
  Spectroscopic Study of the Environments of Gravitational Lens Galaxies}.
  \emph{\apj} 641(1):169--189, \doi{10.1086/500382}, \eprint{astro-ph/0511594}

\bibitem[{{Montiel} et~al.(2014){Montiel}, {Lazkoz}, {Sendra},
  {Escamilla-Rivera}, and {Salzano}}]{montiel2014}
{Montiel} A., {Lazkoz} R., {Sendra} I., {Escamilla-Rivera} C., {Salzano} V.
  (2014) {Nonparametric reconstruction of the cosmic expansion with local
  regression smoothing and simulation extrapolation}. \emph{\prd} 89(4):043007,
  \doi{10.1103/PhysRevD.89.043007}, \eprint{1401.4188}

\bibitem[{{Montiel} et~al.(2021){Montiel}, {Cabrera}, and
  {Hidalgo}}]{montiel21}
{Montiel} A., {Cabrera} J.~I., {Hidalgo} J.~C. (2021) {Improving sampling and
  calibration of gamma-ray bursts as distance indicators}. \emph{\mnras}
  501(3):3515--3526, \doi{10.1093/mnras/staa3926}, \eprint{2003.03387}

\bibitem[{{Mooley} et~al.(2018){Mooley}, {Deller}, {Gottlieb}, {Nakar},
  {Hallinan}, {Bourke}, {Frail}, {Horesh}, {Corsi}, and
  {Hotokezaka}}]{2018Natur.561..355M}
{Mooley} K.~P., {Deller} A.~T., {Gottlieb} O., {Nakar} E., {Hallinan} G.,
  {Bourke} S., {Frail} D.~A., {Horesh} A., et~al. (2018) {Superluminal motion
  of a relativistic jet in the neutron-star merger GW170817}. \emph{\nat}
  561(7723):355--359, \doi{10.1038/s41586-018-0486-3}, \eprint{1806.09693}

\bibitem[{{Moresco}(2015)}]{moresco2015}
{Moresco} M. (2015) {Raising the bar: new constraints on the Hubble parameter
  with cosmic chronometers at z {\textasciitilde} 2}. \emph{\mnras}
  450:L16--L20, \doi{10.1093/mnrasl/slv037}, \eprint{1503.01116}

\bibitem[{{Moresco} and {Marulli}(2017)}]{moresco2017}
{Moresco} M., {Marulli} F. (2017) {Cosmological constraints from a joint
  analysis of cosmic growth and expansion}. \emph{\mnras} 471(1):L82--L86,
  \doi{10.1093/mnrasl/slx112}, \eprint{1705.07903}

\bibitem[{{Moresco} et~al.(2011){Moresco}, {Jimenez}, {Cimatti}, and
  {Pozzetti}}]{moresco2011}
{Moresco} M., {Jimenez} R., {Cimatti} A., {Pozzetti} L. (2011) {Constraining
  the expansion rate of the Universe using low-redshift ellipticals as cosmic
  chronometers}. \emph{\jcap} 2011(3):045, \doi{10.1088/1475-7516/2011/03/045},
  \eprint{1010.0831}

\bibitem[{{Moresco} et~al.(2012{\natexlab{a}}){Moresco}, {Cimatti}, {Jimenez},
  {Pozzetti}, {Zamorani}, {Bolzonella}, {Dunlop}, {Lamareille}, {Mignoli},
  {Pearce}, {Rosati}, {Stern}, {Verde}, and et~al.}]{moresco2012}
{Moresco} M., {Cimatti} A., {Jimenez} R., {Pozzetti} L., {Zamorani} G.,
  {Bolzonella} M., {Dunlop} J., {Lamareille} F., et~al. (2012{\natexlab{a}})
  {Improved constraints on the expansion rate of the Universe up to z \~{} 1.1
  from the spectroscopic evolution of cosmic chronometers}. \emph{\jcap} 8:006,
  \doi{10.1088/1475-7516/2012/08/006}, \eprint{1201.3609}

\bibitem[{{Moresco} et~al.(2012{\natexlab{b}}){Moresco}, {Verde}, {Pozzetti},
  {Jimenez}, and {Cimatti}}]{moresco2012b}
{Moresco} M., {Verde} L., {Pozzetti} L., {Jimenez} R., {Cimatti} A.
  (2012{\natexlab{b}}) {New constraints on cosmological parameters and neutrino
  properties using the expansion rate of the Universe to z \~{} 1.75}.
  \emph{\jcap} 7:053, \doi{10.1088/1475-7516/2012/07/053}, \eprint{1201.6658}

\bibitem[{{Moresco} et~al.(2013){Moresco}, {Pozzetti}, {Cimatti}, {Zamorani},
  {Bolzonella}, {Lamareille}, {Mignoli}, {Zucca}, {Lilly}, {Carollo},
  {Contini}, {Kneib}, {Le F{\`e}vre}, {Mainieri}, {Renzini}, and
  et~al.}]{Moresco2013}
{Moresco} M., {Pozzetti} L., {Cimatti} A., {Zamorani} G., {Bolzonella} M.,
  {Lamareille} F., {Mignoli} M., {Zucca} E., et~al. (2013) {Spot the
  difference. Impact of different selection criteria on observed properties of
  passive galaxies in zCOSMOS-20k sample}. \emph{\aap} 558:A61,
  \doi{10.1051/0004-6361/201321797}, \eprint{1305.1308}

\bibitem[{{Moresco} et~al.(2016{\natexlab{a}}){Moresco}, {Jimenez}, {Verde},
  {Cimatti}, {Pozzetti}, {Maraston}, and {Thomas}}]{moresco2016b}
{Moresco} M., {Jimenez} R., {Verde} L., {Cimatti} A., {Pozzetti} L., {Maraston}
  C., {Thomas} D. (2016{\natexlab{a}}) {Constraining the time evolution of dark
  energy, curvature and neutrino properties with cosmic chronometers}.
  \emph{\jcap} 12:039, \doi{10.1088/1475-7516/2016/12/039}, \eprint{1604.00183}

\bibitem[{{Moresco} et~al.(2016{\natexlab{b}}){Moresco}, {Pozzetti}, {Cimatti},
  {Jimenez}, {Maraston}, {Verde}, {Thomas}, {Citro}, {Tojeiro}, and
  {Wilkinson}}]{moresco2016}
{Moresco} M., {Pozzetti} L., {Cimatti} A., {Jimenez} R., {Maraston} C., {Verde}
  L., {Thomas} D., {Citro} A., et~al. (2016{\natexlab{b}}) {A 6\% measurement
  of the Hubble parameter at z\~{}0.45: direct evidence of the epoch of cosmic
  re-acceleration}. \emph{\jcap} 5:014, \doi{10.1088/1475-7516/2016/05/014},
  \eprint{1601.01701}

\bibitem[{{Moresco} et~al.(2018){Moresco}, {Jimenez}, {Verde}, {Pozzetti},
  {Cimatti}, and {Citro}}]{Moresco2018}
{Moresco} M., {Jimenez} R., {Verde} L., {Pozzetti} L., {Cimatti} A., {Citro} A.
  (2018) {Setting the Stage for Cosmic Chronometers. I. Assessing the Impact of
  Young Stellar Populations on Hubble Parameter Measurements}. \emph{\apj}
  868(2):84, \doi{10.3847/1538-4357/aae829}, \eprint{1804.05864}

\bibitem[{{Moresco} et~al.(2020){Moresco}, {Jimenez}, {Verde}, {Cimatti}, and
  {Pozzetti}}]{Moresco2020}
{Moresco} M., {Jimenez} R., {Verde} L., {Cimatti} A., {Pozzetti} L. (2020)
  {Setting the Stage for Cosmic Chronometers. II. Impact of Stellar Population
  Synthesis Models Systematics and Full Covariance Matrix}. \emph{\apj}
  898(1):82, \doi{10.3847/1538-4357/ab9eb0}, \eprint{2003.07362}

\bibitem[{{Morishita} et~al.(2019){Morishita}, {Abramson}, {Treu}, {Brammer},
  {Jones}, {Kelly}, {Stiavelli}, {Trenti}, {Vulcani}, and
  {Wang}}]{morishita2019}
{Morishita} T., {Abramson} L.~E., {Treu} T., {Brammer} G.~B., {Jones} T.,
  {Kelly} P., {Stiavelli} M., {Trenti} M., et~al. (2019) {Massive Dead Galaxies
  at z$\sim$2 with HST Grism Spectroscopy. I. Star Formation Histories and
  Metallicity Enrichment}. \emph{\apj} 877(2):141,
  \doi{10.3847/1538-4357/ab1d53}, \eprint{1812.06980}

\bibitem[{{Mortlock} et~al.(2019){Mortlock}, {Feeney}, {Peiris}, {Williamson},
  and {Nissanke}}]{2019PhRvD.100j3523M}
{Mortlock} D.~J., {Feeney} S.~M., {Peiris} H.~V., {Williamson} A.~R.,
  {Nissanke} S.~M. (2019) {Unbiased Hubble constant estimation from binary
  neutron star mergers}. \emph{\prd} 100(10):103523,
  \doi{10.1103/PhysRevD.100.103523}, \eprint{1811.11723}

\bibitem[{{Mould} and {Sakai}(2009)}]{mould09}
{Mould} J., {Sakai} S. (2009) {The Extragalactic Distance Scale Without
  Cepheids. II. Surface Brightness Fluctuations}. \emph{\apj}
  694(2):1331--1334, \doi{10.1088/0004-637X/694/2/1331}

\bibitem[{{Muccino} et~al.(2021){Muccino}, {Izzo}, {Luongo}, {Boshkayev},
  {Amati}, {Della Valle}, {Pisani}, and {Zaninoni}}]{muccino21}
{Muccino} M., {Izzo} L., {Luongo} O., {Boshkayev} K., {Amati} L., {Della Valle}
  M., {Pisani} G.~B., {Zaninoni} E. (2021) {Tracing Dark Energy History with
  Gamma-Ray Bursts}. \emph{\apj} 908(2):181, \doi{10.3847/1538-4357/abd254},
  \eprint{2012.03392}

\bibitem[{{Mueller} et~al.(2021){Mueller}, {Rezaie}, {Percival}, {Ross},
  {Ruggeri}, {Seo}, {Gil-Mar{\i}n}, {Bautista}, {Brownstein}, {Dawson}, {de la
  Macorra}, {Palanque-Delabrouille}, {Rossi}, {Schneider}, and
  {Yeche}}]{Mueller:2021tqa}
{Mueller} E.-M., {Rezaie} M., {Percival} W.~J., {Ross} A.~J., {Ruggeri} R.,
  {Seo} H.-J., {Gil-Mar{\i}n} H., {Bautista} J., et~al. (2021) {The clustering
  of galaxies in the completed SDSS-IV extended Baryon Oscillation
  Spectroscopic Survey: Primordial non-Gaussianity in Fourier Space}.
  \emph{arXiv e-prints} arXiv:2106.13725, \eprint{2106.13725}

\bibitem[{{Mukherjee} and {Mukherjee}(2021)}]{mukherjee2021}
{Mukherjee} P., {Mukherjee} A. (2021) {Assessment of the cosmic distance
  duality relation using Gaussian process}. \emph{\mnras} 504(3):3938--3946,
  \doi{10.1093/mnras/stab1054}, \eprint{2104.06066}

\bibitem[{{Mukherjee} and {Wandelt}(2018)}]{Mukherjee:2018ebj}
{Mukherjee} S., {Wandelt} B.~D. (2018) {Beyond the classical distance-redshift
  test: cross-correlating redshift-free standard candles and sirens with
  redshift surveys}. \emph{arXiv e-prints} arXiv:1808.06615,
  \eprint{1808.06615}

\bibitem[{{Mukherjee} et~al.(2021){Mukherjee}, {Wandelt}, {Nissanke}, and
  {Silvestri}}]{2021PhRvD.103d3520M}
{Mukherjee} S., {Wandelt} B.~D., {Nissanke} S.~M., {Silvestri} A. (2021)
  {Accurate precision cosmology with redshift unknown gravitational wave
  sources}. \emph{\prd} 103(4):043520, \doi{10.1103/PhysRevD.103.043520},
  \eprint{2007.02943}

\bibitem[{{Muzzin} et~al.(2013){Muzzin}, {Marchesini}, {Stefanon}, {Franx},
  {McCracken}, {Milvang-Jensen}, {Dunlop}, {Fynbo}, {Brammer}, {Labb{\'e}}, and
  {van Dokkum}}]{muzzin2013}
{Muzzin} A., {Marchesini} D., {Stefanon} M., {Franx} M., {McCracken} H.~J.,
  {Milvang-Jensen} B., {Dunlop} J.~S., {Fynbo} J.~P.~U., et~al. (2013) {The
  Evolution of the Stellar Mass Functions of Star-forming and Quiescent
  Galaxies to z = 4 from the COSMOS/UltraVISTA Survey}. \emph{\apj} 777(1):18,
  \doi{10.1088/0004-637X/777/1/18}, \eprint{1303.4409}

\bibitem[{{Nadathur} and {Crittenden}(2016)}]{Nadathur2016}
{Nadathur} S., {Crittenden} R. (2016) {A Detection of the Integrated
  Sachs-Wolfe Imprint of Cosmic Superstructures Using a Matched-filter
  Approach}. \emph{\apjl} 830(1):L19, \doi{10.3847/2041-8205/830/1/L19},
  \eprint{1608.08638}

\bibitem[{{Nadathur} et~al.(2020){Nadathur}, {Woodfinden}, {Percival},
  {Aubert}, {Bautista}, {Dawson}, {Escoffier}, {Fromenteau}, {Gil-Mar{\'\i}n},
  {Rich}, {Ross}, {Rossi}, {Maga{\~n}a}, {Brownstein}, and
  {Schneider}}]{Nadathur2020}
{Nadathur} S., {Woodfinden} A., {Percival} W.~J., {Aubert} M., {Bautista} J.,
  {Dawson} K., {Escoffier} S., {Fromenteau} S., et~al. (2020) {The completed
  SDSS-IV extended baryon oscillation spectroscopic survey: geometry and growth
  from the anisotropic void-galaxy correlation function in the luminous red
  galaxy sample}. \emph{\mnras} 499(3):4140--4157,
  \doi{10.1093/mnras/staa3074}, \eprint{2008.06060}

\bibitem[{{Nakar} and {Piran}(2005)}]{Nakar05}
{Nakar} E., {Piran} T. (2005) {Outliers to the peak energy-isotropic energy
  relation in gamma-ray bursts}. \emph{\mnras} 360(1):L73--L76,
  \doi{10.1111/j.1745-3933.2005.00049.x}, \eprint{astro-ph/0412232}

\bibitem[{{Nardini} et~al.(2019){Nardini}, {Lusso}, {Risaliti}, {Bisogni},
  {Civano}, {Elvis}, {Fabbiano}, {Gilli}, {Marconi}, {Salvestrini}, and
  {Vignali}}]{nardini2019}
{Nardini} E., {Lusso} E., {Risaliti} G., {Bisogni} S., {Civano} F., {Elvis} M.,
  {Fabbiano} G., {Gilli} R., et~al. (2019) {The most luminous blue quasars at
  3.0 \&lt; z \&lt; 3.3. I. A tale of two X-ray populations}. \emph{\aap}
  632:A109, \doi{10.1051/0004-6361/201936911}, \eprint{1910.04604}

\bibitem[{{Nava} et~al.(2011){Nava}, {Ghirlanda}, {Ghisellini}, and
  {Celotti}}]{Nava11}
{Nava} L., {Ghirlanda} G., {Ghisellini} G., {Celotti} A. (2011) {Fermi/GBM and
  BATSE gamma-ray bursts: comparison of the spectral properties}. \emph{\mnras}
  415(4):3153--3162, \doi{10.1111/j.1365-2966.2011.18928.x}, \eprint{1012.3968}

\bibitem[{{Navarro} et~al.(1997){Navarro}, {Frenk}, and {White}}]{NFW}
{Navarro} J.~F., {Frenk} C.~S., {White} S. D.~M. (1997) {A Universal Density
  Profile from Hierarchical Clustering}. \emph{\apj} 490(2):493--508,
  \doi{10.1086/304888}, \eprint{astro-ph/9611107}

\bibitem[{{Nayyeri} et~al.(2014){Nayyeri}, {Mobasher}, {Hemmati}, {De Barros},
  {Ferguson}, {Wiklind}, {Dahlen}, {Dickinson}, {Giavalisco}, {Fontana},
  {Ashby}, {Barro}, {Guo}, {Hathi}, {Kassin}, {Koekemoer}, {Willner}, {Dunlop},
  {Paris}, and {Targett}}]{nayyeri2014}
{Nayyeri} H., {Mobasher} B., {Hemmati} S., {De Barros} S., {Ferguson} H.~C.,
  {Wiklind} T., {Dahlen} T., {Dickinson} M., et~al. (2014) {A Study of Massive
  and Evolved Galaxies at High Redshift}. \emph{\apj} 794(1):68,
  \doi{10.1088/0004-637X/794/1/68}, \eprint{1408.3684}

\bibitem[{Newburgh et~al.(2016)}]{Newburgh:2016mwi}
Newburgh L., et~al. (2016) {HIRAX: A Probe of Dark Energy and Radio
  Transients}. \emph{Proc SPIE Int Soc Opt Eng} 9906:99065X,
  \doi{10.1117/12.2234286}, \eprint{1607.02059}

\bibitem[{Newburgh et~al.(2014)}]{Newburgh:2014toa}
Newburgh L.~B., et~al. (2014) {Calibrating CHIME, A New Radio Interferometer to
  Probe Dark Energy}. \emph{Proc SPIE Int Soc Opt Eng} 9145:4V,
  \doi{10.1117/12.2056962}, \eprint{1406.2267}

\bibitem[{{Neyrinck}(2008)}]{Neyrinck2008}
{Neyrinck} M.~C. (2008) {ZOBOV: a parameter-free void-finding algorithm}.
  \emph{\mnras} 386:2101--2109, \doi{10.1111/j.1365-2966.2008.13180.x},
  \eprint{0712.3049}

\bibitem[{{Nguyen} et~al.(2020){Nguyen}, {den Brok}, {Seth}, {Davis}, {Greene},
  {Cappellari}, {Jensen}, {Thater}, {Iguchi}, {Imanishi}, {Izumi}, {Nyland},
  {Neumayer}, {Nakanishi}, {Nguyen}, {Tsukui}, {Bureau}, {Onishi}, {Nguyen},
  and {Le}}]{nguyen20}
{Nguyen} D.~D., {den Brok} M., {Seth} A.~C., {Davis} T.~A., {Greene} J.~E.,
  {Cappellari} M., {Jensen} J.~B., {Thater} S., et~al. (2020) {The
  MBHBM$_{{\ensuremath{\star}}}$ Project. I. Measurement of the Central Black
  Hole Mass in Spiral Galaxy NGC 3504 Using Molecular Gas Kinematics}.
  \emph{\apj} 892(1):68, \doi{10.3847/1538-4357/ab77aa}, \eprint{1902.03813}

\bibitem[{{Nielsen} et~al.(2016){Nielsen}, {Guffanti}, and
  {Sarkar}}]{nielsen2016}
{Nielsen} J.~T., {Guffanti} A., {Sarkar} S. (2016) {Marginal evidence for
  cosmic acceleration from Type Ia supernovae}. \emph{Scientific Reports}
  6:35596, \doi{10.1038/srep35596}, \eprint{1506.01354}

\bibitem[{{Nunes} et~al.(2016){Nunes}, {Pan}, and {Saridakis}}]{nunes2016}
{Nunes} R.~C., {Pan} S., {Saridakis} E.~N. (2016) {New constraints on
  interacting dark energy from cosmic chronometers}. \emph{\prd} 94(2):023508,
  \doi{10.1103/PhysRevD.94.023508}, \eprint{1605.01712}

\bibitem[{Offringa et~al.(2010)Offringa, de~Bruyn, Biehl, Zaroubi, Bernardi,
  and Pandey}]{Offringa:2010kb}
Offringa A.~R., de~Bruyn A.~G., Biehl M., Zaroubi S., Bernardi G., Pandey V.~N.
  (2010) {Post-correlation radio frequency interference classification
  methods}. \emph{Mon Not Roy Astron Soc} 405:155--167,
  \doi{10.1111/j.1365-2966.2010.16471.x}, \eprint{1002.1957}

\bibitem[{Oguri(2016)}]{Oguri:2016dgk}
Oguri M. (2016) {Measuring the distance-redshift relation with the
  cross-correlation of gravitational wave standard sirens and galaxies}.
  \emph{Phys Rev D} 93(8):083511, \doi{10.1103/PhysRevD.93.083511},
  \eprint{1603.02356}

\bibitem[{{Oguri} and {Kawano}(2003)}]{Oguri:2003}
{Oguri} M., {Kawano} Y. (2003) {Gravitational lens time delays for distant
  supernovae: breaking the degeneracy between radial mass profiles and the
  Hubble constant}. \emph{\mnras} 338(4):L25--L29,
  \doi{10.1046/j.1365-8711.2003.06290.x}, \eprint{astro-ph/0211499}

\bibitem[{{Oguri} and {Marshall}(2010)}]{Oguri:2010}
{Oguri} M., {Marshall} P.~J. (2010) {Gravitationally lensed quasars and
  supernovae in future wide-field optical imaging surveys}. \emph{\mnras}
  405(4):2579--2593, \doi{10.1111/j.1365-2966.2010.16639.x}, \eprint{1001.2037}

\bibitem[{{Oguri} et~al.(2006){Oguri}, {Inada}, {Pindor}, {Strauss},
  {Richards}, {Hennawi}, {Turner}, {Lupton}, {Schneider}, {Fukugita}, and
  {Brinkmann}}]{SQLS}
{Oguri} M., {Inada} N., {Pindor} B., {Strauss} M.~A., {Richards} G.~T.,
  {Hennawi} J.~F., {Turner} E.~L., {Lupton} R.~H., et~al. (2006) {The Sloan
  Digital Sky Survey Quasar Lens Search. I. Candidate Selection Algorithm}.
  \emph{\aj} 132(3):999--1013, \doi{10.1086/506019}, \eprint{astro-ph/0605571}

\bibitem[{Oh and Mack(2003)}]{Oh:2003jy}
Oh S.~P., Mack K.~J. (2003) {Foregrounds for 21cm observations of neutral gas
  at high redshift}. \emph{Mon Not Roy Astron Soc} 346:871,
  \doi{10.1111/j.1365-2966.2003.07133.x}, \eprint{astro-ph/0302099}

\bibitem[{Olivari et~al.(2018)Olivari, Dickinson, Battye, Ma, Costa,
  Remazeilles, and Harper}]{Olivari:2017bfv}
Olivari L.~C., Dickinson C., Battye R.~A., Ma Y.-Z., Costa A.~A., Remazeilles
  M., Harper S. (2018) {Cosmological parameter forecasts for HI intensity
  mapping experiments using the angular power spectrum}. \emph{Mon Not Roy
  Astron Soc} 473(3):4242--4256, \doi{10.1093/mnras/stx2621},
  \eprint{1707.07647}

\bibitem[{{O'Malley} et~al.(2017){O'Malley}, {Gilligan}, and
  {Chaboyer}}]{OMalley}
{O'Malley} E.~M., {Gilligan} C., {Chaboyer} B. (2017) {Absolute Ages and
  Distances of 22 GCs Using Monte Carlo Main-sequence Fitting}. \emph{\apj}
  838(2):162, \doi{10.3847/1538-4357/aa6574}, \eprint{1703.01915}

\bibitem[{{Onodera} et~al.(2012){Onodera}, {Renzini}, {Carollo}, {Cappellari},
  {Mancini}, {Strazzullo}, {Daddi}, {Arimoto}, {Gobat}, {Yamada}, {McCracken},
  {Ilbert}, {Capak}, {Cimatti}, {Giavalisco}, {Koekemoer}, {Kong}, {Lilly},
  {Motohara}, {Ohta}, {Sanders}, {Scoville}, {Tamura}, and
  {Taniguchi}}]{onodera2012}
{Onodera} M., {Renzini} A., {Carollo} M., {Cappellari} M., {Mancini} C.,
  {Strazzullo} V., {Daddi} E., {Arimoto} N., et~al. (2012) {Deep Near-infrared
  Spectroscopy of Passively Evolving Galaxies at z
  >\raisebox{-0.5ex}\textasciitilde 1.4}. \emph{\apj} 755(1):26,
  \doi{10.1088/0004-637X/755/1/26}, \eprint{1206.1540}

\bibitem[{{Onodera} et~al.(2015){Onodera}, {Carollo}, {Renzini}, {Cappellari},
  {Mancini}, {Arimoto}, {Daddi}, {Gobat}, {Strazzullo}, {Tacchella}, and
  {Yamada}}]{onodera2015}
{Onodera} M., {Carollo} C.~M., {Renzini} A., {Cappellari} M., {Mancini} C.,
  {Arimoto} N., {Daddi} E., {Gobat} R., et~al. (2015) {The Ages, Metallicities,
  and Element Abundance Ratios of Massive Quenched Galaxies at z$\geq$1.6}.
  \emph{\apj} 808:161, \doi{10.1088/0004-637X/808/2/161}, \eprint{1411.5023}

\bibitem[{{Ostriker} and {Steinhardt}(1995)}]{ostriker}
{Ostriker} J.~P., {Steinhardt} P.~J. (1995) {The observational case for a
  low-density Universe with a non-zero cosmological constant}. \emph{\nat}
  377(6550):600--602, \doi{10.1038/377600a0}

\bibitem[{{Pacifici} et~al.(2016){Pacifici}, {Kassin}, {Weiner}, {Holden},
  {Gardner}, {Faber}, {Ferguson}, {Koo}, {Primack}, {Bell}, {Dekel}, {Gawiser},
  {Giavalisco}, {Rafelski}, {Simons}, {Barro}, {Croton}, {Dav{\'e}}, {Fontana},
  {Grogin}, {Koekemoer}, {Lee}, {Salmon}, {Somerville}, and
  {Behroozi}}]{pacifici2016}
{Pacifici} C., {Kassin} S.~A., {Weiner} B.~J., {Holden} B., {Gardner} J.~P.,
  {Faber} S.~M., {Ferguson} H.~C., {Koo} D.~C., et~al. (2016) {The Evolution of
  Star Formation Histories of Quiescent Galaxies}. \emph{\apj} 832(1):79,
  \doi{10.3847/0004-637X/832/1/79}, \eprint{1609.03572}

\bibitem[{{Padilla} et~al.(2005){Padilla}, {Ceccarelli}, and
  {Lambas}}]{Padilla2005}
{Padilla} N.~D., {Ceccarelli} L., {Lambas} D.~G. (2005) {Spatial and dynamical
  properties of voids in a {\ensuremath{\Lambda}} cold dark matter universe}.
  \emph{\mnras} 363(3):977--990, \doi{10.1111/j.1365-2966.2005.09500.x},
  \eprint{astro-ph/0508297}

\bibitem[{Padmanabhan et~al.(2016)Padmanabhan, Choudhury, and
  Refregier}]{Padmanabhan:2015wja}
Padmanabhan H., Choudhury T.~R., Refregier A. (2016) {Modelling the cosmic
  neutral hydrogen from DLAs and 21 cm observations}. \emph{Mon Not Roy Astron
  Soc} 458(1):781--788, \doi{10.1093/mnras/stw353}, \eprint{1505.00008}

\bibitem[{Padmanabhan et~al.(2017)Padmanabhan, Refregier, and
  Amara}]{Padmanabhan:2016fgy}
Padmanabhan H., Refregier A., Amara A. (2017) {A halo model for cosmological
  neutral hydrogen : abundances and clustering H i abundances and clustering}.
  \emph{Mon Not Roy Astron Soc} 469(2):2323--2334, \doi{10.1093/mnras/stx979},
  \eprint{1611.06235}

\bibitem[{Padmanabhan et~al.(2019)Padmanabhan, Refregier, and
  Amara}]{Padmanabhan:2018llf}
Padmanabhan H., Refregier A., Amara A. (2019) {Impact of astrophysics on
  cosmology forecasts for 21 cm surveys}. \emph{Mon Not Roy Astron Soc}
  485(3):4060--4070, \doi{10.1093/mnras/stz683}, \eprint{1804.10627}

\bibitem[{{Padoan} and {Jimenez}(1997)}]{PadoanJimenezLF}
{Padoan} P., {Jimenez} R. (1997) {Ages of Globular Clusters: Breaking the
  Age-Distance Degeneracy with the Luminosity Function}. \emph{\apj}
  475(2):580--583, \doi{10.1086/303582}, \eprint{astro-ph/9603060}

\bibitem[{{Paech} et~al.(2017){Paech}, {Hamaus}, {Hoyle}, {Costanzi},
  {Giannantonio}, {Hagstotz}, {Sauerwein}, and {Weller}}]{Paech2017}
{Paech} K., {Hamaus} N., {Hoyle} B., {Costanzi} M., {Giannantonio} T.,
  {Hagstotz} S., {Sauerwein} G., {Weller} J. (2017) {Cross-correlation of
  galaxies and galaxy clusters in the Sloan Digital Sky Survey and the
  importance of non-Poissonian shot noise}. \emph{\mnras} 470(3):2566--2577,
  \doi{10.1093/mnras/stx1354}, \eprint{1612.02018}

\bibitem[{{Paillas} et~al.(2019){Paillas}, {Cautun}, {Li}, {Cai}, {Padilla},
  {Armijo}, and {Bose}}]{Paillas2019}
{Paillas} E., {Cautun} M., {Li} B., {Cai} Y.-C., {Padilla} N., {Armijo} J.,
  {Bose} S. (2019) {The Santiago-Harvard-Edinburgh-Durham void comparison II:
  unveiling the Vainshtein screening using weak lensing}. \emph{\mnras}
  484(1):1149--1165, \doi{10.1093/mnras/stz022}, \eprint{1810.02864}

\bibitem[{{Paillas} et~al.(2021){Paillas}, {Cai}, {Padilla}, and
  {S{\'a}nchez}}]{Paillas2021}
{Paillas} E., {Cai} Y.-C., {Padilla} N., {S{\'a}nchez} A.~G. (2021)
  {Redshift-space distortions with split densities}. \emph{\mnras}
  505(4):5731--5752, \doi{10.1093/mnras/stab1654}, \eprint{2101.09854}

\bibitem[{Palmese and Kim(2021)}]{Palmese:2020kxn}
Palmese A., Kim A.~G. (2021) {Probing gravity and growth of structure with
  gravitational waves and galaxies\textquoteright{} peculiar velocity}.
  \emph{Phys Rev D} 103(10):103507, \doi{10.1103/PhysRevD.103.103507},
  \eprint{2005.04325}

\bibitem[{{Palmese} et~al.(2020){Palmese}, {deVicente}, {Pereira}, {Annis},
  {Hartley}, {Herner}, {Soares-Santos}, {Crocce}, {Huterer}, {Maga{\~n}a
  Hernandez}, and et~al.}]{2020ApJ...900L..33P}
{Palmese} A., {deVicente} J., {Pereira} M.~E.~S., {Annis} J., {Hartley} W.,
  {Herner} K., {Soares-Santos} M., {Crocce} M., et~al. (2020) {A Statistical
  Standard Siren Measurement of the Hubble Constant from the LIGO/Virgo
  Gravitational Wave Compact Object Merger GW190814 and Dark Energy Survey
  Galaxies}. \emph{\apjl} 900(2):L33, \doi{10.3847/2041-8213/abaeff},
  \eprint{2006.14961}

\bibitem[{{Palmese} et~al.(2021){Palmese}, {Bom}, {Mucesh}, and
  {Hartley}}]{2021arXiv211106445P}
{Palmese} A., {Bom} C.~R., {Mucesh} S., {Hartley} W.~G. (2021) {A standard
  siren measurement of the Hubble constant using gravitational wave events from
  the first three LIGO/Virgo observing runs and the DESI Legacy Survey}.
  \emph{arXiv e-prints} arXiv:2111.06445, \eprint{2111.06445}

\bibitem[{Palmese et~al.(2021)Palmese, Fishbach, Burke, Annis, and
  Liu}]{Palmese:2021wcv}
Palmese A., Fishbach M., Burke C.~J., Annis J.~T., Liu X. (2021) {Do LIGO/Virgo
  Black Hole Mergers Produce AGN Flares? The Case of GW190521 and Prospects for
  Reaching a Confident Association}. \emph{Astrophys J Lett} 914(2):L34,
  \doi{10.3847/2041-8213/ac0883}, \eprint{2103.16069}

\bibitem[{{Pan} et~al.(2012){Pan}, {Vogeley}, {Hoyle}, {Choi}, and
  {Park}}]{Pan2012}
{Pan} D.~C., {Vogeley} M.~S., {Hoyle} F., {Choi} Y.-Y., {Park} C. (2012)
  {Cosmic voids in Sloan Digital Sky Survey Data Release 7}. \emph{\mnras}
  421(2):926--934, \doi{10.1111/j.1365-2966.2011.20197.x}, \eprint{1103.4156}

\bibitem[{{Panchal} et~al.(2020){Panchal}, {Pisani}, and
  {Spergel}}]{Panchal2020}
{Panchal} R.~R., {Pisani} A., {Spergel} D.~N. (2020) {How Do Galaxy Properties
  Affect Void Statistics?} \emph{\apj} 901(1):87,
  \doi{10.3847/1538-4357/abadff}, \eprint{2009.14751}

\bibitem[{{Paraficz} and {Hjorth}(2009)}]{Paraficz:2009}
{Paraficz} D., {Hjorth} J. (2009) {Gravitational lenses as cosmic rulers:
  {\ensuremath{\Omega}}$_{m}$, {\ensuremath{\Omega}}$_{{\ensuremath{\Lambda}}}$
  from time delays and velocity dispersions}. \emph{\aap} 507(3):L49--L52,
  \doi{10.1051/0004-6361/200913307}, \eprint{0910.5823}

\bibitem[{{Paranjape} et~al.(2012){Paranjape}, {Lam}, and
  {Sheth}}]{Paranjape2012}
{Paranjape} A., {Lam} T.~Y., {Sheth} R.~K. (2012) {A hierarchy of voids: more
  ado about nothing}. \emph{\mnras} 420(2):1648--1655,
  \doi{10.1111/j.1365-2966.2011.20154.x}, \eprint{1106.2041}

\bibitem[{{Pardo} et~al.(2018){Pardo}, {Fishbach}, {Holz}, and
  {Spergel}}]{2018JCAP...07..048P}
{Pardo} K., {Fishbach} M., {Holz} D.~E., {Spergel} D.~N. (2018) {Limits on the
  number of spacetime dimensions from GW170817}. \emph{\jcap} 2018(7):048,
  \doi{10.1088/1475-7516/2018/07/048}, \eprint{1801.08160}

\bibitem[{{Park} and {Rozo}(2020)}]{Park2020}
{Park} Y., {Rozo} E. (2020) {Concordance cosmology?} \emph{\mnras}
  499(4):4638--4645, \doi{10.1093/mnras/staa2647}, \eprint{1907.05798}

\bibitem[{{Patiri} et~al.(2006){Patiri}, {Prada}, {Holtzman}, {Klypin}, and
  {Betancort-Rijo}}]{Patiri2006}
{Patiri} S.~G., {Prada} F., {Holtzman} J., {Klypin} A., {Betancort-Rijo} J.
  (2006) {The properties of galaxies in voids}. \emph{\mnras}
  372(4):1710--1720, \doi{10.1111/j.1365-2966.2006.10975.x},
  \eprint{astro-ph/0605703}

\bibitem[{Paul et~al.(2021)Paul, Santos, Townsend, Jarvis, Maddox, Collier,
  Frank, and Taylor}]{Paul:2020ank}
Paul S., Santos M.~G., Townsend J., Jarvis M.~J., Maddox N., Collier J.~D.,
  Frank B.~S., Taylor R. (2021) {H\,i intensity mapping with the MIGHTEE
  survey: power spectrum estimates}. \emph{Mon Not Roy Astron Soc}
  505(2):2039--2050, \doi{10.1093/mnras/stab1089}, \eprint{2009.13550}

\bibitem[{{Paz} et~al.(2013){Paz}, {Lares}, {Ceccarelli}, {Padilla}, and
  {Lambas}}]{Paz2013}
{Paz} D., {Lares} M., {Ceccarelli} L., {Padilla} N., {Lambas} D.~G. (2013)
  {Clues on void evolution-II. Measuring density and velocity profiles on SDSS
  galaxy redshift space distortions}. \emph{\mnras} 436:3480--3491,
  \doi{10.1093/mnras/stt1836}, \eprint{1306.5799}

\bibitem[{{Peebles}(1980)}]{Peebles1980}
{Peebles} P.~J.~E. (1980) {The large-scale structure of the universe}.
  Princeton University Press, Princeton U.S.A.

\bibitem[{{Pe'er}(2015)}]{Peer15}
{Pe'er} A. (2015) {Physics of Gamma-Ray Bursts Prompt Emission}. \emph{Advances
  in Astronomy} 2015:907321, \doi{10.1155/2015/907321}, \eprint{1504.02626}

\bibitem[{{Peng} et~al.(2010){Peng}, {Lilly}, {Kova{\v{c}}}, {Bolzonella},
  {Pozzetti}, {Renzini}, {Zamorani}, {Ilbert}, {Knobel}, and et~al.}]{peng10}
{Peng} Y.-j., {Lilly} S.~J., {Kova{\v{c}}} K., {Bolzonella} M., {Pozzetti} L.,
  {Renzini} A., {Zamorani} G., {Ilbert} O., et~al. (2010) {Mass and Environment
  as Drivers of Galaxy Evolution in SDSS and zCOSMOS and the Origin of the
  Schechter Function}. \emph{\apj} 721(1):193--221,
  \doi{10.1088/0004-637X/721/1/193}, \eprint{1003.4747}

\bibitem[{{Percival} et~al.(2001){Percival}, {Baugh}, {Bland-Hawthorn},
  {Bridges}, {Cannon}, {Cole}, {Colless}, {Collins}, {Couch}, {Dalton}, {De
  Propris}, {Driver}, {Efstathiou}, {Ellis}, {Frenk}, {Glazebrook}, {Jackson},
  {Lahav}, {Lewis}, {Lumsden}, {Maddox}, {Moody}, {Norberg}, {Peacock},
  {Peterson}, {Sutherland}, and {Taylor}}]{percival2001}
{Percival} W.~J., {Baugh} C.~M., {Bland-Hawthorn} J., {Bridges} T., {Cannon}
  R., {Cole} S., {Colless} M., {Collins} C., et~al. (2001) {The 2dF Galaxy
  Redshift Survey: the power spectrum and the matter content of the Universe}.
  \emph{\mnras} 327(4):1297--1306, \doi{10.1046/j.1365-8711.2001.04827.x},
  \eprint{astro-ph/0105252}

\bibitem[{{Perico} et~al.(2019){Perico}, {Voivodic}, {Lima}, and
  {Mota}}]{Perico2019}
{Perico} E. L.~D., {Voivodic} R., {Lima} M., {Mota} D.~F. (2019) {Cosmic voids
  in modified gravity scenarios}. \emph{\aap} 632:A52,
  \doi{10.1051/0004-6361/201935949}

\bibitem[{{Perlick} and {Tsupko}(2022)}]{perlick2022}
{Perlick} V., {Tsupko} O.~Y. (2022) {Calculating black hole shadows: Review of
  analytical studies}. \emph{\physrep} 947:1--39,
  \doi{10.1016/j.physrep.2021.10.004}, \eprint{2105.07101}

\bibitem[{{Perlmutter} et~al.(1998){Perlmutter}, {Aldering}, {della Valle},
  {Deustua}, {Ellis}, {Fabbro}, {Fruchter}, {Goldhaber}, {Groom}, {Hook},
  {Kim}, {Kim}, {Knop}, {Lidman}, {McMahon}, {Nugent}, {Pain}, {Panagia},
  {Pennypacker}, {Ruiz-Lapuente}, {Schaefer}, and {Walton}}]{perlmutter1998}
{Perlmutter} S., {Aldering} G., {della Valle} M., {Deustua} S., {Ellis} R.~S.,
  {Fabbro} S., {Fruchter} A., {Goldhaber} G., et~al. (1998) {Discovery of a
  supernova explosion at half the age of the Universe}. \emph{\nat}
  391(6662):51--54, \doi{10.1038/34124}, \eprint{astro-ph/9712212}

\bibitem[{{Perlmutter} et~al.(1999){Perlmutter}, {Aldering}, {Goldhaber},
  {Knop}, {Nugent}, {Castro}, {Deustua}, {Fabbro}, {Goobar}, {Groom}, {Hook},
  {Kim}, et~al., and {Project}}]{perlmutter1999}
{Perlmutter} S., {Aldering} G., {Goldhaber} G., {Knop} R.~A., {Nugent} P.,
  {Castro} P.~G., {Deustua} S., {Fabbro} S., et~al. (1999) {Measurements of
  {\ensuremath{\Omega}} and {\ensuremath{\Lambda}} from 42 High-Redshift
  Supernovae}. \emph{\apj} 517(2):565--586, \doi{10.1086/307221},
  \eprint{astro-ph/9812133}

\bibitem[{{Peterson} et~al.(2009){Peterson}, {Aleksan}, {Ansari}, {Bandura},
  {Bond}, {Bunton}, {Carlson}, {Chang}, {DeJongh}, {Dobbs}, {Dodelson},
  {Darhmaoui}, {Gnedin}, {Halpern}, {Hogan}, {Le Goff}, {Liu}, {Legrouri},
  {Loeb}, {Loudiyi}, {Magneville}, {Marriner}, {McGinnis}, {McWilliams},
  {Moniez}, {Palanque-Delabruille}, {Pasquinelli}, {Pen}, {Rich}, {Scarpine},
  {Seo}, {Sigurdson}, {Seljak}, {Stebbins}, {Steffen}, {Stoughton}, {Timbie},
  {Vallinotto}, and {Teche}}]{Peterson:2009ka}
{Peterson} J.~B., {Aleksan} R., {Ansari} R., {Bandura} K., {Bond} D., {Bunton}
  J., {Carlson} K., {Chang} T.-C., et~al. (2009) {21-cm Intensity Mapping}. In:
  astro2010: The Astronomy and Astrophysics Decadal Survey, vol 2010, p 234,
  \eprint{0902.3091}

\bibitem[{{Petrillo} et~al.(2017){Petrillo}, {Tortora}, {Chatterjee},
  {Vernardos}, {Koopmans}, {Verdoes Kleijn}, {Napolitano}, {Covone},
  {Schneider}, {Grado}, and {McFarland}}]{Petrillo2017}
{Petrillo} C.~E., {Tortora} C., {Chatterjee} S., {Vernardos} G., {Koopmans}
  L.~V.~E., {Verdoes Kleijn} G., {Napolitano} N.~R., {Covone} G., et~al. (2017)
  {Finding strong gravitational lenses in the Kilo Degree Survey with
  Convolutional Neural Networks}. \emph{\mnras} 472(1):1129--1150,
  \doi{10.1093/mnras/stx2052}, \eprint{1702.07675}

\bibitem[{{Philcox} et~al.(2020){Philcox}, {Massara}, and
  {Spergel}}]{Philcox2020}
{Philcox} O. H.~E., {Massara} E., {Spergel} D.~N. (2020) {What does the marked
  power spectrum measure? Insights from perturbation theory}. \emph{\prd}
  102(4):043516, \doi{10.1103/PhysRevD.102.043516}, \eprint{2006.10055}

\bibitem[{{Phillips}(1993)}]{phillips1993}
{Phillips} M.~M. (1993) {The Absolute Magnitudes of Type IA Supernovae}.
  \emph{\apjl} 413:L105, \doi{10.1086/186970}

\bibitem[{{Pietrzy{\'n}ski} et~al.(2019){Pietrzy{\'n}ski}, {Graczyk},
  {Gallenne}, {Gieren}, {Thompson}, {Pilecki}, {Karczmarek}, {G{\'o}rski},
  {Suchomska}, {Taormina}, {Zgirski}, {Wielg{\'o}rski}, {Ko{\l}aczkowski},
  {Konorski}, {Villanova}, {Nardetto}, {Kervella}, {Bresolin}, {Kudritzki},
  {Storm}, {Smolec}, and {Narloch}}]{pietrzynski19}
{Pietrzy{\'n}ski} G., {Graczyk} D., {Gallenne} A., {Gieren} W., {Thompson}
  I.~B., {Pilecki} B., {Karczmarek} P., {G{\'o}rski} M., et~al. (2019) {A
  distance to the Large Magellanic Cloud that is precise to one per cent}.
  \emph{\nat} 567(7747):200--203, \doi{10.1038/s41586-019-0999-4},
  \eprint{1903.08096}

\bibitem[{{Piotto} et~al.(2002){Piotto}, {King}, {Djorgovski}, {Sosin},
  {Zoccali}, {Saviane}, {De Angeli}, {Riello}, {Recio-Blanco}, {Rich},
  {Meylan}, and {Renzini}}]{piotto02}
{Piotto} G., {King} I.~R., {Djorgovski} S.~G., {Sosin} C., {Zoccali} M.,
  {Saviane} I., {De Angeli} F., {Riello} M., et~al. (2002) {HST color-magnitude
  diagrams of 74 galactic globular clusters in the HST F439W and F555W bands}.
  \emph{\aap} 391:945--965, \doi{10.1051/0004-6361:20020820},
  \eprint{astro-ph/0207124}

\bibitem[{{Pisani} et~al.(2014){Pisani}, {Lavaux}, {Sutter}, and
  {Wandelt}}]{Pisani2014}
{Pisani} A., {Lavaux} G., {Sutter} P.~M., {Wandelt} B.~D. (2014) {Real-space
  density profile reconstruction of stacked voids}. \emph{\mnras}
  443(4):3238--3250, \doi{10.1093/mnras/stu1399}, \eprint{1306.3052}

\bibitem[{{Pisani} et~al.(2015{\natexlab{a}}){Pisani}, {Sutter}, {Hamaus},
  {Alizadeh}, {Biswas}, {Wandelt}, and {Hirata}}]{Pisani2015a}
{Pisani} A., {Sutter} P.~M., {Hamaus} N., {Alizadeh} E., {Biswas} R., {Wandelt}
  B.~D., {Hirata} C.~M. (2015{\natexlab{a}}) {Counting voids to probe dark
  energy}. \emph{\prd} 92(8):083531, \doi{10.1103/PhysRevD.92.083531},
  \eprint{1503.07690}

\bibitem[{{Pisani} et~al.(2015{\natexlab{b}}){Pisani}, {Sutter}, and
  {Wandelt}}]{Pisani2015b}
{Pisani} A., {Sutter} P.~M., {Wandelt} B.~D. (2015{\natexlab{b}}) {Mastering
  the effects of peculiar velocities in cosmic voids}. \emph{ArXiv e-prints}
  \eprint{1506.07982}

\bibitem[{{Pisani} et~al.(2019){Pisani}, {Massara}, {Spergel}, {Alonso},
  {Baker}, {Cai}, {Cautun}, {Davies}, {Demchenko}, {Dor{\'e}}, {Goulding},
  {Habouzit}, {Hamaus}, {Hawken}, {Hirata}, {Ho}, {Jain}, {Kreisch}, {Marulli},
  {Padilla}, {Pollina}, {Sahl{\'e}n}, {Sheth}, {Somerville}, {Szapudi}, {van de
  Weygaert}, {Villaescusa-Navarro}, {Wandelt}, and {Wang}}]{Pisani2019}
{Pisani} A., {Massara} E., {Spergel} D.~N., {Alonso} D., {Baker} T., {Cai}
  Y.-C., {Cautun} M., {Davies} C., et~al. (2019) {Cosmic voids: a novel probe
  to shed light on our Universe}. \emph{\textnormal{BAAS}} 51(3):40,
  \eprint{1903.05161}

\bibitem[{{Planck Collaboration} et~al.(2014{\natexlab{a}}){Planck
  Collaboration}, {Ade}, {Aghanim}, {Armitage-Caplan}, {Arnaud}, {Ashdown},
  {Atrio-Barandela}, {Aumont}, {Baccigalupi}, {Banday}, {Barreiro}, {Bartlett},
  and et~al.}]{Planck14}
{Planck Collaboration}, {Ade} P.~A.~R., {Aghanim} N., {Armitage-Caplan} C.,
  {Arnaud} M., {Ashdown} M., {Atrio-Barandela} F., {Aumont} J., et~al.
  (2014{\natexlab{a}}) {Planck 2013 results. XVI. Cosmological parameters}.
  \emph{\aap} 571:A16, \doi{10.1051/0004-6361/201321591}, \eprint{1303.5076}

\bibitem[{{Planck Collaboration} et~al.(2014{\natexlab{b}}){Planck
  Collaboration}, {Ade}, {Aghanim}, {Armitage-Caplan}, {Arnaud}, {Ashdown},
  {Atrio-Barandela}, {Aumont}, {Baccigalupi}, {Banday}, {Barreiro}, {Bartlett},
  {Bartolo}, and et~al.}]{Planck2014}
{Planck Collaboration}, {Ade} P.~A.~R., {Aghanim} N., {Armitage-Caplan} C.,
  {Arnaud} M., {Ashdown} M., {Atrio-Barandela} F., {Aumont} J., et~al.
  (2014{\natexlab{b}}) {Planck 2013 results. XIX. The integrated Sachs-Wolfe
  effect}. \emph{\aap} 571:A19, \doi{10.1051/0004-6361/201321526},
  \eprint{1303.5079}

\bibitem[{{Planck Collaboration} et~al.(2020{\natexlab{a}}){Planck
  Collaboration}, {Aghanim}, {Akrami}, {Ashdown}, {Aumont}, {Baccigalupi},
  {Ballardini}, {Banday}, {Barreiro}, {Bartolo}, {Basak}, {Battye}, {Benabed},
  {Bernard}, {Bersanelli}, {Bielewicz}, {Bock}, {Bond}, and
  et~al.}]{planck2018}
{Planck Collaboration}, {Aghanim} N., {Akrami} Y., {Ashdown} M., {Aumont} J.,
  {Baccigalupi} C., {Ballardini} M., {Banday} A.~J., et~al.
  (2020{\natexlab{a}}) {Planck 2018 results. VI. Cosmological parameters}.
  \emph{\aap} 641:A6, \doi{10.1051/0004-6361/201833910}, \eprint{1807.06209}

\bibitem[{{Planck Collaboration} et~al.(2020{\natexlab{b}}){Planck
  Collaboration}, {Aghanim}, {Akrami}, {Ashdown}, {Aumont}, {Baccigalupi},
  {Ballardini}, {Banday}, {Barreiro}, {Bartolo}, {Basak}, {Battye}, {Benabed},
  {Bernard}, {Bersanelli}, {Bielewicz}, {Bock}, {Bond}, and
  https://www.overleaf.com/4448358379csmrynjykryg}]{planck20}
{Planck Collaboration}, {Aghanim} N., {Akrami} Y., {Ashdown} M., {Aumont} J.,
  {Baccigalupi} C., {Ballardini} M., {Banday} A.~J., et~al.
  (2020{\natexlab{b}}) {Planck 2018 results. VI. Cosmological parameters}.
  \emph{\aap} 641:A6, \doi{10.1051/0004-6361/201833910}, \eprint{1807.06209}

\bibitem[{{Planck Collaboration} et~al.(2020{\natexlab{c}}){Planck
  Collaboration}, {Aghanim}, {Akrami}, {Ashdown}, {Aumont}, {Baccigalupi},
  {Ballardini}, {Banday}, {Barreiro}, {Bartolo}, {Basak}, {Battye}, {Benabed},
  {Bernard}, {Bersanelli}, {Bielewicz}, {Bock}, and
  https://www.overleaf.com/4448358379csmrynjykryg}]{Planck2020}
{Planck Collaboration}, {Aghanim} N., {Akrami} Y., {Ashdown} M., {Aumont} J.,
  {Baccigalupi} C., {Ballardini} M., {Banday} A.~J., et~al.
  (2020{\natexlab{c}}) {Planck 2018 results. VI. Cosmological parameters}.
  \emph{\aap} 641:A6, \doi{10.1051/0004-6361/201833910}, \eprint{1807.06209}

\bibitem[{{Planck Collaboration} et~al.(2015){Planck Collaboration}, Ade,
  Aghanim, Arnaud et~al.}]{PlanckCollaboration2015}
{Planck Collaboration} P. A.~R., Ade P. A.~R., Aghanim N., Arnaud M., et~al.
  (2015) {Planck 2015 results. XIII. Cosmological parameters}. \emph{Astronomy
  {\&} Astrophysics} 594:A13, \doi{10.1051/0004-6361/201525830},
  \eprint{1502.01589}

\bibitem[{{Platen} et~al.(2007){Platen}, {van de Weygaert}, and
  {Jones}}]{Platen2007}
{Platen} E., {van de Weygaert} R., {Jones} B. J.~T. (2007) {A cosmic watershed:
  the WVF void detection technique}. \emph{\mnras} 380(2):551--570,
  \doi{10.1111/j.1365-2966.2007.12125.x}, \eprint{0706.2788}

\bibitem[{{Pollina} et~al.(2016){Pollina}, {Baldi}, {Marulli}, and
  {Moscardini}}]{Pollina2016}
{Pollina} G., {Baldi} M., {Marulli} F., {Moscardini} L. (2016) {Cosmic voids in
  coupled dark energy cosmologies: the impact of halo bias}. \emph{\mnras}
  455(3):3075--3085, \doi{10.1093/mnras/stv2503}, \eprint{1506.08831}

\bibitem[{{Pollina} et~al.(2017){Pollina}, {Hamaus}, {Dolag}, {Weller},
  {Baldi}, and {Moscardini}}]{Pollina2017}
{Pollina} G., {Hamaus} N., {Dolag} K., {Weller} J., {Baldi} M., {Moscardini} L.
  (2017) {On the linearity of tracer bias around voids}. \emph{\mnras}
  469(1):787--799, \doi{10.1093/mnras/stx785}, \eprint{1610.06176}

\bibitem[{{Pollina} et~al.(2019){Pollina}, {Hamaus}, {Paech}, {Dolag},
  {Weller}, {S{\'a}nchez}, {Rykoff}, {Jain}, {Abbott}, {Allam}, {Avila},
  {Bernstein}, {Bertin}, et~al., and {DES Collaboration}}]{Pollina2019}
{Pollina} G., {Hamaus} N., {Paech} K., {Dolag} K., {Weller} J., {S{\'a}nchez}
  C., {Rykoff} E.~S., {Jain} B., et~al. (2019) {On the relative bias of void
  tracers in the Dark Energy Survey}. \emph{\mnras} 487(2):2836--2852,
  \doi{10.1093/mnras/stz1470}, \eprint{1806.06860}

\bibitem[{{Porqueres} et~al.(2019){Porqueres}, {Jasche}, {Lavaux}, and
  {En{\ss}lin}}]{Porqueres2019}
{Porqueres} N., {Jasche} J., {Lavaux} G., {En{\ss}lin} T. (2019) {Inferring
  high-redshift large-scale structure dynamics from the
  Lyman-{\ensuremath{\alpha}} forest}. \emph{\aap} 630:A151,
  \doi{10.1051/0004-6361/201936245}, \eprint{1907.02973}

\bibitem[{{Postman} et~al.(2012){Postman}, {Coe}, {Ben{\'\i}tez}, {Bradley},
  {Broadhurst}, {Donahue}, {Ford}, {Graur}, {Graves}, {Jouvel}, {Koekemoer},
  and et~al.}]{Postman:2012}
{Postman} M., {Coe} D., {Ben{\'\i}tez} N., {Bradley} L., {Broadhurst} T.,
  {Donahue} M., {Ford} H., {Graur} O., et~al. (2012) {The Cluster Lensing and
  Supernova Survey with Hubble: An Overview}. \emph{\apjs} 199(2):25,
  \doi{10.1088/0067-0049/199/2/25}, \eprint{1106.3328}

\bibitem[{Pourtsidou and Metcalf(2014)}]{Pourtsidou:2013hea}
Pourtsidou A., Metcalf R.~B. (2014) {Weak lensing with 21cm intensity mapping
  at $z \sim 2-3$}. \emph{Mon Not Roy Astron Soc} 439:L36--L40,
  \doi{10.1093/mnrasl/slt175}, \eprint{1311.4484}

\bibitem[{Pourtsidou et~al.(2016)Pourtsidou, Bacon, Crittenden, and
  Metcalf}]{Pourtsidou:2015mia}
Pourtsidou A., Bacon D., Crittenden R., Metcalf R.~B. (2016) {Prospects for
  clustering and lensing measurements with forthcoming intensity mapping and
  optical surveys}. \emph{Mon Not Roy Astron Soc} 459(1):863--870,
  \doi{10.1093/mnras/stw658}, \eprint{1509.03286}

\bibitem[{Pourtsidou et~al.(2017)Pourtsidou, Bacon, and
  Crittenden}]{Pourtsidou:2016dzn}
Pourtsidou A., Bacon D., Crittenden R. (2017) {HI and cosmological constraints
  from intensity mapping, optical and CMB surveys}. \emph{Mon Not Roy Astron
  Soc} 470(4):4251--4260, \doi{10.1093/mnras/stx1479}, \eprint{1610.04189}

\bibitem[{{Pozzetti} and {Mannucci}(2000)}]{pozzetti2000}
{Pozzetti} L., {Mannucci} F. (2000) {Extremely red galaxies: age and dust
  degeneracy solved?} \emph{\mnras} 317(1):L17--L21,
  \doi{10.1046/j.1365-8711.2000.03829.x}, \eprint{astro-ph/0006430}

\bibitem[{{Pozzetti} et~al.(2010){Pozzetti}, {Bolzonella}, {Zucca}, {Zamorani},
  {Lilly}, {Renzini}, {Moresco}, {Mignoli}, {Cassata}, {Tasca}, {Lamareille},
  {Maier}, and et~al.}]{pozzetti2010}
{Pozzetti} L., {Bolzonella} M., {Zucca} E., {Zamorani} G., {Lilly} S.,
  {Renzini} A., {Moresco} M., {Mignoli} M., et~al. (2010) {zCOSMOS - 10k-bright
  spectroscopic sample. The bimodality in the galaxy stellar mass function:
  exploring its evolution with redshift}. \emph{\aap} 523:A13,
  \doi{10.1051/0004-6361/200913020}, \eprint{0907.5416}

\bibitem[{{Predehl}(2012)}]{2012SPIE.8443E..1RP}
{Predehl} P. (2012) {eROSITA}. In: \procspie, Society of Photo-Optical
  Instrumentation Engineers (SPIE) Conference Series, vol 8443, p 84431R,
  \doi{10.1117/12.925843}

\bibitem[{{Punturo} et~al.(2010){Punturo}, {Abernathy}, {Acernese}, {Allen},
  {Andersson}, {Arun}, {Barone}, {Barr}, {Barsuglia}, {Beker}, {Beveridge},
  {Birindelli}, {Bose}, {Bosi}, {Braccini}, {Bradaschia}, {Bulik}, {Calloni},
  {Cella}, {Chassande Mottin}, and et~al.}]{Punturo:2010}
{Punturo} M., {Abernathy} M., {Acernese} F., {Allen} B., {Andersson} N., {Arun}
  K., {Barone} F., {Barr} B., et~al. (2010) {The Einstein Telescope: a
  third-generation gravitational wave observatory}. \emph{Classical and Quantum
  Gravity} 27(19):194002, \doi{10.1088/0264-9381/27/19/194002}

\bibitem[{Qin et~al.(2019)Qin, Howlett, and Staveley-Smith}]{Qin:2019axr}
Qin F., Howlett C., Staveley-Smith L. (2019) {The redshift-space momentum power
  spectrum \textendash{} II. Measuring the growth rate from the combined 2MTF
  and 6dFGSv surveys}. \emph{Mon Not Roy Astron Soc} 487(4):5235--5247,
  \doi{10.1093/mnras/stz1576}, \eprint{1906.02874}

\bibitem[{{Quartin} and {Amendola}(2010)}]{quartin2010}
{Quartin} M., {Amendola} L. (2010) {Distinguishing between void models and dark
  energy with cosmic parallax and redshift drift}. \emph{\prd} 81(4):043522,
  \doi{10.1103/PhysRevD.81.043522}, \eprint{0909.4954}

\bibitem[{Quartin et~al.(2014)Quartin, Marra, and Amendola}]{Quartin:2013moa}
Quartin M., Marra V., Amendola L. (2014) {Accurate Weak Lensing of Standard
  Candles. II. Measuring sigma8 with Supernovae}. \emph{PhysRev} D89:023009,
  \doi{10.1103/PhysRevD.89.023009}, \eprint{1307.1155}

\bibitem[{Quartin et~al.(2022)Quartin, Amendola, and Moraes}]{Quartin:2021dmr}
Quartin M., Amendola L., Moraes B. (2022) {The 6~\texttimes{}~2pt method:
  supernova velocities meet multiple tracers}. \emph{Mon Not Roy Astron Soc}
  512(2):2841--2853, \doi{10.1093/mnras/stac571}, \eprint{2111.05185}

\bibitem[{{Raichoor} et~al.(2021){Raichoor}, {de Mattia}, {Ross}, {Zhao},
  {Alam}, {Avila}, {Bautista}, {Brinkmann}, and et~al.}]{raichoor2021}
{Raichoor} A., {de Mattia} A., {Ross} A.~J., {Zhao} C., {Alam} S., {Avila} S.,
  {Bautista} J., {Brinkmann} J., et~al. (2021) {The completed SDSS-IV extended
  Baryon Oscillation Spectroscopic Survey: large-scale structure catalogues and
  measurement of the isotropic BAO between redshift 0.6 and 1.1 for the
  Emission Line Galaxy Sample}. \emph{\mnras} 500(3):3254--3274,
  \doi{10.1093/mnras/staa3336}, \eprint{2007.09007}

\bibitem[{{Raimondo}(2009)}]{raimondo09}
{Raimondo} G. (2009) {Joint Analysis of Near-Infrared Properties and Surface
  Brightness Fluctuations of Large Magellanic Cloud Star Clusters}. \emph{\apj}
  700(2):1247--1261, \doi{10.1088/0004-637X/700/2/1247}, \eprint{0907.1408}

\bibitem[{{Raimondo} et~al.(2005){Raimondo}, {Brocato}, {Cantiello}, and
  {Capaccioli}}]{raimondo05}
{Raimondo} G., {Brocato} E., {Cantiello} M., {Capaccioli} M. (2005) {New
  Optical and Near-Infrared Surface Brightness Fluctuation Models. II. Young
  and Intermediate-Age Stellar Populations}. \emph{\aj} 130(6):2625--2646,
  \doi{10.1086/497591}, \eprint{astro-ph/0509020}

\bibitem[{{Rakavy} et~al.(1967){Rakavy}, {Shaviv}, and
  {Zinamon}}]{1967ApJ...150..131R}
{Rakavy} G., {Shaviv} G., {Zinamon} Z. (1967) {Carbon and Oxygen Burning Stars
  and Pre-Supernova Models}. \emph{\apj} 150:131, \doi{10.1086/149318}

\bibitem[{{R{\"a}s{\"a}nen} et~al.(2015){R{\"a}s{\"a}nen}, {Bolejko}, and
  {Finoguenov}}]{Rasanen:2014mca}
{R{\"a}s{\"a}nen} S., {Bolejko} K., {Finoguenov} A. (2015) {New Test of the
  Friedmann-Lema{\^\i}tre-Robertson-Walker Metric Using the Distance Sum Rule}.
  \emph{\prl} 115(10):101301, \doi{10.1103/PhysRevLett.115.101301},
  \eprint{1412.4976}

\bibitem[{{Rathna Kumar} et~al.(2013){Rathna Kumar}, {Tewes}, {Stalin},
  {Courbin}, {Asfandiyarov}, {Meylan}, {Eulaers}, {Prabhu}, {Magain}, {Van
  Winckel}, and {Ehgamberdiev}}]{Rathna2013}
{Rathna Kumar} S., {Tewes} M., {Stalin} C.~S., {Courbin} F., {Asfandiyarov} I.,
  {Meylan} G., {Eulaers} E., {Prabhu} T.~P., et~al. (2013) {COSMOGRAIL: the
  COSmological MOnitoring of GRAvItational Lenses. XIV. Time delay of the
  doubly lensed quasar SDSS J1001+5027}. \emph{\aap} 557:A44,
  \doi{10.1051/0004-6361/201322116}, \eprint{1306.5105}

\bibitem[{{Ratsimbazafy} et~al.(2017){Ratsimbazafy}, {Loubser}, {Crawford},
  {Cress}, {Bassett}, {Nichol}, and {V{\"a}is{\"a}nen}}]{ratsimbazafy2017}
{Ratsimbazafy} A.~L., {Loubser} S.~I., {Crawford} S.~M., {Cress} C.~M.,
  {Bassett} B.~A., {Nichol} R.~C., {V{\"a}is{\"a}nen} P. (2017) {Age-dating
  Luminous Red Galaxies observed with the Southern African Large Telescope}.
  \emph{ArXiv e-prints} \eprint{1702.00418}

\bibitem[{{Read} et~al.(2007){Read}, {Saha}, and {Macci{\`o}}}]{Read:2007}
{Read} J.~I., {Saha} P., {Macci{\`o}} A.~V. (2007) {Radial Density Profiles of
  Time-Delay Lensing Galaxies}. \emph{\apj} 667(2):645--654,
  \doi{10.1086/520714}, \eprint{0704.3267}

\bibitem[{{Reed} et~al.(2015){Reed}, {Schneider}, {Smith}, {Potter}, {Stadel},
  and {Moore}}]{Reed2015}
{Reed} D.~S., {Schneider} A., {Smith} R.~E., {Potter} D., {Stadel} J., {Moore}
  B. (2015) {The same with less: the cosmic web of warm versus cold dark matter
  dwarf galaxies}. \emph{\mnras} 451(4):4413--4423,
  \doi{10.1093/mnras/stv1233}, \eprint{1410.1541}

\bibitem[{{Refsdal}(1964)}]{Refsdal:1964}
{Refsdal} S. (1964) {On the possibility of determining Hubble's parameter and
  the masses of galaxies from the gravitational lens effect}. \emph{\mnras}
  128:307, \doi{10.1093/mnras/128.4.307}

\bibitem[{{Reichart} et~al.(2001){Reichart}, {Lamb}, {Fenimore},
  {Ramirez-Ruiz}, {Cline}, and {Hurley}}]{Reichart01}
{Reichart} D.~E., {Lamb} D.~Q., {Fenimore} E.~E., {Ramirez-Ruiz} E., {Cline}
  T.~L., {Hurley} K. (2001) {A Possible Cepheid-like Luminosity Estimator for
  the Long Gamma-Ray Bursts}. \emph{\apj} 552(1):57--71, \doi{10.1086/320434},
  \eprint{astro-ph/0004302}

\bibitem[{{Reitze} et~al.(2019){Reitze}, {Adhikari}, {Ballmer}, {Barish},
  {Barsotti}, {Billingsley}, {Brown}, {Chen}, {Coyne}, {Eisenstein}, {Evans},
  {Fritschel}, {Hall}, {Lazzarini}, {Lovelace}, {Read}, {Sathyaprakash},
  {Shoemaker}, {Smith}, {Torrie}, {Vitale}, {Weiss}, {Wipf}, and
  {Zucker}}]{Reitze:2019}
{Reitze} D., {Adhikari} R.~X., {Ballmer} S., {Barish} B., {Barsotti} L.,
  {Billingsley} G., {Brown} D.~A., {Chen} Y., et~al. (2019) {Cosmic Explorer:
  The U.S. Contribution to Gravitational-Wave Astronomy beyond LIGO}. In:
  Bulletin of the American Astronomical Society, vol~51, p~35,
  \eprint{1907.04833}

\bibitem[{{Renzi} and {Martinelli}(2022)}]{renzi2022}
{Renzi} F., {Martinelli} M. (2022) {Climbing out of the shadows: building the
  distance ladder with black hole images}. \emph{arXiv e-prints}
  arXiv:2205.03396, \eprint{2205.03396}

\bibitem[{{Renzi} et~al.(2021){Renzi}, {Hogg}, and {Giar{\`e}}}]{renzi2021}
{Renzi} F., {Hogg} N.~B., {Giar{\`e}} W. (2021) {The resilience of the
  Etherington--Hubble relation}. \emph{arXiv e-prints} arXiv:2112.05701,
  \eprint{2112.05701}

\bibitem[{{Renzini}(2006)}]{renzini2006}
{Renzini} A. (2006) {Stellar Population Diagnostics of Elliptical Galaxy
  Formation}. \emph{\araa} 44(1):141--192,
  \doi{10.1146/annurev.astro.44.051905.092450}, \eprint{astro-ph/0603479}

\bibitem[{{Reyes} and {Escamilla-Rivera}(2021)}]{reyes2021}
{Reyes} M., {Escamilla-Rivera} C. (2021) {Improving data-driven
  model-independent reconstructions and updated constraints on dark energy
  models from Horndeski cosmology}. \emph{\jcap} 2021(7):048,
  \doi{10.1088/1475-7516/2021/07/048}, \eprint{2104.04484}

\bibitem[{{Ricciardelli} et~al.(2014{\natexlab{a}}){Ricciardelli}, {Cava},
  {Varela}, and {Quilis}}]{Ricciardelli2014b}
{Ricciardelli} E., {Cava} A., {Varela} J., {Quilis} V. (2014{\natexlab{a}})
  {The star formation activity in cosmic voids}. \emph{\mnras}
  445(4):4045--4054, \doi{10.1093/mnras/stu2061}, \eprint{1410.0023}

\bibitem[{{Ricciardelli} et~al.(2014{\natexlab{b}}){Ricciardelli}, {Quilis},
  and {Varela}}]{Ricciardelli2014a}
{Ricciardelli} E., {Quilis} V., {Varela} J. (2014{\natexlab{b}}) {On the
  universality of void density profiles}. \emph{\mnras} 440(1):601--609,
  \doi{10.1093/mnras/stu307}, \eprint{1402.2976}

\bibitem[{{Riess} et~al.(1998){Riess}, {Filippenko}, {Challis}, {Clocchiatti},
  {Diercks}, {Garnavich}, {Gilliland}, {Hogan}, {Jha}, {Kirshner},
  {Leibundgut}, {Phillips}, {Reiss}, {Schmidt}, {Schommer}, {Smith},
  {Spyromilio}, {Stubbs}, {Suntzeff}, and {Tonry}}]{riess1998}
{Riess} A.~G., {Filippenko} A.~V., {Challis} P., {Clocchiatti} A., {Diercks}
  A., {Garnavich} P.~M., {Gilliland} R.~L., {Hogan} C.~J., et~al. (1998)
  {Observational Evidence from Supernovae for an Accelerating Universe and a
  Cosmological Constant}. \emph{\aj} 116(3):1009--1038, \doi{10.1086/300499},
  \eprint{astro-ph/9805201}

\bibitem[{{Riess} et~al.(2011){Riess}, {Macri}, {Casertano}, {Lampeit},
  {Ferguson}, {Filippenko}, {Jha}, {Li}, {Chornock}, and
  {Silverman}}]{Riess2011}
{Riess} A.~G., {Macri} L., {Casertano} S., {Lampeit} H., {Ferguson} H.~C.,
  {Filippenko} A.~V., {Jha} S.~W., {Li} W., et~al. (2011) {Erratum: ``A 3\%
  Solution: Determination of the Hubble Constant with the Hubble Space
  Telescope and Wide Field Camera 3'' <A
  href=``/abs/2011ApJ...730..119R''>(2011, ApJ, 730, 119)</A>}. \emph{\apj}
  732(2):129, \doi{10.1088/0004-637X/732/2/129}

\bibitem[{{Riess} et~al.(2016){Riess}, {Macri}, {Hoffmann}, {Scolnic},
  {Casertano}, {Filippenko}, {Tucker}, {Reid}, {Jones}, {Silverman},
  {Chornock}, {Challis}, {Yuan}, {Brown}, and {Foley}}]{Riess2016}
{Riess} A.~G., {Macri} L.~M., {Hoffmann} S.~L., {Scolnic} D., {Casertano} S.,
  {Filippenko} A.~V., {Tucker} B.~E., {Reid} M.~J., et~al. (2016) {A 2.4\%
  Determination of the Local Value of the Hubble Constant}. \emph{\apj}
  826(1):56, \doi{10.3847/0004-637X/826/1/56}, \eprint{1604.01424}

\bibitem[{{Riess} et~al.(2018){Riess}, {Rodney}, {Scolnic}, {Shafer},
  {Strolger}, {Ferguson}, {Postman}, {Graur}, {Maoz}, {Jha}, {Mobasher},
  {Casertano}, {Hayden}, {Molino}, {Hjorth}, {Garnavich}, {Jones}, {Kirshner},
  {Koekemoer}, {Grogin}, {Brammer}, {Hemmati}, {Dickinson}, {Challis}, {Wolff},
  {Clubb}, {Filippenko}, {Nayyeri}, {U}, {Koo}, {Faber}, {Kocevski}, {Bradley},
  and {Coe}}]{riess2018b}
{Riess} A.~G., {Rodney} S.~A., {Scolnic} D.~M., {Shafer} D.~L., {Strolger}
  L.-G., {Ferguson} H.~C., {Postman} M., {Graur} O., et~al. (2018) {Type Ia
  Supernova Distances at Redshift >1.5 from the Hubble Space Telescope
  Multi-cycle Treasury Programs: The Early Expansion Rate}. \emph{\apj}
  853(2):126, \doi{10.3847/1538-4357/aaa5a9}, \eprint{1710.00844}

\bibitem[{{Riess} et~al.(2019){Riess}, {Casertano}, {Yuan}, {Macri}, and
  {Scolnic}}]{Riess:2019}
{Riess} A.~G., {Casertano} S., {Yuan} W., {Macri} L.~M., {Scolnic} D. (2019)
  {Large Magellanic Cloud Cepheid Standards Provide a 1\% Foundation for the
  Determination of the Hubble Constant and Stronger Evidence for Physics beyond
  {\ensuremath{\Lambda}}CDM}. \emph{\apj} 876(1):85,
  \doi{10.3847/1538-4357/ab1422}, \eprint{1903.07603}

\bibitem[{{Riess} et~al.(2020){Riess}, {Yuan}, {Casertano}, {Macri}, and
  {Scolnic}}]{Riess2020}
{Riess} A.~G., {Yuan} W., {Casertano} S., {Macri} L.~M., {Scolnic} D. (2020)
  {The Accuracy of the Hubble Constant Measurement Verified through Cepheid
  Amplitudes}. \emph{\apjl} 896(2):L43, \doi{10.3847/2041-8213/ab9900},
  \eprint{2005.02445}

\bibitem[{{Riess} et~al.(2021{\natexlab{a}}){Riess}, {Casertano}, {Yuan},
  {Bowers}, {Macri}, {Zinn}, and {Scolnic}}]{Riess:2021}
{Riess} A.~G., {Casertano} S., {Yuan} W., {Bowers} J.~B., {Macri} L., {Zinn}
  J.~C., {Scolnic} D. (2021{\natexlab{a}}) {Cosmic Distances Calibrated to 1\%
  Precision with Gaia EDR3 Parallaxes and Hubble Space Telescope Photometry of
  75 Milky Way Cepheids Confirm Tension with {\ensuremath{\Lambda}}CDM}.
  \emph{\apjl} 908(1):L6, \doi{10.3847/2041-8213/abdbaf}, \eprint{2012.08534}

\bibitem[{{Riess} et~al.(2021{\natexlab{b}}){Riess}, {Yuan}, {Macri},
  {Scolnic}, {Brout}, {Casertano}, {Jones}, {Murakami}, {Breuval}, {Brink},
  {Filippenko}, {Hoffmann}, {Jha}, {Kenworthy}, {Mackenty}, {Stahl}, and
  {Zheng}}]{Riess2021b}
{Riess} A.~G., {Yuan} W., {Macri} L.~M., {Scolnic} D., {Brout} D., {Casertano}
  S., {Jones} D.~O., {Murakami} Y., et~al. (2021{\natexlab{b}}) {A
  Comprehensive Measurement of the Local Value of the Hubble Constant with 1
  km/s/Mpc Uncertainty from the Hubble Space Telescope and the SH0ES Team}.
  \emph{arXiv e-prints} arXiv:2112.04510, \eprint{2112.04510}

\bibitem[{Riess et~al.(2004)}]{Riess:2004nr}
Riess A.~G., et~al. (2004) {Type Ia Supernova Discoveries at z>1 From the
  Hubble Space Telescope: Evidence for Past Deceleration and Constraints on
  Dark Energy Evolution}. \emph{Astrophys J} 607:665--687,
  \eprint{astro-ph/0402512}

\bibitem[{{Risaliti} and {Lusso}(2015)}]{rl15}
{Risaliti} G., {Lusso} E. (2015) {A Hubble Diagram for Quasars}. \emph{\apj}
  815:33, \doi{10.1088/0004-637X/815/1/33}, \eprint{1505.07118}

\bibitem[{{Risaliti} and {Lusso}(2019)}]{rl19}
{Risaliti} G., {Lusso} E. (2019) {Cosmological constraints from the Hubble
  diagram of quasars at high redshifts}. \emph{Nature Astronomy} p 195,
  \doi{10.1038/s41550-018-0657-z}

\bibitem[{{Rodney} et~al.(2021){Rodney}, {Brammer}, {Pierel}, {Richard},
  {Toft}, {O'Connor}, {Akhshik}, and {Whitaker}}]{Rodney:2021}
{Rodney} S.~A., {Brammer} G.~B., {Pierel} J. D.~R., {Richard} J., {Toft} S.,
  {O'Connor} K.~F., {Akhshik} M., {Whitaker} K.~E. (2021) {A gravitationally
  lensed supernova with an observable two-decade time delay}. \emph{Nature
  Astronomy} \doi{10.1038/s41550-021-01450-9}, \eprint{2106.08935}

\bibitem[{{Romanowsky} and {Kochanek}(1999)}]{Romanowsky:1999}
{Romanowsky} A.~J., {Kochanek} C.~S. (1999) {Constraints on H$_{0}$ from the
  Central Velocity Dispersions of Lens Galaxies}. \emph{\apj} 516(1):18--26,
  \doi{10.1086/307077}, \eprint{astro-ph/9805080}

\bibitem[{{Ronconi} and {Marulli}(2017)}]{Ronconi2017}
{Ronconi} T., {Marulli} F. (2017) {Cosmological exploitation of cosmic void
  statistics. New numerical tools in the CosmoBolognaLib to extract
  cosmological constraints from the void size function}. \emph{\aap} 607:A24,
  \doi{10.1051/0004-6361/201730852}, \eprint{1703.07848}

\bibitem[{{Ronconi} et~al.(2019){Ronconi}, {Contarini}, {Marulli}, {Baldi}, and
  {Moscardini}}]{Ronconi2019}
{Ronconi} T., {Contarini} S., {Marulli} F., {Baldi} M., {Moscardini} L. (2019)
  {Cosmic voids uncovered - first-order statistics of depressions in the biased
  density field}. \emph{\mnras} 488(4):5075--5084, \doi{10.1093/mnras/stz2115},
  \eprint{1902.04585}

\bibitem[{{Ruffini} et~al.(2014){Ruffini}, {Izzo}, {Muccino}, {Pisani},
  {Rueda}, {Wang}, {Barbarino}, {Bianco}, {Enderli}, and
  {Kovacevic}}]{Ruffini2014}
{Ruffini} R., {Izzo} L., {Muccino} M., {Pisani} G.~B., {Rueda} J.~A., {Wang}
  Y., {Barbarino} C., {Bianco} C.~L., et~al. (2014) {Induced gravitational
  collapse at extreme cosmological distances: the case of GRB 090423}.
  \emph{\aap} 569:A39, \doi{10.1051/0004-6361/201423457}, \eprint{1404.1840}

\bibitem[{{Ruiz} et~al.(2015){Ruiz}, {Paz}, {Lares}, {Luparello}, {Ceccarelli},
  and {Lambas}}]{Ruiz2015}
{Ruiz} A.~N., {Paz} D.~J., {Lares} M., {Luparello} H.~E., {Ceccarelli} L.,
  {Lambas} D.~G. (2015) {Clues on void evolution - III. Structure and dynamics
  in void shells}. \emph{\mnras} 448(2):1471--1482, \doi{10.1093/mnras/stv019},
  \eprint{1501.02120}

\bibitem[{{Rusu} et~al.(2017){Rusu}, {Fassnacht}, {Sluse}, {Hilbert}, {Wong},
  {Huang}, {Suyu}, {Collett}, {Marshall}, {Treu}, and {Koopmans}}]{Rusu:2017}
{Rusu} C.~E., {Fassnacht} C.~D., {Sluse} D., {Hilbert} S., {Wong} K.~C.,
  {Huang} K.-H., {Suyu} S.~H., {Collett} T.~E., et~al. (2017) {H0LiCOW - III.
  Quantifying the effect of mass along the line of sight to the gravitational
  lens HE 0435-1223 through weighted galaxy counts}. \emph{\mnras}
  467(4):4220--4242, \doi{10.1093/mnras/stx285}, \eprint{1607.01047}

\bibitem[{{Rusu} et~al.(2020){Rusu}, {Wong}, {Bonvin}, {Sluse}, {Suyu},
  {Fassnacht}, {Chan}, {Hilbert}, {Auger}, {Sonnenfeld}, {Birrer}, {Courbin},
  {Treu}, {Chen}, {Halkola}, {Koopmans}, {Marshall}, and {Shajib}}]{Rusu:2020}
{Rusu} C.~E., {Wong} K.~C., {Bonvin} V., {Sluse} D., {Suyu} S.~H., {Fassnacht}
  C.~D., {Chan} J. H.~H., {Hilbert} S., et~al. (2020) {H0LiCOW XII. Lens mass
  model of WFI2033-4723 and blind measurement of its time-delay distance and
  H$_{0}$}. \emph{\mnras} 498(1):1440--1468, \doi{10.1093/mnras/stz3451},
  \eprint{1905.09338}

\bibitem[{{Ryden}(1995)}]{Ryden1995}
{Ryden} B.~S. (1995) {Measuring Q 0 from the Distortion of Voids in Redshift
  Space}. \emph{\apj} 452:25, \doi{10.1086/176277}, \eprint{astro-ph/9506028}

\bibitem[{{Ryden} and {Melott}(1996)}]{Ryden1996}
{Ryden} B.~S., {Melott} A.~L. (1996) {Voids in Real Space and in Redshift
  Space}. \emph{\apj} 470:160, \doi{10.1086/177857}, \eprint{astro-ph/9510108}

\bibitem[{{Sacchi} et~al.(2022){Sacchi}, {Risaliti}, {Signorini}, {Lusso},
  {Nardini}, {Bargiacchi}, {Bisogni}, {Civano}, {Elvis}, {Fabbiano}, {Gilli},
  {Trefoloni}, and {Vignali}}]{sacchi2022arXiv}
{Sacchi} A., {Risaliti} G., {Signorini} M., {Lusso} E., {Nardini} E.,
  {Bargiacchi} G., {Bisogni} S., {Civano} F., et~al. (2022) {Quasars as
  high-redshift standard candles}. \emph{arXiv e-prints} arXiv:2206.13528,
  \eprint{2206.13528}

\bibitem[{{Sadler} et~al.(2020){Sadler}, {Moss}, {Allison}, {Mahony},
  {Whiting}, {Johnston}, {Ellison}, {Lagos}, and {Koribalski}}]{sadler2020}
{Sadler} E.~M., {Moss} V.~A., {Allison} J.~R., {Mahony} E.~K., {Whiting} M.~T.,
  {Johnston} H.~M., {Ellison} S.~L., {Lagos} C. d.~P., et~al. (2020) {A
  successful search for intervening 21 cm H I absorption in galaxies at 0.4 < z
  <1.0 with the Australian square kilometre array pathfinder (ASKAP)}.
  \emph{\mnras} 499(3):4293--4311, \doi{10.1093/mnras/staa2390},
  \eprint{2007.05648}

\bibitem[{{Saha} and {Williams}(2006)}]{Saha:2006}
{Saha} P., {Williams} L. L.~R. (2006) {Gravitational Lensing Model
  Degeneracies: Is Steepness All-Important?} \emph{\apj} 653(2):936--941,
  \doi{10.1086/508798}, \eprint{astro-ph/0608496}

\bibitem[{{Sahl{\'e}n}(2019)}]{Sahlen2019}
{Sahl{\'e}n} M. (2019) {Cluster-void degeneracy breaking: Neutrino properties
  and dark energy}. \emph{\prd} 99(6):063525, \doi{10.1103/PhysRevD.99.063525},
  \eprint{1807.02470}

\bibitem[{{Sahl{\'e}n} and {Silk}(2018)}]{Sahlen2018}
{Sahl{\'e}n} M., {Silk} J. (2018) {Cluster-void degeneracy breaking: Modified
  gravity in the balance}. \emph{\prd} 97(10):103504,
  \doi{10.1103/PhysRevD.97.103504}, \eprint{1612.06595}

\bibitem[{{Sahl{\'e}n} et~al.(2016){Sahl{\'e}n}, {Zubeld{\'\i}a}, and
  {Silk}}]{Sahlen2016}
{Sahl{\'e}n} M., {Zubeld{\'\i}a} {\'I}., {Silk} J. (2016) {Cluster-Void
  Degeneracy Breaking: Dark Energy, Planck, and the Largest Cluster and Void}.
  \emph{\apjl} 820(1):L7, \doi{10.3847/2041-8205/820/1/L7}, \eprint{1511.04075}

\bibitem[{{Sahlholdt} et~al.(2019){Sahlholdt}, {Feltzing}, {Lindegren}, and
  {Church}}]{Lund}
{Sahlholdt} C.~L., {Feltzing} S., {Lindegren} L., {Church} R.~P. (2019)
  {Benchmark ages for the Gaia benchmark stars}. \emph{\mnras} 482(1):895--920,
  \doi{10.1093/mnras/sty2732}, \eprint{1810.02829}

\bibitem[{{Sakamoto} et~al.(2011){Sakamoto}, {Barthelmy}, {Baumgartner},
  {Cummings}, {Fenimore}, {Gehrels}, {Krimm}, {Markwardt}, {Palmer}, {Parsons},
  {Sato}, {Stamatikos}, {Tueller}, {Ukwatta}, and {Zhang}}]{Sakamoto11}
{Sakamoto} T., {Barthelmy} S.~D., {Baumgartner} W.~H., {Cummings} J.~R.,
  {Fenimore} E.~E., {Gehrels} N., {Krimm} H.~A., {Markwardt} C.~B., et~al.
  (2011) {The Second Swift Burst Alert Telescope Gamma-Ray Burst Catalog}.
  \emph{\apjs} 195(1):2, \doi{10.1088/0067-0049/195/1/2}, \eprint{1104.4689}

\bibitem[{Sakstein and Jain(2017)}]{Sakstein2017}
Sakstein J., Jain B. (2017) {Implications of the Neutron Star Merger GW170817
  for Cosmological Scalar-Tensor Theories}. \emph{Physical Review Letters}
  119(25):251303, \doi{10.1103/PhysRevLett.119.251303}

\bibitem[{{Salpeter}(1955)}]{salpeter1955}
{Salpeter} E.~E. (1955) {The Luminosity Function and Stellar Evolution.}
  \emph{\apj} 121:161, \doi{10.1086/145971}

\bibitem[{Saltelli(2002)}]{saltelli:2002}
Saltelli A. (2002) Making best use of model evaluations to compute sensitivity
  indices. \emph{Computer Physics Communications} 145(2):280--297,
  \urlprefix\url{https://doi.org/10.1016/S0010-4655(02)00280-1}

\bibitem[{Saltelli et~al.(2010)Saltelli, Annoni, Azzini, Campolongo, Ratto, and
  Tarantola}]{saltelli:2010}
Saltelli A., Annoni P., Azzini I., Campolongo F., Ratto M., Tarantola S. (2010)
  Variance based sensitivity analysis of model output. design and estimator for
  the total sensitivity index. \emph{Computer Physics Communications}
  181(2):259--270, \doi{https://doi.org/10.1016/j.cpc.2009.09.018},
  \urlprefix\url{https://www.sciencedirect.com/science/article/pii/S0010465509003087}

\bibitem[{{Salvestrini} et~al.(2019){Salvestrini}, {Risaliti}, {Bisogni},
  {Lusso}, and {Vignali}}]{salvestrini2019}
{Salvestrini} F., {Risaliti} G., {Bisogni} S., {Lusso} E., {Vignali} C. (2019)
  {Quasars as standard candles II. The non-linear relation between UV and X-ray
  emission at high redshifts}. \emph{\aap} 631:A120,
  \doi{10.1051/0004-6361/201935491}, \eprint{1909.12309}

\bibitem[{{S{\'a}nchez}(2020)}]{Sanchez2020}
{S{\'a}nchez} A.~G. (2020) {Arguments against using h$^{-1}$ Mpc units in
  observational cosmology}. \emph{\prd} 102(12):123511,
  \doi{10.1103/PhysRevD.102.123511}, \eprint{2002.07829}

\bibitem[{{S{\'a}nchez} et~al.(2017){S{\'a}nchez}, {Clampitt}, {Kovacs},
  {Jain}, {Garc{\'\i}a-Bellido}, {Nadathur}, {Gruen}, {Hamaus}, {Huterer},
  {Vielzeuf}, {Amara}, et~al., and {DES Collaboration}}]{Sanchez2017}
{S{\'a}nchez} C., {Clampitt} J., {Kovacs} A., {Jain} B., {Garc{\'\i}a-Bellido}
  J., {Nadathur} S., {Gruen} D., {Hamaus} N., et~al. (2017) {Cosmic voids and
  void lensing in the Dark Energy Survey Science Verification data}.
  \emph{\mnras} 465(1):746--759, \doi{10.1093/mnras/stw2745},
  \eprint{1605.03982}

\bibitem[{{S{\'a}nchez-Bl{\'a}zquez} et~al.(2006){S{\'a}nchez-Bl{\'a}zquez},
  {Peletier}, {Jim{\'e}nez-Vicente}, {Cardiel}, {Cenarro},
  {Falc{\'o}n-Barroso}, {Gorgas}, {Selam}, and
  {Vazdekis}}]{sanchezblazquez2006}
{S{\'a}nchez-Bl{\'a}zquez} P., {Peletier} R.~F., {Jim{\'e}nez-Vicente} J.,
  {Cardiel} N., {Cenarro} A.~J., {Falc{\'o}n-Barroso} J., {Gorgas} J., {Selam}
  S., et~al. (2006) {Medium-resolution Isaac Newton Telescope library of
  empirical spectra}. \emph{\mnras} 371(2):703--718,
  \doi{10.1111/j.1365-2966.2006.10699.x}, \eprint{astro-ph/0607009}

\bibitem[{{Sandage}(1962)}]{sandage1962}
{Sandage} A. (1962) {The Change of Redshift and Apparent Luminosity of Galaxies
  due to the Deceleration of Selected Expanding Universes.} \emph{ApJ} 136:319,
  \doi{10.1086/147385}

\bibitem[{{Sandage} and {Schwarzschild}(1952)}]{Sandage}
{Sandage} A.~R., {Schwarzschild} M. (1952) {Inhomogeneous Stellar Models. II.
  Models with Exhausted Cores in Gravitational Contraction.} \emph{\apj}
  116:463, \doi{10.1086/145638}

\bibitem[{Vargas~dos Santos et~al.(2019)Vargas~dos Santos, Quartin, and
  Reis}]{VargasdosSantos:2019ovq}
Vargas~dos Santos M., Quartin M., Reis R. R.~R. (2019) {On the cosmological
  performance of photometric classified supernovae with machine learning}.
  \emph{MNRAS} \doi{10.1093/mnras/staa1968}, \eprint{1908.04210}

\bibitem[{Santos et~al.(2005)Santos, Cooray, and Knox}]{Santos:2004ju}
Santos M.~G., Cooray A., Knox L. (2005) {Multifrequency analysis of 21 cm
  fluctuations from the era of reionization}. \emph{Astrophys J} 625:575--587,
  \doi{10.1086/429857}, \eprint{astro-ph/0408515}

\bibitem[{Santos et~al.(2015)}]{Santos:2015gra}
Santos M.~G., et~al. (2015) {Cosmology from a SKA HI intensity mapping survey}.
  \emph{PoS} AASKA14:019, \doi{10.22323/1.215.0019}, \eprint{1501.03989}

\bibitem[{Santos et~al.(2017)}]{Santos:2017qgq}
Santos M.~G., et~al. (2017) {MeerKLASS: MeerKAT Large Area Synoptic Survey}.
  In: {MeerKAT Science}: {On the Pathway to the SKA}, \eprint{1709.06099}

\bibitem[{{Santos-da-Costa} et~al.(2015){Santos-da-Costa}, {Busti}, and
  {Holanda}}]{santos2015}
{Santos-da-Costa} S., {Busti} V.~C., {Holanda} R. F.~L. (2015) {Two new tests
  to the distance duality relation with galaxy clusters}. \emph{\jcap}
  2015(10):061, \doi{10.1088/1475-7516/2015/10/061}, \eprint{1506.00145}

\bibitem[{{Sapone} et~al.(2014){Sapone}, {Majerotto}, and
  {Nesseris}}]{sapone2014}
{Sapone} D., {Majerotto} E., {Nesseris} S. (2014) {Curvature versus distances:
  Testing the FLRW cosmology}. \emph{\prd} 90(2):023012,
  \doi{10.1103/PhysRevD.90.023012}, \eprint{1402.2236}

\bibitem[{{Saracco} et~al.(2019){Saracco}, {La Barbera}, {Gargiulo},
  {Mannucci}, {Marchesini}, {Nonino}, and {Ciliegi}}]{saracco2019}
{Saracco} P., {La Barbera} F., {Gargiulo} A., {Mannucci} F., {Marchesini} D.,
  {Nonino} M., {Ciliegi} P. (2019) {Age, metallicity, and star formation
  history of spheroidal galaxies in cluster at $z\sim1.2$}. \emph{\mnras}
  484(2):2281--2295, \doi{10.1093/mnras/sty3509}, \eprint{1901.01595}

\bibitem[{{Sathyaprakash} et~al.(2010){Sathyaprakash}, {Schutz}, and {Van Den
  Broeck}}]{2010CQGra..27u5006S}
{Sathyaprakash} B.~S., {Schutz} B.~F., {Van Den Broeck} C. (2010) {Cosmography
  with the Einstein Telescope}. \emph{Classical and Quantum Gravity}
  27(21):215006, \doi{10.1088/0264-9381/27/21/215006}, \eprint{0906.4151}

\bibitem[{{Scelfo} et~al.(2022){Scelfo}, {Spinelli}, {Raccanelli}, {Boco},
  {Lapi}, and {Viel}}]{Scelfo:2021fqe}
{Scelfo} G., {Spinelli} M., {Raccanelli} A., {Boco} L., {Lapi} A., {Viel} M.
  (2022) {Gravitational waves {\texttimes} HI intensity mapping: cosmological
  and astrophysical applications}. \emph{\jcap} 2022(1):004,
  \doi{10.1088/1475-7516/2022/01/004}, \eprint{2106.09786}

\bibitem[{{Schlegel} et~al.(1998){Schlegel}, {Finkbeiner}, and
  {Davis}}]{schlegel98}
{Schlegel} D.~J., {Finkbeiner} D.~P., {Davis} M. (1998) {Maps of Dust Infrared
  Emission for Use in Estimation of Reddening and Cosmic Microwave Background
  Radiation Foregrounds}. \emph{\apj} 500:525, \doi{10.1086/305772},
  \eprint{arXiv:astro-ph/9710327}

\bibitem[{{Schmidt} et~al.(1998){Schmidt}, {Suntzeff}, {Phillips}, {Schommer},
  {Clocchiatti}, {Kirshner}, {Garnavich}, {Challis}, {Leibundgut},
  {Spyromilio}, {Riess}, {Filippenko}, {Hamuy}, {Smith}, {Hogan}, {Stubbs},
  {Diercks}, {Reiss}, {Gilliland}, {Tonry}, {Maza}, {Dressler}, {Walsh}, and
  {Ciardullo}}]{schmidt1998}
{Schmidt} B.~P., {Suntzeff} N.~B., {Phillips} M.~M., {Schommer} R.~A.,
  {Clocchiatti} A., {Kirshner} R.~P., {Garnavich} P., {Challis} P., et~al.
  (1998) {The High-Z Supernova Search: Measuring Cosmic Deceleration and Global
  Curvature of the Universe Using Type IA Supernovae}. \emph{\apj}
  507(1):46--63, \doi{10.1086/306308}, \eprint{astro-ph/9805200}

\bibitem[{{Schneider}(1985)}]{Schneider:1985}
{Schneider} P. (1985) {A new formulation of gravitational lens theory,
  time-delay, and Fermat's principle}. \emph{\aap} 143(2):413--420

\bibitem[{{Schneider} and {Sluse}(2013)}]{Schneider:2013}
{Schneider} P., {Sluse} D. (2013) {Mass-sheet degeneracy, power-law models and
  external convergence: Impact on the determination of the Hubble constant from
  gravitational lensing}. \emph{\aap} 559:A37,
  \doi{10.1051/0004-6361/201321882}, \eprint{1306.0901}

\bibitem[{{Schneider} and {Sluse}(2014)}]{Schneider:2014}
{Schneider} P., {Sluse} D. (2014) {Source-position transformation: an
  approximate invariance in strong gravitational lensing}. \emph{\aap}
  564:A103, \doi{10.1051/0004-6361/201322106}, \eprint{1306.4675}

\bibitem[{{Schneider} et~al.(1992){Schneider}, {Ehlers}, and
  {Falco}}]{Schneider:1992}
{Schneider} P., {Ehlers} J., {Falco} E.~E. (1992) {Gravitational Lenses}.
  Springer-Verlag Berlin Heidelberg New York, \doi{10.1007/978-3-662-03758-4}

\bibitem[{{Schreiber} et~al.(2018){Schreiber}, {Glazebrook}, {Nanayakkara},
  {Kacprzak}, {Labb{\'e}}, {Oesch}, {Yuan}, {Tran}, {Papovich}, {Spitler}, and
  {Straatman}}]{schreiber2018}
{Schreiber} C., {Glazebrook} K., {Nanayakkara} T., {Kacprzak} G.~G.,
  {Labb{\'e}} I., {Oesch} P., {Yuan} T., {Tran} K.~V., et~al. (2018) {Near
  infrared spectroscopy and star-formation histories of 3 {\ensuremath{\leq}} z
  {\ensuremath{\leq}} 4 quiescent galaxies}. \emph{\aap} 618:A85,
  \doi{10.1051/0004-6361/201833070}, \eprint{1807.02523}

\bibitem[{{Schuster} et~al.(2019){Schuster}, {Hamaus}, {Pisani}, {Carbone},
  {Kreisch}, {Pollina}, and {Weller}}]{Schuster2019}
{Schuster} N., {Hamaus} N., {Pisani} A., {Carbone} C., {Kreisch} C.~D.,
  {Pollina} G., {Weller} J. (2019) {The bias of cosmic voids in the presence of
  massive neutrinos}. \emph{\jcap} 2019(12):055,
  \doi{10.1088/1475-7516/2019/12/055}, \eprint{1905.00436}

\bibitem[{{Schutz}(1986)}]{1986Natur.323..310S}
{Schutz} B.~F. (1986) {Determining the Hubble constant from gravitational wave
  observations}. \emph{\nat} 323(6086):310--311, \doi{10.1038/323310a0}

\bibitem[{{Schwarzschild}(1979)}]{Schwarzschild:1979}
{Schwarzschild} M. (1979) {A numerical model for a triaxial stellar system in
  dynamical equilibrium.} \emph{\apj} 232:236--247, \doi{10.1086/157282}

\bibitem[{{Scolnic} et~al.(2018){Scolnic}, {Jones}, {Rest}, {Pan}, {Chornock},
  {Foley}, {Huber}, {Kessler}, {Narayan}, {Riess}, {Rodney}, {Berger}, {Brout},
  and et~al.}]{scolnic2018}
{Scolnic} D.~M., {Jones} D.~O., {Rest} A., {Pan} Y.~C., {Chornock} R., {Foley}
  R.~J., {Huber} M.~E., {Kessler} R., et~al. (2018) {The Complete Light-curve
  Sample of Spectroscopically Confirmed SNe Ia from Pan-STARRS1 and
  Cosmological Constraints from the Combined Pantheon Sample}. \emph{\apj}
  859(2):101, \doi{10.3847/1538-4357/aab9bb}, \eprint{1710.00845}

\bibitem[{Scovacricchi et~al.(2017)Scovacricchi, Nichol, Macaulay, and
  Bacon}]{Scovacricchi:2016ylt}
Scovacricchi D., Nichol R.~C., Macaulay E., Bacon D. (2017) {Measuring weak
  lensing correlations of Type Ia Supernovae}. \emph{Mon Not Roy Astron Soc}
  465(3):2862--2872, \doi{10.1093/mnras/stw2878}, \eprint{1611.01315}

\bibitem[{{Secco} et~al.(2021){Secco}, {Samuroff}, {Krause}, {Jain}, {Blazek},
  {Raveri}, {Campos}, {Amon}, {Chen}, {Doux}, {Choi}, {Gruen}, {Bernstein}, and
  et~al.}]{Secco2021}
{Secco} L.~F., {Samuroff} S., {Krause} E., {Jain} B., {Blazek} J., {Raveri} M.,
  {Campos} A., {Amon} A., et~al. (2021) {Dark Energy Survey Year 3 Results:
  Cosmology from Cosmic Shear and Robustness to Modeling Uncertainty}.
  \emph{arXiv e-prints} arXiv:2105.13544, \eprint{2105.13544}

\bibitem[{{Seikel} et~al.(2012){Seikel}, {Yahya}, {Maartens}, and
  {Clarkson}}]{seikel2012}
{Seikel} M., {Yahya} S., {Maartens} R., {Clarkson} C. (2012) {Using H(z) data
  as a probe of the concordance model}. \emph{\prd} 86(8):083001,
  \doi{10.1103/PhysRevD.86.083001}, \eprint{1205.3431}

\bibitem[{Seljak(2009)}]{Seljak:2009}
Seljak U. (2009) {Extracting primordial non-gaussianity without cosmic
  variance}. \emph{Phys Rev Lett} 102:021302,
  \doi{10.1103/PhysRevLett.102.021302}, \eprint{0807.1770}

\bibitem[{Seo and Eisenstein(2003)}]{Seo:2003pu}
Seo H.-J., Eisenstein D.~J. (2003) {Probing dark energy with baryonic acoustic
  oscillations from future large galaxy redshift surveys}. \emph{Astrophys J}
  598:720--740, \doi{10.1086/379122}, \eprint{astro-ph/0307460}

\bibitem[{Seo et~al.(2010)Seo, Dodelson, Marriner, Mcginnis, Stebbins,
  Stoughton, and Vallinotto}]{Seo:2009fq}
Seo H.-J., Dodelson S., Marriner J., Mcginnis D., Stebbins A., Stoughton C.,
  Vallinotto A. (2010) {A ground-based 21cm Baryon acoustic oscillation
  survey}. \emph{Astrophys J} 721:164--173, \doi{10.1088/0004-637X/721/1/164},
  \eprint{0910.5007}

\bibitem[{{Shajib} et~al.(2018){Shajib}, {Treu}, and {Agnello}}]{Shajib:2018}
{Shajib} A.~J., {Treu} T., {Agnello} A. (2018) {Improving time-delay
  cosmography with spatially resolved kinematics}. \emph{\mnras}
  473(1):210--226, \doi{10.1093/mnras/stx2302}, \eprint{1709.01517}

\bibitem[{{Shajib} et~al.(2020){Shajib}, {Birrer}, {Treu}, {Agnello},
  {Buckley-Geer}, {Chan}, {Christensen}, {Lemon}, {Lin}, {Millon}, {Poh},
  {Rusu}, {Sluse}, {Spiniello}, {Chen}, {Collett}, {Courbin}, {Fassnacht}, and
  et~al.}]{Shajib:2020}
{Shajib} A.~J., {Birrer} S., {Treu} T., {Agnello} A., {Buckley-Geer} E.~J.,
  {Chan} J.~H.~H., {Christensen} L., {Lemon} C., et~al. (2020) {STRIDES: a 3.9
  per cent measurement of the Hubble constant from the strong lens system DES
  J0408-5354}. \emph{\mnras} 494(4):6072--6102, \doi{10.1093/mnras/staa828},
  \eprint{1910.06306}

\bibitem[{{Shajib} et~al.(2021){Shajib}, {Treu}, {Birrer}, and
  {Sonnenfeld}}]{Shajib:2021slacs}
{Shajib} A.~J., {Treu} T., {Birrer} S., {Sonnenfeld} A. (2021) {Dark matter
  haloes of massive elliptical galaxies at z {\ensuremath{\sim}} 0.2 are well
  described by the Navarro-Frenk-White profile}. \emph{\mnras}
  503(2):2380--2405, \doi{10.1093/mnras/stab536}, \eprint{2008.11724}

\bibitem[{{Shakura} and {Sunyaev}(1973)}]{ss1973}
{Shakura} N.~I., {Sunyaev} R.~A. (1973) {Reprint of 1973A\&A....24..337S. Black
  holes in binary systems. Observational appearance.} \emph{\aap} 500:33--51

\bibitem[{Shaw et~al.(2015)Shaw, Sigurdson, Sitwell, Stebbins, and
  Pen}]{Shaw:2014khi}
Shaw J.~R., Sigurdson K., Sitwell M., Stebbins A., Pen U.-L. (2015) {Coaxing
  cosmic 21 cm fluctuations from the polarized sky using m-mode analysis}.
  \emph{Phys Rev D} 91(8):083514, \doi{10.1103/PhysRevD.91.083514},
  \eprint{1401.2095}

\bibitem[{{Sheth}(2005)}]{Sheth2005}
{Sheth} R.~K. (2005) {The halo-model description of marked statistics}.
  \emph{\mnras} 364(3):796--806, \doi{10.1111/j.1365-2966.2005.09609.x},
  \eprint{astro-ph/0511772}

\bibitem[{{Sheth} and {van de Weygaert}(2004)}]{Sheth2004}
{Sheth} R.~K., {van de Weygaert} R. (2004) {A hierarchy of voids: much ado
  about nothing}. \emph{\mnras} 350:517--538,
  \doi{10.1111/j.1365-2966.2004.07661.x}, \eprint{astro-ph/0311260}

\bibitem[{{Shim} et~al.(2021){Shim}, {Park}, {Kim}, and {Hwang}}]{Shim2021}
{Shim} J., {Park} C., {Kim} J., {Hwang} H.~S. (2021) {Identification of Cosmic
  Voids as Massive Cluster Counterparts}. \emph{\apj} 908(2):211,
  \doi{10.3847/1538-4357/abd0f6}, \eprint{2012.03511}

\bibitem[{{Shirokov} et~al.(2020){Shirokov}, {Sokolov}, {Lovyagin}, {Amati},
  {Baryshev}, {Sokolov}, and {Gorokhov}}]{Shirokov20}
{Shirokov} S.~I., {Sokolov} I.~V., {Lovyagin} N.~Y., {Amati} L., {Baryshev}
  Y.~V., {Sokolov} V.~V., {Gorokhov} V.~L. (2020) {High-redshift long gamma-ray
  bursts Hubble diagram as a test of basic cosmological relations}.
  \emph{\mnras} 496(2):1530--1544, \doi{10.1093/mnras/staa1548},
  \eprint{2006.00981}

\bibitem[{{Si} et~al.(2018){Si}, {Qi}, {Xue}, {Liu}, {Wu}, {Yi}, {Tang}, {Zou},
  {Wang}, and {Wang}}]{Si18}
{Si} S.-K., {Qi} Y.-Q., {Xue} F.-X., {Liu} Y.-J., {Wu} X., {Yi} S.-X., {Tang}
  Q.-W., {Zou} Y.-C., et~al. (2018) {The Three-parameter Correlations About the
  Optical Plateaus of Gamma-Ray Bursts}. \emph{\apj} 863(1):50,
  \doi{10.3847/1538-4357/aad08a}, \eprint{1807.00241}

\bibitem[{{Simon} et~al.(2005){Simon}, {Verde}, and {Jimenez}}]{simon2005}
{Simon} J., {Verde} L., {Jimenez} R. (2005) {Constraints on the redshift
  dependence of the dark energy potential}. \emph{\prd} 71(12):123001,
  \doi{10.1103/PhysRevD.71.123001}, \eprint{astro-ph/0412269}

\bibitem[{{SKA Cosmology SWG}(2020)}]{SKARedBook2018}
{SKA Cosmology SWG} (2020) {Cosmology with Phase 1 of the Square Kilometre
  Array: Red Book 2018: Technical specifications and performance forecasts}.
  \emph{Publ Astron Soc Austral} 37:e007, \doi{10.1017/pasa.2019.51},
  \eprint{1811.02743}

\bibitem[{{Skrutskie} et~al.(2006){Skrutskie}, {Cutri}, {Stiening}, {Weinberg},
  {Schneider}, {Carpenter}, {Beichman}, {Capps}, {Chester}, {Elias}, {Huchra},
  {Liebert}, {Lonsdale}, {Monet}, {Price}, {Seitzer}, {Jarrett}, {Kirkpatrick},
  {Gizis}, {Howard}, {Evans}, {Fowler}, {Fullmer}, {Hurt}, {Light}, {Kopan},
  {Marsh}, {McCallon}, {Tam}, {Van Dyk}, and {Wheelock}}]{2006AJ....131.1163S}
{Skrutskie} M.~F., {Cutri} R.~M., {Stiening} R., {Weinberg} M.~D., {Schneider}
  S., {Carpenter} J.~M., {Beichman} C., {Capps} R., et~al. (2006) {The Two
  Micron All Sky Survey (2MASS)}. \emph{The Astrophysical Journal}
  131(2):1163--1183, \doi{10.1086/498708}

\bibitem[{{Slosar} et~al.(2019){Slosar}, {Ahmed}, {Alonso}, {Amin}, {Arena},
  {Bandura}, {Battaglia}, {Blazek}, {Bull}, {Castorina}, {Chang}, {Connor},
  {Dav{\'e}}, {Dvorkin}, {van Engelen}, {Ferraro}, and et~al.}]{PUMA:2019jwd}
{Slosar} A., {Ahmed} Z., {Alonso} D., {Amin} M.~A., {Arena} E.~J., {Bandura}
  K., {Battaglia} N., {Blazek} J., et~al. (2019) {Packed Ultra-wideband Mapping
  Array (PUMA): A Radio Telescope for Cosmology and Transients}. In: Bulletin
  of the American Astronomical Society, vol~51, p~53, \eprint{1907.12559}

\bibitem[{{Sluse} et~al.(2003){Sluse}, {Surdej}, {Claeskens},
  {Hutsem{\'e}kers}, {Jean}, {Courbin}, {Nakos}, {Billeres}, and
  {Khmil}}]{Sluse:2003}
{Sluse} D., {Surdej} J., {Claeskens} J.~F., {Hutsem{\'e}kers} D., {Jean} C.,
  {Courbin} F., {Nakos} T., {Billeres} M., et~al. (2003) {A quadruply imaged
  quasar with an optical Einstein ring candidate: 1RXS J113155.4-123155}.
  \emph{\aap} 406:L43--L46, \doi{10.1051/0004-6361:20030904},
  \eprint{astro-ph/0307345}

\bibitem[{{Sluse} et~al.(2017{\natexlab{a}}){Sluse}, {Sonnenfeld}, {Rumbaugh},
  {Rusu}, {Fassnacht}, {Treu}, {Suyu}, {Wong}, {Auger}, {Bonvin}, {Collett},
  {Courbin}, {Hilbert}, {Koopmans}, {Marshall}, {Meylan}, {Spiniello}, and
  {Tewes}}]{Sluse2017}
{Sluse} D., {Sonnenfeld} A., {Rumbaugh} N., {Rusu} C.~E., {Fassnacht} C.~D.,
  {Treu} T., {Suyu} S.~H., {Wong} K.~C., et~al. (2017{\natexlab{a}}) {H0LiCOW -
  II. Spectroscopic survey and galaxy-group identification of the strong
  gravitational lens system HE 0435-1223}. \emph{\mnras} 470(4):4838--4857,
  \doi{10.1093/mnras/stx1484}, \eprint{1607.00382}

\bibitem[{{Sluse} et~al.(2017{\natexlab{b}}){Sluse}, {Sonnenfeld}, {Rumbaugh},
  {Rusu}, {Fassnacht}, {Treu}, {Suyu}, {Wong}, {Auger}, {Bonvin}, {Collett},
  {Courbin}, {Hilbert}, {Koopmans}, {Marshall}, {Meylan}, {Spiniello}, and
  {Tewes}}]{Sluse:2017}
{Sluse} D., {Sonnenfeld} A., {Rumbaugh} N., {Rusu} C.~E., {Fassnacht} C.~D.,
  {Treu} T., {Suyu} S.~H., {Wong} K.~C., et~al. (2017{\natexlab{b}}) {H0LiCOW -
  II. Spectroscopic survey and galaxy-group identification of the strong
  gravitational lens system HE 0435-1223}. \emph{\mnras} 470(4):4838--4857,
  \doi{10.1093/mnras/stx1484}, \eprint{1607.00382}

\bibitem[{{Sluse} et~al.(2019){Sluse}, {Rusu}, {Fassnacht}, {Sonnenfeld},
  {Richard}, {Auger}, {Coccato}, {Wong}, {Suyu}, {Treu}, {Agnello}, {Birrer},
  and et~al.}]{Sluse:2019}
{Sluse} D., {Rusu} C.~E., {Fassnacht} C.~D., {Sonnenfeld} A., {Richard} J.,
  {Auger} M.~W., {Coccato} L., {Wong} K.~C., et~al. (2019) {H0LiCOW - X.
  Spectroscopic/imaging survey and galaxy-group identification around the
  strong gravitational lens system WFI 2033-4723}. \emph{\mnras}
  490(1):613--633, \doi{10.1093/mnras/stz2483}, \eprint{1905.08800}

\bibitem[{{Smoot} et~al.(1992){Smoot}, {Bennett}, {Kogut}, {Wright}, {Aymon},
  {Boggess}, {Cheng}, {de Amici}, {Gulkis}, {Hauser}, {Hinshaw}, {Jackson},
  {Janssen}, {Kaita}, {Kelsall}, {Keegstra}, {Lineweaver}, {Loewenstein},
  {Lubin}, {Mather}, {Meyer}, {Moseley}, {Murdock}, {Rokke}, {Silverberg},
  {Tenorio}, {Weiss}, and {Wilkinson}}]{COBE}
{Smoot} G.~F., {Bennett} C.~L., {Kogut} A., {Wright} E.~L., {Aymon} J.,
  {Boggess} N.~W., {Cheng} E.~S., {de Amici} G., et~al. (1992) {Structure in
  the COBE Differential Microwave Radiometer First-Year Maps}. \emph{\apjl}
  396:L1, \doi{10.1086/186504}

\bibitem[{Soares et~al.(2021)Soares, Cunnington, Pourtsidou, and
  Blake}]{Soares:2020zaq}
Soares P.~S., Cunnington S., Pourtsidou A., Blake C. (2021) {Power spectrum
  multipole expansion for HI intensity mapping experiments: unbiased parameter
  estimation}. \emph{Mon Not Roy Astron Soc} 502(2):2549--2564,
  \doi{10.1093/mnras/stab027}, \eprint{2008.12102}

\bibitem[{{Soares} et~al.(2021){Soares}, {Watkinson}, {Cunnington}, and
  {Pourtsidou}}]{Soares:2021ohm}
{Soares} P.~S., {Watkinson} C.~A., {Cunnington} S., {Pourtsidou} A. (2021)
  {Gaussian Process Regression for foreground removal in HI intensity mapping
  experiments}. \emph{\mnras} \doi{10.1093/mnras/stab2594}, \eprint{2105.12665}

\bibitem[{{Soares-Santos} et~al.(2019){Soares-Santos}, {Palmese}, {Hartley},
  {Annis}, {Garcia-Bellido}, {Lahav}, {Doctor}, {Fishbach}, {Holz}, {Lin}, and
  et~al.}]{2019ApJ...876L...7S}
{Soares-Santos} M., {Palmese} A., {Hartley} W., {Annis} J., {Garcia-Bellido}
  J., {Lahav} O., {Doctor} Z., {Fishbach} M., et~al. (2019) {First Measurement
  of the Hubble Constant from a Dark Standard Siren using the Dark Energy
  Survey Galaxies and the LIGO/Virgo Binary-Black-hole Merger GW170814}.
  \emph{\apjl} 876(1):L7, \doi{10.3847/2041-8213/ab14f1}, \eprint{1901.01540}

\bibitem[{Sobol'(2001)}]{sobol:2001}
Sobol' I. (2001) Global sensitivity indices for nonlinear mathematical models
  and their monte carlo estimates. \emph{Mathematics and Computers in
  Simulation} 55(1):271--280,
  \doi{https://doi.org/10.1016/S0378-4754(00)00270-6},
  \urlprefix\url{https://www.sciencedirect.com/science/article/pii/S0378475400002706},
  the Second IMACS Seminar on Monte Carlo Methods

\bibitem[{{Soderblom}(2010)}]{Soderblom}
{Soderblom} D.~R. (2010) {The Ages of Stars}. \emph{\araa} 48:581--629,
  \doi{10.1146/annurev-astro-081309-130806}, \eprint{1003.6074}

\bibitem[{{Solomon} and {Stojkovic}(2022)}]{ss2022}
{Solomon} R., {Stojkovic} D. (2022) {Variability in quasar light curves: using
  quasars as standard candles}. \emph{\jcap} 2022(4):060,
  \doi{10.1088/1475-7516/2022/04/060}, \eprint{2110.03671}

\bibitem[{{Soltis} et~al.(2021){Soltis}, {Casertano}, and {Riess}}]{soltis21}
{Soltis} J., {Casertano} S., {Riess} A.~G. (2021) {The Parallax of
  {\ensuremath{\omega}} Centauri Measured from Gaia EDR3 and a Direct,
  Geometric Calibration of the Tip of the Red Giant Branch and the Hubble
  Constant}. \emph{\apjl} 908(1):L5, \doi{10.3847/2041-8213/abdbad},
  \eprint{2012.09196}

\bibitem[{{Sonnenfeld}(2018)}]{Sonnenfeld:2018}
{Sonnenfeld} A. (2018) {On the choice of lens density profile in time delay
  cosmography}. \emph{\mnras} 474(4):4648--4659, \doi{10.1093/mnras/stx3105},
  \eprint{1710.05925}

\bibitem[{{Soucail} et~al.(2004){Soucail}, {Kneib}, and {Golse}}]{Soucail:2004}
{Soucail} G., {Kneib} J.~P., {Golse} G. (2004) {Multiple-images in the cluster
  lens Abell 2218: Constraining the geometry of the Universe?} \emph{\aap}
  417:L33--L37, \doi{10.1051/0004-6361:20040077}, \eprint{astro-ph/0402658}

\bibitem[{{Spergel} et~al.(2015){Spergel}, {Gehrels}, {Baltay}, {Bennett},
  {Breckinridge}, {Donahue}, {Dressler}, {Gaudi}, {Greene}, {Guyon}, {Hirata},
  {Kalirai}, and et~al.}]{Spergel:2015}
{Spergel} D., {Gehrels} N., {Baltay} C., {Bennett} D., {Breckinridge} J.,
  {Donahue} M., {Dressler} A., {Gaudi} B.~S., et~al. (2015) {Wide-Field
  InfrarRed Survey Telescope-Astrophysics Focused Telescope Assets WFIRST-AFTA
  2015 Report}. \emph{arXiv e-prints} arXiv:1503.03757, \eprint{1503.03757}

\bibitem[{Spinelli et~al.(2020)Spinelli, Zoldan, De~Lucia, Xie, and
  Viel}]{Spinelli:2019smg}
Spinelli M., Zoldan A., De~Lucia G., Xie L., Viel M. (2020) {The atomic
  Hydrogen content of the post-reionization Universe}. \emph{Mon Not Roy Astron
  Soc} 493(4):5434--5455, \doi{10.1093/mnras/staa604}, \eprint{1909.02242}

\bibitem[{{Spinelli} et~al.(2022){Spinelli}, {Carucci}, {Cunnington}, {Harper},
  {Irfan}, {Fonseca}, {Pourtsidou}, and {Wolz}}]{Spinelli:2021emp}
{Spinelli} M., {Carucci} I.~P., {Cunnington} S., {Harper} S.~E., {Irfan} M.~O.,
  {Fonseca} J., {Pourtsidou} A., {Wolz} L. (2022) {SKAO H I intensity mapping:
  blind foreground subtraction challenge}. \emph{\mnras} 509(2):2048--2074,
  \doi{10.1093/mnras/stab3064}, \eprint{2107.10814}

\bibitem[{{Spinrad} et~al.(1997){Spinrad}, {Dey}, {Stern}, {Dunlop}, {Peacock},
  {Jimenez}, and {Windhorst}}]{Spinrad}
{Spinrad} H., {Dey} A., {Stern} D., {Dunlop} J., {Peacock} J., {Jimenez} R.,
  {Windhorst} R. (1997) {LBDS 53W091: An Old, Red Galaxy at z = 1.552}.
  \emph{\apj} 484(2):581--601, \doi{10.1086/304381}, \eprint{astro-ph/9702233}

\bibitem[{{Spolyar} et~al.(2013){Spolyar}, {Sahl{\'e}n}, and
  {Silk}}]{Spolyar2013}
{Spolyar} D., {Sahl{\'e}n} M., {Silk} J. (2013) {Topology and Dark Energy:
  Testing Gravity in Voids}. \emph{Physical Review Letters} 111(24):241103,
  \doi{10.1103/PhysRevLett.111.241103}, \eprint{1304.5239}

\bibitem[{{Stark} et~al.(2015){Stark}, {Font-Ribera}, {White}, and
  {Lee}}]{Stark2015}
{Stark} C.~W., {Font-Ribera} A., {White} M., {Lee} K.-G. (2015) {Finding
  high-redshift voids using Lyman {\ensuremath{\alpha}} forest tomography}.
  \emph{\mnras} 453(4):4311--4323, \doi{10.1093/mnras/stv1868},
  \eprint{1504.03290}

\bibitem[{{Stefanon} et~al.(2013){Stefanon}, {Marchesini}, {Rudnick},
  {Brammer}, and {Whitaker}}]{stefanon2013}
{Stefanon} M., {Marchesini} D., {Rudnick} G.~H., {Brammer} G.~B., {Whitaker}
  K.~E. (2013) {What are the Progenitors of Compact, Massive, Quiescent
  Galaxies at z = 2.3? The Population of Massive Galaxies at z > 3 from NMBS
  and CANDELS}. \emph{\apj} 768(1):92, \doi{10.1088/0004-637X/768/1/92},
  \eprint{1301.7063}

\bibitem[{{Steinhardt} et~al.(2020){Steinhardt}, {Jauzac}, {Acebron}, {Atek},
  {Capak}, {Davidzon}, {Eckert}, {Harvey}, {Koekemoer}, {Lagos}, and
  et~al.}]{Steinhardt:2020}
{Steinhardt} C.~L., {Jauzac} M., {Acebron} A., {Atek} H., {Capak} P.,
  {Davidzon} I., {Eckert} D., {Harvey} D., et~al. (2020) {The BUFFALO HST
  Survey}. \emph{\apjs} 247(2):64, \doi{10.3847/1538-4365/ab75ed},
  \eprint{2001.09999}

\bibitem[{{Stern} et~al.(2010){Stern}, {Jimenez}, {Verde}, {Kamionkowski}, and
  {Stanford}}]{stern2010}
{Stern} D., {Jimenez} R., {Verde} L., {Kamionkowski} M., {Stanford} S.~A.
  (2010) {Cosmic chronometers: constraining the equation of state of dark
  energy. I: H(z) measurements}. \emph{\jcap} 2:008,
  \doi{10.1088/1475-7516/2010/02/008}, \eprint{0907.3149}

\bibitem[{{Stern} et~al.(2021){Stern}, {Djorgovski}, {Krone-Martins}, {Sluse},
  {Delchambre}, {Ducourant}, {Teixeira}, {Surdej}, {Boehm}, {den Brok},
  {Dobie}, {Drake}, {Galluccio}, {Graham}, {Jalan}, {Kl{\"u}ter}, {Le Campion},
  {Mahabal}, {Mignard}, {Murphy}, {Nierenberg}, {Scarano}, {Simon}, {Slezak},
  {Spindola-Duarte}, and {Wambsganss}}]{Stern21}
{Stern} D., {Djorgovski} S.~G., {Krone-Martins} A., {Sluse} D., {Delchambre}
  L., {Ducourant} C., {Teixeira} R., {Surdej} J., et~al. (2021) {Gaia GraL:
  Gaia DR2 Gravitational Lens Systems. VI. Spectroscopic Confirmation and
  Modeling of Quadruply Imaged Lensed Quasars}. \emph{\apj} 921(1):42,
  \doi{10.3847/1538-4357/ac0f04}, \eprint{2012.10051}

\bibitem[{{Stopyra} et~al.(2021){Stopyra}, {Peiris}, and
  {Pontzen}}]{Stopyra2021}
{Stopyra} S., {Peiris} H.~V., {Pontzen} A. (2021) {How to build a catalogue of
  linearly evolving cosmic voids}. \emph{\mnras} 500(3):4173--4180,
  \doi{10.1093/mnras/staa3587}, \eprint{2007.14395}

\bibitem[{{Straatman} et~al.(2014){Straatman}, {Labb{\'e}}, {Spitler}, {Allen},
  {Altieri}, {Brammer}, {Dickinson}, {van Dokkum}, {Inami}, {Glazebrook},
  {Kacprzak}, {Kawinwanichakij}, {Kelson}, {McCarthy}, {Mehrtens}, {Monson},
  {Murphy}, {Papovich}, {Persson}, {Quadri}, {Rees}, {Tomczak}, {Tran}, and
  {Tilvi}}]{straatman2014}
{Straatman} C. M.~S., {Labb{\'e}} I., {Spitler} L.~R., {Allen} R., {Altieri}
  B., {Brammer} G.~B., {Dickinson} M., {van Dokkum} P., et~al. (2014) {A
  Substantial Population of Massive Quiescent Galaxies at z
  \raisebox{-0.5ex}\textasciitilde 4 from ZFOURGE}. \emph{\apjl} 783(1):L14,
  \doi{10.1088/2041-8205/783/1/L14}, \eprint{1312.4952}

\bibitem[{{Sun} et~al.(2021{\natexlab{a}}){Sun}, {Goetz}, {Kissel},
  {Betzwieser}, {Karki}, {Bhattacharjee}, {Covas}, {Datrier}, {Kandhasamy},
  {Lecoeuche}, and et~al.}]{2021arXiv210700129S}
{Sun} L., {Goetz} E., {Kissel} J.~S., {Betzwieser} J., {Karki} S.,
  {Bhattacharjee} D., {Covas} P.~B., {Datrier} L. E.~H., et~al.
  (2021{\natexlab{a}}) {Characterization of systematic error in Advanced LIGO
  calibration in the second half of O3}. \emph{arXiv e-prints}
  arXiv:2107.00129, \eprint{2107.00129}

\bibitem[{{Sun} et~al.(2021{\natexlab{b}}){Sun}, {Jiao}, and {Zhang}}]{sun2021}
{Sun} W., {Jiao} K., {Zhang} T.-J. (2021{\natexlab{b}}) {Influence of the
  Bounds of the Hyperparameters on the Reconstruction of the Hubble Constant
  with the Gaussian Process}. \emph{\apj} 915(2):123,
  \doi{10.3847/1538-4357/ac05b8}

\bibitem[{{Sutter} et~al.(2012{\natexlab{a}}){Sutter}, {Lavaux}, {Wandelt}, and
  {Weinberg}}]{Sutter2012b}
{Sutter} P.~M., {Lavaux} G., {Wandelt} B.~D., {Weinberg} D.~H.
  (2012{\natexlab{a}}) {A First Application of the Alcock-Paczynski Test to
  Stacked Cosmic Voids}. \emph{\apj} 761:187,
  \doi{10.1088/0004-637X/761/2/187}, \eprint{1208.1058}

\bibitem[{{Sutter} et~al.(2012{\natexlab{b}}){Sutter}, {Lavaux}, {Wandelt}, and
  {Weinberg}}]{Sutter2012a}
{Sutter} P.~M., {Lavaux} G., {Wandelt} B.~D., {Weinberg} D.~H.
  (2012{\natexlab{b}}) {A Public Void Catalog from the SDSS DR7 Galaxy Redshift
  Surveys Based on the Watershed Transform}. \emph{\apj} 761:44,
  \doi{10.1088/0004-637X/761/1/44}, \eprint{1207.2524}

\bibitem[{{Sutter} et~al.(2014{\natexlab{a}}){Sutter}, {Elahi}, {Falck},
  {Onions}, {Hamaus}, {Knebe}, {Srisawat}, and {Schneider}}]{Sutter2014c}
{Sutter} P.~M., {Elahi} P., {Falck} B., {Onions} J., {Hamaus} N., {Knebe} A.,
  {Srisawat} C., {Schneider} A. (2014{\natexlab{a}}) {The life and death of
  cosmic voids}. \emph{\mnras} 445:1235--1244, \doi{10.1093/mnras/stu1845},
  \eprint{1403.7525}

\bibitem[{{Sutter} et~al.(2014{\natexlab{b}}){Sutter}, {Lavaux}, {Hamaus},
  {Wandelt}, {Weinberg}, and {Warren}}]{Sutter2014a}
{Sutter} P.~M., {Lavaux} G., {Hamaus} N., {Wandelt} B.~D., {Weinberg} D.~H.,
  {Warren} M.~S. (2014{\natexlab{b}}) {Sparse sampling, galaxy bias, and
  voids}. \emph{\mnras} 442:462--471, \doi{10.1093/mnras/stu893},
  \eprint{1309.5087}

\bibitem[{{Sutter} et~al.(2014{\natexlab{c}}){Sutter}, {Lavaux}, {Wandelt},
  {Weinberg}, {Warren}, and {Pisani}}]{Sutter2014}
{Sutter} P.~M., {Lavaux} G., {Wandelt} B.~D., {Weinberg} D.~H., {Warren} M.~S.,
  {Pisani} A. (2014{\natexlab{c}}) {Voids in the SDSS DR9: observations,
  simulations, and the impact of the survey mask}. \emph{\mnras}
  442(4):3127--3137, \doi{10.1093/mnras/stu1094}, \eprint{1310.7155}

\bibitem[{{Sutter} et~al.(2014{\natexlab{d}}){Sutter}, {Pisani}, {Wandelt}, and
  {Weinberg}}]{Sutter2014b}
{Sutter} P.~M., {Pisani} A., {Wandelt} B.~D., {Weinberg} D.~H.
  (2014{\natexlab{d}}) {A measurement of the Alcock-Paczy{\'n}ski effect using
  cosmic voids in the SDSS}. \emph{\mnras} 443:2983--2990,
  \doi{10.1093/mnras/stu1392}, \eprint{1404.5618}

\bibitem[{{Sutter} et~al.(2015){Sutter}, {Lavaux}, {Hamaus}, {Pisani},
  {Wandelt}, {Warren}, {Villaescusa-Navarro}, {Zivick}, {Mao}, and
  {Thompson}}]{Sutter2015}
{Sutter} P.~M., {Lavaux} G., {Hamaus} N., {Pisani} A., {Wandelt} B.~D.,
  {Warren} M., {Villaescusa-Navarro} F., {Zivick} P., et~al. (2015) {VIDE: The
  Void IDentification and Examination toolkit}. \emph{Astronomy and Computing}
  9:1--9, \doi{10.1016/j.ascom.2014.10.002}, \eprint{1406.1191}

\bibitem[{{Suyu} et~al.(2006){Suyu}, {Marshall}, {Hobson}, and
  {Blandford}}]{Suyu:2006}
{Suyu} S.~H., {Marshall} P.~J., {Hobson} M.~P., {Blandford} R.~D. (2006) {A
  Bayesian analysis of regularized source inversions in gravitational lensing}.
  \emph{\mnras} 371(2):983--998, \doi{10.1111/j.1365-2966.2006.10733.x},
  \eprint{astro-ph/0601493}

\bibitem[{{Suyu} et~al.(2009){Suyu}, {Marshall}, {Blandford}, {Fassnacht},
  {Koopmans}, {McKean}, and {Treu}}]{Suyu:2009}
{Suyu} S.~H., {Marshall} P.~J., {Blandford} R.~D., {Fassnacht} C.~D.,
  {Koopmans} L.~V.~E., {McKean} J.~P., {Treu} T. (2009) {Dissecting the
  Gravitational Lens B1608+656. I. Lens Potential Reconstruction}. \emph{\apj}
  691(1):277--298, \doi{10.1088/0004-637X/691/1/277}, \eprint{0804.2827}

\bibitem[{{Suyu} et~al.(2010){Suyu}, {Marshall}, {Auger}, {Hilbert},
  {Blandford}, {Koopmans}, {Fassnacht}, and {Treu}}]{Suyu:2010}
{Suyu} S.~H., {Marshall} P.~J., {Auger} M.~W., {Hilbert} S., {Blandford} R.~D.,
  {Koopmans} L.~V.~E., {Fassnacht} C.~D., {Treu} T. (2010) {Dissecting the
  Gravitational lens B1608+656. II. Precision Measurements of the Hubble
  Constant, Spatial Curvature, and the Dark Energy Equation of State}.
  \emph{\apj} 711(1):201--221, \doi{10.1088/0004-637X/711/1/201},
  \eprint{0910.2773}

\bibitem[{{Suyu} et~al.(2013){Suyu}, {Auger}, {Hilbert}, {Marshall}, {Tewes},
  {Treu}, {Fassnacht}, {Koopmans}, {Sluse}, {Blandford}, {Courbin}, and
  {Meylan}}]{Suyu:2013}
{Suyu} S.~H., {Auger} M.~W., {Hilbert} S., {Marshall} P.~J., {Tewes} M., {Treu}
  T., {Fassnacht} C.~D., {Koopmans} L.~V.~E., et~al. (2013) {Two Accurate
  Time-delay Distances from Strong Lensing: Implications for Cosmology}.
  \emph{\apj} 766(2):70, \doi{10.1088/0004-637X/766/2/70}, \eprint{1208.6010}

\bibitem[{{Suyu} et~al.(2014){Suyu}, {Treu}, {Hilbert}, {Sonnenfeld}, {Auger},
  {Blandford}, {Collett}, {Courbin}, {Fassnacht}, {Koopmans}, {Marshall},
  {Meylan}, {Spiniello}, and {Tewes}}]{Suyu:2014}
{Suyu} S.~H., {Treu} T., {Hilbert} S., {Sonnenfeld} A., {Auger} M.~W.,
  {Blandford} R.~D., {Collett} T., {Courbin} F., et~al. (2014) {Cosmology from
  Gravitational Lens Time Delays and Planck Data}. \emph{\apjl} 788(2):L35,
  \doi{10.1088/2041-8205/788/2/L35}, \eprint{1306.4732}

\bibitem[{{Suyu} et~al.(2017){Suyu}, {Bonvin}, {Courbin}, {Fassnacht}, {Rusu},
  {Sluse}, {Treu}, {Wong}, {Auger}, {Ding}, {Hilbert}, {Marshall}, {Rumbaugh},
  {Sonnenfeld}, {Tewes}, {Tihhonova}, {Agnello}, {Blandford}, {Chen},
  {Collett}, {Koopmans}, {Liao}, {Meylan}, and {Spiniello}}]{Suyu:2017}
{Suyu} S.~H., {Bonvin} V., {Courbin} F., {Fassnacht} C.~D., {Rusu} C.~E.,
  {Sluse} D., {Treu} T., {Wong} K.~C., et~al. (2017) {H0LiCOW - I. H$_{0}$
  Lenses in COSMOGRAIL's Wellspring: program overview}. \emph{\mnras}
  468(3):2590--2604, \doi{10.1093/mnras/stx483}, \eprint{1607.00017}

\bibitem[{{Svensson} and {Zdziarski}(1994)}]{sz1994}
{Svensson} R., {Zdziarski} A.~A. (1994) {Black Hole Accretion Disks with
  Coronae}. \emph{\apj} 436:599, \doi{10.1086/174934}

\bibitem[{{Swetz} et~al.(2011){Swetz}, {Ade}, {Amiri}, {Appel}, {Battistelli},
  {Burger}, {Chervenak}, {Devlin}, {Dicker}, {Doriese}, {D{\"u}nner},
  {Essinger-Hileman}, {Fisher}, {Fowler}, {Halpern}, {Hasselfield}, {Hilton},
  {Hincks}, {Irwin}, {Jarosik}, {Kaul}, {Klein}, {Lau}, {Limon}, {Marriage},
  {Marsden}, {Martocci}, {Mauskopf}, {Moseley}, {Netterfield}, {Niemack},
  {Nolta}, {Page}, {Parker}, {Staggs}, {Stryzak}, {Switzer}, {Thornton},
  {Tucker}, {Wollack}, and {Zhao}}]{ACT}
{Swetz} D.~S., {Ade} P.~A.~R., {Amiri} M., {Appel} J.~W., {Battistelli} E.~S.,
  {Burger} B., {Chervenak} J., {Devlin} M.~J., et~al. (2011) {Overview of the
  Atacama Cosmology Telescope: Receiver, Instrumentation, and Telescope
  Systems}. \emph{\apjs} 194(2):41, \doi{10.1088/0067-0049/194/2/41},
  \eprint{1007.0290}

\bibitem[{Switzer et~al.(2013)Switzer, Masui, Bandura, Calin, Chang, Chen, Li,
  Liao, Natarajan, Pen, and et~al.}]{Switzer_2013}
Switzer E.~R., Masui K.~W., Bandura K., Calin L.-M., Chang T.-C., Chen X.-L.,
  Li Y.-C., Liao Y.-W., et~al. (2013) Determination of z$\sim$0.8 neutral
  hydrogen fluctuations using the 21cm intensity mapping autocorrelation.
  \emph{Mon Not Roy Astron Soc: Letters} 434(1):L46--L50,
  \doi{10.1093/mnrasl/slt074},
  \urlprefix\url{http://dx.doi.org/10.1093/mnrasl/slt074}

\bibitem[{Switzer et~al.(2015)Switzer, Chang, Masui, Pen, and
  Voytek}]{Switzer_2015}
Switzer E.~R., Chang T.-C., Masui K.~W., Pen U.-L., Voytek T.~C. (2015)
  Interpreting the unresolved intensity of cosmologically redshifted line
  radiation. \emph{The Astrophysical Journal} 815(1):51,
  \doi{10.1088/0004-637x/815/1/51},
  \urlprefix\url{http://dx.doi.org/10.1088/0004-637X/815/1/51}

\bibitem[{{Takada} et~al.(2014){Takada}, {Ellis}, {Chiba}, {Greene}, {Aihara},
  {Arimoto}, {Bundy}, {Cohen}, {Dor{\'e}}, {Graves}, {Gunn}, {Heckman},
  {Hirata}, {Ho}, {Kneib}, {Le F{\`e}vre}, {Lin}, {More}, {Murayama}, {Nagao},
  {Ouchi}, {Seiffert}, {Silverman}, {Sodr{\'e}}, {Spergel}, {Strauss}, {Sugai},
  {Suto}, {Takami}, and {Wyse}}]{Takada:2014}
{Takada} M., {Ellis} R.~S., {Chiba} M., {Greene} J.~E., {Aihara} H., {Arimoto}
  N., {Bundy} K., {Cohen} J., et~al. (2014) {Extragalactic science, cosmology,
  and Galactic archaeology with the Subaru Prime Focus Spectrograph}.
  \emph{\pasj} 66(1):R1, \doi{10.1093/pasj/pst019}, \eprint{1206.0737}

\bibitem[{{Tananbaum} et~al.(1979){Tananbaum}, {Avni}, {Branduardi}, {Elvis},
  {Fabbiano}, {Feigelson}, {Giacconi}, {Henry}, {Pye}, {Soltan}, and
  {Zamorani}}]{avnitananbaum79}
{Tananbaum} H., {Avni} Y., {Branduardi} G., {Elvis} M., {Fabbiano} G.,
  {Feigelson} E., {Giacconi} R., {Henry} J.~P., et~al. (1979) {X-ray studies of
  quasars with the Einstein Observatory}. \emph{\apjl} 234:L9--L13,
  \doi{10.1086/183100}

\bibitem[{{Tavasoli}(2021)}]{Tavasoli2021}
{Tavasoli} S. (2021) {Void Galaxy Distribution: A Challenge for
  {\ensuremath{\Lambda}}CDM}. \emph{\apjl} 916(2):L24,
  \doi{10.3847/2041-8213/ac1357}

\bibitem[{{Taylor} and {Gair}(2012)}]{2012PhRvD..86b3502T}
{Taylor} S.~R., {Gair} J.~R. (2012) {Cosmology with the lights off: Standard
  sirens in the Einstein Telescope era}. \emph{\prd} 86(2):023502,
  \doi{10.1103/PhysRevD.86.023502}, \eprint{1204.6739}

\bibitem[{{Taylor} et~al.(2012){Taylor}, {Gair}, and
  {Mandel}}]{2012PhRvD..85b3535T}
{Taylor} S.~R., {Gair} J.~R., {Mandel} I. (2012) {Cosmology using advanced
  gravitational-wave detectors alone}. \emph{\prd} 85(2):023535,
  \doi{10.1103/PhysRevD.85.023535}, \eprint{1108.5161}

\bibitem[{Tegmark(1997)}]{Tegmark:1996qs}
Tegmark M. (1997) {How to make maps from CMB data without losing information}.
  \emph{Astrophys J Lett} 480:L87--L90, \doi{10.1086/310631},
  \eprint{astro-ph/9611130}

\bibitem[{{Terlevich} et~al.(2015){Terlevich}, {Terlevich}, {Melnick},
  {Ch{\'a}vez}, {Plionis}, {Bresolin}, and {Basilakos}}]{terlevich2015}
{Terlevich} R., {Terlevich} E., {Melnick} J., {Ch{\'a}vez} R., {Plionis} M.,
  {Bresolin} F., {Basilakos} S. (2015) {On the road to precision cosmology with
  high-redshift H II galaxies}. \emph{\mnras} 451(3):3001--3010,
  \doi{10.1093/mnras/stv1128}, \eprint{1505.04376}

\bibitem[{{Tewes} et~al.(2013{\natexlab{a}}){Tewes}, {Courbin}, and
  {Meylan}}]{Tewes:2013b}
{Tewes} M., {Courbin} F., {Meylan} G. (2013{\natexlab{a}}) {COSMOGRAIL: the
  COSmological MOnitoring of GRAvItational Lenses. XI. Techniques for time
  delay measurement in presence of microlensing}. \emph{\aap} 553:A120,
  \doi{10.1051/0004-6361/201220123}, \eprint{1208.5598}

\bibitem[{{Tewes} et~al.(2013{\natexlab{b}}){Tewes}, {Courbin}, {Meylan},
  {Kochanek}, {Eulaers}, {Cantale}, {Mosquera}, {Magain}, {Van Winckel},
  {Sluse}, {Cataldi}, {V{\"o}r{\"o}s}, and {Dye}}]{Tewes2013b}
{Tewes} M., {Courbin} F., {Meylan} G., {Kochanek} C.~S., {Eulaers} E.,
  {Cantale} N., {Mosquera} A.~M., {Magain} P., et~al. (2013{\natexlab{b}})
  {COSMOGRAIL: the COSmological MOnitoring of GRAvItational Lenses. XIII. Time
  delays and 9-yr optical monitoring of the lensed quasar RX J1131-1231}.
  \emph{\aap} 556:A22, \doi{10.1051/0004-6361/201220352}, \eprint{1208.6009}

\bibitem[{{Tewes} et~al.(2013{\natexlab{c}}){Tewes}, {Courbin}, {Meylan},
  {Kochanek}, {Eulaers}, {Cantale}, {Mosquera}, {Magain}, {Van Winckel},
  {Sluse}, {Cataldi}, {V{\"o}r{\"o}s}, and {Dye}}]{Tewes:2013a}
{Tewes} M., {Courbin} F., {Meylan} G., {Kochanek} C.~S., {Eulaers} E.,
  {Cantale} N., {Mosquera} A.~M., {Magain} P., et~al. (2013{\natexlab{c}})
  {COSMOGRAIL: the COSmological MOnitoring of GRAvItational Lenses. XIII. Time
  delays and 9-yr optical monitoring of the lensed quasar RX J1131-1231}.
  \emph{\aap} 556:A22, \doi{10.1051/0004-6361/201220352}, \eprint{1208.6009}

\bibitem[{{The LIGO Scientific Collaboration} et~al.(2021{\natexlab{a}}){The
  LIGO Scientific Collaboration}, {the Virgo Collaboration}, {the KAGRA
  Collaboration}, {Abbott}, {Abbott}, {Acernese}, {Ackley}, {Adams},
  {Adhikari}, {Adhikari}, and et~al.}]{2021arXiv211103606T}
{The LIGO Scientific Collaboration}, {the Virgo Collaboration}, {the KAGRA
  Collaboration}, {Abbott} R., {Abbott} T.~D., {Acernese} F., {Ackley} K.,
  {Adams} C., et~al. (2021{\natexlab{a}}) {GWTC-3: Compact Binary Coalescences
  Observed by LIGO and Virgo During the Second Part of the Third Observing
  Run}. \emph{arXiv e-prints} arXiv:2111.03606, \eprint{2111.03606}

\bibitem[{{The LIGO Scientific Collaboration} et~al.(2021{\natexlab{b}}){The
  LIGO Scientific Collaboration}, {the Virgo Collaboration}, {the KAGRA
  Collaboration}, {Abbott}, {Abbott}, {Acernese}, {Ackley}, {Adams},
  {Adhikari}, {Adhikari}, and et~al.}]{2021arXiv211103634T}
{The LIGO Scientific Collaboration}, {the Virgo Collaboration}, {the KAGRA
  Collaboration}, {Abbott} R., {Abbott} T.~D., {Acernese} F., {Ackley} K.,
  {Adams} C., et~al. (2021{\natexlab{b}}) {The population of merging compact
  binaries inferred using gravitational waves through GWTC-3}. \emph{arXiv
  e-prints} arXiv:2111.03634, \eprint{2111.03634}

\bibitem[{{The LIGO Scientific Collaboration} et~al.(2021{\natexlab{c}}){The
  LIGO Scientific Collaboration}, {the Virgo Collaboration}, {the KAGRA
  Collaboration}, {Abbott}, {Abe}, {Acernese}, {Ackley}, {Adhikari},
  {Adhikari}, and et~al.}]{LIGO2021}
{The LIGO Scientific Collaboration}, {the Virgo Collaboration}, {the KAGRA
  Collaboration}, {Abbott} R., {Abe} H., {Acernese} F., {Ackley} K., {Adhikari}
  N., et~al. (2021{\natexlab{c}}) {Constraints on the cosmic expansion history
  from GWTC-3}. \emph{arXiv e-prints} arXiv:2111.03604, \eprint{2111.03604}

\bibitem[{{The LSST Dark Energy Science Collaboration} et~al.(2018){The LSST
  Dark Energy Science Collaboration}, {Mandelbaum}, {Eifler}, {Hlo{\v{z}}ek},
  {Collett}, {Gawiser}, {Scolnic}, {Alonso}, {Awan}, {Biswas}, {Blazek},
  {Burchat}, {Chisari}, {Dell'Antonio}, {Digel}, {Frieman}, {Goldstein},
  {Hook}, {Ivezi{\'c}}, {Kahn}, {Kamath}, {Kirkby}, {Kitching}, {Krause},
  {Leget}, {Marshall}, {Meyers}, {Miyatake}, {Newman}, {Nichol}, {Rykoff},
  {Sanchez}, {Slosar}, {Sullivan}, and {Troxel}}]{Mandelbaum:2018ouv}
{The LSST Dark Energy Science Collaboration}, {Mandelbaum} R., {Eifler} T.,
  {Hlo{\v{z}}ek} R., {Collett} T., {Gawiser} E., {Scolnic} D., {Alonso} D.,
  et~al. (2018) {The LSST Dark Energy Science Collaboration (DESC) Science
  Requirements Document}. \emph{arXiv e-prints} arXiv:1809.01669,
  \eprint{1809.01669}

\bibitem[{{Thomas} et~al.(2010){Thomas}, {Maraston}, {Schawinski}, {Sarzi}, and
  {Silk}}]{thomas2010}
{Thomas} D., {Maraston} C., {Schawinski} K., {Sarzi} M., {Silk} J. (2010)
  {Environment and self-regulation in galaxy formation}. \emph{\mnras}
  404:1775--1789, \doi{10.1111/j.1365-2966.2010.16427.x}, \eprint{0912.0259}

\bibitem[{{Thomas} et~al.(2011){Thomas}, {Maraston}, and
  {Johansson}}]{thomas2011}
{Thomas} D., {Maraston} C., {Johansson} J. (2011) {Flux-calibrated stellar
  population models of Lick absorption-line indices with variable element
  abundance ratios}. \emph{\mnras} 412(4):2183--2198,
  \doi{10.1111/j.1365-2966.2010.18049.x}, \eprint{1010.4569}

\bibitem[{{Tihhonova} et~al.(2018){Tihhonova}, {Courbin}, {Harvey}, {Hilbert},
  {Rusu}, {Fassnacht}, {Bonvin}, {Marshall}, {Meylan}, {Sluse}, {Suyu}, {Treu},
  and {Wong}}]{Tihhonova:2018}
{Tihhonova} O., {Courbin} F., {Harvey} D., {Hilbert} S., {Rusu} C.~E.,
  {Fassnacht} C.~D., {Bonvin} V., {Marshall} P.~J., et~al. (2018) {H0LiCOW
  VIII. A weak-lensing measurement of the external convergence in the field of
  the lensed quasar HE 0435-1223}. \emph{\mnras} 477(4):5657--5669,
  \doi{10.1093/mnras/sty1040}, \eprint{1711.08804}

\bibitem[{{Tihhonova} et~al.(2020){Tihhonova}, {Courbin}, {Harvey}, {Hilbert},
  {Peel}, {Rusu}, {Fassnacht}, {Bonvin}, {Marshall}, {Meylan}, {Sluse}, {Suyu},
  {Treu}, and {Wong}}]{Tihhonova:2020}
{Tihhonova} O., {Courbin} F., {Harvey} D., {Hilbert} S., {Peel} A., {Rusu}
  C.~E., {Fassnacht} C.~D., {Bonvin} V., et~al. (2020) {H0LiCOW - XI. A weak
  lensing measurement of the external convergence in the field of the lensed
  quasar B1608+656 using HST and Subaru deep imaging}. \emph{\mnras}
  498(1):1406--1419, \doi{10.1093/mnras/staa1436}, \eprint{2005.12295}

\bibitem[{{Titov} et~al.(2011){Titov}, {Lambert}, and {Gontier}}]{titov2011}
{Titov} O., {Lambert} S.~B., {Gontier} A.~M. (2011) {VLBI measurement of the
  secular aberration drift}. \emph{\aap} 529:A91,
  \doi{10.1051/0004-6361/201015718}, \eprint{1009.3698}

\bibitem[{{Tojeiro} et~al.(2007){Tojeiro}, {Heavens}, {Jimenez}, and
  {Panter}}]{tojeiro2007}
{Tojeiro} R., {Heavens} A.~F., {Jimenez} R., {Panter} B. (2007) {Recovering
  galaxy star formation and metallicity histories from spectra using VESPA}.
  \emph{\mnras} 381(3):1252--1266, \doi{10.1111/j.1365-2966.2007.12323.x},
  \eprint{0704.0941}

\bibitem[{{Tonry} and {Schneider}(1988)}]{ts88}
{Tonry} J., {Schneider} D.~P. (1988) {A new technique for measuring
  extragalactic distances}. \emph{The Astrophysical Journal} 96:807--815,
  \doi{10.1086/114847}

\bibitem[{{Tonry}(1997)}]{tonry97}
{Tonry} J.~L. (1997) {The SBF Survey of Galaxy Distances}. In: {Livio} M.,
  {Donahue} M., {Panagia} N. (eds) The Extragalactic Distance Scale, p 297

\bibitem[{{Tonry} et~al.(1990){Tonry}, {Ajhar}, and {Luppino}}]{tal90}
{Tonry} J.~L., {Ajhar} E.~A., {Luppino} G.~A. (1990) {Observations of
  surface-brightness fluctuations in Virgo}. \emph{The Astrophysical Journal}
  100:1416, \doi{10.1086/115606}

\bibitem[{{Tonry} et~al.(2000){Tonry}, {Blakeslee}, {Ajhar}, and
  {Dressler}}]{tonry00}
{Tonry} J.~L., {Blakeslee} J.~P., {Ajhar} E.~A., {Dressler} A. (2000) {The
  Surface Brightness Fluctuation Survey of Galaxy Distances. II. Local and
  Large-Scale Flows}. \emph{\apj} 530(2):625--651, \doi{10.1086/308409},
  \eprint{astro-ph/9907062}

\bibitem[{{Tonry} et~al.(2001){Tonry}, {Dressler}, {Blakeslee}, {Ajhar},
  {Fletcher}, {Luppino}, {Metzger}, and {Moore}}]{tonry01}
{Tonry} J.~L., {Dressler} A., {Blakeslee} J.~P., {Ajhar} E.~A., {Fletcher}
  A.~B., {Luppino} G.~A., {Metzger} M.~R., {Moore} C.~B. (2001) {The SBF Survey
  of Galaxy Distances. IV. SBF Magnitudes, Colors, and Distances}. \emph{\apj}
  546:681, \doi{10.1086/318301}

\bibitem[{{Torri} et~al.(2004){Torri}, {Meneghetti}, {Bartelmann},
  {Moscardini}, {Rasia}, and {Tormen}}]{Torri:2004}
{Torri} E., {Meneghetti} M., {Bartelmann} M., {Moscardini} L., {Rasia} E.,
  {Tormen} G. (2004) {The impact of cluster mergers on arc statistics}.
  \emph{\mnras} 349(2):476--490, \doi{10.1111/j.1365-2966.2004.07508.x},
  \eprint{astro-ph/0310898}

\bibitem[{{Treu} and {Koopmans}(2002)}]{Treu:2002}
{Treu} T., {Koopmans} L.~V.~E. (2002) {The internal structure of the lens
  PG1115+080: breaking degeneracies in the value of the Hubble constant}.
  \emph{\mnras} 337(2):L6--L10, \doi{10.1046/j.1365-8711.2002.06107.x},
  \eprint{astro-ph/0210002}

\bibitem[{{Treu} and {Koopmans}(2004)}]{Treu:2004}
{Treu} T., {Koopmans} L. V.~E. (2004) {Massive Dark Matter Halos and Evolution
  of Early-Type Galaxies to z \raisebox{-0.5ex}\textasciitilde 1}. \emph{\apj}
  611(2):739--760, \doi{10.1086/422245}, \eprint{astro-ph/0401373}

\bibitem[{{Treu} and {Marshall}(2016)}]{Treu:2016}
{Treu} T., {Marshall} P.~J. (2016) {Time delay cosmography}. \emph{\aapr}
  24(1):11, \doi{10.1007/s00159-016-0096-8}, \eprint{1605.05333}

\bibitem[{{Tr{\"o}ster} et~al.(2021){Tr{\"o}ster}, {Asgari}, {Blake},
  {Cataneo}, {Heymans}, {Hildebrandt}, {Joachimi}, {Lin}, {S{\'a}nchez},
  {Wright}, {Bilicki}, {Bose}, {Crocce}, {Dvornik}, {Erben}, {Giblin},
  {Glazebrook}, {Hoekstra}, {Joudaki}, {Kannawadi}, {K{\"o}hlinger}, {Kuijken},
  {Lidman}, {Lombriser}, {Mead}, {Parkinson}, {Shan}, {Wolf}, and
  {Xia}}]{Troster2021}
{Tr{\"o}ster} T., {Asgari} M., {Blake} C., {Cataneo} M., {Heymans} C.,
  {Hildebrandt} H., {Joachimi} B., {Lin} C.-A., et~al. (2021) {KiDS-1000
  Cosmology: Constraints beyond flat {\ensuremath{\Lambda}}CDM}. \emph{\aap}
  649:A88, \doi{10.1051/0004-6361/202039805}, \eprint{2010.16416}

\bibitem[{{Tsaprazi} et~al.(2021){Tsaprazi}, {Jasche}, {Goobar}, {Peiris},
  {Andreoni}, {Coughlin}, {Fremling}, {Graham}, {Kasliwal}, {Kulkarni},
  {Mahabal} et~al.}]{Tsaprazi2021}
{Tsaprazi} E., {Jasche} J., {Goobar} A., {Peiris} H.~V., {Andreoni} I.,
  {Coughlin} M.~W., {Fremling} C.~U., {Graham} M.~J., et~al. (2021) {The
  large-scale environment of thermonuclear and core-collapse supernovae}.
  \emph{} arXiv:2109.02651, \eprint{2109.02651}

\bibitem[{{Tsupko} et~al.(2020){Tsupko}, {Fan}, and
  {Bisnovatyi-Kogan}}]{tsupko2020}
{Tsupko} O.~Y., {Fan} Z., {Bisnovatyi-Kogan} G.~S. (2020) {Black hole shadow as
  a standard ruler in cosmology}. \emph{Classical and Quantum Gravity}
  37(6):065016, \doi{10.1088/1361-6382/ab6f7d}, \eprint{1905.10509}

\bibitem[{{Unruh} et~al.(2017){Unruh}, {Schneider}, and {Sluse}}]{Unruh:2017}
{Unruh} S., {Schneider} P., {Sluse} D. (2017) {Ambiguities in gravitational
  lens models: the density field from the source position transformation}.
  \emph{\aap} 601:A77, \doi{10.1051/0004-6361/201629048}, \eprint{1606.04321}

\bibitem[{{Vagnozzi} et~al.(2020){Vagnozzi}, {Bambi}, and
  {Visinelli}}]{vagnozzi2020}
{Vagnozzi} S., {Bambi} C., {Visinelli} L. (2020) {Concerns regarding the use of
  black hole shadows as standard rulers}. \emph{Classical and Quantum Gravity}
  37(8):087001, \doi{10.1088/1361-6382/ab7965}, \eprint{2001.02986}

\bibitem[{{Vagnozzi} et~al.(2021){Vagnozzi}, {Loeb}, and
  {Moresco}}]{vagnozzi2021}
{Vagnozzi} S., {Loeb} A., {Moresco} M. (2021) {Eppur {\`e} piatto? The Cosmic
  Chronometers Take on Spatial Curvature and Cosmic Concordance}. \emph{\apj}
  908(1):84, \doi{10.3847/1538-4357/abd4df}, \eprint{2011.11645}

\bibitem[{Vagnozzi et~al.(2021)Vagnozzi, Pacucci, and Loeb}]{sunny}
Vagnozzi S., Pacucci F., Loeb A. (2021) Implications for the hubble tension
  from the ages of the oldest astrophysical objects. \eprint{2105.10421}

\bibitem[{{Valcin} et~al.(2020){Valcin}, {Bernal}, {Jimenez}, {Verde}, and
  {Wandelt}}]{Valcin1}
{Valcin} D., {Bernal} J.~L., {Jimenez} R., {Verde} L., {Wandelt} B.~D. (2020)
  {Inferring the age of the universe with globular clusters}. \emph{\jcap}
  2020(12):002, \doi{10.1088/1475-7516/2020/12/002}, \eprint{2007.06594}

\bibitem[{{Valcin} et~al.(2021){Valcin}, {Jimenez}, {Verde}, {Bernal}, and
  {Wandelt}}]{Valcin2}
{Valcin} D., {Jimenez} R., {Verde} L., {Bernal} J.~L., {Wandelt} B.~D. (2021)
  {The age of the Universe with globular clusters: reducing systematic
  uncertainties}. \emph{\jcap} 2021(8):017,
  \doi{10.1088/1475-7516/2021/08/017}, \eprint{2102.04486}

\bibitem[{{van de Weygaert} and {van Kampen}(1993)}]{vdWeygaert1993}
{van de Weygaert} R., {van Kampen} E. (1993) {Voids in Gravitational
  Instability Scenarios - Part One - Global Density and Velocity Fields in an
  Einstein - De-Sitter Universe}. \emph{\mnras} 263:481,
  \doi{10.1093/mnras/263.2.481}

\bibitem[{{van de Weygaert} et~al.(2011){van de Weygaert}, {Kreckel}, {Platen},
  {Beygu}, {van Gorkom}, {van der Hulst}, {Arag{\'o}n-Calvo}, {Peebles},
  {Jarrett}, {Rhee}, {Kova{\v{c}}}, and {Yip}}]{vdWeygaert2011}
{van de Weygaert} R., {Kreckel} K., {Platen} E., {Beygu} B., {van Gorkom}
  J.~H., {van der Hulst} J.~M., {Arag{\'o}n-Calvo} M.~A., {Peebles} P.~J.~E.,
  et~al. (2011) {The Void Galaxy Survey}. \emph{Astrophysics and Space Science
  Proceedings} 27:17, \doi{10.1007/978-3-642-20285-8\_3}, \eprint{1101.4187}

\bibitem[{{van Dokkum} et~al.(2018){van Dokkum}, {Danieli}, {Cohen},
  {Romanowsky}, and {Conroy}}]{vdk18}
{van Dokkum} P., {Danieli} S., {Cohen} Y., {Romanowsky} A.~J., {Conroy} C.
  (2018) {The Distance of the Dark Matter Deficient Galaxy NGC 1052-DF2}.
  \emph{\apjl} 864(1):L18, \doi{10.3847/2041-8213/aada4d}, \eprint{1807.06025}

\bibitem[{{van Dokkum} et~al.(2000){van Dokkum}, {Franx}, {Fabricant},
  {Illingworth}, and {Kelson}}]{vandokkum2000}
{van Dokkum} P.~G., {Franx} M., {Fabricant} D., {Illingworth} G.~D., {Kelson}
  D.~D. (2000) {Hubble Space Telescope Photometry and Keck Spectroscopy of the
  Rich Cluster MS 1054-03: Morphologies, Butcher-Oemler Effect, and the
  Color-Magnitude Relation at Z = 0.83}. \emph{\apj} 541(1):95--111,
  \doi{10.1086/309402}, \eprint{astro-ph/0002507}

\bibitem[{{Vanden Berk} et~al.(2001)}]{vandenberk2001}
{Vanden Berk} D.~E., et~al. (2001) {Composite Quasar Spectra from the Sloan
  Digital Sky Survey}. \emph{\aj} 122:549--564, \doi{10.1086/321167},
  \eprint{arXiv:astro-ph/0105231}

\bibitem[{{Vandenberg} et~al.(1996){Vandenberg}, {Bolte}, and
  {Stetson}}]{Bolte+}
{Vandenberg} D.~A., {Bolte} M., {Stetson} P.~B. (1996) {The Age of the Galactic
  Globular Cluster System}. \emph{\araa} 34:461--510,
  \doi{10.1146/annurev.astro.34.1.461}

\bibitem[{{Vanderlinde} et~al.(2019){Vanderlinde}, {Liu}, {Gaensler}, {Bond},
  {Hinshaw}, {Ng}, {Chiang}, {Stairs}, {Brown}, {Sievers}, {Mena}, {Smith},
  {Bandura}, {Masui}, {Spekkens}, {Belostotski}, {Dobbs}, {Turok}, {Boyle},
  {Rupen}, {Landecker}, {Pen}, and {Kaspi}}]{Vanderlinde:2019tjt}
{Vanderlinde} K., {Liu} A., {Gaensler} B., {Bond} D., {Hinshaw} G., {Ng} C.,
  {Chiang} C., {Stairs} I., et~al. (2019) {The Canadian Hydrogen Observatory
  and Radio-transient Detector (CHORD)}. In: Canadian Long Range Plan for
  Astronomy and Astrophysics White Papers, vol 2020, p~28,
  \doi{10.5281/zenodo.3765414}, \eprint{1911.01777}

\bibitem[{{Vavry{\v{c}}uk} and {Kroupa}(2020)}]{vavrycuk2020}
{Vavry{\v{c}}uk} V., {Kroupa} P. (2020) {The failure of testing for cosmic
  opacity via the distance-duality relation}. \emph{\mnras} 497(1):378--388,
  \doi{10.1093/mnras/staa1936}, \eprint{2007.10472}

\bibitem[{Vazdekis et~al.(2015)Vazdekis, Coelho, Cassisi, Ricciardelli,
  Falc{\'{o}}n-Barroso, S{\'{a}}nchez-Bl{\'{a}}zquez, Barbera, Beasley, and
  Pietrinferni}]{vazdekis2015}
Vazdekis A., Coelho P., Cassisi S., Ricciardelli E., Falc{\'{o}}n-Barroso J.,
  S{\'{a}}nchez-Bl{\'{a}}zquez P., Barbera F.~L., Beasley M.~A., et~al. (2015)
  Evolutionary stellar population synthesis with {MILES} {\textendash} {II}.
  scaled-solar and $\alpha$-enhanced models. \emph{Monthly Notices of the Royal
  Astronomical Society} 449(2):1177--1214, \doi{10.1093/mnras/stv151}

\bibitem[{{Vazdekis} et~al.(2016){Vazdekis}, {Koleva}, {Ricciardelli},
  {R{\"o}ck}, and {Falc{\'o}n-Barroso}}]{vazdekis2016}
{Vazdekis} A., {Koleva} M., {Ricciardelli} E., {R{\"o}ck} B.,
  {Falc{\'o}n-Barroso} J. (2016) {UV-extended E-MILES stellar population
  models: young components in massive early-type galaxies}. \emph{\mnras}
  463(4):3409--3436, \doi{10.1093/mnras/stw2231}, \eprint{1612.01187}

\bibitem[{{Verde} et~al.(2014){Verde}, {Protopapas}, and
  {Jimenez}}]{protopapas2014}
{Verde} L., {Protopapas} P., {Jimenez} R. (2014) {The expansion rate of the
  intermediate universe in light of Planck}. \emph{Physics of the Dark
  Universe} 5:307--314, \doi{10.1016/j.dark.2014.09.003}, \eprint{1403.2181}

\bibitem[{{Verde} et~al.(2019){Verde}, {Treu}, and {Riess}}]{verde2019}
{Verde} L., {Treu} T., {Riess} A.~G. (2019) {Tensions between the early and
  late Universe}. \emph{Nature Astronomy} 3:891--895,
  \doi{10.1038/s41550-019-0902-0}, \eprint{1907.10625}

\bibitem[{{Verza} et~al.(2019){Verza}, {Pisani}, {Carbone}, {Hamaus}, and
  {Guzzo}}]{Verza2019}
{Verza} G., {Pisani} A., {Carbone} C., {Hamaus} N., {Guzzo} L. (2019) {The void
  size function in dynamical dark energy cosmologies}. \emph{\jcap}
  2019(12):040, \doi{10.1088/1475-7516/2019/12/040}, \eprint{1906.00409}

\bibitem[{{Vielzeuf} et~al.(2021){Vielzeuf}, {Kov{\'a}cs}, {Demirbozan},
  {Fosalba}, {Baxter}, {Hamaus}, {Huterer}, {Miquel}, {Nadathur}, {Pollina},
  {S{\'a}nchez}, et~al., and {DES Collaboration}}]{Vielzeuf2021}
{Vielzeuf} P., {Kov{\'a}cs} A., {Demirbozan} U., {Fosalba} P., {Baxter} E.,
  {Hamaus} N., {Huterer} D., {Miquel} R., et~al. (2021) {Dark Energy Survey
  Year 1 results: the lensing imprint of cosmic voids on the cosmic microwave
  background}. \emph{\mnras} 500(1):464--480, \doi{10.1093/mnras/staa3231},
  \eprint{1911.02951}

\bibitem[{{Vijaykumar} et~al.(2020){Vijaykumar}, {Saketh}, {Kumar}, {Ajith},
  and {Choudhury}}]{2020arXiv200501111V}
{Vijaykumar} A., {Saketh} M.~V.~S., {Kumar} S., {Ajith} P., {Choudhury} T.~R.
  (2020) {Probing the large scale structure using gravitational-wave
  observations of binary black holes}. \emph{arXiv e-prints} arXiv:2005.01111,
  \eprint{2005.01111}

\bibitem[{Viljoen et~al.(2020)Viljoen, Fonseca, and Maartens}]{Viljoen:2020efi}
Viljoen J.-A., Fonseca J., Maartens R. (2020) {Constraining the growth rate by
  combining multiple future surveys}. \emph{JCAP} 09:054,
  \doi{10.1088/1475-7516/2020/09/054}, \eprint{2007.04656}

\bibitem[{Villaescusa-Navarro et~al.(2014)Villaescusa-Navarro, Viel, Datta, and
  Choudhury}]{Villaescusa-Navarro:2014cma}
Villaescusa-Navarro F., Viel M., Datta K.~K., Choudhury T.~R. (2014) {Modeling
  the neutral hydrogen distribution in the post-reionization Universe:
  intensity mapping}. \emph{JCAP} 09:050, \doi{10.1088/1475-7516/2014/09/050},
  \eprint{1405.6713}

\bibitem[{Villaescusa-Navarro et~al.(2015)Villaescusa-Navarro, Bull, and
  Viel}]{Villaescusa-Navarro:2015cca}
Villaescusa-Navarro F., Bull P., Viel M. (2015) {Weighing neutrinos with cosmic
  neutral hydrogen}. \emph{Astrophys J} 814(2):146,
  \doi{10.1088/0004-637X/814/2/146}, \eprint{1507.05102}

\bibitem[{Villaescusa-Navarro et~al.(2017)Villaescusa-Navarro, Alonso, and
  Viel}]{Villaescusa-Navarro:2016kbz}
Villaescusa-Navarro F., Alonso D., Viel M. (2017) {Baryonic acoustic
  oscillations from 21 cm intensity mapping: the Square Kilometre Array case}.
  \emph{Mon Not Roy Astron Soc} 466(3):2736--2751, \doi{10.1093/mnras/stw3224},
  \eprint{1609.00019}

\bibitem[{Villaescusa-Navarro et~al.(2018)}]{Villaescusa-Navarro:2018vsg}
Villaescusa-Navarro F., et~al. (2018) {Ingredients for 21 cm Intensity
  Mapping}. \emph{Astrophys J} 866(2):135, \doi{10.3847/1538-4357/aadba0},
  \eprint{1804.09180}

\bibitem[{Villar et~al.(2020)}]{Villar:2020epn}
Villar V.~A., et~al. (2020) {SuperRAENN: A Semisupervised Supernova Photometric
  Classification Pipeline Trained on Pan-STARRS1 Medium-Deep Survey
  Supernovae}. \emph{Astrophys J} 905(2):94, \doi{10.3847/1538-4357/abc6fd},
  \eprint{2008.04921}

\bibitem[{{Vitale} and {Chen}(2018)}]{2018PhRvL.121b1303V}
{Vitale} S., {Chen} H.-Y. (2018) {Measuring the Hubble Constant with Neutron
  Star Black Hole Mergers}. \emph{\prl} 121(2):021303,
  \doi{10.1103/PhysRevLett.121.021303}, \eprint{1804.07337}

\bibitem[{{Vito} et~al.(2019){Vito}, {Brandt}, {Bauer}, {Calura}, {Gilli},
  {Luo}, {Shemmer}, {Vignali}, {Zamorani}, {Brusa}, {Civano}, {Comastri}, and
  {Nanni}}]{vito2019}
{Vito} F., {Brandt} W.~N., {Bauer} F.~E., {Calura} F., {Gilli} R., {Luo} B.,
  {Shemmer} O., {Vignali} C., et~al. (2019) {The X-ray properties of z \&gt; 6
  quasars: no evident evolution of accretion physics in the first Gyr of the
  Universe}. \emph{\aap} 630:A118, \doi{10.1051/0004-6361/201936217},
  \eprint{1908.09849}

\bibitem[{{Voivodic} et~al.(2017){Voivodic}, {Lima}, {Llinares}, and
  {Mota}}]{Voivodic2017}
{Voivodic} R., {Lima} M., {Llinares} C., {Mota} D.~F. (2017) {Modeling void
  abundance in modified gravity}. \emph{\prd} 95(2):024018,
  \doi{10.1103/PhysRevD.95.024018}, \eprint{1609.02544}

\bibitem[{{Voivodic} et~al.(2020){Voivodic}, {Rubira}, and
  {Lima}}]{Voivodic2020}
{Voivodic} R., {Rubira} H., {Lima} M. (2020) {The Halo Void (Dust) Model of
  large scale structure}. \emph{\jcap} 2020(10):033,
  \doi{10.1088/1475-7516/2020/10/033}, \eprint{2003.06411}

\bibitem[{{von Marttens} et~al.(2019){von Marttens}, {Marra}, {Casarini},
  {Gonzalez}, and {Alcaniz}}]{vonmarttens2019}
{von Marttens} R., {Marra} V., {Casarini} L., {Gonzalez} J.~E., {Alcaniz} J.
  (2019) {Null test for interactions in the dark sector}. \emph{\prd}
  99(4):043521, \doi{10.1103/PhysRevD.99.043521}, \eprint{1812.02333}

\bibitem[{{Vuissoz} et~al.(2007){Vuissoz}, {Courbin}, {Sluse}, {Meylan},
  {Ibrahimov}, {Asfandiyarov}, {Stoops}, {Eigenbrod}, {Le Guillou}, {van
  Winckel}, and {Magain}}]{Vuissoz2007}
{Vuissoz} C., {Courbin} F., {Sluse} D., {Meylan} G., {Ibrahimov} M.,
  {Asfandiyarov} I., {Stoops} E., {Eigenbrod} A., et~al. (2007) {COSMOGRAIL:
  the COSmological MOnitoring of GRAvItational Lenses. V. The time delay in
  SDSS J1650+4251}. \emph{\aap} 464(3):845--851,
  \doi{10.1051/0004-6361:20065823}, \eprint{astro-ph/0606317}

\bibitem[{{Vuissoz} et~al.(2008){Vuissoz}, {Courbin}, {Sluse}, {Meylan},
  {Chantry}, {Eulaers}, {Morgan}, {Eyler}, {Kochanek}, {Coles}, {Saha},
  {Magain}, and {Falco}}]{Vuissoz2008}
{Vuissoz} C., {Courbin} F., {Sluse} D., {Meylan} G., {Chantry} V., {Eulaers}
  E., {Morgan} C., {Eyler} M.~E., et~al. (2008) {COSMOGRAIL: the COSmological
  MOnitoring of GRAvItational Lenses. VII. Time delays and the Hubble constant
  from WFI J2033-4723}. \emph{\aap} 488(2):481--490,
  \doi{10.1051/0004-6361:200809866}, \eprint{0803.4015}

\bibitem[{{Wagner-Kaiser} et~al.(2017){Wagner-Kaiser}, {Sarajedini}, {von
  Hippel}, {Stenning}, {van Dyk}, {Jeffery}, {Robinson}, {Stein}, {Anderson},
  and {Jefferys}}]{BayesianGC}
{Wagner-Kaiser} R., {Sarajedini} A., {von Hippel} T., {Stenning} D.~C., {van
  Dyk} D.~A., {Jeffery} E., {Robinson} E., {Stein} N., et~al. (2017) {The ACS
  survey of Galactic globular clusters - XIV. Bayesian single-population
  analysis of 69 globular clusters}. \emph{\mnras} 468(1):1038--1055,
  \doi{10.1093/mnras/stx544}, \eprint{1702.08856}

\bibitem[{{Walsh} et~al.(1979){Walsh}, {Carswell}, and {Weymann}}]{Walsh:1979}
{Walsh} D., {Carswell} R.~F., {Weymann} R.~J. (1979) {0957+561 A, B: twin
  quasistellar objects or gravitational lens?} \emph{\nat} 279:381--384,
  \doi{10.1038/279381a0}

\bibitem[{{Wang} et~al.(2015){Wang}, {Dai}, and {Liang}}]{Wang15}
{Wang} F.~Y., {Dai} Z.~G., {Liang} E.~W. (2015) {Gamma-ray burst cosmology}.
  \emph{\nar} 67:1--17, \doi{10.1016/j.newar.2015.03.001}, \eprint{1504.00735}

\bibitem[{{Wang} et~al.(2021){Wang}, {Hu}, {Zhang}, and {Dai}}]{Wang2021}
{Wang} F.~Y., {Hu} J.~P., {Zhang} G.~Q., {Dai} Z.~G. (2021) {Standardized long
  gamma-ray bursts as a cosmic distance indicator}. \emph{arXiv e-prints}
  arXiv:2106.14155, \eprint{2106.14155}

\bibitem[{Wang et~al.(2021)}]{Wang:2020lkn}
Wang J., et~al. (2021) {H\,i intensity mapping with MeerKAT: calibration
  pipeline for multidish autocorrelation observations}. \emph{Mon Not Roy
  Astron Soc} 505(3):3698--3721, \doi{10.1093/mnras/stab1365},
  \eprint{2011.13789}

\bibitem[{{Wang} et~al.(2020){Wang}, {Zhao}, {Zhang}, and
  {Zhang}}]{2020JCAP...11..012W}
{Wang} L.-F., {Zhao} Z.-W., {Zhang} J.-F., {Zhang} X. (2020) {A preliminary
  forecast for cosmological parameter estimation with gravitational-wave
  standard sirens from TianQin}. \emph{\jcap} 2020(11):012,
  \doi{10.1088/1475-7516/2020/11/012}, \eprint{1907.01838}

\bibitem[{{Wang} et~al.(2018){Wang}, {Luo}, {Shen}, {Hou}, {Kong}, {Song},
  {Zhang}, {Wu}, {Cao}, {Hou}, {Wang}, {Zhang}, and {Zhao}}]{wang2018}
{Wang} L.-L., {Luo} A.~L., {Shen} S.-Y., {Hou} W., {Kong} X., {Song} Y.-H.,
  {Zhang} J.-N., {Wu} H., et~al. (2018) {Spectral classification and composites
  of galaxies in LAMOST DR4}. \emph{\mnras} 474(2):1873--1885,
  \doi{10.1093/mnras/stx2798}, \eprint{1710.10611}

\bibitem[{{Wang} et~al.(2016){Wang}, {Elbaz}, {Schreiber}, {Pannella}, {Shu},
  {Willner}, {Ashby}, {Huang}, {Fontana}, {Dekel}, {Daddi}, {Ferguson},
  {Dunlop}, {Ciesla}, {Koekemoer}, {Giavalisco}, {Boutsia}, {Finkelstein},
  {Juneau}, {Barro}, {Koo}, {Micha{\l}owski}, {Orellana}, {Lu}, {Castellano},
  {Bourne}, {Buitrago}, {Santini}, {Faber}, {Hathi}, {Lucas}, and
  {P{\'e}rez-Gonz{\'a}lez}}]{wang2016}
{Wang} T., {Elbaz} D., {Schreiber} C., {Pannella} M., {Shu} X., {Willner}
  S.~P., {Ashby} M.~L.~N., {Huang} J.~S., et~al. (2016) {Infrared Color
  Selection of Massive Galaxies at z > 3}. \emph{\apj} 816(2):84,
  \doi{10.3847/0004-637X/816/2/84}, \eprint{1512.02656}

\bibitem[{{Wang} et~al.(2019){Wang}, {Dickinson}, {Hillenbrand}, {Robberto},
  {Armus}, {Ballardini}, {Barkhouser}, {Bartlett}, {Behroozi}, {Benjamin},
  {Brinchmann}, {Chary}, {Chuang}, {Cimatti}, {Conroy}, {Content}, {Daddi},
  {Donahue}, {Dore}, {Eisenhardt}, {Ferguson}, {Faisst}, {Fraser},
  {Glazebrook}, {Gorjian}, {Helou}, {Hirata}, {Hudson}, {Kirkpatrick},
  {Malhotra}, {Mei}, {Moscardini}, {Newman}, {Ninkov}, {Orsi}, {Ressler},
  {Rhoads}, {Rhodes}, {Ryan}, {Samushia}, {Scarlata}, {Scolnic}, {Seiffert},
  {Shapley}, {Smee}, {Valentino}, {Vorobiev}, and {Wechsler}}]{Atlasmission}
{Wang} Y., {Dickinson} M., {Hillenbrand} L., {Robberto} M., {Armus} L.,
  {Ballardini} M., {Barkhouser} R., {Bartlett} J., et~al. (2019) {ATLAS Probe:
  Breakthrough Science of Galaxy Evolution, Cosmology, Milky Way, and the Solar
  System}. \emph{arXiv e-prints} arXiv:1909.00070, \eprint{1909.00070}

\bibitem[{{Watson} et~al.(2011){Watson}, {Denney}, {Vestergaard}, and
  {Davis}}]{watson2011}
{Watson} D., {Denney} K.~D., {Vestergaard} M., {Davis} T.~M. (2011) {A New
  Cosmological Distance Measure Using Active Galactic Nuclei}. \emph{\apjl}
  740(2):L49, \doi{10.1088/2041-8205/740/2/L49}, \eprint{1109.4632}

\bibitem[{{Webb} et~al.(2020){Webb}, {Coriat}, {Traulsen}, {Ballet}, {Motch},
  {Carrera}, {Koliopanos}, {Authier}, {de la Calle}, {Ceballos}, {Colomo},
  {Chuard}, {Freyberg}, {Garcia}, {Kolehmainen}, {Lamer}, {Lin}, {Maggi},
  {Michel}, {Page}, {Page}, {Perea-Calderon}, {Pineau}, {Rodriguez}, {Rosen},
  {Santos Lleo}, {Saxton}, {Schwope}, {Tom{\'a}s}, {Watson}, and
  {Zakardjian}}]{nb2020}
{Webb} N.~A., {Coriat} M., {Traulsen} I., {Ballet} J., {Motch} C., {Carrera}
  F.~J., {Koliopanos} F., {Authier} J., et~al. (2020) {The XMM-Newton
  serendipitous survey. IX. The fourth XMM-Newton serendipitous source
  catalogue}. \emph{\aap} 641:A136, \doi{10.1051/0004-6361/201937353},
  \eprint{2007.02899}

\bibitem[{{Wei}(2010)}]{Wei10}
{Wei} H. (2010) {Observational constraints on cosmological models with the
  updated long gamma-ray bursts}. \emph{\jcap} 2010(8):020,
  \doi{10.1088/1475-7516/2010/08/020}, \eprint{1004.4951}

\bibitem[{{Wei} and {Melia}(2020)}]{wm2020}
{Wei} J.-J., {Melia} F. (2020) {Model-independent Distance Calibration and
  Curvature Measurement Using Quasars and Cosmic Chronometers}. \emph{\apj}
  888(2):99, \doi{10.3847/1538-4357/ab5e7d}, \eprint{1912.00668}

\bibitem[{{Wei} and {Wu}(2017)}]{Wei17}
{Wei} J.-J., {Wu} X.-F. (2017) {Gamma-ray burst cosmology: Hubble diagram and
  star formation history}. \emph{International Journal of Modern Physics D}
  26(2):1730002, \doi{10.1142/S0218271817300026}, \eprint{1607.01550}

\bibitem[{Weltman et~al.(2020)}]{Weltman:2018zrl}
Weltman A., et~al. (2020) {Fundamental physics with the Square Kilometre
  Array}. \emph{Publ Astron Soc Austral} 37:e002, \doi{10.1017/pasa.2019.42},
  \eprint{1810.02680}

\bibitem[{{White}(2016)}]{White2016}
{White} M. (2016) {A marked correlation function for constraining modified
  gravity models}. \emph{\jcap} 2016(11):057,
  \doi{10.1088/1475-7516/2016/11/057}, \eprint{1609.08632}

\bibitem[{{White} and {Padmanabhan}(2017)}]{White2017}
{White} M., {Padmanabhan} N. (2017) {Matched filtering with interferometric 21
  cm experiments}. \emph{\mnras} 471(1):1167--1180,
  \doi{10.1093/mnras/stx1682}, \eprint{1705.09669}

\bibitem[{{Wiklind} et~al.(2008){Wiklind}, {Dickinson}, {Ferguson},
  {Giavalisco}, {Mobasher}, {Grogin}, and {Panagia}}]{wiklind2008}
{Wiklind} T., {Dickinson} M., {Ferguson} H.~C., {Giavalisco} M., {Mobasher} B.,
  {Grogin} N.~A., {Panagia} N. (2008) {A Population of Massive and Evolved
  Galaxies at z gtrsim 5}. \emph{\apj} 676(2):781--806, \doi{10.1086/524919},
  \eprint{0710.0406}

\bibitem[{{Wilkinson} et~al.(2017){Wilkinson}, {Maraston}, {Goddard}, {Thomas},
  and {Parikh}}]{wilkinson2017}
{Wilkinson} D.~M., {Maraston} C., {Goddard} D., {Thomas} D., {Parikh} T. (2017)
  {FIREFLY (Fitting IteRativEly For Likelihood analYsis): a full spectral
  fitting code}. \emph{\mnras} 472(4):4297--4326, \doi{10.1093/mnras/stx2215},
  \eprint{1711.00865}

\bibitem[{{Williams} et~al.(2009){Williams}, {Quadri}, {Franx}, {van Dokkum},
  and {Labb{\'e}}}]{williams2009}
{Williams} R.~J., {Quadri} R.~F., {Franx} M., {van Dokkum} P., {Labb{\'e}} I.
  (2009) {Detection of Quiescent Galaxies in a Bicolor Sequence from Z = 0-2}.
  \emph{\apj} 691(2):1879--1895, \doi{10.1088/0004-637X/691/2/1879},
  \eprint{0806.0625}

\bibitem[{{Willingale} et~al.(2007){Willingale}, {O'Brien}, {Osborne}, {Godet},
  {Page}, {Goad}, {Burrows}, {Zhang}, {Rol}, {Gehrels}, and
  {Chincarini}}]{Willingale2007}
{Willingale} R., {O'Brien} P.~T., {Osborne} J.~P., {Godet} O., {Page} K.~L.,
  {Goad} M.~R., {Burrows} D.~N., {Zhang} B., et~al. (2007) {Testing the
  Standard Fireball Model of Gamma-Ray Bursts Using Late X-Ray Afterglows
  Measured by Swift}. \emph{\apj} 662(2):1093--1110, \doi{10.1086/517989},
  \eprint{astro-ph/0612031}

\bibitem[{{Wilson} and {Bean}(2021)}]{Wilson2021}
{Wilson} C., {Bean} R. (2021) {Testing f (R ) gravity with scale dependent
  cosmic void velocity profiles}. \emph{\prd} 104(2):023512,
  \doi{10.1103/PhysRevD.104.023512}, \eprint{2012.05925}

\bibitem[{{Wilson} et~al.(2016){Wilson}, {Zabludoff}, {Ammons}, {Momcheva},
  {Williams}, and {Keeton}}]{Wilson2016}
{Wilson} M.~L., {Zabludoff} A.~I., {Ammons} S.~M., {Momcheva} I.~G., {Williams}
  K.~A., {Keeton} C.~R. (2016) {A Spectroscopic Survey of the Fields of 28
  Strong Gravitational Lenses: the Group Catalog}. \emph{\apj} 833(2):194,
  \doi{10.3847/1538-4357/833/2/194}, \eprint{1710.09908}

\bibitem[{{Witzemann} et~al.(2019){Witzemann}, {Alonso}, {Fonseca}, and
  {Santos}}]{Witzemann:2018cdx}
{Witzemann} A., {Alonso} D., {Fonseca} J., {Santos} M.~G. (2019) {Simulated
  multitracer analyses with H I intensity mapping}. \emph{\mnras}
  485(4):5519--5531, \doi{10.1093/mnras/stz778}, \eprint{1808.03093}

\bibitem[{{Wojtak} et~al.(2016){Wojtak}, {Powell}, and {Abel}}]{Wojtak2016}
{Wojtak} R., {Powell} D., {Abel} T. (2016) {Voids in cosmological simulations
  over cosmic time}. \emph{\mnras} 458(4):4431--4442,
  \doi{10.1093/mnras/stw615}, \eprint{1602.08541}

\bibitem[{{Wojtak} et~al.(2019){Wojtak}, {Hjorth}, and {Gall}}]{Wojtak:2019}
{Wojtak} R., {Hjorth} J., {Gall} C. (2019) {Magnified or multiply imaged? -
  Search strategies for gravitationally lensed supernovae in wide-field
  surveys}. \emph{\mnras} 487(3):3342--3355, \doi{10.1093/mnras/stz1516},
  \eprint{1903.07687}

\bibitem[{Wolz et~al.(2014)Wolz, Abdalla, Blake, Shaw, Chapman, and
  Rawlings}]{Wolz:2013wna}
Wolz L., Abdalla F.~B., Blake C., Shaw J.~R., Chapman E., Rawlings S. (2014)
  {The effect of foreground subtraction on cosmological measurements from
  Intensity Mapping}. \emph{Mon Not Roy Astron Soc} 441(4):3271--3283,
  \doi{10.1093/mnras/stu792}, \eprint{1310.8144}

\bibitem[{Wolz et~al.(2017{\natexlab{a}})Wolz, Blake, and
  Wyithe}]{Wolz:2017rlw}
Wolz L., Blake C., Wyithe J. S.~B. (2017{\natexlab{a}}) {Determining the HI
  content of galaxies via intensity mapping cross-correlations}. \emph{Mon Not
  Roy Astron Soc} 470(3):3220--3226, \doi{10.1093/mnras/stx1388},
  \eprint{1703.08268}

\bibitem[{{Wolz} et~al.(2022){Wolz}, {Pourtsidou}, {Masui}, {Chang},
  {Bautista}, {M{\"u}ller}, {Avila}, {Bacon}, {Percival}, {Cunnington},
  {Anderson}, {Chen}, {Kneib}, {Li}, {Liao}, {Pen}, {Peterson}, {Rossi},
  {Schneider}, {Yadav}, and {Zhao}}]{Wolz:2021ofa}
{Wolz} L., {Pourtsidou} A., {Masui} K.~W., {Chang} T.-C., {Bautista} J.~E.,
  {M{\"u}ller} E.-M., {Avila} S., {Bacon} D., et~al. (2022) {H I constraints
  from the cross-correlation of eBOSS galaxies and Green Bank Telescope
  intensity maps}. \emph{\mnras} 510(3):3495--3511,
  \doi{10.1093/mnras/stab3621}, \eprint{2102.04946}

\bibitem[{Wolz et~al.(2017{\natexlab{b}})}]{Wolz:2015lwa}
Wolz L., et~al. (2017{\natexlab{b}}) {Erasing the Milky Way: new cleaning
  technique applied to GBT intensity mapping data}. \emph{Mon Not Roy Astron
  Soc} 464(4):4938--4949, \doi{10.1093/mnras/stw2556}, \eprint{1510.05453}

\bibitem[{{Wong} et~al.(2017){Wong}, {Suyu}, {Auger}, {Bonvin}, {Courbin},
  {Fassnacht}, {Halkola}, {Rusu}, {Sluse}, {Sonnenfeld}, {Treu}, {Collett},
  {Hilbert}, {Koopmans}, {Marshall}, and {Rumbaugh}}]{Wong:2017}
{Wong} K.~C., {Suyu} S.~H., {Auger} M.~W., {Bonvin} V., {Courbin} F.,
  {Fassnacht} C.~D., {Halkola} A., {Rusu} C.~E., et~al. (2017) {H0LiCOW - IV.
  Lens mass model of HE 0435-1223 and blind measurement of its time-delay
  distance for cosmology}. \emph{\mnras} 465(4):4895--4913,
  \doi{10.1093/mnras/stw3077}, \eprint{1607.01403}

\bibitem[{{Wong} et~al.(2020){Wong}, {Suyu}, {Chen}, {Rusu}, {Millon}, {Sluse},
  {Bonvin}, {Fassnacht}, {Taubenberger}, {Auger}, {Birrer}, {Chan}, {Courbin},
  {Hilbert}, {Tihhonova}, {Treu}, {Agnello}, {Ding}, {Jee}, {Komatsu},
  {Shajib}, {Sonnenfeld}, {Blandford}, {Koopmans}, {Marshall}, and
  {Meylan}}]{Wong:2020}
{Wong} K.~C., {Suyu} S.~H., {Chen} G. C.~F., {Rusu} C.~E., {Millon} M., {Sluse}
  D., {Bonvin} V., {Fassnacht} C.~D., et~al. (2020) {H0LiCOW {\textendash}
  XIII. A 2.4 per cent measurement of H$_{0}$ from lensed quasars:
  5.3{\ensuremath{\sigma}} tension between early- and late-Universe probes}.
  \emph{\mnras} 498(1):1420--1439, \doi{10.1093/mnras/stz3094},
  \eprint{1907.04869}

\bibitem[{{Worthey}(1993)}]{worthey93}
{Worthey} G. (1993) {The dependence of the brightness fluctuation distance
  indicator on stellar population age and metallicity}. \emph{\apj}
  409:530--536, \doi{10.1086/172684}

\bibitem[{{Worthey}(1994)}]{worthey1994}
{Worthey} G. (1994) {Comprehensive Stellar Population Models and the
  Disentanglement of Age and Metallicity Effects}. \emph{\apjs} 95:107,
  \doi{10.1086/192096}

\bibitem[{{Worthey} and {Ottaviani}(1997)}]{worthey1997}
{Worthey} G., {Ottaviani} D.~L. (1997) {H{\ensuremath{\gamma}} and
  H{\ensuremath{\delta}} Absorption Features in Stars and Stellar Populations}.
  \emph{\apjs} 111(2):377--386, \doi{10.1086/313021}

\bibitem[{{Wright} et~al.(2010){Wright}, {Eisenhardt}, {Mainzer}, {Ressler},
  {Cutri}, {Jarrett}, {Kirkpatrick}, {Padgett}, {McMillan}, {Skrutskie},
  {Stanford}, {Cohen}, {Walker}, {Mather}, {Leisawitz}, {Gautier}, {McLean},
  {Benford}, {Lonsdale}, {Blain}, {Mendez}, {Irace}, {Duval}, {Liu}, {Royer},
  {Heinrichsen}, {Howard}, {Shannon}, {Kendall}, {Walsh}, {Larsen}, {Cardon},
  {Schick}, {Schwalm}, {Abid}, {Fabinsky}, {Naes}, and {Tsai}}]{write2010}
{Wright} E.~L., {Eisenhardt} P. R.~M., {Mainzer} A.~K., {Ressler} M.~E.,
  {Cutri} R.~M., {Jarrett} T., {Kirkpatrick} J.~D., {Padgett} D., et~al. (2010)
  {The Wide-field Infrared Survey Explorer (WISE): Mission Description and
  Initial On-orbit Performance}. \emph{The Astrophysical Journal}
  140(6):1868--1881, \doi{10.1088/0004-6256/140/6/1868}, \eprint{1008.0031}

\bibitem[{{Wu} et~al.(2021){Wu}, {Li}, {Zuo}, {Chen}, {Das}, {Marriner},
  {Oxholm}, {Phan}, {Stebbins}, {Timbie}, {Ansari}, {Campagne}, {Chen}, {Cong},
  {Huang}, {Kwak}, {Li}, {Liu}, {Liu}, {Niu}, {Osinga}, {Perdereau},
  {Peterson}, {Podczerwinski}, {Shi}, {Siebert}, {Sun}, {Tian}, {Tucker},
  {Wang}, {Wang}, {Wang}, {Wu}, {Xu}, {Yu}, {Yu}, {Zhang}, {Zhang}, and
  {Zhu}}]{Wu:2020jwm}
{Wu} F., {Li} J., {Zuo} S., {Chen} X., {Das} S., {Marriner} J.~P., {Oxholm}
  T.~M., {Phan} A., et~al. (2021) {The Tianlai dish pathfinder array: design,
  operation, and performance of a prototype transit radio interferometer}.
  \emph{\mnras} 506(3):3455--3482, \doi{10.1093/mnras/stab1802},
  \eprint{2011.05946}

\bibitem[{{Wu} et~al.(2012){Wu}, {Vanden Berk}, {Grupe}, {Koch}, {Gelbord},
  {Schneider}, {Gronwall}, {Wesolowski}, and {Porterfield}}]{wu2012}
{Wu} J., {Vanden Berk} D., {Grupe} D., {Koch} S., {Gelbord} J., {Schneider}
  D.~P., {Gronwall} C., {Wesolowski} S., et~al. (2012) {A Quasar Catalog with
  Simultaneous UV, Optical, and X-Ray Observations by Swift}. \emph{\apjs}
  201(2):10, \doi{10.1088/0067-0049/201/2/10}, \eprint{1203.6071}

\bibitem[{{Wucknitz}(2002)}]{Wucknitz:2002}
{Wucknitz} O. (2002) {Degeneracies and scaling relations in general power-law
  models for gravitational lenses}. \emph{\mnras} 332(4):951--961,
  \doi{10.1046/j.1365-8711.2002.05426.x}, \eprint{astro-ph/0202376}

\bibitem[{{Wucknitz} et~al.(2021){Wucknitz}, {Spitler}, and
  {Pen}}]{wucknitz2021}
{Wucknitz} O., {Spitler} L.~G., {Pen} U.~L. (2021) {Cosmology with
  gravitationally lensed repeating fast radio bursts}. \emph{\aap} 645:A44,
  \doi{10.1051/0004-6361/202038248}, \eprint{2004.11643}

\bibitem[{Wuensche(2019)}]{Wuensche:2018alk}
Wuensche C.~A. (2019) {The BINGO telescope: a new instrument exploring the new
  21 cm cosmology window}. \emph{J Phys Conf Ser} 1269(1):012002,
  \doi{10.1088/1742-6596/1269/1/012002}, \eprint{1803.01644}

\bibitem[{Wyithe(2008)}]{Wyithe:2008th}
Wyithe S. (2008) {A Method to Measure the Mass of Damped Ly-alpha Absorber Host
  Galaxies Using Fluctuations in 21cm Emission}. \emph{Mon Not Roy Astron Soc}
  388:1889, \doi{10.1111/j.1365-2966.2008.13546.x}, \eprint{0804.1624}

\bibitem[{{Xu} et~al.(2016){Xu}, {Sluse}, {Schneider}, {Springel},
  {Vogelsberger}, {Nelson}, and {Hernquist}}]{Xu:2016}
{Xu} D., {Sluse} D., {Schneider} P., {Springel} V., {Vogelsberger} M., {Nelson}
  D., {Hernquist} L. (2016) {Lens galaxies in the Illustris simulation:
  power-law models and the bias of the Hubble constant from time delays}.
  \emph{\mnras} 456(1):739--755, \doi{10.1093/mnras/stv2708},
  \eprint{1507.07937}

\bibitem[{{Xu} et~al.(2021){Xu}, {Tang}, {Geng}, {Wang}, {Wang}, {Kuerban}, and
  {Huang}}]{Xu2021}
{Xu} F., {Tang} C.-H., {Geng} J.-J., {Wang} F.-Y., {Wang} Y.-Y., {Kuerban} A.,
  {Huang} Y.-F. (2021) {X-Ray Plateaus in Gamma-Ray Burst Afterglows and Their
  Application in Cosmology}. \emph{\apj} 920(2):135,
  \doi{10.3847/1538-4357/ac158a}, \eprint{2012.05627}

\bibitem[{{Yang} et~al.(2015){Yang}, {Neyrinck}, {Arag{\'o}n-Calvo}, {Falck},
  and {Silk}}]{Yang2015}
{Yang} L.~F., {Neyrinck} M.~C., {Arag{\'o}n-Calvo} M.~A., {Falck} B., {Silk} J.
  (2015) {Warmth elevating the depths: shallower voids with warm dark matter}.
  \emph{\mnras} 451:3606--3614, \doi{10.1093/mnras/stv1087}, \eprint{1411.5029}

\bibitem[{{Yang} et~al.(2019){Yang}, {Pan}, {Di Valentino}, {Saridakis}, and
  {Chakraborty}}]{yang2019}
{Yang} W., {Pan} S., {Di Valentino} E., {Saridakis} E.~N., {Chakraborty} S.
  (2019) {Observational constraints on one-parameter dynamical dark-energy
  parametrizations and the H$_{0}$ tension}. \emph{\prd} 99(4):043543,
  \doi{10.1103/PhysRevD.99.043543}, \eprint{1810.05141}

\bibitem[{{Ye} and {Fishbach}(2021)}]{2021arXiv210314038Y}
{Ye} C., {Fishbach} M. (2021) {Cosmology with Standard Sirens at Cosmic Noon}.
  \emph{arXiv e-prints} arXiv:2103.14038, \eprint{2103.14038}

\bibitem[{{Y{\i}ld{\i}r{\i}m} et~al.(2020){Y{\i}ld{\i}r{\i}m}, {Suyu}, and
  {Halkola}}]{Yildirim:2020}
{Y{\i}ld{\i}r{\i}m} A., {Suyu} S.~H., {Halkola} A. (2020) {Time-delay
  cosmographic forecasts with strong lensing and JWST stellar kinematics}.
  \emph{\mnras} 493(4):4783--4807, \doi{10.1093/mnras/staa498},
  \eprint{1904.07237}

\bibitem[{{Y{\i}ld{\i}r{\i}m} et~al.(2021){Y{\i}ld{\i}r{\i}m}, {Suyu}, {Chen},
  and {Komatsu}}]{Yildirim:2021}
{Y{\i}ld{\i}r{\i}m} A., {Suyu} S.~H., {Chen} G.~C.~F., {Komatsu} E. (2021)
  {TDCOSMO VIII: Cosmological distance measurements in light of the mass-sheet
  degeneracy -- forecasts from strong lensing and IFU stellar kinematics}.
  \emph{arXiv e-prints} arXiv:2109.14615, \eprint{2109.14615}

\bibitem[{{Yonetoku} et~al.(2004){Yonetoku}, {Murakami}, {Nakamura},
  {Yamazaki}, {Inoue}, and {Ioka}}]{Yonetoku04}
{Yonetoku} D., {Murakami} T., {Nakamura} T., {Yamazaki} R., {Inoue} A.~K.,
  {Ioka} K. (2004) {Gamma-Ray Burst Formation Rate Inferred from the Spectral
  Peak Energy-Peak Luminosity Relation}. \emph{\apj} 609(2):935--951,
  \doi{10.1086/421285}, \eprint{astro-ph/0309217}

\bibitem[{{You} et~al.(2021){You}, {Zhu}, {Ashton}, {Thrane}, and
  {Zhu}}]{2021ApJ...908..215Y}
{You} Z.-Q., {Zhu} X.-J., {Ashton} G., {Thrane} E., {Zhu} Z.-H. (2021)
  {Standard-siren Cosmology Using Gravitational Waves from Binary Black Holes}.
  \emph{\apj} 908(2):215, \doi{10.3847/1538-4357/abd4d4}, \eprint{2004.00036}

\bibitem[{{Young} et~al.(2010){Young}, {Elvis}, and {Risaliti}}]{young2010}
{Young} M., {Elvis} M., {Risaliti} G. (2010) {The X-ray Energy Dependence of
  the Relation Between Optical and X-ray Emission in Quasars}. \emph{\apj}
  708:1388--1397, \doi{10.1088/0004-637X/708/2/1388}, \eprint{0911.0474}

\bibitem[{{Yu} et~al.(2014){Yu}, {Zhang}, and {Pen}}]{yu2014}
{Yu} H.-R., {Zhang} T.-J., {Pen} U.-L. (2014) {Method for Direct Measurement of
  Cosmic Acceleration by 21-cm Absorption Systems}. \emph{Phys\ Rev\ Lett}
  113(4):041303, \doi{10.1103/PhysRevLett.113.041303}, \eprint{1311.2363}

\bibitem[{{Zamorani} et~al.(1981){Zamorani}, {Henry}, {Maccacaro}, {Tananbaum},
  {Soltan}, {Avni}, {Liebert}, {Stocke}, {Strittmatter}, {Weymann}, {Smith},
  and {Condon}}]{zamorani81}
{Zamorani} G., {Henry} J.~P., {Maccacaro} T., {Tananbaum} H., {Soltan} A.,
  {Avni} Y., {Liebert} J., {Stocke} J., et~al. (1981) {X-ray studies of quasars
  with the Einstein Observatory. II}. \emph{\apj} 245:357--374,
  \doi{10.1086/158815}

\bibitem[{{Zaninoni} et~al.(2014){Zaninoni}, {Bernardini}, {Margutti}, and
  {Amati}}]{Zaninoni2014}
{Zaninoni} E., {Bernardini} M.~G., {Margutti} R., {Amati} L. (2014) {Ten years
  of Swift: a universal scaling for short and long gamma-ray bursts
  (EX,iso-Egamma,iso-Epk)}. In: Proceedings of Swift: 10 Years of Discovery
  (SWIFT 10, p 120

\bibitem[{{Zeldovich} et~al.(1982){Zeldovich}, {Einasto}, and
  {Shandarin}}]{Zeldovich1982}
{Zeldovich} I.~B., {Einasto} J., {Shandarin} S.~F. (1982) {Giant voids in the
  Universe}. \emph{\nat} 300(5891):407--413, \doi{10.1038/300407a0}

\bibitem[{{Zeldovich}(1970)}]{Zeldovich1970}
{Zeldovich} Y.~B. (1970) {Reprint of 1970A\&A.....5...84Z. Gravitational
  instability: an approximate theory for large density perturbations.}
  \emph{\aap} 500:13--18

\bibitem[{{Zhai} et~al.(2017){Zhai}, {Tinker}, {Hahn}, {Seo}, {Blanton},
  {Tojeiro}, {Camacho}, {Lima}, {Carnero Rosell}, {Sobreira}, {da Costa},
  {Bautista}, {Brownstein}, {Comparat}, {Dawson}, {Newman}, {Prakash},
  {Roman-Lopes}, and {Schneider}}]{Zhai2017}
{Zhai} Z., {Tinker} J.~L., {Hahn} C., {Seo} H.-J., {Blanton} M.~R., {Tojeiro}
  R., {Camacho} H.~O., {Lima} M., et~al. (2017) {The Clustering of Luminous Red
  Galaxies at z {\ensuremath{\sim}} 0.7 from EBOSS and BOSS Data}. \emph{\apj}
  848(2):76, \doi{10.3847/1538-4357/aa8eee}, \eprint{1607.05383}

\bibitem[{{Zhang}(2014)}]{Zhang14}
{Zhang} B. (2014) {Gamma-Ray Burst Prompt Emission}. \emph{International
  Journal of Modern Physics D} 23(2):1430002, \doi{10.1142/S021827181430002X},
  \eprint{1402.7022}

\bibitem[{Zhang and {M{\'e}sz{\'a}ros}(2002)}]{Zhang02}
Zhang B., {M{\'e}sz{\'a}ros} P. (2002) {An Analysis of Gamma-Ray Burst Spectral
  Break Models}. \emph{\apj} 581(2):1236--1247, \doi{10.1086/344338},
  \eprint{astro-ph/0206158}

\bibitem[{{Zhang} et~al.(2021){Zhang}, {Zhu}, {Wu}, {Yu}, {Jiang}, {Yue},
  {Huang}, and {Hao}}]{zhang2021}
{Zhang} B., {Zhu} M., {Wu} Z.-Z., {Yu} Q.-Z., {Jiang} P., {Yue} Y.-L., {Huang}
  M.-L., {Hao} Q.-L. (2021) {Extragalactic H I 21-cm absorption line
  observations with the Five-hundred-meter Aperture Spherical radio Telescope}.
  \emph{\mnras} 503(4):5385--5396, \doi{10.1093/mnras/stab754},
  \eprint{2103.06573}

\bibitem[{{Zhang} et~al.(2014){Zhang}, {Zhang}, {Yuan}, {Liu}, {Zhang}, and
  {Sun}}]{zhang2014}
{Zhang} C., {Zhang} H., {Yuan} S., {Liu} S., {Zhang} T.-J., {Sun} Y.-C. (2014)
  {Four new observational H(z) data from luminous red galaxies in the Sloan
  Digital Sky Survey data release seven}. \emph{Research in Astronomy and
  Astrophysics} 14:1221-1233, \doi{10.1088/1674-4527/14/10/002},
  \eprint{1207.4541}

\bibitem[{{Zhang} et~al.(2020){Zhang}, {Li}, {Liu}, {Spergel}, {Kreisch},
  {Pisani}, and {Wandelt}}]{Zhang2020}
{Zhang} G., {Li} Z., {Liu} J., {Spergel} D.~N., {Kreisch} C.~D., {Pisani} A.,
  {Wandelt} B.~D. (2020) {Void halo mass function: A promising probe of
  neutrino mass}. \emph{\prd} 102(8):083537, \doi{10.1103/PhysRevD.102.083537},
  \eprint{1910.07553}

\bibitem[{{Zhao} et~al.(2016){Zhao}, {Tao}, {Liang}, {Kitaura}, and
  {Chuang}}]{Zhao2016}
{Zhao} C., {Tao} C., {Liang} Y., {Kitaura} F.-S., {Chuang} C.-H. (2016) {DIVE
  in the cosmic web: voids with Delaunay triangulation from discrete matter
  tracer distributions}. \emph{\mnras} 459(3):2670--2680,
  \doi{10.1093/mnras/stw660}, \eprint{1511.04299}

\bibitem[{{Zhao} and {Xia}(2021)}]{2021EPJC...81..694Z}
{Zhao} D., {Xia} J.-Q. (2021) {Constraining the anisotropy of the Universe with
  the X-ray and UV fluxes of quasars}. \emph{European Physical Journal C}
  81(8):694, \doi{10.1140/epjc/s10052-021-09491-0}, \eprint{2105.03965}

\bibitem[{{Zhao} et~al.(2017){Zhao}, {Raveri}, {Pogosian}, {Wang},
  {Crittenden}, {Handley}, {Percival}, {Beutler}, {Brinkmann}, {Chuang},
  {Cuesta}, {Eisenstein}, {Kitaura}, {Koyama}, {L'Huillier}, {Nichol}, {Pieri},
  {Rodriguez-Torres}, {Ross}, {Rossi}, {S{\'a}nchez}, {Shafieloo}, {Tinker},
  {Tojeiro}, {Vazquez}, and {Zhang}}]{zhao2017}
{Zhao} G.-B., {Raveri} M., {Pogosian} L., {Wang} Y., {Crittenden} R.~G.,
  {Handley} W.~J., {Percival} W.~J., {Beutler} F., et~al. (2017) {Dynamical
  dark energy in light of the latest observations}. \emph{Nature Astronomy}
  1:627--632, \doi{10.1038/s41550-017-0216-z}, \eprint{1701.08165}

\bibitem[{{Zhao} et~al.(2020{\natexlab{a}}){Zhao}, {Zhang}, {Zhang}, {Liang},
  {Luan}, {Zhou}, {Yi}, {Wang}, and {Zhang}}]{Zhao20}
{Zhao} W., {Zhang} J.-C., {Zhang} Q.-X., {Liang} J.-T., {Luan} X.-H., {Zhou}
  Q.-Q., {Yi} S.-X., {Wang} F.-F., et~al. (2020{\natexlab{a}}) {Statistical
  Study of Gamma-Ray Bursts with Jet Break Features in Multiwavelength
  Afterglow Emissions}. \emph{\apj} 900(2):112, \doi{10.3847/1538-4357/aba43a},
  \eprint{2009.03550}

\bibitem[{{Zhao} et~al.(2020{\natexlab{b}}){Zhao}, {Wang}, {Zhang}, and
  {Zhang}}]{2020SciBu..65.1340Z}
{Zhao} Z.-W., {Wang} L.-F., {Zhang} J.-F., {Zhang} X. (2020{\natexlab{b}})
  {Prospects for improving cosmological parameter estimation with
  gravitational-wave standard sirens from Taiji}. \emph{Science Bulletin}
  65(16):1340--1348, \doi{10.1016/j.scib.2020.04.032}, \eprint{1912.11629}

\bibitem[{Zheng et~al.(2017)Zheng, Tegmark, Dillon, Liu, Neben, Jonas, Reich,
  Reich, Kim, and Neben}]{Zheng:2016lul}
Zheng H., Tegmark M., Dillon J.~S., Liu A., Neben A., Jonas J., Reich P., Reich
  W., et~al. (2017) {An improved model of diffuse galactic radio emission from
  10 MHz to 5 THz}. \emph{Mon Not Roy Astron Soc} 464(3):3486--3497,
  \doi{10.1093/mnras/stw2525}, \eprint{1605.04920}

\bibitem[{{Zheng} et~al.(2017){Zheng}, {Xue}, {Brandt}, {Li}, {Paolillo},
  {Yang}, {Zhu}, {Luo}, {Sun}, {Hughes}, {Bauer}, {Vito}, {Wang}, {Liu},
  {Vignali}, and {Shu}}]{zheng2017}
{Zheng} X.~C., {Xue} Y.~Q., {Brandt} W.~N., {Li} J.~Y., {Paolillo} M., {Yang}
  G., {Zhu} S.~F., {Luo} B., et~al. (2017) {Deepest View of AGN X-Ray
  Variability with the 7 Ms Chandra Deep Field-South Survey}. \emph{\apj}
  849(2):127, \doi{10.3847/1538-4357/aa9378}, \eprint{1710.04358}

\bibitem[{{Zitrin} et~al.(2014){Zitrin}, {Redlich}, and
  {Broadhurst}}]{Zitrin:2014}
{Zitrin} A., {Redlich} M., {Broadhurst} T. (2014) {Consistent Use of Type Ia
  Supernovae Highly Magnified by Galaxy Clusters to Constrain the Cosmological
  Parameters}. \emph{\apj} 789(1):51, \doi{10.1088/0004-637X/789/1/51},
  \eprint{1311.5224}

\bibitem[{{Zivick} et~al.(2015){Zivick}, {Sutter}, {Wandelt}, {Li}, and
  {Lam}}]{Zivick2015}
{Zivick} P., {Sutter} P.~M., {Wandelt} B.~D., {Li} B., {Lam} T.~Y. (2015)
  {Using cosmic voids to distinguish f(R) gravity in future galaxy surveys}.
  \emph{\mnras} 451:4215--4222, \doi{10.1093/mnras/stv1209}, \eprint{1411.5694}

\bibitem[{{Zucca} et~al.(2009){Zucca}, {Bardelli}, {Bolzonella}, {Zamorani},
  {Ilbert}, {Pozzetti}, {Mignoli}, {Kova{\v{c}}}, {Lilly}, {Tresse}, {Tasca},
  {Cassata}, and et~al.}]{zucca2009}
{Zucca} E., {Bardelli} S., {Bolzonella} M., {Zamorani} G., {Ilbert} O.,
  {Pozzetti} L., {Mignoli} M., {Kova{\v{c}}} K., et~al. (2009) {The zCOSMOS
  survey: the role of the environment in the evolution of the luminosity
  function of different galaxy types}. \emph{\aap} 508(3):1217--1234,
  \doi{10.1051/0004-6361/200912665}, \eprint{0909.4674}

\bibitem[{Zumalacarregui and Seljak(2018)}]{Zumalacarregui:2017qqd}
Zumalacarregui M., Seljak U. (2018) {Limits on stellar-mass compact objects as
  dark matter from gravitational lensing of type Ia supernovae}. \emph{Phys Rev
  Lett} 121(14):141101, \doi{10.1103/PhysRevLett.121.141101},
  \eprint{1712.02240}

\end{thebibliography}

\section*{Affiliations}

$^1$ Dipartimento di Fisica e Astronomia ``Augusto Righi'', Alma Mater Studiorum Universit\`{a} di Bologna, via Piero Gobetti 93/2, I-40129 Bologna, Italy\\
$^2$ INAF - Osservatorio di Astrofisica e Scienza dello Spazio di Bologna, via Piero Gobetti 93/3, I-40129 Bologna, Italy\\
$^3$ Institute of Theoretical Physics, Heidelberg University, Philosophenweg 16, 69120 Heidelberg, Germany\\
$^4$ Kavli Institute for Particle Astrophysics and Cosmology and Department of Physics, Stanford University, Stanford, CA 94305, USA\\
$^5$ SLAC National Accelerator Laboratory, Menlo Park, CA, 94025, USA\\
$^6$ NSF’s National Optical-Infrared Astronomy Research Laboratory (NOIRLab), Tucson, AZ 85719, USA\\
$^7$ INAF Osservatorio Astr. d'Abruzzo, via Maggini, I-64100, Teramo, Italy\\
$^8$ INAF - Osservatorio Astrofisico di Arcetri, Largo E. Fermi 5, I-50125, Firenze, Italy\\
$^{9}$ Center for Astrophysics and Space Astronomy, Department of Astrophysical and Planetary Sciences, University of Colorado, 389 UCB, Boulder, CO 80309-0389, USA\\
$^{10}$ INAF-Capodimonte, Naples, Italy \& ESO, Garching, Germany\\
$^{11}$ Center for Interdisciplinary Exploration and Research in Astrophysics (CIERA) and Department of Physics and Astronomy, Northwestern University, 1800 Sherman Ave, Evanston, IL 60201, USA\\
$^{12}$ Dipartimento di Fisica, Universit\`a degli Studi di Milano, via Celoria 16, I-20133 Milano, Italy\\
$^{13}$ INAF - IASF Milano, via A. Corti 12, I-20133 Milano, Italy\\
$^{14}$ Universitäts-Sternwarte München, Fakultät für Physik, Ludwig-Maximilians-Universität München, Scheinerstrasse~1, 81679 München, Germany\\
$^{15}$ Department of Physics, University of Chicago, Chicago, IL 60637, USA\\
$^{16}$ Department of Astronomy and Astrophysics, Enrico Fermi Institute, and Kavli Institute for Cosmological Physics, University of Chicago, Chicago, IL 60637, USA\\
$^{17}$ DARK, Niels Bohr Institute, University of Copenhagen, Lyngbyvej 2, DK-2100 Copenhagen, Denmark\\
$^{18}$ ICCUB, University of Barcelona, Barcelona 08028, Spain\\
$^{19}$ Institucio Catalana de Recerca i Estudis Avancats, Passeig Lluis Companys 23, Barcelona 08010, Spain\\
$^{20}$ Dipartimento di Fisica e Astronomia, Università di Firenze, via G. Sansone 1, 50019 Sesto Fiorentino, Firenze, Italy\\
$^{21}$ Istituto Nazionale Di Fisica Nucleare, viale Berti Pichat 6/2, I-40127 Bologna, Italy\\
$^{22}$ Dipartimento di Fisica, Università degli Studi di Napoli Federico II, Compl. Univ. Monte S. Angelo, 80126 Naples, Italy\\
$^{23}$ Istituto Nazionale Di Fisica Nucleare, Sez. di Napoli, Compl. Univ. Monte S. Angelo, Edificio 6, via Cinthia, 80126 - Napoli, Italy\\
$^{24}$ Department of Astrophysical Sciences, Princeton University, Peyton Hall, 4 Ivy Lane, Princeton, NJ 08544, USA\\
$^{25}$ Center for Computational Astrophysics, Flatiron Institute, 162 5th Avenue, New York, NY, 10010, USA\\
$^{26}$ The Cooper Union for the Advancement of Science and Art, 41 Cooper Square, New York, NY 10003, USA\\
$^{27}$ Higgs Centre for Theoretical Physics, School of Physics and Astronomy, The University of Edinburgh, Edinburgh EH9 3FD, UK \\
$^{28}$ Institute for Astronomy, The University of Edinburgh, Royal Observatory, Edinburgh EH9 3HJ, UK\\
$^{29}$ Department of Physics \& Astronomy, University of the Western Cape, Cape Town 7535, South Africa\\
$^{30}$ Instituto de Física, Universidade Federal do Rio de Janeiro, 21941-972, Rio de Janeiro, RJ, Brazil\\
$^{31}$ Observatório do Valongo, Universidade Federal do Rio de Janeiro, 20080-090, Rio de Janeiro, RJ, Brazil\\
$^{32}$ Dipartimento di Fisica e Scienze della Terra, Università degli Studi di Ferrara, Via Saragat 1, I-44122 Ferrara, Italy

\end{document}